\documentclass[%
  manyauthors,
  nocleardouble,
  COMPASS,
  american]{cernphprep}

\usepackage{multicol}  %
\usepackage{float}     %
\usepackage{packages}

\makeatletter
\let\@fnsymbol\@alph
\makeatother

\makeatletter
\g@addto@macro\bfseries{\boldmath}
\makeatother

\graphicspath{%
  {./},%
  {./tikz/},%
  {./plots/}
}

\usepackage{macros}

\FPeval{\plotHeightRatioOneDTwoD}{550/511}           %
\FPeval{\widthOneD}{1/(1+\plotHeightRatioOneDTwoD)}  %
\newlength{\totalPlotWidth}
\newlength{\twoPlotWidth}
\newlength{\twoPlotWidthTwoD}
\newlength{\twoPlotSpacing}
\newlength{\twoPlotWidthOneD}
\newlength{\twoPlotSpacingOneD}
\FPeval{\widthOneD}{1/(2+\plotHeightRatioOneDTwoD)}  %
\newlength{\threePlotWidth}
\newlength{\threePlotWidthTwoD}
\newlength{\threePlotSpacing}
\begin{document}

%
%

%
%
%
%
%

\begin{titlepage}
  \PHnumber{2015--233}
  \PHdate{
  \vspace*{-1.1\baselineskip}
  \begin{tabular}[c]{r@{}}
  August 31, 2015\\
  rev.\ Feb.\ 3, 2017
  \end{tabular}
  }
  %
  %
  %
%
%

\title{Resonance Production and $\pi\pi$ $S$-wave in \reaction at \SI{190}{\GeVc}}
   \makeatletter
  \ShortTitle{\@title}
  \makeatother
  %

  %
  %
  %
%
%

\begin{abstract}
  The COMPASS collaboration has collected the currently largest data
  set on diffractively produced \threePi final states using a negative
  pion beam of \SI{190}{\GeVc} momentum impinging on a stationary
  proton target.  This data set allows for a systematic partial-wave
  analysis in 100~bins of three-pion mass,
  \SIvalRange{0.5}{\mThreePi}{2.5}{\GeVcc}, and in 11~bins of the
  reduced four-momentum transfer squared,
  \SIvalRange{0.1}{\tpr}{1.0}{\GeVcsq}.  This two-dimensional analysis
  offers sensitivity to genuine one-step resonance production, \ie the
  production of a state followed by its decay, as well as to more
  complex dynamical effects in nonresonant $3\pi$ production.  In this
  paper, we present detailed studies on selected $3\pi$ partial waves
  with $\JPC = 0^{-+}$, $1^{++}$, $2^{-+}$, $2^{++}$, and $4^{++}$.
  In these waves, we observe the well-known ground-state mesons as
  well as a new narrow axial-vector meson \PaOne[1420] decaying into
  $\PfZero\,\pi$.  In addition, we present the results of a novel
  method to extract the amplitude of the \twoPi subsystem with
  $\IGJPC = 0^+\,0^{++}$ in various partial waves from the \threePi
  data.  Evidence is found for correlation of the \PfZero and
  \PfZero[1500] appearing as intermediate \twoPi isobars in the decay
  of the known \Ppi[1800] and \PpiTwo[1880].
\end{abstract}
   \vfill\vfill
  \begin{flushleft}
    Keywords:
    %
%
experimental results, magnetic spectrometer;
hadron spectroscopy, meson, light;
CERN Lab;
CERN SPS;
COMPASS;
beam, pi-, 190 GeV/c;
pi-, hadroproduction, meson resonance;
pi-, diffraction, dissociation;
pi-, multiple production, (pi+ 2pi-);
target, hydrogen;
pi- p, inelastic scattering, exclusive reaction;
pi- p --> p pi+ 2pi-;
partial-wave analysis; isobar model; hadronic decay, amplitude analysis;
mass spectrum, (pi+ 2pi-);
spin, density matrix;
momentum transfer dependence, slope;
data analysis method;
mass spectrum, (pi+ pi-);
(2pi) s-wave;
scalar meson, isoscalar;
pseudoscalar meson, isovector;
vector meson, isovector;
axial-vector meson, isovector;
tensor meson;
f0(500); rho(770); f0(980); f2(1270); f0(1500); rho3(1690);
a1(1260); a2(1320); a1(1420); pi2(1670); pi(1800); pi2(1880); a4(2040)
   \end{flushleft}
  \Submitted{(submitted to Physical Review D)
  }
\end{titlepage}

{\pagestyle{empty}
  %
%
%
\section*{The COMPASS Collaboration}
\label{app:collab}
\renewcommand\labelenumi{\textsuperscript{\theenumi}~}
\renewcommand\theenumi{\arabic{enumi}}
\begin{flushleft}
C.~Adolph\Irefn{erlangen},
R.~Akhunzyanov\Irefn{dubna}, 
M.G.~Alexeev\Irefn{turin_u},
G.D.~Alexeev\Irefn{dubna}, 
A.~Amoroso\Irefnn{turin_u}{turin_i},
V.~Andrieux\Irefn{saclay},
V.~Anosov\Irefn{dubna}, 
W.~Augustyniak\Irefn{warsaw},
A.~Austregesilo\Irefn{munichtu},
C.D.R.~Azevedo\Irefn{aveiro}
% ...
\end{flushleft}
%
%
\begin{Authlist}
\item [{\makebox[2mm][l]{\textsuperscript{\#}}}] Corresponding author
\item \Idef{turin_p}{University of Eastern Piedmont, 15100 Alessandria, Italy}
\item \Idef{aveiro}{University of Aveiro, Department of Physics, 3810-193 Aveiro, Portugal} 
\item \Idef{bochum}{Universit\"at Bochum, Institut f\"ur Experimentalphysik, 44780 Bochum, Germany\Arefs{l}\Arefs{s}}
\item \Idef{bonniskp}{Universit\"at Bonn, Helmholtz-Institut f\"ur  Strahlen- und Kernphysik, 53115 Bonn, Germany\Arefs{l}}
\item \Idef{bonnpi}{Universit\"at Bonn, Physikalisches Institut, 53115 Bonn, Germany\Arefs{l}}
\item \Idef{brno}{Institute of Scientific Instruments, AS CR, 61264 Brno, Czech Republic\Arefs{m}}
\item \Idef{calcutta}{Matrivani Institute of Experimental Research \& Education, Calcutta-700 030, India\Arefs{n}}
\item \Idef{dubna}{Joint Institute for Nuclear Research, 141980 Dubna, Moscow region, Russia\Arefs{o}}
\item \Idef{erlangen}{Universit\"at Erlangen--N\"urnberg, Physikalisches Institut, 91054 Erlangen, Germany\Arefs{l}}
\item \Idef{freiburg}{Universit\"at Freiburg, Physikalisches Institut, 79104 Freiburg, Germany\Arefs{l}\Arefs{s}}
\item \Idef{cern}{CERN, 1211 Geneva 23, Switzerland}
\item \Idef{liberec}{Technical University in Liberec, 46117 Liberec, Czech Republic\Arefs{m}}
\item \Idef{lisbon}{LIP, 1000-149 Lisbon, Portugal\Arefs{p}}
\item \Idef{mainz}{Universit\"at Mainz, Institut f\"ur Kernphysik, 55099 Mainz, Germany\Arefs{l}}
\item \Idef{miyazaki}{University of Miyazaki, Miyazaki 889-2192, Japan\Arefs{q}}
\item \Idef{moscowlpi}{Lebedev Physical Institute, 119991 Moscow, Russia}
\item \Idef{munichtu}{Technische Universit\"at M\"unchen, Physik Department, 85748 Garching, Germany\Arefs{l}\Arefs{r}}
\item \Idef{nagoya}{Nagoya University, 464 Nagoya, Japan\Arefs{q}}
\item \Idef{praguecu}{Charles University in Prague, Faculty of Mathematics and Physics, 18000 Prague, Czech Republic\Arefs{m}}
\item \Idef{praguectu}{Czech Technical University in Prague, 16636 Prague, Czech Republic\Arefs{m}}
\item \Idef{protvino}{State Scientific Center Institute for High Energy Physics of National Research Center `Kurchatov Institute', 142281 Protvino, Russia}
\item \Idef{saclay}{CEA IRFU/SPhN Saclay, 91191 Gif-sur-Yvette, France\Arefs{s}}
\item \Idef{taipei}{Academia Sinica, Institute of Physics, Taipei, 11529 Taiwan}
\item \Idef{telaviv}{Tel Aviv University, School of Physics and Astronomy, 69978 Tel Aviv, Israel\Arefs{t}}
\item \Idef{triest_u}{University of Trieste, Department of Physics, 34127 Trieste, Italy}
\item \Idef{triest_i}{Trieste Section of INFN, 34127 Trieste, Italy}
\item \Idef{triest_ictp}{Abdus Salam ICTP, 34151 Trieste, Italy}
\item \Idef{turin_u}{University of Turin, Department of Physics, 10125 Turin, Italy}
\item \Idef{turin_i}{Torino Section of INFN, 10125 Turin, Italy}
\item \Idef{illinois}{University of Illinois at Urbana-Champaign, Department of Physics, Urbana, IL 61801-3080, U.S.A.}   
\item \Idef{warsaw}{National Centre for Nuclear Research, 00-681 Warsaw, Poland\Arefs{u} }
\item \Idef{warsawu}{University of Warsaw, Faculty of Physics, 02-093 Warsaw, Poland\Arefs{u} }
\item \Idef{warsawtu}{Warsaw University of Technology, Institute of Radioelectronics, 00-665 Warsaw, Poland\Arefs{u} }
\item \Idef{yamagata}{Yamagata University, Yamagata, 992-8510 Japan\Arefs{q} }
\end{Authlist}
%
%
\renewcommand\theenumi{\alph{enumi}}
\begin{Authlist}
\item [{\makebox[2mm][l]{\textsuperscript{*}}}] Deceased
\item \Adef{a}{Also at Instituto Superior T\'ecnico, Universidade de Lisboa, Lisbon, Portugal}
\item \Adef{b}{Also at Department of Physics, Pusan National University, Busan 609-735, Republic of Korea and at Physics Department, Brookhaven National Laboratory, Upton, NY 11973, U.S.A. }
\item \Adef{r}{Supported by the DFG cluster of excellence `Origin and Structure of the Universe' (www.universe-cluster.de)}
\item \Adef{d}{Also at Chubu University, Kasugai, Aichi, 487-8501 Japan\Arefs{q}}
\item \Adef{e}{Also at KEK, 1-1 Oho, Tsukuba, Ibaraki, 305-0801 Japan}
\item \Adef{f}{Present address: Universit\"at Bonn, Helmholtz-Institut f\"ur Strahlen- und Kernphysik, 53115 Bonn, Germany}
\item \Adef{g}{Also at Moscow Institute of Physics and Technology, Moscow Region, 141700, Russia}
\item \Adef{v}{Supported by Presidential grant NSh - 999.2014.2}
\item \Adef{h}{Present address: RWTH Aachen University, III. Physikalisches Institut, 52056 Aachen, Germany}
\item \Adef{i}{Present address: Uppsala University, Box 516, SE-75120 Uppsala, Sweden}
\item \Adef{c}{Supported by the DFG Research Training Group Programme 1102  ``Physics at Hadron Accelerators''}
%
%
\item \Adef{l}{Supported by the German Bundesministerium f\"ur Bildung und Forschung}
\item \Adef{s}{Supported by EU FP7 (HadronPhysics3, Grant Agreement number 283286)}
\item \Adef{m}{Supported by Czech Republic MEYS Grant LG13031}
\item \Adef{n}{Supported by SAIL (CSR), Govt.\ of India}
\item \Adef{o}{Supported by CERN-RFBR Grant 12-02-91500}
\item \Adef{p}{\raggedright Supported by the Portuguese FCT - Funda\c{c}\~{a}o para a Ci\^{e}ncia e Tecnologia, COMPETE and QREN,
 Grants CERN/FP 109323/2009, 116376/2010, 123600/2011 and CERN/FIS-NUC/0017/2015}
\item \Adef{q}{Supported by the MEXT and the JSPS under the Grants No.18002006, No.20540299 and No.18540281; Daiko Foundation and Yamada Foundation}
\item \Adef{t}{Supported by the Israel Academy of Sciences and Humanities}
\item \Adef{u}{Supported by the Polish NCN Grant DEC-2011/01/M/ST2/02350}
%
%
%
\end{Authlist}

   \clearpage
}

\setcounter{page}{1}
 \tableofcontents

\setlist{noitemsep}
\setlist{nolistsep}
\setlist[enumerate]{label=\textit{\roman*})}

\clearpage
%
%
%

\section{Introduction}
\label{sec:introduction}

In this paper, we report on the results of a partial-wave analysis of
the \threePi system produced by a \SI{190}{\GeVc} $\pi^-$ beam
impinging on a liquid-hydrogen target.  The reaction of interest is
diffractive dissociation of a $\pi^-$ into a \threePi system,
\begin{equation}
  \reaction,
  \label{eq:reaction}
\end{equation}
with $p_\text{recoil}$ denoting the recoiling target proton.  The data
for this analysis were recorded with the COMPASS experiment at the
CERN SPS in 2008.

Despite many decades of research in hadron spectroscopy, the
excitation spectrum of light mesons, which are made of $u$, $d$, and
$s$ quarks, is still only partially known.  In the framework of the
simple constituent-quark model using
$\text{SU(3)}_\text{flavor}\, \otimes\, \text{SU(2)}_\text{spin}\,
\otimes\, \text{SU(3)}_\text{color}$
symmetry, a number of frequently observed states are commonly
interpreted in terms of orbital and radial excitations of
quark-antiquark ground-state mesons, \ie they are assigned to the
multiplets resulting from the symmetry.  Some of these assignments are
still disputed, as \eg the isovector mesons \Prho[1450], \Prho[1700],
\Ppi[1300], and \Ppi[1800]~\cite{klempt:2007cp}, as well as the whole
sector of scalar mesons~\cite{PDG_scalars:2012}.  In addition, a
number of extra states have been found, which cannot be accommodated by
the constituent-quark model.  These extra states appear in mass ranges
where quark-model states have already been identified, \eg the
\PpiTwo[1880] which is close to the \PpiTwo ground state.  Other
observed states seem to have peculiar decay modes or decay widths that
do not fit well into the general pattern.  Searching for new states
beyond the constituent-quark model, attempts have been made to
establish the existence of gluonic degrees of freedom.  The
fingerprints are expected to be so-called \emph{exotic} spin quantum
numbers\footnote{\JPC quantum numbers that are forbidden for \qqbar in
  the nonrelativistic limit.} or decay branching ratios, which could
identify them as \emph{hybrids}~\cite{Meyer:2015eta,Meyer:2010ku},
\emph{glueballs}~\cite{Ochs:2013gi,Crede:2008vw}, or \emph{tetra-quark
  systems}~\cite{klempt:2007cp}.  Potential candidates are \eg \PpiOne[1600], \Ppi[1800],
\PpiTwo[1880] or \PfZero[1500], \PfZero[1710] or \PfZero[980],
\PaZero, \PfOne[1420], respectively.

The COMPASS collaboration has already studied properties of isovector
$3\pi$ resonances~\cite{adolph:2014mup,Alekseev:2009aa} in the mass
range between \SIlist{1.1;2.1}{\GeVcc} using a lead target.  In this
paper, isovector mesons decaying into three charged pions are studied
using a hydrogen target with the emphasis on \one production
kinematics, \two separation of nonresonant processes, \three search
for new and excited mesons, and \four on properties of the \pipiSW
amplitude.  This paper is the first in a planned series of
publications to present precision studies revisiting all quantum
numbers accessible in reaction~\eqref{eq:reaction} up to total spin
$J = 6$.  The analysis is limited to states belonging to the family of
$\pi_J$ and $a_J$. In addition, the large data set allows to apply a
novel method for investigating isoscalar states, which occur as \twoPi
subsystems in the decays of isovector mesons.

The Particle Data Group (PDG)~\cite{Agashe:2014kda} lists a total of
eleven well-established isovector states with masses below
\SI{2.1}{\GeVcc} (see \cref{tab:PDG_mesons_2014}), where only the
\PaZero* states do not decay into $3\pi$ due to parity conservation.
The widths of the \PaZero*, \PaTwo*, and \PaFour* ground states have
values of about \SI{10}{\percent} of their mass values, while the
\PaOne is much broader.  Pionic excitations are typically broader with
values of their width being about \SIrange{15}{20}{\percent} of their
mass values.  In addition, the table contains a number of less
well-established states.  Even for some established states, properties
such as mass and width are poorly determined, \eg for the \PaOne as
the lightest \PaOne* state, the reported widths vary between
\SIlist{250;600}{\MeVcc}.  Another example is the inconsistency in the
mass measurements of \Ppi[1800], where experimental results cluster
around two different mean values.  This has lead to speculations on
the existence of two states, one being an ordinary meson and the other
one a hybrid.  Extensive discussions of the light-meson sector are
found in \refsCite{klempt:2007cp,Brambilla:2014jmp}.

\begin{table}[htbp]
  \centering
  \renewcommand{\arraystretch}{1.2}
  \caption{Resonance parameters of $a_J$ and $\pi_J$ mesons in the mass
    region below \SI{2.1}{\GeVcc} as given in PDG~\cite{Agashe:2014kda}.
    Note that due to parity conservation the $a_0$ states cannot decay into
    \threePi.}
  \label{tab:PDG_mesons_2014}
  \begin{small}
    \begin{tabular}[t]{lcll}
      \toprule
      \textbf{Particle} &
      \textbf{\JPC} &
      \textbf{Mass [\si{\MeVcc}]} &
      \textbf{Width [\si{\MeVcc}]} \\
      \midrule

      \addlinespace[1mm]
      \multicolumn{4}{c}{\textbf{Established states}} \\
      \addlinespace[1mm]

      \PaZero       & $0^{++}$ & $980 \pm 20$        & $50~\text{to}~100$ \\
      \PaOne        & $1^{++}$ & $1230 \pm 40$       & $250~\text{to}~600$ \\
      \PaTwo        & $2^{++}$ & $1318.3^{+0.5}_{-0.6}$ & $107 \pm 5$ \\
      \PaZero[1450] & $0^{++}$ & $1474 \pm 19$       & $265 \pm 13$ \\
      \PaFour       & $4^{++}$ & $1996^{+10}_{-9}$     & $255^{+28}_{-24}$ \\

      \midrule

      \Ppi[1300]    & $0^{-+}$ & $1300 \pm 100$   & $200~\text{to}~600$ \\
      \PpiOne[1400] & $1^{-+}$ & $1354 \pm 25$    & $330 \pm 35 $ \\
      \PpiOne[1600] & $1^{-+}$ & $1662^{+8}_{-9}$   & $ 241 \pm 40$ \\
      \PpiTwo       & $2^{-+}$ & $1672.2 \pm 3.0$ & $260 \pm 9$ \\
      \Ppi[1800]    & $0^{-+}$ & $1812 \pm 12$    & $208 \pm 12$ \\
      \PpiTwo[1880] & $2^{-+}$ & $1895 \pm 16$    & $235 \pm 34$ \\

      \midrule

      \addlinespace[1mm]
      \multicolumn{4}{c}{\textbf{States omitted from summary table}} \\
      \addlinespace[1mm]

      \PaOne[1640] & $1^{++}$ & $1647 \pm 22$ & $254 \pm 27$ \\
      \PaTwo[1700] & $2^{++}$ & $1732 \pm 16$ & $194 \pm 40$ \\

      \midrule

      \PpiTwo[2100] & $2^{-+}$ & $2090 \pm 29$ & $625 \pm 50$ \\

      \midrule

      \addlinespace[1mm]
      \multicolumn{4}{c}{\textbf{Further states}} \\
      \addlinespace[1mm]

      \PaThree[1875] & $3^{++}$ & $1874 \pm 43 \pm 96$   & $385 \pm 121 \pm 114$ \\
      \PaOne[1930]   & $1^{++}$ & $1930^{+30}_{-70}$       & $155 \pm 45$ \\
      \PaTwo[1950]   & $2^{++}$ & $1950^{+30}_{-70}$       & $180^{+30}_{-70}$ \\
      \PaTwo[1990]   & $2^{++}$ & $2050 \pm 10 \pm 40$   & $190 \pm 22 \pm 100$ \\[-1ex]
                     &         & $2003 \pm 10 \pm 19$   & $249 \pm 23 \pm 32$ \\
      \PaZero[2020]  & $0^{++}$ & $2025 \pm 30$          & $330 \pm 75$ \\
      \PaTwo[2030]   & $2^{++}$ & $2030 \pm 20$          & $205 \pm 30$ \\
      \PaThree[2030] & $3^{++}$ & $2031 \pm 12$          & $150 \pm 18$ \\
      \PaOne[2095]   & $1^{++}$ & $2096 \pm 17 \pm 121 $ & $451 \pm 41 \pm 81$ \\

      \midrule

      \PpiTwo[2005] & $2^{-+}$ & $1974 \pm 14 \pm 83$ & $341 \pm 61 \pm 139$ \\[-1ex]
                    &         & $2005 \pm 15$        & $200 \pm 40$ \\
      \PpiOne[2015] & $1^{-+}$ & $2014 \pm 20 \pm 16$ & $230 \pm 32 \pm 73$ \\[-1ex]
                    &         & $2001 \pm 30 \pm 92$ & $333 \pm 52 \pm 49$ \\
      \Ppi[2070]    & $0^{-+}$ & $2070 \pm 35$        & $310^{+100}_{-50}$ \\

      \midrule

      \PX[1775] & $\text{?}^{-+}$       & $1763 \pm 20$ & $192 \pm 60$ \\[-1ex]
                &                      & $1787 \pm 18$ & $118 \pm 60$ \\
      \PX[2000] & $\text{?}^{\text{?}+}$ & $1964 \pm 35$ & $225 \pm 50$ \\[-1ex]
                &                      & $\sim 2100$   & $\sim 500$ \\[-1ex]
                &                      & $2214 \pm 15$ & $355 \pm 21$ \\[-1ex]
                &                      & $2080 \pm 40$ & $340 \pm 80$ \\

      \bottomrule
    \end{tabular}
  \end{small}
\end{table}

The partial-wave analysis of the $3\pi$ system has a long
history~\cite{klempt:2007cp}.  The technique of partial-wave analysis
(PWA) of $3\pi$ systems was established by the work of Ascoli
\etal~\cite{Ascoli:1970xi,Ascoli:1973nj} in 1968.  The CERN-Munich
collaboration
(ACCMOR)~\cite{daum:1979ix,Daum:1979sx,Daum:1979iv,Daum:1980ay}
further developed this method and measured significant contributions
from partial waves up to $J = 2$, without including spin-exotic waves.
The largest data set used so far, which is the basis of several
publications on the $3\pi$ final state, was obtained and analyzed by
the BNL~E852
collaboration~\cite{adams:1998ff,Chung:2002pu,Dzierba:2005jg}.  They
have studied reaction~\eqref{eq:reaction} at beam momenta of
\SI{18}{\GeVc} and observed significant waves with $\JPC = 0^{-+}$,
$1^{++}$, $2^{++}$, and $2^{-+}$ quantum numbers.  In addition, they
have detected a $1^{-+}$ spin-exotic wave in the $\Prho\,\pi$ decay
channel with significant fluctuation in intensity depending on the
number of partial waves used, \ie with a considerable model
dependence.  Also the VES experiment has large data sets, the analysis
of which was published mostly in conference proceedings, see
\eg~\refsCite{amelin:1995gu,Khokhlov:2000tk,Kachaev:2001jj,Khokhlov:2012nn}.

\begin{figure}[htbp]
  \centering
  \includegraphics[scale=1]{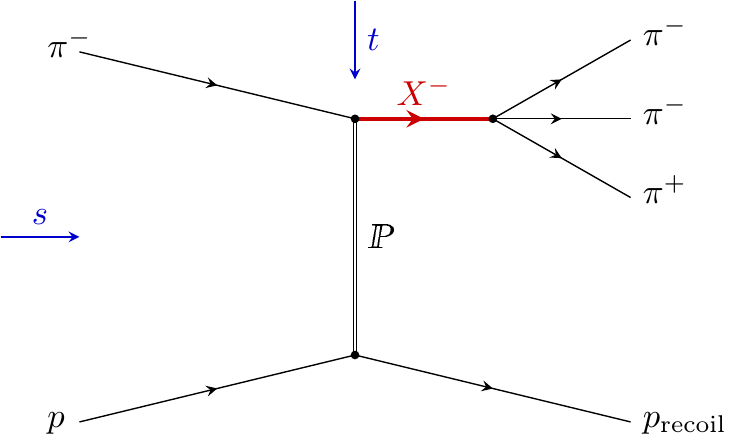}
  \caption{Diffractive dissociation of a beam pion on a target proton
    into the three-pion final state.  The figure shows the excitation
    of an intermediate resonance $X^-$ via Pomeron exchange and its
    subsequent decay into $3\pi$.}
  \label{fig:diffractive_dissociation_3pi}
\end{figure}

As illustrated in \cref{fig:diffractive_dissociation_3pi}, for
reaction~\eqref{eq:reaction} at \SI{190}{\GeV} beam energy, the strong
interaction can be described by the exchange of a quasi-particle
called Pomeron,~\Ppom, which is a flavorless glueball-like object that
accounts for diffractive dissociation and most of the two-body elastic
scattering ~\cite{Donnachie:2002en}.  The Regge trajectory
$\alpha_\Ppom(t)$ of the Pomeron determines the elastic scattering
amplitude
\begin{equation}
  A(s, t) \propto s^{\alpha_\Ppom(t)}.
  \label{eq:pomeron1}
\end{equation}
Here, $s$ is the squared center-of-mass energy, $t$ the squared
four-momentum transferred between beam particle and target nucleon,
and
\begin{equation}
 \alpha_\Ppom(t) = 1 + \epsilon_\Ppom + \alpha'_\Ppom\, t,
 \label{eq:pomeron2}
\end{equation}
where $0.081 \lesssim \epsilon_\Ppom \lesssim 0.112$ and
$\alpha'_\Ppom \approx \SI{0.25}{\perGeVcsq}$~\cite{Donnachie:2002en}.
The Pomeron is an even-signature Regge trajectory with
$\JPC = 2^{++}$, $4^{++}$, $6^{++}$, \ldots, and its first Regge pole
corresponds to a flavorless hadron with $\JPC = 2^{++}$ and a mass of
about \SI{1.9}{\GeVcc}.  The parameter $\alpha'_\Ppom$ modifies the
dependence of the differential cross section on the four-momentum
transfer. \Cref{eq:pomeron1} implies a dependence of the cross section
on $t$ as
\begin{equation}
 \od{\sigma}{t} \propto e^{{-b}\, t}.
 \label{eq:sigma}
\end{equation}
The slope parameter $b$ is given by
\begin{equation}
 b = b_0 + 4 \alpha'_{\Ppom}\, \ln \sqrt{\frac{s}{s_0}},
 \label{eq:slope}
\end{equation}
where $b_0$ is a generic slope parameter and the unknown scale
parameter $s_0$ is usually taken to be \SI{1}{GeV^2}.  The reduced
four-momentum transfer squared is
\begin{equation}
 \tpr \equiv \tabs - \tmin \geq 0,
 ~\text{where}~
 \tmin \approx \rBrk{\frac{\mThreePi^2 - m_\pi^2}{2 \Abs{\vec{p}_\text{beam}}}}^2
 \label{eq:tmin}
\end{equation}
is the minimum momentum transfer needed to excite the beam particle to
a mass \mThreePi, which is the invariant mass of the $3\pi$ final
state.  The beam momentum $\vec{p}_\text{beam}$ is measured in the
laboratory frame.  For the $3\pi$ mass range of
\SIrange{0.5}{2.5}{\GeVcc} considered in this analysis, typical values
of \tmin are well below \SI{e-3}{\GeVcsq}.  Different production
mechanisms, \ie different exchange particles, can lead to different
slopes~$b$.  The existence of concurrent exchange processes thus
results in a more complex form of the \tpr dependence with coherently
and/or incoherently overlapping exponentials.  The \tpr range for this
analysis is \SIrange{0.1}{1.0}{\GeVcsq}.

Studies of diffractive dissociation of pions, see
\eg~\refsCite{Alekseev:2009aa,Dzierba:2005jg,Kachaev:2001jj,Daum:1980ay},
reveal the existence of nonresonant background processes such as the
Deck effect~\cite{Deck:1964hm}.  These processes exhibit strongly
mass-dependent production amplitudes that occur in the same partial
waves as the resonances under study.  In particular, the analyses
presented in~\refsCite{Daum:1980ay,Dzierba:2005jg} showed the
importance of the kinematic variable \tpr in a partial-wave analysis
and illustrated the power of accounting for the difference in the \tpr
dependence of the reaction mechanisms and also of the different
resonances.  In this work, we take advantage of the large size of our
data sample and develop this approach further in order to better
disentangle resonant and nonresonant components.

In the case of Pomeron exchange, the partial waves induced by a pion
beam can be assessed as follows: the $\pi^-$ is an isovector
pseudoscalar with negative $G$ parity and the Pomeron is assumed to be
an isoscalar $C = +1$ object, so that the partial waves all have
$\IG = 1^-$.  Possible \JPC quantum numbers\footnote{Although the $C$
  parity is not defined for a charged system, it is customary to quote
  the \JPC quantum numbers of the corresponding neutral partner state
  in the isospin multiplet. The $C$ parity can be generalized to the
  $G$ parity $G \equiv C\, e^{i \pi I_y}$, a multiplicative quantum
  number, which is defined for the non-strange states of a meson
  multiplet.} of partial waves are listed in \cref{tab:allowedJpc} for
the lowest values of the relative orbital angular momentum $\ell$
between the beam particle and a $\JPC = 2^{++}$ Pomeron as an example.
As we will demonstrate in this paper, almost all partial waves listed
in \cref{tab:allowedJpc} are indeed observed in our data.  Higher-spin
waves with $J \geq 5$ contribute significantly only at masses above
\SI{2}{\GeVcc}.  The table includes spin-exotic partial waves such as
$\JPC = 1^{-+}$, $3^{-+}$, and $5^{-+}$.  The present paper focuses on
non-exotic spin quantum numbers with the emphasis on known states.
They are extracted from the data by partial-wave methods that contain
an \apriori unknown dependence on \tpr, which is extracted from the
data.

\begin{table}[tbp]
  \centering
  \renewcommand{\arraystretch}{1.2}
  \caption{List of allowed \JPC quantum numbers for $X$ assuming
    that it is produced in the interaction of a $\JPC = 0^{-+}$ beam
    pion and a $2^{++}$ Pomeron as an example, with relative orbital
    angular momentum $\ell$ between the two.}
  \label{tab:allowedJpc}
  \begin{tabular}{cl}
    \toprule
    \textbf{$\ell$} &
    \textbf{\JPC of $X$} \\
    \midrule
    0      & $2^{-+}$ \\
    1      & $1^{++}$, $2^{++}$, $3^{++}$ \\
    2      & $0^{-+}$, $1^{-+}$, $2^{-+}$, $3^{-+}$, $4^{-+}$ \\
    3      & $1^{++}$, $2^{++}$, $3^{++}$, $4^{++}$, $5^{++}$ \\
    4      & $2^{-+}$, $3^{-+}$, $4^{-+}$, $5^{-+}$, $6^{-+}$ \\
    \vdots & \\
    \bottomrule
  \end{tabular}
\end{table}

The work related to this topic is subdivided into two publications,
owing to the large amount of material and various, in parts novel
analysis techniques used.  This paper contains details on the
experiment in \cref{sec:experimental_setup} and a description of the
basic event selection criteria in
\cref{sec:event_selection_trigger,sec:event_selection_cuts}, where we
also present the general features of our data set and the overall
kinematic distributions for both \mThreePi and \tpr.
\Cref{sec:pwa_method} contains a detailed description of our analysis
method and the PWA model used.  For clarity, we include a rather
extensive mathematical description summarizing the work of many
authors, who laid the basis for our analysis (see \eg\
\refsCite{Jacob:1959at,chung:1971ri,Hansen:1973gb,Chung:1974fq,Herndon:1973yn,Richman:1984gh,Salgado:2013dja}).
In this scheme, the analysis follows a two-step procedure described
in~\refCite{Salgado:2013dja}.  In the first step, a PWA is performed
in bins of \mThreePi and \tpr.  The results of this so-called
\emph{mass-independent} fit are presented and discussed in
\cref{sec:results_pwa_massindep,sec:tprim_dependence}.  In these and
the following sections, the focus lies on $3\pi$ resonances with
masses below \SI{2.1}{\GeVcc}.  The discussion on \tpr dependences
includes the kinematic distributions and \JPC-resolved \tpr spectra.
In \cref{sec:results_free_pipi_s_wave}, we present a novel approach
that allows us to investigate the amplitude of \twoPi subsystems in
the decay process.  In particular, we address the topic of the scalar
sector containing \PfZero* mesons and its complicated relation to
\pipiSW scattering.  The relation of \PfZero[980] and \PfZero[1500]
mesons to \pipi scattering will be demonstrated.  In this paper, all
error bars shown in the figures represent statistical uncertainties
only.  Systematic effects are discussed in
\cref{sec:pwa_massindep_systematic_studies,sec:appendix_syst_studies_massindep}.
In \cref{sec:conclusions}, we conclude by summarizing the findings
based on qualitative arguments.  The appendices contain details about
more technical issues.

The analysis methods and results presented in this paper will serve as
a basis for further publications that will be dedicated to individual
partial waves.  In the second step of the analysis, physics parameters
will be extracted from the data presented in this paper by performing
a fit that models the resonance amplitudes and the amplitudes of
nonresonant processes. This involves simultaneous fitting to many
partial-wave amplitudes in all bins of \tpr.  Such a
\emph{mass-dependent} fit, which will allow us to extract the \tpr
dependences of various components, \ie resonant and nonresonant
contributions for individual partial waves as well as resonance
parameters for the mesonic states observed with different \JPC, will
be described in a forthcoming paper~\cite{COMPASS_3pi_mass_dep_fit}.
 %
%
%

\section{Experimental Setup and Event Selection}
\label{sec:setup_and_event_selection}

\subsection{COMPASS Setup}
\label{sec:experimental_setup}

\begin{figure*}
  \centering
  \includegraphics[scale=1]{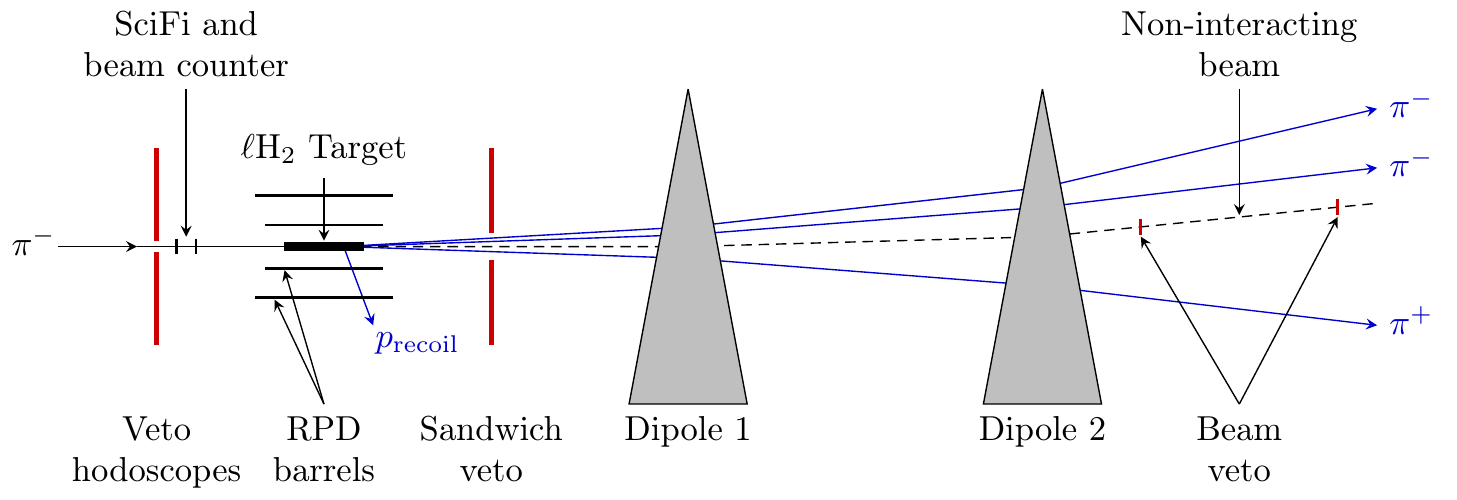}
  \caption{\colorPlot Simplified scheme of the diffractive trigger.
    The main components are the beam trigger, which selects beam
    particles, and the RPD, which triggers on slow charged particles
    leaving the target.  The veto system (red) rejects uninteresting
    events and consists of three parts: The veto hodoscopes, the
    sandwich, and the beam veto.}
  \label{fig:trigger_scheme_DT0}
\end{figure*}

The COMPASS spectrometer, which is described in general in
\refCite{Abbon:2007pq}, is situated at the CERN SPS.  The setup used
for the measurement presented here is explained in more detail in
\refCite{Abbon:2014aex}.  COMPASS uses secondary hadron and tertiary
muon beams that are produced by the \SI{400}{\GeVc} SPS proton beam
impinging on a \SI{50}{\centi\meter} long beryllium target.  The
measurement described in this paper is based on data recorded during
the 2008 COMPASS run.  The beam was tuned to deliver negatively
charged hadrons of \SI{190}{\GeVc} momentum passing through a pair of
beam Cherenkov detectors (CEDARs) for beam particle identification.
The beam impinged on a \SI{40}{\centi\meter} long liquid-hydrogen
target with an intensity of \num{5e7}~particles per SPS spill
(\SI{10}{\second} extraction with a repetition time of
\SI{45}{\second}).  At the target, the hadronic component of the beam
consisted of \SI{96.8}{\percent}~$\pi^-$, \SI{2.4}{\percent}~$K^-$,
and \SI{0.8}{\percent}~\Ppbar.  In addition, the beam contained about
\SI{1}{\percent} $\mu^-$ and an even smaller amount of electrons.

The target was surrounded by a Recoil-Proton Detector~(RPD) consisting
of two concentric, inner and outer, barrels of scintillators with
12~and 24~azimuthal segments, respectively.  Recoil protons emerging
from diffraction-like reactions must carry momenta of at least
\SI{270}{\MeVc} in order to traverse the target containment and to be
detected in the two RPD rings.  This leads to a minimum detectable
squared four-momentum transfer \tpr of about \SI{0.07}{\GeVcsq}.

Incoming beam particles and outgoing reaction products that emerge in
the forward region were detected by a set of silicon micro-strip
detector stations, each consisting of two double-sided detector
modules that were arranged to view four projections.  Particles
emerging in the forward direction were momentum-analyzed by the
two-stage magnetic spectrometer with a wide angular acceptance of
$\pm \SI{180}{\milli\radian}$.  Both spectrometer stages are each
composed of a bending magnet, charged-particle tracking,
electromagnetic and hadronic calorimetry, and muon identification.
Particles in the momentum range between \SIlist{2.5;50}{\GeVc} and
passing through the Ring-Imaging Cherenkov (RICH) detector in the
first stage can be identified as pion, kaon, or proton.  The
experiment offers large acceptance and high reconstruction efficiency
over a wide range of three-pion mass \mThreePi and squared
four-momentum transfer \tpr.

\subsection{Hardware Trigger}
\label{sec:event_selection_trigger}

A minimum-bias trigger, the so-called \emph{diffractive trigger}
(DT0)~\cite{Abbon:2014aex,Bernhard:2014}, was used to preselect events
with interacting beam particles and a recoiling proton emerging from
the target.  The trigger elements are shown schematically in
\cref{fig:trigger_scheme_DT0}.  The DT0 trigger is a coincidence of
three independent trigger signals: \one the \emph{beam trigger}, \two
the \emph{recoil-proton trigger}, and \three the veto signal.
Incoming beam particles are selected by the beam trigger requiring a
signal in one plane of the scintillating-fiber detector (SciFi) in
coincidence with a hit in the beam counter, which is a scintillator
disc of \SI{32}{\milli\meter} diameter and \SI{4}{\milli\meter}
thickness.  Both beam-trigger elements are located upstream of the
target.  The proton trigger selects events with protons recoiling from
the target. It features target pointing and discrimination of protons
from other charged particles by measuring the energy loss in each ring
of the RPD.  The veto signal has three sub-components.  The \emph{veto
  hodoscopes} reject incoming beam particles with trajectories far
from the nominal one.  Similarly, the \emph{sandwich} scintillation
detector that is positioned downstream close to the target, vetoes
events with particles leaving the target area outside of the
geometrical acceptance of the spectrometer.  Lastly, the \emph{beam
  veto}, two scintillator discs of \SI{35}{\milli\meter} diameter and
\SI{5}{\milli\meter} thickness positioned between the second analyzing
magnet and the second electromagnetic calorimeter, vetoes signals from
non-interacting beam particles.  Events recorded with the diffractive
trigger can be regarded as good candidates for diffractive
dissociation reactions.

\subsection{Event Selection}
\label{sec:event_selection_cuts}

\begin{figure}[tbp]
  \centering
  \subfloat[][]{%
    \label{fig:exclusivity_Esum}%
    \includegraphics[width=\twoPlotWidth]{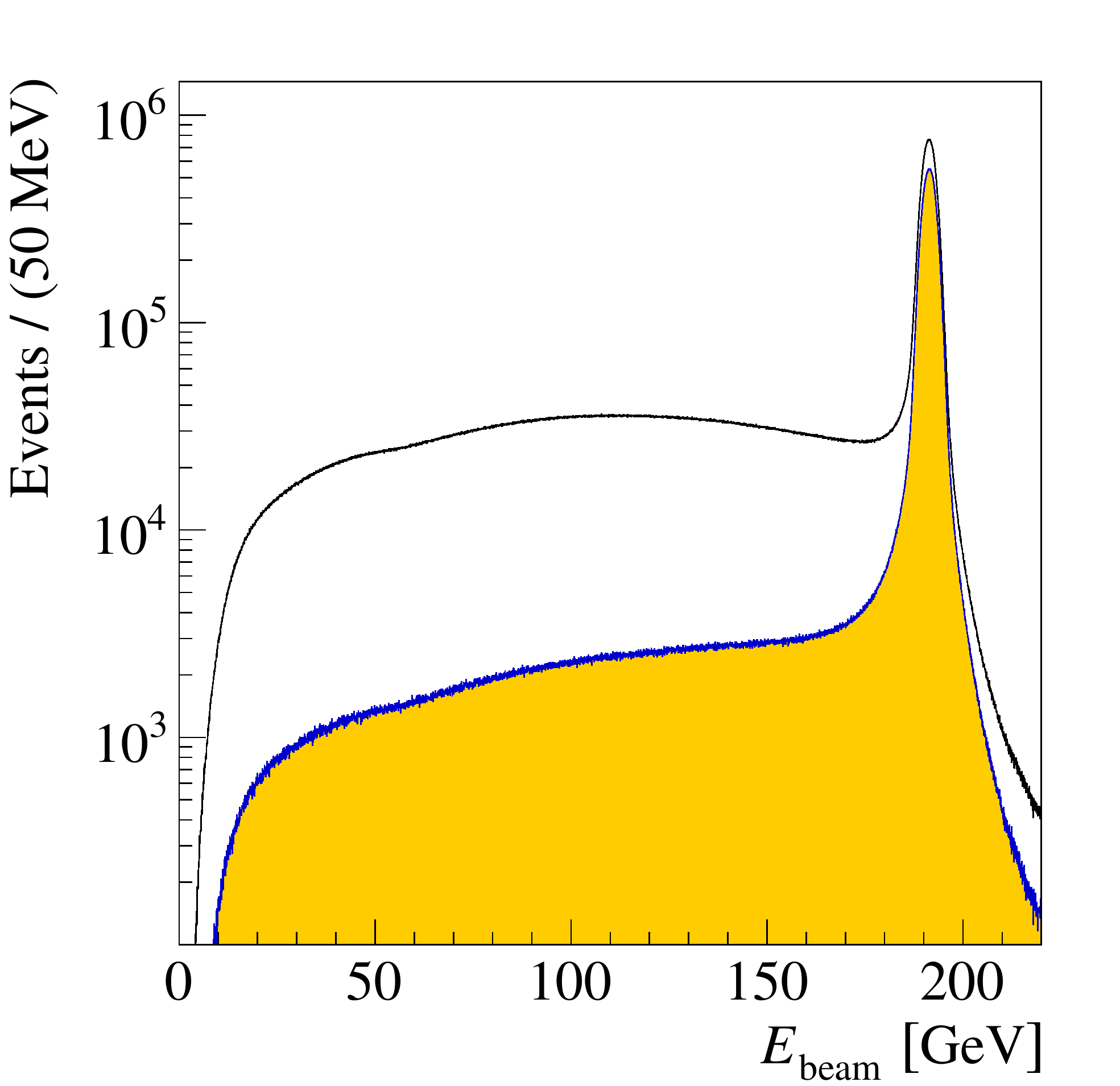}%
  }%
  \newLineOrHspace{\twoPlotSpacing}%
  \subfloat[][]{%
    \label{fig:exclusivity_Esum_zoom}%
    \includegraphics[width=\twoPlotWidth]{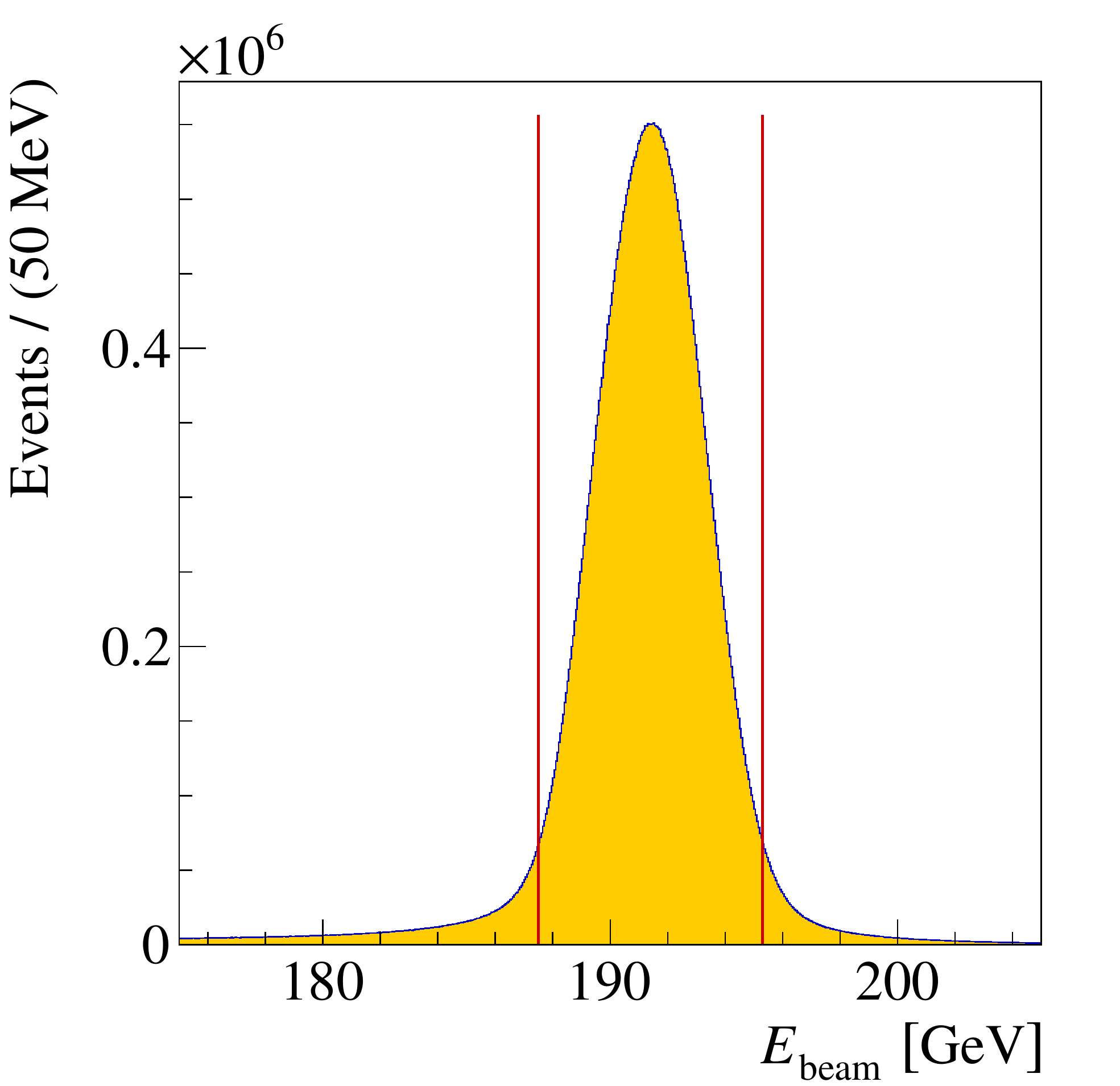}%
  }%
  \caption{Panels~(a) and~(b) show the reconstructed beam energy
    $E_\text{beam}$ after selection cuts (filled histograms).  The
    open histogram in~(a) represents the energy distribution without
    the RPD information.  In the zoomed view~(b), the vertical red
    lines indicate the accepted range.}
  \label{fig:exclusivity}
\end{figure}

The analysis is based on a data set of about \num[round-mode =
figures,round-precision = 2]{6.367916817e9}~events selected by the
hardware trigger (see \cref{sec:event_selection_trigger}).  The event
selection aims at a clean sample of exclusive \reaction events (see
\cref{fig:diffractive_dissociation_3pi}) and consists of the following
criteria (see \refCite{Haas:2014bzm} for more details):
\begin{enumerate}
\item A vertex is required to be formed by the beam particle and three
  charged outgoing tracks with a total charge sum of~$-1$. The vertex
  must be located within the fiducial volume of the liquid-hydrogen
  target.
\item Momentum conservation is applied by requiring exactly one recoil
  particle detected in the RPD that is back-to-back with the outgoing
  \threePi system in the plane transverse to the beam (transverse
  momentum balance).  This suppresses contributions from
    double-diffractive processes, in which also the target proton is
    excited.
\item The beam energy $E_\text{beam}$, which is calculated from the
  energy and momentum of the three outgoing particles corrected for
  the target recoil, must be within a window of $\pm \SI{3.78}{\GeV}$
  around the nominal beam energy, which corresponds to two standard
  deviations (see \cref{fig:exclusivity_Esum_zoom}).
\end{enumerate}

\begin{figure}[tbp]
  \centering
  \subfloat[][]{%
    \label{fig:CP_veto_m3pi}%
    \includegraphics[width=\twoPlotWidth]{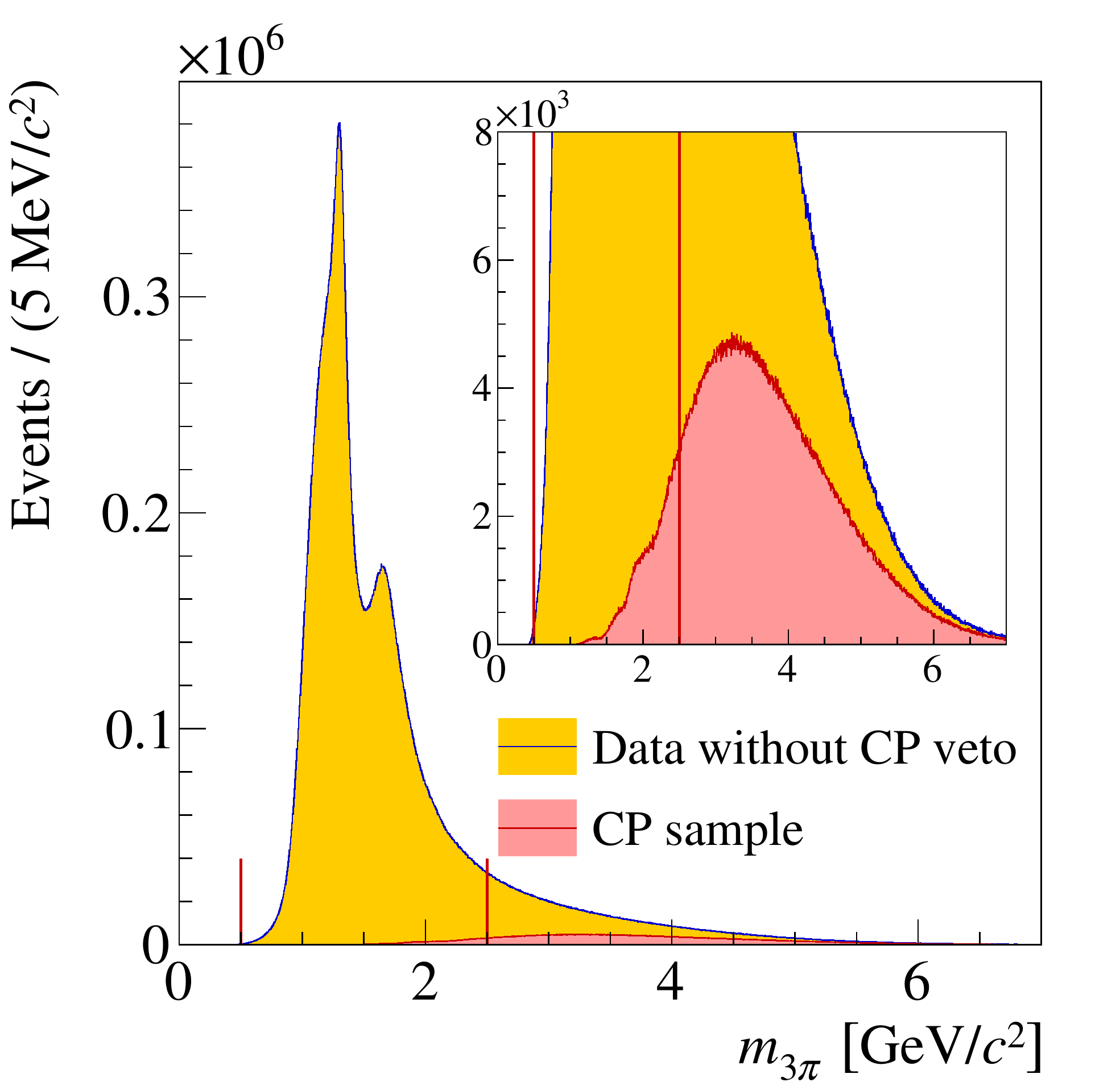}%
  }%
  \newLineOrHspace{\twoPlotSpacing}%
  \subfloat[][]{%
    \includegraphics[width=\twoPlotWidth]{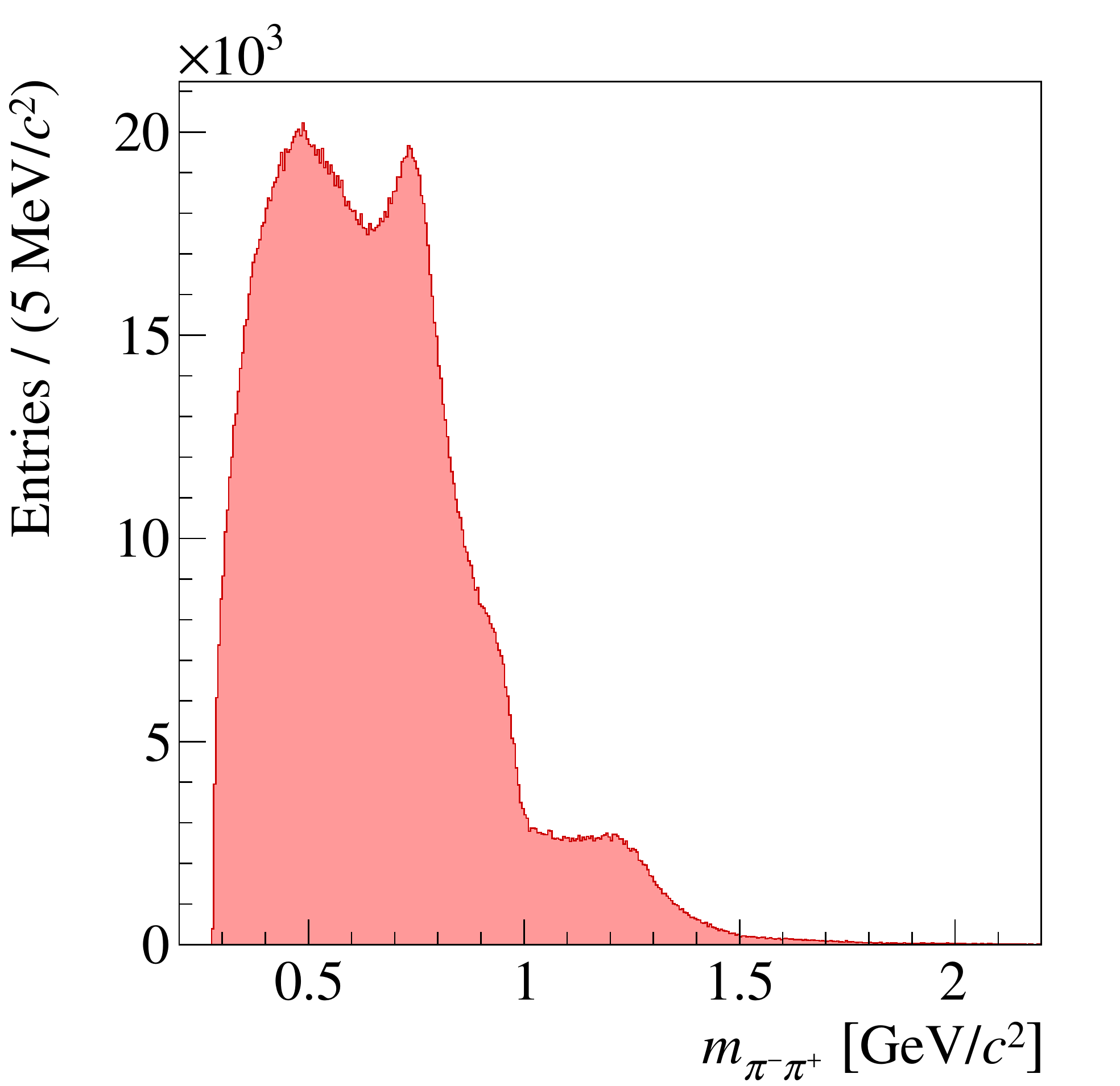}%
  }%
  \caption{\colorPlot Effect of the central-production (CP) veto.
    Panel~(a): \threePi invariant mass spectrum without the
    central-production veto (yellow histogram) together with the
    sample that is removed by the central-production veto (red
    histogram).  The inset shows the same histogram with magnified
    ordinate scale.  Note that the partial-wave analysis is performed
    only in the mass region of
    \SIvalRange{0.5}{\mThreePi}{2.5}{\GeVcc} indicated by the vertical
    red lines.  Panel~(b): \twoPi invariant mass distribution (two
    entries per event) of the sample that is cut away.}
  \label{fig:CP_veto}
\end{figure}

A number of additional selection criteria is applied in order to
reject background events originating from other processes.  Events are
disregarded if the incoming beam particle is identified by the two
beam Cherenkov detectors (CEDARs) as a kaon.  This suppresses
kaon-beam induced events, like \eg
$K^- + p \to K^-\pi^-\pi^+ + p_\text{recoil}$.  If at least one of the
three forward-going particles is identified by the RICH detector as a
kaon, proton, electron, muon, or noise, the event is rejected, thereby
suppressing events such as \eg
$\pi^- + p \to \pi^-K^-K^+ + p_\text{recoil}$.  In order to reject
background events stemming from the central-production reaction
$\pi^- + p \to \pi^-_\text{fast} + \twoPi + p_\text{recoil}$, in which
no three-pion resonances are formed, the faster $\pi^-$ in the event
is required to have a Feynman-$x$ below 0.9 defined in the overall
center-of-mass frame.  The rapidity difference between the faster
$\pi^-$ and the remaining \twoPi pair is limited to the range from
\numrange{2.7}{4.5}.  \Cref{fig:CP_veto} shows the \mThreePi and
\mTwoPi distributions of the sample that is cut away.

\begin{figure*}[tbp]
  \centering
  \subfloat[][]{%
    \includegraphics[width=\twoPlotWidth]{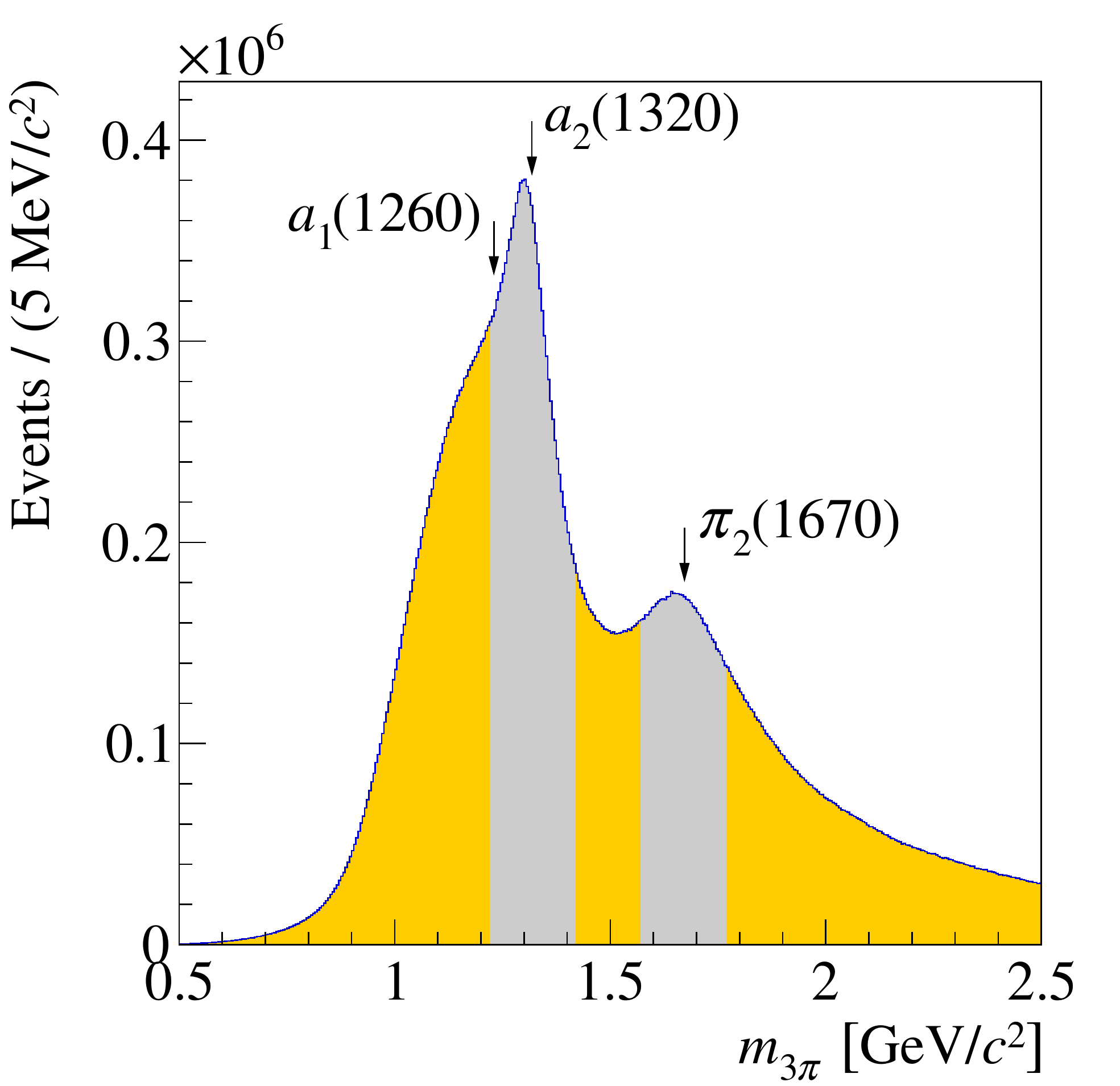}%
    \label{fig:mass_spectrum_3pi}%
  }%
  \hspace*{\twoPlotSpacing}%
  \subfloat[][]{%
    \includegraphics[width=\twoPlotWidth]{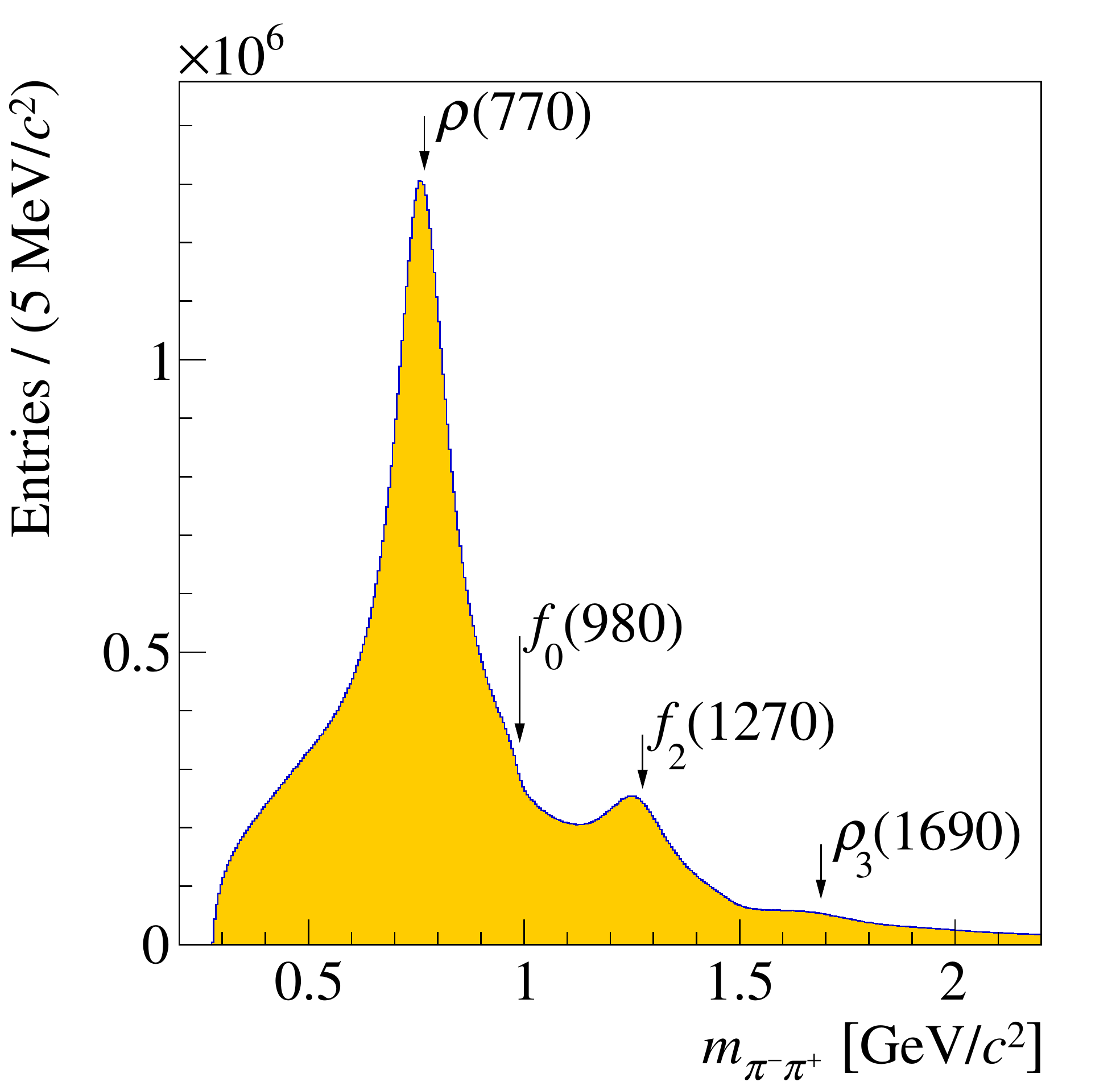}%
    \label{fig:mass_spectrum_2pi}%
  }%
  \\
  \subfloat[][]{%
    \includegraphics[width=\twoPlotWidthTwoD]{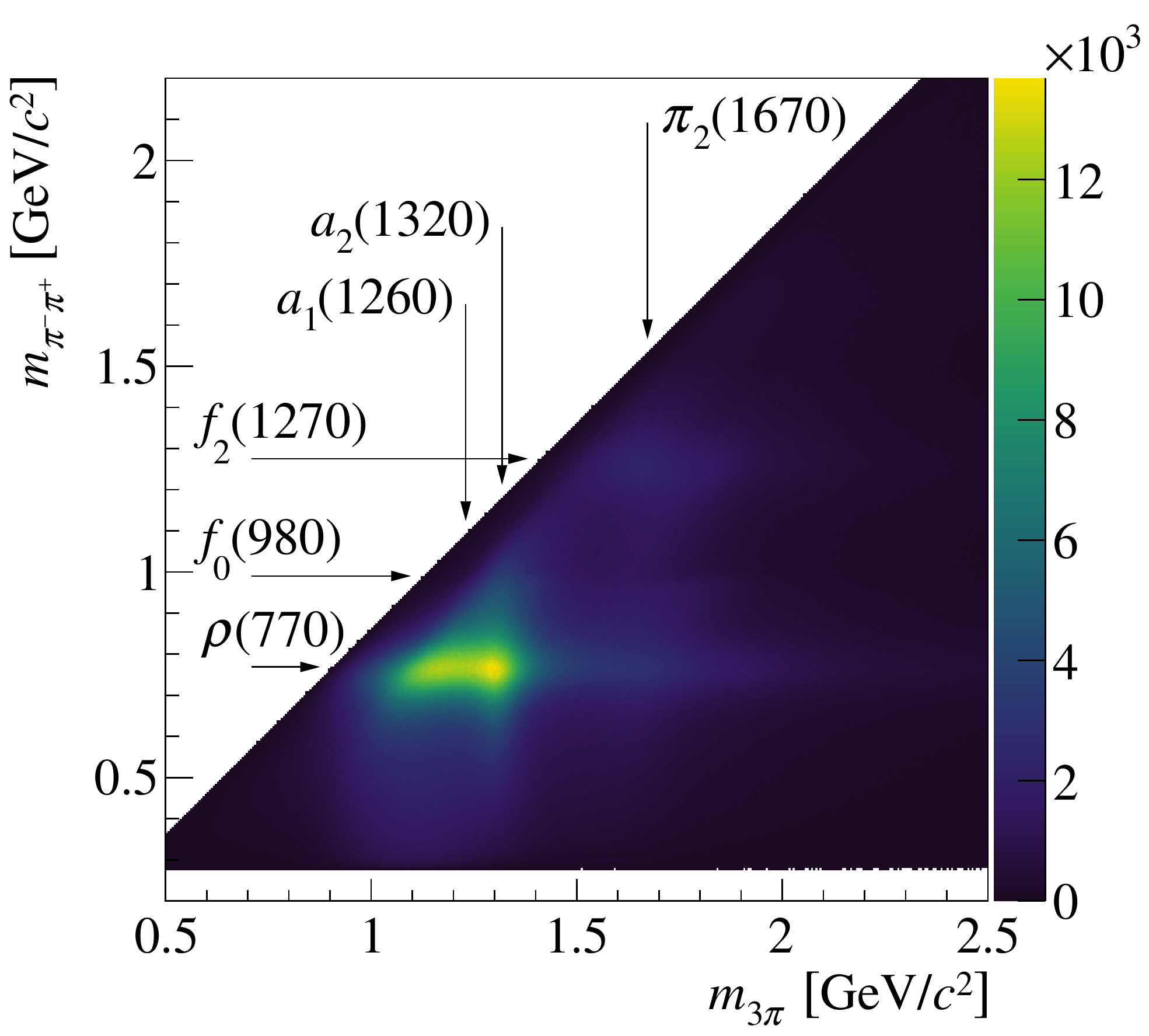}%
    \label{3pi_vs_2pi_mass}%
  }%
  \subfloat[][]{%
    \includegraphics[width=\twoPlotWidth]{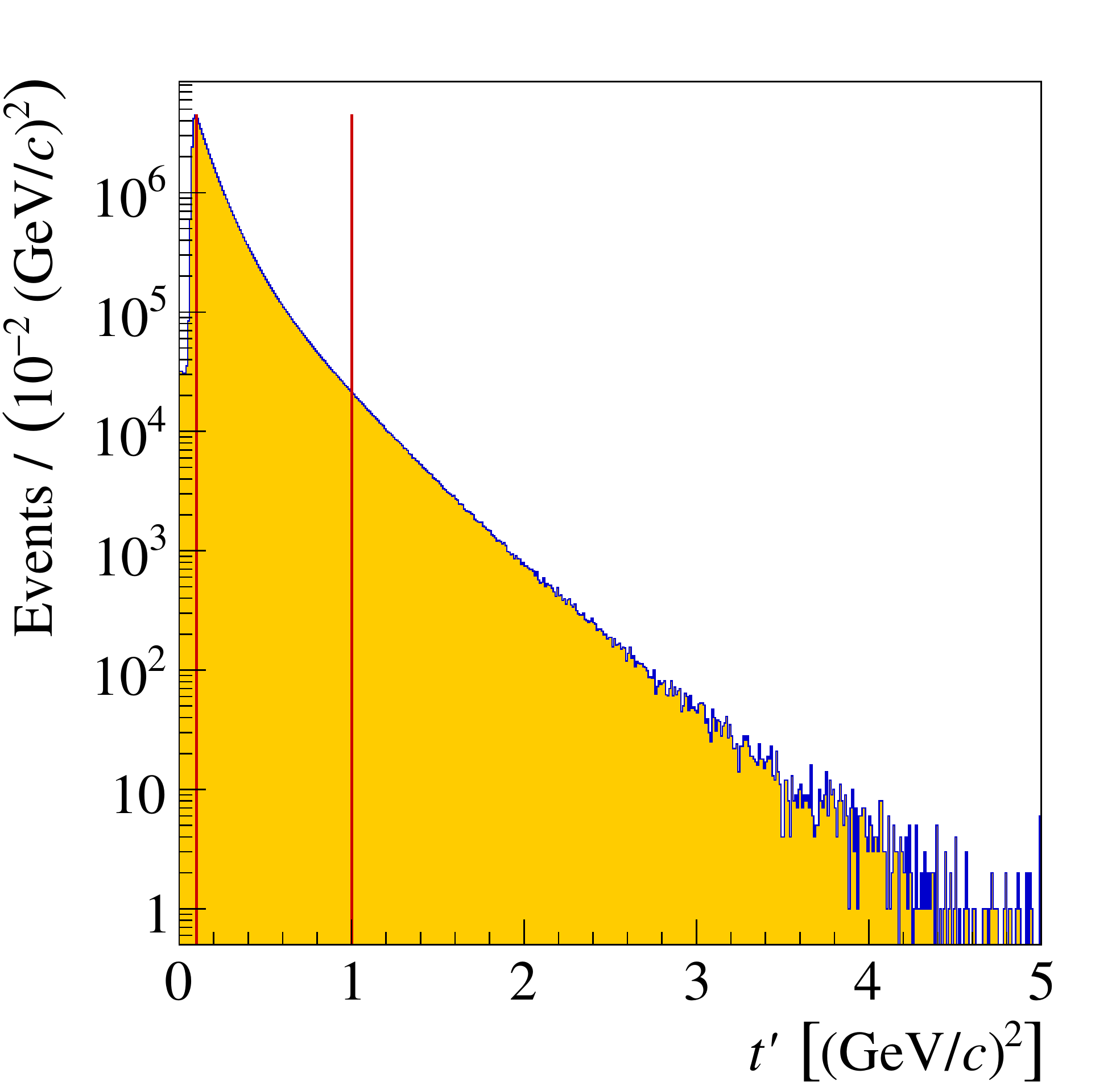}%
    \label{fig:t_prime_2008_log}%
  }%
  \caption{\colorPlot Final event sample after all selection cuts:
    (a)~invariant mass spectrum of \threePi in the range used in this
    analysis (see vertical lines in \cref{fig:CP_veto_m3pi}),
    (b)~\twoPi mass distribution, (c)~correlation of the two, (d)~\tpr
    distribution with vertical lines indicating the range of \tpr
    values used in this analysis.  The histograms in (b) and (c) have
    two entries per event. The labels indicate the position of major
    $3\pi$ and $2\pi$ resonances, the gray shaded areas in~(a) the
    mass regions used to generate the Dalitz plots in
    \cref{fig:dalitz_plots}.}
  \label{fig:mass_spectra}
\end{figure*}

After all cuts, the data sample consists of \num[round-mode =
figures,round-precision = 2]{45.927632e6}~events in the analyzed
kinematic region of three-pion mass,
\SIvalRange{0.5}{\mThreePi}{2.5}{\GeVcc}, and four-momentum transfer
squared, \SIvalRange{0.1}{\tpr}{1.0}{\GeVcsq}.
\Cref{fig:mass_spectrum_3pi,fig:mass_spectrum_2pi} show for all
selected events the mass spectrum of \threePi and of the two \twoPi
combinations.  The known pattern of resonances \PaOne, \PaTwo, and
\PpiTwo is seen in the $3\pi$ system as well as \Prho, \PfZero,
\PfTwo, and \PrhoThree in the \twoPi subsystem.  From
\cref{3pi_vs_2pi_mass}, the correlation of the resonances in the
\threePi system and in the \twoPi subsystem is clearly visible.  This
correlation is the basis of our analysis model described in
\cref{sec:pwa_method}.  The \tpr spectrum is shown in
\cref{fig:t_prime_2008_log}.

A Monte Carlo simulation has shown that for the reaction under study,
the $3\pi$ mass resolution of the spectrometer varies between
\SI{5.4}{\MeVcc} at small \mThreePi (in the range from
\SIrange{0.5}{1.0}{\GeVcc}) and \SI{15.5}{\MeVcc} at large \mThreePi
(in the range from \SIrange{2.0}{2.5}{\GeVcc}), respectively.  The
\tpr resolution as obtained from the reconstructed $3\pi$ final state
ranges between \SIlist{7e-3;20e-3}{\GeVcsq} depending on the \mThreePi
and \tpr region.  The resolution of the reconstructed beam energy
$E_\text{beam}$ is smaller than the intrinsic energy spread of the
beam and varies between \SIlist{0.6;0.9}{\GeV}.  The position of the
primary interaction vertex along the beam axis is reconstructed with a
resolution of approximately \SI{6}{mm} at small and \SI{1.5}{mm} at
large \mThreePi.  The overall \emph{detection efficiency}, which
includes detector acceptance, reconstruction efficiency, and event
selection, is estimated for isotropically distributed (phase-space)
\threePi events.  Integrated over the analyzed kinematic region, it is
\SI{49}{\percent} on average.  More details are found in
\cref{sec:acc_res} and \refCite{Haas:2014bzm}.
 %
%
%

\section{Partial-Wave Analysis Method}
\label{sec:pwa_method}

The goal of the analysis described in this paper is to extract the
resonances contributing to the reaction \reaction and to determine
their quantum numbers from the observed kinematic distributions of the
outgoing \threePi system.  This is accomplished using partial-wave
analysis techniques.  The basic assumption is that resonances dominate
the $3\pi$ intermediate states $X^-$ produced in the scattering
process, so that the $X^-$ production can be treated independently of
the $X^-$ decay (see \cref{fig:diffractive_dissociation_3pi}).  The
amplitude for a certain intermediate state $X^-$ is therefore
factorized into two terms: \one the \emph{transition amplitude}
$\mathcal{T}$ describing the production of a state $X^-$ with specific
quantum numbers and \two the \emph{decay amplitude} $\Psi$ that
describes the decay of the $X^-$ state into a particular \threePi
final state.  For fixed beam energy, the measured kinematic
distribution of the final-state particles depends on the $3\pi$
invariant mass \mThreePi, the four-momentum transfer squared \tpr, and
a set of five additional phase-space variables denoted as $\tau$,
which fully describe the three-body decay and are defined below.

\subsection{Isobar Model}
\label{sec:isobar_model}

\begin{figure}[tbp]
  \centering
  \subfloat[][]{%
    \includegraphics[width=\twoPlotWidth]{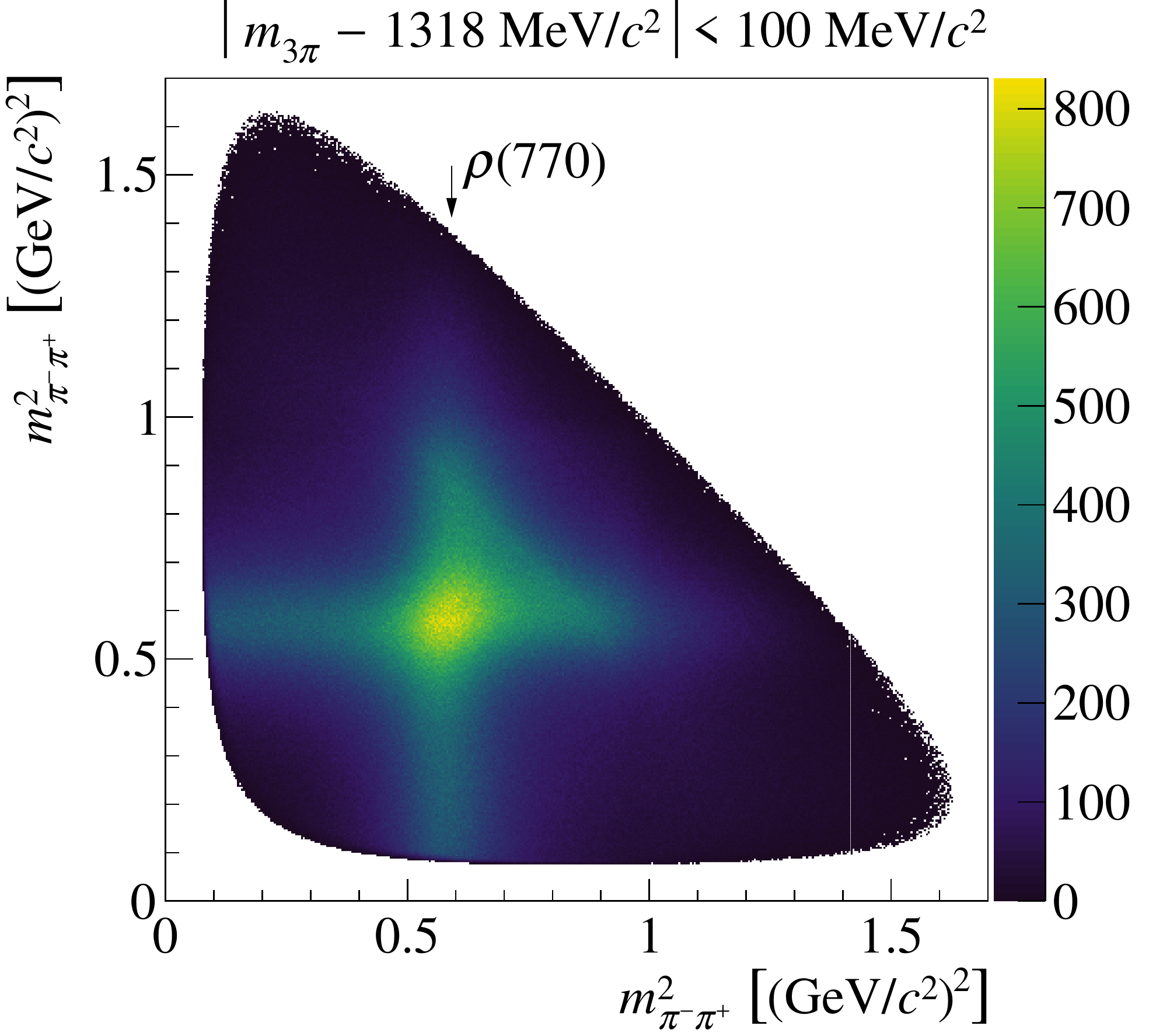}%
  }%
  \newLineOrHspace{\twoPlotSpacing}%
  \subfloat[][]{%
    \includegraphics[width=\twoPlotWidth]{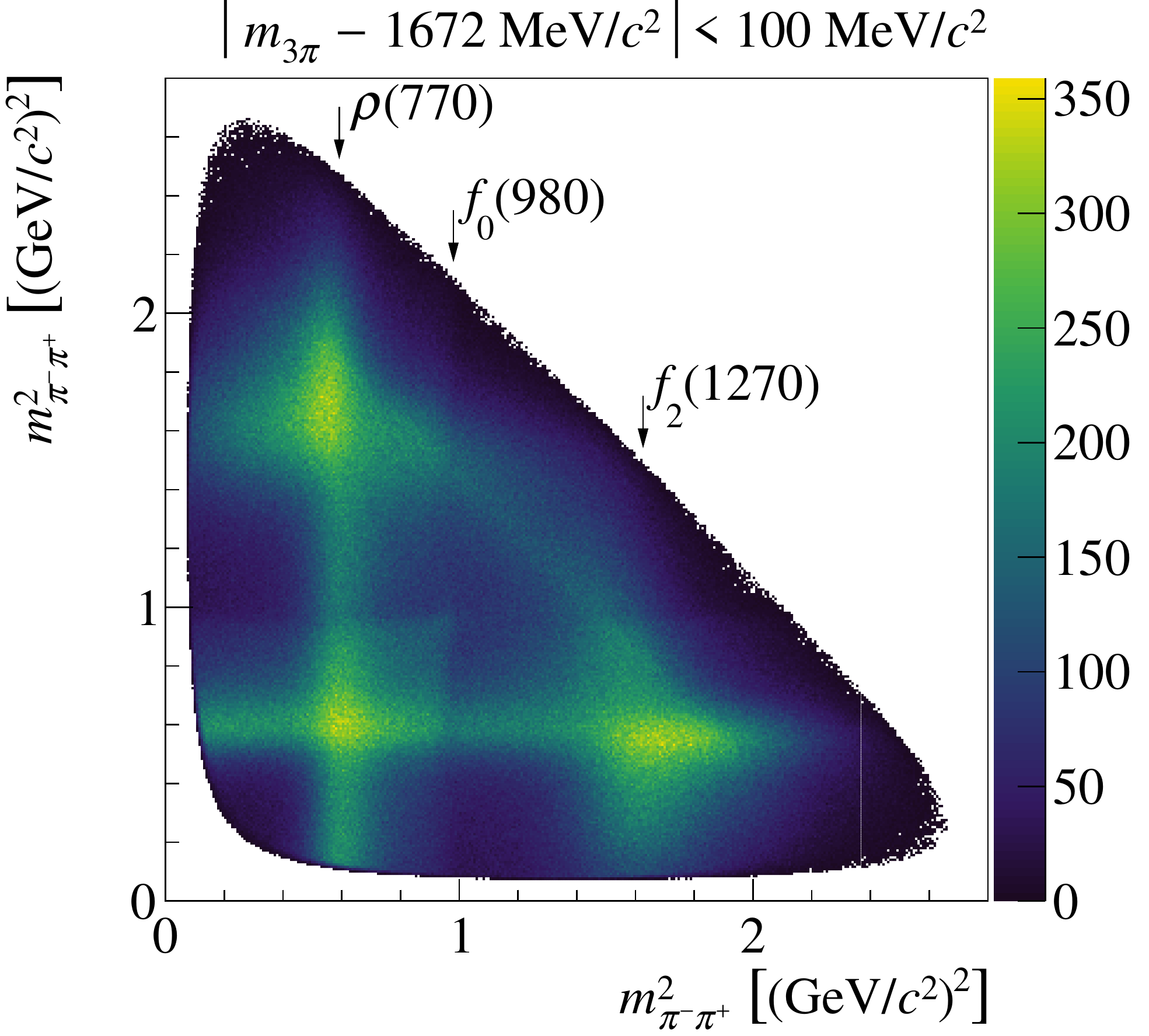}%
  }%
  \caption{\colorPlot (a)~Dalitz plot in the mass regions of the
    \PaTwo, which also includes the \PaOne, (b) around the \PpiTwo.
    The used $3\pi$ mass regions are indicated in
    \cref{fig:mass_spectrum_3pi}.  The dominant $\Prho\,\pi$ decays
    of \PaOne and \PaTwo are clearly visible.  The \PpiTwo region
    exhibits $\Prho\,\pi$, $\PfTwo\,\pi$, and $\PfZero[980]\,\pi$
    decay modes.}
  \label{fig:dalitz_plots}
\end{figure}

In order to illustrate the isobar ansatz, we give in
\cref{fig:dalitz_plots} two examples for Dalitz plots for two
different regions of \mThreePi.  In the $3\pi$ mass region around
\PaTwo, which also includes contributions from \PaOne, we see a
dominant contribution of the \Prho in the \twoPi subsystem, while for
values of \mThreePi around \PpiTwo several $2\pi$ resonances
contribute, \ie \Prho, \PfZero, and \PfTwo.

\begin{figure}[tbp]
  \centering
  \includegraphics[scale=1]{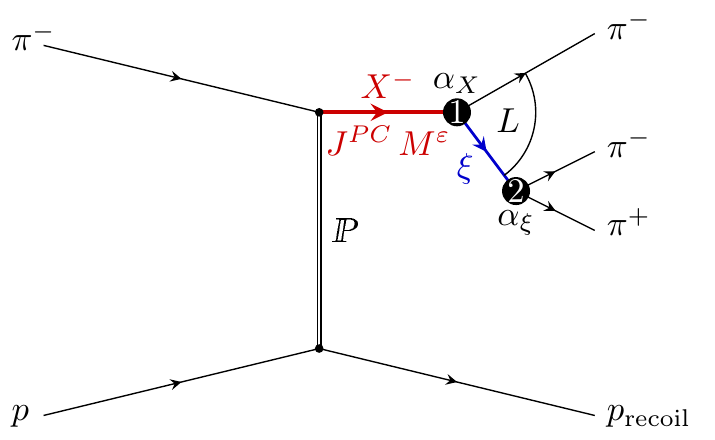}
  \caption{The decay of $X^-$, as described in the isobar model, is
    assumed to proceed via an intermediate \twoPi state $\xi$, the
    so-called isobar.}
  \label{fig:3pi_reaction_isobar}
\end{figure}

Because of the strong contribution of resonances in the \twoPi
subsystem, the three-body decay amplitude
$\widetilde{\Psi}(\tau, \mThreePi)$ is factorized into two two-body
decay terms (see \cref{fig:3pi_reaction_isobar}).  This factorization
is known as the \emph{isobar model}\footnote{An early detailed
  discussion can be found in \refCite{Herndon:1973yn}.}  and the
introduced intermediate \twoPi state $\xi$ is called the
\emph{isobar}.  In the first two-body decay, $X^- \to \xi^0 + \pi^-$,
a relative orbital angular momentum $L$ is involved in the decay.  The
decay amplitude $\widetilde{\Psi}(\tau, \mThreePi)$ completely
describes the kinematic distribution of the three outgoing pions for
particular quantum numbers of $X^-$ and for a particular isobar
channel with a given $L$.

\begin{figure*}[tbp]
  \centering
  \includegraphics[scale=1]{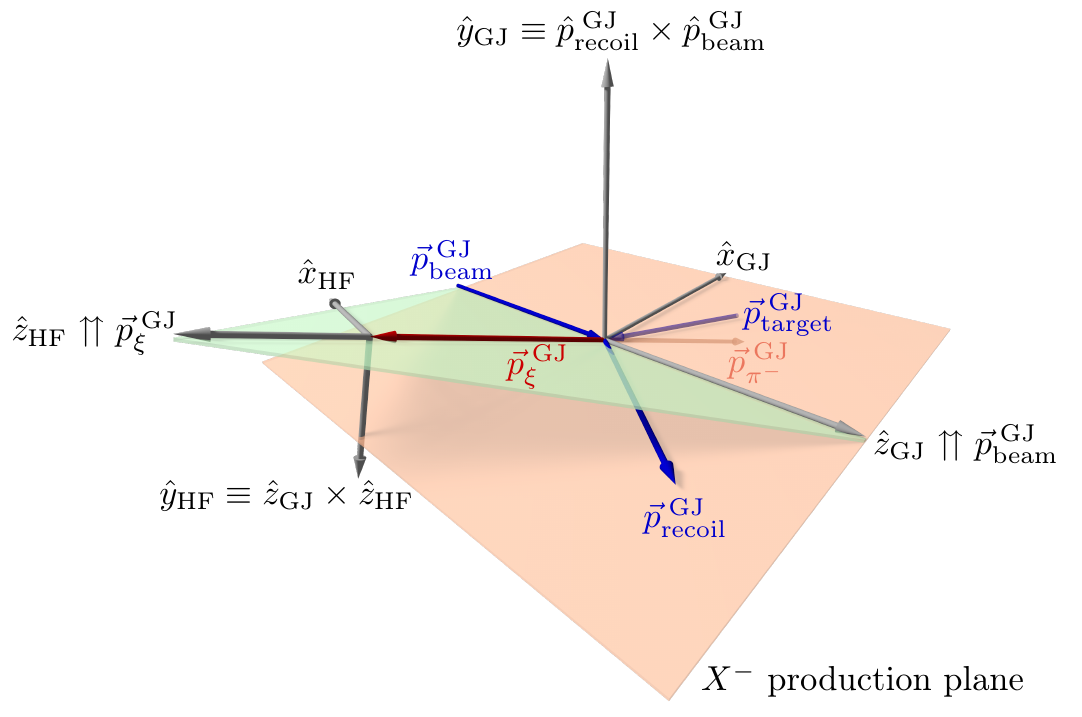}
  \caption{\colorPlot Definition of the Gottfried-Jackson
      reference frame~(GJ) in the $X$ rest system and of the helicity
      reference frame~(HF) in the $\xi^0$ rest system as they
      are
    used to analyze the angular distributions of the decays $X^- \to
    \xi^0 + \pi^-$ and $\xi^0 \to \pi^- + \pi^+$,
    respectively.  Unit vectors are indicated by a circumflex.}
  \label{fig:coordsys}
\end{figure*}

The two subsequent two-body decays are described in different
right-handed coordinate systems, \ie the Gottfried-Jackson and the
helicity reference frame (see \cref{fig:coordsys}).  The
Gottfried-Jackson~(GJ) frame is used to describe the angular
distribution of the decay of the intermediate state $X^-$ into the
isobar $\xi$ and the bachelor pion.  It is constructed in the $X^-$
rest system, in which the direction of the beam particle defines the
\zGJ axis and the \yGJ axis is oriented along the normal to the
production plane ($\yGJv \equiv \hat{p}_\text{beam}^\text{\,lab}
\times \hat{p}_X^\text{\,lab} = \hat{p}_\text{recoil}^\text{\;GJ}
\times \hat{p}_\text{beam}^\text{\;GJ}$, where unit vectors are
indicated by a circumflex).  In this system, the momenta of the isobar and the
bachelor pion are back to back, so that the two-body decay $X^- \to
\xi^0 + \pi^-$ is described by the polar angle \thetaGJ and the
azimuthal angle \phiGJ of the isobar, the latter being also referred
to as Treiman-Yang angle.

For the decay of the isobar $\xi$ into \twoPi, the helicity reference
system~(HF) is used to describe the angular distribution.  This frame
is constructed by boosting from the Gottfried-Jackson system into the
$\xi$ rest frame.  The \zHF axis is taken along the original direction
of the isobar and $\yHFv \equiv \zGJv \times \zHFv$.  The two pions
are emitted back to back, so that the $\xi^0 \to \twoPi$ decay is
described by the polar angle \thetaHF and the azimuthal angle \phiHF
of the negative pion.

\begin{figure*}[tbp]
  \centering
  \newsavebox{\tempbox}%
  \sbox{\tempbox}{%
    \shortstack{%
      \includegraphics[width=0.26\textwidth]{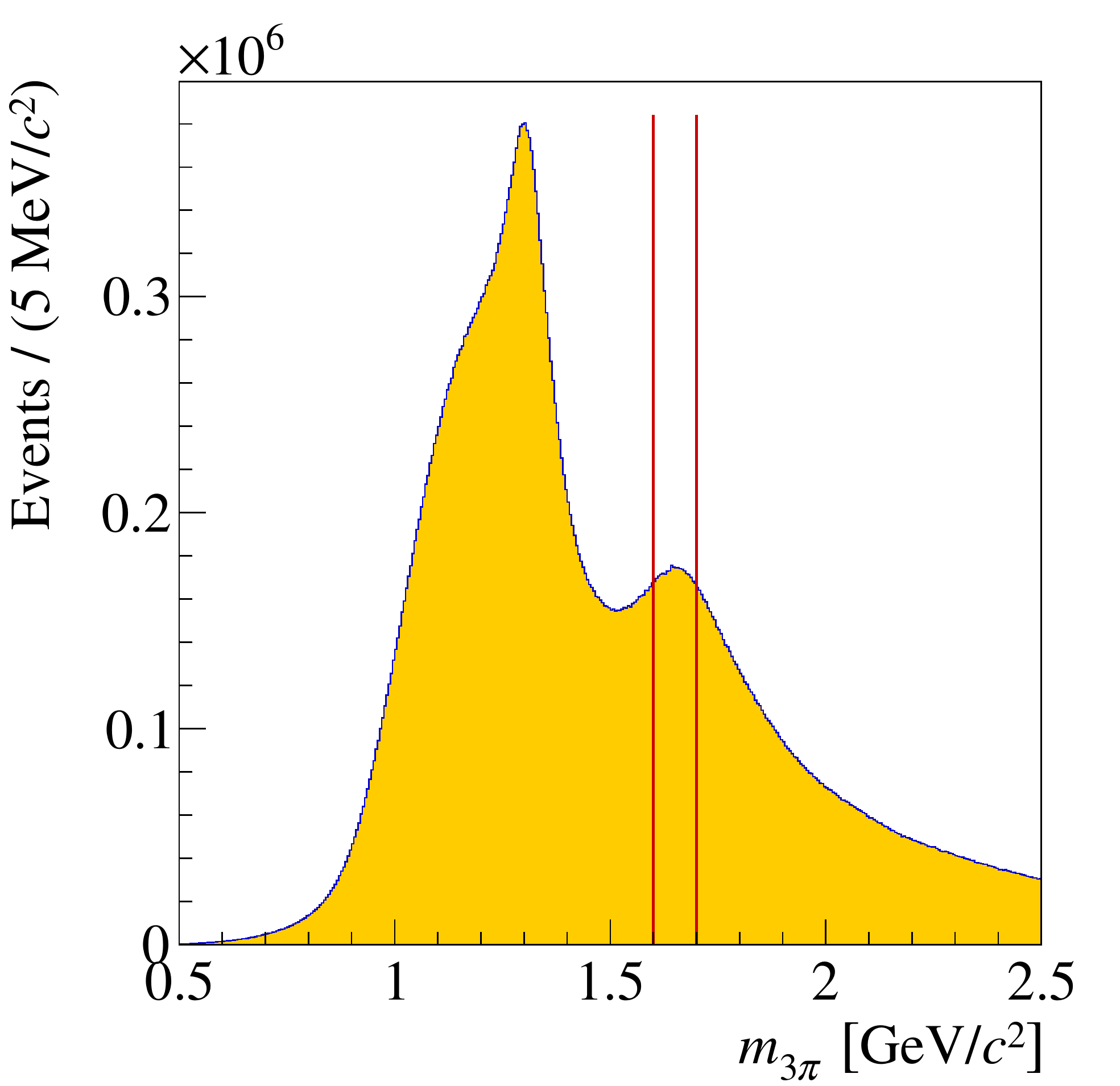}%
      \\%
      \includegraphics[width=0.26\textwidth]{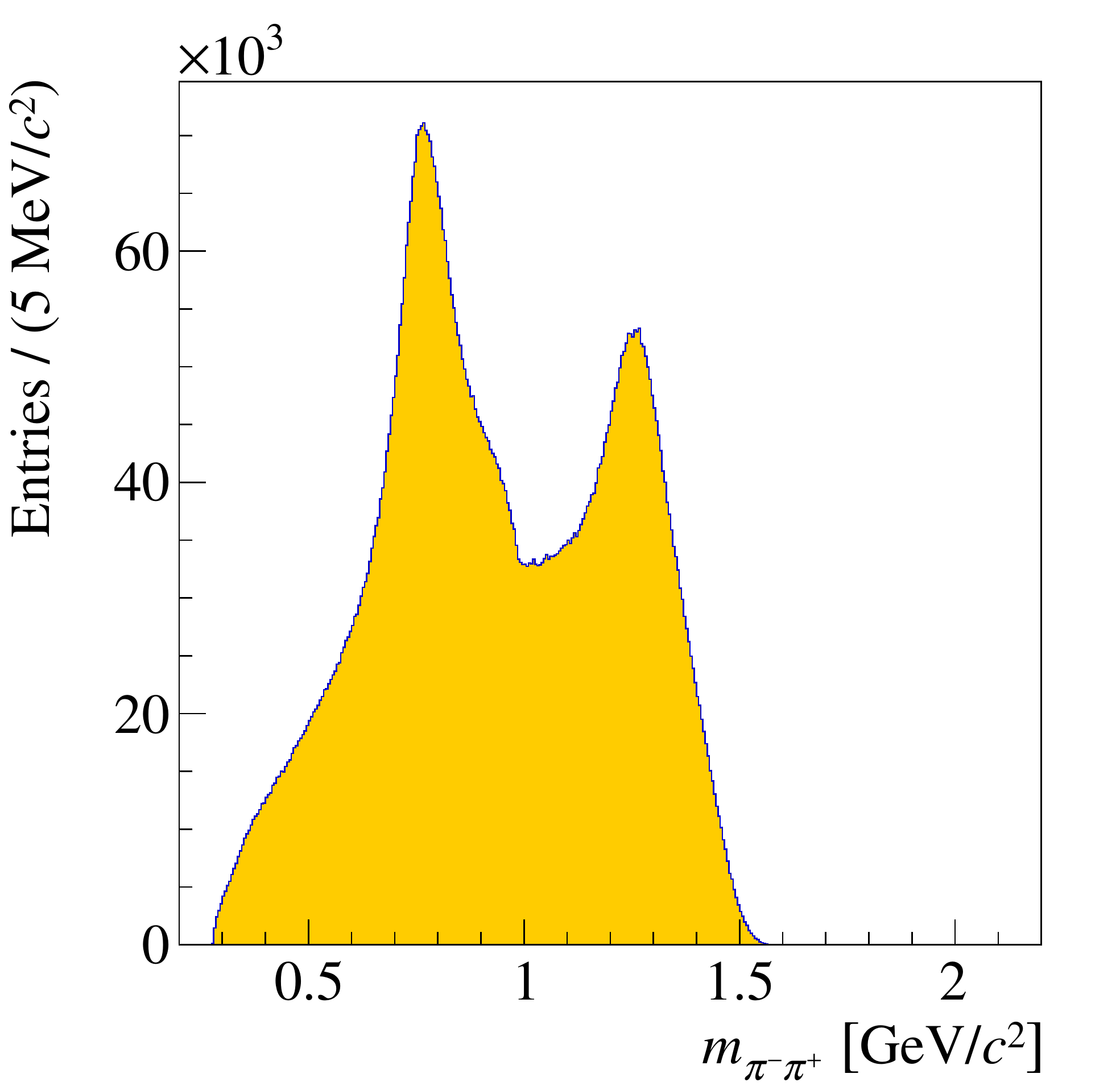}%
    }%
  }%
  \subfloat{%
    \usebox{\tempbox}%
  }%
  \subfloat{%
    \vbox to \ht\tempbox{%
      \vfil
      \hbox{%
        \includegraphics[width=0.37\textwidth]{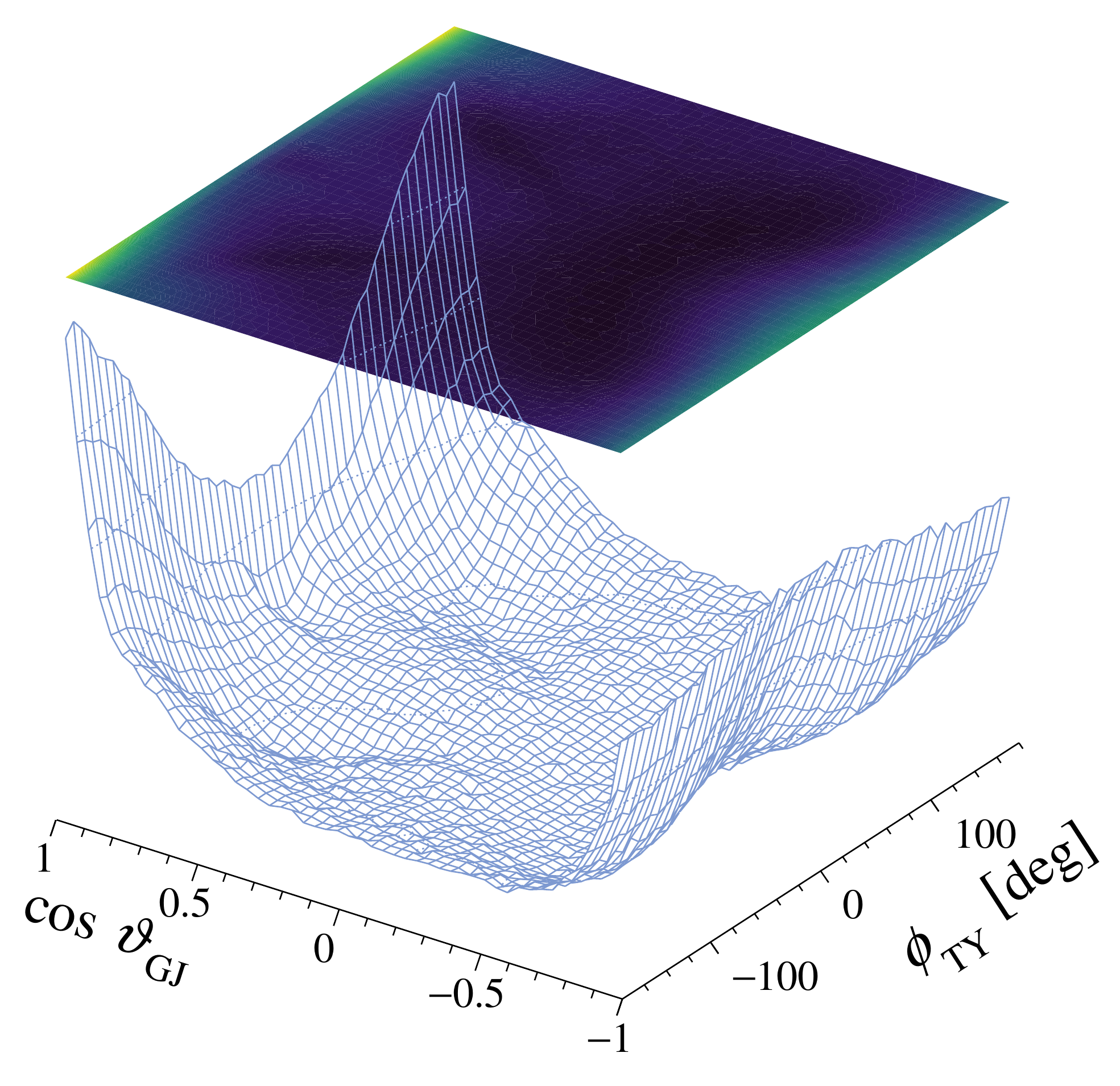}%
      }%
      \vfil
    }%
  }%
  \subfloat{%
    \vbox to \ht\tempbox{%
      \vfil
      \hbox{%
        \includegraphics[width=0.37\textwidth]{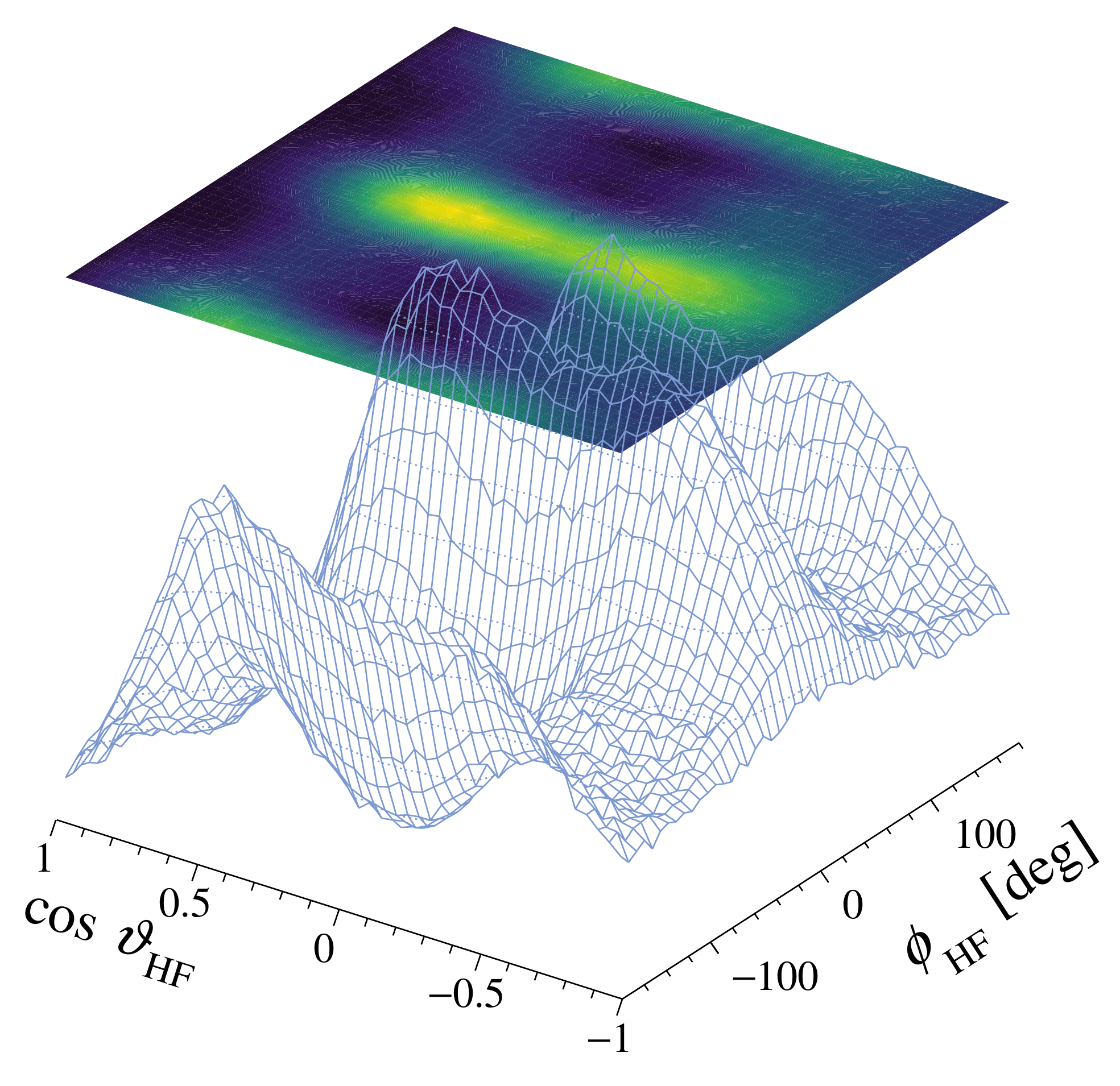}%
      }%
      \vfil
    }%
  }%
  \caption{\colorPlot Example of a $3\pi$ angular distribution
    observed in the mass region
    \SIvalRange{1.6}{\mThreePi}{1.7}{\GeVcc} around the \PpiTwo
    indicated by vertical red lines in the upper left panel.  The main
    decay of this resonance is through the \PfTwo isobar, which is a
    $\JPC = 2^{++}$ state, decaying into \twoPi in a $D$-wave.  The
    \PfTwo and the bachelor pion are in a relative $S$ or
    $D$-wave.}
  \label{fig:example_distribution}
\end{figure*}

For illustration, \cref{fig:example_distribution} shows the observed, \ie
acceptance-uncorrected angular distributions in the two reference
systems for events around the \PpiTwo mass region.  The main decay of
this resonance is through the \PfTwo isobar, which is a
$\JPC = 2^{++}$ state decaying into \twoPi in a relative $D$-wave in
the helicity reference frame.  The \PfTwo and the bachelor pion are
emitted in a relative $S$ or $D$-wave in the Gottfried-Jackson
coordinate system.  Note that the shown distribution is complicated by
the fact that other decay modes of the \PpiTwo as well as decays of
other $3\pi$ resonances with different angular distributions interfere
with the $\PpiTwo \to \PfTwo\, \pi^-$ decay.

\subsection{Parametrization of Decay Amplitudes}
\label{sec:isobar_model_amplitude}

In the helicity
formalism~\cite{Jacob:1959at,chung:1971ri,Richman:1984gh}, the
amplitude $\mathcal{A}_M^R$ for a two-body decay of a state $R$ with
spin $J$ into particles~1 and~2 can be factorized into a dynamic part
$f_{\lambda_1\, \lambda_2}^J(m_R; m_1, m_2)$ that describes the mass
dependence and an angular part.  The latter is related to the rotation
between the rest frame of the parent system $R$, in which its spin
projection $M$ is defined, and the helicity frame used to define the
daughter spin states, which are given by the helicities
$\lambda_{1, 2}$.  The rotation is described by the Wigner
$D$-function.  In addition, there are two Clebsch-Gordan coefficients
arising in the decay $R \to 1 + 2$: \one for the coupling of the spins
$J_{1,2}$ of the daughter particles to the total intrinsic spin $S$
and \two for the coupling of the relative orbital angular momentum
$L_{1 2}$ between the daughter particles with $S$ to $J$.  As the
orbital angular momentum $L_{1 2}$ in the decay is by definition
perpendicular to the quantization axis in the helicity formalism, its
$z$ projection vanishes.

The amplitude $\mathcal{A}_\lambda^\xi$ for the two-body decay of the
isobar $\xi$ with spin $J_\xi$ and helicity $\lambda$ into \twoPi is
given by
\begin{multlineOrEq}
  \label{eq:decay_amplitude_isobar}
  \mathcal{A}_\lambda^\xi(\thetaHF, \phiHF, m_\xi) \newLineOrNot
  = D^{J_\xi \text{*}}_{\lambda\, 0}(\phiHF, \thetaHF, 0)\, f_{0\, 0}^{J_\xi}(m_\xi; m_\pi, m_\pi),
\end{multlineOrEq}
with $m_\xi$ being the \twoPi invariant mass.  The dynamic part
factorizes into several components:
\begin{multlineOrEq}
  \label{eq:decay_amplitude_isobar_dyn}
  f_{0\, 0}^{J_\xi}(m_\xi; m_\pi, m_\pi) \newLineOrNot
  = \underbrace{\vphantom{F_{J_\xi}}\sqrt{2J_\xi + 1}}_{\text{normalization}}
  \underbrace{\alpha_\xi\, F_{J_\xi}(m_\xi; m_\pi, m_\pi)\, \Delta_\xi(m_\xi; m_\pi, m_\pi)}_{\text{dynamics}}.
\end{multlineOrEq}
Here, the fact was already used that pions are spinless isospin-1
particles.  Therefore, the $L$-$S$ coupling Clebsch-Gordan coefficient
is unity and the orbital angular momentum $L_\xi$ in the decay is
identical to the spin $J_\xi$ of the isobar.  The coupling amplitude
$\alpha_\xi$ describes the strength of the decay and is usually
unknown.  Parametrizations for the barrier factor $F_{J_\xi}$ and the
isobar line shape $\Delta_\xi$ are discussed in
\cref{sec:pwa_method_isobar_parametrization}.

The amplitude $\mathcal{A}_M^X$ for the two-body decay of $X^-$ into
the isobar $\xi$ and the bachelor pion is constructed by summing over
the helicity $\lambda$ of the intermediate isobar:
\begin{multlineOrEq}
  \label{eq:decay_amplitude_X}
  \mathcal{A}_M^X(\thetaGJ, \phiGJ, \mThreePi) \newLineOrNot
  = \sum_\lambda D^{J \text{*}}_{M\, \lambda}(\phiGJ, \thetaGJ, 0)\, f_{\lambda\, 0}^J(\mThreePi; m_\xi, m_\pi).
\end{multlineOrEq}
Taking into account the quantum numbers of the bachelor pion the
dynamic part of the amplitude reads:
\begin{multlineOrEq}
  \label{eq:decay_amplitude_X_dyn}
  f_{\lambda\, 0}^J(\mThreePi; m_\xi, m_\pi) \newLineOrNot
  = \underbrace{\vphantom{J_\xi}\sqrt{2L + 1}}_{\text{normalization}}
  \underbrace{\clebsch{L}{0}{J_\xi}{\lambda}{J}{\lambda}}_{\substack{\text{$L$-$S$ coupling} \\ \text{Clebsch-Gordan}}}\,
  \underbrace{\alpha_X\, F_L(\mThreePi; m_\xi, m_\pi)}_{\text{dynamics}}.
\end{multlineOrEq}
This is the nonrelativistic $L$-$S$ coupling scheme as introduced by
Jacob and Wick in \refCite{Jacob:1959at}, which is equivalent to the
nonrelativistic Zemach tensors~\cite{Zemach:1965z,Zemach:1965zz}.
Relativistic corrections as worked out in \refCite{chung:2007nn} are
not applied.  The results presented here are therefore comparable to
those of previous analyses.  The relativistic corrections are expected
to become important for large breakup momenta in the
$X^- \to \xi^0 + \pi^-$ decay and will be studied in detail in a
future analysis.

In \cref{eq:decay_amplitude_X_dyn}, again an unknown coupling
amplitude $\alpha_X$ appears.  Note that the line shape
$\Delta_X(\mThreePi)$ of the $X^-$ is unknown.  It is actually the
goal of the analysis to extract it from the data.  This is achieved by
setting $\Delta_X$ to unity so that it does not appear in the above
formula and by performing the analysis in narrow bins of \mThreePi,
thereby neglecting the \mThreePi dependence within each bin.

Combining \cref{eq:decay_amplitude_isobar,eq:decay_amplitude_X} yields
the $X^-$ decay amplitude
\begin{multlineOrEq}
  \label{eq:decay_amplitude_before_sym}
  \psi_{i, j}(\overbrace{\thetaHF, \phiHF, m_\xi, \thetaGJ, \phiGJ}^{\displaystyle \hspace{0.75em} \equiv \tau}; \mThreePi) \newLineOrNot
  = \sum_\lambda D^{J \text{*}}_{M\, \lambda}(\phiGJ, \thetaGJ, 0)\, f_{\lambda\, 0}^J(\mThreePi; m_\xi, m_\pi)\,
  \newLineOrNot[-1.8ex] \mathcal{A}_\lambda^\xi(\thetaHF, \phiHF, m_\xi).
\end{multlineOrEq}
However, the above amplitude does not yet have the correct Bose
symmetry under exchange of the two indistinguishable $\pi^-$ in the
final state.  The symmetrized amplitude is
\begin{multlineOrEq}
  \label{eq:decay_amplitude}
  \widetilde{\Psi}_{i, j}(\tau_{13}, \tau_{23}; \mThreePi) \newLineOrNot
  = \frac{1}{\sqrt{2}} \sBrk[1]{\psi_{i, j}(\tau_{13}; \mThreePi) + \psi_{i, j}(\tau_{23}; \mThreePi)}
\end{multlineOrEq}
where $\tau_{13}$ and $\tau_{23}$ are the sets of phase-space
variables calculated for the two possible \twoPi combinations of the
$\pi_1^-\pi_2^-\pi_3^+$ system.  \Cref{eq:decay_amplitude} takes
correctly into account the self-interference due to the
particle-exchange symmetry.  For better readability, we will use the
simplified notation $\widetilde{\Psi}_{i, j}(\tau; \mThreePi)$ in the
text below.

The $X^-$ decay amplitude $\widetilde{\Psi}_{i, j}$ is uniquely
defined by two indices: a)~the set of $X^-$ quantum numbers (isospin
$I$, $G$ parity, spin $J$, parity $P$, $C$ parity, and the spin
projection $M$), represented here by the index
$i \equiv (\IG, \JPC, M)$, and b)~by the $X^-$ decay mode enumerated
by $j \equiv (\xi, L)$.  In this way we can describe the decay of a
diffractively produced intermediate state $X^-$ with mass \mThreePi
decaying into a \twoPi isobar $\xi$ and a bachelor $\pi^-$.

\subsection{Partial-Wave Decomposition}
\label{sec:pwa_decomp}

The intensity distribution $\mathcal{I}(\mThreePi, \tpr, \tau)$ of the
final-state particles is written as a truncated series of partial
waves denoted by the indices $i$ and $j$, which represent certain
quantum number combinations as discussed in
\cref{sec:isobar_model_amplitude}.  The strengths and phases, with
which the different intermediate states $X^-$ are produced, are
described by the production amplitudes
$\widetilde{\mathcal{T}}_i^{r \refl}(\mThreePi, \tpr)$.  They depend
on the production kinematics and on the set $i = (\IG, \JPC, M)$ of
the $X^-$ quantum numbers.  Together with the decay amplitudes from
\cref{eq:decay_amplitude}, the intensity is written as the coherent
sum over the different intermediate $X^-$ states represented by $i$
and the different $X^-$ decay modes enumerated by $j$:
\begin{multlineOrEq}
  \label{eq:intensity_ansatz}
  \mathcal{I}(\tau; \mThreePi, \tpr) \newLineOrNot
  = \sum_{\refl = \pm 1} \sum_{r = 1}^{N_r^\refl} \Abs[3]{\sum_i \widetilde{\mathcal{T}}_i^{r \refl}(\mThreePi, \tpr)
  \sum_j \widetilde{\Psi}_{i, j}^\refl(\tau; \mThreePi)}^2.
\end{multlineOrEq}
In the above formula, two additional indices, the so-called
reflectivity \refl and the rank index $r$, are introduced, which are
both summed over incoherently.  Before discussing these two indices,
we transform \cref{eq:intensity_ansatz} further.

In the helicity formalism, the isobar-model decay amplitudes are
calculable up to the unknown couplings $\alpha_\xi$ and $\alpha_X$,
which appear at each decay vertex and were introduced in
\cref{sec:isobar_model_amplitude} [see
\cref{eq:decay_amplitude_isobar_dyn,eq:decay_amplitude_X_dyn,fig:3pi_reaction_isobar}].
Assuming that these couplings do not depend on the kinematics, these
unknowns can be be pulled out of the decay amplitude in
\cref{eq:decay_amplitude} and absorbed into the production amplitudes
by the following redefinitions:
\begin{align}
  \widebar{\Psi}\vphantom{\Psi}_{i, j}^{\refl}(\tau; \mThreePi)
  \alignOrNot\equiv
  \frac{\widetilde{\Psi}_{i, j}^{\refl}(\tau; \mThreePi)}{\alpha_\xi\, \alpha_X}
  \\ \intertext{and}%
  \widebar{\mathcal{T}}\vphantom{\mathcal{T}}_{i, j}^{r \refl}(\mThreePi, \tpr)
  \alignOrNot\equiv
  \alpha_\xi\, \alpha_X\, \widetilde{\mathcal{T}}_i^{r \refl}(\mThreePi, \tpr).
\end{align}
Note that now the amplitudes
$\widebar{\mathcal{T}}\vphantom{\mathcal{T}}_{i, j}^{r \refl}$ carry
not only information about the production of the state~$i$, but also
about its coupling to a certain decay channel $j$.  Therefore, we will
refer to the
$\widebar{\mathcal{T}}\vphantom{\mathcal{T}}_{i, j}^{r \refl}$ as
\emph{transition amplitudes} in the rest of the text.  We introduce
the index
\begin{equation}
  \label{eq:wave_notation}
  a \equiv (i, j).
\end{equation}
This notation represents a certain partial wave and contains all
information about the production as well as the decay (see
\cref{sec:isobar_model_amplitude}).  With these modifications, we
rewrite the expression for the intensity:
\begin{multlineOrEq}
  \label{eq:intensity_ansatz_couplings}
  \mathcal{I}(\tau; \mThreePi, \tpr) \newLineOrNot
  = \sum_{\refl = \pm1} \sum_{r = 1}^{N_r^\refl} \Abs[3]{\sum_a
  \widebar{\mathcal{T}}\vphantom{\mathcal{T}}_a^{r \refl}(\mThreePi, \tpr)\,
  \widebar{\Psi}\vphantom{\Psi}_a^\refl(\tau; \mThreePi)}^2.
\end{multlineOrEq}

It is convenient to introduce the \emph{spin-density matrix}
\begin{equation}
  \label{eq:spin_density_definition}
  \widebar{\varrho}_{a b}^\refl(\mThreePi, \tpr)
  = \sum_{r = 1}^{N_r^\refl} \widebar{\mathcal{T}}\vphantom{\mathcal{T}}_a^{r \refl}\,
  \widebar{\mathcal{T}}\vphantom{\mathcal{T}}_b^{r \refl \text{*}},
\end{equation}
which represents the full information that is obtainable about $X^-$.
The diagonal elements of $\widebar{\varrho}$ are proportional to the
partial-wave intensities and the off-diagonal entries to the
interference terms.

There are several effects that could lead to deviations from full
coherence of the intermediate states, \eg spin-flip and spin-non-flip
processes or the excitation of baryon resonances.  Also, performing
the analysis over a large range of four-momentum transfer without
taking into account the different \tpr dependences of the intermediate
states may appear like incoherence (see \cref{sec:tprim_dependence}).
One way of including these incoherences is the introduction of the
additional rank index~$r$ for the transition amplitudes, which is
summed over incoherently [see \cref{eq:intensity_ansatz_couplings}].
The parameter~$N_r$ is called the \emph{rank} of the spin-density
matrix.

The constraints due to parity conservation in the production process
are directly taken into account by working in the so-called
\emph{reflectivity basis}, where positive and negative values for the
spin projection $M$ are combined to yield amplitudes characterized by
$M \geq 0$ and an additional quantum number $\refl = \pm 1$, called
reflectivity.  This is achieved by replacing the $D$-function in the
$X^- \to \xi^0 + \pi^-$ two-body decay amplitude of
\cref{eq:decay_amplitude_X} by
\begin{multlineOrEq}
  \label{eq:d_func_refl_basis}
  \prescript{\refl\!}{}{D}^J_{M\, \lambda}(\phiGJ, \thetaGJ, 0) \newLineOrNot
  \ifMultiColumnLayout{\shoveleft}{}
  \equiv c(M)\, \bigg[ D^J_{(+M)\, \lambda}(\phiGJ, \thetaGJ, 0) \newLineOrNot
    - \refl\, P\, (-)^{J - M}\, D^J_{(-M)\, \lambda}(\phiGJ, \thetaGJ, 0) \bigg],
\end{multlineOrEq}
with $\refl = \pm 1$, $M \geq 0$, and the normalization factor
\begin{equation}
  c(M) =
  \begin{cases}
    1 / 2 & \text{for $M = 0$}, \\
    1 / \sqrt{2} & \text{otherwise}.
  \end{cases}
\end{equation}

The reflectivity is the eigenvalue of reflection through the $X$
production plane.  In the high-energy limit, \refl corresponds to the
naturality of the exchange in the scattering process, such that
$\refl = +1$ ($-1$) corresponds to natural spin-parity of the
exchanged Reggeon, \ie $\JP = 1^-$ or $2^+$ or $3^-$ \ldots (unnatural
spin-parity: $\JP = 0^-$ or $1^+$ or $2^-$ \ldots) transfer to the
beam particle.  Expressing the amplitudes in the reflectivity basis
brings the spin-density matrix into a block-diagonal form \wrt
\refl~\cite{Chung:1974fq}.  Hence states with different
reflectivities, \ie those produced by Regge-trajectories with
different naturalities, do not interfere and are thus summed up
incoherently [see \cref{eq:intensity_ansatz_couplings}].  In general,
the rank $N_r$ of the spin-density matrix may be different in the two
reflectivity sectors, \ie $N_r^\refl$.

Finally, we introduce the phase-space-normalized decay amplitudes
$\Psi_a^\refl(\tau; \mThreePi)$ as
\begin{equation}
  \label{eq:decay_amplitude_norm}
  \Psi_a^\refl(\tau; \mThreePi) \equiv \frac{\widebar{\Psi}\vphantom{\Psi}_a^\refl(\tau; \mThreePi)}
  {\sqrt{\int\! \dif{\varphi_3(\tau')} \Abs[1]{\widebar{\Psi}\vphantom{\Psi}_a^\refl(\tau'; \mThreePi)}^2}},
\end{equation}
where $\dif{\varphi_3(\tau')}$ is the differential three-body
phase-space element.  This normalizes the transition amplitudes via
\begin{multlineOrEq}
  \label{eq:spin_density}
  \mathcal{T}_a^{r \refl}(\mThreePi, \tpr) \newLineOrNot
  \equiv \widebar{\mathcal{T}}\vphantom{\mathcal{T}}_a^{r \refl}(\mThreePi, \tpr)\,
  \sqrt{\tint\! \dif{\varphi_3(\tau')} \Abs[1]{\widebar{\Psi}\vphantom{\Psi}_a^\refl(\tau'; \mThreePi)}^2}
\end{multlineOrEq}
with
\begin{equation}
  \varrho_{a b}^\refl(\mThreePi, \tpr)
  = \sum_{r = 1}^{N_r^\refl} \mathcal{T}_a^{r \refl}\, \mathcal{T}_b^{r \refl \text{*}},
\end{equation}
such that the partial-wave intensities, which are the diagonal
elements of the spin-density matrix in \cref{eq:spin_density}, are
given in terms of number of events that would be observed in a perfect
detector.

The goal of the partial-wave analysis is to extract the unknown
transition amplitudes $\mathcal{T}_a^{r \refl}(\mThreePi, \tpr)$ from
the data, because they contain information about the intermediate
$3\pi$ resonances.  Since the mass dependence of the transition
amplitudes is unknown, the event sample is divided into \mThreePi bins
much narrower than the width of typical hadronic resonances.  Within
each mass bin, the \mThreePi dependence is assumed to be negligible,
so that the $\mathcal{T}_a^{r \refl}$ only depend on \tpr.

Also the \tpr dependence of the transition amplitudes is \apriori
unknown.  In previous analyses it was often assumed that the \mThreePi
and \tpr dependences factorize and the \tpr dependence was modeled by
real functions $g_a^\refl(\tpr)$.  These functions were extracted from
the analyzed data set by integrating over wide \mThreePi ranges, often
only for groups of waves.  The COMPASS data, however, exhibit a
complicated and significant correlation of the \tpr and \mThreePi
dependences (see \cref{sec:tprim_dependence}), which renders this
approach inapplicable.  As it will be shown in
\cref{sec:results_pwa_massindep_major_waves}, this is mainly due to
different production processes (resonance production and nonresonant
processes, like \eg the Deck process~\cite{Deck:1964hm}), which
contribute with amplitudes that may have very different dependences on
\tpr.  Therefore, the partial-wave decomposition is performed for each
\mThreePi bin independently in different slices of \tpr (see
\cref{sec:pwa_massindep_model,tab:t-bins}).  Within a \tpr bin, the
transition amplitude is assumed to be independent of \tpr.  Taking out
the explicit assumptions about the \tpr dependences by virtue of our
large data set is an advantage compared to most previous analyses
(\eg~\cite{Alekseev:2009aa}).

For a given bin in \mThreePi and \tpr, the intensity has thus a
simpler form as it depends only on the five phase-space variables
$\tau$:
\begin{equation}
  \label{eq:intensity_bin}
  \mathcal{I}(\tau)
  = \sum_{\refl = \pm 1} \sum_{r = 1}^{N_r^\refl}
  \Abs[3]{\sum_a \mathcal{T}_a^{r \refl}\, \Psi_a^\refl(\tau)}^2 + \,\mathcal{I}_\text{flat},
\end{equation}
with the transition amplitudes appearing as constants.  Here, we
introduced an additional incoherently added wave that is isotropic in
$\tau$ and from now on is referred to as \emph{flat wave}.  The
purpose of this wave is to absorb intensity of events with three
uncorrelated pions in the final state, \eg non-exclusive background.
The flat wave is always part of the wave set, even if not mentioned
explicitly.

\subsection{Maximum-Likelihood Method}
\label{sec:maxlikelihood_method}

The transition amplitudes $\mathcal{T}_a^{r \refl}$ are determined for
each bin in \mThreePi and \tpr by fitting the model intensity
$\mathcal{I}(\tau)$ of \cref{eq:intensity_bin} to the measured $\tau$
distribution.  The fit is based on an extended likelihood function
constructed from the probabilities to observe the $N$ measured events
with phase-space coordinates $\tau_i$:
\begin{equation}
  \label{eq:likelihood_function_ansatz}
  \mathcal{L} =
  \underbrace{\frac{\widebar{N}^N}{N!\vphantom{\int}}\, e^{-\widebar{N}}}_{\substack{\text{Poisson} \\
      \text{probability}}}\,
  \prod_{i = 1}^{N} \underbrace{\frac{\mathcal{I}(\tau_i)}
  {\int\! \dif{\varphi_3(\tau)}\, \eta(\tau)\, \mathcal{I}(\tau)}}_\text{Probability for event $i$}.
\end{equation}
Here, $\eta(\tau)$ is the detection efficiency and
$\dif{\varphi_3(\tau)}$ the differential three-body phase-space
element.  The expected number of events $\widebar{N}$ in the detector
is given by the normalization integral
\begin{equation}
  \label{eq:expected_ev_nmb}
  \widebar{N} = \int\! \dif{\varphi_3(\tau)}\, \eta(\tau)\, \mathcal{I}(\tau).
\end{equation}
By this integral, the detection efficiency is taken into account in
the fit model, thereby avoiding the binning of the data, which would
be impractical given the high dimensionality of the intensity
distribution.

Inserting \cref{eq:expected_ev_nmb} into
\cref{eq:likelihood_function_ansatz} and dropping all constant terms
as well as taking the logarithm, the expression reads
\begin{multline}
  \label{eq:likelihood_function_amp}
  \ln \mathcal{L} = \sum_{i = 1}^{N} \ln \sBrk[3]{%
    \sum_{\refl = \pm1} \sum_{r = 1}^{N_r^\refl}
    \Abs[3]{\sum_a \mathcal{T}_a^{r \refl}\, \Psi_a^\refl(\tau_i)}^2 + \,\mathcal{I}_\text{flat}} \\
  - \sum_{\refl = \pm1} \sum_{r = 1}^{N_r^\refl}
  \sum_{a, b} \mathcal{T}_a^{r \refl}\, \mathcal{T}_b^{r \refl *}\,
  \underbrace{\int\! \dif{\varphi_3(\tau)}\, \eta(\tau)\,
    \Psi_a^\refl(\tau)\, \Psi_b^{\refl *}(\tau)}_{\displaystyle \equiv I_{a b}^\refl} \newLineOrNot
  - \mathcal{I}_\text{flat}\, \underbrace{\int\! \dif{\varphi_3(\tau)}\, \eta(\tau)}_{\displaystyle \equiv I_\text{flat}}.
\end{multline}
Here, the complex-valued integral matrix $I_{a b}^\refl$, which is
independent of the transition amplitudes, is calculated using Monte
Carlo methods.  The same is true for the real-valued integral
$I_\text{flat}$ for the isotropic flat wave.

In every individual $(\mThreePi, \tpr)$ bin, the transition amplitudes
$\mathcal{T}_a^{r \refl}$ are determined by maximizing the likelihood
function of \cref{eq:likelihood_function_amp}, which allows the
determination of the spin-density matrix elements
\begin{equation}
  \label{eq:spin_density_norm}
  \varrho_{a b}^\refl
  = \sum_{r = 1}^{N_r^\refl} \mathcal{T}_a^{r \refl}\, \mathcal{T}_b^{r \refl \text{*}}.
\end{equation}
Setting the detection efficiency $\eta(\tau) = 1$ in
\cref{eq:expected_ev_nmb} gives the expected acceptance-corrected
number of events:
\begin{multlineOrEq}
  \begin{aligned}
    N_\text{corr}
    \alignOrNot= \int\! \dif{\varphi_3(\tau)}\, \mathcal{I}(\tau) \newLineOrNot
    \alignOrNot= \sum_{\refl = \pm1} \sum_{a, b} \varrho_{a b}^\refl\,
    \int\! \dif{\varphi_3(\tau)}\, \Psi_a^\refl(\tau)\, \Psi_b^{\refl *}(\tau)
  \end{aligned} \newLineOrNot
  + \mathcal{I}_\text{flat}\, \int\! \dif{\varphi_3(\tau)}.
\end{multlineOrEq}
Using the fact that the decay amplitudes $\Psi_a^\refl(\tau)$ are
normalized via \cref{eq:decay_amplitude_norm} and that
$\varrho_{a b}^\refl$ is hermitian, the expression can be rewritten as
\begin{multlineOrEq}
  \label{eq:expected_ev_nmb_corr}
  N_\text{corr} = \sum_{\refl = \pm1} \bigg\{ \sum_a \underbrace{\varrho_{a a}^\refl \vphantom{\rBrk[2]{\int}}}_{\mathclap{\displaystyle \text{Intensities}}} \newLineOrNot
  \ifMultiColumnLayout{\mbox{}\hfill}{}
    + \sum_{a < b} \underbrace{2 \operatorname{Re}\!\!\sBrk[2]{\varrho_{a b}^\refl\,
  \int\! \dif{\varphi_3(\tau)}\, \Psi_a^\refl(\tau)\, \Psi_b^{\refl *}(\tau)}}_{\displaystyle \text{Overlaps}} \bigg\} \newLineOrNot
  + \mathcal{I}_\text{flat}\, \int\! \dif{\varphi_3(\tau)}.
\end{multlineOrEq}
From this equation, the interpretation of the spin-density matrix
elements becomes obvious.  The diagonal elements $\varrho_{a a}^\refl$
are the \emph{partial wave intensities}, \ie the expected
acceptance-corrected number of events in wave $a$.  The
\emph{overlaps} are the respective number of events that exhibit
interference between waves $a$ and $b$.  Limiting the summation in
\cref{eq:expected_ev_nmb_corr} to a subset of partial waves yields the
expected acceptance-corrected number of events in these waves
including all interferences.  Such sums will be denoted as
\emph{coherent sums} of partial waves in the following text.

The procedure described in this section is referred to as
\emph{mass-independent} fit.  It is worth stressing that fits in
different kinematic bins are independent of each other.  The fit model
of \cref{eq:intensity_bin} does not contain any assumptions about
possible $3\pi$ resonances.  They will be extracted in a second
analysis step from the \mThreePi dependence of the spin-density
matrix.  This so-called \emph{mass-dependent} fit will be described in
a forthcoming paper~\cite{COMPASS_3pi_mass_dep_fit}.
 %
%
%

\section{Partial-Wave Decomposition in Bins of \mThreePi and \tpr}
\label{sec:results_pwa_massindep}

In principle, the partial-wave expansion in \cref{eq:intensity_bin}
includes an infinite number of waves.  In practice, the expansion
series has to be truncated.  This means that one has to define a
\emph{wave set} that describes the data sufficiently well, without
introducing too many free parameters.

Since the intermediate state $X^-$ decays into a system of three
charged pions, the $G$ parity of $X^-$ is $-1$ and the isospin
$I = 1$, ignoring flavor-exotic states with $I > 1$.  The number of
possible partial waves is largely determined by the maximum allowed
spin~$J$ of $X^-$, the maximum allowed orbital angular momentum~$L$ in
the decay of the $X^-$ to the isobar and the bachelor $\pi^-$, and the
choice of the isobars.  Since there are no known resonances in the
flavor-exotic $\pi^-\pi^-$ channel, we choose to include only \twoPi
isobars.  We include \pipiS, \Prho, \PfZero[980], \PfTwo,
\PfZero[1500], and \PrhoThree as isobars into the fit model.  This
selection is based on the features observed in the \twoPi invariant
mass spectrum in \cref{fig:mass_spectrum_2pi,fig:dalitz_plots} and on
findings of previous
experiments~\cite{adams:1998ff,Chung:2002pu,Dzierba:2005jg,amelin:1995gu,Kachaev:2001jj}.

\subsection{Isobar Parametrization}
\label{sec:pwa_method_isobar_parametrization}

In this section, we present the parametrizations of the mass-dependent
amplitudes of the six isobars chosen above, which enter the analysis
via
\cref{eq:decay_amplitude_isobar,eq:decay_amplitude_X,eq:decay_amplitude_before_sym}
and are summarized in \Cref{tab:isobar_param}.

\begin{table}[tbp]
  \centering
  \renewcommand{\arraystretch}{1.5}
  \caption{Overview of the isobar parametrizations used in the
    partial-wave analysis.}
  \label{tab:isobar_param}
  \begin{tabularx}{\linewidth}{cXp{0.38\linewidth}}
    \toprule
    \textbf{Isobar} &
    \textbf{Formula} &
    \textbf{Parameters} \\
    \midrule

    \pipiS &
    $M$~solution from \refCite{au:1986vs} \newline (see \cref{fig:pipiS_M_param}) &
    see text and \namecref{tab:isobar_param}~1 in \refCite{au:1986vs} \\

    \Prho &
    \cref{eq:relBW} with \cref{eq:massDepWidthRho} &
    $m_0 = \SI{768.5}{\MeVcc}$ \newline
    $\Gamma_0 = \SI{150.7}{\MeVcc}$ \\

    \PfZero[980] &
    \cref{eq:f0980_flatte} (see \refCite{Ablikim:2004wn}) &
    $m_0 = \SI{965}{\MeVcc}$ \newline
    $g_\pipi = \SI{0.165}{\GeVccsq}$ \newline
    $g_\KKbar / g_\pipi = 4.21$ \\

    \PfTwo &
    \cref{eq:relBW} with \cref{eq:massDepWidth} &
    $m_0 = \SI{1275.4}{\MeVcc}$ \newline
    $\Gamma_0 = \SI{185.2}{\MeVcc}$ \\

    \PfZero[1500] &
    \cref{eq:relBW} with \cref{eq:constWidth} &
    $m_0 = \SI{1507}{\MeVcc}$ \newline
    $\Gamma_0 = \SI{109}{\MeVcc}$ \\

    \PrhoThree &
    \cref{eq:relBWrho3} &
    $m_0 = \SI{1690}{\MeVcc}$ \newline
    $\Gamma_0 = \SI{190}{\MeVcc}$ \\

    \bottomrule
  \end{tabularx}
\end{table}

In most cases, the \twoPi isobar resonances are described using a
relativistic Breit-Wigner amplitude~\cite{Breit:1936zzb}
\begin{equation}
  \label{eq:relBW}
  \Delta_\text{BW}(m; m_0, \Gamma_0) = \frac{m_0\, \Gamma_0}{m_0^2 - m^2 - i\, m_0\, \Gamma(m)},
\end{equation}
where $m_0$ and $\Gamma_0$ are mass and width of the resonance.  For a
single two-body decay channel, the mass-dependent width $\Gamma(m)$ is
given by
\begin{equation}
  \label{eq:massDepWidth}
  \Gamma(m) = \Gamma_{0}\, \frac{m_0}{m}\, \frac{q}{q_{0}}\, \frac{F_\ell^2(q)}{F_\ell^2(q_{0})}.
\end{equation}
By applying \cref{eq:massDepWidth}, we assume that the isobar decays
predominantly into two pions and neglect other decay modes.  Here,
$q(m)$ is the momentum of $\pi^-$ and $\pi^+$ in the rest frame of the
isobar with mass $m$.  At the nominal resonance mass, the breakup
momentum is given by $q_0 = q(m_0)$.  By $F_\ell(q)$ we denote the
Blatt-Weisskopf barrier factors~\cite{Blatt:1952}, which appear also
in \cref{eq:decay_amplitude_isobar_dyn} and take into account the
centrifugal-barrier effect caused by the orbital angular
momentum~$\ell$ in the isobar decay.\footnote{For the decay of the
  isobar into two spinless particles, $\ell$ is given by the spin
  $J_\xi$ of the isobar.}  We use the parametrization of von~Hippel
and Quigg~\cite{VonHippel:1972fg}, where
\begin{align}
  F_0^2(q) &= 1, \\
  F_1^2(q) &= \frac{2 z}{z + 1}, \\
  F_2^2(q) &= \frac{13 z^2}{z^2 + 3 z + 9}, \\
  F_3^2(q) &= \frac{277 z^3}{z^3 + 6 z^2 + 45 z + 225}, \\
  F_4^2(q) &= \frac{\num{12746} z^4}{z^4 + 10 z^3 + 135 z^2 + 1575 z + 11025},
\end{align}
\ifMultiColumnLayout{\begin{widetext}}{}
  \begin{align}
    F_5^2(q) &= \frac{\num{998881} z^5}{z^5 + 15 z^4 + 315 z^3 + \num{6300} z^2 + \num{99225} z + \num{893025}},~\text{and} \\
    F_6^2(q) &= \frac{\num{118394977} z^6}{z^6 + 21 z^5 + 630 z^4 + \num{18900} z^3 + \num{496125} z^2 + \num{9823275} z + \num{108056025}}.
  \end{align}
\ifMultiColumnLayout{\end{widetext}}{}
Here, $z \equiv (q / q_R)^2$ with the range parameter
$q_R = \SI{202.4}{\MeVc}$ that corresponds to an assumed strong interaction
range of \SI{1}{\femto\meter}.\footnote{Instead of the original
  normalization of the barrier factors such that $F_\ell(q) \to 1$ for
  $q \to \infty$, von~Hippel and Quigg modified the normalization in a
  way that $F_\ell(q) = 1$ for $z = 1$.}  For small breakup momenta
$q \approx 0$, the amplitude behaves like $F_\ell(q) \propto q^\ell$.

The description of the \Prho isobar is slightly
improved by modifying \cref{eq:massDepWidth} as shown in
\refsCite{Pisut:1968zza,Bowler:1987bj}:
\begin{equation}
  \label{eq:massDepWidthRho}
  \Gamma(m) = \Gamma_{0}\, \frac{q}{q_{0}}\, \frac{F_\ell^2(q)}{F_\ell^2(q_{0})}.
\end{equation}

For the \PrhoThree isobar, a slightly modified Breit-Wigner
amplitude is used:
\begin{equation}
  \label{eq:relBWrho3}
  \Delta_{\PrhoThree}(m; m_0, \Gamma_0) = \frac{\sqrt{m\, m_0}\, \Gamma_0}{m_0^2 - m^2 - i\, m_0\, \Gamma_0}.
\end{equation}
Since the \twoPi decay mode of the \PrhoThree is not
dominant, a constant total width is used.

The most difficult sector is that of the scalar isobars with
$\JPC = 0^{++}$, which consists of several overlapping \PfZero*
resonances.  In this analysis, we consider three independent isobar
amplitudes that have quite different properties.  A broad component
with slow phase motion, which we denote by \pipiS, dominates the mass
spectrum from low to intermediate two-pion masses.  This component
interferes with the narrow \PfZero[980].  In elastic \pipi scattering,
this interference is destructive, so that the \PfZero[980] appears as
a pronounced dip.  However, in \threePi decays, the \pipiSW subsystem
behaves differently.  As will be shown in
\cref{sec:results_free_pipi_s_wave}, the relative phase between the
two components depends on the quantum numbers of the $3\pi$
intermediate state and on its mass.  In order to give the model the
freedom to adjust the couplings of the various $3\pi$ states to the
$\pipiS\, \pi$ and $\PfZero[980]\, \pi$ decay modes separately, the
broad \pipiSW component and the \PfZero[980] are treated as
independent isobars.  Similarly, the \PfZero[1500] is included using a
Breit-Wigner amplitude [see \cref{eq:relBW}] with constant width
\begin{equation}
  \label{eq:constWidth}
  \Gamma(m) = \Gamma_{0}.
\end{equation}

The Breit-Wigner amplitude is not able to describe the \PfZero[980]
well as it peaks close to the \KKbar threshold.  Therefore, this
isobar is described by a Flatt\'e parametrization~\cite{Flatte:1976xv}
that takes into account the coupling to the \pipi and \KKbar decay
channels:
\begin{multlineOrEq}
  \label{eq:f0980_flatte}
  \Delta_\text{Flatt\'e}(m; m_0, g_\pipi, g_\KKbar) \newLineOrNot
  = \frac{1}{m_0^2 - m^2 - i \rBrk{\varphi_2^\pipi\, g_\pipi + \varphi_2^\KKbar\, g_\KKbar}}.
\end{multlineOrEq}
Here, $\varphi_2^i = 2 q_i / m$ is the two-body phase space for the
two decay channels $i = \pipi$, \KKbar with the respective breakup
momenta $q_i(m)$, which become complex-valued below threshold.  The
values for the couplings $g_\pipi$ and $g_\KKbar$ as well as that for
the mass $m_0$ are given in \cref{tab:isobar_param} as determined by
the BES experiment from a partial-wave analysis of $J/\psi$ decays
into $\phi\, \twoPi$ and $\phi\, K^- K^+$~\cite{Ablikim:2004wn}.

The parametrization of the broad \pipiSW component is the most
complicated one.  It is based on the parametrization of the \pipiSW
from \refCite{au:1986vs}, which was extracted from \pipi elastic
scattering data.  We modify the so-called $M$~solution (see
\namecref{tab:isobar_param}~1 in \refCite{au:1986vs}) as suggested by
the VES collaboration~\cite{kachaev:pipiS}.  In order to remove the
\PfZero[980] from the amplitude, the parameters $f_1^1$, $f_2^1$,
$f_1^3$, $c_{11}^4$, and $c_{22}^4$ as well as the diagonal elements
of the $M$ matrix in \namecref{eq:f0980_flatte}~(3.20) of
\refCite{au:1986vs} are set to zero.  \Cref{fig:pipiS_M_param} shows
the resulting \pipiS amplitude ($\mathcal{T}_{11}$ of
\namecref{eq:f0980_flatte}~(3.15) in \refCite{au:1986vs}).  It has a
broad intensity distribution that extends to two-pion masses of about
\SI{1.5}{\GeVcc} accompanied by a slow phase motion.

\begin{figure}[tbp]
  \centering
  \subfloat[][]{%
    \includegraphics[width=\twoPlotWidth]{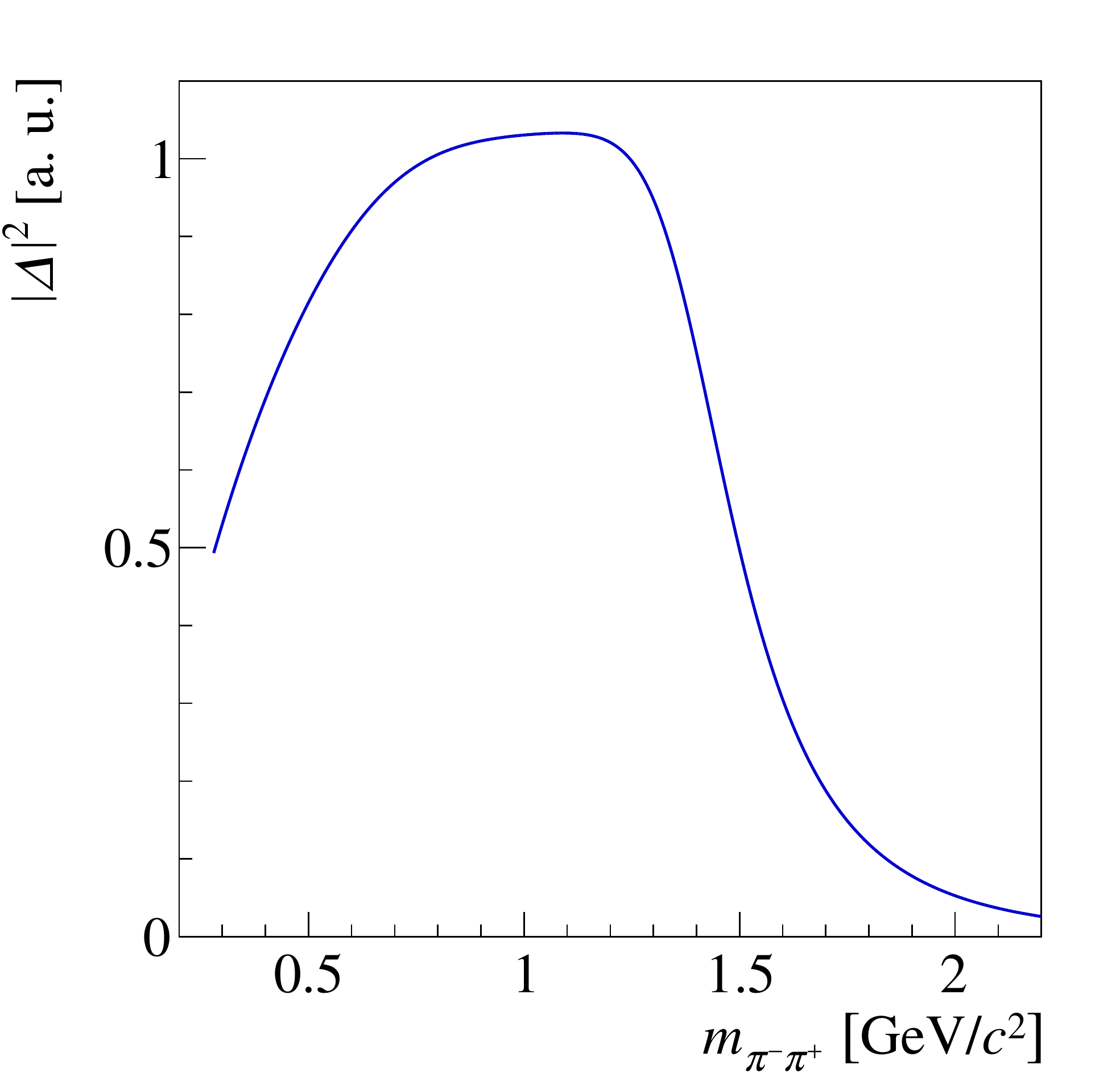}%
  }%
  \newLineOrHspace{\twoPlotSpacing}%
  \subfloat[][]{%
    \includegraphics[width=\twoPlotWidth]{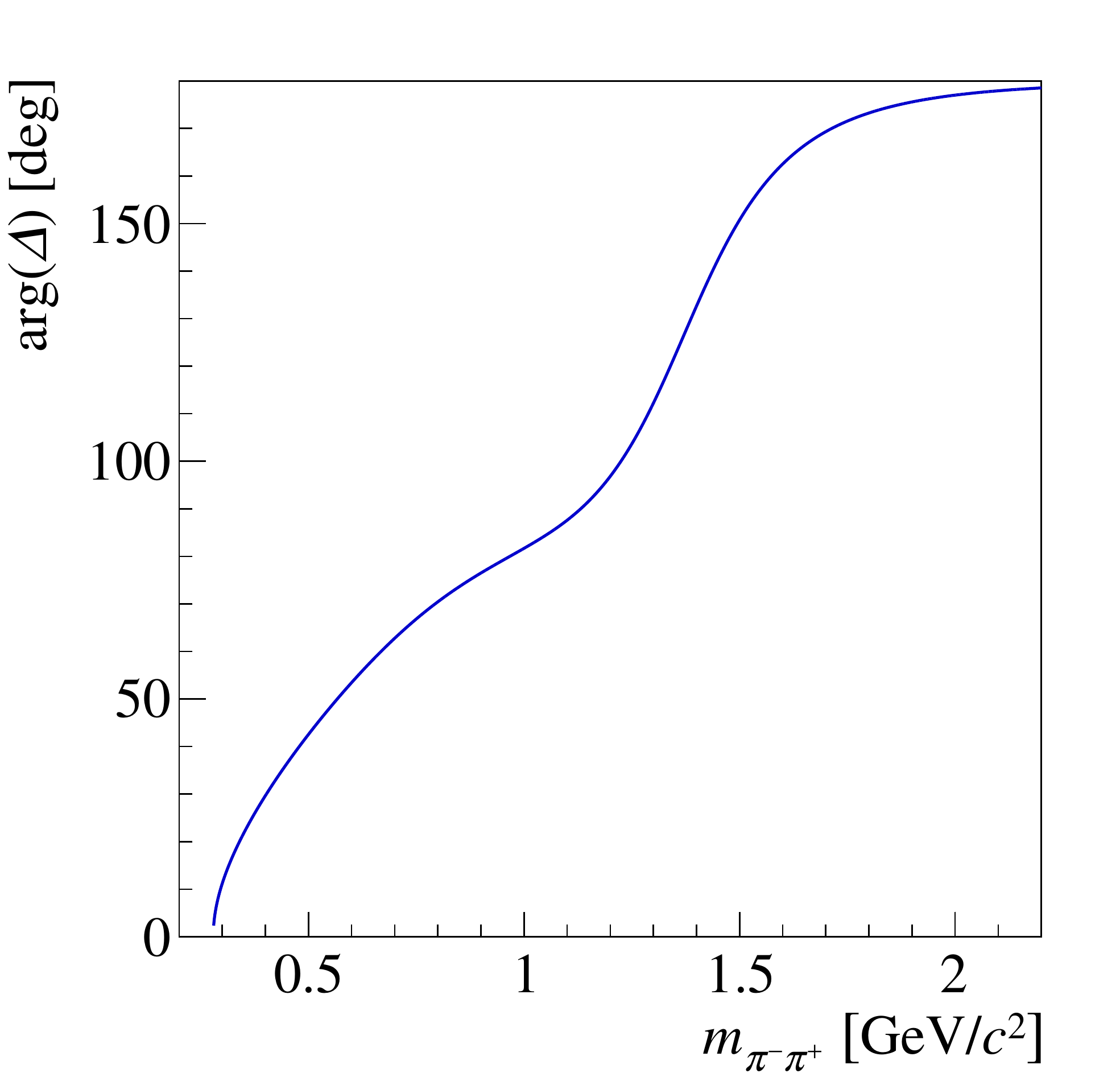}%
  }%
  \\
  \subfloat[][]{%
    \includegraphics[width=\twoPlotWidth]{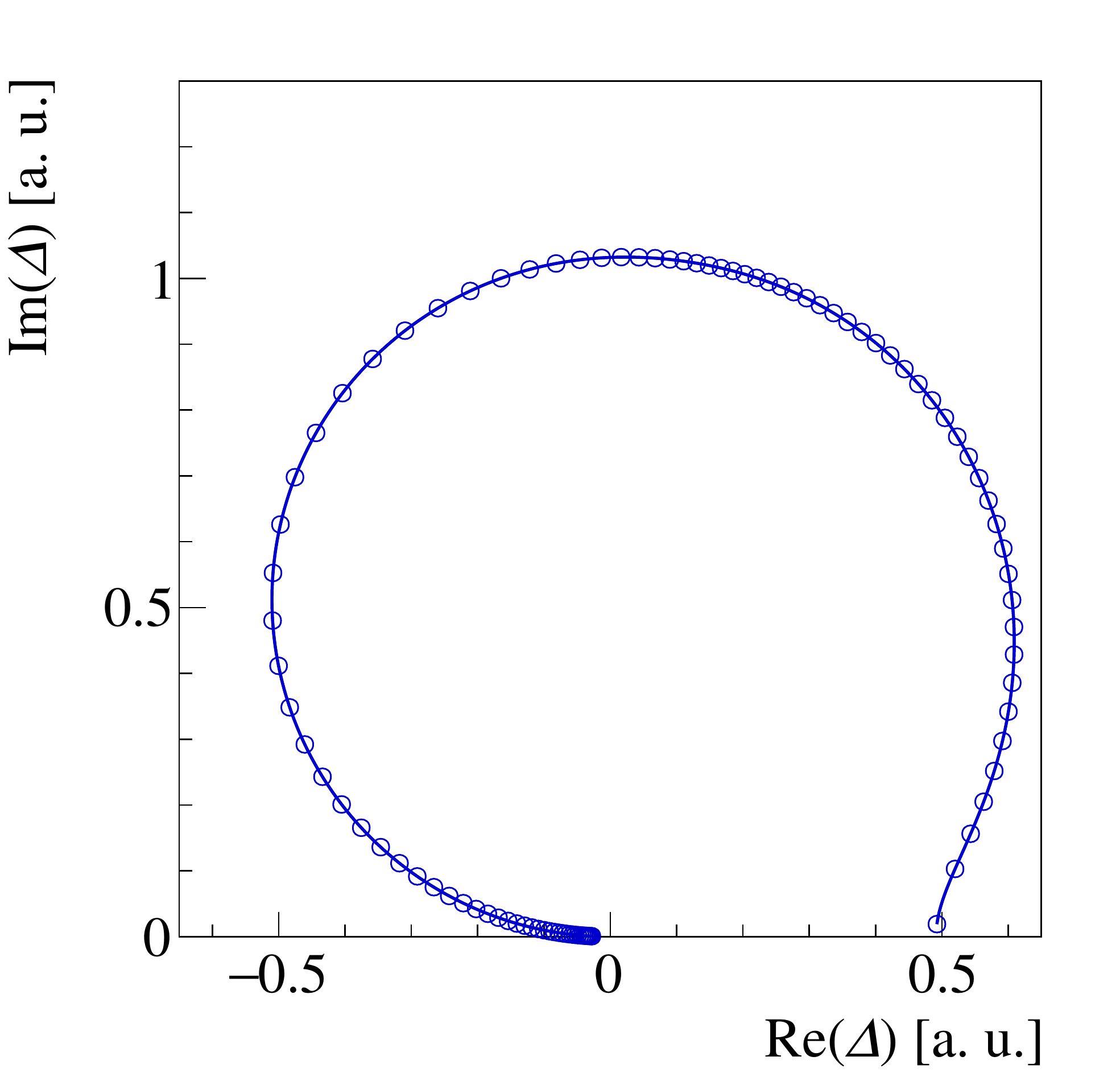}%
  }%
  \caption{Parametrization of the \pipiS isobar amplitude based on the
    $M$~solution described in \refCite{au:1986vs}.  Panel~(a) shows
    the intensity, (b)~the phase, and (c)~the corresponding \Argand.
    The open circles in the latter are evenly spaced in \mTwoPi in
    \SI{20}{\MeVcc} intervals.  Arbitrary units are denoted by
    \enquote{a.~u.}.}
  \label{fig:pipiS_M_param}
\end{figure}

\subsection{Fit Model}
\label{sec:pwa_massindep_model}
When using the isobar model, we have in principle to take into account
all observed \twoPi correlations.  In accordance with the \twoPi
invariant mass spectrum shown in \cref{fig:mass_spectrum_2pi} and with
analyses by previous experiments, we include \pipiS, \Prho,
\PfZero[980], \PfTwo, \PfZero[1500], and \PrhoThree as isobars into
the fit model.  Based on these six isobars, we have constructed a set
of partial waves that consists of 88~waves in total, \ie 80~waves with
reflectivity $\refl = +1$, seven waves with $\refl = -1$ and one
noninterfering flat wave representing three uncorrelated pions.  This
wave set has been derived from a larger set of 128~waves, which
includes mainly positive-reflectivity partial waves with spin
$J \leq 6$, orbital angular momentum $L \leq 6$, and spin projection
$M = \numlist{0;1;2}$.  Omission of structureless waves with relative
intensities below approximately \num{e-3} yields the 88~partial waves
that are used in this analysis and given in \cref{tab:waveset} in
\cref{sec:appendix_wave_set}.

The wave set includes waves with spin-exotic $\JPC = 1^{-+}$ and
$3^{-+}$.  These waves have intensities significantly different from
zero.  They contribute \SI{1.8}{\percent} and \SI{0.1}{\percent},
respectively, to the observed intensity.  Removing the three
$1^{-+}$ waves from the fit model\footnote{This reduces the number of free parameters in the PWA
fit by 6.} decreases the log-likelihood value,
summed over the 11~\tpr bins,
by more than \SI{4000}{units} in the $3\pi$ mass range from
\SIrange{1.1}{1.7}{\GeVcc}.  If instead the two $3^{-+}$ waves are
removed,\footnote{This reduces the number of free parameters in the PWA
fit by 4.} the log-likelihood value, summed over the 11~\tpr bins,
decreases by \SI{200}{units} in the $3\pi$ mass range from
\SIrange{1.1}{1.7}{\GeVcc}.  The spin-exotic waves will
not be discussed any further in this paper.

In the construction of the wave set, problems may arise when more than
one isobar with the same \JPC quantum numbers and a broad overlap of
their mass functions are used simultaneously, causing considerable
overlap between the corresponding decay amplitudes.  Such cases have
to be treated with great care as the fit tends to become unstable.  In
our fit model, this applies to the $0^{++}$ isobars.  Here, the broad
\pipiS, the narrow \PfZero[980], and the \PfZero[1500] do have
considerable overlap.  Because of the narrowness of the \PfZero[980],
the fit is able to separate it well from the broad \pipiS, as it is
demonstrated in \cref{sec:results_free_pipi_s_wave}.  In contrast, the
inclusion of several waves with $\PfZero[1500]\,\pi$ decay modes tends
to destabilize the fit.  Therefore, the 88-wave model includes only
one $\PfZero[1500]\,\pi$ wave.  We decided to include the
\wave{0}{-+}{0}{+}{\PfZero[1500]}{S} wave for
$\mThreePi > \SI{1.7}{\GeVcc}$ in order to study a potential signal
for the decay $\Ppi[1800] \to \PfZero[1500] + \pi$.  The
parametrizations used for the line shapes of the isobars are based on
prior knowledge and were described in
\cref{sec:pwa_method_isobar_parametrization}.  The effect of isobars
with uncertain line shapes may lead to spurious results and is
addressed by systematic studies discussed in
\cref{sec:pwa_massindep_systematic_studies,sec:appendix_syst_studies_massindep_isobar_param}.
We also apply an extended analysis method, which partly removes the
model bias due to the isobar parametrizations.  Results are presented
in \cref{sec:results_free_pipi_s_wave}.

The likelihood function to be maximized in the fit with the production
amplitudes as free parameters is built according to
\cref{eq:likelihood_function_amp}.  Using such a large wave set to fit
the three-pion system, we have to be concerned about stability of the
results, which in turn may be influenced by correlations and cross
talk of partial waves.  In order to reduce such effects, different
subsets of the 88~waves are used, which grow in size with increasing
three-pion mass.  High-spin waves and waves with heavy isobars are
typically omitted from the wave set in the region of low \mThreePi.
This has two reasons: first, the intensity of such waves is expected
to vanish at low \mThreePi, and secondly, they would artificially
contribute to ambiguities since the phase space at low masses appears
to be too small to find a unique solution.  A disadvantage of
introducing the mass thresholds for particular waves are possible
discontinuities induced in the mass dependence of other partial waves.
Therefore, such thresholds have to be placed as low as possible.  In
our analysis, thresholds were applied to~27 of the 88~partial waves.
The threshold values, which were carefully tuned in order to reduce
artificial structures, are listed in \cref{tab:waveset} in
\cref{sec:appendix_wave_set}.

The partial-wave analysis is performed independently in
100~equidistant \mThreePi bins with a width of \SI{20}{\MeVcc}, each
of which is subdivided into eleven non-equidistant \tpr bins (see
\cref{tab:t-bins}). The \tpr bins are chosen such that, except for the
two highest \tpr bins, each bin contains approximately the same number
of events.  Within each of these \num{1100} bins, the transition
amplitudes $\mathcal{T}_\alpha^{r \refl}(\mThreePi, \tpr)$ in
\cref{eq:intensity_bin} are assumed to be constant.
\Cref{fig:t_vs_m_binning} illustrates the correlation of \tpr and
\mThreePi, where the subdivision into bins of \tpr is indicated by
horizontal lines.

\begin{figure}[tbp]
  \centering
  \includegraphics[width=\twoPlotWidthTwoD]{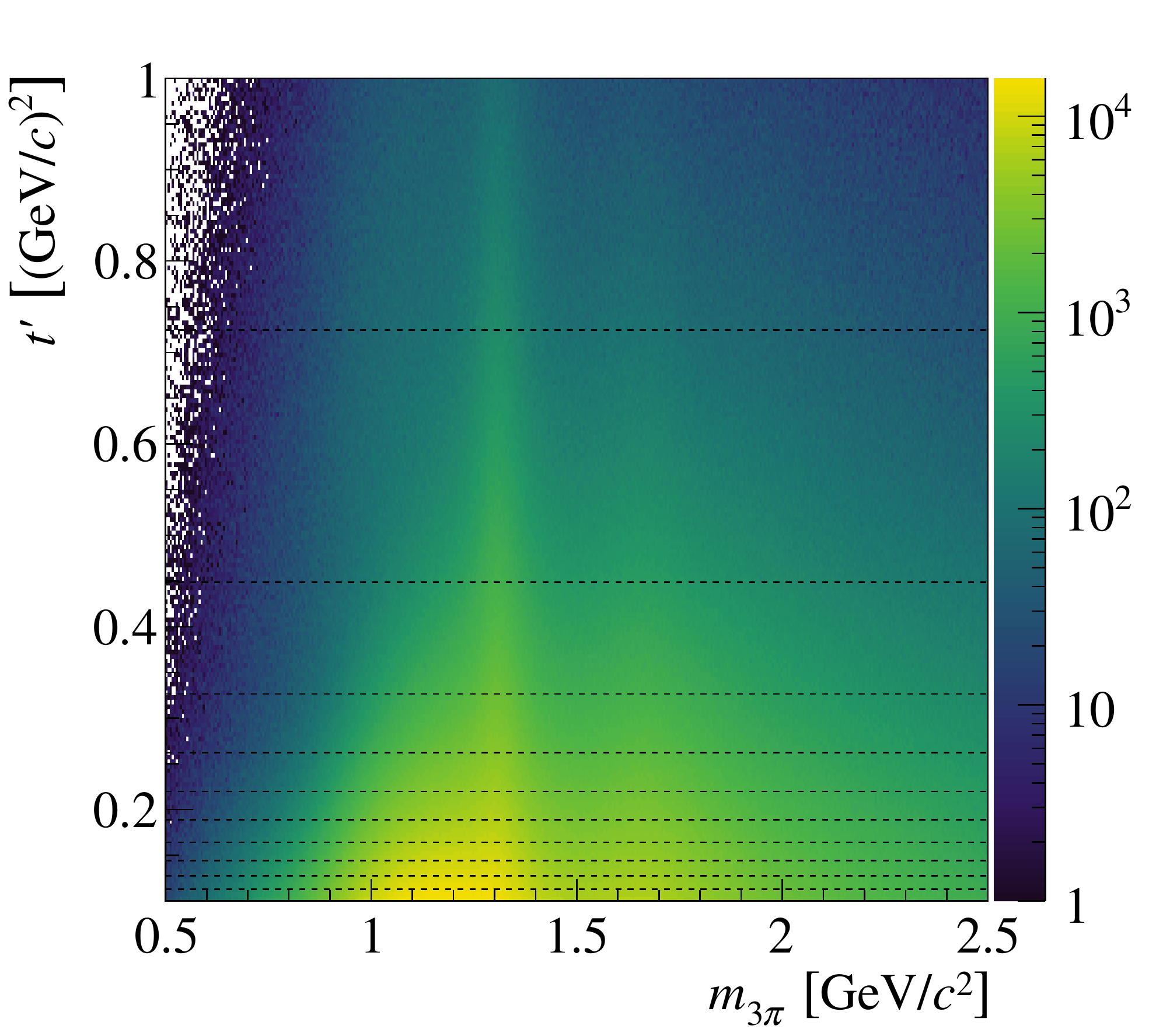}
  \caption{\colorPlot Correlation of the reduced four-momentum
    transfer squared \tpr and the invariant mass \mThreePi of the
    $3\pi$ system in the analyzed kinematic region.  The partial-wave
    analysis is performed independently in 100~equidistant \mThreePi
    bins with a width of \SI{20}{\MeVcc}, each of which is subdivided
    into eleven non-equidistant \tpr bins.  The latter are indicated
    by the dashed horizontal lines.  The numerical values for the \tpr
    bins are listed in \cref{tab:t-bins}.}
  \label{fig:t_vs_m_binning}
\end{figure}

\begin{table*}[htbp]
  \sisetup{%
    round-mode = places,
    round-precision = 3
  }
  \caption{Borders of the eleven non-equidistant \tpr bins, in which
    the partial-wave analysis is performed (see
    \cref{fig:t_vs_m_binning}).  The intervals are chosen such that
    each bin contains approximately \num[round-mode =
    places, round-precision = 1]{4.6e6} events.  Only
    the last range from \SIrange{0.448588}{1.000000}{\GeVcsq} is
    subdivided further into two bins.}
  \label{tab:t-bins}
  \renewcommand{\arraystretch}{1.2}
  \newcolumntype{Z}{%
    >{\Makebox[0pt][c]\bgroup}%
    c%
    <{\egroup}%
  }
  \setlength{\tabcolsep}{0pt}  %
  \ifMultiColumnLayout{%
    \begin{tabular}{l@{\extracolsep{12pt}}c@{\extracolsep{6pt}}Z*{10}{cZ}c}
      \toprule
      \textbf{Bin} && 1 && 2 && 3 && 4 && 5 && 6 && 7 && 8 && 9 && 10 && 11 & \\
      \midrule
      \textbf{\tpr [\si{\GeVcsq}]} &
      \num{0.100000} &&
      \num{0.112853} &&
      \num{0.127471} &&
      \num{0.144385} &&
      \num{0.164401} &&
      \num{0.188816} &&
      \num{0.219907} &&
      \num{0.262177} &&
      \num{0.326380} &&
      \num{0.448588} &&
      \num{0.724294} &&
      \num{1.000000} \\
      \bottomrule
    \end{tabular}%
  }{%
    \begin{tabular}{l@{\extracolsep{12pt}}c@{\extracolsep{6pt}}Z*{4}{cZ}cc}
      \toprule
      \textbf{Bin} && 1 && 2 && 3 && 4 && 5 && 6 \\
      \midrule
      \textbf{\tpr [\si{\GeVcsq}]} &
      \num{0.100000} &&
      \num{0.112853} &&
      \num{0.127471} &&
      \num{0.144385} &&
      \num{0.164401} &&
      \num{0.188816} \\
      \bottomrule
    \end{tabular}
    \\[3ex]
    \begin{tabular}{l@{\extracolsep{12pt}}c@{\extracolsep{6pt}}Z*{4}{cZ}c}
      \toprule
      \textbf{Bin} && 7 && 8 && 9 && 10 && 11 & \\
      \midrule
      \textbf{\tpr [\si{\GeVcsq}]} &
      \num{0.219907} &&
      \num{0.262177} &&
      \num{0.326380} &&
      \num{0.448588} &&
      \num{0.724294} &&
      \num{1.000000} \\
      \bottomrule
    \end{tabular}%
  }
\end{table*}

In the present analysis, we have limited ourselves to a rank-1
spin-density matrix for the waves in the positive-reflectivity sector,
which is dominated by Pomeron exchange (see also discussion in
\cref{sec:pwa_decomp}).  The smallest rank, $N_r^{(\refl = +1)} = 1$,
is sufficient mainly because the analysis is performed in narrow bins
of \tpr, where the relative phases of the partial waves do not vary
significantly.  As part of the systematic studies, a fit with rank~2
was investigated, which shows enhanced artificial structures as well
as increased instabilities (see
\cref{sec:pwa_massindep_systematic_studies,sec:appendix_syst_studies_massindep_rank}).
For the negative-reflectivity waves, which can be produced by the
exchange of various Reggeons [\eg \PbOne], we use
$N_r^{(\refl = -1)} = 2$.

\subsection{Selected Partial Waves with Spin Projections $M = \numlist{0;1;2}$}
\label{sec:results_pwa_massindep_major_waves}

In this section, we present the result of the fit in bins of \mThreePi
and \tpr for 18~selected waves with positive reflectivity as listed in
\cref{tab:selected-waves}.  The waves are selected partly in view of
the \emph{mass-dependent} fit that will be described in a forthcoming
paper, in which all resonance parameters determined by the fit will be
presented~\cite{COMPASS_3pi_mass_dep_fit}.  This selection includes
waves with spin projections $M = \numlist{0;1;2}$ that either have
large intensities or exhibit clear signals of well-established
resonances or even unexpected signals.  In addition, we have selected
large waves with \pipiS and \PfZero[980] isobars, which are related to
the detailed study of the \twoPi subsystem with $\JPC = 0^{++}$
presented in \cref{sec:results_free_pipi_s_wave}.  The amplitudes of
17~out of the 18~selected waves are found to be practically
insensitive to systematic effects arising from the remaining waves, in
particular to the truncation of the partial-wave expansion series in
\cref{eq:intensity_bin} (see
\cref{sec:pwa_massindep_systematic_studies,sec:appendix_syst_studies_massindep}).
The intensity distributions of the remaining 69~waves are shown in the
Supplemental
Material\ifMultiColumnLayout{~\cite{paper1_supplemental_material}}{ in
  \cref{sec:additional_waves}}.

The \emph{total intensity} of all partial waves is defined as the total number of
acceptance-corrected events as given by \cref{eq:expected_ev_nmb_corr}.
The \emph{relative intensity} of a given partial wave, as listed in
\cref{tab:selected-waves}, is defined as the ratio of its intensity
integral over the analyzed mass range to the integral of the
total intensity.  This value is in general different from
the contribution of a wave to the total intensity, owing to
interference effects between the waves.  Therefore, the relative intensities
of all 88~partial waves add up to \SI{105.3}{\percent} instead of
\SI{100}{\percent}.  However, self-interference due to Bose
symmetrization is included via \cref{eq:decay_amplitude}.

\begin{table*}[htbp]
  \sisetup{%
    round-mode = places,
    round-precision = 1
  }
  \centering
  \renewcommand{\arraystretch}{1.2}
  \caption{Waves selected for presentation in this paper out of the much
    larger pool of 88~waves used in the \emph{mass-independent} fit (see
    \cref{tab:waveset} in \cref{sec:appendix_wave_set}).  The partial
    waves with \pipiS and \PfZero isobars at the bottom of the table
    will be discussed in \cref{sec:results_free_pipi_s_wave}.  The
    intensities are evaluated as a sum over the 11~\tpr bins and are
    normalized to the total number of acceptance-corrected events.  They do not include
    interference effects between the waves.}
  \label{tab:selected-waves}
  \begin{tabular}{cccSl}
    \toprule
    \textbf{\JPC} &
    \textbf{\Mrefl} &
    \textbf{Isobaric decay} &
    \textbf{Relative intensity [\si{\percent}]} &
    \textbf{Shown in} \\
    \midrule

    \onePP  & $0^+$ & $\Prho\,\Ppi\,S$  & \num{32.65} & \cref{fig:a1_total_m0,fig:a1_t_bin_low,fig:a1_t_bin_high} \\
    \onePP  & $1^+$ & $\Prho\,\Ppi\,S$  & \num{4.10}  & \cref{fig:a1_total_m1} \\
    \onePP  & $0^+$ & $\PfTwo\,\Ppi\,P$ & \num{0.44}  & \cref{fig:a1P_t_bin_low,fig:a1P_t_bin_high,fig:a1_f2_pi_P_total_m0} \\[1.2ex]

    \twoPP  & $1^+$ & $\Prho\,\Ppi\,D$  & \num{7.66}  & \cref{fig:a2_total_m1_rho,fig:a2_t_bin_low,fig:a2_t_bin_high} \\
    \twoPP  & $2^+$ & $\Prho\,\Ppi\,D$  & \num{0.33}  & \cref{fig:a2_total_m2} \\
    \twoPP  & $1^+$ & $\PfTwo\,\Ppi\,P$ & \num{0.48}  & \cref{fig:a2_total_m1_f2,fig:a2P_t_bin_low,fig:a2P_t_bin_high} \\[1.2ex]

    \twoMP  & $0^+$ & $\Prho\,\Ppi\,F$  & \num{2.19}  & \cref{fig:pi2F_t_bin_low,fig:pi2F_t_bin_high,fig:pi2_total_rho} \\
    \twoMP  & $0^+$ & $\PfTwo\,\Ppi\,S$ & \num{6.72}  & \cref{fig:pi2_total_m0,fig:pi2_t_bin_low,fig:pi2_t_bin_high,fig:pi2_total_m0_2} \\
    \twoMP  & $1^+$ & $\PfTwo\,\Ppi\,S$ & \num{0.87}  & \cref{fig:pi2_total_m1,fig:pi2_total_m1_2} \\
    \twoMP  & $0^+$ & $\PfTwo\,\Ppi\,D$ & \num{0.91}  & \cref{fig:pi2D_total_m0} \\[1.2ex]

    \fourPP & $1^+$ & $\Prho\,\Ppi\,G$  & \num{0.76}  & \cref{fig:a4_total_m1_rho,fig:a4_t_bin_low,fig:a4_t_bin_high} \\
    \fourPP & $1^+$ & $\PfTwo\,\Ppi\,F$ & \num{0.18}  & \cref{fig:a4_total_m1_f2,fig:a4F_t_bin_low,fig:a4F_t_bin_high} \\[1.2ex]

    \midrule

    \zeroMP & $0^+$ & $\pipiS\,\Ppi\,S$  & \num{7.96} & \cref{fig:0mp_pipiS,fig:0mp_pipiS_t_bin_low,fig:0mp_pipiS_t_bin_high} \\
    \zeroMP & $0^+$ & $\PfZero\,\Ppi\,S$ & \num{2.44} & \cref{fig:0mp_f0980} \\[1.2ex]

    \onePP  & $0^+$ & $\pipiS\,\Ppi\,P$  & \num{4.07} & \cref{fig:1pp_pipiS} \\
    \onePP  & $0^+$ & $\PfZero\,\Ppi\,P$ & \num{0.25} & \cref{fig:1pp_f0980} \\[1.2ex]

    \twoMP  & $0^+$ & $\pipiS\,\Ppi\,D$  & \num{2.96} & \cref{fig:2mp_pipiS} \\
    \twoMP  & $0^+$ & $\PfZero\,\Ppi\,D$ & \num{0.55} & \cref{fig:2mp_f0980} \\

    \bottomrule

    & & \hfill \textbf{Intensity Sum} & 75.8 & \\  %

  \end{tabular}
\end{table*}

If not indicated otherwise, the wave intensities shown in the figures
below are the sum of the intensities over the individual \tpr bins.
They will be referred to as \emph{\tpr-summed} intensities in the text
that follows.  The percent numbers given in the mass spectra are the
relative intensities of the particular partial wave shown.  In
addition, we show for some waves the intensity distribution in
individual \tpr bins.  While mass and width of resonances do not
depend on the production kinematics, coherent nonresonant
contributions may vary in shape and phase with \tpr.  This may lead to
significant \tpr-dependent shifts of mass peaks.  Examples of such
effects are discussed below.

As shown in \cref{fig:neg_refl_wave_sum}, waves with negative
reflectivity, which correspond to unnatural-parity exchange processes,
contribute only \SI{2.2}{\percent} to the total intensity.  The
dominance of natural-parity exchange processes is expected at COMPASS
energies because the Pomeron contribution is considered to be
dominant.  Therefore, we are only taking into account
positive-reflectivity partial waves in the following.

\begin{figure}[tbp]
  \centering
  \subfloat[][]{%
    \includegraphics[width=\twoPlotWidth]{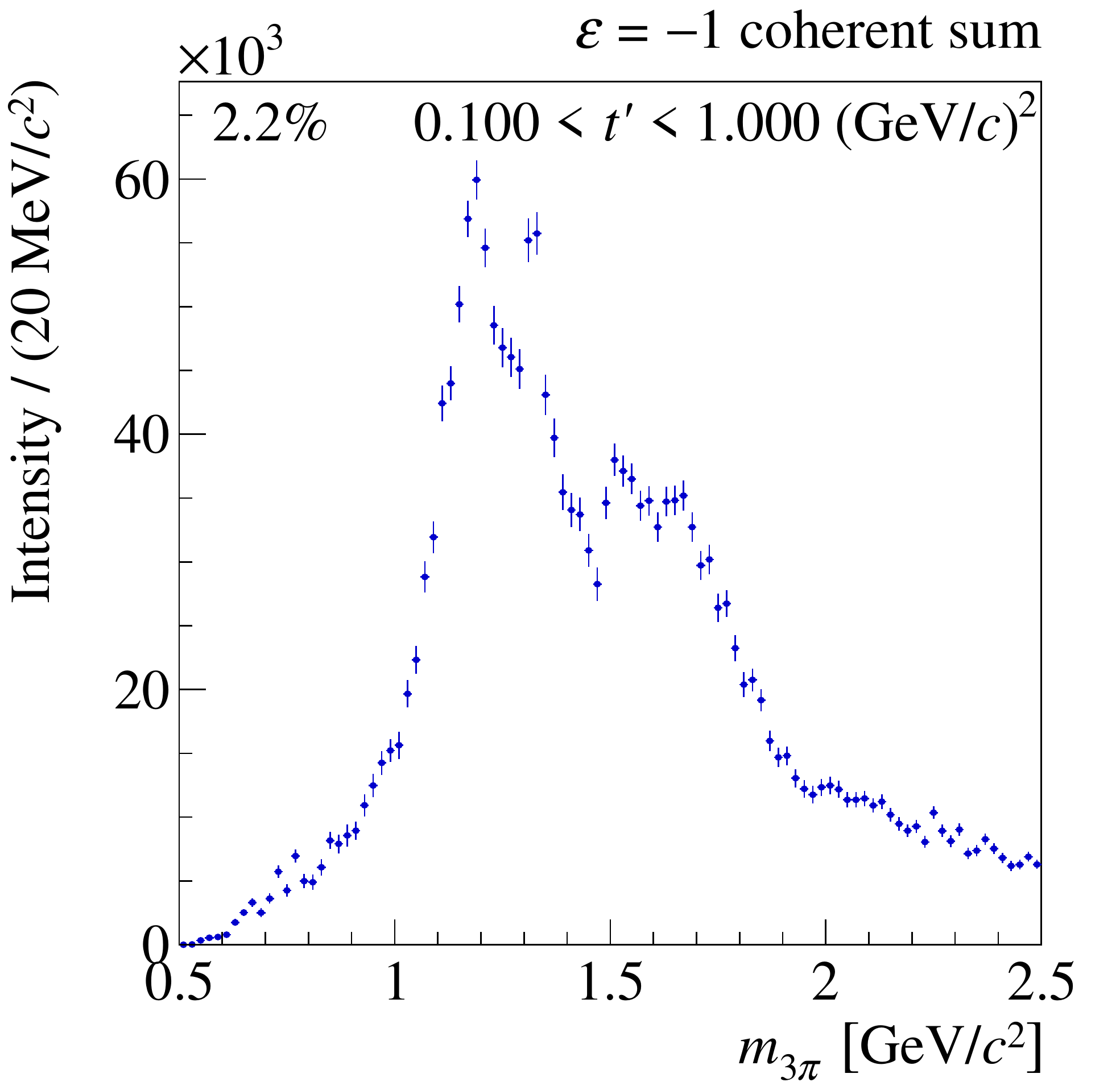}%
  }%
  \newLineOrHspace{\twoPlotSpacing}%
  \subfloat[][]{%
    \includegraphics[width=\twoPlotWidth]{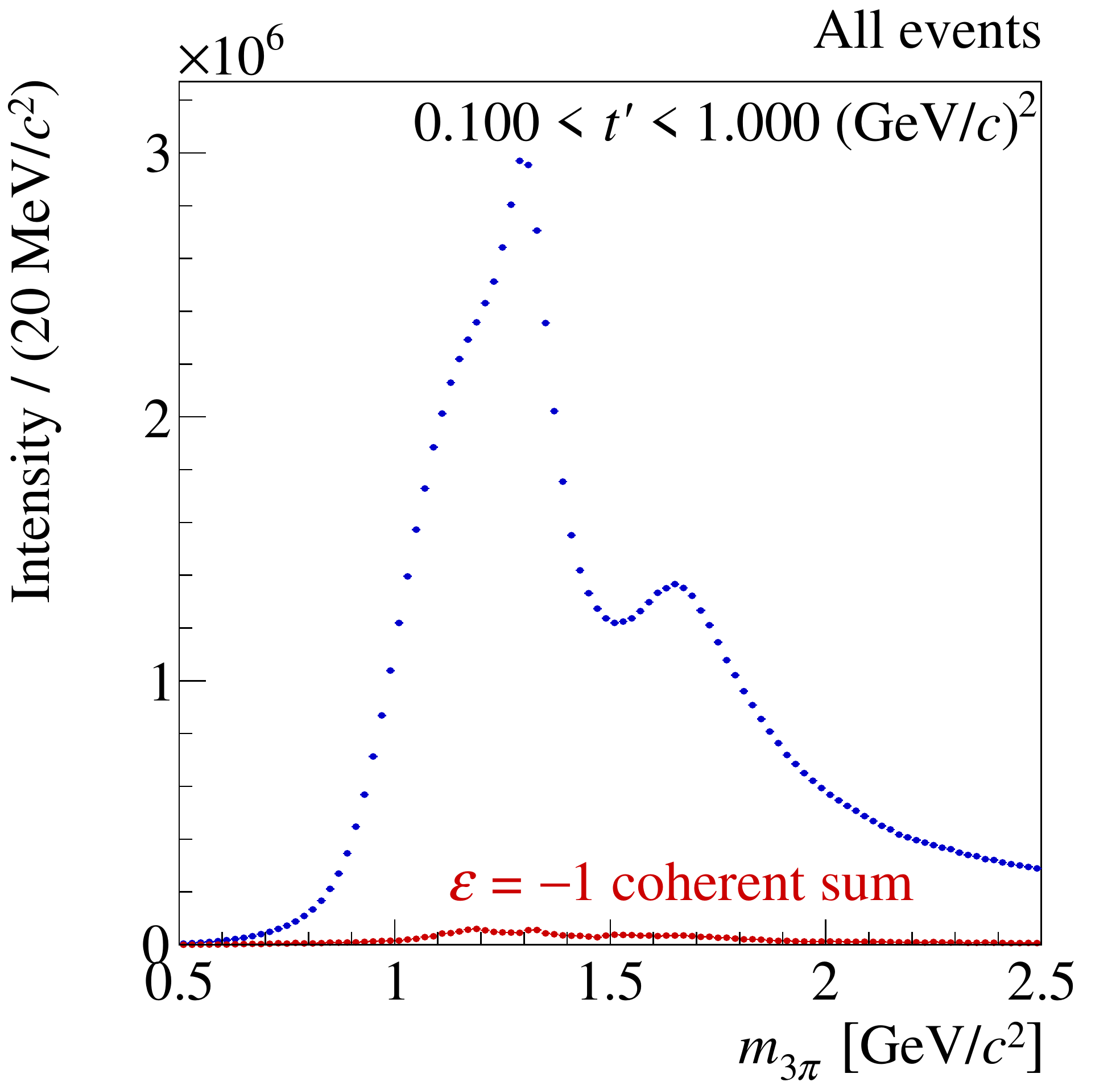}%
  }%
  \caption{The \tpr-summed intensity of the coherent sum of all
    negative-reflectivity waves~(a) and, for comparison, together with
    the total intensity of all partial waves~(b).}
  \label{fig:neg_refl_wave_sum}
\end{figure}

The incoherent isotropic flat wave turns out to contribute about
\SI{3.1}{\percent} to the total observed intensity (see
\cref{fig:flat_wave}).  This magnitude is roughly consistent with the
background level that one expects from extrapolating the non-exclusive
background component visible in \cref{fig:exclusivity_Esum_zoom} into
the signal region.

\begin{figure}[tbp]
  \centering
  \subfloat[][]{%
    \includegraphics[width=\twoPlotWidth]{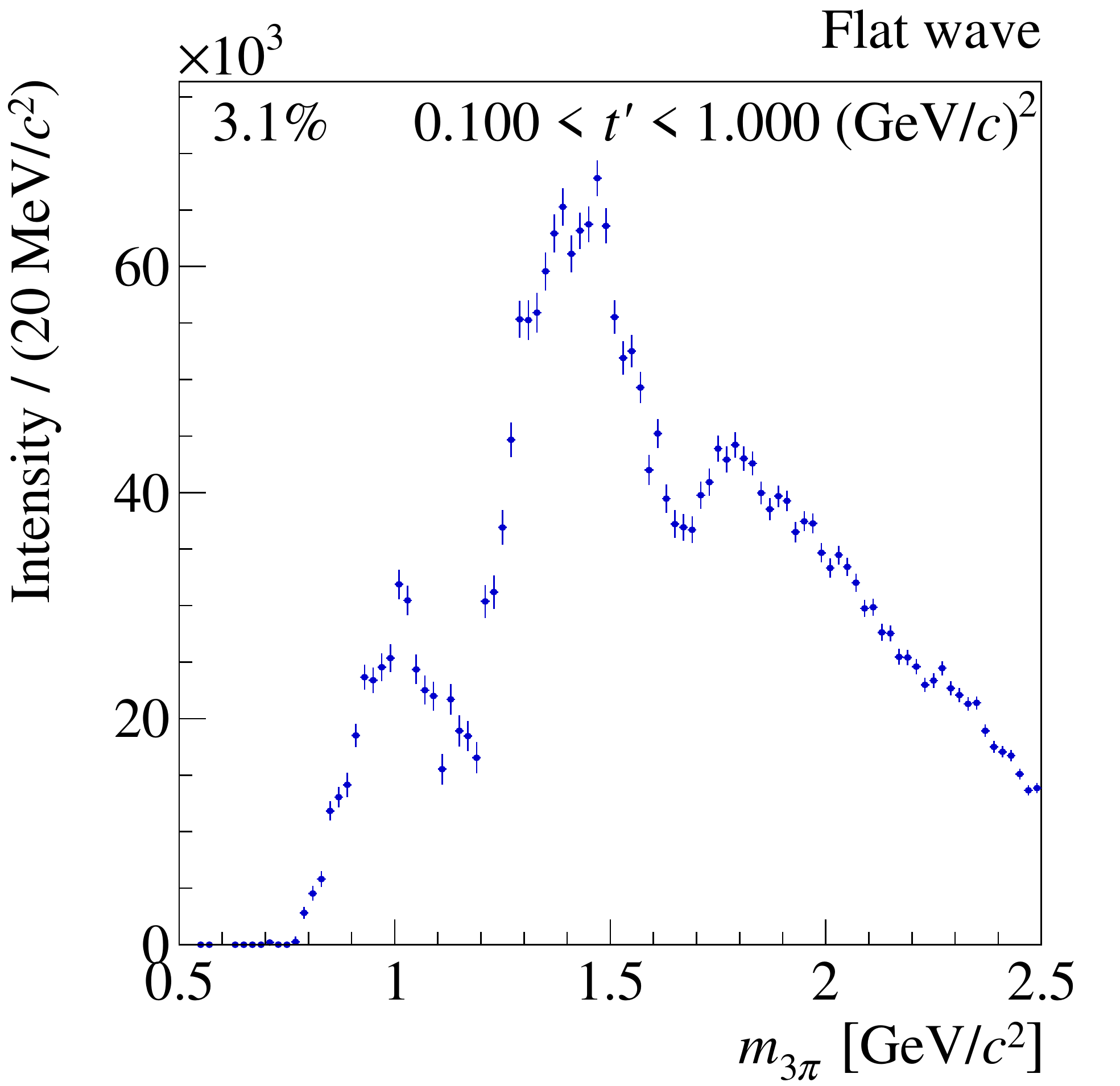}%
  }%
  \newLineOrHspace{\twoPlotSpacing}%
  \subfloat[][]{%
    \includegraphics[width=\twoPlotWidth]{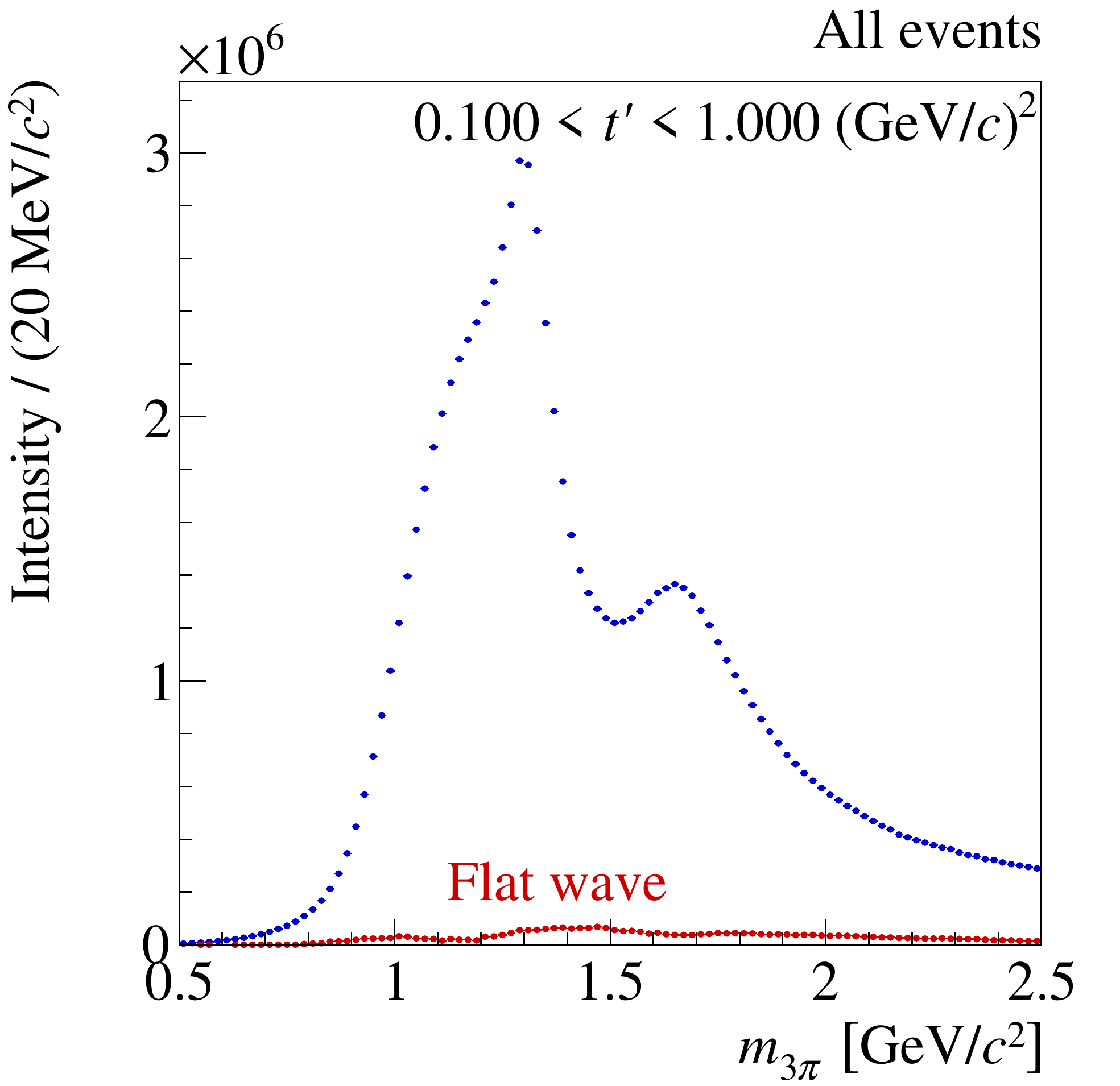}%
  }%
  \caption{The \tpr-summed intensity of the flat wave~(a) and, for
    comparison, together with the total intensity of all partial
    waves~(b).}
  \label{fig:flat_wave}
\end{figure}

\Cref{fig:major_waves_m0} shows the \tpr-summed intensities of two
major waves with spin projection $M = 0$, \ie the
\wave{1}{++}{0}{+}{\Prho}{S} and \wave{2}{-+}{0}{+}{\PfTwo}{S} waves.
Both exhibit clear peaks corresponding to the \PaOne and the \PpiTwo
resonances.

\begin{figure}[tbp]
  \centering
  \subfloat[][]{%
    \label{fig:a1_total_m0}%
    \includegraphics[width=\twoPlotWidth]{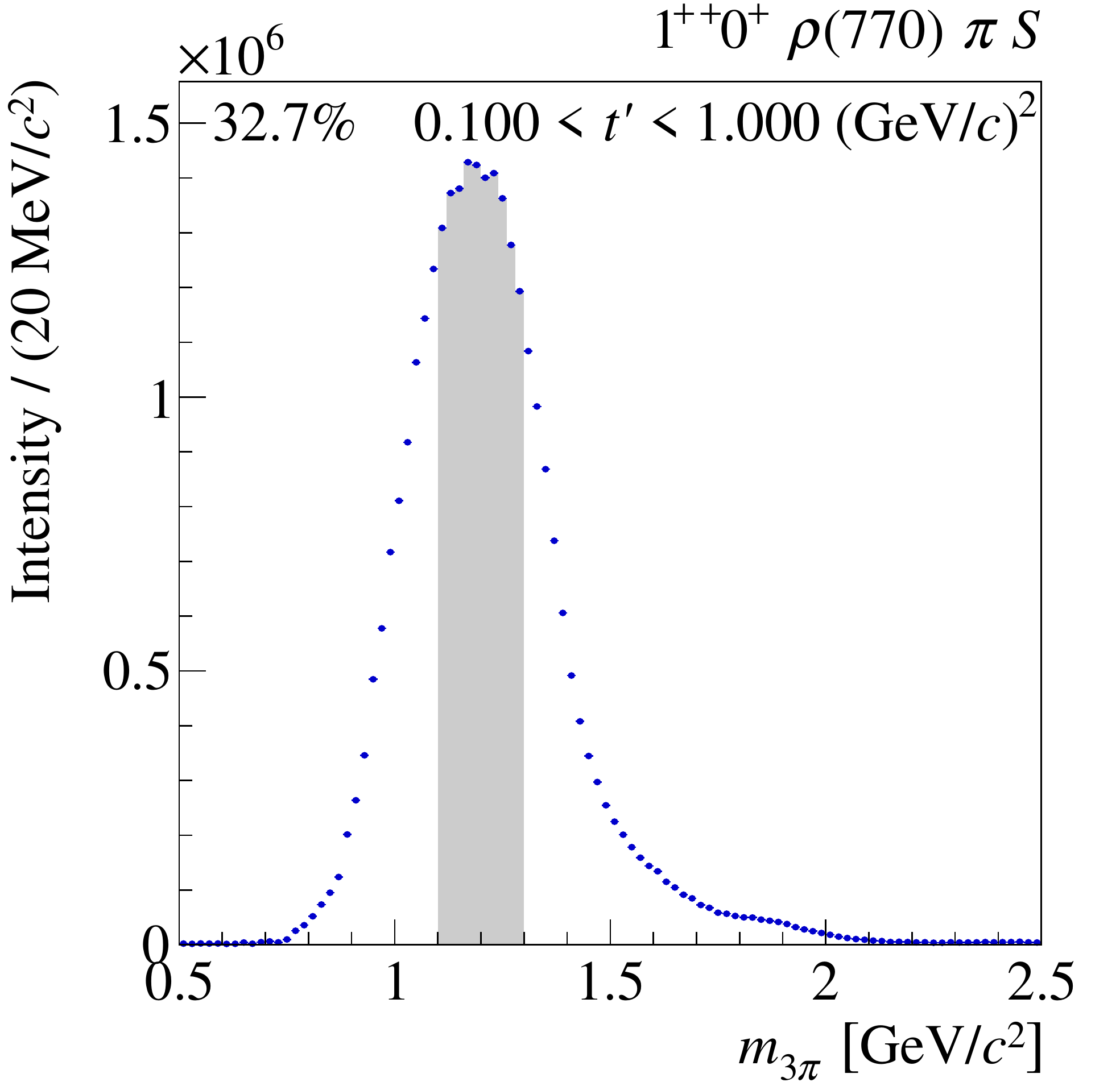}%
  }%
  \newLineOrHspace{\twoPlotSpacing}%
  \subfloat[][]{%
    \label{fig:pi2_total_m0}%
    \includegraphics[width=\twoPlotWidth]{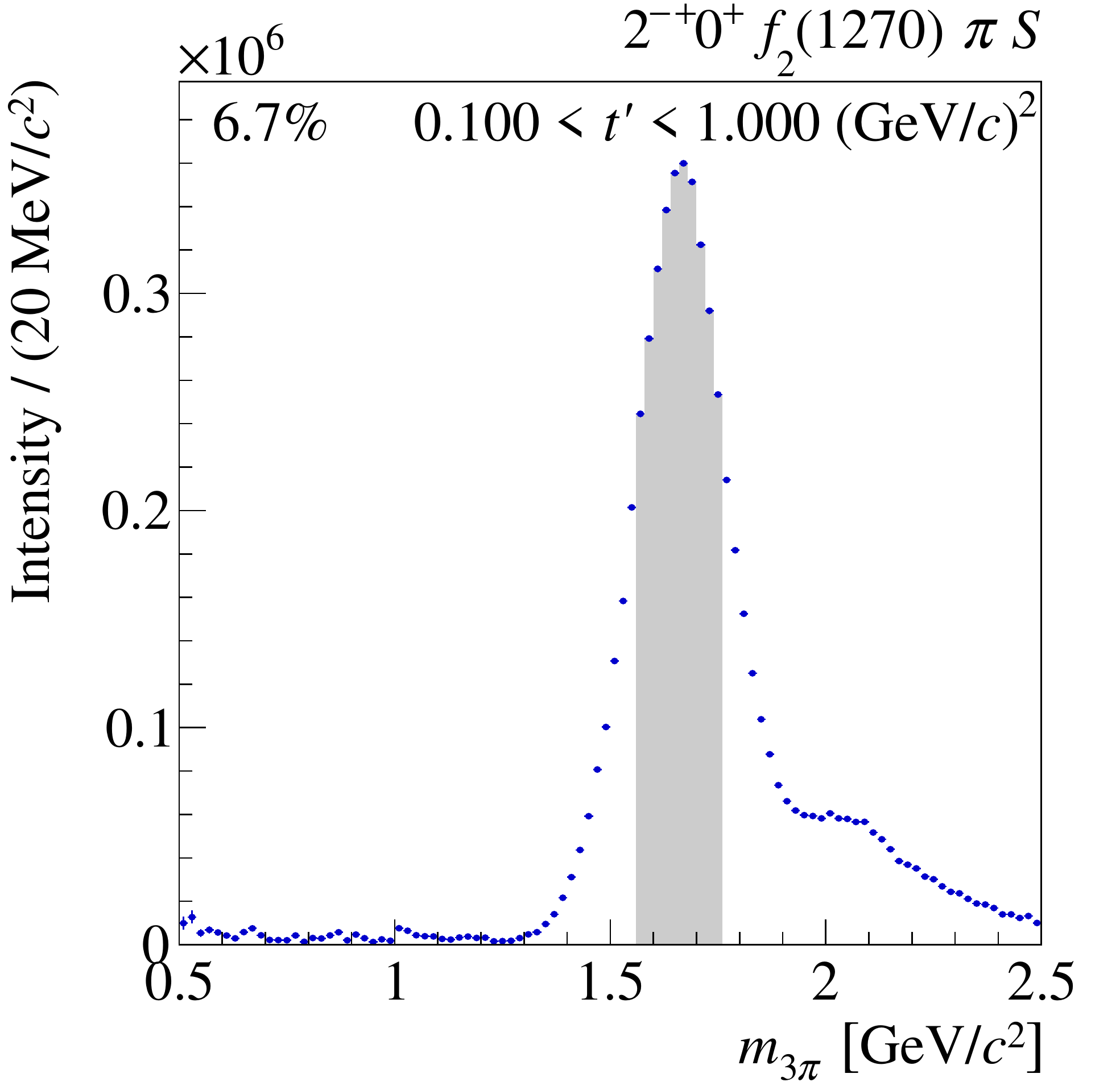}%
  }%
  \caption{The \tpr-summed intensities of major waves with spin
    projection $M = 0$ showing in
      \protect\subref{fig:a1_total_m0} the \PaOne
    and in \protect\subref{fig:pi2_total_m0} the
    \PpiTwo.  The shaded regions indicate the mass intervals that
    are integrated over to generate the \tpr spectra (see
    \cref{fig:a1_total_m0_t}).}
  \label{fig:major_waves_m0}
\end{figure}

Selecting spin projection $M = 1$, we have access to the
\wave{1}{++}{1}{+}{\Prho}{S}, \wave{2}{++}{1}{+}{\Prho}{D},
\wave{2}{++}{1}{+}{\PfTwo}{P}, \wave{2}{-+}{1}{+}{\PfTwo}{S}, as well
as to the \wave{4}{++}{1}{+}{\Prho}{G} and
\wave{4}{++}{1}{+}{\PfTwo}{F} waves, as shown in
\cref{fig:major_waves_m1}.  The intensity maxima can be identified
with the well-known resonances \PaOne, \PaTwo, \PpiTwo, and \PaFour.
Comparing \cref{fig:a1_total_m1,fig:pi2_total_m1} to
\cref{fig:a1_total_m0,fig:pi2_total_m0}, respectively, a suppression
of intensities for waves with $M = 1$ by about one order of magnitude
as compared to $M = 0$ can clearly be observed.

\begin{figure*}[htbp]
  \centering
  \subfloat[][]{%
    \label{fig:a1_total_m1}%
    \includegraphics[width=\twoPlotWidth]{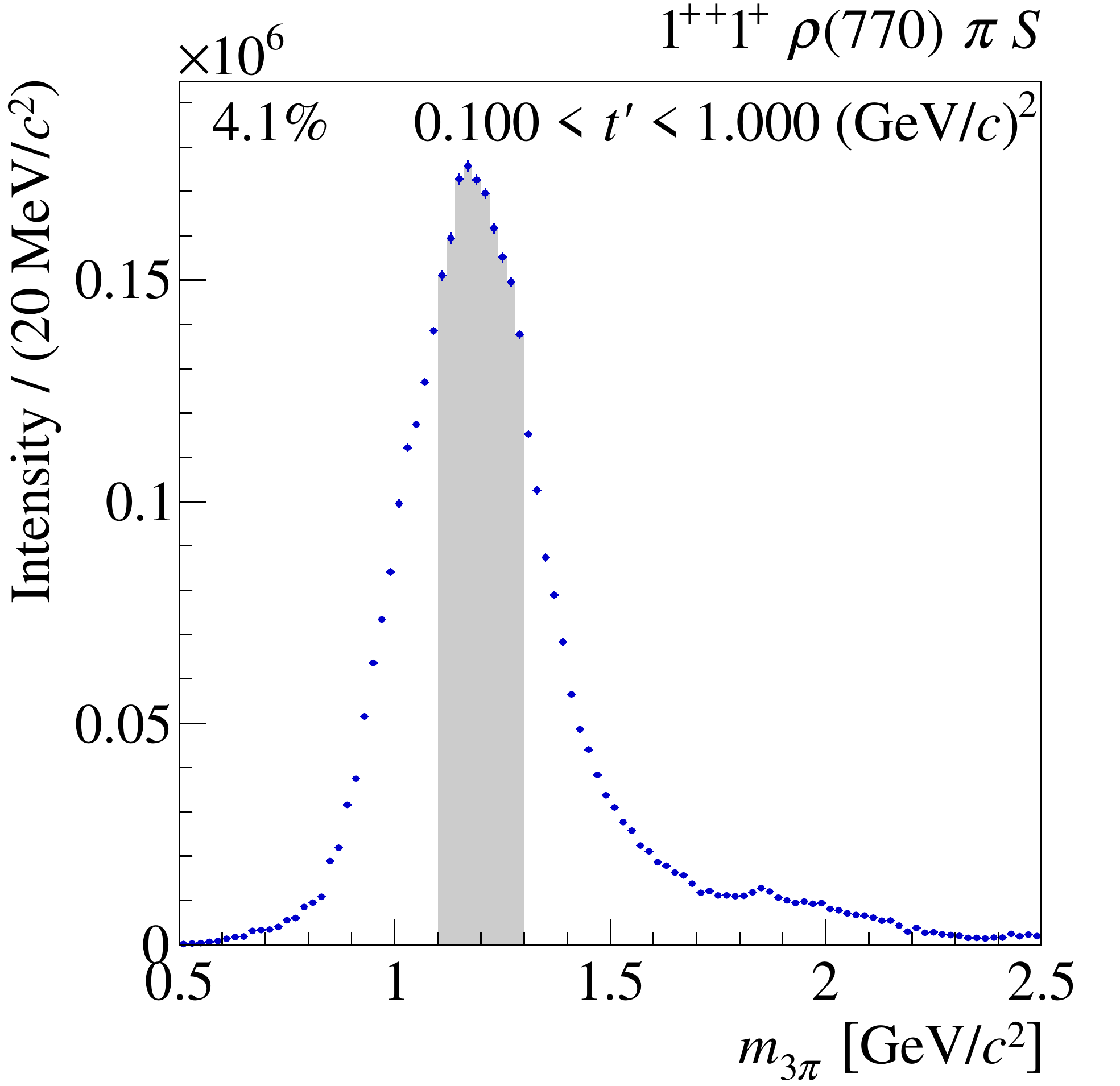}%
  }%
  \hspace*{\twoPlotSpacing}
  \subfloat[][]{%
    \label{fig:a2_total_m1_rho}%
    \includegraphics[width=\twoPlotWidth]{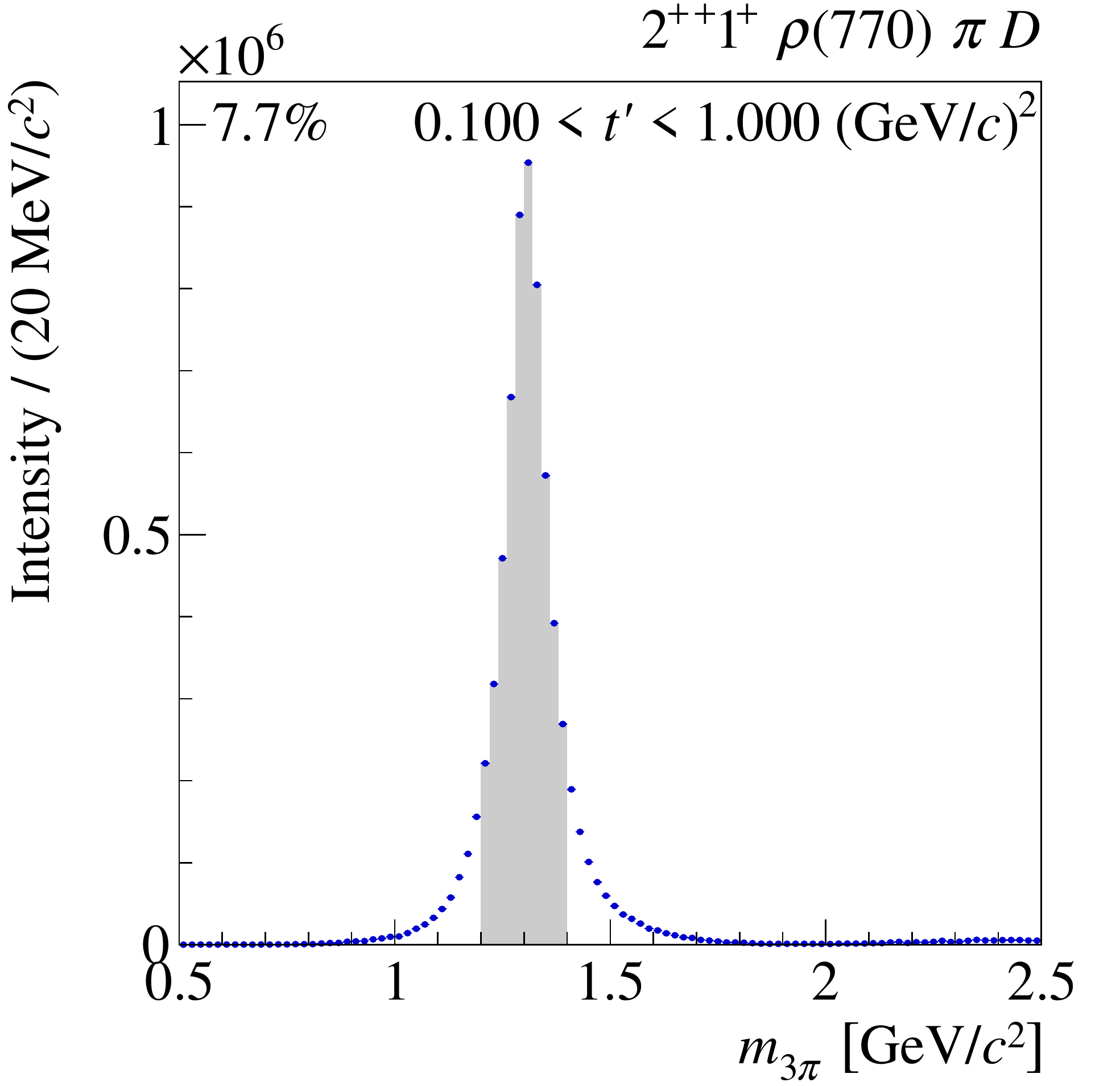}%
  }%
  \\
  \subfloat[][]{%
    \label{fig:a2_total_m1_f2}%
    \includegraphics[width=\twoPlotWidth]{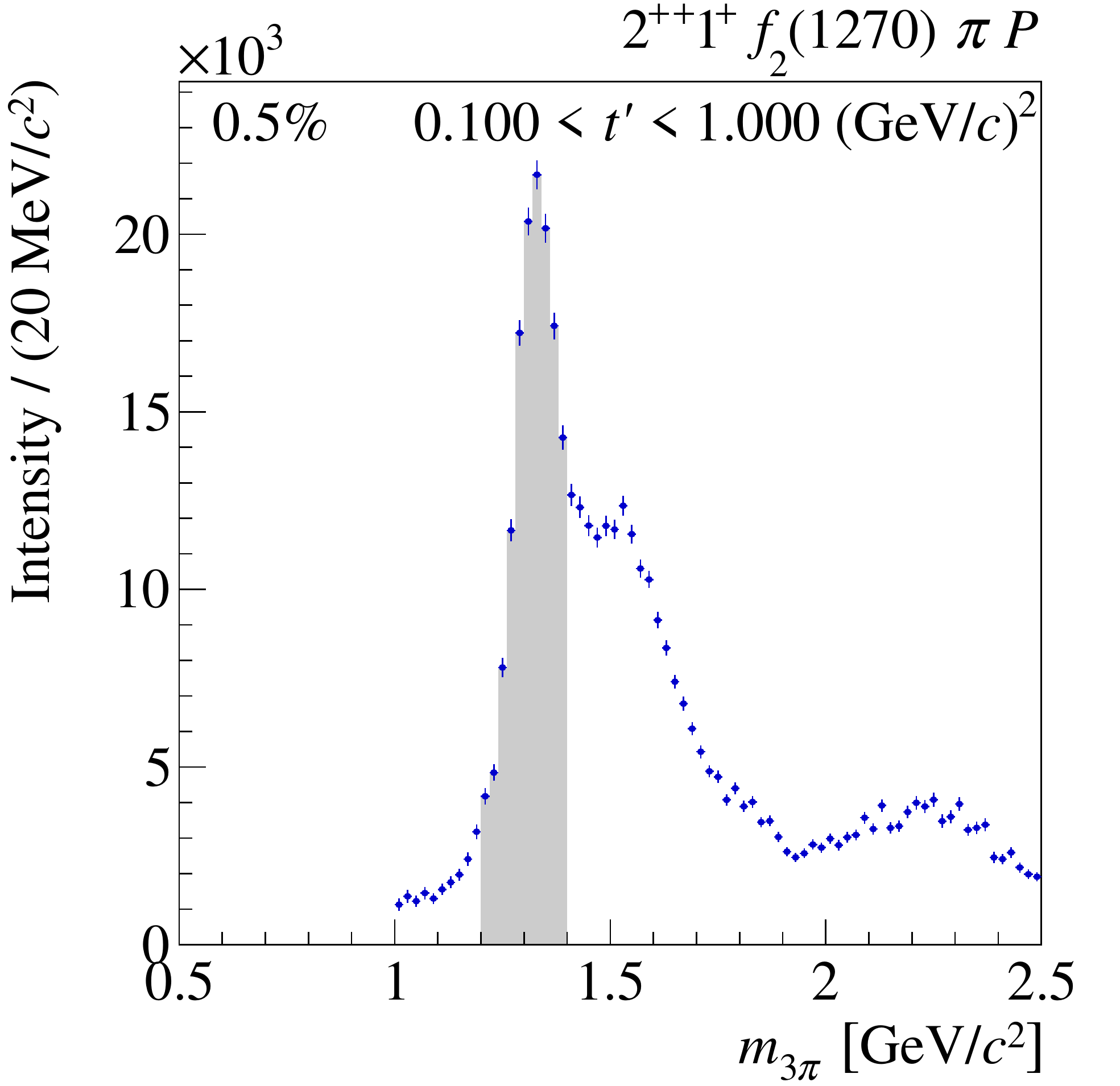}%
  }%
  \hspace*{\twoPlotSpacing}
  \subfloat[][]{%
    \label{fig:pi2_total_m1}%
    \includegraphics[width=\twoPlotWidth]{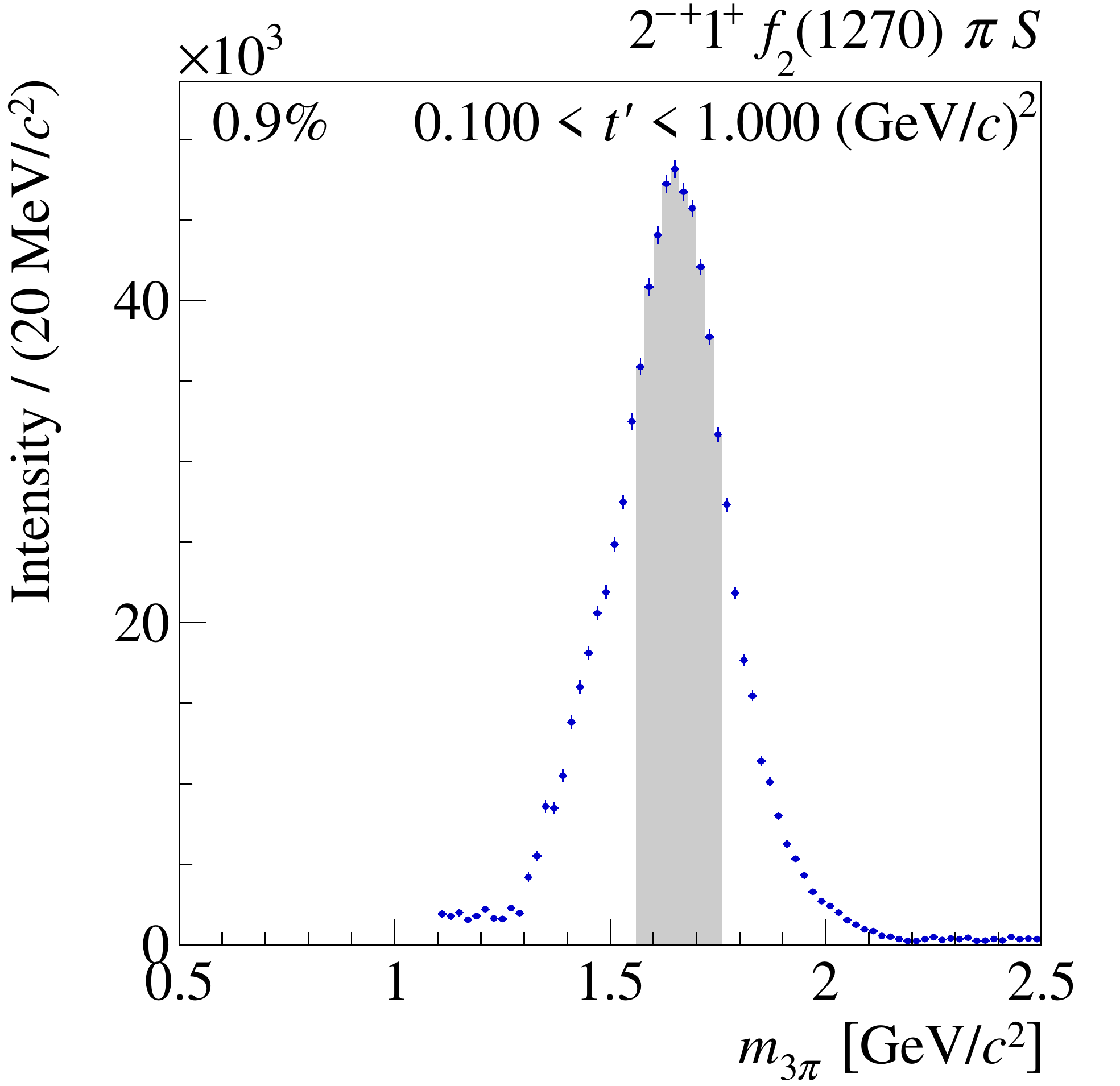}%
  }%
  \\
  \subfloat[][]{%
    \label{fig:a4_total_m1_rho}%
    \includegraphics[width=\twoPlotWidth]{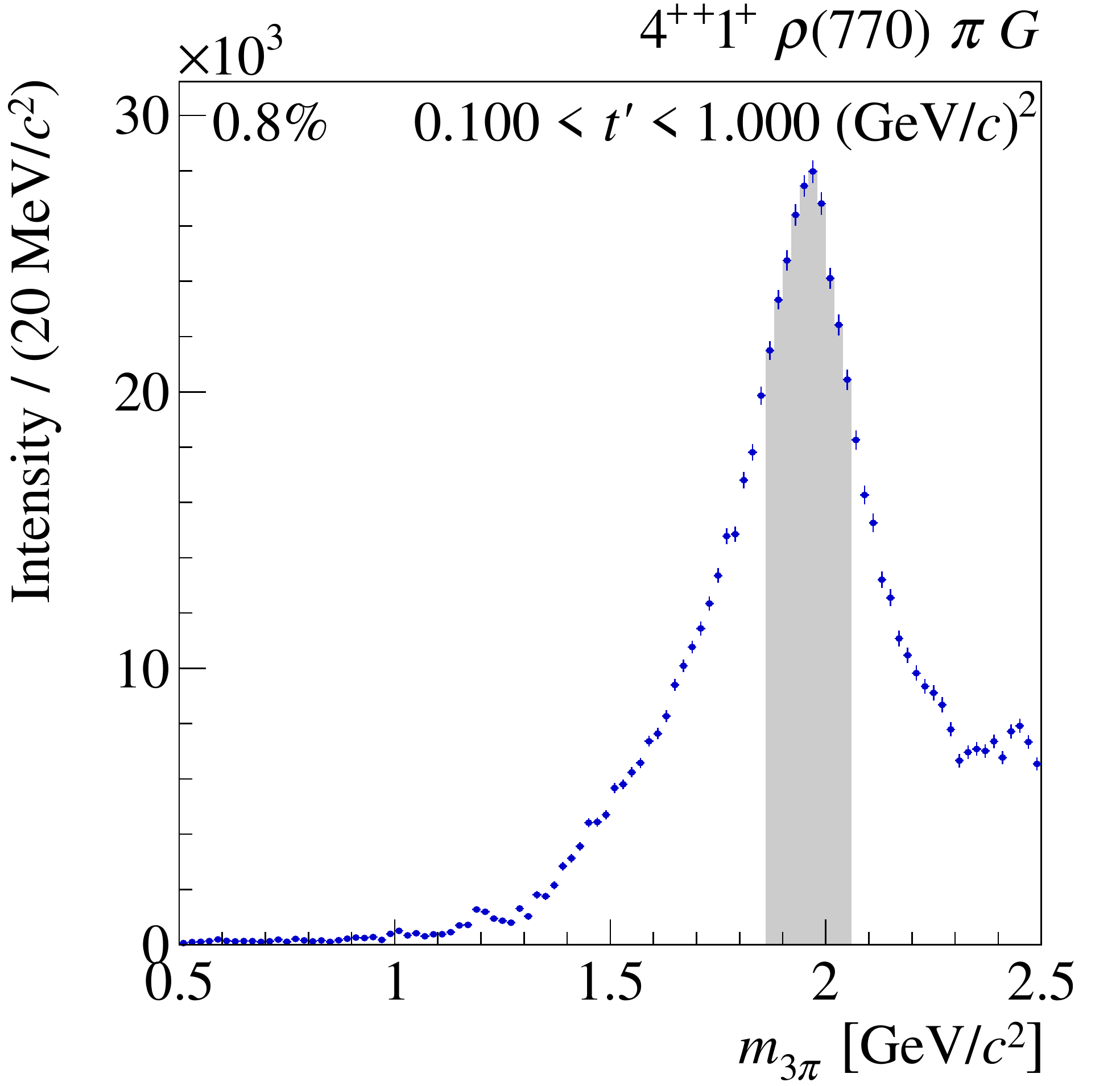}%
  }%
  \hspace*{\twoPlotSpacing}
  \subfloat[][]{%
    \label{fig:a4_total_m1_f2}%
    \includegraphics[width=\twoPlotWidth]{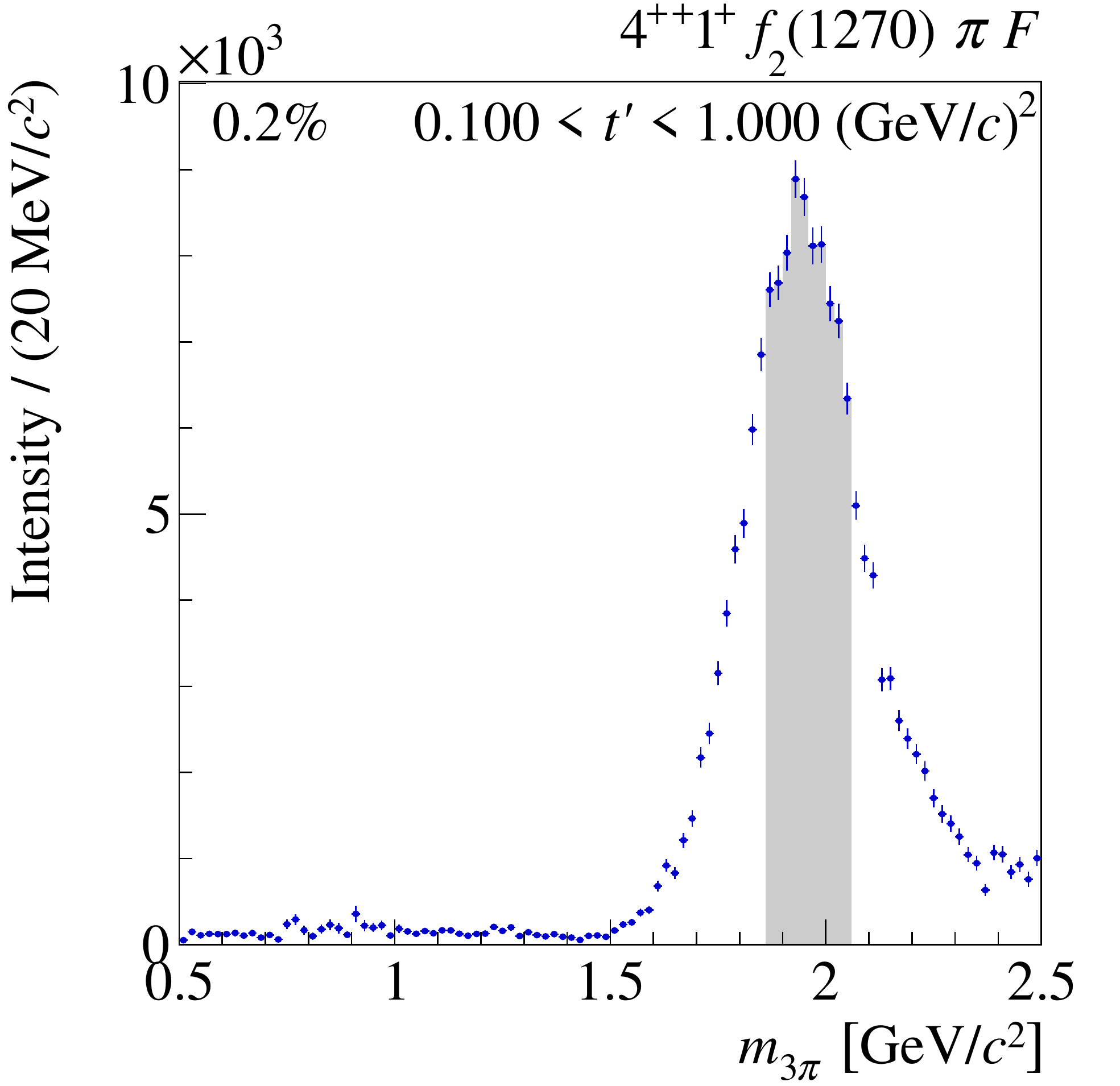}%
  }%
  \caption{The \tpr-summed intensity of waves with spin projection
    $M = 1$ showing in \protect\subref{fig:a1_total_m1} the \PaOne, in
      \protect\subref{fig:a2_total_m1_rho} and \protect\subref{fig:a2_total_m1_f2} the \PaTwo, in \protect\subref{fig:pi2_total_m1}
      the \PpiTwo, and in \protect\subref{fig:a4_total_m1_rho} and \protect\subref{fig:a4_total_m1_f2} the
    \PaFour.  The shaded regions indicate the mass intervals that
    are integrated over to generate the \tpr spectra (see
    \cref{fig:t_a1_m1,fig:t_a2_m1,fig:t_a2_m1b,fig:t_dependence_4pp}).}
  \label{fig:major_waves_m1}
\end{figure*}

Clear evidence is obtained for an $M = 2$ component of the
\wave{2}{++}{\!\!}{}{\Prho}{D} wave (\cref{fig:a2_total_m2}). Its
relative intensity with respect to the $M = 1$ wave
(\cref{fig:a2_total_m1_rho}) is about \SI{5}{\percent}.  This is in
good agreement with our result for the $2^{++}$ wave in the
$\eta\,\pi$ final state, which is dominated by the
\PaTwo~\cite{Adolph:2014rpp}.  In the analyzed range of
\SIvalRange{0.1}{\tpr}{1.0}{\GeVcsq}, the observed suppression is
twice as large as the suppression of $M = 1$ versus $M = 0$ waves.

\begin{figure}[tbp]
  \centering
  \subfloat[][]{%
    \label{fig:a2_total_m2}%
    \includegraphics[width=\twoPlotWidth]{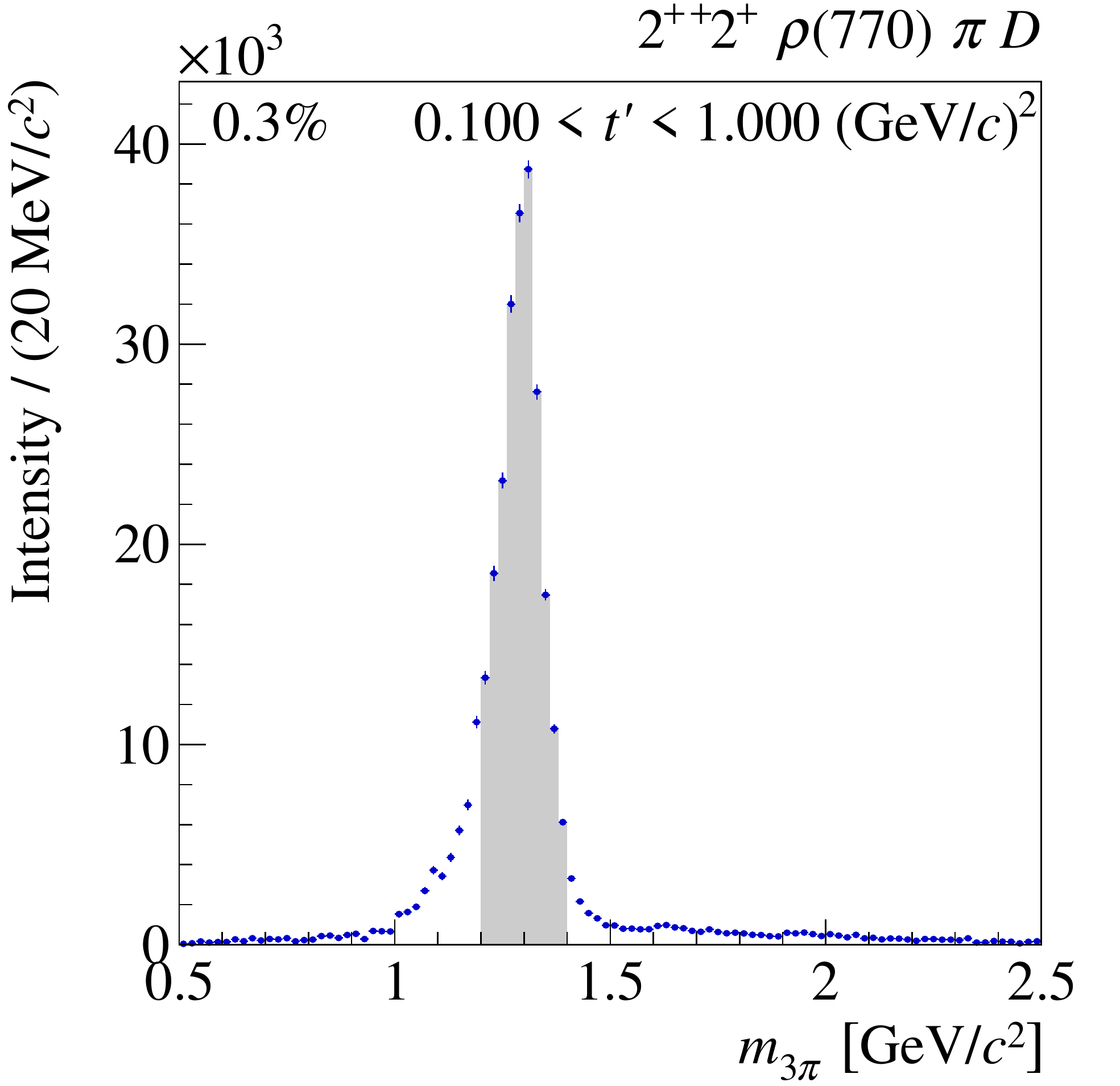}%
  }%
  \newLineOrHspace{\twoPlotSpacing}%
  \subfloat[][]{%
    \label{fig:a1_f2_pi_P_total_m0}%
    \includegraphics[width=\twoPlotWidth]{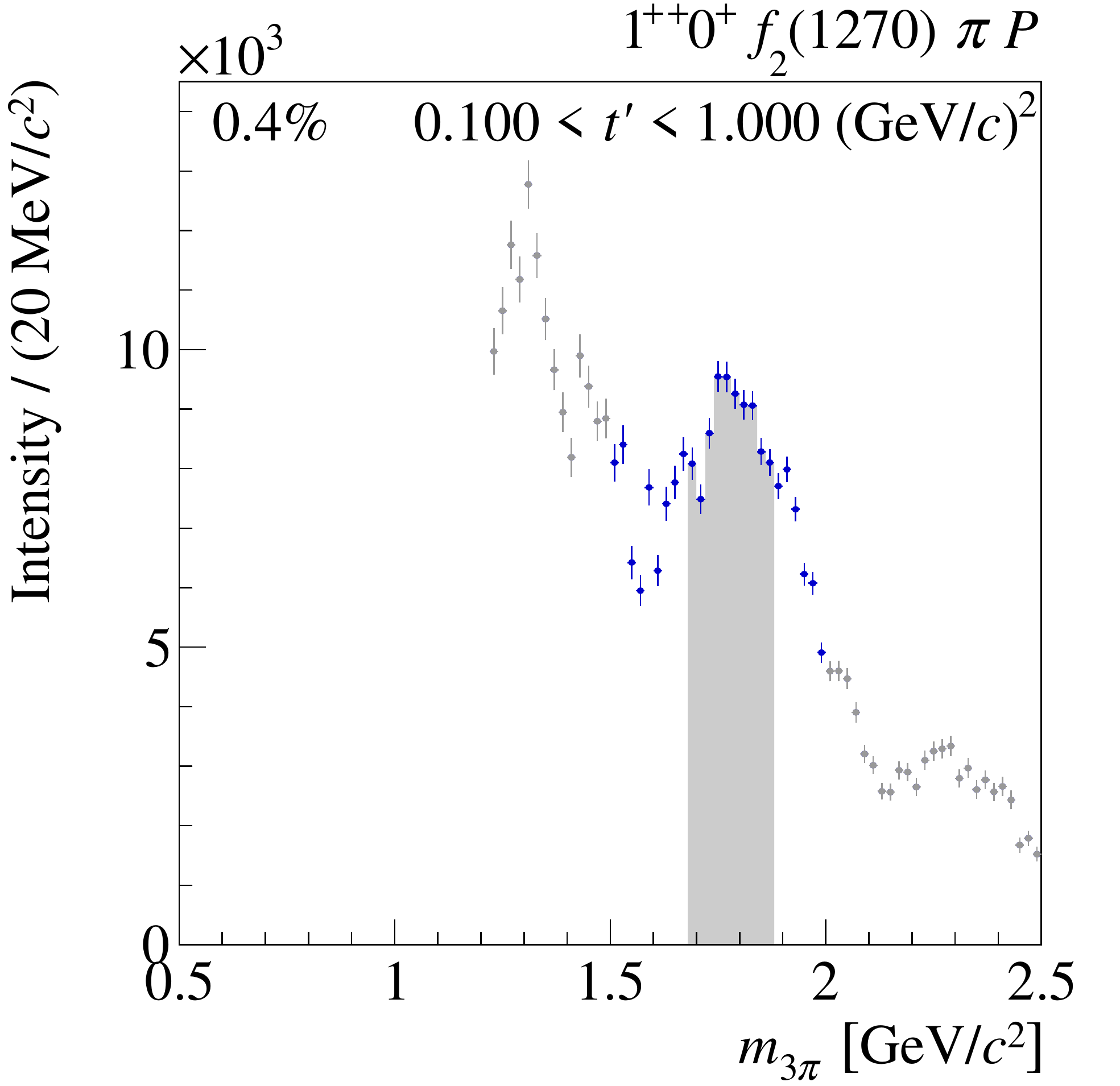}%
  }%
  \caption{Panel (a): \tpr-summed intensity of
    \wave{2}{++}{2}{+}{\Prho}{D} wave with spin projection $M = 2$ and
    the \PaTwo peak.  Panel~(b): \tpr-summed intensity of
    \wave{1}{++}{0}{+}{\PfTwo}{P} wave with spin projection $M =
    0$.  In this wave, the mass regions below
    \SI{1.5}{\GeVcc} and above \SI{2.0}{\GeVcc} (shown by gray
    points) are sensitive to the truncation of the partial-wave
    expansion series (see \cref{sec:pwa_massindep_systematic_studies}).  In both waves,
    the shaded regions indicate the mass intervals that are
    integrated over to generate the \tpr spectra (see
    \cref{fig:t_a2_m2,fig:t_a2_m2}).}
  \label{fig:a2_a1_totals}
\end{figure}

Nonresonant and resonant contributions are expected to follow
different production paths with possibly different dependences on
\tpr.  In order to investigate possible nonresonant contributions, we
show in \cref{fig:major_waves_1_t_bins,fig:major_waves_2_t_bins} the
intensities of four selected waves for two intervals of \tpr, \ie
\SIvalRange{0.100}{\tpr}{0.113}{\GeVcsq} and
\SIvalRange{0.449}{\tpr}{0.724}{\GeVcsq}, which represent regions of
low and high \tpr in this analysis.  When comparing these two regions,
the shapes of the \PaTwo and \PaFour resonances in the
\wave{2}{++}{1}{+}{\Prho}{D} and \wave{4}{++}{1}{+}{\Prho}{G} waves,
respectively, stay largely unaltered.  In contrast, we observe that
the peak in the \wave{1}{++}{0}{+}{\Prho}{S} wave, which presumably
contains the \PaOne, significantly shifts towards higher masses with
increasing \tpr.  A similar but less strong effect is observed for the
\PpiTwo peak in the \wave{2}{-+}{0}{+}{\PfTwo}{S} wave.  This shows
that the peak structures in the latter two partial waves are not only
due to ordinary resonances but are distorted by nonresonant
contributions.  The Deck process proposed in~\refCite{Deck:1964hm} and
illustrated in \cref{fig:deck_process} may provide an explanation for
the \tpr-dependent nonresonant contributions observed in the $1^{++}$
and $2^{-+}$ waves.  The \tpr dependence of the shape of the
\wave{1}{++}{0}{+}{\Prho}{S} mass spectrum was already observed by the
ACCMOR collaboration~\cite{Daum:1979sx,Daum:1980ay} and our results
confirm their findings.

\begin{figure*}[htbp]
  \centering
  \subfloat[][]{%
    \label{fig:a1_t_bin_low}%
    \includegraphics[width=\twoPlotWidth]{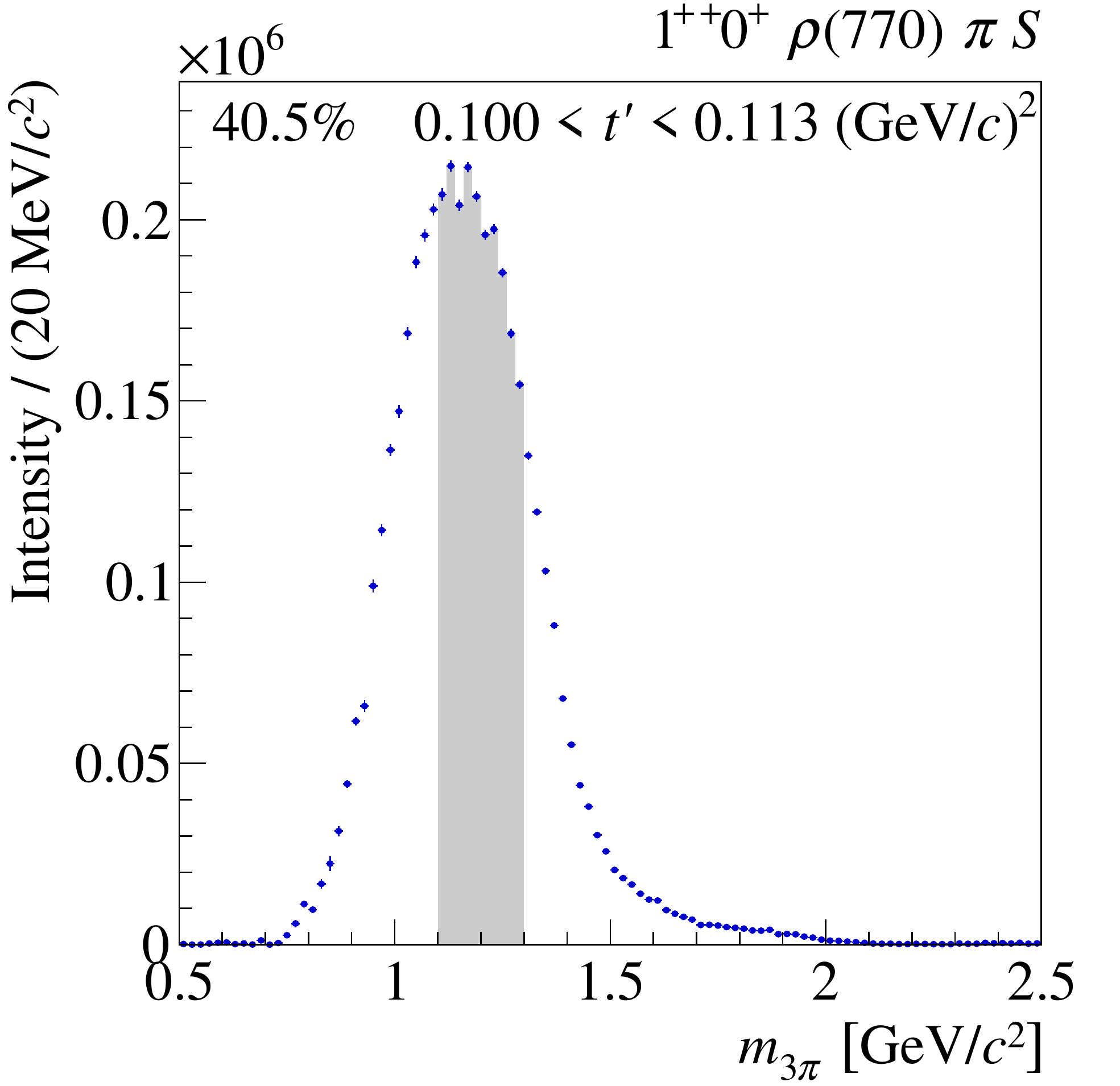}%
  }%
  \hspace*{\twoPlotSpacing}
  \subfloat[][]{%
    \label{fig:a2_t_bin_low}%
    \includegraphics[width=\twoPlotWidth]{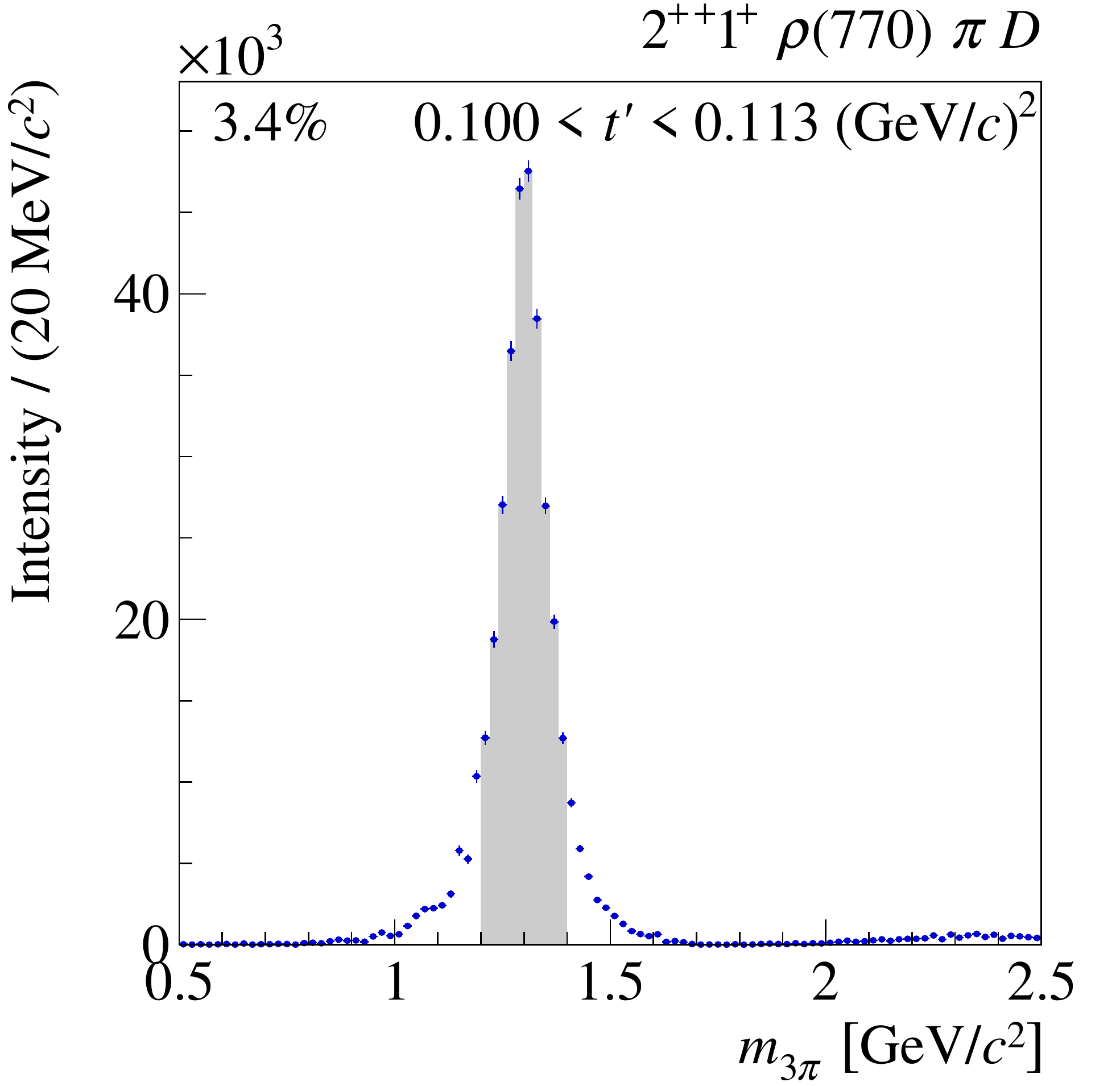}%
  }%
  \\
  \subfloat[][]{%
    \label{fig:a1_t_bin_high}%
    \includegraphics[width=\twoPlotWidth]{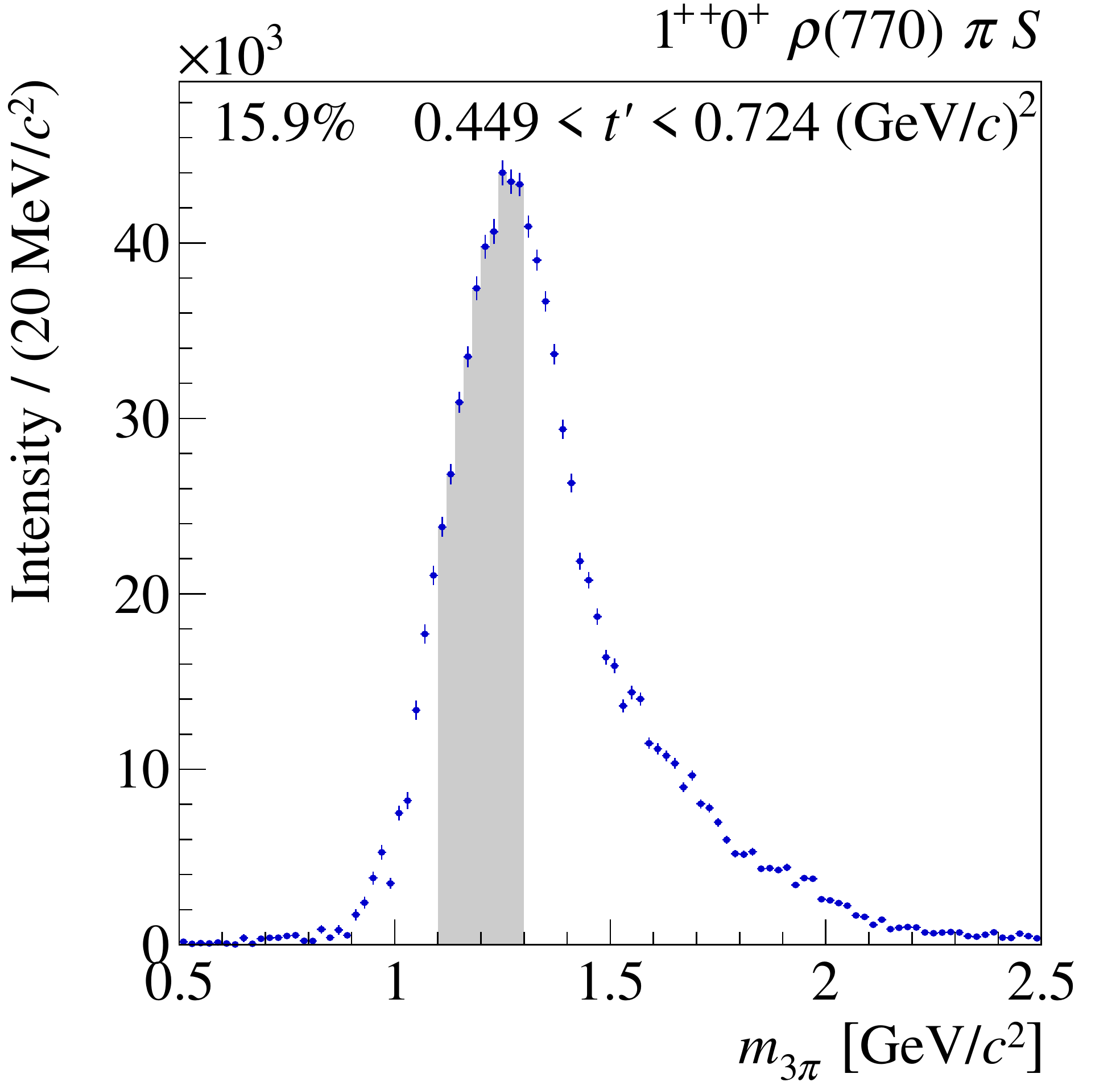}%
  }%
  \hspace*{\twoPlotSpacing}
  \subfloat[][]{%
    \label{fig:a2_t_bin_high}%
    \includegraphics[width=\twoPlotWidth]{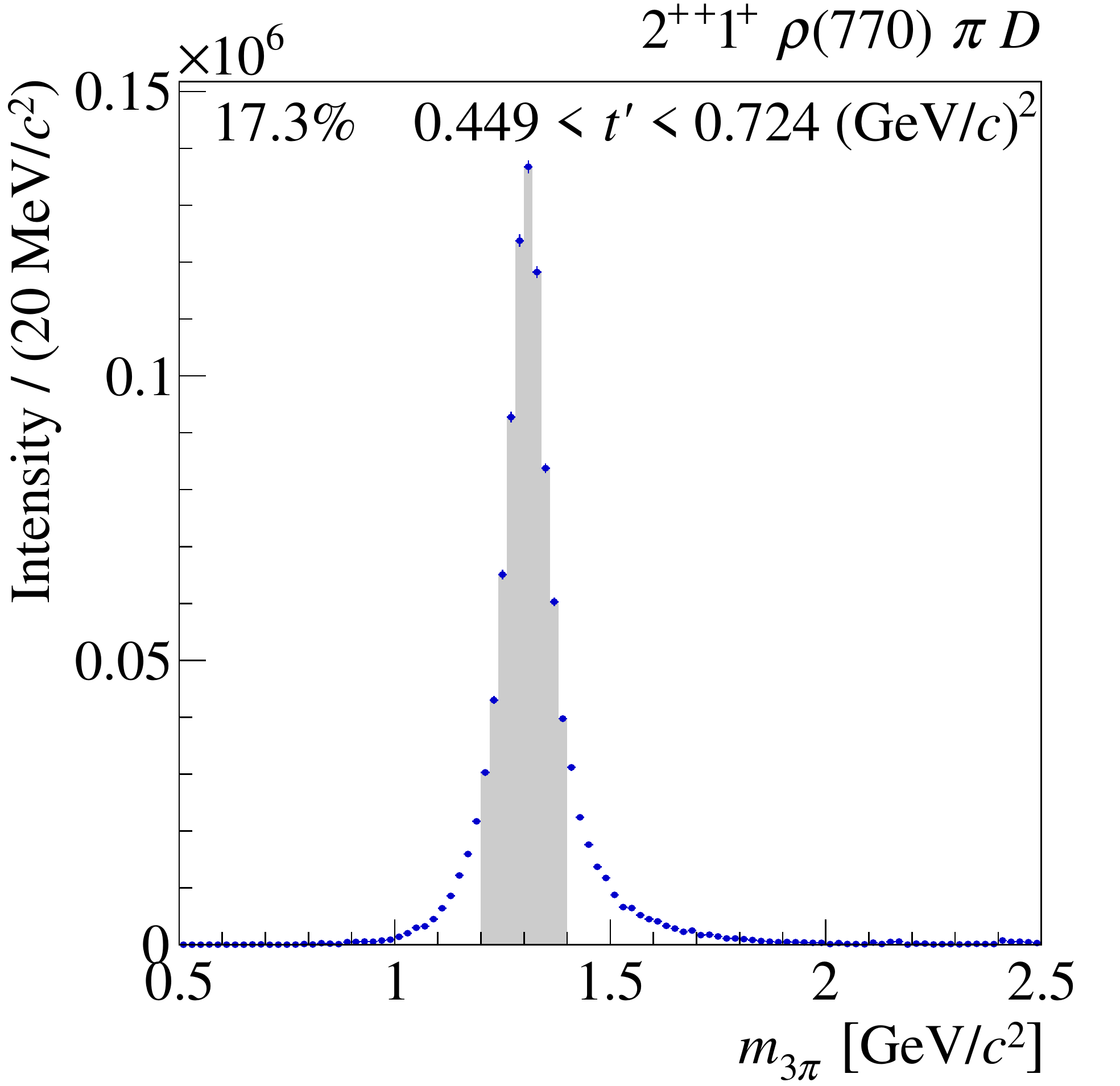}%
  }%
  \caption{The intensities of the \wave{1}{++}{0}{+}{\Prho}{S} and
    \wave{2}{++}{1}{+}{\Prho}{D} waves in two different \tpr regions.
    Upper row: low \tpr; Lower row: high \tpr. The shaded regions
    indicate the mass intervals that are integrated over to generate
    the \tpr spectra.}
  \label{fig:major_waves_1_t_bins}
\end{figure*}

\begin{figure*}[htbp]
  \centering
  \subfloat[][]{%
    \label{fig:pi2_t_bin_low}%
    \includegraphics[width=\twoPlotWidth]{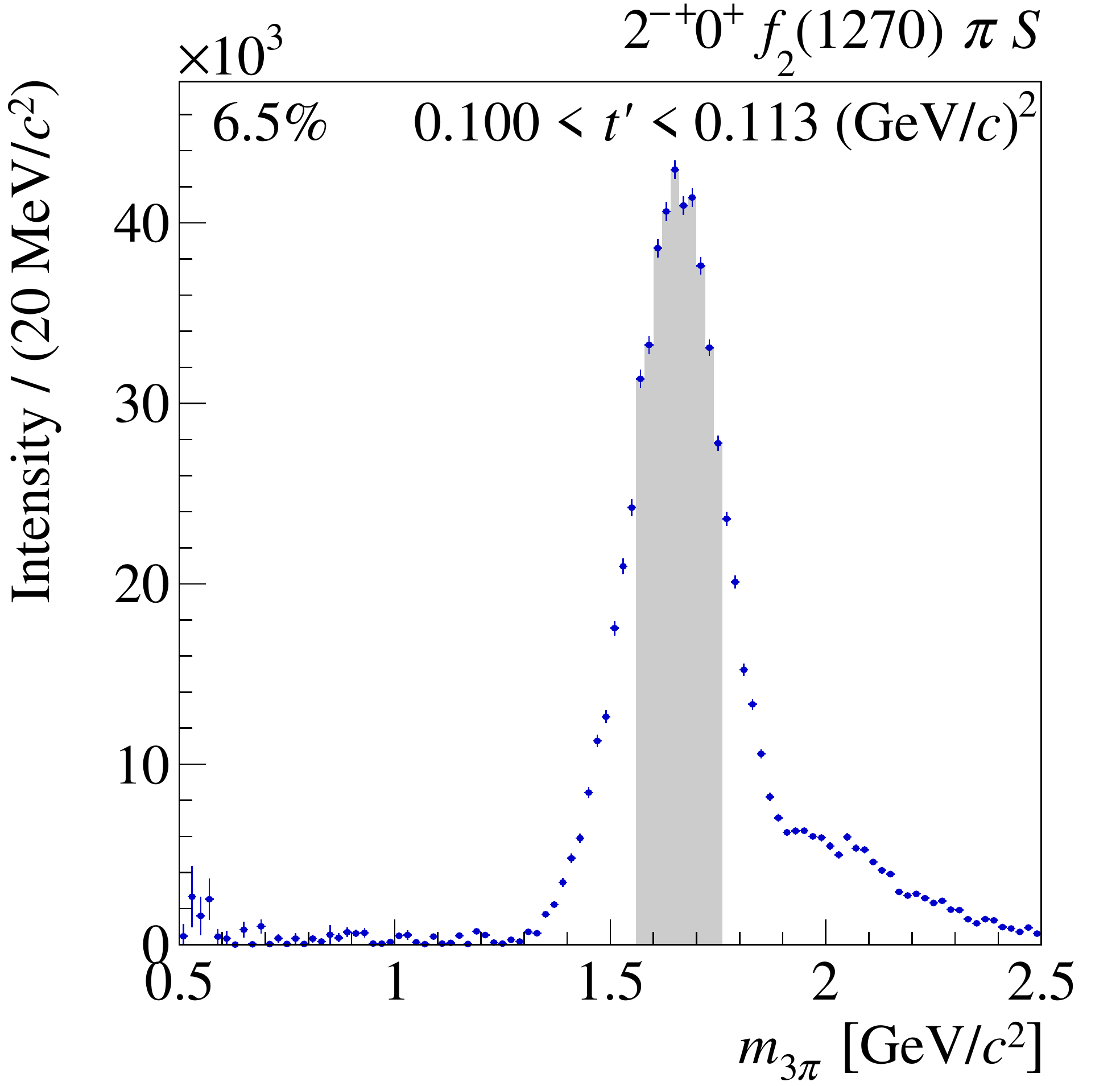}%
  }%
  \hspace*{\twoPlotSpacing}
  \subfloat[][]{%
    \label{fig:a4_t_bin_low}%
    \includegraphics[width=\twoPlotWidth]{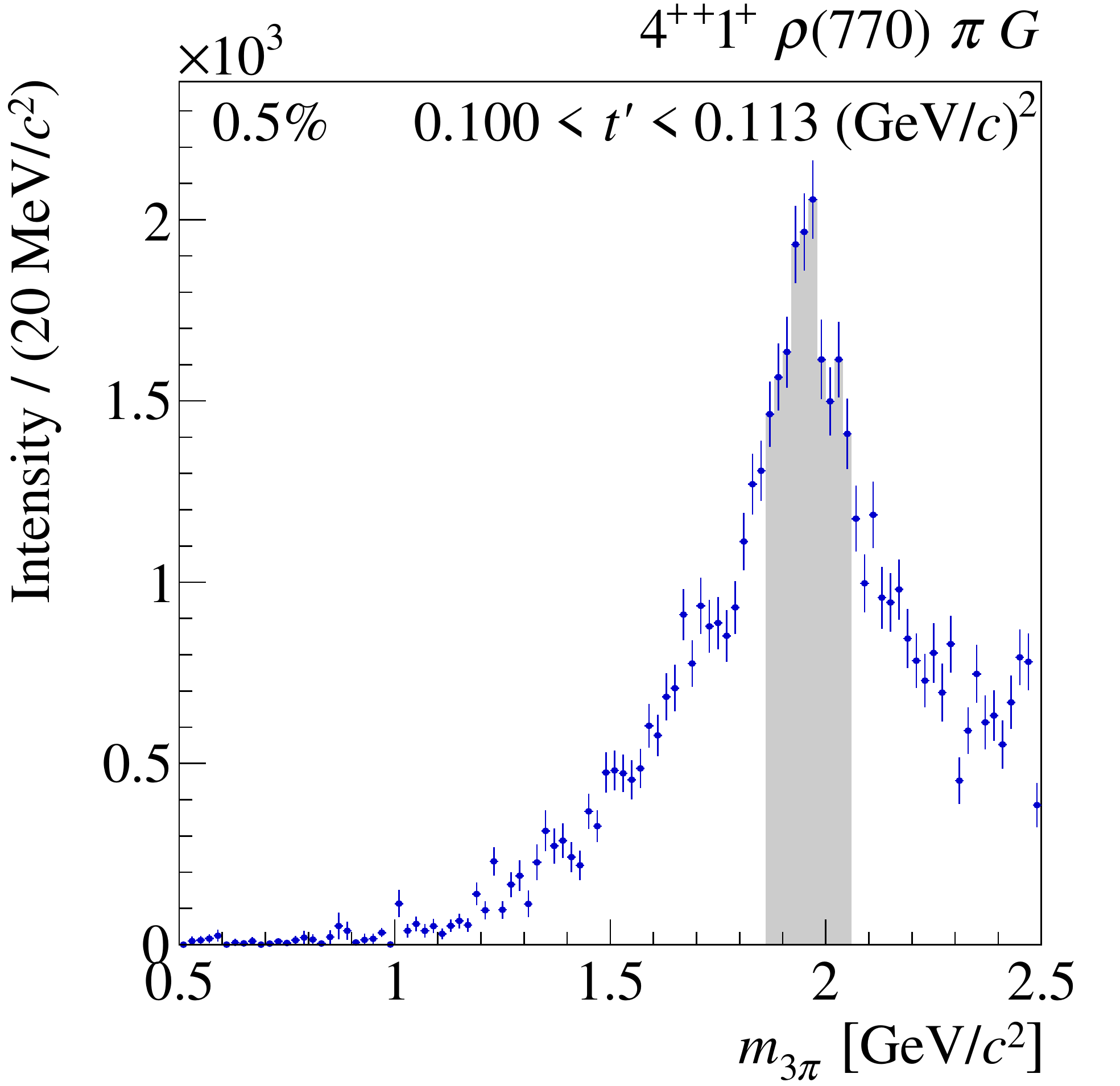}%
  }%
  \\
  \subfloat[][]{%
    \label{fig:pi2_t_bin_high}%
    \includegraphics[width=\twoPlotWidth]{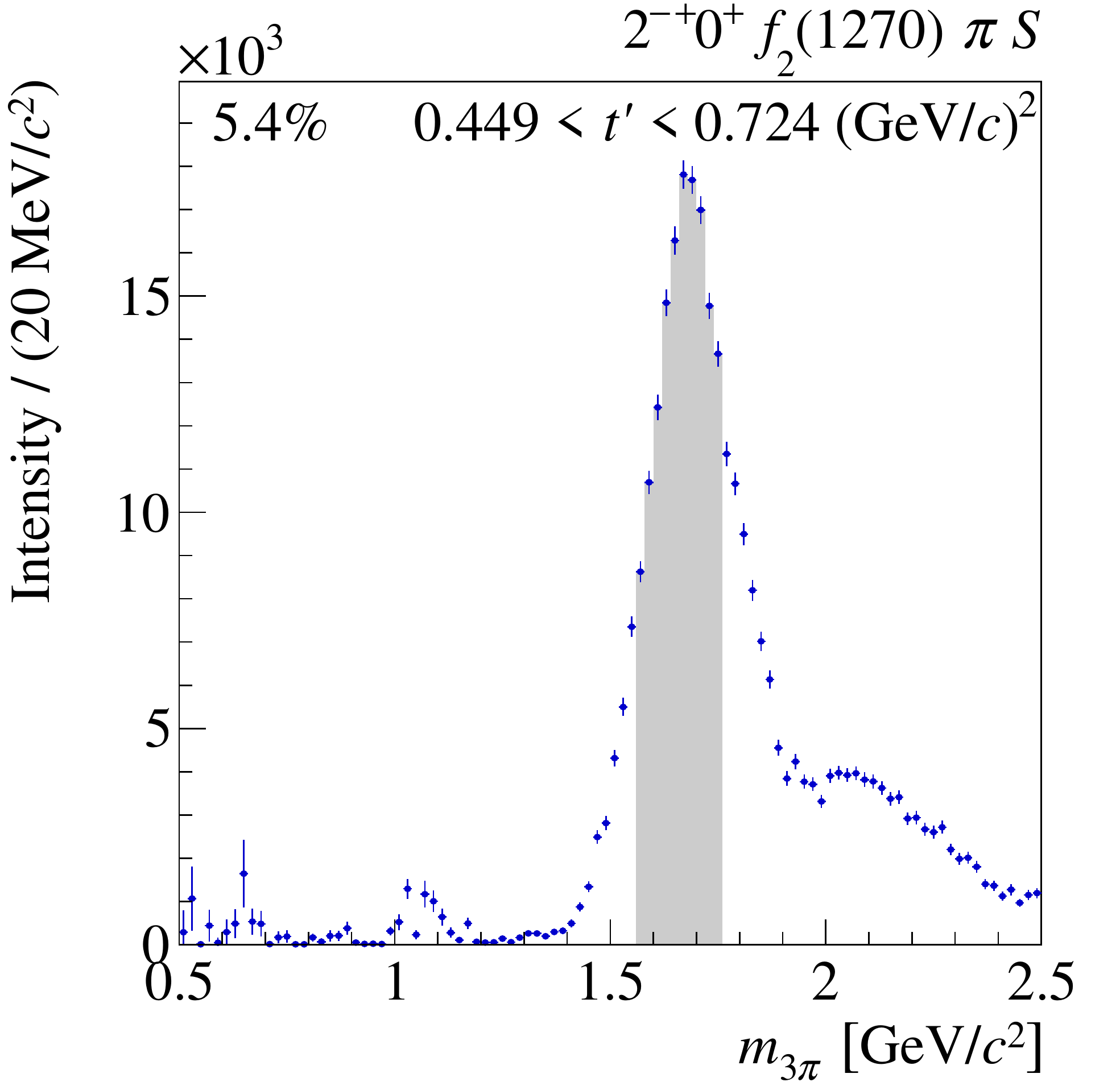}%
  }%
  \hspace*{\twoPlotSpacing}
  \subfloat[][]{%
    \label{fig:a4_t_bin_high}%
    \includegraphics[width=\twoPlotWidth]{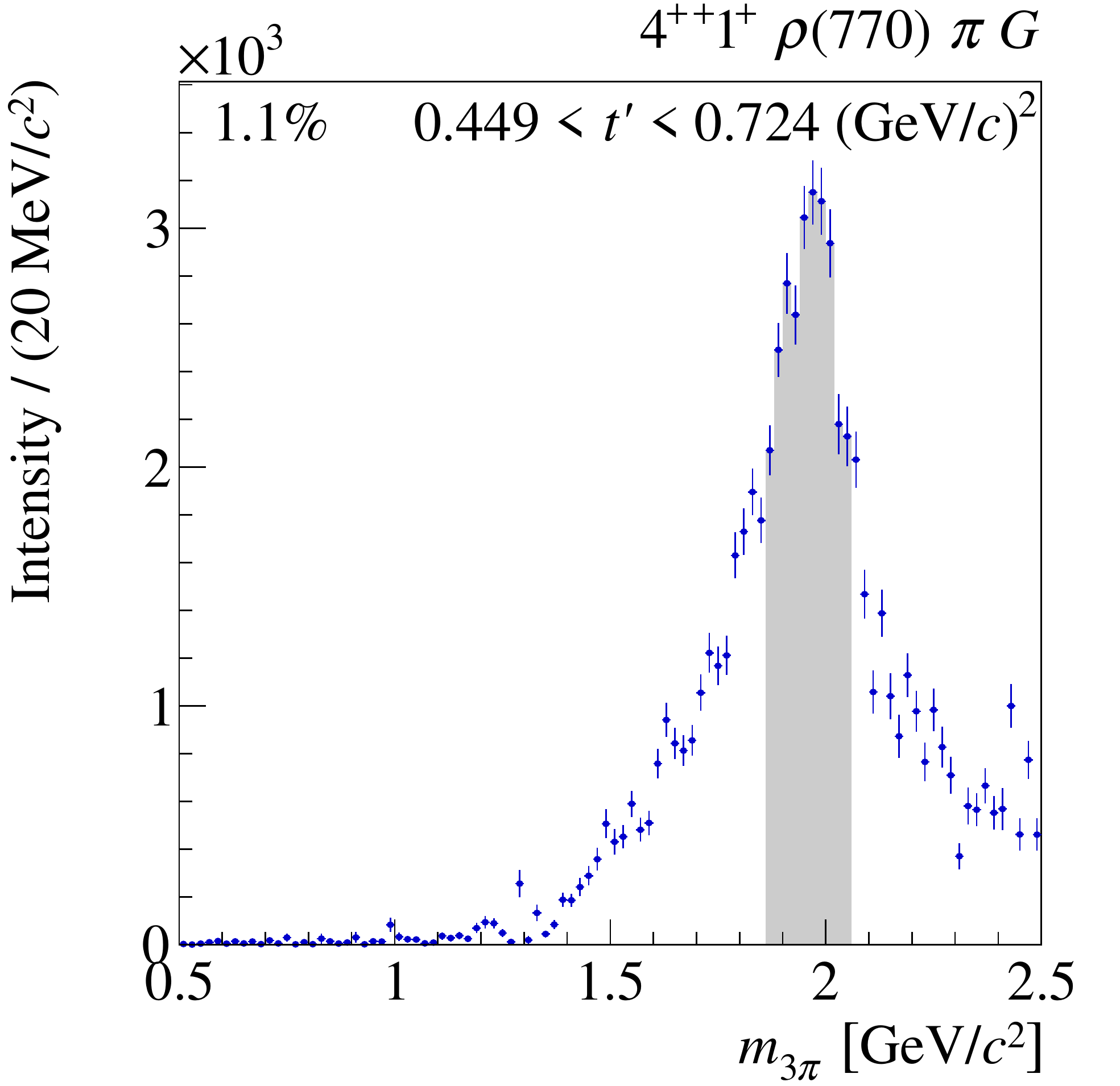}%
  }%
  \caption{Same as \cref{fig:major_waves_1_t_bins}, but for the
    \wave{2}{-+}{0}{+}{\PfTwo}{S} and \wave{4}{++}{1}{+}{\Prho}{G}
    waves.}
  \label{fig:major_waves_2_t_bins}
\end{figure*}

\begin{figure}[tbp]
  \centering
  \includegraphics[scale=1]{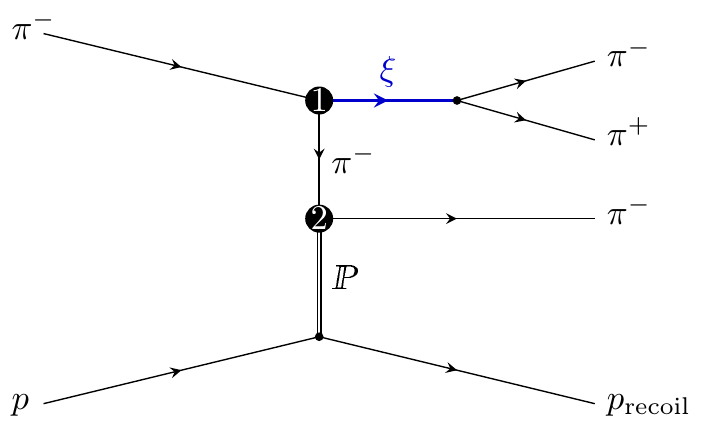}
  \caption{Example for a nonresonant production process for the $3\pi$
    final state as proposed by Deck~\cite{Deck:1964hm}.}
  \label{fig:deck_process}
\end{figure}

We show in \cref{fig:minor_waves_1_t_bins,fig:minor_waves_2_t_bins}
the same \tpr regions for the small-intensity waves
\wave{1}{++}{0}{+}{\PfTwo}{P}, \wave{2}{++}{1}{+}{\PfTwo}{P},
\wave{2}{-+}{0}{+}{\Prho}{F}, and \wave{4}{++}{1}{+}{\PfTwo}{F}.  All
waves show a pronounced dependence of the mass spectrum on \tpr.  In
contrast to the $1^{++}\,\Prho\,\pi\,S$ wave, the \PaOne cannot be
clearly identified in the $\PfTwo\,\pi\,P$ wave.  Instead, the latter
wave shows a broad enhancement around \SI{1.8}{\GeVcc} (see also
\cref{fig:a1_f2_pi_P_total_m0}).  In the $2^{++}\,\PfTwo\,\pi\,P$
wave, the \PaTwo exhibits a high-mass shoulder, which is particularly
pronounced at large values of \tpr, although it is clearly
identifiable also at low \tpr.  Such a high-mass shoulder also becomes
prominent for the \PpiTwo in the $\Prho\,\pi\,F$ wave, for which the
spectrum exhibits a richer structure than for the
\wave{2}{-+}{0}{+}{\PfTwo}{S} wave.

\begin{figure*}[htbp]
  \centering
  \subfloat[][]{%
    \label{fig:a1P_t_bin_low}%
    \includegraphics[width=\twoPlotWidth]{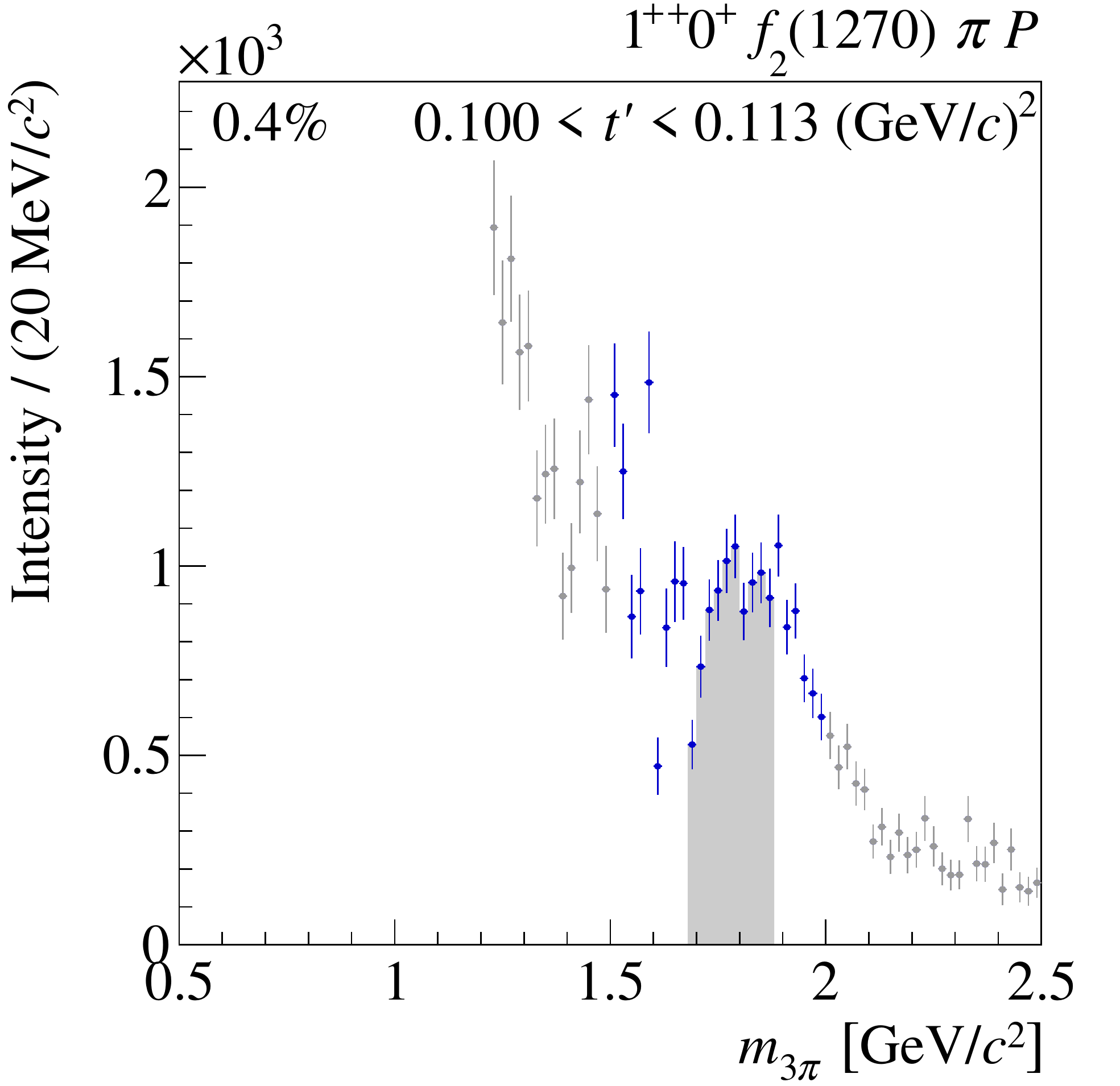}%
  }%
  \hspace*{\twoPlotSpacing}
  \subfloat[][]{%
    \label{fig:a2P_t_bin_low}%
    \includegraphics[width=\twoPlotWidth]{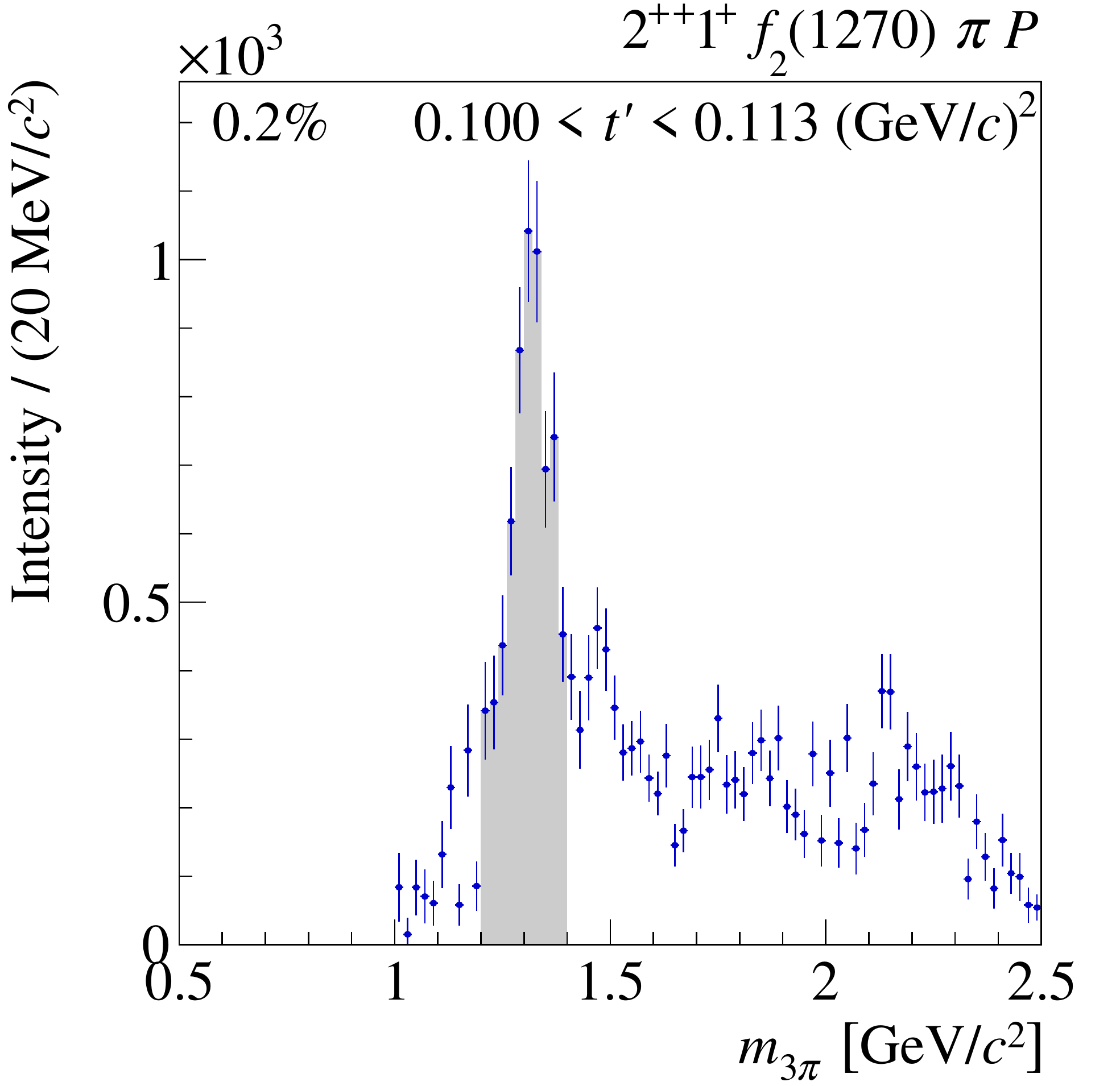}%
  }%
  \\
  \subfloat[][]{%
    \label{fig:a1P_t_bin_high}%
    \includegraphics[width=\twoPlotWidth]{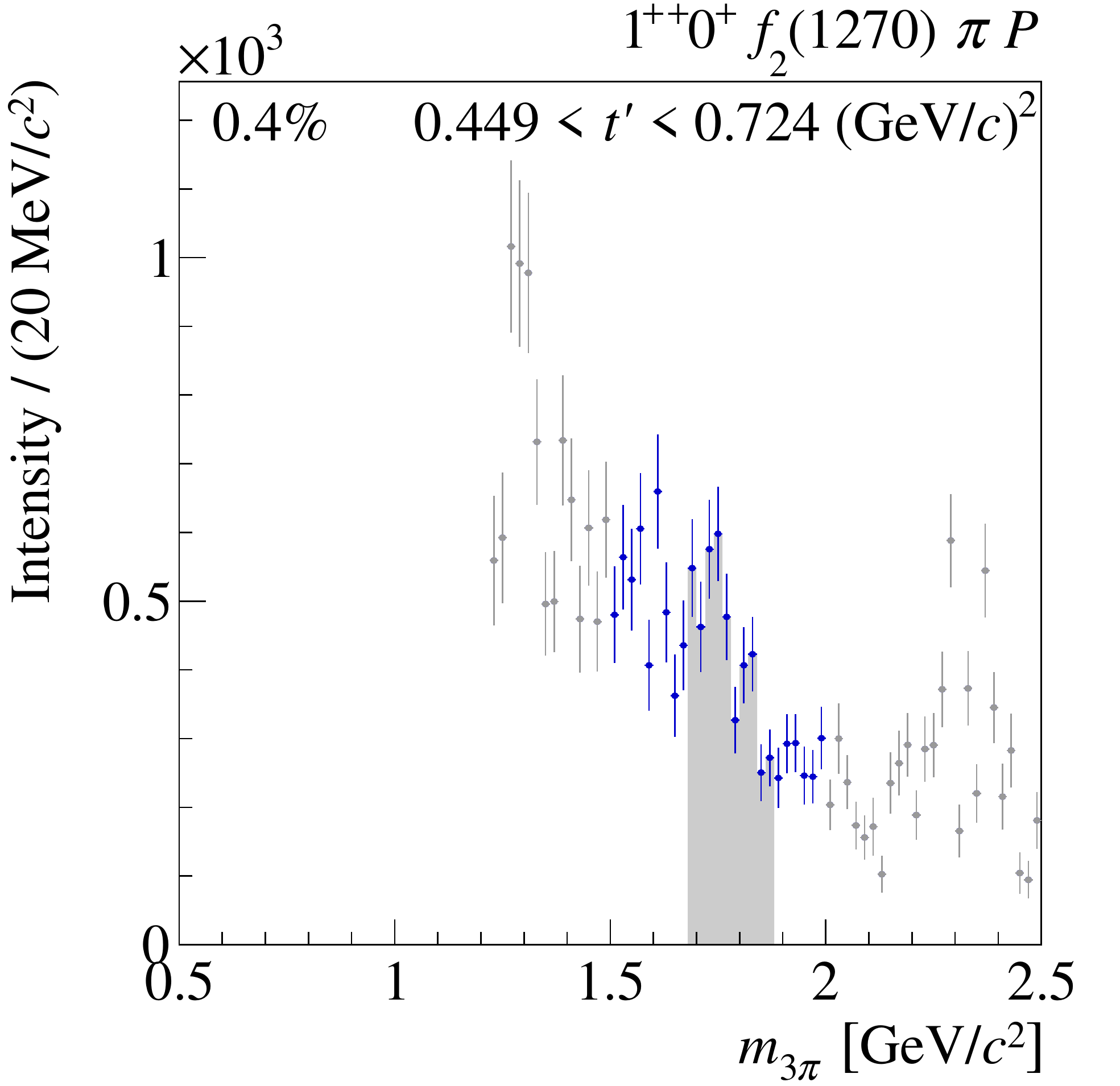}%
  }%
  \hspace*{\twoPlotSpacing}
  \subfloat[][]{%
    \label{fig:a2P_t_bin_high}%
    \includegraphics[width=\twoPlotWidth]{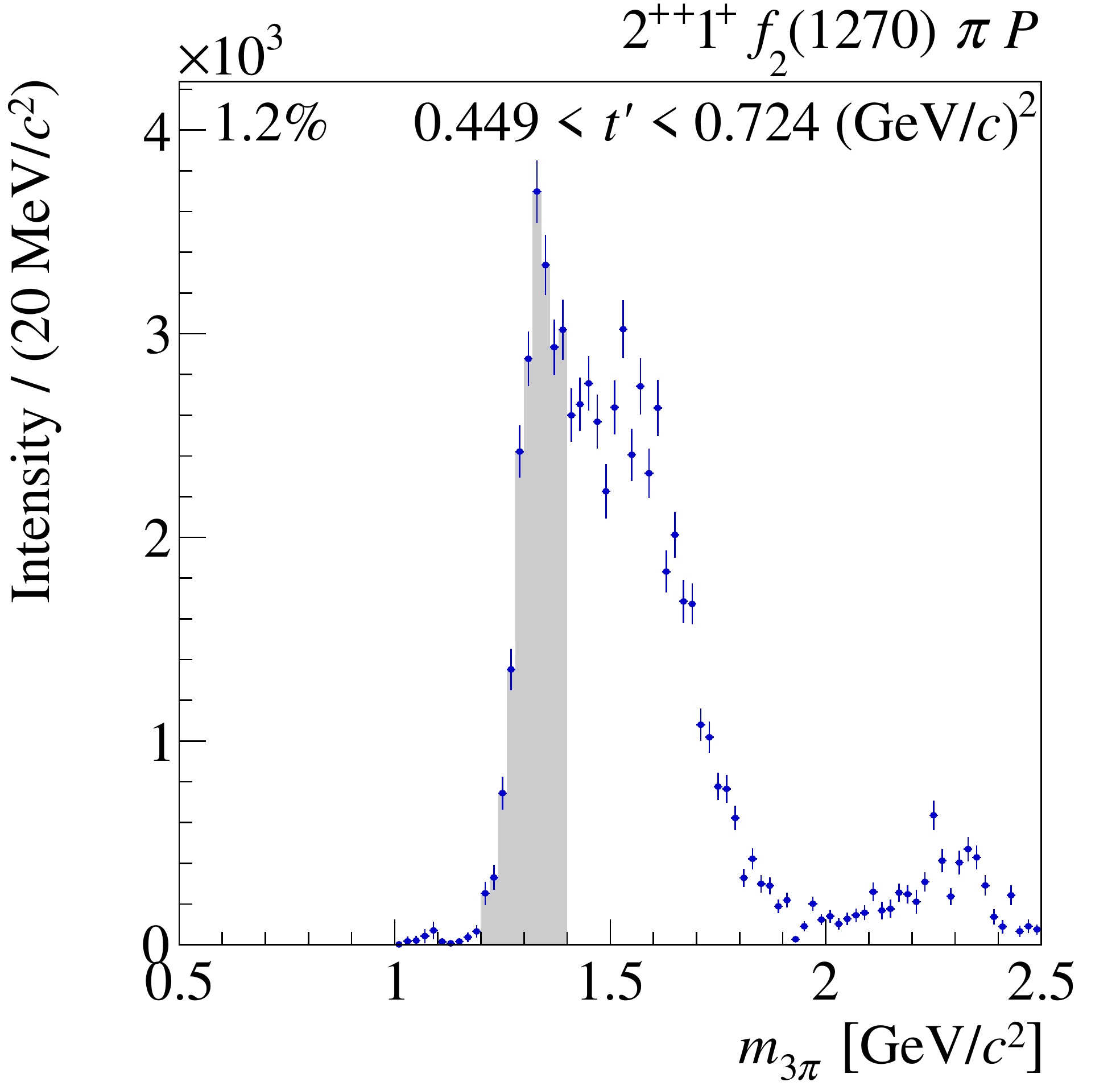}%
  }%
  \caption{Same as \cref{fig:major_waves_1_t_bins}, but for the
    \wave{1}{++}{0}{+}{\PfTwo}{P} and \wave{2}{++}{1}{+}{\PfTwo}{P}
    waves.  In the former wave, the mass regions below
    \SI{1.5}{\GeVcc} and above \SI{2.0}{\GeVcc} (shown by gray
    points) are sensitive to the truncation of the partial-wave
    expansion series (see \cref{sec:pwa_massindep_systematic_studies}).}
  \label{fig:minor_waves_1_t_bins}
\end{figure*}

\begin{figure*}[htbp]
  \centering
  \subfloat[][]{%
    \label{fig:pi2F_t_bin_low}%
    \includegraphics[width=\twoPlotWidth]{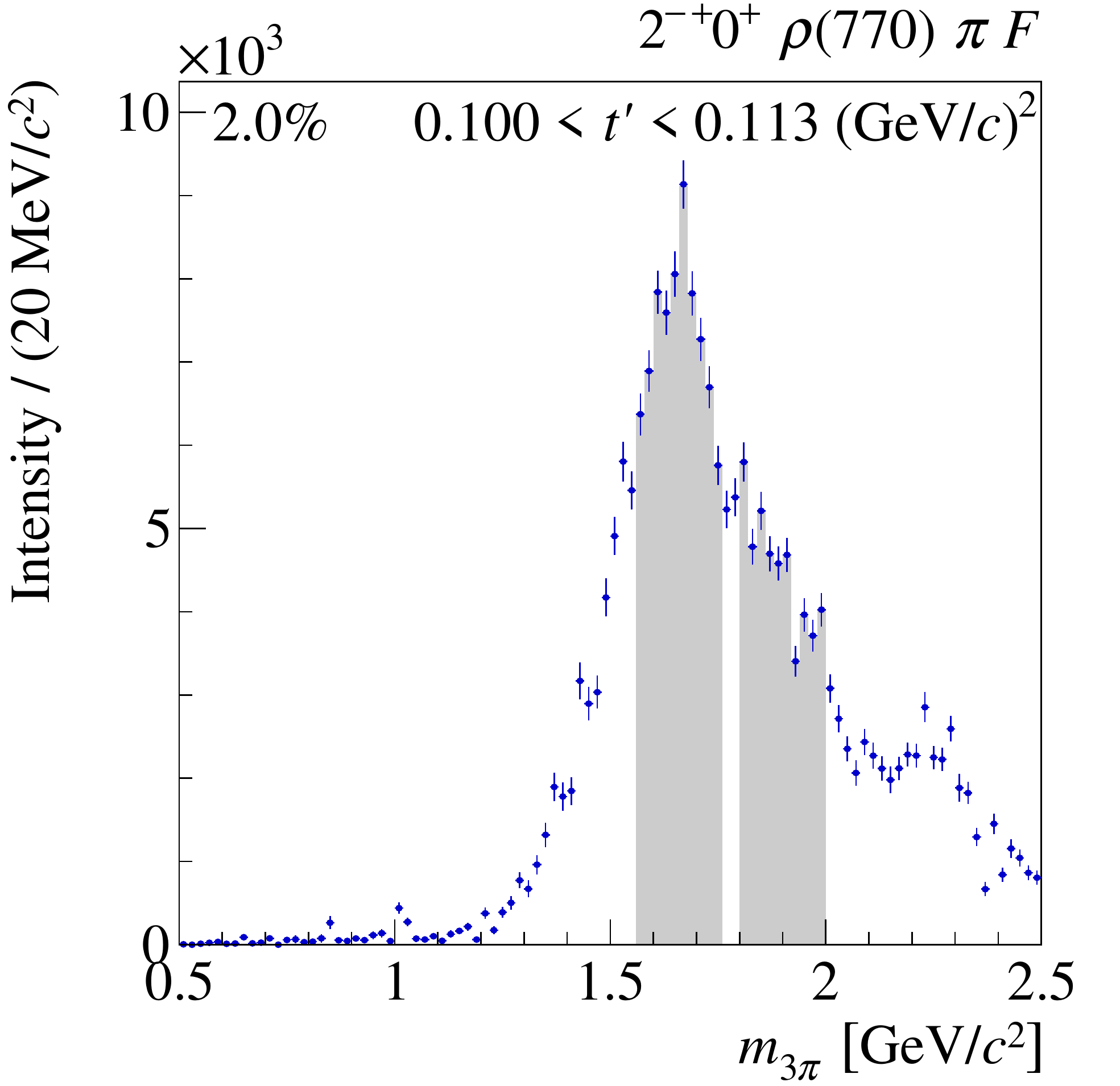}%
  }%
  \hspace*{\twoPlotSpacing}
  \subfloat[][]{%
    \label{fig:a4F_t_bin_low}%
    \includegraphics[width=\twoPlotWidth]{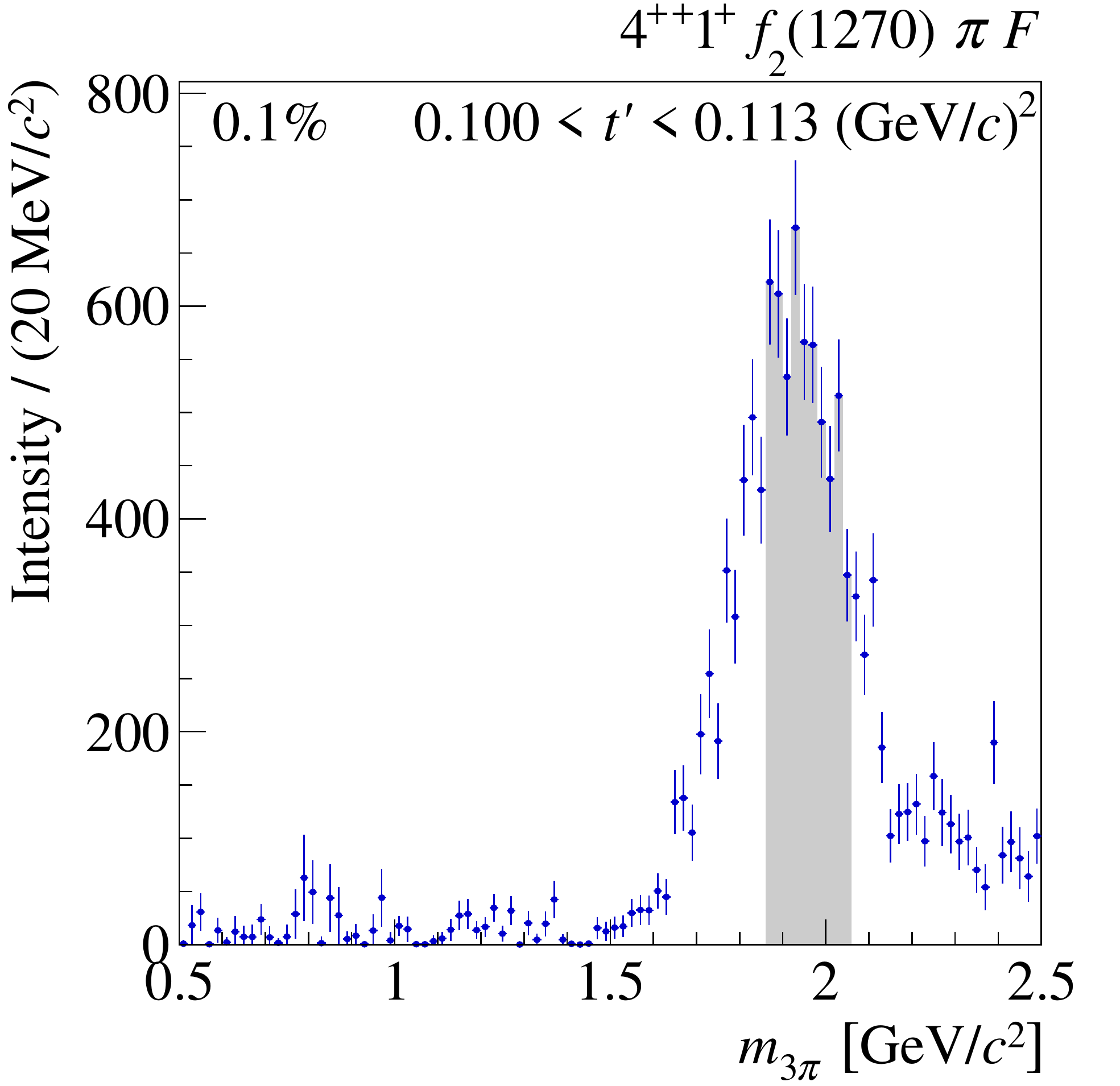}%
  }%
  \\
  \subfloat[][]{%
    \label{fig:pi2F_t_bin_high}%
    \includegraphics[width=\twoPlotWidth]{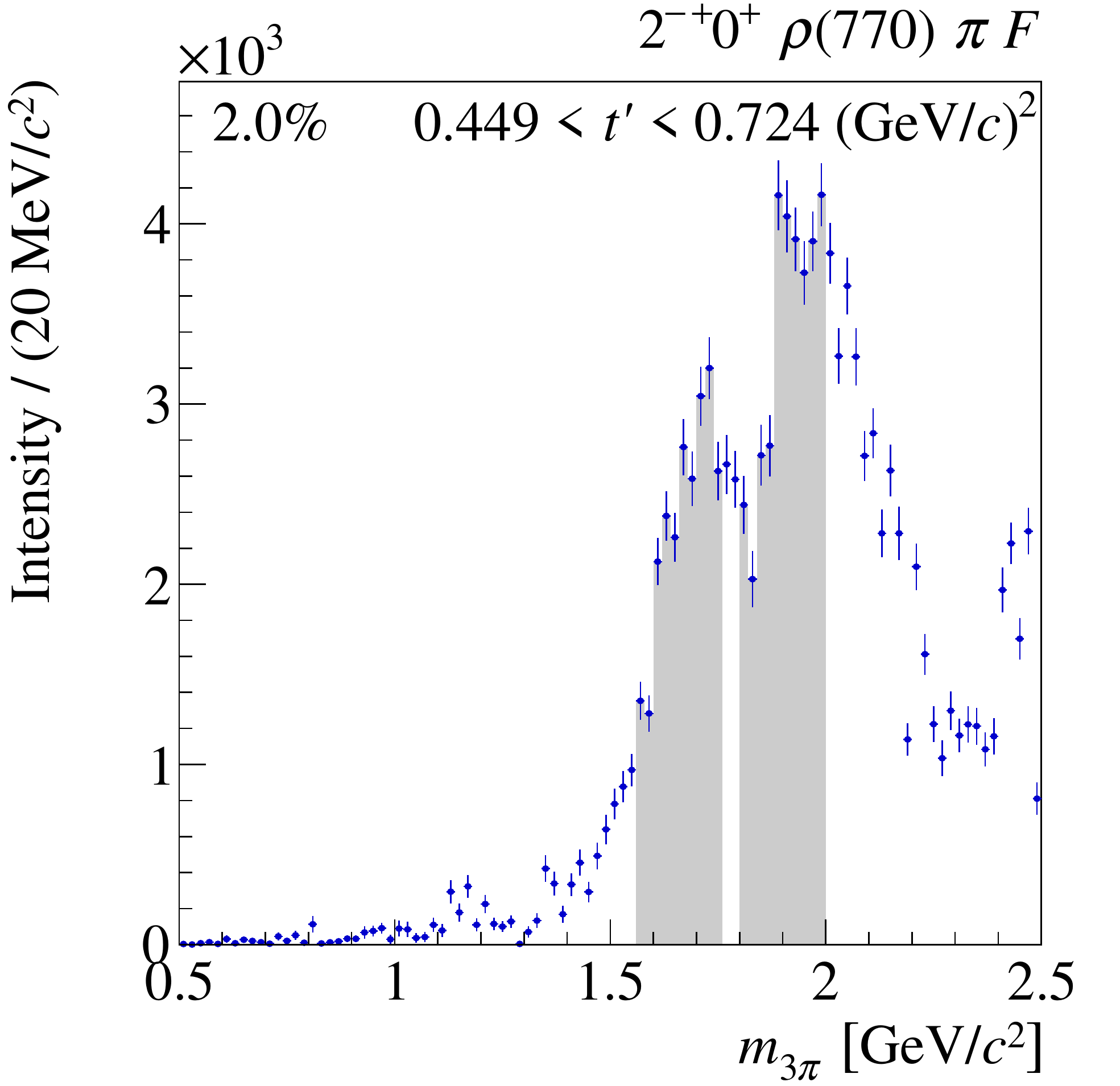}%
  }%
  \hspace*{\twoPlotSpacing}
  \subfloat[][]{%
    \label{fig:a4F_t_bin_high}%
    \includegraphics[width=\twoPlotWidth]{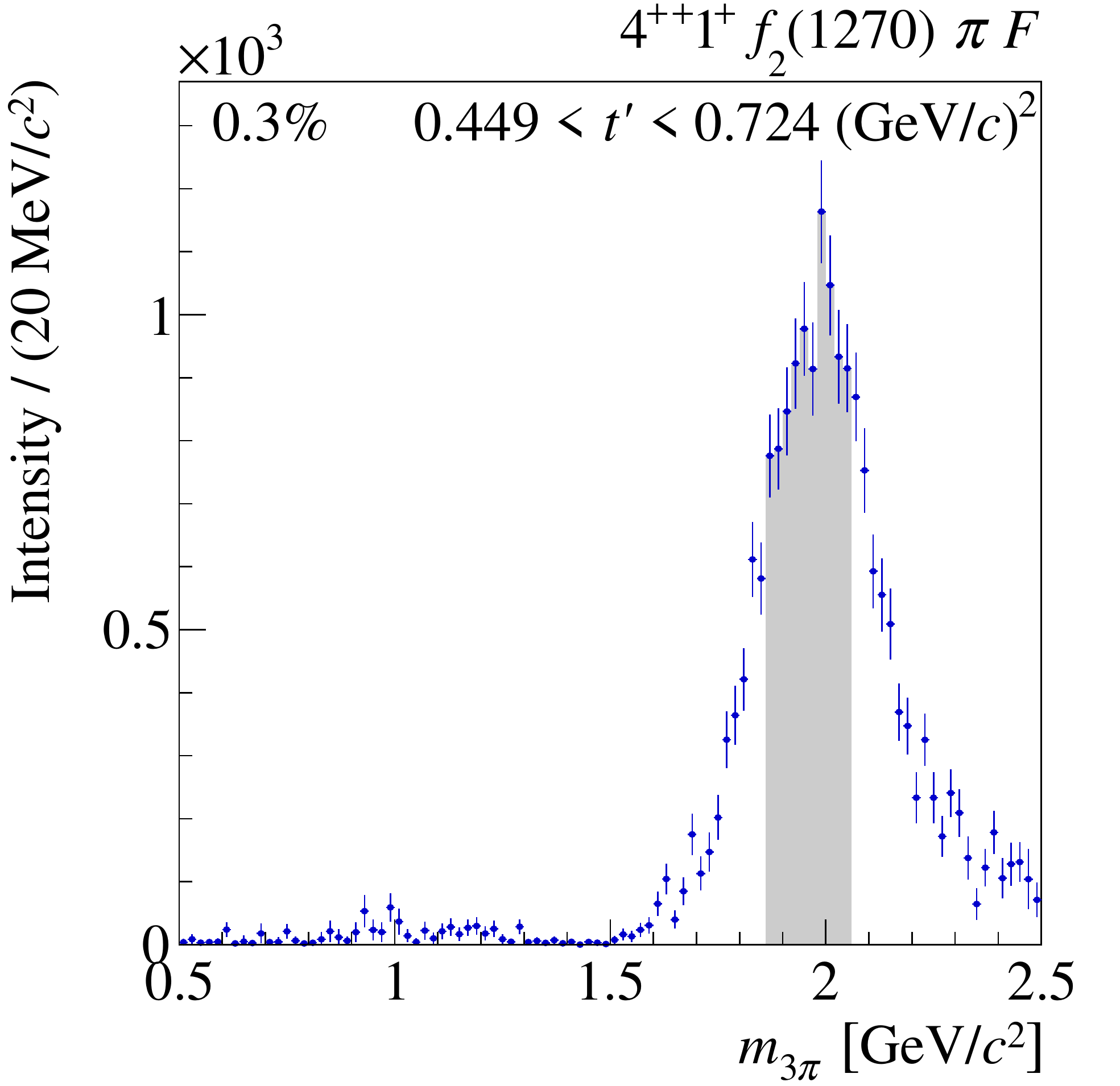}%
  }%
  \caption{Same as \cref{fig:major_waves_1_t_bins}, but for the
    \wave{2}{-+}{0}{+}{\Prho}{F} and \wave{4}{++}{1}{+}{\PfTwo}{F}
    waves.}
 \label{fig:minor_waves_2_t_bins}
\end{figure*}

The selective effect of the orbital angular momentum~$L$ in the decay
is clearly demonstrated in
\cref{fig:pi2_total_m0_2,fig:pi2D_total_m0}, which show the
\wave{2}{-+}{0}{+}{\PfTwo}{} waves with $L = 0$ and $L = 2$.  The
\PpiTwo dominates the $S$-wave, while the \PpiTwo[1880] favors the
$D$-wave.  The \PpiTwo[1880] is considerably lighter than the expected
radial excitation of the \PpiTwo ground state and has been rated as a
viable hybrid-meson candidate~\cite{klempt:2007cp,barnes:1996ff}.
However, a dominance of $S$ over $D$-wave $\PfTwo\,\pi$ decay modes
was predicted for hybrid mesons by model
calculations~\cite{close:1994hc,page:1998gz}.  This is at variance
with the present observation of a prevailing $D$-wave decay of the
\PpiTwo[1880].  The existence of the \PpiTwo[1880] was questioned
by~\refCite{Dudek:2006ud}, which explains it as an interference of the
\PpiTwo ground state with the nonresonant Deck process causing an
apparent shift of the peak position.  This might be counterargued by
the observation of two peaks in the \wave{2}{-+}{0}{+}{\Prho}{F} wave
(see \cref{fig:pi2F_t_bin_low,fig:pi2F_t_bin_high,fig:pi2_total_rho}).
Studying additional decay modes and the \tpr dependence of the mass
spectra should resolve this issue.

\begin{figure*}[htbp]
  \centering
  \subfloat[][]{%
    \label{fig:pi2_total_m0_2}%
    \includegraphics[width=\twoPlotWidth]{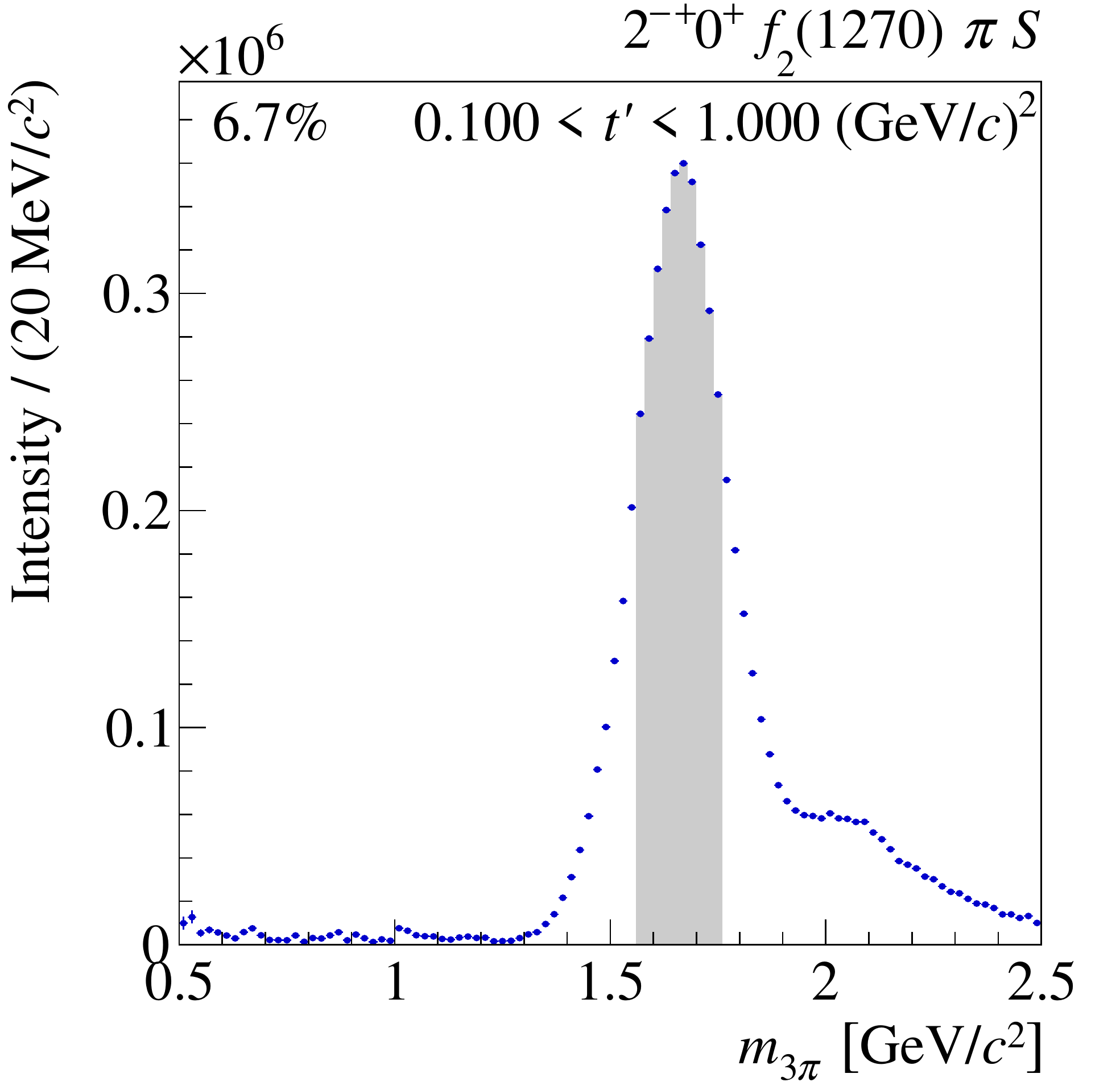}%
  }%
  \hspace*{\twoPlotSpacing}
  \subfloat[][]{%
    \label{fig:pi2D_total_m0}%
    \includegraphics[width=\twoPlotWidth]{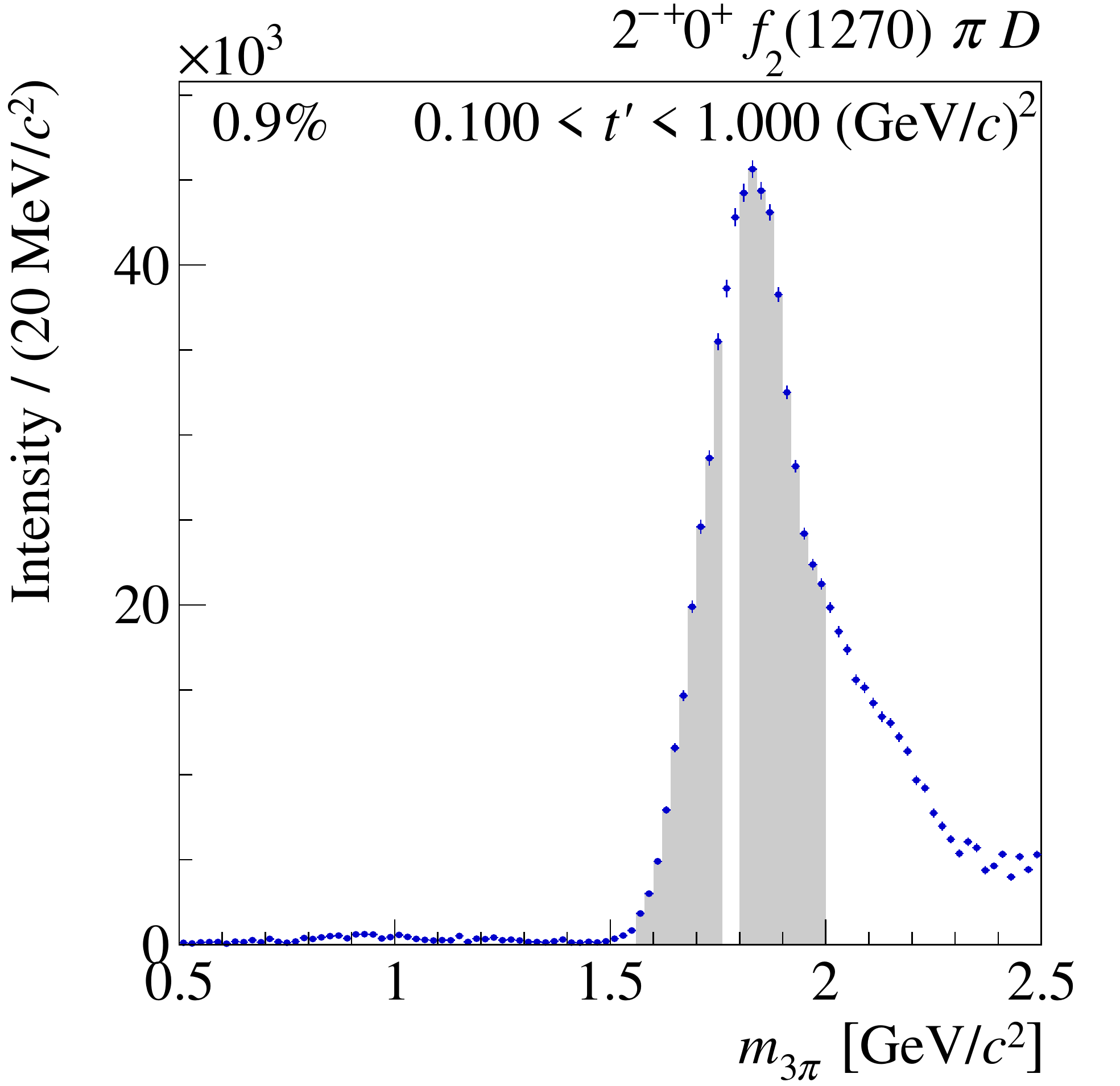}%
  }%
  \\
  \subfloat[][]{%
    \label{fig:pi2_total_m1_2}%
    \includegraphics[width=\twoPlotWidth]{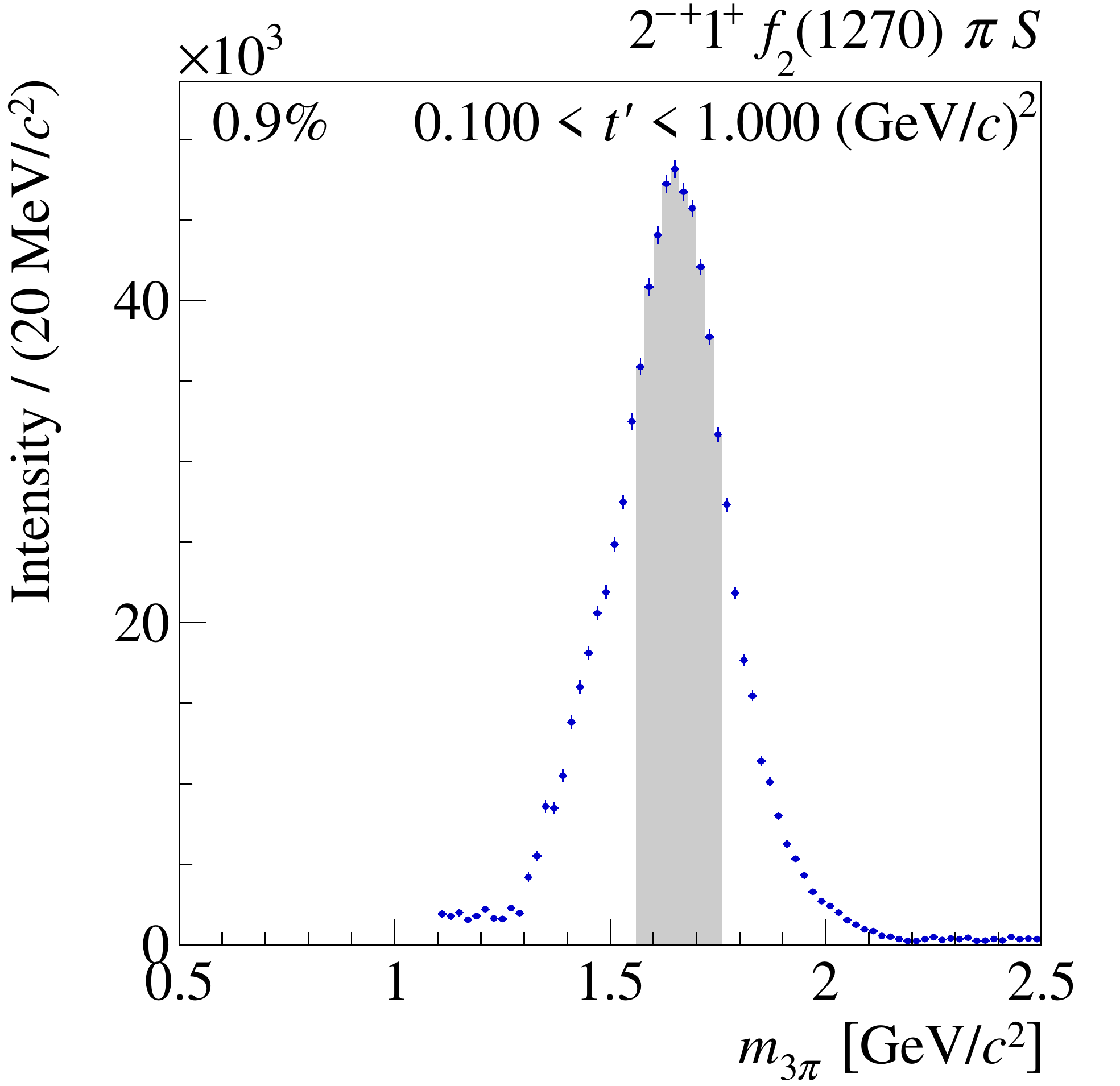}%
  }%
  \hspace*{\twoPlotSpacing}
  \subfloat[][]{%
    \label{fig:pi2_total_rho}%
    \includegraphics[width=\twoPlotWidth]{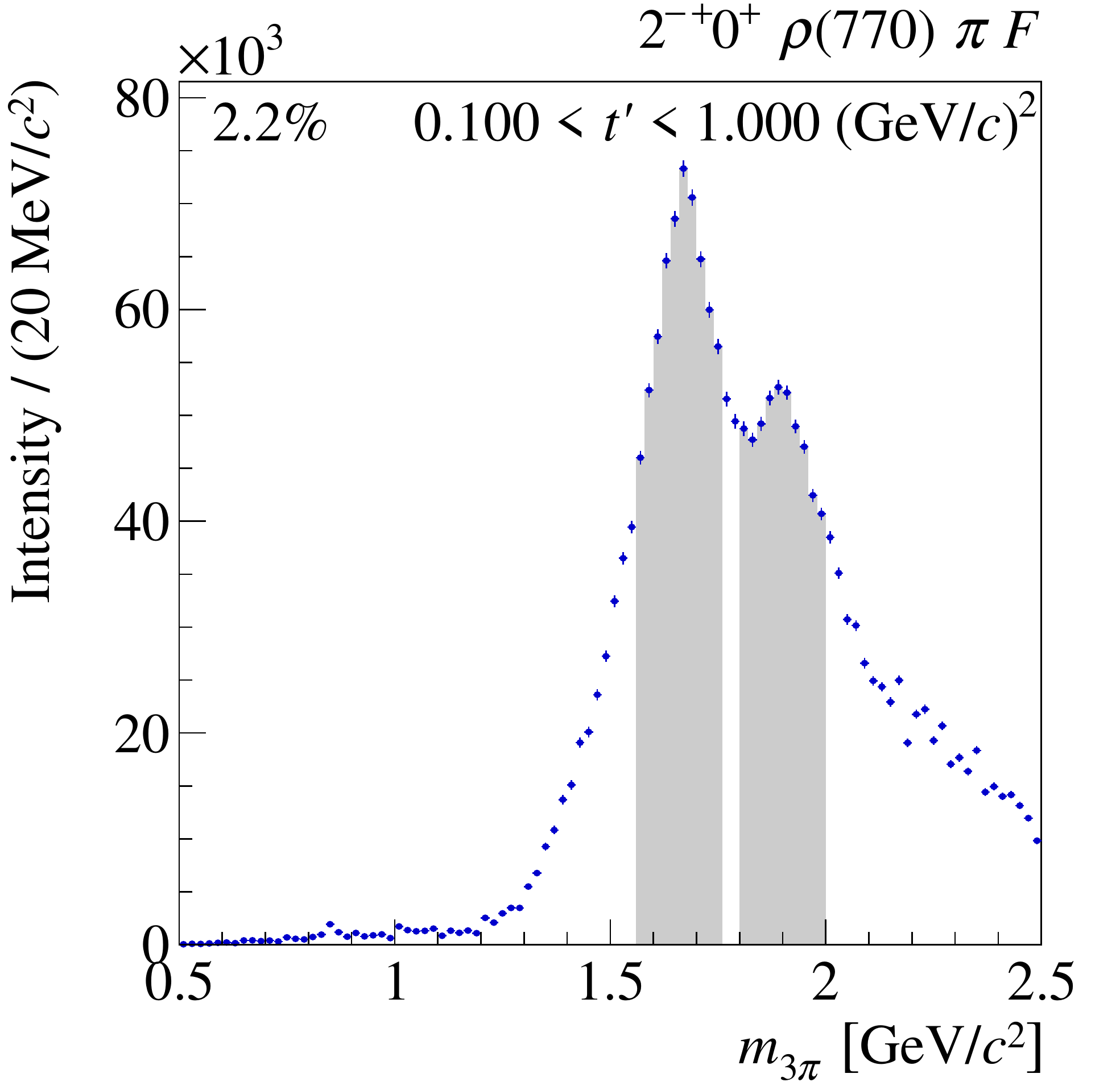}%
  }%
  \caption{The \tpr-summed intensities of
    \wave{2}{-+}{M}{+}{\PfTwo}{L} waves in panels~(a), (b), and~(c)
    and of \wave{2}{-+}{0}{+}{\Prho}{F} wave in panel~(d).  The \PpiTwo dominates
    the waves shown in \protect\subref{fig:pi2_total_m0_2} and
    \protect\subref{fig:pi2_total_m1_2}, the \PpiTwo[1880] the one
    shown in \protect\subref{fig:pi2D_total_m0}.  Both resonances
    appear in the wave shown in \protect\subref{fig:pi2_total_rho}.
    Upper row: comparison
    of same decay mode but different orbital angular momenta $L = 0$
    and 2 in the decay; Left column: comparison of same decay mode but
    different spin projections $M = 0$ and 1.  The shaded regions
    indicate the mass intervals that are integrated over to generate
    the \tpr spectra discussed in \cref{sec:tprim_partial_waves}.}
  \label{fig:angular_momentum_effect}
\end{figure*}

\subsection{Partial Waves with \PfZero and broad \pipiSW Isobars}
\label{sec:results_pwa_massindep_pipiS_waves}

As discussed in the previous section, the shape of the peak in the
\PaOne region in the \wave{1}{++}{0}{+}{\Prho}{S} wave and that in the
\PpiTwo region in the $2^{-+}$ waves change as a function of \tpr (see
\cref{fig:a1_t_bin_low,fig:a1_t_bin_high,fig:pi2_t_bin_low,fig:pi2_t_bin_high}).
A possible explanation for this behavior is the Deck process.  We have
therefore investigated partial waves that are expected to have small
contributions from the Deck process.  Owing to the nature and small
width of the \PfZero[980], this is in particular true for
$\PfZero[980]\,\pi$ partial waves.  Only a few meson resonances have
been observed to decay via \PfZero, such as
$\Ppi[1800] \to \PfZero \pi$, $\Pphi \to \PfZero \gamma$,
$\Pphi[2170] \to \PfZero \Pphi*$, and $\Peta[1405] \to \PfZero \eta$,
where the latter is a sub-threshold decay.  Among these, the
\Ppi[1800] is the only isovector state and thus accessible in the
$3\pi$ final state.  Because of its small width, the \PfZero accounts
for only a small fraction of the full \pipiSW.  It is easily separated
from the broad \pipiSW structure, which is shown in
\cref{fig:pipiS_M_param}.  Compared to the positive-reflectivity waves
containing the \Prho isobar, those with the \PfZero are suppressed by
a factor of approximately 20.  \Cref{fig:f0980_spin_totals} shows the
\tpr-summed intensity of the coherent sum of all partial waves with an
\PfZero isobar and positive reflectivity, which amounts to a relative
intensity of \SI{3.3}{\percent}.

\begin{figure}[tbp]
  \centering
  \includegraphics[width=\twoPlotWidth]{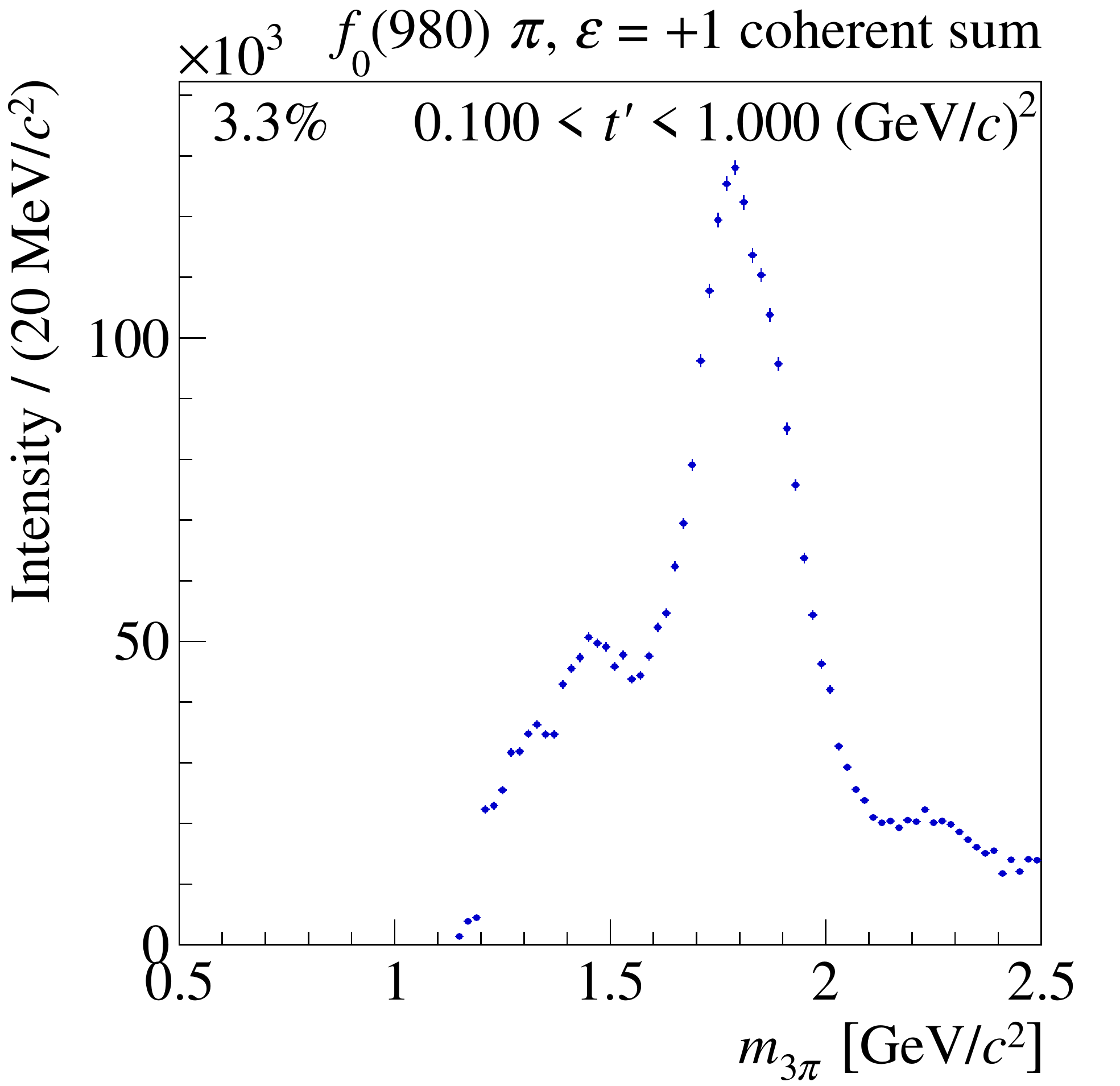}
  \caption{Intensity of the coherent sum of all $\PfZero[980]\,\pi$
    waves with positive reflectivity, summed over all \tpr bins.}
  \label{fig:f0980_spin_totals}
\end{figure}

The intensity distribution of the \wave{0}{-+}{0}{+}{\PfZero}{S} wave
is dominated by the \Ppi[1800] peak (see \cref{fig:0mp_f0980}).  The
more complicated mass spectrum of the \wave{2}{-+}{0}{+}{\PfZero}{D}
wave is shown in \cref{fig:2mp_f0980}.  This wave, which should
contain signals of \PpiTwo[1670] and \PpiTwo[1880], is characterized
by pronounced destructive interference around
$\mThreePi = \SI{1.8}{\GeVcc}$.

\begin{figure}[tbp]
  \centering
  \subfloat[][]{%
    \label{fig:0mp_pipiS_t_bin_low}%
    \includegraphics[width=\twoPlotWidth]{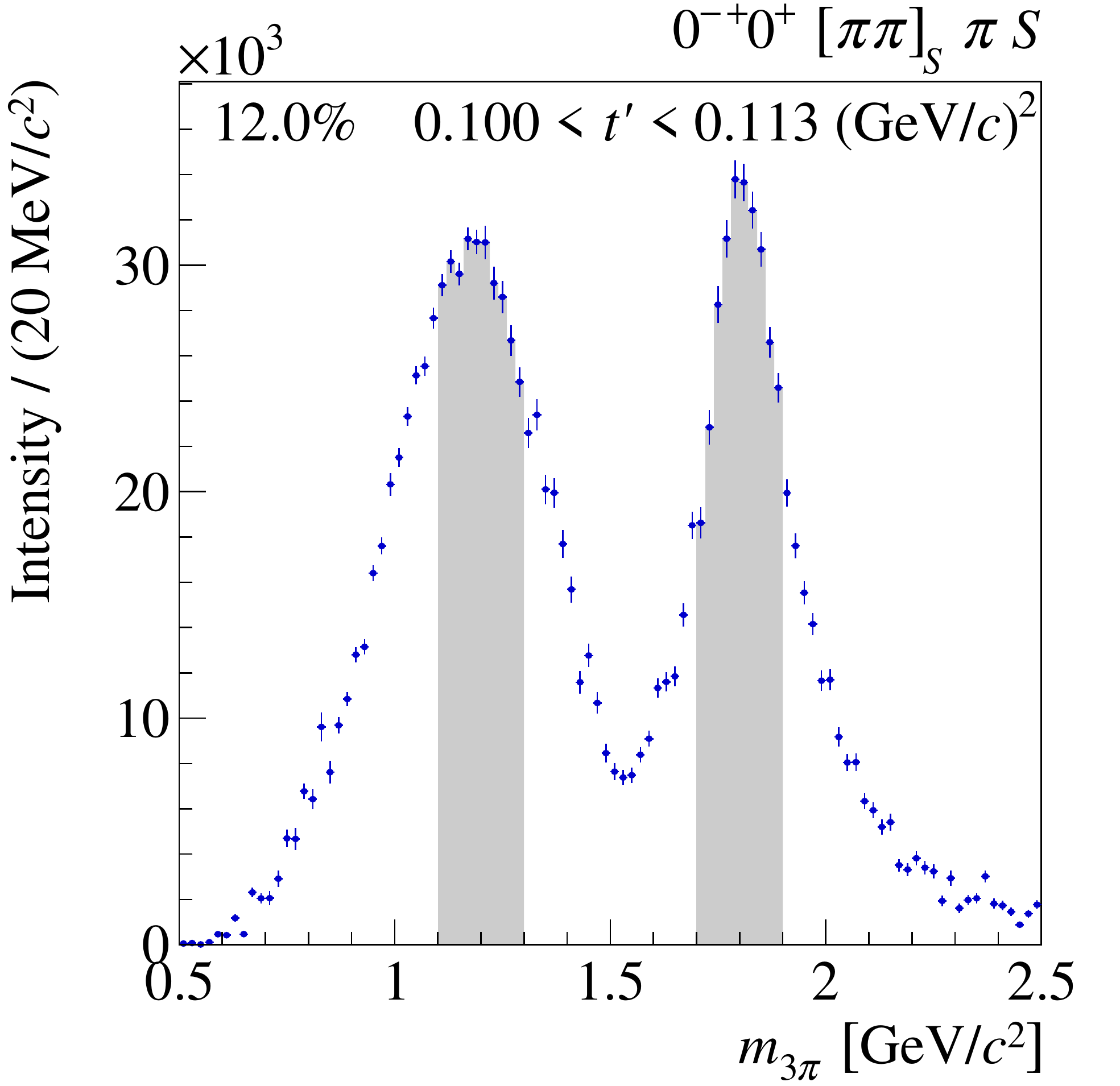}%
  }%
  \newLineOrHspace{\twoPlotSpacing}%
  \subfloat[][]{%
    \label{fig:0mp_pipiS_t_bin_high}%
    \includegraphics[width=\twoPlotWidth]{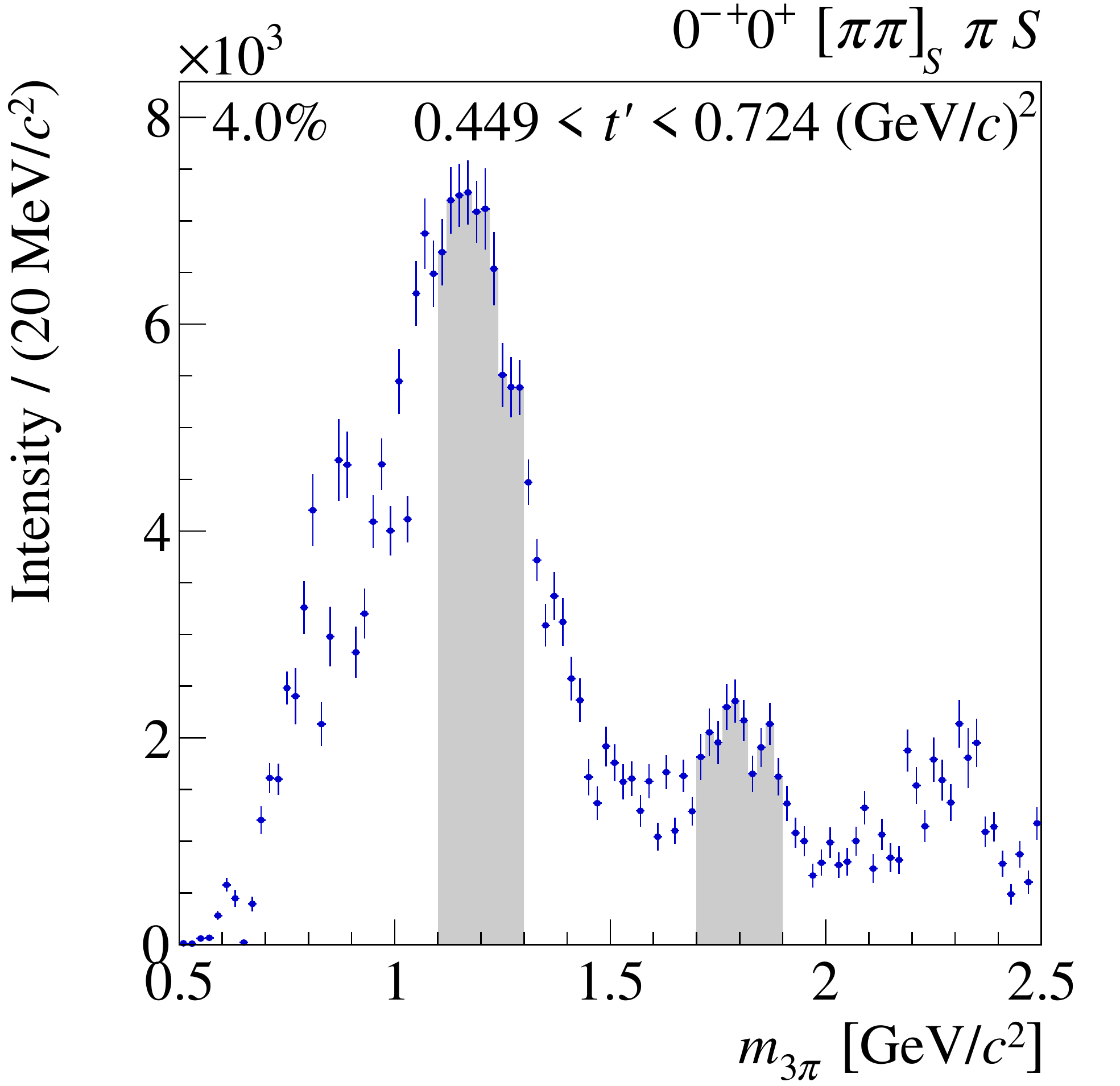}%
  }%
  \caption{Intensity of the \wave{0}{-+}{0}{+}{\pipiS}{S} wave in two
    different \tpr regions.  (a)~low \tpr; (b)~high \tpr. The shaded
    regions indicate the mass intervals that are integrated over to
    generate the \tpr spectra.}
  \label{fig:0mp_pipiS_t_bins}
\end{figure}

\begin{figure*}[htbp]
  \centering
  \subfloat[][]{%
    \label{fig:0mp_f0980}%
    \includegraphics[width=\twoPlotWidth]{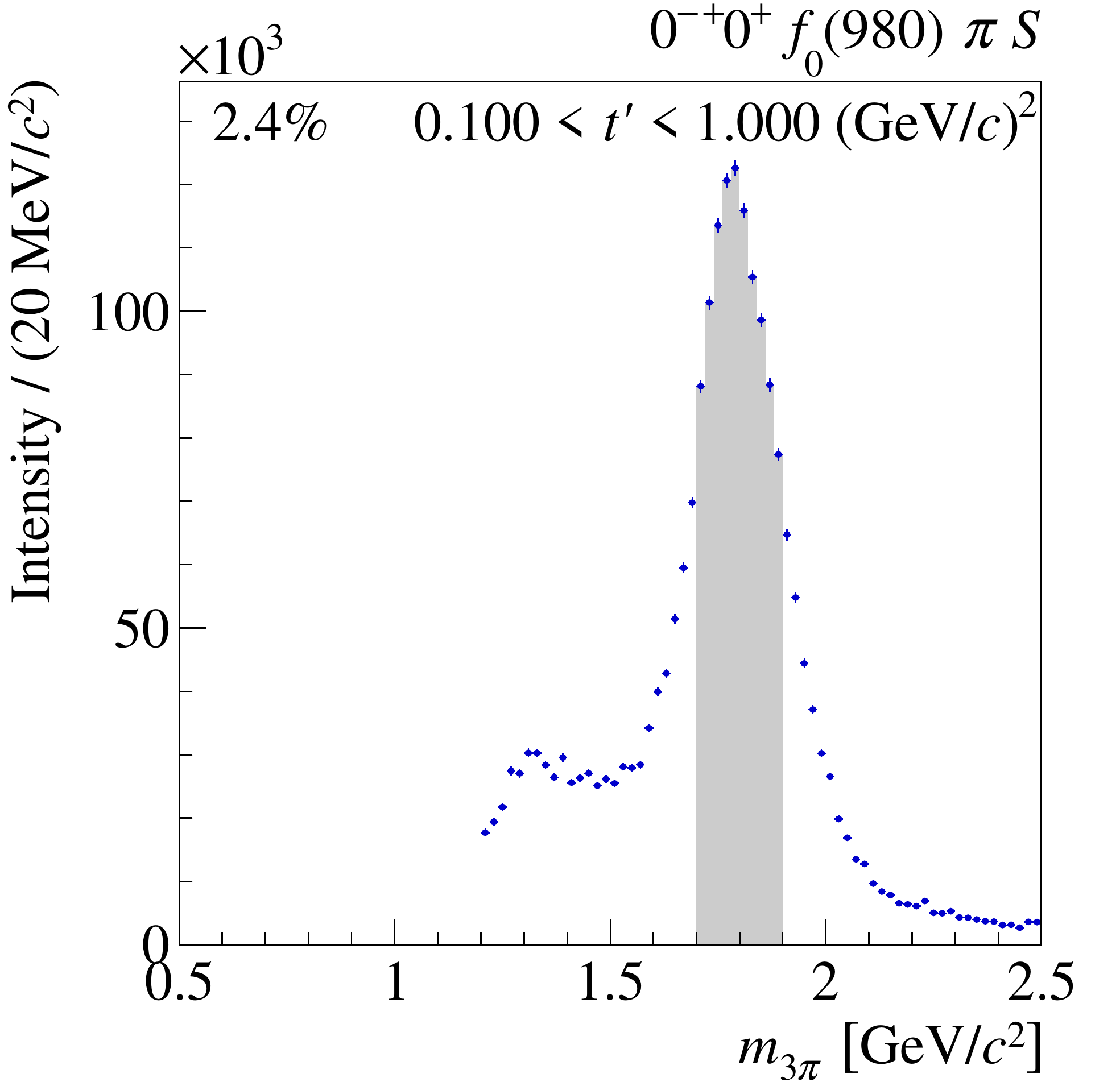}%
  }%
  \hspace*{\twoPlotSpacing}
  \subfloat[][]{%
    \label{fig:0mp_pipiS}%
    \includegraphics[width=\twoPlotWidth]{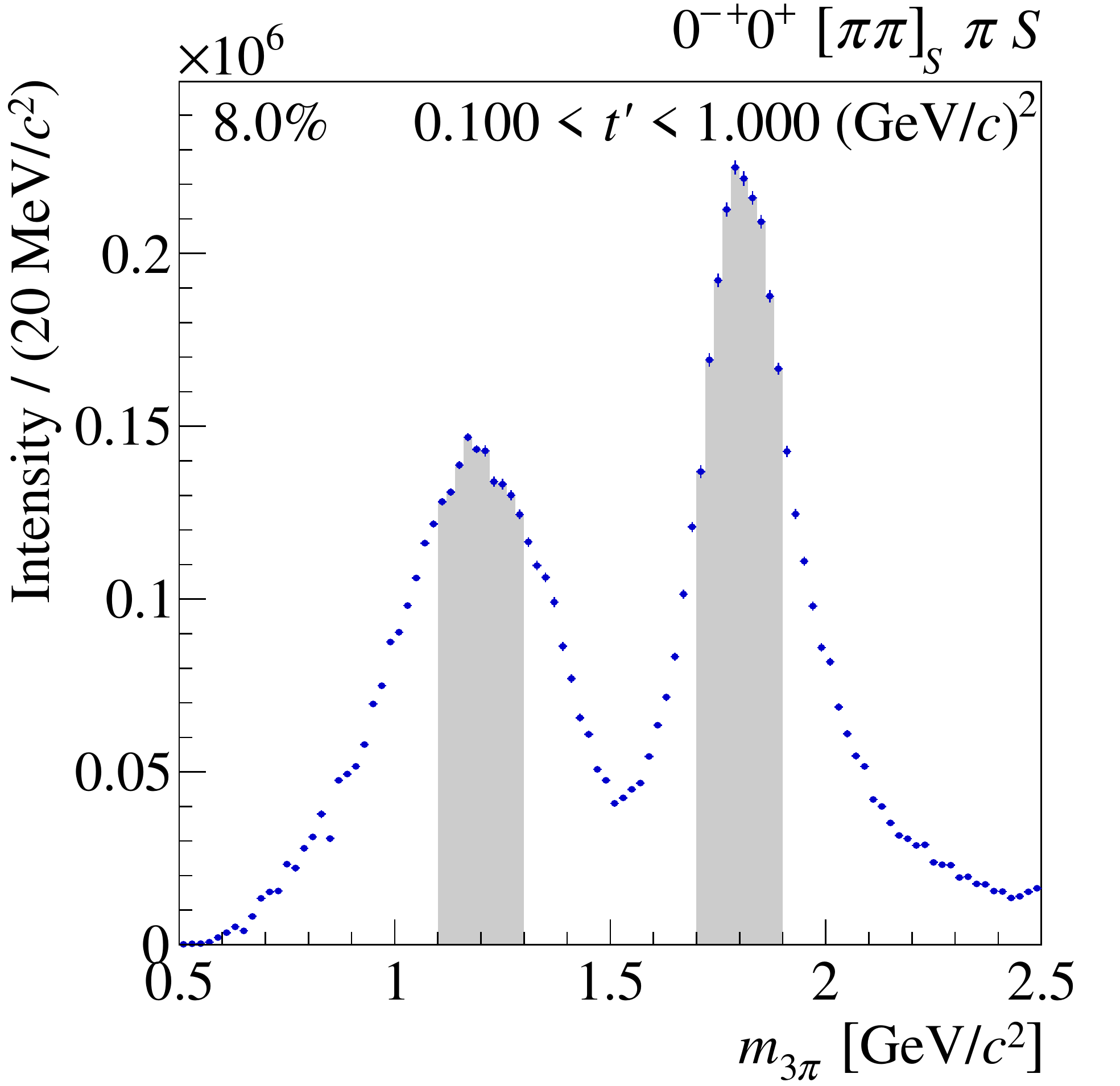}%
  }%
  \\
  \subfloat[][]{%
    \label{fig:2mp_f0980}%
    \includegraphics[width=\twoPlotWidth]{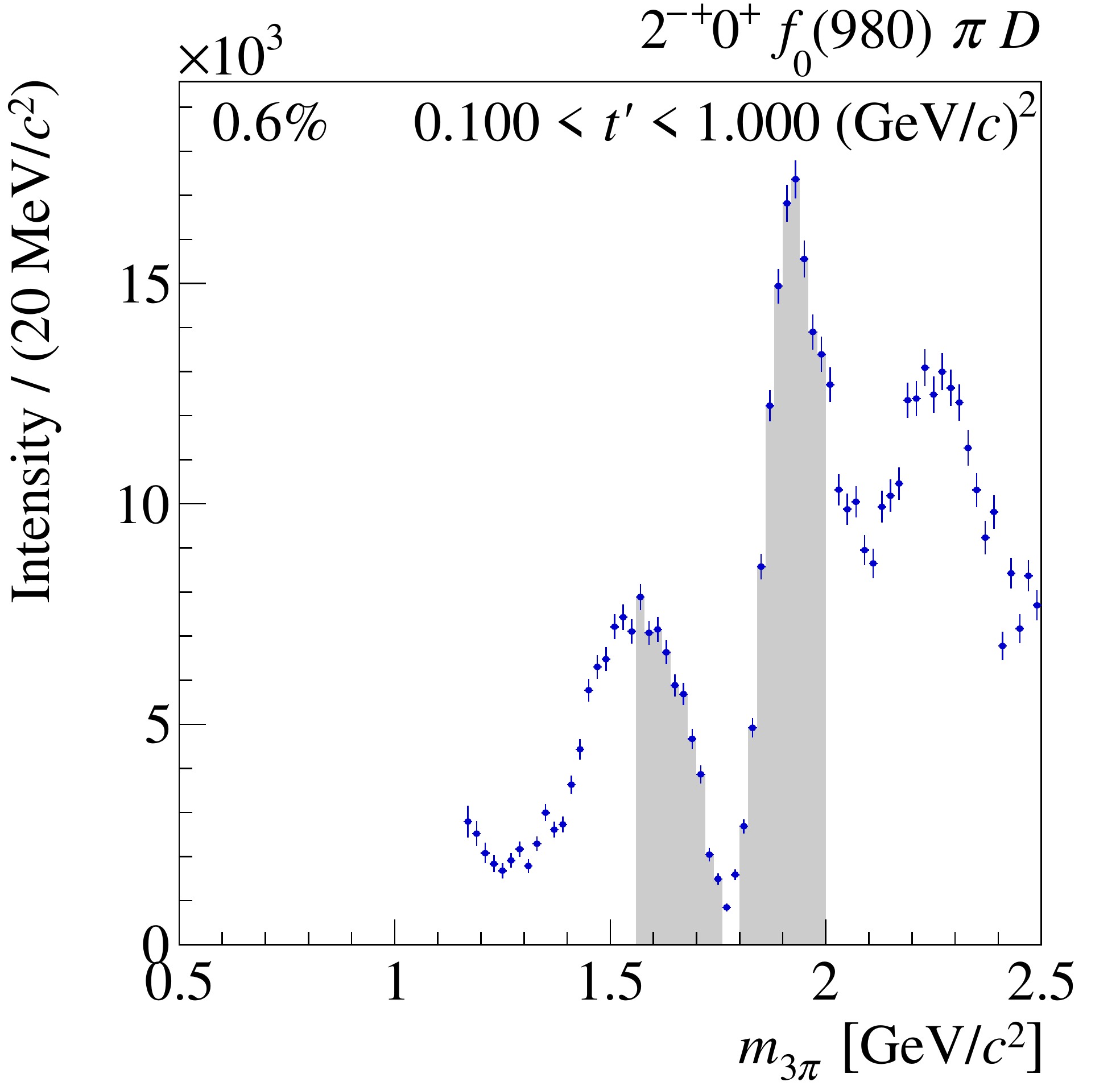}%
  }%
  \hspace*{\twoPlotSpacing}
  \subfloat[][]{%
    \label{fig:2mp_pipiS}%
    \includegraphics[width=\twoPlotWidth]{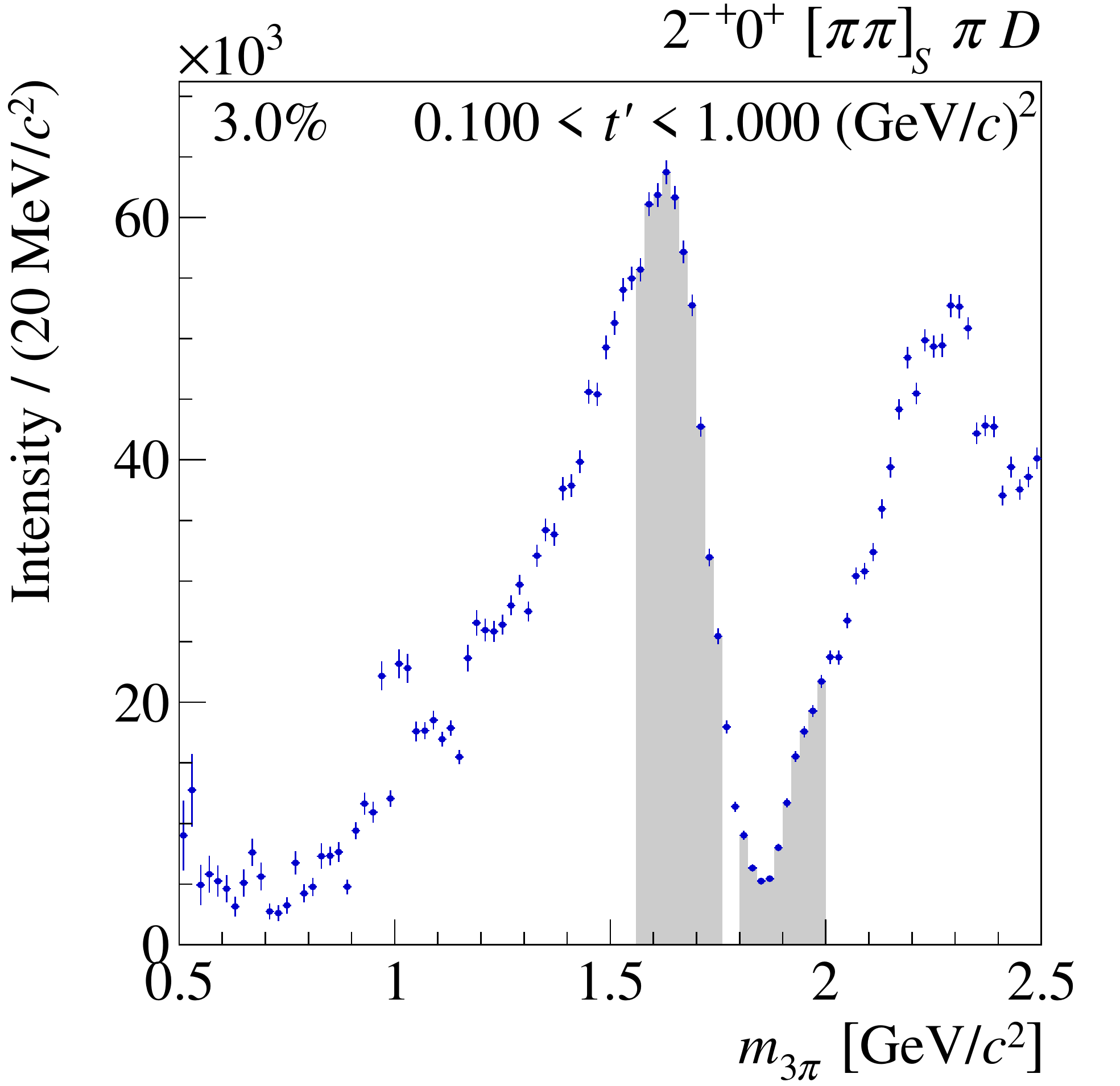}%
  }%
  \\
  \subfloat[][]{%
    \label{fig:1pp_f0980}%
    \includegraphics[width=\twoPlotWidth]{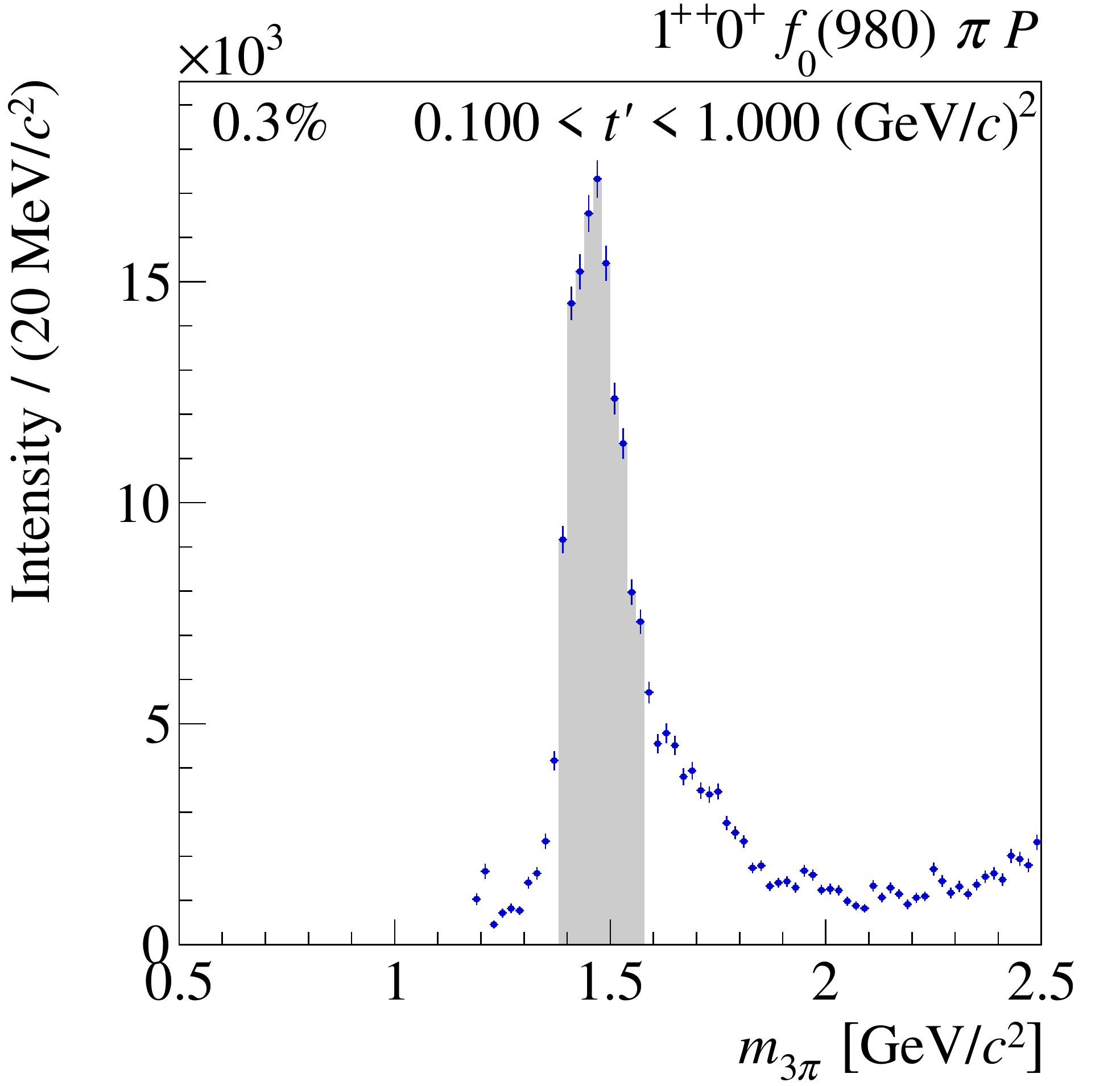}%
  }%
  \hspace*{\twoPlotSpacing}
  \subfloat[][]{%
    \label{fig:1pp_pipiS}%
    \includegraphics[width=\twoPlotWidth]{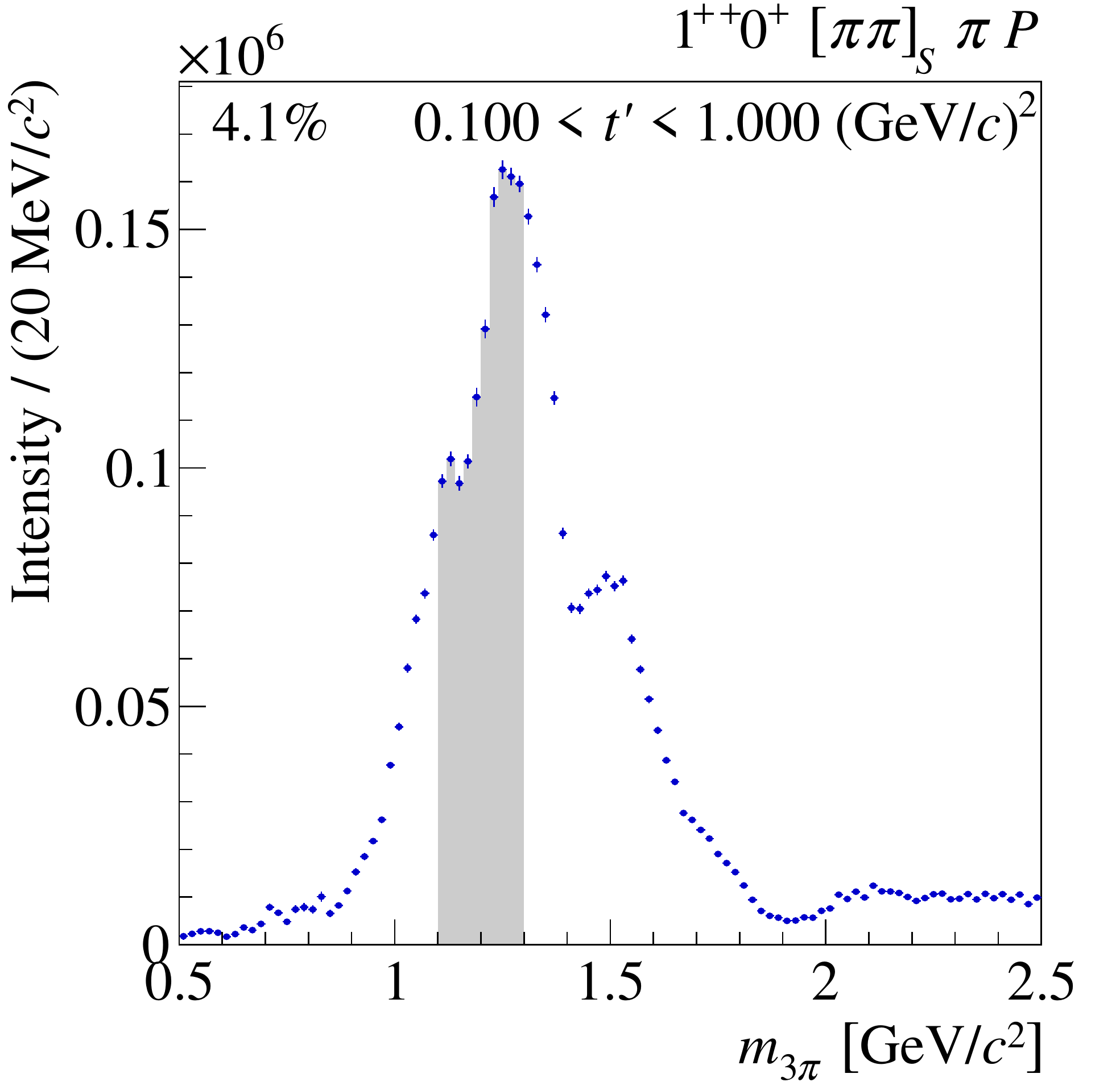}%
  }%
  \caption{The \tpr-summed intensity for selected waves with \pipiSW
    isobars.  Left column: waves with the narrow \PfZero isobar; Right
    column: waves with the broad \pipiS isobar.  The $\JPC = 0^{-+}$
    waves in the top row show the \Ppi[1800].  The structure at
    \SI{1.2}{\GeVcc} is probably mainly of nonresonant origin.  The
    $2^{-+}$ intensities in the center row exhibit a complicated
    destructive interference pattern around \SI{1.8}{\GeVcc}.  The
    bottom row shows an enhancement in the region of the \PaOne in the
    \wave{1}{++}{0}{+}{\pipiS}{P} wave and a new state, the
    \PaOne[1420], in \wave{1}{++}{0}{+}{\PfZero}{P}.  The shaded
    regions indicate the mass intervals that are integrated over to
    generate the \tpr spectra (see
    \cref{fig:t_dependence_PIPIS_pi_0mp,fig:t_dependence_PIPIS_pi_1pp}).}
  \label{fig:pipi_s-wave}
\end{figure*}

The \wave{1}{++}{0}{+}{\PfZero}{P} wave is shown in
\cref{fig:1pp_f0980}. It exhibits the new axial-vector meson
\PaOne[1420].  The resonance features of this signal were presented in
\refCite{Adolph:2015pws}.  It should be noted that the intensity of
this wave corresponds to only about \SI{0.3}{\percent} of the total
number of events.  Since the signal is very small, we conducted
several systematic studies that will be discussed in
\cref{sec:pwa_massindep_systematic_studies}.

The same partial waves discussed above are shown in the right column
of \cref{fig:pipi_s-wave} for the broad component of the \pipiSW as
isobar, which is parametrized as described in
\cref{sec:pwa_method_isobar_parametrization} and denoted by \pipiS.
The intensity spectrum of the \wave{0}{-+}{0}{+}{\pipiS}{S} wave (see
\cref{fig:0mp_pipiS}) exhibits two pronounced maxima and differs
considerably from that of the corresponding $\PfZero[980]\,\pi$ wave
in \cref{fig:0mp_f0980}.  The maximum at \SI{1.8}{\GeVcc} corresponds
to the \Ppi[1800], but we also observe a broad structure around
\SI{1.2}{\GeVcc}, which could contain the \Ppi[1300].  As it will be
discussed in \cref{sec:tprim_partial_waves}, the latter structure has
a very distinct dependence on \tpr with a minimum around
$\tpr \approx \SI{0.35}{\GeVcsq}$.  \Cref{fig:0mp_pipiS_t_bins} shows
as an example the intensity spectrum in two \tpr bins.  At high \tpr,
the \Ppi[1800] peak nearly vanishes and the structure around
\SI{1.2}{\GeVcc} is shifted towards lower \mThreePi.  A more detailed
analysis discussed in
\cref{sec:results_free_pipi_s_wave_int_correlations} indicates that in
addition to interference effects with the \Ppi[1800] also nonresonant
processes seem to contribute to the \SI{1.2}{\GeVcc} mass region.

Another resonance that was observed to couple to the \pipiS isobar is
the \PpiTwo~\cite{Agashe:2014kda}, whose main decay mode into
$\PfTwo\,\pi$ is discussed above in
\cref{sec:results_pwa_massindep_major_waves}.  The mass spectrum of
the \wave{2}{-+}{0}{+}{\pipiS}{D} wave is shown in
\cref{fig:2mp_pipiS} and exhibits marked destructive interference
effects at masses around \SI{1.8}{\GeVcc}, similar to the ones
observed in the corresponding wave with the \PfZero decay mode in
\cref{fig:2mp_f0980}.

The \wave{1}{++}{0}{+}{\pipiS}{P} wave is even more difficult to
interpret (see \cref{fig:1pp_pipiS}). A significant signal is observed
in the region of \PaOne.  However, as it will be shown in
\cref{sec:tprim_partial_waves} and
\cref{sec:results_free_pipi_s_wave}, this structure exhibits a strong
\tpr dependence, which is a signature for significant nonresonant
contributions.

Comparing the $\PfZero[980]\,\pi$ and $\pipiS\,\pi$ decay modes, the
latter are obviously more affected by nonresonant contributions.  We
will discuss the sector of partial waves with \pipiSW isobars again in
\cref{sec:results_free_pipi_s_wave} in the context of an extended
analysis.

\subsection{Comparison of Fit Result and Real Data}
\label{sec:pwa_massindep_weighted_mc}

In order to estimate the goodness of the \emph{mass-independent} fit,
three-pion phase-space Monte Carlo events, which were processed
through the detector simulation and reconstruction chain and satisfied
the selection criteria, were weighted with the intensity distribution
of the fit model [see \cref{eq:intensity_bin}].  For a good fit,
distributions obtained from these weighted Monte Carlo events are
expected to approximate the real data.

For fixed values of \mThreePi and \tpr, the phase space of the three
final-state particles is five-dimensional.  Therefore, we can show
only projections in certain kinematic regions.  For the comparison we
use the same five kinematic variables that also enter in the decay
amplitudes (see \cref{sec:isobar_model_amplitude}), \ie \cosThetaGJ
and \phiGJ of the isobar in the Gottfried-Jackson frame, the isobar
mass $m_\xi = \mTwoPi$, and \cosThetaHF and \phiHF of the $\pi^-$ in
the helicity frame.

\Crefrange{fig:wMC_low_t_overview}{fig:wMC_high_t_low_mX} show as examples
the distributions of the kinematic variables in various regions of
\mThreePi and \tpr.  These kinematic regions contain different
resonant and nonresonant contributions leading to different shapes of
the angular distributions and the isobar mass spectrum.

\begin{figure*}[htbp]
  \centering
  \includegraphics[width=\totalPlotWidth]{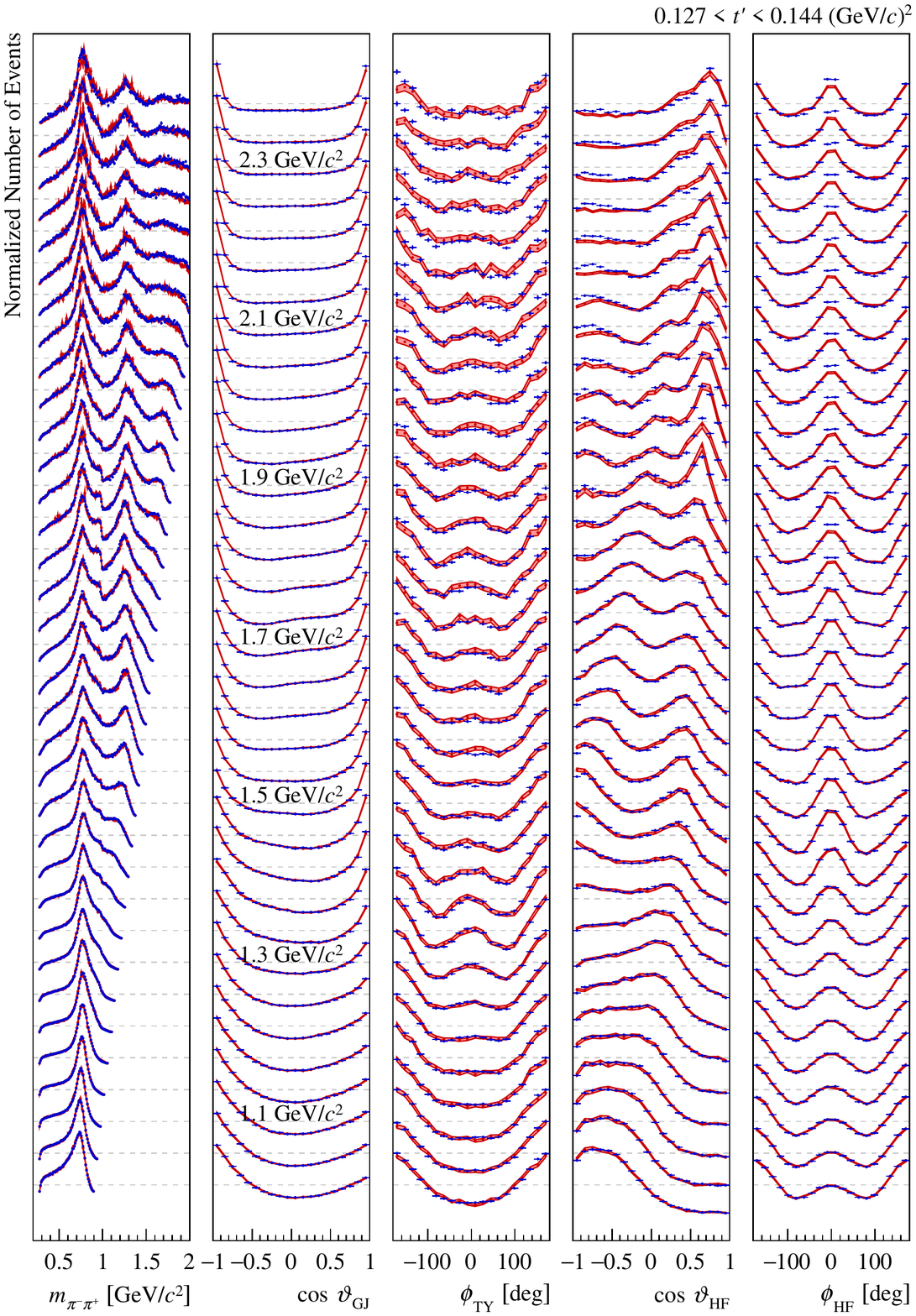}
  \caption{\colorPlot Distributions of the five phase-space variables
    used to calculate the decay amplitudes shown for different $3\pi$
    mass bins in the region \SIvalRange{0.127}{\tpr}{0.144}{\GeVcsq}.
    Each distribution is shown for real data (blue points) and for
    weighted Monte Carlo events (red bands), which
    are generated according to the fit result.
    Each distribution is normalized to its maximum deviation from its
    average $y$~value.  Along the ordinate, the average $y$~values for the distributions
    (indicated by gray lines) are shifted equidistantly
    with respect to one another.  The $3\pi$ mass ranges from
    \SIrange{1.0}{2.4}{\GeVcc} and is subdivided into \SI{40}{\MeVcc}
    wide bins.  The central values of selected $3\pi$ mass bins are
    given as labels in the \cosThetaGJ distribution.}
  \label{fig:wMC_low_t_overview}
\end{figure*}

\begin{figure*}[htbp]
  \centering
  \subfloat[][]{%
    \label{fig:wMC_low_t_massX}%
    \includegraphics[width=\twoPlotWidthOneD]{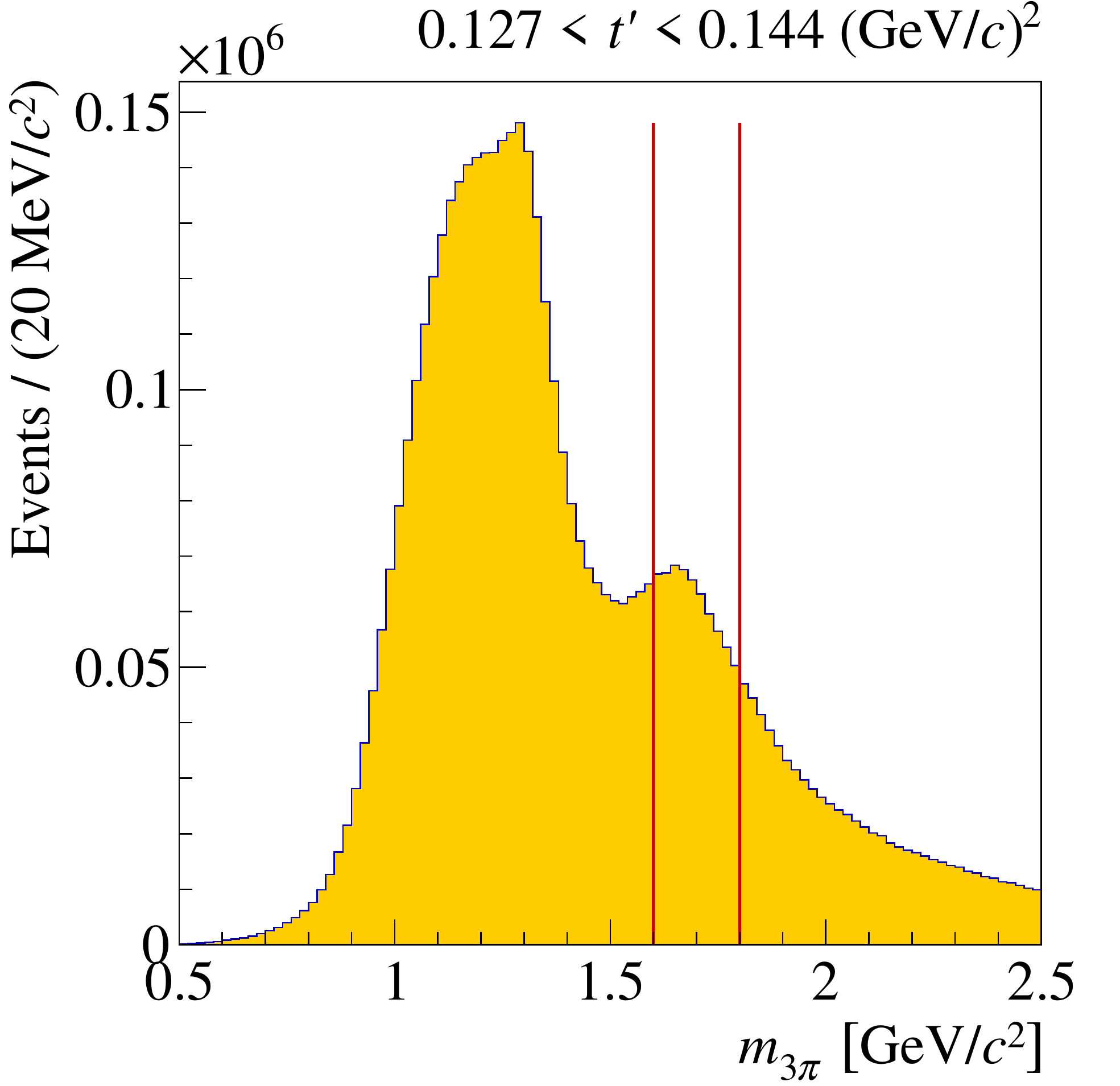}%
  }%
  \hspace*{\twoPlotSpacingOneD}
  \subfloat[][]{%
    \label{fig:wMC_low_t_isobar}%
    \includegraphics[width=\twoPlotWidthOneD]{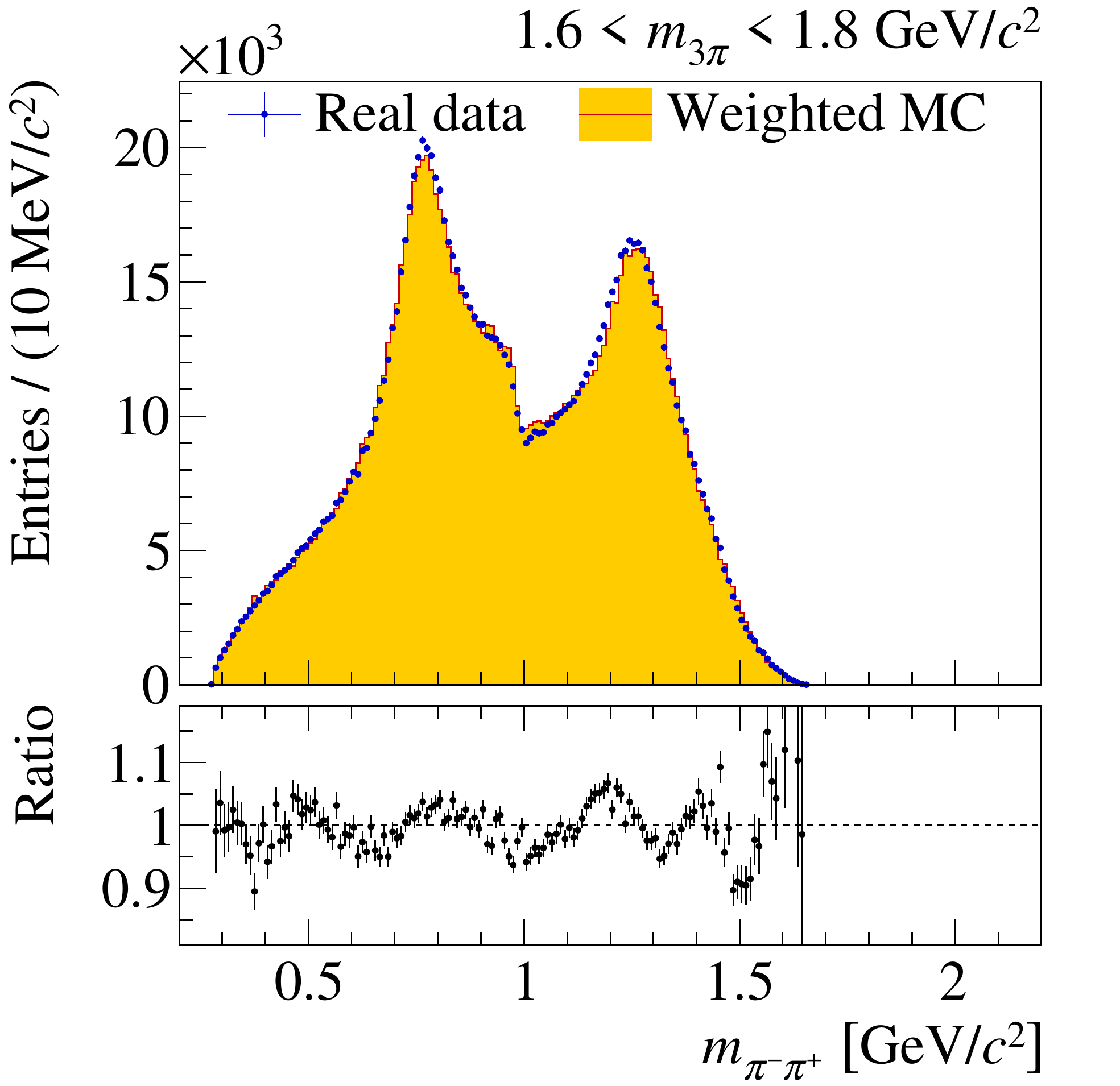}%
  }%
  \hspace*{\twoPlotSpacingOneD}\hspace*{0.5em}
  \\
  \subfloat[][]{%
    \label{fig:wMC_low_t_GJ_data}%
    \includegraphics[width=\twoPlotWidthTwoD]{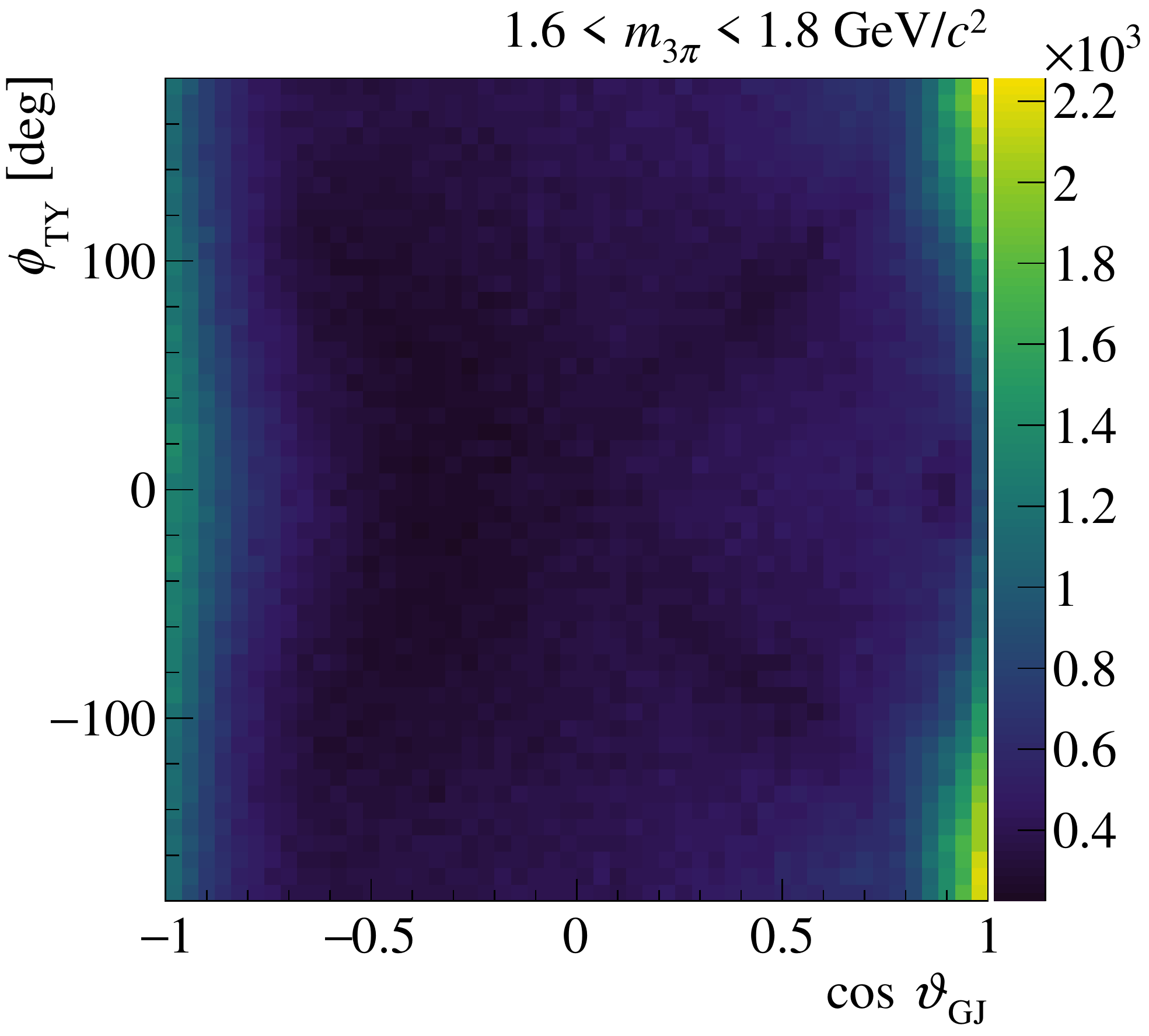}%
  }%
  \subfloat[][]{%
    \label{fig:wMC_low_t_HF_data}%
    \includegraphics[width=\twoPlotWidthTwoD]{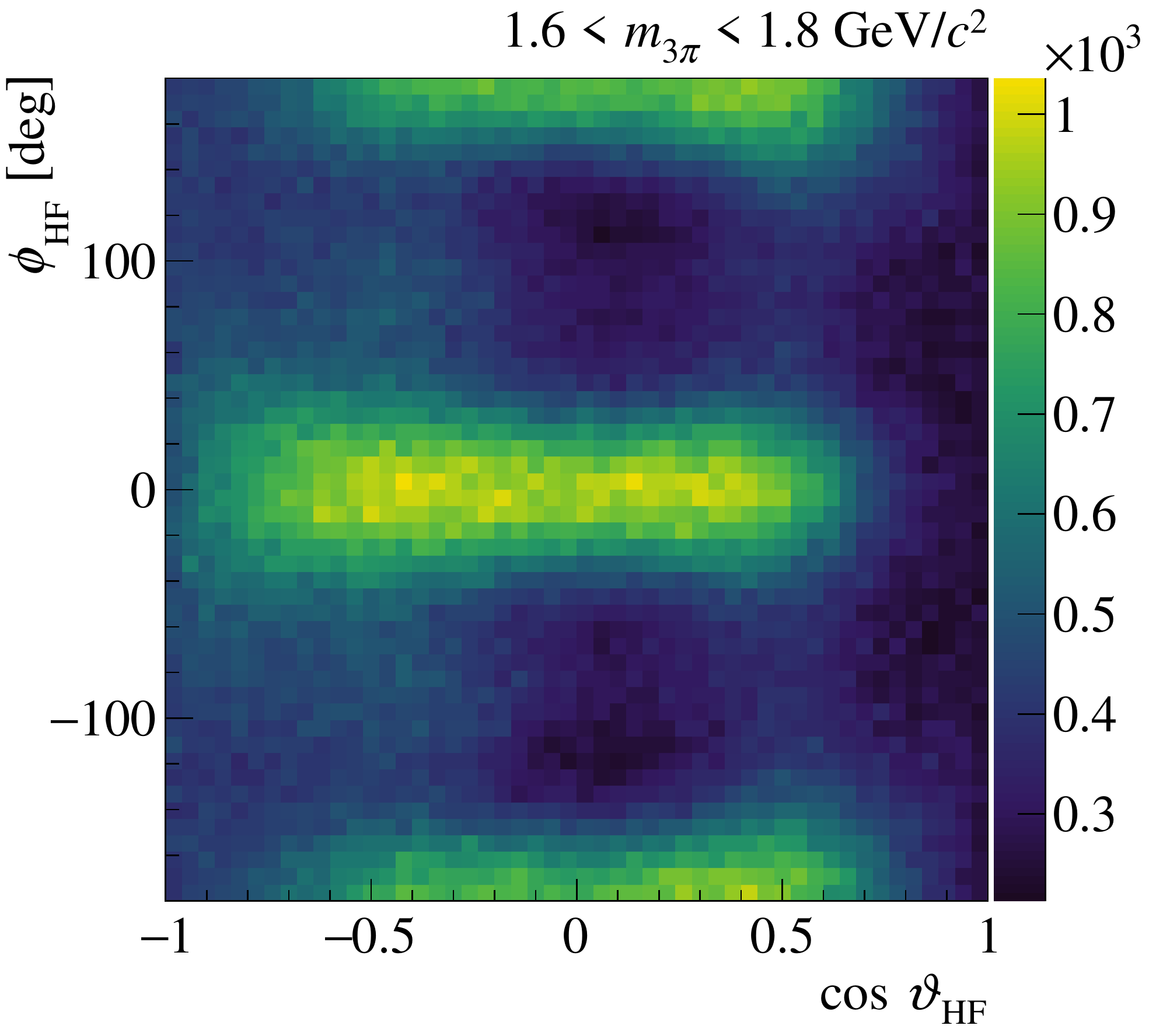}%
  }%
  \\
  \subfloat[][]{%
    \label{fig:wMC_low_t_ratio_GJ}%
    \includegraphics[width=\twoPlotWidthTwoD]{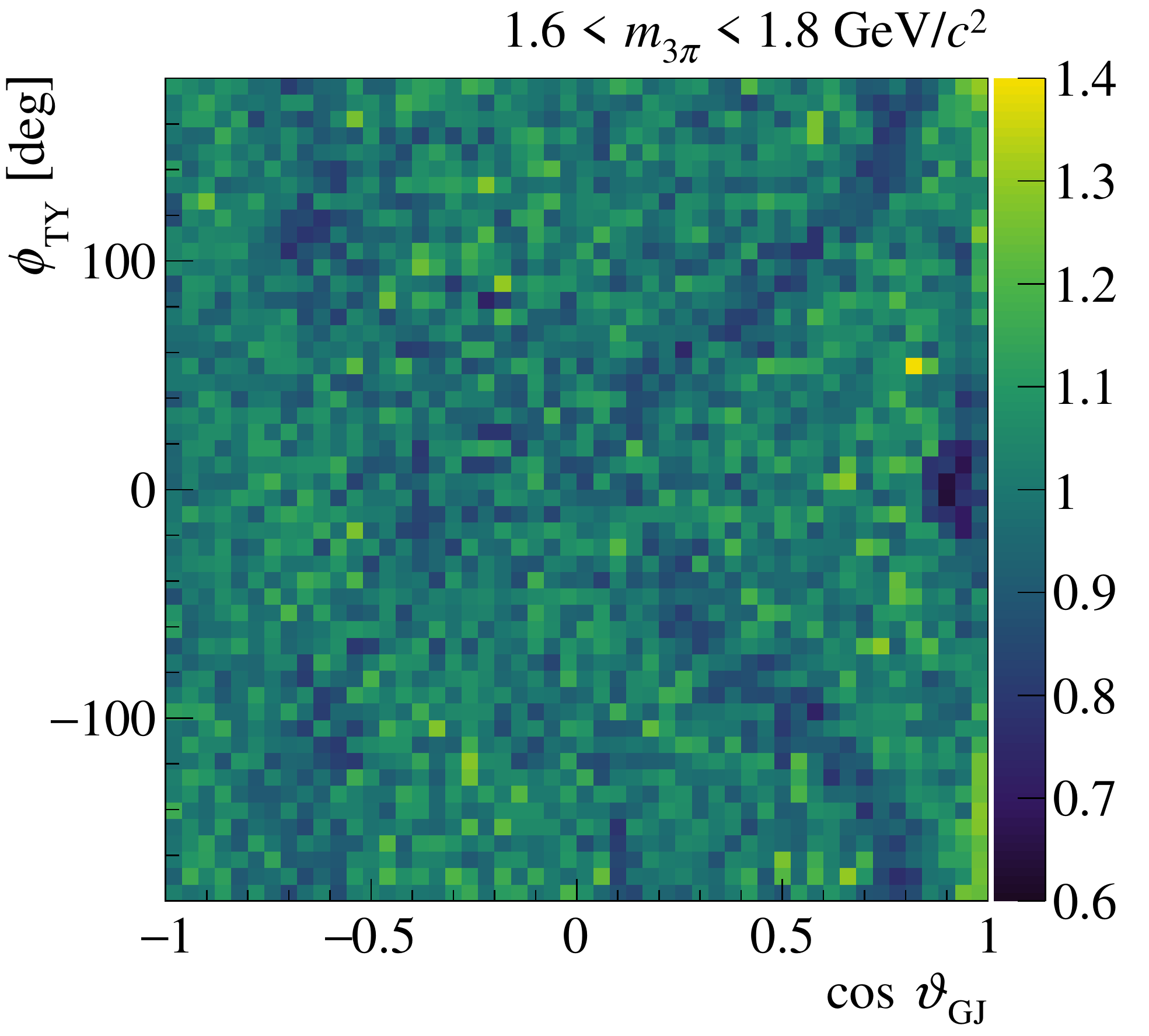}%
  }%
  \subfloat[][]{%
    \label{fig:wMC_low_t_ratio_HF}%
    \includegraphics[width=\twoPlotWidthTwoD]{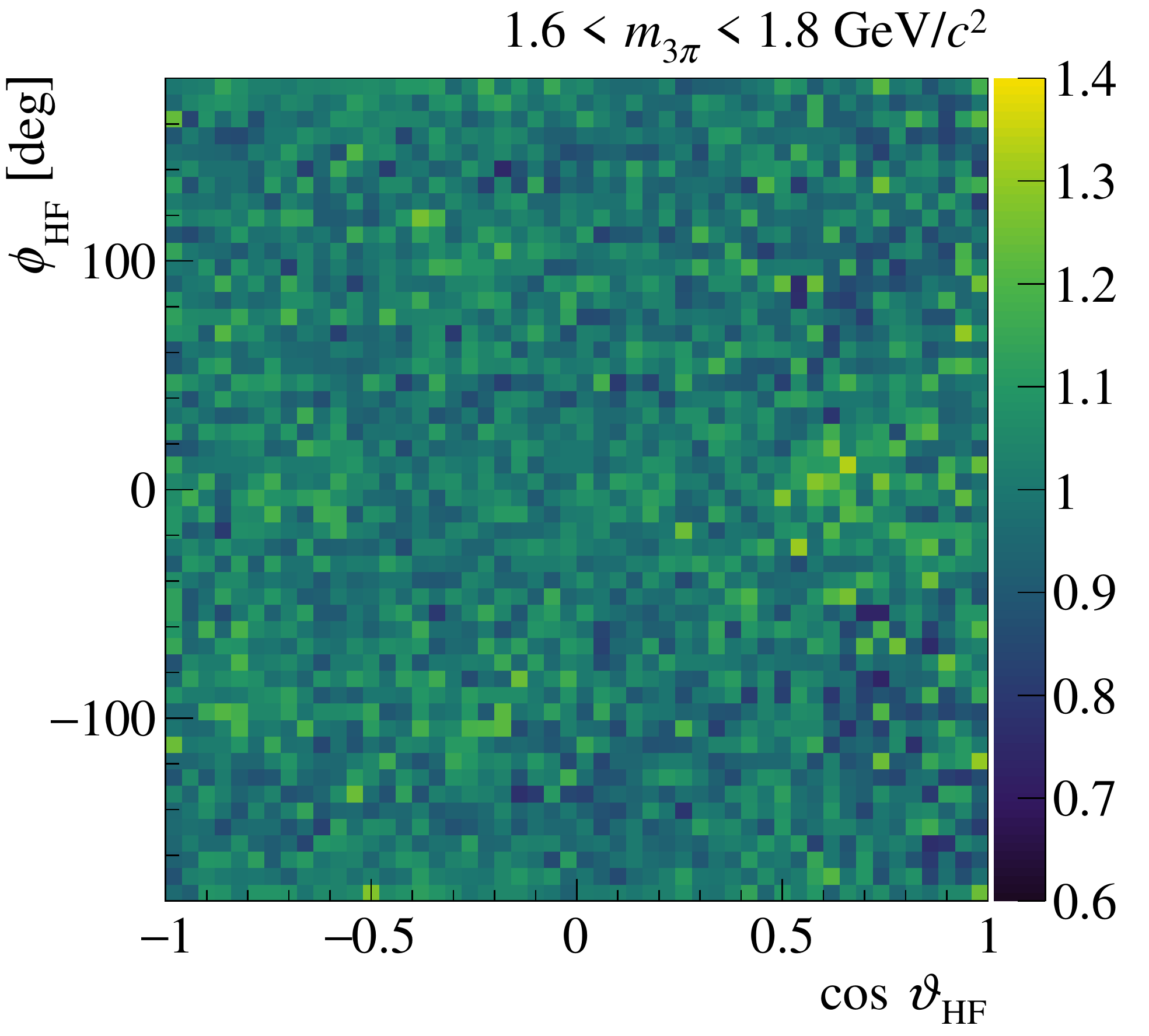}%
  }%
  \caption{\colorPlot Comparison of kinematic distributions of
    weighted Monte Carlo events, generated according to the fit
    result, with the corresponding real-data distributions in the
    low-\tpr region \SIvalRange{0.127}{\tpr}{0.144}{\GeVcsq}.
    Panel~(a) shows the acceptance-corrected $3\pi$ invariant mass
    distribution.  The other panels show kinematic distributions in
    the mass interval \SIvalRange{1.6}{\mThreePi}{1.8}{\GeVcc} around
    the \PpiTwo, which is indicated by vertical red lines in~(a).
    (b)~invariant mass spectrum of the \twoPi subsystem;
    (c)~distribution of the Gottfried-Jackson angles for real data;
    (e)~ratio of the real-data distribution in~(c) and that of the
    weighted Monte Carlo.  Panels~(d) and~(f) show the respective
    distributions for the helicity angles.  Note that (b), (c), and
    (d) have two entries per event.}
  \label{fig:wMC_low_t_high_mX}
\end{figure*}

\begin{figure*}[htbp]
  \centering
  \includegraphics[width=\totalPlotWidth]{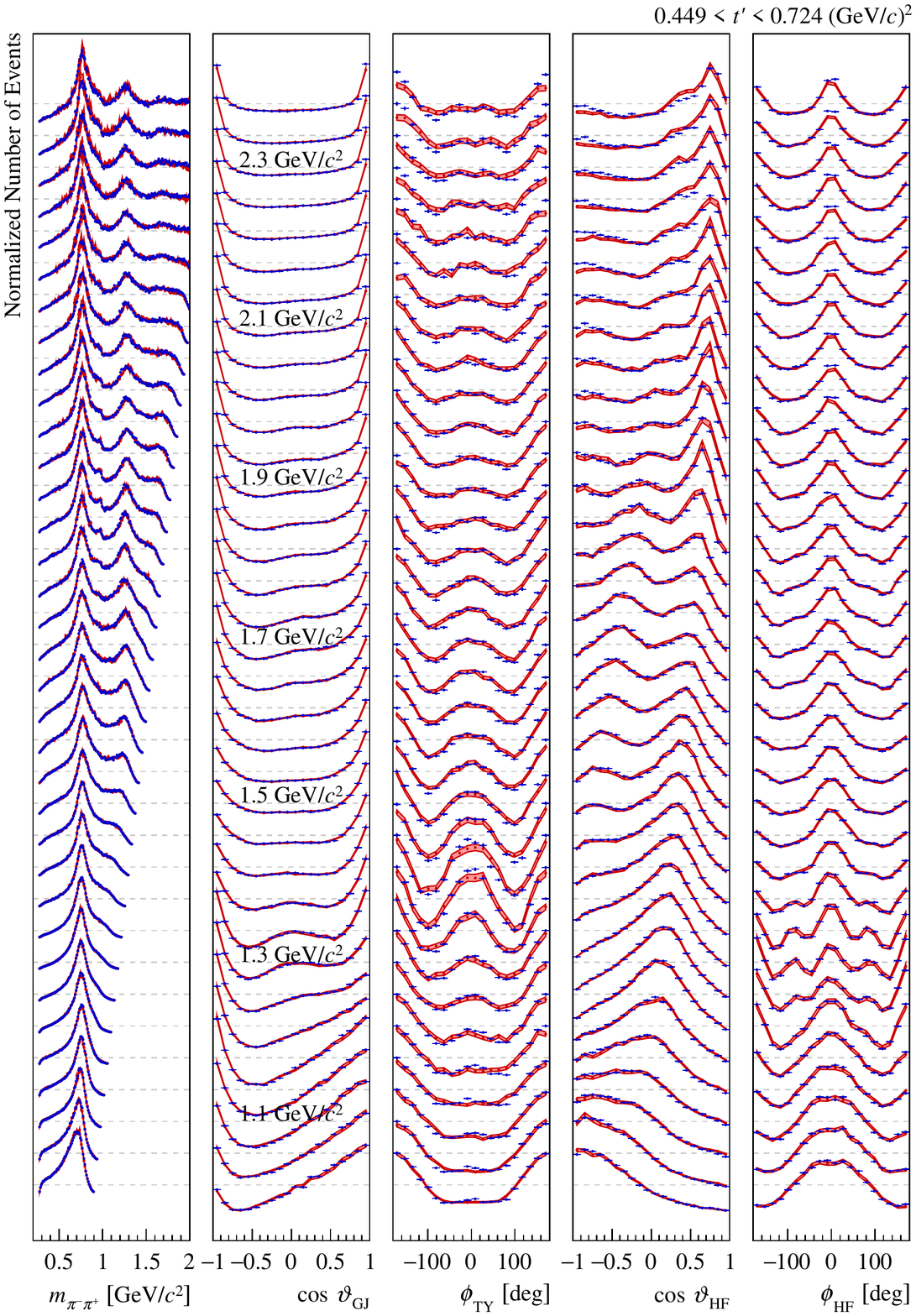}
  \caption{\colorPlot Same as \cref{fig:wMC_low_t_overview}, but for
    the high-\tpr region \SIvalRange{0.449}{\tpr}{0.724}{\GeVcsq}.}
  \label{fig:wMC_high_t_overview}
\end{figure*}

\begin{figure*}[htbp]
  \centering
  \subfloat[][]{%
    \label{fig:wMC_high_t_massX}%
    \includegraphics[width=\twoPlotWidthOneD]{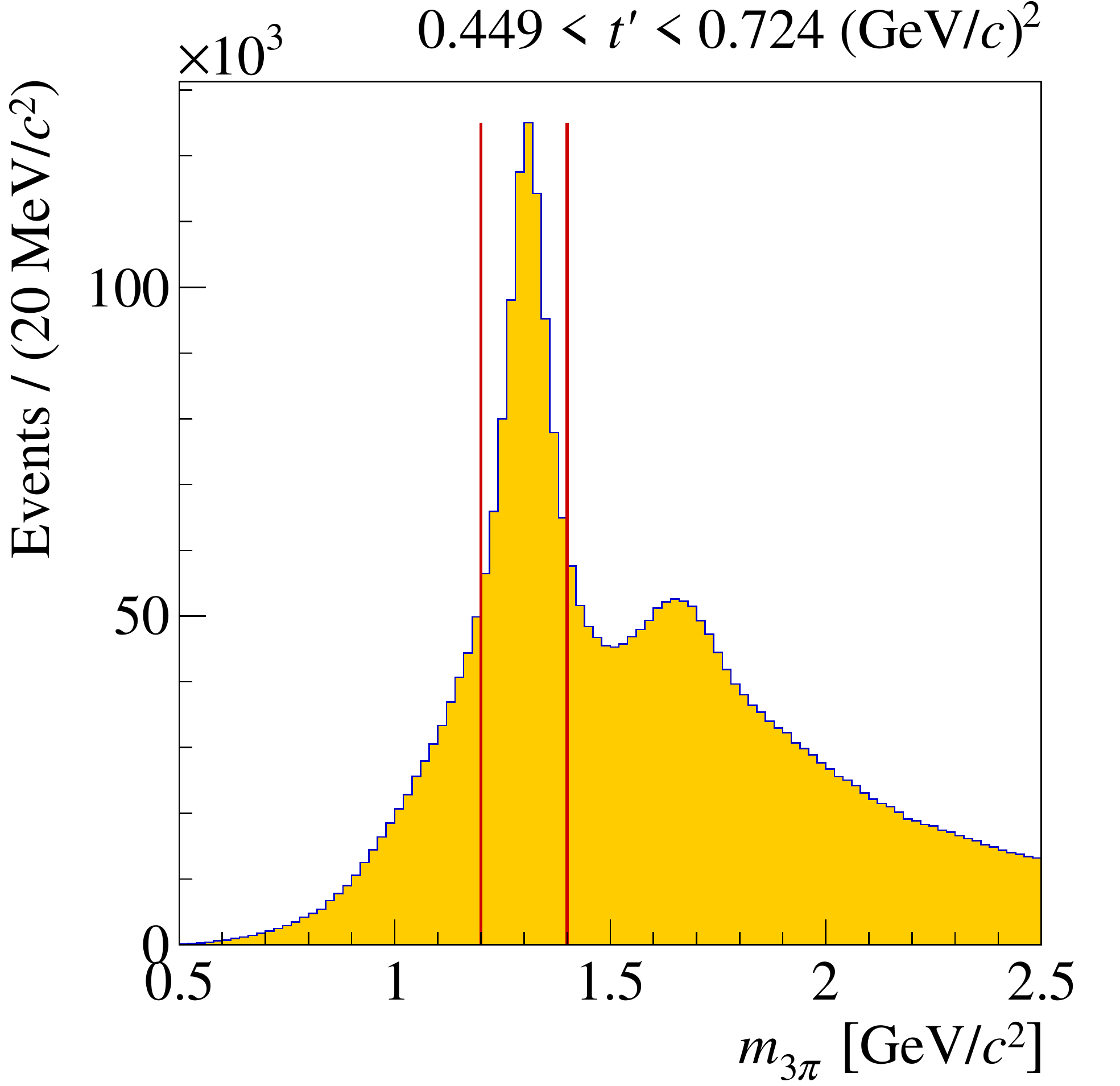}%
  }%
  \hspace*{\twoPlotSpacingOneD}
  \subfloat[][]{%
    \label{fig:wMC_high_t_isobar}%
    \includegraphics[width=\twoPlotWidthOneD]{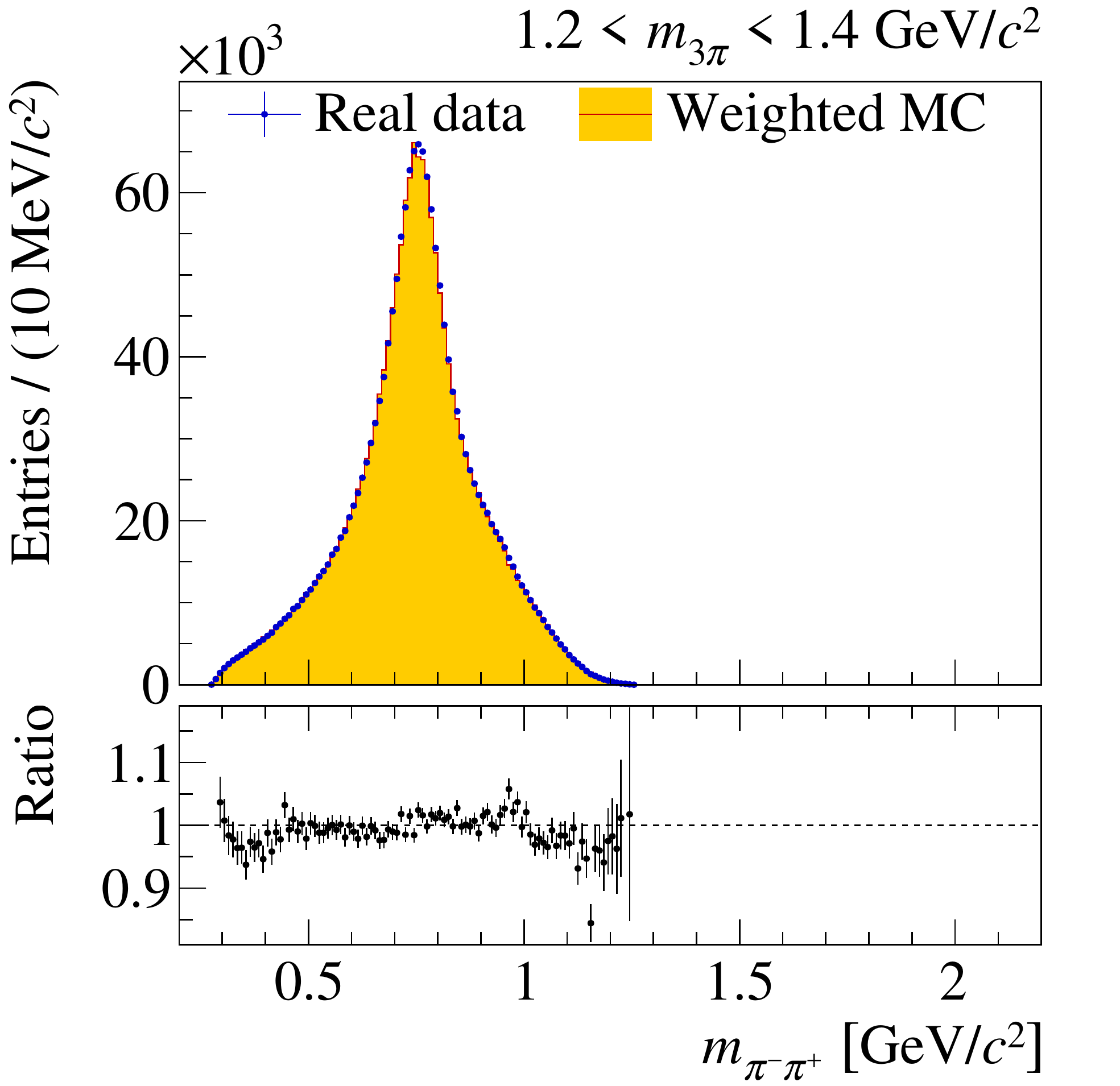}%
  }%
  \hspace*{\twoPlotSpacingOneD}\hspace*{0.5em}
  \\
  \subfloat[][]{%
    \label{fig:wMC_high_t_GJ_data}%
    \includegraphics[width=\twoPlotWidthTwoD]{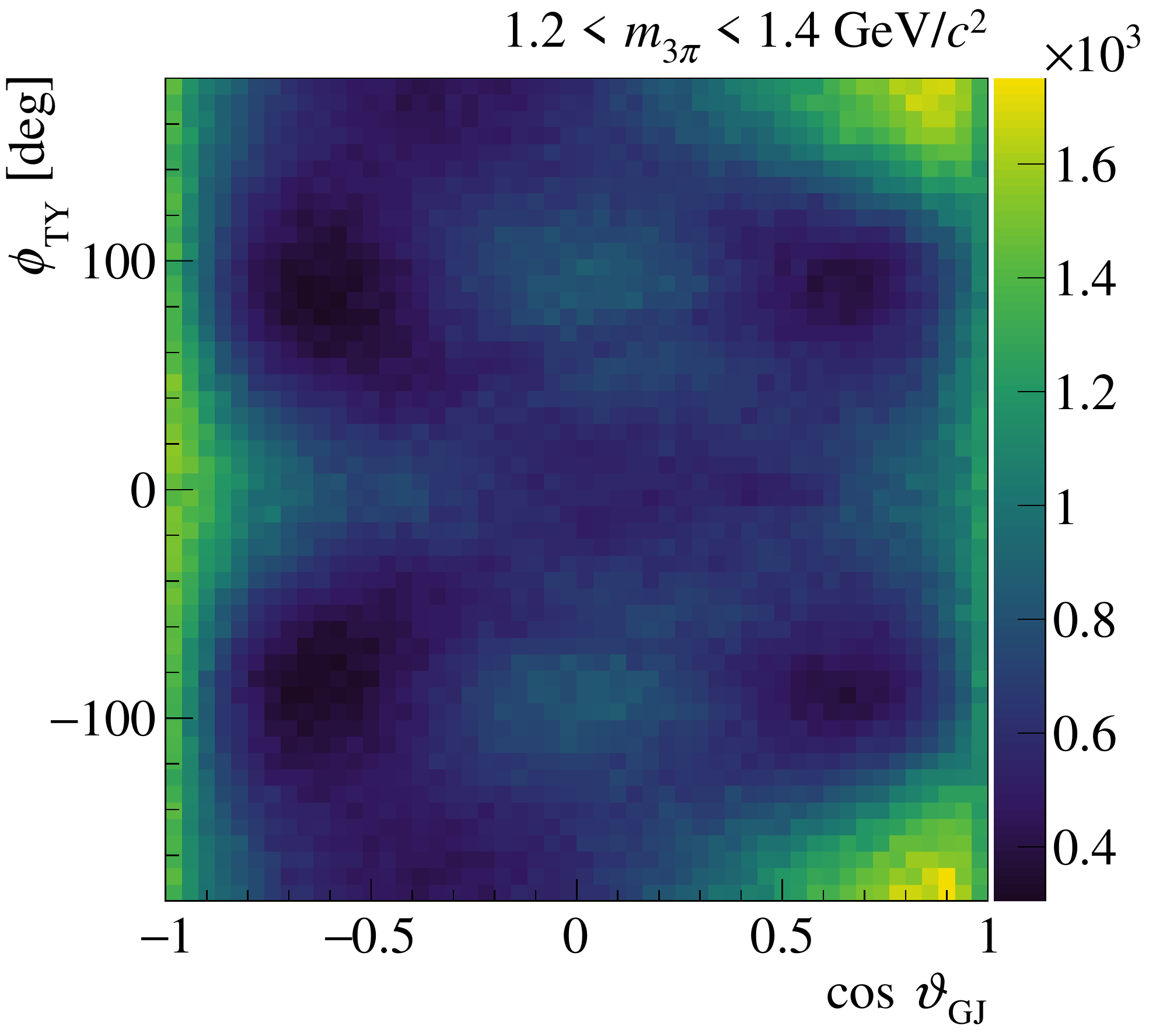}%
  }%
  \subfloat[][]{%
    \label{fig:wMC_high_t_HF_data}%
    \includegraphics[width=\twoPlotWidthTwoD]{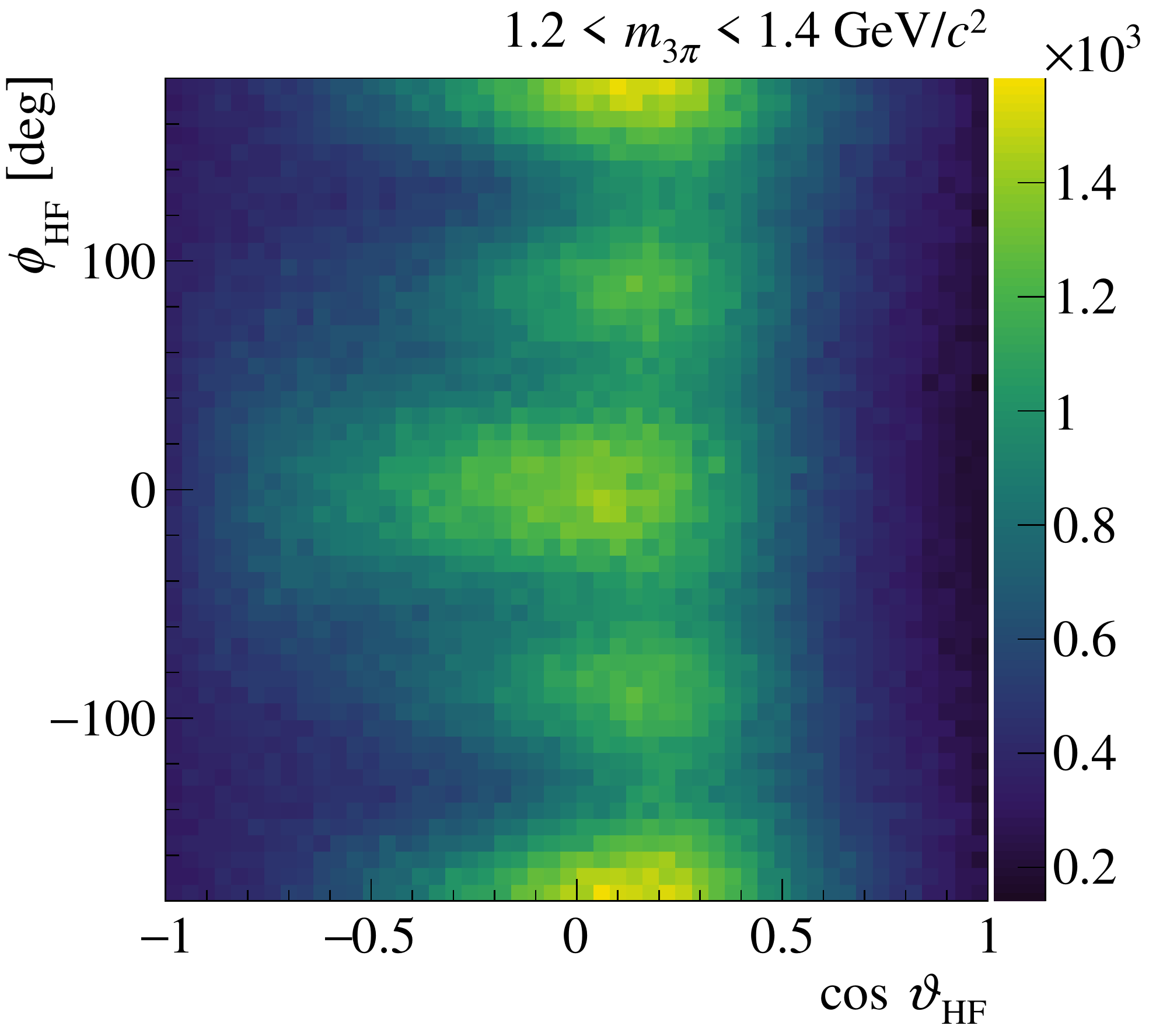}%
  }%
  \\
  \subfloat[][]{%
    \label{fig:wMC_high_t_ratio_GJ}%
    \includegraphics[width=\twoPlotWidthTwoD]{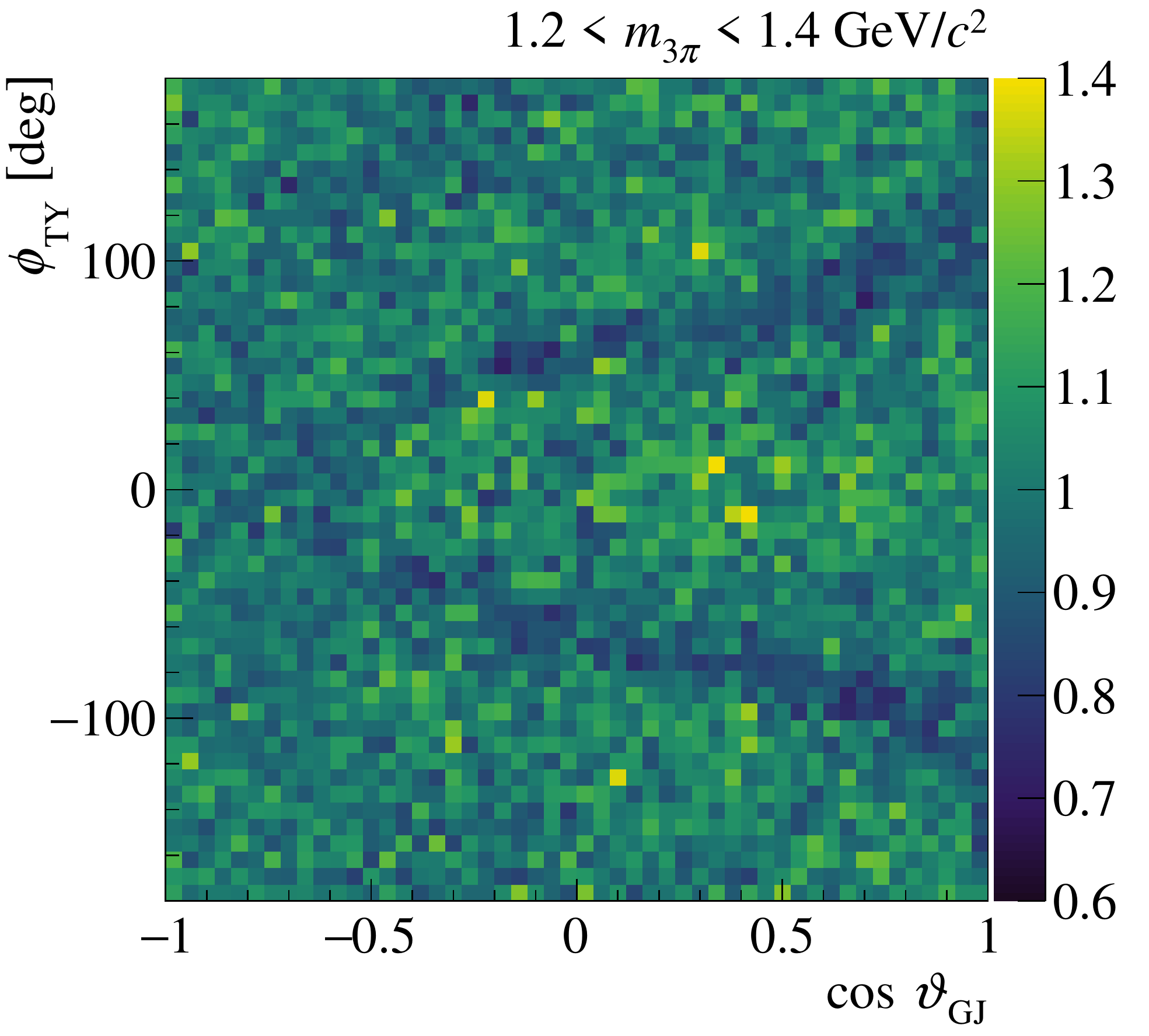}%
  }%
  \subfloat[][]{%
    \label{fig:wMC_high_t_ratio_HF}%
    \includegraphics[width=\twoPlotWidthTwoD]{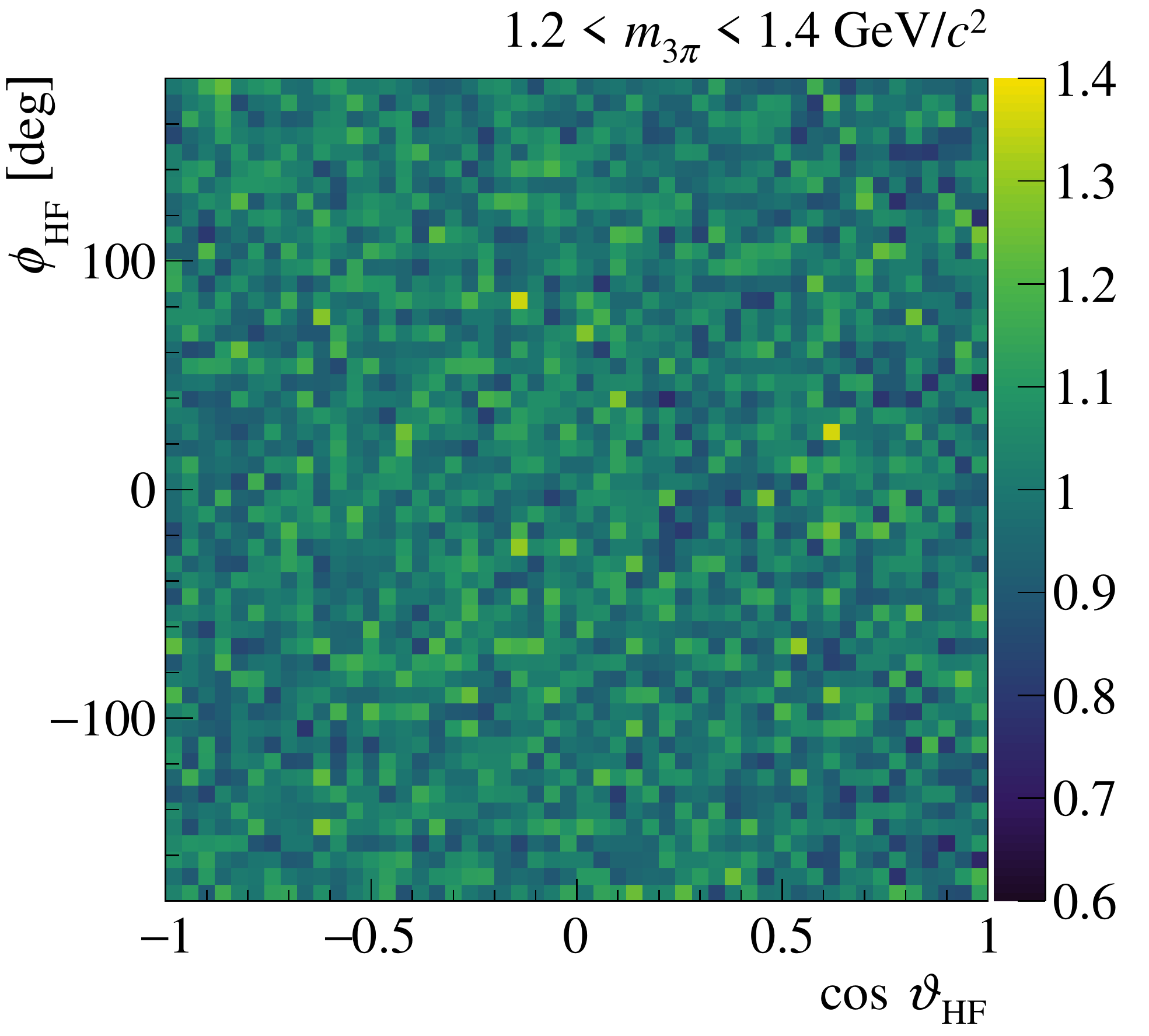}%
  }%
  \caption{\colorPlot Same as \cref{fig:wMC_low_t_high_mX}, but for
    the high-\tpr region \SIvalRange{0.449}{\tpr}{0.724}{\GeVcsq} and
    the mass slice \SIvalRange{1.2}{\mThreePi}{1.4}{\GeVcc} around the
    \PaTwo.}
  \label{fig:wMC_high_t_low_mX}
\end{figure*}

In general, the agreement between the weighted Monte Carlo and the
real-data events is very good, in particular at intermediate $3\pi$
masses. At larger \mThreePi, we observe small deviations concerning
the description of the \PfZero and \PfTwo isobars (see
\cref{fig:wMC_low_t_isobar}) as well as small localized differences in
the angular distribution in the Gottfried-Jackson frame (see
\cref{fig:wMC_low_t_ratio_GJ,fig:wMC_high_t_ratio_GJ}).

\subsection{Systematic Studies}
\label{sec:pwa_massindep_systematic_studies}

Given the high precision of the data, statistical uncertainties are
negligibly small in most cases, \ie systematic uncertainties are
dominant.  We have performed several tests to assess the stability of
the result of the mass-independent fit.  Here, we summarize the
findings of these studies; more details can be found in
\cref{sec:appendix_syst_studies_massindep}.

Possible effects from processes, in which the target proton
is excited, are expected to be negligible.  Due to Pomeron
dominance, target excitations will be mostly $N^*$.  The
recoil-proton trigger and the momentum-conservation criterion
applied in the event selection (see
\cref{sec:event_selection_trigger,sec:event_selection_cuts})
suppress such events on average by about an order of magnitude.  The
remaining contributions consist
predominantly of low-mass $N^*$ produced at large \tpr.  In
diffractive reactions, target and beam vertex factorize, so that
these events are expected to have only little effect on the production of the
three-pion final state.  As we assume for these events the proton
mass for the mass of the recoiling particle, the calculated values
of \tpr would be slightly shifted by values comparable to the \tpr
resolution.  The same is true for the reconstructed beam energy
$E_\text{beam}$.

In order to reduce the probability for the fit to converge to a local
maximum, the likelihood fit is repeated in each kinematic bin in
\mThreePi and \tpr \num{30}~times with random starting values for the
transition amplitudes $\mathcal{T}_a^{r \refl}$ in
\cref{eq:likelihood_function_amp}.  From these 30~fits, the one with
the highest likelihood is selected in order to generate the results
presented in
\cref{sec:results_pwa_massindep_major_waves,sec:results_pwa_massindep_pipiS_waves}.
In the $3\pi$ mass range above about \SI{1}{\GeVcc}, the fits reliably
yield a single solution.  Only a few fits are trapped in local maxima
with significantly lower likelihood.  In contrast, for mass bins below
about \SI{1}{\GeVcc}, the fits find multiple local maxima that deviate
from each other only by a few units of log-likelihood.  We attribute
this behavior to the fact that, due to the smaller phase-space volume
at low \mThreePi, mainly the low-mass tails of the isobars contribute
to the decay amplitudes.  Therefore, it is harder to distinguish
partial waves with different isobars.  Since we do not expect any
$3\pi$ resonances below \SI{1}{\GeVcc}, no efforts were made to
resolve these ambiguous solutions.

We have studied how the truncation of the partial-wave
expansion series in \cref{eq:intensity_bin} influences the
intensities of the 18~partial waves discussed in this paper (see
\cref{tab:selected-waves}) by comparing to a fit with a reduced set
of only 53~partial waves~\cite{haas:2011rj}.  Except for one wave,
the intensities exhibit only relatively small changes, which
typically affect the high-mass regions.  The intensity of the
\wave{1}{++}{0}{+}{\PfTwo}{P} wave changes significantly in the mass
regions above \SI{2.0}{\GeVcc} and below \SI{1.5}{\GeVcc}, the latter
of which is attributed to model leakage from the
  $\Prho\,\pi$ $S$-wave decay of the \PaOne.  However,
the region around the enhancement at \SI{1.8}{\GeVcc} (marked by the
shaded region in \cref{fig:a1_f2_pi_P_total_m0}) is only slightly
affected.

As mentioned in \cref{sec:isobar_model_amplitude}, we do not
apply relativistic corrections to the decay amplitudes in the
partial-wave analysis.  First studies show that the effect on the
shapes of the selected 18~waves is small.

 In order to study the effect of the rank of the spin-density matrix,
the data are fit with a rank-2 spin-density matrix instead of the
rank-1 for the positive-reflectivity sector.  The most striking
feature of the rank-2 fit, which has nearly twice the number of free
parameters, is that the flat wave practically disappears.  In
addition, intensity is shifted from the negative into the
positive-reflectivity sector.  However, the fit shows substantial
instabilities and artificial structures in the \mThreePi region
between \SIlist{1.0;1.3}{\GeVcc} in some partial waves.  We therefore
conclude that the rank-1 fit offers a better description of the data
using significantly less parameters.  For more details see
\cref{sec:appendix_syst_studies_massindep_rank}.

Omitting the waves with negative reflectivity from the wave set feeds
intensity mostly into the flat wave, but causes little change of
positive-reflectivity waves (see
\cref{sec:appendix_syst_studies_massindep_neg_refl}).  This means that
in the given range of four-momentum transfer, the positive and
negative-reflectivity sectors do not mutually influence each other and
are well-separated by the fit as opposed to the case of very small
$\tpr < \SI{e-3}{\GeVcsq}$~\cite{adolph:2014mup}.

The isobar parametrizations (see
\cref{sec:pwa_method_isobar_parametrization}) are an important input
for the PWA model.
The \Prho is the dominant isobar.  For most of the waves in
\cref{tab:selected-waves}, the intensity distribution is not sensitive
to the details of the \Prho parametrization or to small changes of the
used parameter values.  In contrast, the region around the \PaOne mass
in the \wave{1}{++}{0}{+}{\pipiS}{P} wave and the \PaTwo signal in the
\wave{2}{++}{1}{+}{\PfTwo}{P} wave change significantly (see
\cref{sec:appendix_syst_studies_massindep_isobar_param}).  Both seem to
be contaminated by model leakage due to the imperfect description of
the \Prho.
 The dependence of the PWA result on the parametrization of
the \PfZero[980] and that of the broad \pipiSW component is studied
as well.
Using a simple $S$-wave Breit-Wigner amplitude [\cref{eq:relBW} with
\cref{eq:constWidth}] instead of the Flatt\'e parametrization
[\cref{eq:f0980_flatte}] for the \PfZero[980] reduces the height of
the intensity peaks of the resonances decaying into
$\PfZero[980]\,\pi$ by about a factor of two (see
\cref{sec:appendix_syst_studies_massindep_isobar_param}).  However,
the shapes of the peaks in these partial waves remain unchanged.  In
the mass region above \SI{1.3}{\GeVcc}, the fit with the \PfZero[980]
Flatt\'e parametrization has a significantly higher likelihood than the
one with the $S$-wave Breit-Wigner amplitude.  This indicates that the
data are better described by the Flatt\'e parametrization.

The influence of the parametrization used for the broad component of
the \pipiSW on the PWA result is studied by performing a fit with an
alternative description of the mass-dependence of the isobar
amplitude.  Instead of the modified $M$~solution from
\refCite{au:1986vs} (see
\cref{sec:pwa_method_isobar_parametrization}), the $K_1$~solution from
\refCite{au:1986vs} with the \PfZero[980] pole subtracted using a
simple $S$-wave Breit-Wigner amplitude is used.  This parametrization
was originally used by the VES experiment~\cite{amelin:1995gu}.  The
Breit-Wigner for the \PfZero[980] is similar to the one employed in
the \PfZero[980] study described above.  In order to be consistent,
the Breit-Wigner amplitude is also used for all waves with the
\PfZero[980] isobar.  The observed variations in the fit result are
small.

Performing the PWA on a data sample without the CEDAR, RICH, and
central-production vetos described in \cref{sec:event_selection_cuts}
shows that the result is not very sensitive to backgrounds from kaon
diffraction, kaon pairs in the final state, and central-production
reactions (see
\cref{sec:appendix_syst_studies_massindep_selection_cuts}).  The
partial-wave intensities scale approximately with the number of
events, only the relative intensity of the flat wave increases.  It is
very unlikely that the peak around $\mThreePi = \SI{1.4}{\GeVcc}$ in
the \wave{1}{++}{0}{+}{\PfZero}{P} wave is caused by kaon-induced
reactions or that it stems from kaonic final states misinterpreted as
pionic ones.  The CEDAR and RICH vetos applied in the event selection
reduce such contaminations considerably.  Further studies show that
the signal is not correlated with these cuts.

In summary, the PWA fits converge reliably for $3\pi$ masses above
about \SI{1}{\GeVcc}.  The shapes of clear resonance peaks are stable
\wrt changes of the PWA model.  However, in some cases the height of
the intensity peaks is sensitive to the isobar parametrization.  This
issue will mostly be resolved for the \pipiSW isobars by applying a
method introduced in \cref{sec:results_free_pipi_s_wave}, by which the
isobar amplitudes are extracted from the data.
 %
%
%

\section{\tpr Dependences}
\label{sec:tprim_dependence}

\Cref{fig:mass_spectrum_lowt,fig:mass_spectrum_hight} illustrate how
the shape of the measured three-pion invariant mass spectrum changes with \tpr,
while \cref{fig:tprim_spectrum_lowM,fig:tprim_spectrum_highM} show the
change of the measured \tpr distribution with $3\pi$ mass.  It is apparent that
the \tpr spectrum strongly depends on \mThreePi.  This observation has
motivated us to perform the PWA in bins of \tpr instead of weighting
the partial waves with \tpr-dependent model functions, which has been
the conventional approach.

\begin{figure*}[htbp]
  \captionsetup[subfigure]{captionskip=4pt}
  \centering
  \subfloat[\mThreePi spectrum for low values of \tpr.]{%
    \label{fig:mass_spectrum_lowt}%
    \includegraphics[width=\twoPlotWidth]{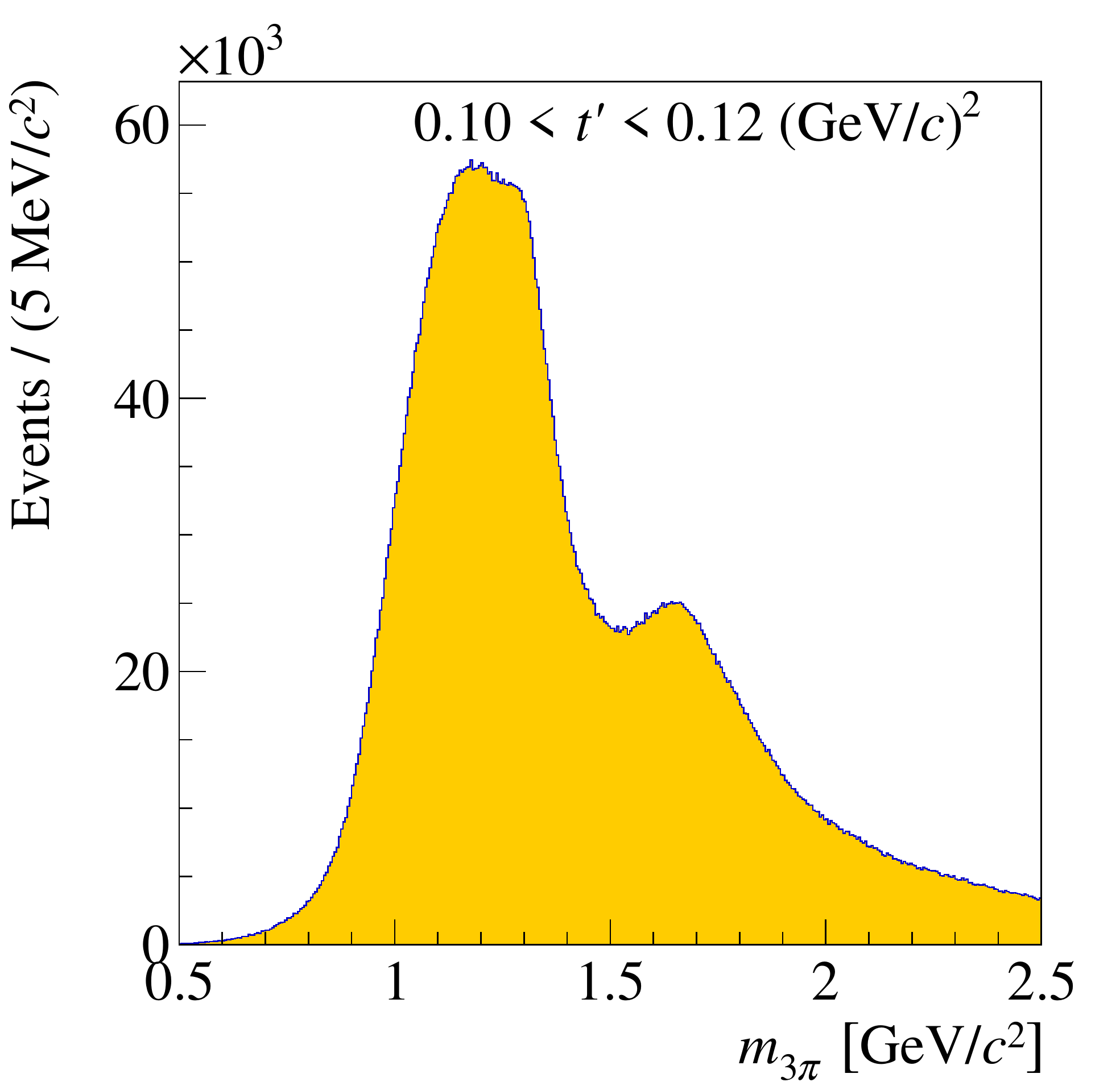}%
  }%
  \hspace*{\twoPlotSpacing}
  \subfloat[\mThreePi spectrum for high values of \tpr.]{%
    \label{fig:mass_spectrum_hight}%
    \includegraphics[width=\twoPlotWidth]{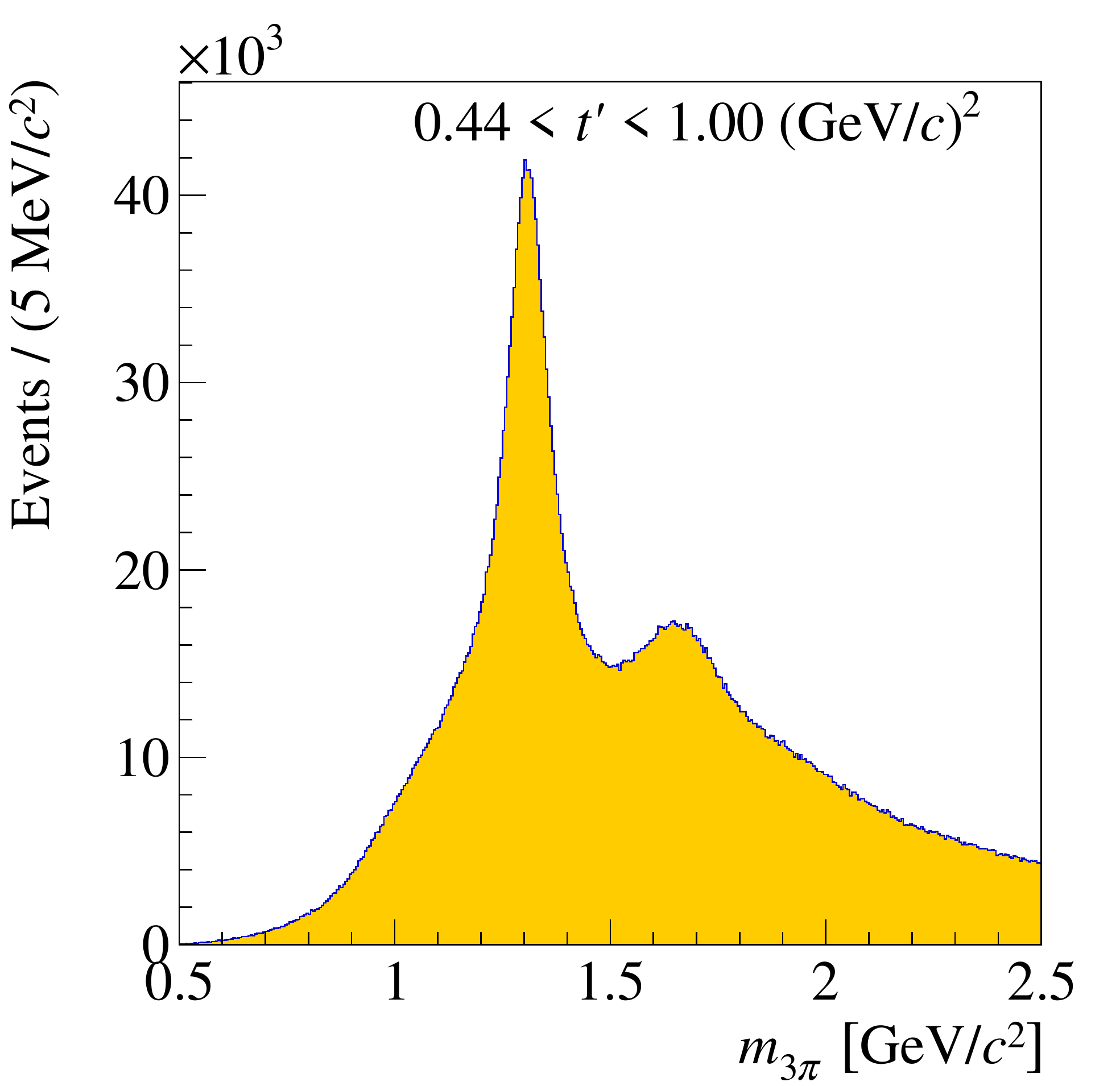}%
  }%
  \\
  \subfloat[\tpr spectrum for low values of \mThreePi.]{%
    \label{fig:tprim_spectrum_lowM}%
    \includegraphics[width=\twoPlotWidth]{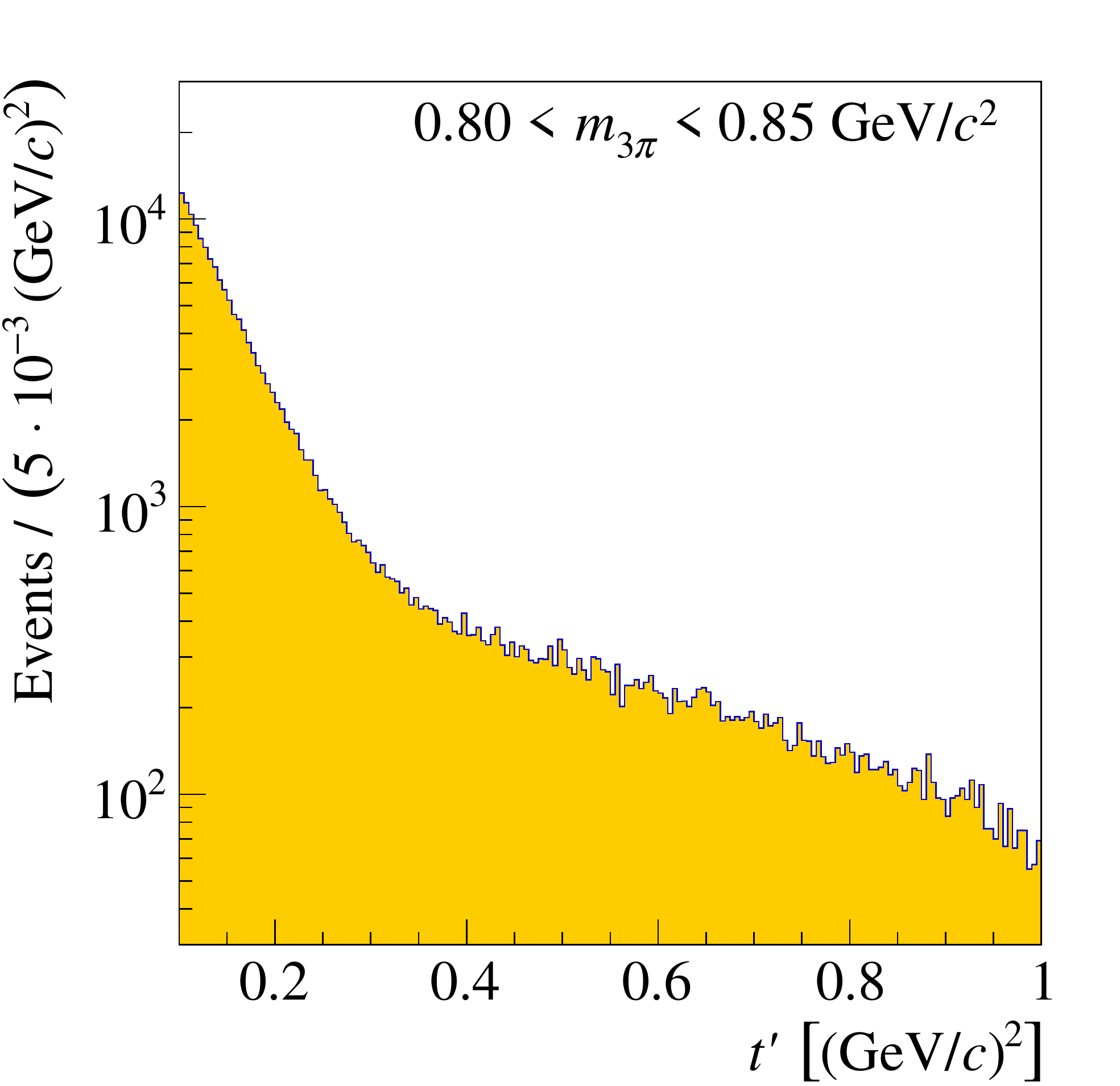}%
  }%
  \hspace*{\twoPlotSpacing}
  \subfloat[\tpr spectrum for medium values of \mThreePi.]{%
    \label{fig:tprim_spectrum_highM}%
    \includegraphics[width=\twoPlotWidth]{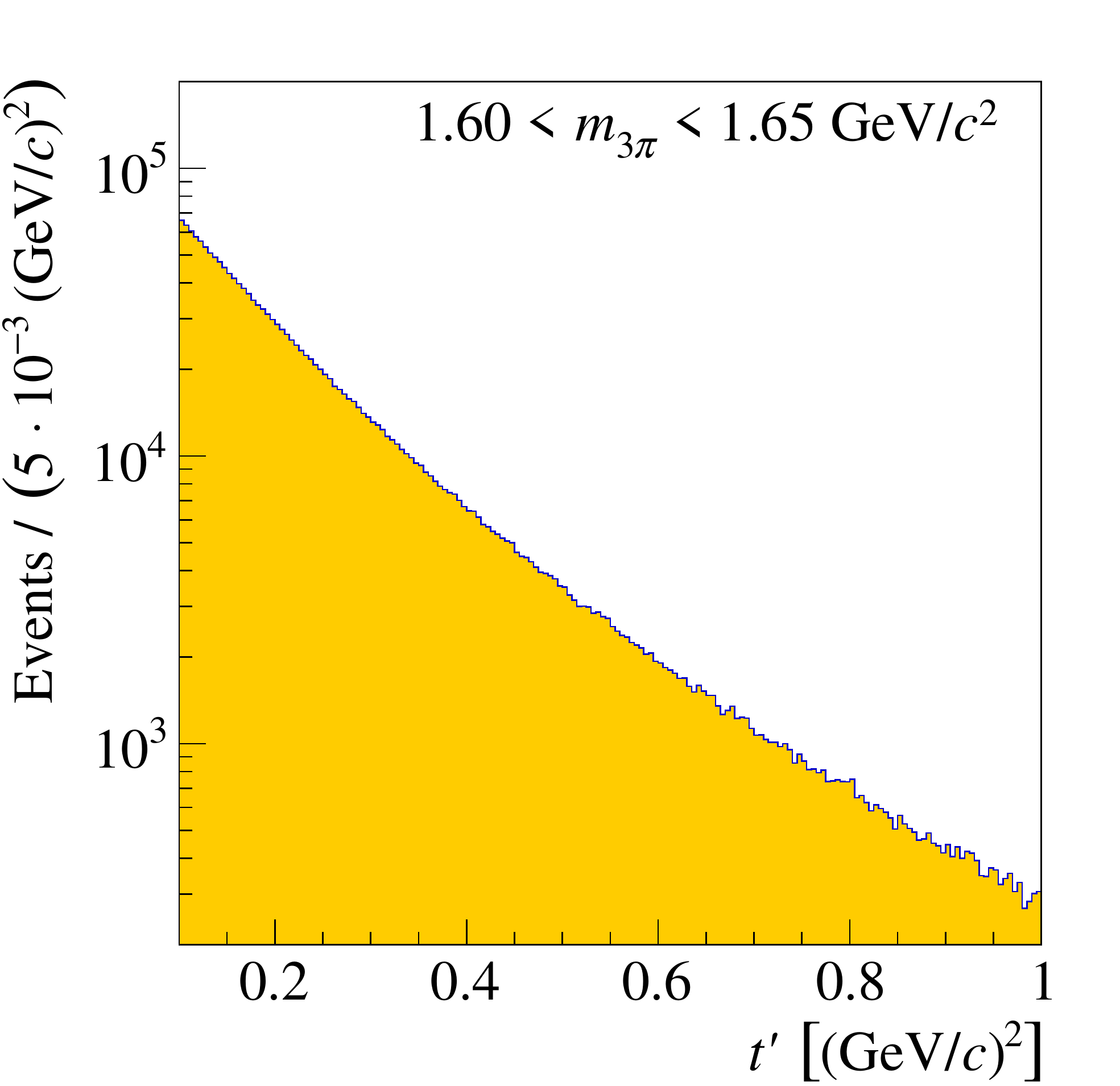}%
  }%
  \caption{The \tpr dependence of the measured $3\pi$ invariant mass spectrum
    and vice versa.}
  \label{fig:mass_spectrum_different_t}
\end{figure*}

In this chapter, we will elaborate on the details of the observed \tpr
distributions.  We will start with global spectra, from which we
determine the \tpr dependence as a function of \mThreePi.  Such an
analysis comes closest to the traditional description of high-energy
reactions in terms of Regge exchange.  Using the results from the
\emph{mass-independent} fit outlined in
\cref{sec:results_pwa_massindep}, we can in addition separate the
contributions of various partial waves to the \tpr spectrum.
Comparing different $3\pi$ mass regions that are either dominated by
well-established resonances or by nonresonant contributions, various
patterns become apparent.

\subsection{Overall \tpr Dependence}
\label{sec:global_tprim_dependences}

The first extensive study of the $3\pi$ mass dependence of the \tpr
spectrum was performed by the ACCMOR
collaboration~\cite{daum:1979ix,Daum:1980ay}.  They investigated the
reaction $\pi^- + p \to {\threePi} + p$ at \SIlist{63;94}{\GeVc}
incoming pion momentum and determined the \tpr dependence as a
function of \mThreePi in the range $\tpr < \SI{1.0}{\GeVcsq}$.  The
\tpr dependence was parametrized for each \SI{50}{\MeVcc} wide $3\pi$
mass bin by two exponentials:
\begin{multlineOrEq}
  \label{eq:two_exponentials}
  \dod{N}{\tpr}(\tpr; \mThreePi) =   A_1(\mThreePi)\, e^{-b_1(\mThreePi)\, \tpr} \newLineOrNot
                                   + A_2(\mThreePi)\, e^{-b_2(\mThreePi)\, \tpr},
\end{multlineOrEq}
with real-valued parameters $A_i$.  The ACCMOR collaboration observed
that the two slope parameters $b_{1,2}$ are different at small values
of \mThreePi and that they vary significantly up to values of
\mThreePi of about \SI{1.2}{\GeVcc}.  This marks the onset of
resonance production.  Beyond this mass value, the slope values of
$b_1 \approx \SI{12}{\perGeVcsq}$ and $b_2 \approx \SI{5}{\perGeVcsq}$
stay almost constant (see open circles in
\cref{fig:t-slopes_vs_mass}).

\begin{figure}[tbp]
  \centering
  \subfloat[][]{%
    \label{fig:t-slopes_vs_mass}%
    \includegraphics[width=\twoPlotWidth]{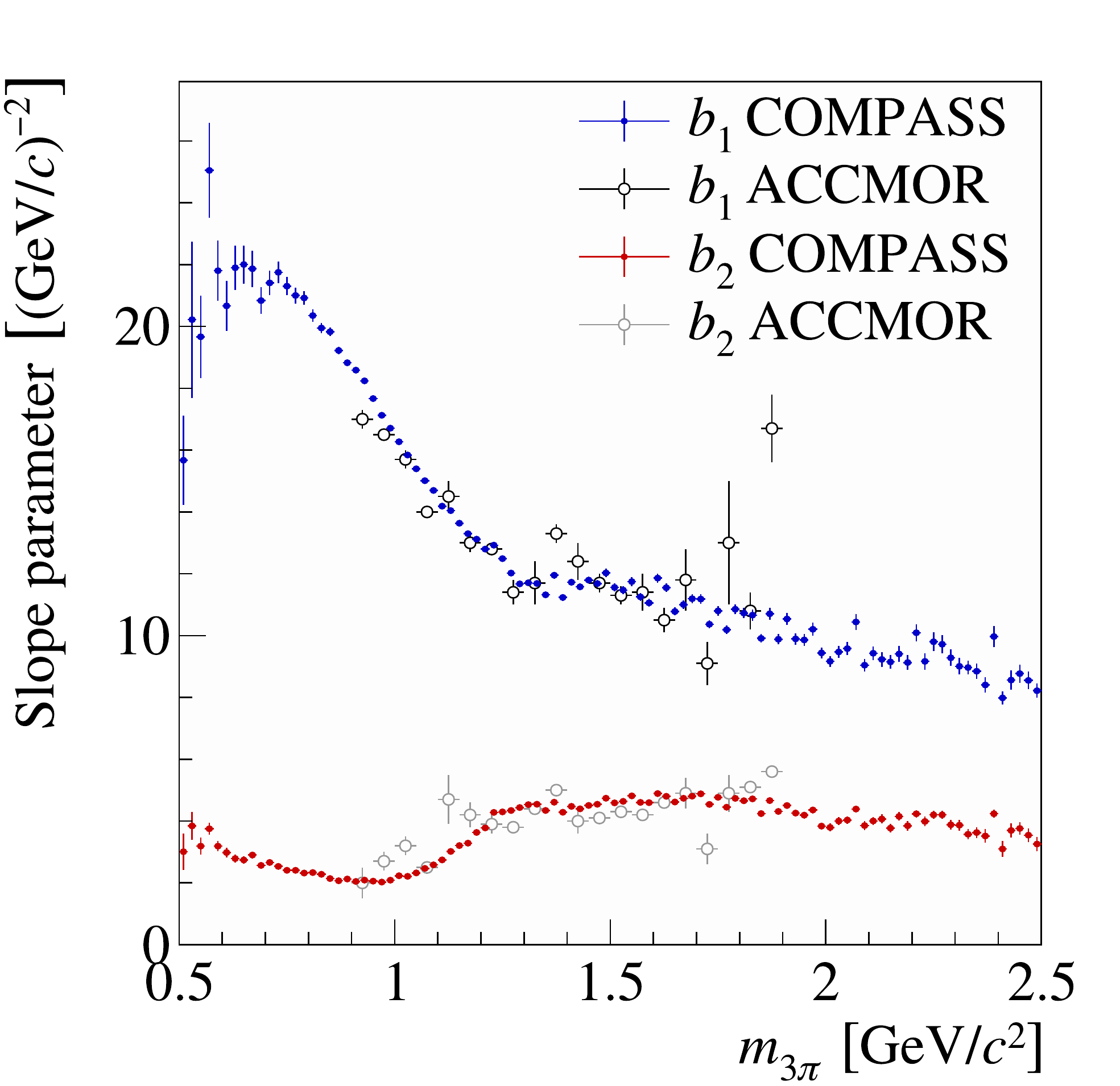}%
  }%
  \newLineOrHspace{\twoPlotSpacing}%
  \subfloat[][]{%
    \label{fig:t-slopes_ratio_vs_mass}%
    \includegraphics[width=\twoPlotWidth]{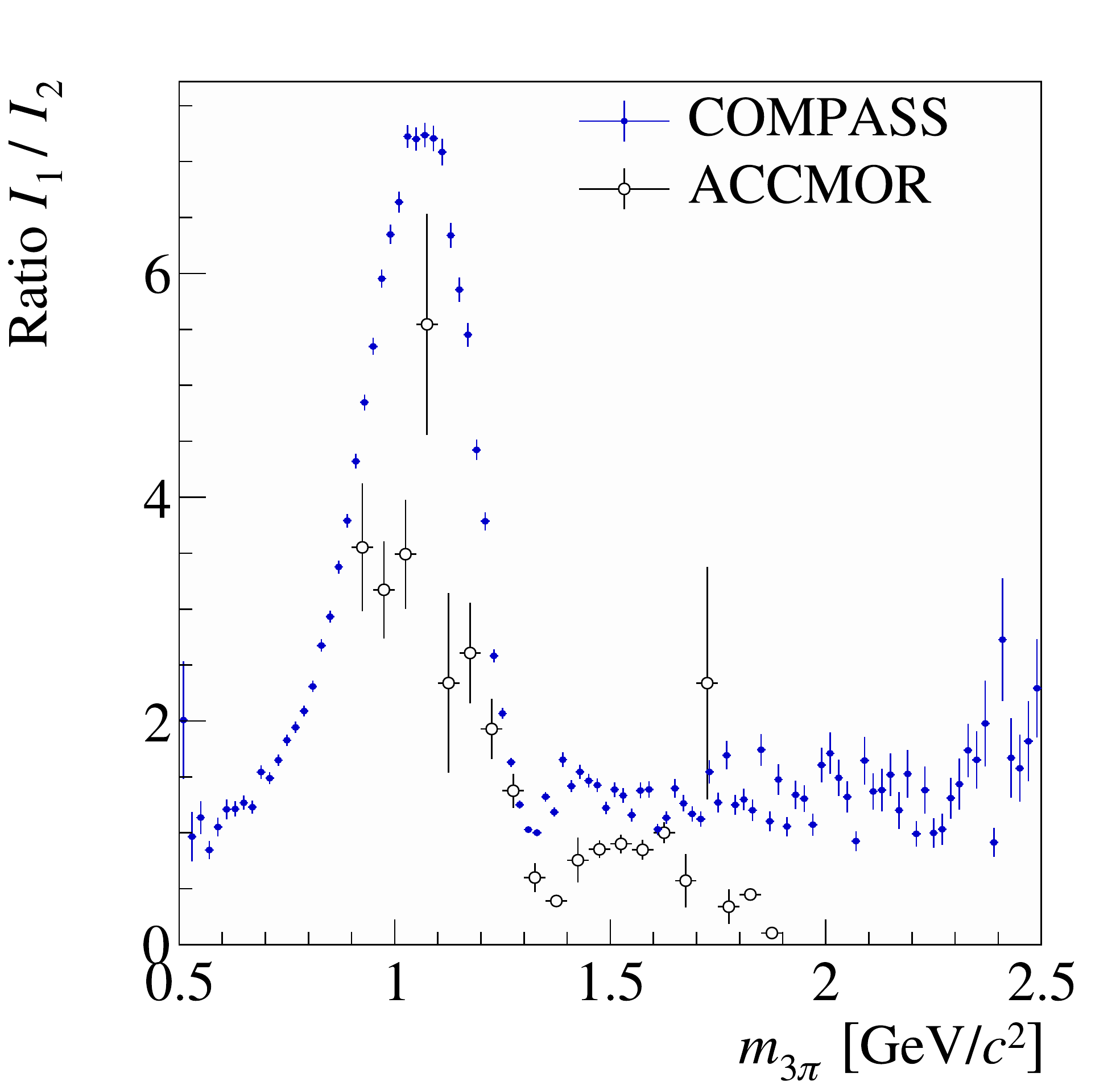}%
  }%
  \caption{\colorPlot Result of a fit to the \tpr dependence of events
    from the diffractive-dissociation reaction
    $\pi^- + p \to {\threePi} + p$, as measured by COMPASS (filled
    circles) and by the ACCMOR collaboration~\cite{Daum:1980ay} (open
    circles).  For each \mThreePi bin, the \tpr spectrum was fit using
    a double-exponential model [see \cref{eq:two_exponentials}].
    Panel~(a) shows the mass dependence of the two slope parameters
    $b_1$ (upper points) and $b_2$ (lower points).  Note the extended
    mass range of the present measurement as compared to the ACCMOR
    data.  Panel~(b) shows the ratio $I_1 / I_2$ of the integrated
    exponential contributions, see
    \cref{eq:two_exponentials_integrals}.}
  \label{fig:t-slopes_overall}
\end{figure}

We perform the same study on the present data in a wider $3\pi$ mass
range and using finer \mThreePi bins of \SI{20}{\MeVcc} width.  In
each mass bin, the acceptance-corrected \tpr spectrum, which is
obtained from the mass-independent fit in 11~\tpr bins, is fit by
\cref{eq:two_exponentials}.  The result is shown as filled circles in
\cref{fig:t-slopes_vs_mass}.  The general pattern and also the
absolute values for the slope parameters agree nicely with the ACCMOR
results.  We observe a strong dependence of both slope parameters on
\mThreePi.  In the region $\mThreePi < \SI{1.0}{\GeVcc}$ below the
resonances, only a few partial waves contribute significantly to the
spectrum, \ie the \wave{1}{++}{0}{+}{\Prho}{S} (\SI{57.1}{\percent}),
\wave{0}{-+}{0}{+}{\pipiS}{S} (\SI{12.8}{\percent}),
\wave{1}{++}{1}{+}{\Prho}{S} (\SI{8.3}{\percent}),
\wave{1}{++}{0}{+}{\pipiS}{P} (\SI{4.2}{\percent}), and
\wave{0}{-+}{0}{+}{\Prho}{P} (\SI{4.1}{\percent}) waves.  In this
region, the parameter~$b_1$ representing the steeper component shows a
rapid drop with increasing three-pion mass.  The parameter $b_2$
representing the shallower component exhibits less variation.  Its
mass dependence shows a dip by \SI{2}{\perGeVcsq} at around
$\mThreePi = \SI{1.0}{\GeVcc}$.  Approaching the mass region of \PaOne
and \PaTwo, above approximately \SI{1.3}{\GeVcc}, the \mThreePi
dependence of $b_1$ and $b_2$ changes abruptly: $b_1$ drops much
slower, decreasing from about \SI{12}{\perGeVcsq} at \SI{1.3}{\GeVcc}
to \SI{8}{\perGeVcsq} at \SI{2.5}{\GeVcc}, whereas $b_2$ stays nearly
constant at about \SI{4}{\perGeVcsq} over the same mass range.

\Cref{fig:t-slopes_ratio_vs_mass} shows the ratio of the
contributions
\begin{equation}
  \label{eq:two_exponentials_integrals}
  I_1 \equiv A_1 \int_{t'_\text{min}}^{t'_\text{max}}\! \dif{\tpr}\, e^{-b_1\, \tpr}
  ~\text{and}~
  I_2 \equiv A_2 \int_{t'_\text{min}}^{t'_\text{max}}\! \dif{\tpr}\, e^{-b_2\, \tpr}
\end{equation}
of the two exponentials from \cref{eq:two_exponentials}, integrated
from $t'_\text{min} = \SI{0.1}{\GeVcsq}$ to
$t'_\text{max} = \SI{1.0}{\GeVcsq}$.  As observed for the slope
parameters, the regions below and above the resonances show very
different behavior.  Below the resonance region, the component with
the steep slope~$b_1$ dominates and its contribution reaches a maximum
at $3\pi$ masses of approximately \SI{1.0}{\GeVcc}.  From there, it
drops quickly with a shallow minimum around
$\mThreePi = \SI{1.3}{\GeVcc}$.  This dip is presumably caused by the
onset of the \twoPP waves with $M = 1$.  Above about \SI{1.3}{\GeVcc},
the relative contributions of the two exponentials only depend weakly
on \mThreePi with almost equal relative weights for the two terms.
The ACCMOR data show a qualitatively similar behavior.  The agreement
with the present data is, however, not as good as that observed for
the slopes.

To our knowledge, the complicated mass dependence of the \tpr spectra
described above is not well understood.  In the region around
\SI{1.3}{\GeVcc}, nonresonant processes are known to play an important
role.  Most available calculations describe these processes as the
dissociation of the beam pion into the isobar~$\xi^0$ and the bachelor
$\pi^-$, followed by diffractive scattering of one of the beam
fragments (typically the $\pi^-$) off the target proton (see
\cref{fig:deck_process}).  These calculations focus mainly on the
$3\pi$ mass dependence and are based on \pipi and $\pi p$
elastic-scattering
data~\cite{Deck:1964hm,Ascoli:1974sp,Ascoli:1974hi,Daum:1980ay,Dudek:2006ud}.
The more elaborate \emph{three-component Deck
  model}~\cite{CohenTannoudji:1976tj,CohenTannoudji:1977xh,Antunes:1984cy}
describes the reaction $\pi^- + p \to \xi^0\, \pi^- + p_\text{recoil}$
by including $\xi$ as well as $\pi$ exchanges in addition to direct
production of $\xi^0\,\pi^-$ via Pomeron exchange.  In general, such
nonresonant processes exhibit a dependence on \tpr that is different
from that of resonant production.  Interferences between resonant and
nonresonant processes may in addition modify the \tpr spectra.  The
three-component Deck model describes the correlations between the
$\xi\pi$ invariant mass, the slope of the \tpr spectrum, and the
\cosThetaGJ distribution in detail and predicts the existence of
interference minima in the \tpr spectra.  As it will be shown in the
following section, the \tpr spectra of some partial waves exhibit such
kind of minima in certain $3\pi$ mass regions.

\subsection{\tpr Dependences of Individual Partial Waves}
\label{sec:tprim_partial_waves}

Using the partial-wave decomposition of the mass spectrum from the
\emph{mass-independent} fit as presented in
\cref{sec:results_pwa_massindep_major_waves}, we can now study the
\tpr dependence of the intensity of individual partial waves in
different mass regions.  The selected mass regions are indicated by
shaded bands in the spectral distributions shown in
\cref{sec:results_pwa_massindep}.  The corresponding \tpr spectra are
obtained by integrating the partial wave intensities over those mass
regions.  The integrated intensities are presented using horizontal
bars, the lengths of which represent the widths of the given \tpr
bins.  Blue horizontal lines represent the central values.  The height
of the gray horizontal bars corresponds to the statistical uncertainty
of the intensity.

We compare waves with the same isobars and angular momentum in the
decay but with different spin projections~$M$.
\Cref{fig:t_dependence_m0} gives an example for the
\wave{1}{++}{M}{+}{\Prho}{S} intensities with $M = \numlist{0;1}$
integrated over the range \SIvalRange{1.1}{\mThreePi}{1.3}{\GeVcc},
which covers part of the \PaOne.  \Cref{fig:t_a2_m1,fig:t_a2_m2} show
the analog comparison for \wave{2}{++}{M}{+}{\Prho}{D} waves with
$M = \numlist{1;2}$.

\begin{figure}[tbp]
  \centering
  \subfloat[][]{%
    \label{fig:a1_total_m0_t}%
    \includegraphics[width=\twoPlotWidth]{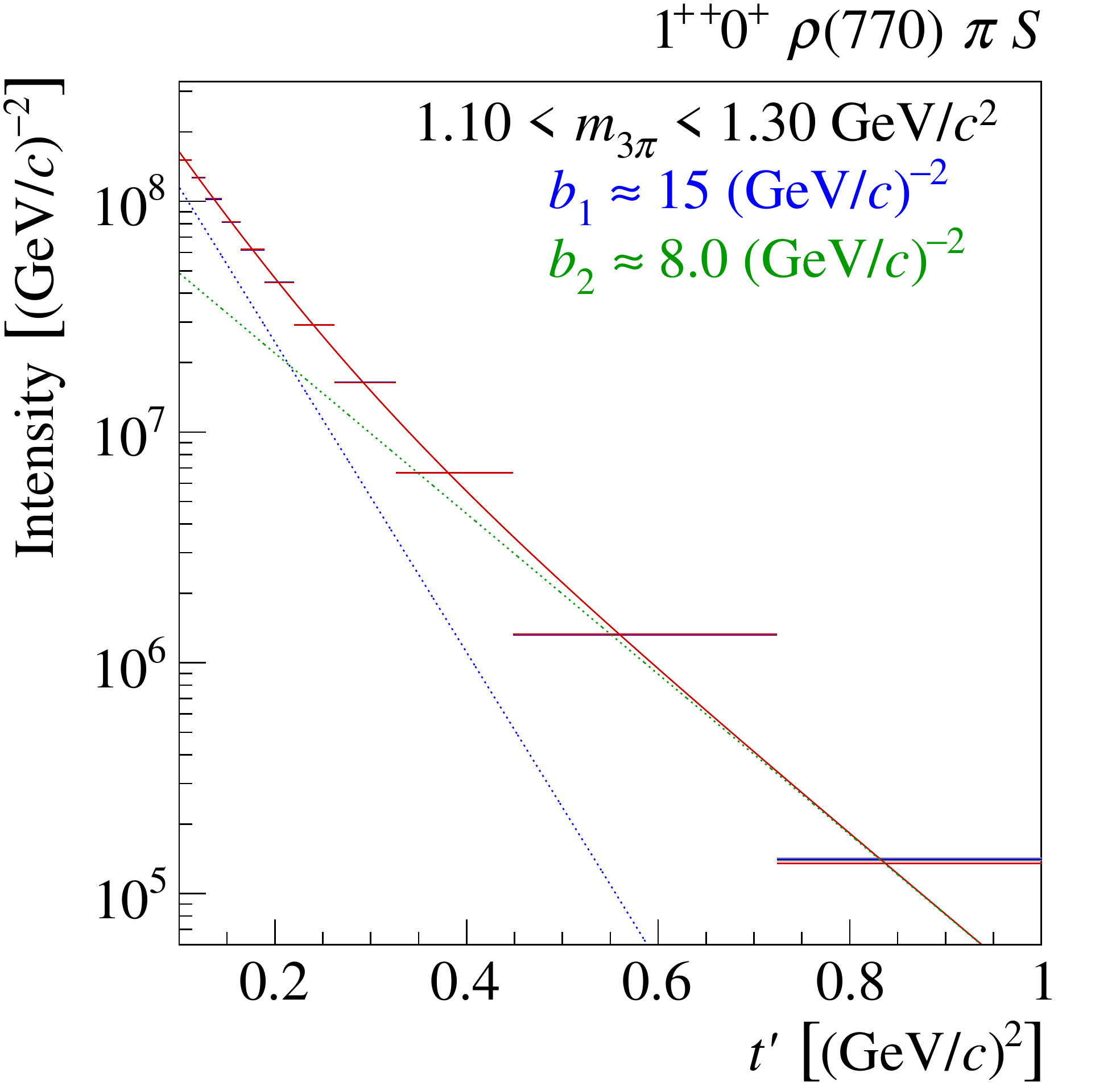}%
  }%
  \newLineOrHspace{\twoPlotSpacing}%
  \subfloat[][]{%
    \label{fig:t_a1_m1}%
    \includegraphics[width=\twoPlotWidth]{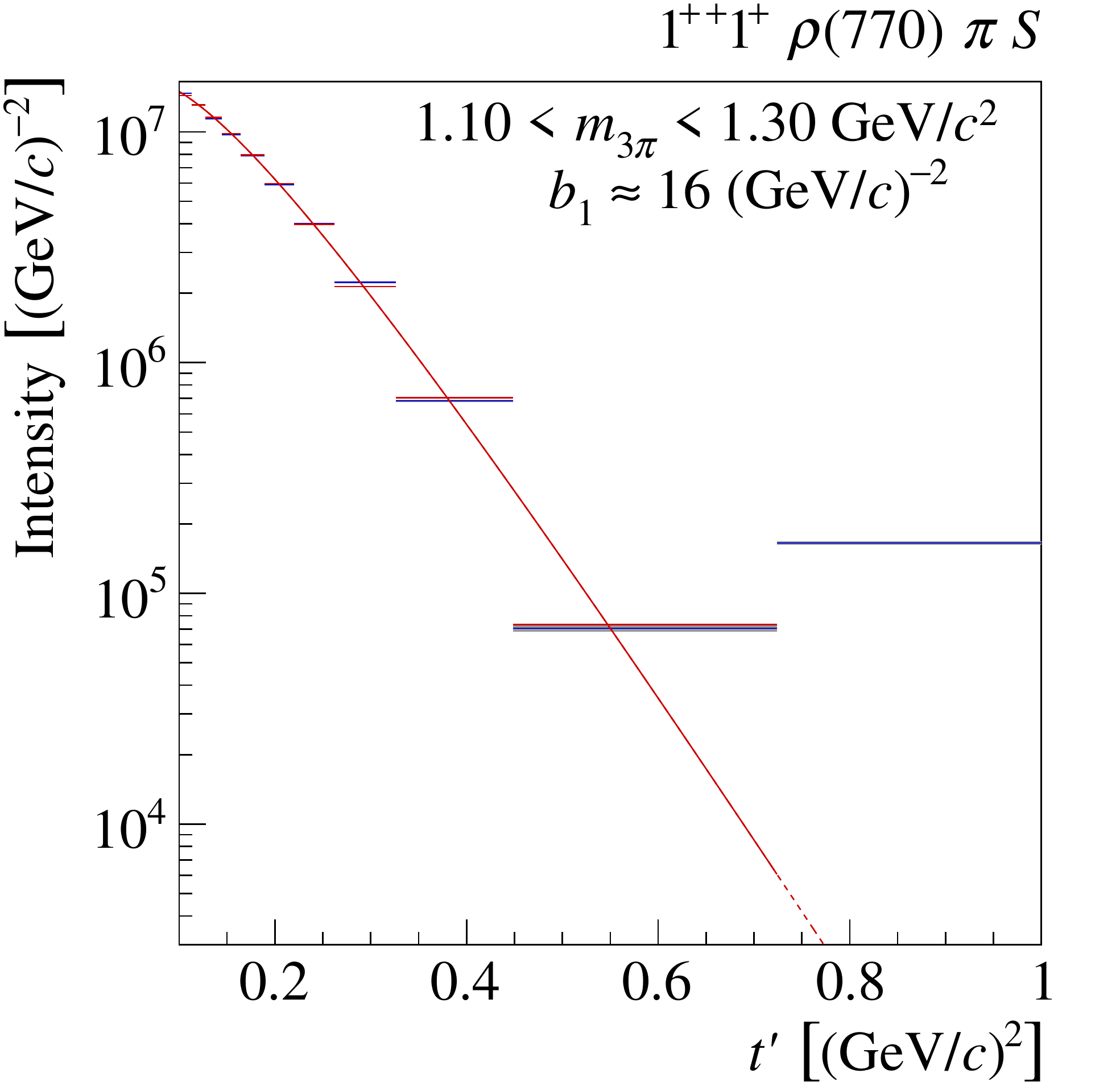}%
  }%
  \caption{\colorPlot The \tpr dependence of the intensity of the
    \wave{1}{++}{M}{+}{\Prho}{S} waves with spin projections
    $M = 0$~(a) and $M = 1$~(b), integrated over the mass region
    around the \PaOne as indicated by the shaded regions in
    \cref{fig:a1_total_m0,fig:a1_total_m1}.  The solid red curve
    in~(a) represents a double-exponential fit using
    \cref{eq:tprim-dependence}, the one in~(b) a single-exponential
    fit with parameter $A_2 = 0$.  In~(a), the two exponential
    components are shown by the dotted lines.  In~(b), the fitted \tpr
    range is indicated by the solid curve; the extrapolation is shown
    as dashed curve.  See text for details.}
  \label{fig:t_dependence_m0}
\end{figure}

Alternatively, waves with same quantum numbers but different isobars
can be compared, again keeping the mass interval fixed.
\Cref{fig:t_dependence_2pp} shows this for the $2^{++}\,1^+$ waves
with the \Prho isobar and $L = 2$ along with waves with the \PfTwo
isobar and $L = 1$, both in the mass region
\SIvalRange{1.2}{\mThreePi}{1.4}{\GeVcc} around the \PaTwo.  A
comparison of the \tpr spectra of the \wave{4}{++}{1}{+}{\Prho}{G} and
$\PfTwo\,\pi\,F$ waves for \SIvalRange{1.86}{\mThreePi}{2.06}{\GeVcc}
is given by \cref{fig:t_dependence_4pp}.

\begin{figure}[tbp]
  \centering
  \subfloat[][]{%
    \label{fig:t_a2_m1}%
    \includegraphics[width=\ifMultiColumnLayout{\twoPlotWidth}{\threePlotWidth}]{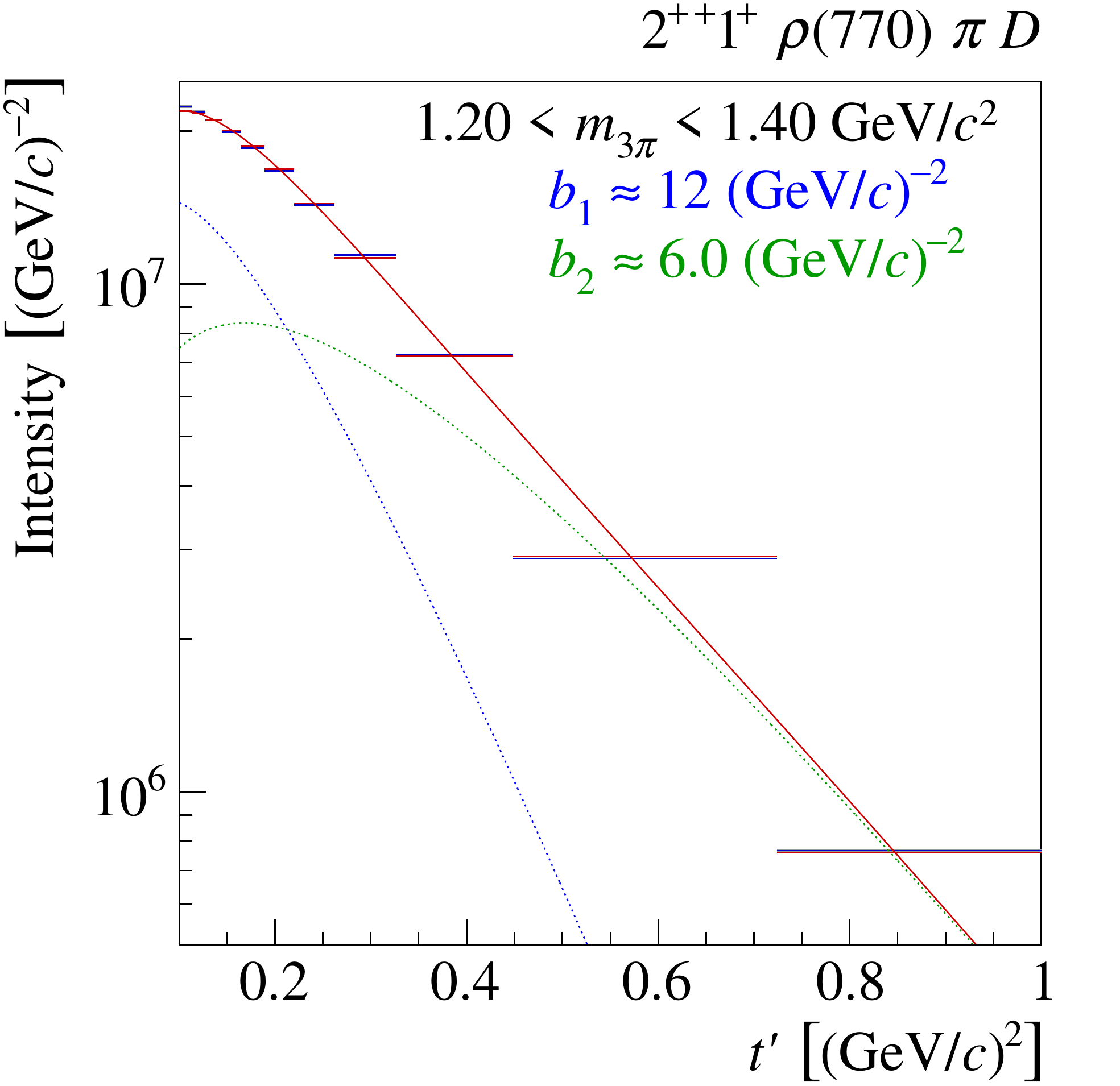}%
  }%
  \newLineOrHspace{\threePlotSpacing}%
  \subfloat[][]{%
    \label{fig:t_a2_m1b}%
    \includegraphics[width=\ifMultiColumnLayout{\twoPlotWidth}{\threePlotWidth}]{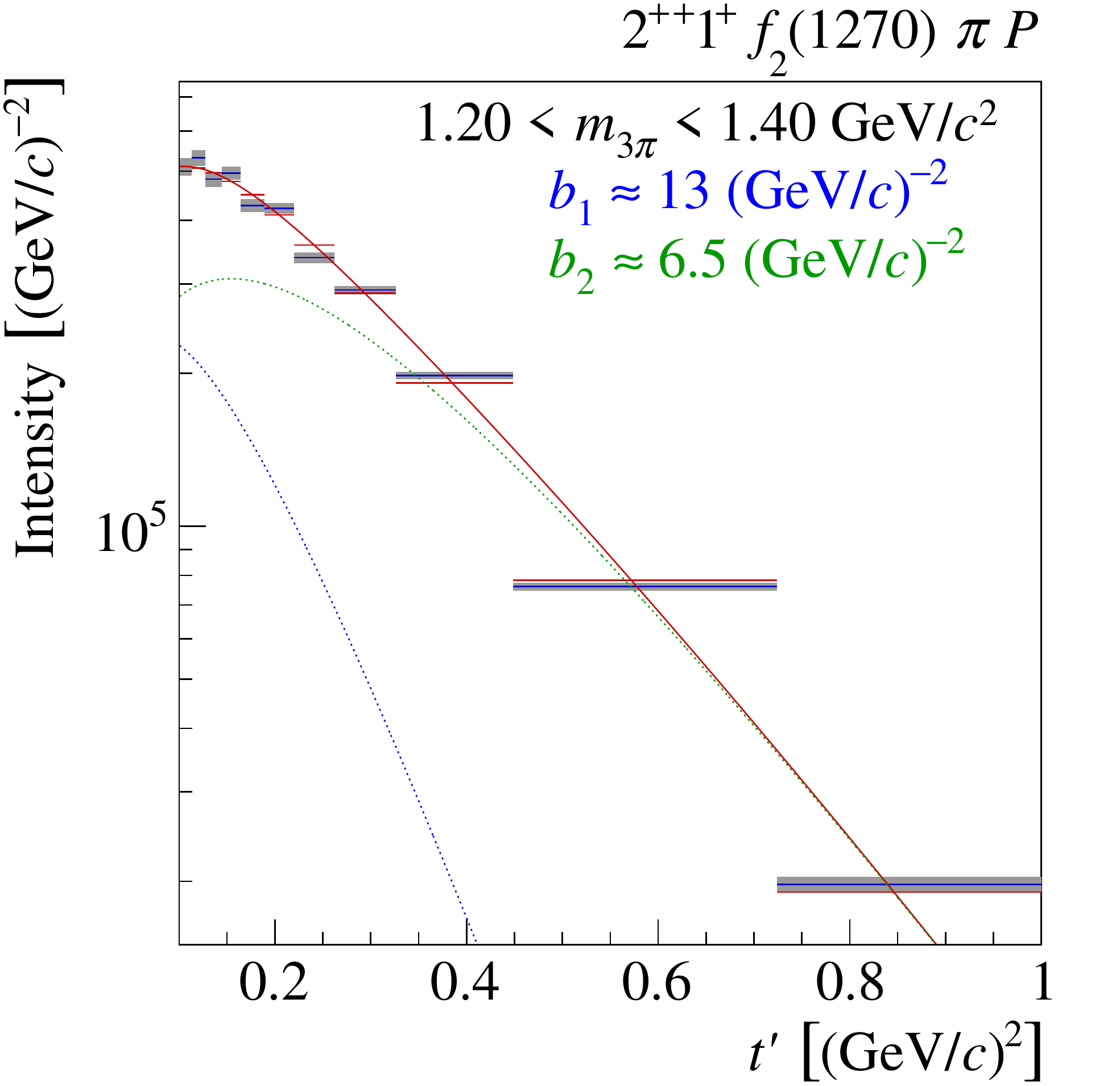}%
  }%
  \newLineOrHspace{\threePlotSpacing}%
  \subfloat[][]{%
    \label{fig:t_a2_m2}%
    \includegraphics[width=\ifMultiColumnLayout{\twoPlotWidth}{\threePlotWidth}]{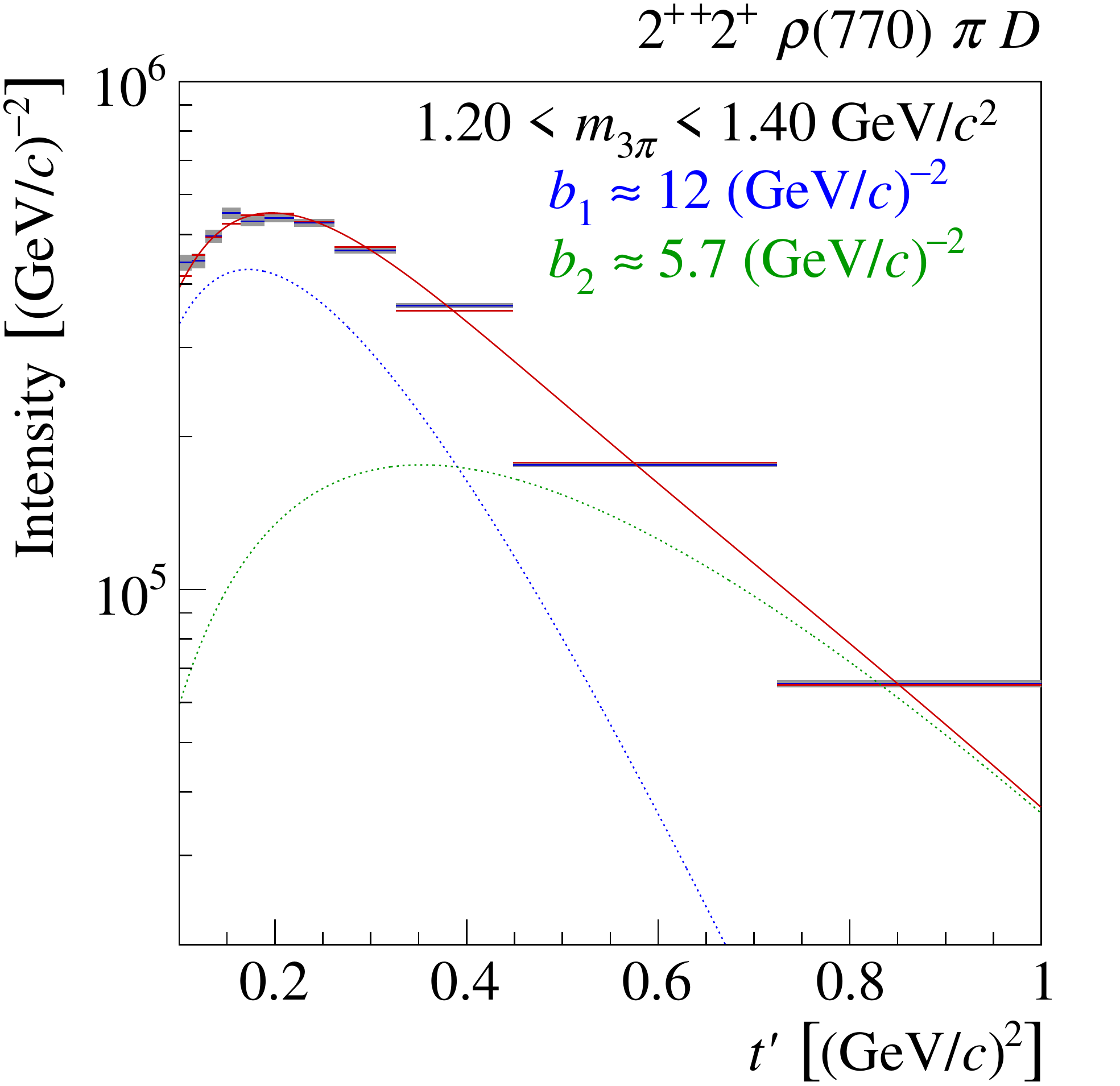}%
  }%
  \caption{\colorPlot The \tpr dependence of the intensity of three
    $2^{++}$ waves with different isobars and different
    spin-projections~$M$, integrated over the mass region around the
    \PaTwo as indicated by the shaded regions in
    \cref{fig:a2_total_m1_rho,fig:a2_total_m1_f2,fig:a2_total_m2}.
    The solid red curves represent double-exponential fits using
    \cref{eq:tprim-dependence}; the fit components are shown as dotted
    curves.  See text for details.}
  \label{fig:t_dependence_2pp}
\end{figure}

\begin{figure}[tbp]
  \centering
  \subfloat[][]{%
    \label{fig:t_a4_m1a}%
    \includegraphics[width=\twoPlotWidth]{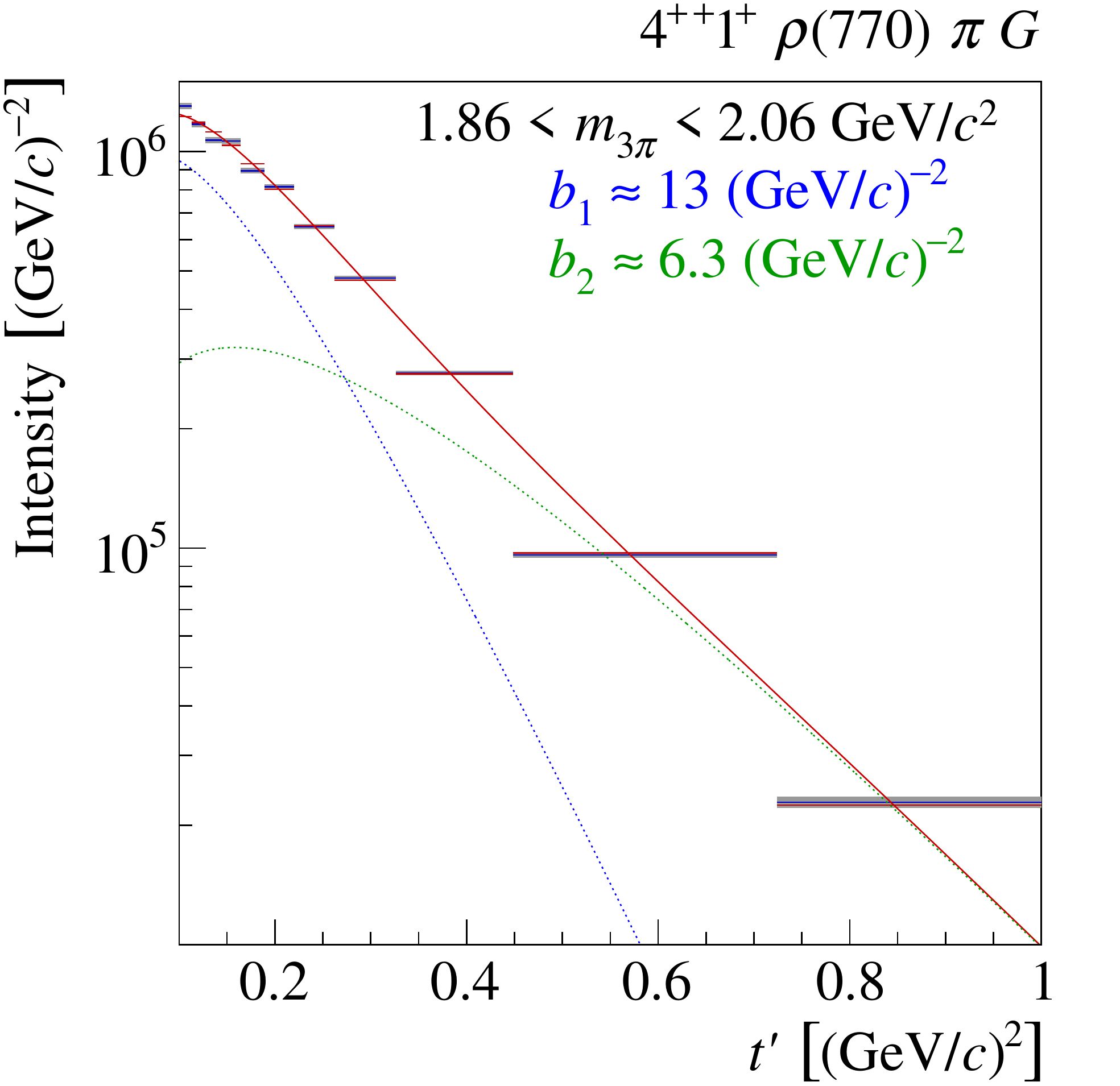}%
  }%
  \newLineOrHspace{\twoPlotSpacing}%
  \subfloat[][]{%
    \label{fig:t_a4_m1b}%
    \includegraphics[width=\twoPlotWidth]{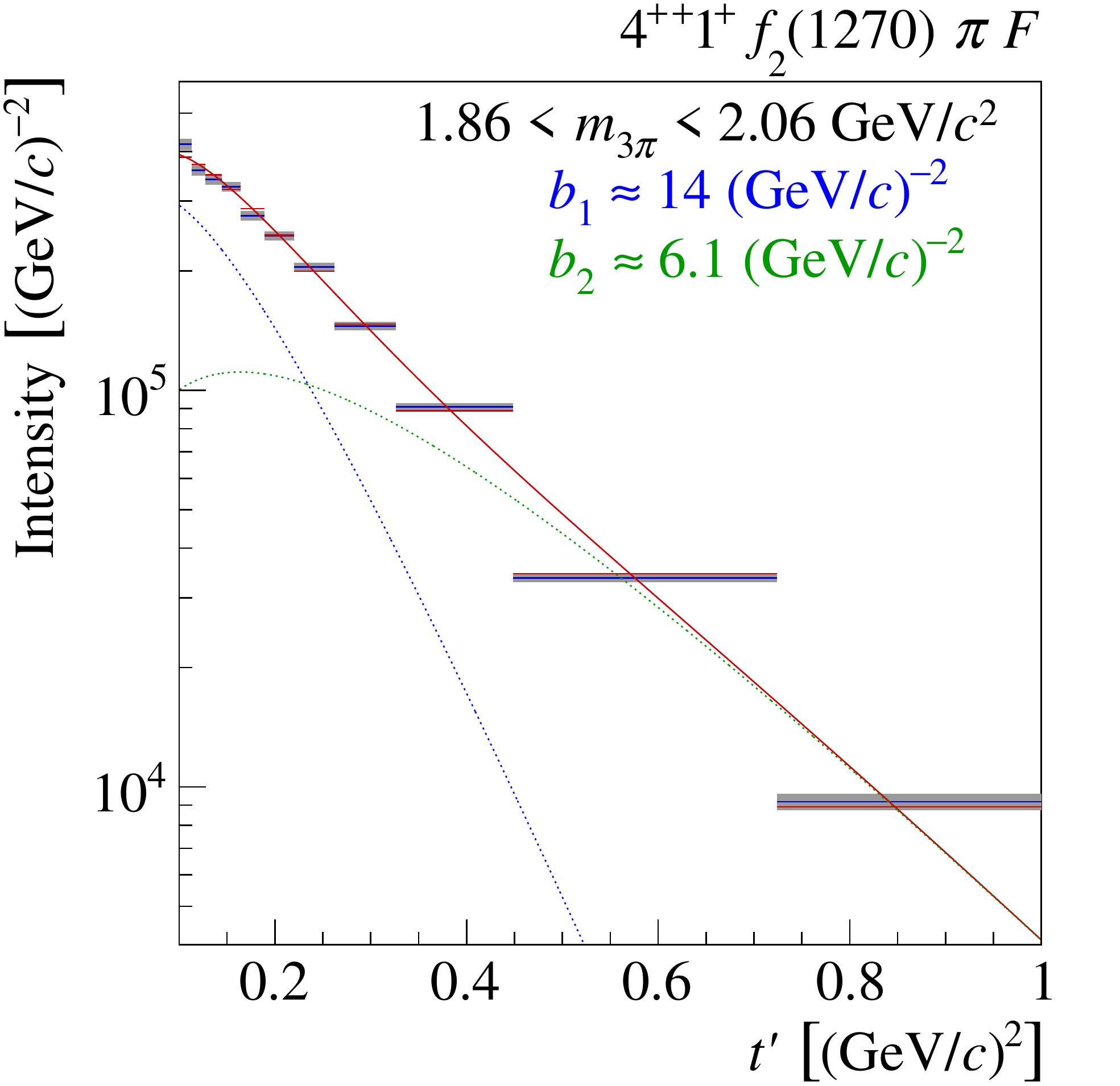}%
  }%
  \caption{\colorPlot The \tpr dependence of the intensity of the
    \wave{4}{++}{1}{+}{\Prho}{G} and $\PfTwo\,\pi\,F$ waves,
    integrated over the mass interval around the \PaFour as indicated
    by the shaded regions in
    \cref{fig:a4_total_m1_rho,fig:a4_total_m1_f2}.  The solid red
    curves represent double-exponential fits using
    \cref{eq:tprim-dependence}; the fit components are shown as dotted
    curves.  See text for details.}
  \label{fig:t_dependence_4pp}
\end{figure}

We may also compare the \tpr spectra of the same partial wave in
different mass intervals.  In \cref{fig:t_dependence_PIPIS_pi_0mp},
this is shown for the \wave{0}{-+}{0}{+}{\pipiS}{S} wave using the
peak regions \SIvalRange{1.1}{\mThreePi}{1.3}{\GeVcc} and
\SIvalRange{1.7}{\mThreePi}{1.9}{\GeVcc}.  These mass intervals
contain the low-mass part of a potential \Ppi[1300] contribution and
the peak region of the \Ppi[1800], respectively.

\begin{figure}[tbp]
  \centering
  \subfloat[][]{%
    \label{fig:t_pi_S_m1}%
    \includegraphics[width=\twoPlotWidth]{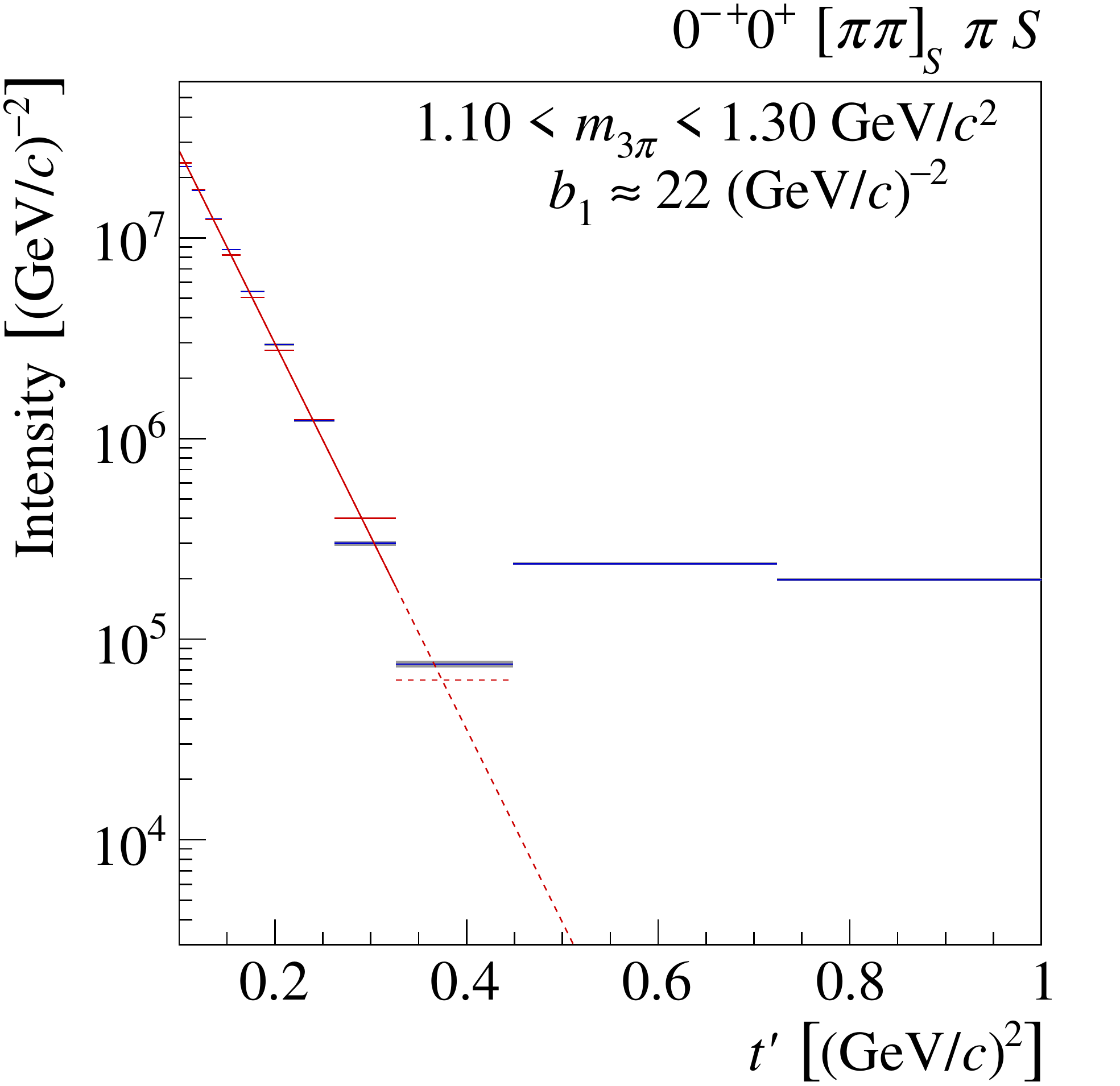}%
  }%
  \newLineOrHspace{\twoPlotSpacing}%
  \subfloat[][]{%
    \label{fig:t_pi_S_m1_2}%
    \includegraphics[width=\twoPlotWidth]{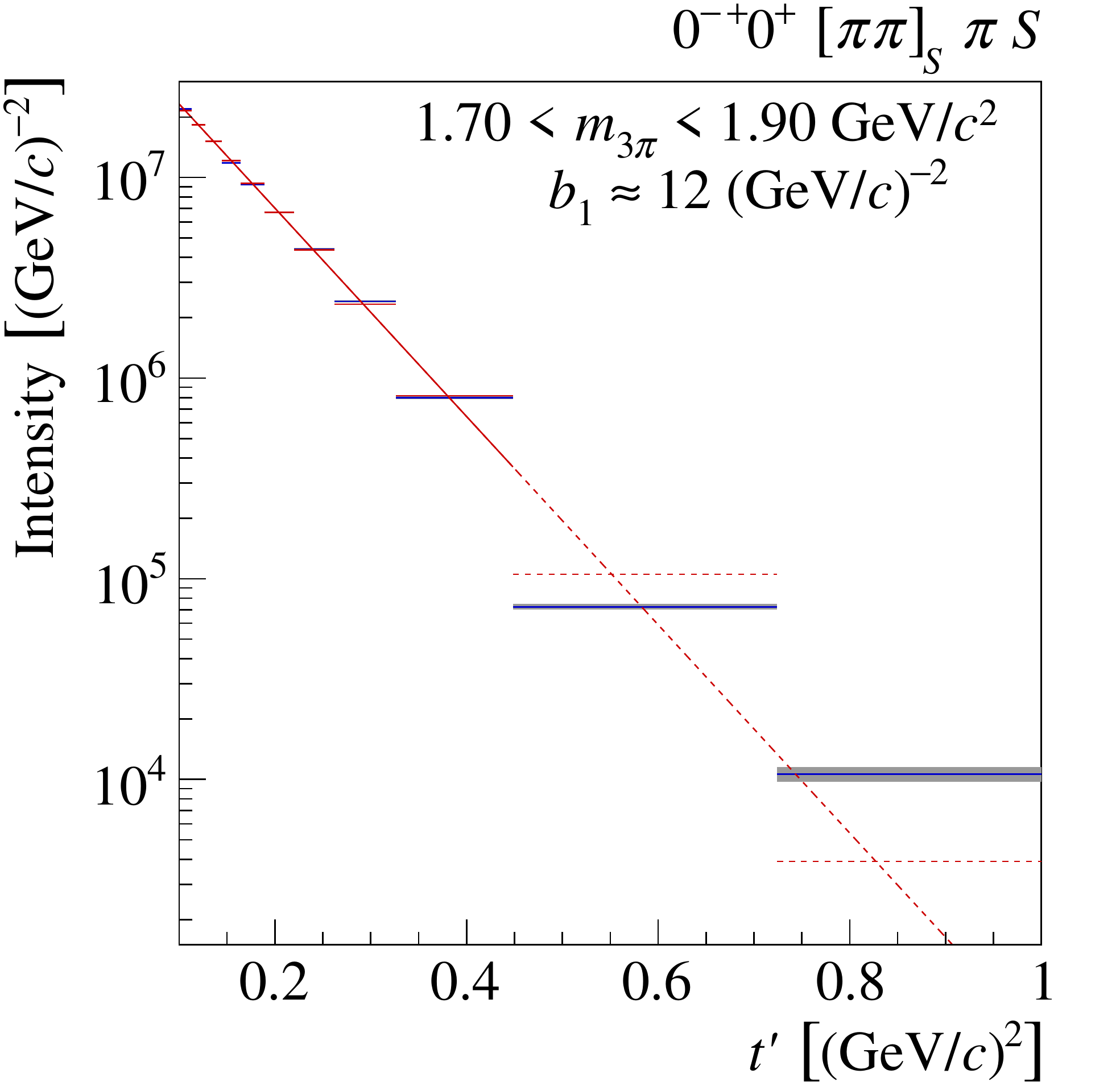}%
  }%
  \caption{\colorPlot The \tpr dependence of the intensity of the
    \wave{0}{-+}{0}{+}{\pipiS}{S} wave in two different mass regions,
    which correspond to the two peaks that are indicated by the shaded
    bands in \cref{fig:0mp_pipiS}.  The red curves represent
    single-exponential fits using \cref{eq:tprim-dependence} with
    $A_2 = 0$.  The fitted \tpr ranges are indicated by the solid
    curves; the extrapolations are shown as dashed curves.  See text
    for details.}
  \label{fig:t_dependence_PIPIS_pi_0mp}
\end{figure}

From the above figures, we can see that at low \tpr, all waves have a
large single-exponential component.  In addition, we can distinguish
three characteristically different patterns in the \tpr spectra:
\one~for about half the spectra, the single exponential dominates the
full measured \tpr range; \two~many waves show larger deviations at
higher \tpr, suggesting additional components; \three~a few waves
exhibit a minimum in the \tpr region between \SIlist{0.3;0.6}{\GeVcsq}
(see \eg \cref{fig:t_pi_S_m1,fig:t_a1_m1}).  The position of such
minima is far below the diffractive minima observed in elastic $\pi p$
scattering.

Our general ansatz for the description of the observed \tpr spectra is
a sum of two terms, each containing an exponential function multiplied
by an $M$-dependent term\footnote{Given by the forward limit of the
  Wigner $D$-functions (see
  \refCite{perl:1974high}).}\ifMultiColumnLayout{\nocite{perl:1974high}}{}
of the form $(\tpr)^M$ with $M \geq 0$:
\begin{equation}
	\label{eq:tprim-dependence}
	\od{N}{\tpr}(\tpr) = (\tpr)^M\, \sBrk{A_1\, e^{-b_1\, \tpr} + A_2\, e^{-b_2\, \tpr}}.
\end{equation}
Here, the $A_i$ are real-valued parameters. The above formula is not
able to describe the behavior of waves that show minima in their
\tpr distribution.

For each partial wave, one or two specific $3\pi$ mass ranges are
selected, which cover known resonances.  However, the \tpr dependence
of the intensity in these mass ranges still reflects the $3\pi$ system
as a whole in a given partial wave, with both resonant and nonresonant
contributions.  Bin migration effects due to the limited \tpr
resolution of the apparatus are not corrected for.  However, in the
analyzed range the \tpr resolution is better than \SI{0.02}{\GeVcsq}
(see \cref{sec:event_selection_cuts}), which renders the observed \tpr
spectra only slightly shallower than the true ones.

We perform two kinds of fits: single-slope fits, where the
parameter~$A_2$ in \cref{eq:tprim-dependence} is set to zero, and
double-slope fits, where all four parameters are left free.  For cases
where the \tpr spectra exhibit more than one component, the range of
the single-slope fits is limited to lower \tpr values.  Since
\cref{eq:tprim-dependence} is not able to describe the dip structures
appearing in some \tpr spectra, those distributions are fit only with
a single exponential.  Details on the fit results are summarized in
\cref{tab:mass_independent:single-slope-parameters,tab:mass_independent:double-slope-parameters}
that show the ranges in \mThreePi, \tpr and the resulting slope
parameters, the intensity ratio of the two components within the fit
range, and the fit quality in terms of $\chi^2/\text{ndf}$.  It should
be noted that $\chi^2$ is calculated using the integrals of the model
function over the respective \tpr bins.  About half of the spectra
require a description with two slopes.  For spectra that can be
described by a single slope only, the double-slope fit results in a
second component having a very small relative weight.  For these
cases, the values are omitted from
\cref{tab:mass_independent:double-slope-parameters}.

Because of the high precision of the data, statistical uncertainties
on the extracted slope parameters are negligible and therefore the
uncertainties are mostly of systematic nature.  The values of the
slope parameters depend, among other things, on the choice of
boundaries of the mass interval and the fit range in \tpr.  Given the
complex interplay between resonant and nonresonant components, which
can only be disentangled later at the stage of the mass-dependent
fit~\cite{COMPASS_3pi_mass_dep_fit}, we have not attempted to quantify
the systematic uncertainties.  We therefore quote the slope parameters
rounded to two-digit precision and do not give the respective
uncertainties.  In the figures, the fit functions are represented by
red curves.  For the double-exponential fits, the full \tpr range from
\SIrange{0.1}{1.0}{\GeVcsq} is used.  In contrast, the
single-exponential fits are performed using narrower \tpr ranges,
which are chosen individually for each partial wave and mass region
(see \cref{tab:mass_independent:single-slope-parameters}).  In this
case, the fit ranges are indicated by solid red curves, while the
extrapolations to the full \tpr range are shown as dashed red curves.
In every \tpr bin, the integral of the fit function, which enters the
$\chi^2$ function to be minimized, is shown as a red horizontal line,
while the blue line represents the data, so that their difference
directly indicates the fit quality.  For the double-exponential fits,
the two components are shown in addition as dotted curves: blue for
the steep component and green for the shallow one.  In the following,
we shall discuss the observed characteristics for each \JPC sector.
\begin{table*}[tbp]
  \centering
  \renewcommand{\arraystretch}{1.2}
  \caption{Slope parameters~$b_1$ from a single-exponential fit to the
    \tpr spectra in the given \tpr ranges.  The listed \mThreePi
    intervals, over which the intensity is integrated, cover the peak
    regions of the different partial waves.}
  \label{tab:mass_independent:single-slope-parameters}
  \begin{tabular}{lccSS}
    \toprule
    \textbf{Partial Wave} &
    \textbf{\mThreePi Range} &
    \textbf{\tpr Range} &
    {\textbf{$b_1$}} &
    {\textbf{$\chi^2 / \text{ndf}$}} \\
    & \textbf{[\si{\GeVcc}]} &
    \textbf{[\si{\GeVcsq}]} &
    \multicolumn{1}{c}{\textbf{[\si{\perGeVcsq}]}} & {} \\
    \midrule

    \wave{1}{++}{0}{+}{\Prho}{S}  & $[1.10, 1.30]$ & $[0.100, 0.326]$ & 12  & 120 \\
    \wave{1}{++}{1}{+}{\Prho}{S}  & $[1.10, 1.30]$ & $[0.100, 0.724]$ & 16  & 6.6 \\
    \wave{1}{++}{0}{+}{\PfTwo}{P} & $[1.68, 1.88]$ & $[0.100, 1.000]$ & 8.4 & 4.3 \\[1.2ex]

    \wave{2}{++}{1}{+}{\Prho}{D}  & $[1.20, 1.40]$ & $[0.100, 0.326]$ & 8.9 & 37  \\
    \wave{2}{++}{2}{+}{\Prho}{D}  & $[1.20, 1.40]$ & $[0.164, 0.724]$ & 8.5 & 18  \\
    \wave{2}{++}{1}{+}{\PfTwo}{P} & $[1.20, 1.40]$ & $[0.127, 0.724]$ & 7.5 & 5.3 \\[1.2ex]

    \wave{2}{-+}{0}{+}{\Prho}{F}  & $[1.56, 1.76]$ & $[0.113, 0.724]$ & 9.2 & 5.1 \\
    \wave{2}{-+}{0}{+}{\PfTwo}{S} & $[1.56, 1.76]$ & $[0.100, 0.326]$ & 9.8 & 30  \\
    \wave{2}{-+}{1}{+}{\PfTwo}{S} & $[1.56, 1.76]$ & $[0.113, 0.724]$ & 6.3 & 3.5 \\
    \wave{2}{-+}{0}{+}{\PfTwo}{D} & $[1.56, 1.76]$ & $[0.113, 1.000]$ & 7.8 & 2.7 \\[1.2ex]

    \wave{2}{-+}{0}{+}{\Prho}{F}  & $[1.80, 2.00]$ & $[0.113, 0.724]$ & 7.2 & 3.5 \\
    \wave{2}{-+}{0}{+}{\PfTwo}{D} & $[1.80, 2.00]$ & $[0.113, 0.724]$ & 8.4 & 14 \\[1.2ex]

    \wave{4}{++}{1}{+}{\Prho}{G}  & $[1.86, 2.06]$ & $[0.164, 0.724]$ & 8.8 & 25  \\
    \wave{4}{++}{1}{+}{\PfTwo}{F} & $[1.86, 2.06]$ & $[0.164, 0.724]$ & 8.4 & 11  \\[1.2ex]

    \midrule

    \addlinespace[1mm]
    \multicolumn{5}{c}{\textbf{Waves with \PfZero isobar}} \\
    \addlinespace[1mm]

    \wave{0}{-+}{0}{+}{\PfZero}{S} & $[1.70, 1.90]$ & $[0.100, 0.724]$ & 11  & 5.6 \\[1.2ex]

    \wave{1}{++}{0}{+}{\PfZero}{P} & $[1.38, 1.58]$ & $[0.100, 0.724]$ & 11  & 2.1 \\[1.2ex]

    \wave{2}{-+}{0}{+}{\PfZero}{D} & $[1.56, 1.76]$ & $[0.100, 0.724]$ & 8.4 & 4.2 \\
    \wave{2}{-+}{0}{+}{\PfZero}{D} & $[1.80, 2.00]$ & $[0.100, 0.724]$ & 7.3 & 4.9 \\[1.2ex]

    \midrule

    \addlinespace[1mm]
    \multicolumn{5}{c}{\textbf{Waves with \pipiS isobar}} \\
    \addlinespace[1mm]

    \wave{0}{-+}{0}{+}{\pipiS}{S} & $[1.10, 1.30]$ & $[0.100, 0.326]$ & 22 & 55  \\
    \wave{0}{-+}{0}{+}{\pipiS}{S} & $[1.70, 1.90]$ & $[0.100, 0.449]$ & 12 & 4.1 \\[1.2ex]

    \wave{1}{++}{0}{+}{\pipiS}{P} & $[1.10, 1.30]$ & $[0.100, 0.449]$ & 13 & 11  \\[1.2ex]

    \wave{2}{-+}{0}{+}{\pipiS}{D} & $[1.56, 1.76]$ & $[0.100, 0.724]$ & 11 & 35  \\
    \wave{2}{-+}{0}{+}{\pipiS}{D} & $[1.80, 2.00]$ & $[0.100, 0.724]$ & 11 & 5.8 \\

    \bottomrule
  \end{tabular}
\end{table*}

\begin{table*}[tbp]
  \centering
  \renewcommand{\arraystretch}{1.2}
  \caption{Same as in \cref{tab:mass_independent:single-slope-parameters},
    but for the double-exponential fit over the full \tpr range of
    \SIvalRange{0.1}{\tpr}{1.0}{\GeVcsq}.  The given intensity ratio is
    the ratio of the integrals $I_1 / I_2$ of the exponentials with
    slopes $b_{1, 2}$.  Partial waves with \tpr spectra, which exhibit a
    clear dip structure (marked with $^\dagger$) or which are already
    well described by a single slope as shown in
    \cref{tab:mass_independent:single-slope-parameters} (marked by
    $^\ddagger$) are not fit with the double-exponential
    model.}
  \label{tab:mass_independent:double-slope-parameters}
  \begin{tabular}{lcSSSS}
    \toprule
    \textbf{Partial Wave} &
    \textbf{\mThreePi Range} &
    {\textbf{$b_1$}} &
    {\textbf{$b_2$}} &
    {\textbf{Intensity}} &
    {\textbf{$\chi^2 / \text{ndf}$}} \\
    & \textbf{[\si{\GeVcc}]} &
    \multicolumn{1}{c}{\textbf{[\si{\perGeVcsq}]}} &
    \multicolumn{1}{c}{\textbf{[\si{\perGeVcsq}]}} &
    {\textbf{ratio $I_1 / I_2$}} & {} \\
    \midrule

    \wave{1}{++}{0}{+}{\Prho}{S}  & $[1.10, 1.30]$ & 15    & 8.0   & 1.2   & 6.9   \\
    \wave{1}{++}{1}{+}{\Prho}{S}$^\dagger$  & $[1.10, 1.30]$ & {---} & {---} & {---} & {---} \\  %
    \wave{1}{++}{0}{+}{\PfTwo}{P}$^\ddagger$ & $[1.68, 1.88]$ & {---} & {---} & {---} & {---} \\[1.2ex]  %

    \wave{2}{++}{1}{+}{\Prho}{D}  & $[1.20, 1.40]$ & 12    & 6.0   & 0.69  & 7.3   \\
    \wave{2}{++}{2}{+}{\Prho}{D}  & $[1.20, 1.40]$ & 12    & 5.7   & 1.2   & 2.2   \\
    \wave{2}{++}{1}{+}{\PfTwo}{P} & $[1.20, 1.40]$ & 13    & 6.5   & 0.27  & 3.1   \\[1.2ex]

    \wave{2}{-+}{0}{+}{\Prho}{F}$^\ddagger$  & $[1.56, 1.76]$ & {---} & {---} & {---} & {---} \\  %
    \wave{2}{-+}{0}{+}{\PfTwo}{S} & $[1.56, 1.76]$ & 14    & 7.0   & 0.71  & 4.1   \\
    \wave{2}{-+}{1}{+}{\PfTwo}{S}$^\ddagger$ & $[1.56, 1.76]$ & {---} & {---} & {---} & {---} \\  %
    \wave{2}{-+}{0}{+}{\PfTwo}{D}$^\ddagger$ & $[1.56, 1.76]$ & {---} & {---} & {---} & {---} \\[1.2ex]  %

    \wave{2}{-+}{0}{+}{\Prho}{F}$^\ddagger$  & $[1.80, 2.00]$ & {---} & {---} & {---} & {---} \\
    \wave{2}{-+}{0}{+}{\PfTwo}{D} & $[1.80, 2.00]$ & 11    & 5.9   & 1.5   & 0.47  \\[1.2ex]

    \wave{4}{++}{1}{+}{\Prho}{G}  & $[1.86, 2.06]$ & 13    & 6.3   & 1.1   & 4.1   \\
    \wave{4}{++}{1}{+}{\PfTwo}{F} & $[1.86, 2.06]$ & 14    & 6.1   & 0.84  & 1.9   \\[1.2ex]

    \midrule

    \addlinespace[1mm]
    \multicolumn{6}{c}{\textbf{Waves with \PfZero isobar}} \\
    \addlinespace[1mm]

    \wave{0}{-+}{0}{+}{\PfZero}{S}$^\ddagger$ & $[1.70, 1.90]$ & {---} & {---} & {---} & {---} \\[1.2ex]  %

    \wave{1}{++}{0}{+}{\PfZero}{P}$^\ddagger$ & $[1.38, 1.58]$ & {---} & {---} & {---} & {---} \\[1.2ex]  %

    \wave{2}{-+}{0}{+}{\PfZero}{D}$^\ddagger$ & $[1.56, 1.76]$ & {---} & {---} & {---} & {---} \\         %
    \wave{2}{-+}{0}{+}{\PfZero}{D}$^\ddagger$ & $[1.80, 2.00]$ & {---} & {---} & {---} & {---} \\[1.2ex]  %

    \midrule

    \addlinespace[1mm]
    \multicolumn{6}{c}{\textbf{Waves with \pipiS isobar}} \\
    \addlinespace[1mm]

    \wave{0}{-+}{0}{+}{\pipiS}{S}$^\dagger$ & $[1.10, 1.30]$ & {---} & {---} & {---} & {---} \\  %
    \wave{0}{-+}{0}{+}{\pipiS}{S}$^\ddagger$ & $[1.70, 1.90]$ & {---} & {---} & {---} & {---} \\[1.2ex]  %

    \wave{1}{++}{0}{+}{\pipiS}{P}$^\ddagger$ & $[1.10, 1.30]$ & {---} & {---} & {---} & {---} \\[1.2ex]  %

    \wave{2}{-+}{0}{+}{\pipiS}{D} & $[1.56, 1.76]$ & 16    & 7.3   & 1.2   & 2.5   \\
    \wave{2}{-+}{0}{+}{\pipiS}{D}$^\ddagger$ & $[1.80, 2.00]$ & {---} & {---} & {---} & {---} \\  %

    \bottomrule
  \end{tabular}
\end{table*}

\paragraph{$\JPC = 0^{-+}$:}
We study the $0^{-+}$ waves containing \PfZero and \pipiS as isobars.
The intensity spectrum of the \wave{0}{-+}{0}{+}{\pipiS}{S} wave shown
in \cref{fig:0mp_pipiS} exhibits two pronounced maxima and differs
strongly from the one for the corresponding wave with a \PfZero isobar
shown in \cref{fig:0mp_f0980}.  The higher-lying maximum in the
$\pipiS\,\pi\,S$ decay mode corresponds to the \Ppi[1800] and exhibits
a slope parameter of $b \approx \SI{12}{\perGeVcsq}$ (see
\cref{fig:t_pi_S_m1_2}), similar to that for the $\PfZero\,\pi\,S$
decay mode.  This is in agreement with the expectation that a
resonance should have the same slope parameter independent of its
decay mode.  In both cases, the \tpr spectra are purely exponential.
In contrast, the \tpr spectrum corresponding to the broad structure in
the \SIrange{1.1}{1.3}{\GeVcc} mass range around the elusive
\Ppi[1300] exhibits a pronounced intensity minimum around
\SI{0.35}{\GeVcsq} and a second maximum around \SI{0.6}{\GeVcsq} (see
\cref{fig:t_pi_S_m1}).  This behavior suggests that different
production processes are interfering and is similar to predictions by
the three-component Deck model~\cite{Antunes:1984cy}.  The
single-exponential fit to the low-\tpr region results in an
exceptionally steep slope of $b \approx \SI{22}{\perGeVcsq}$.  The
strikingly different \tpr dependences of the \Ppi[1300] and \Ppi[1800]
mass regions are further illustrated by \cref{fig:0mp_pipiS_t_bins}
and in \cref{sec:results_free_pipi_s_wave_int_correlations}.

\begin{figure}[tbp]
  \centering
  \subfloat[][]{%
    \label{fig:t_pi_S_m1_1pp}%
    \includegraphics[width=\twoPlotWidth]{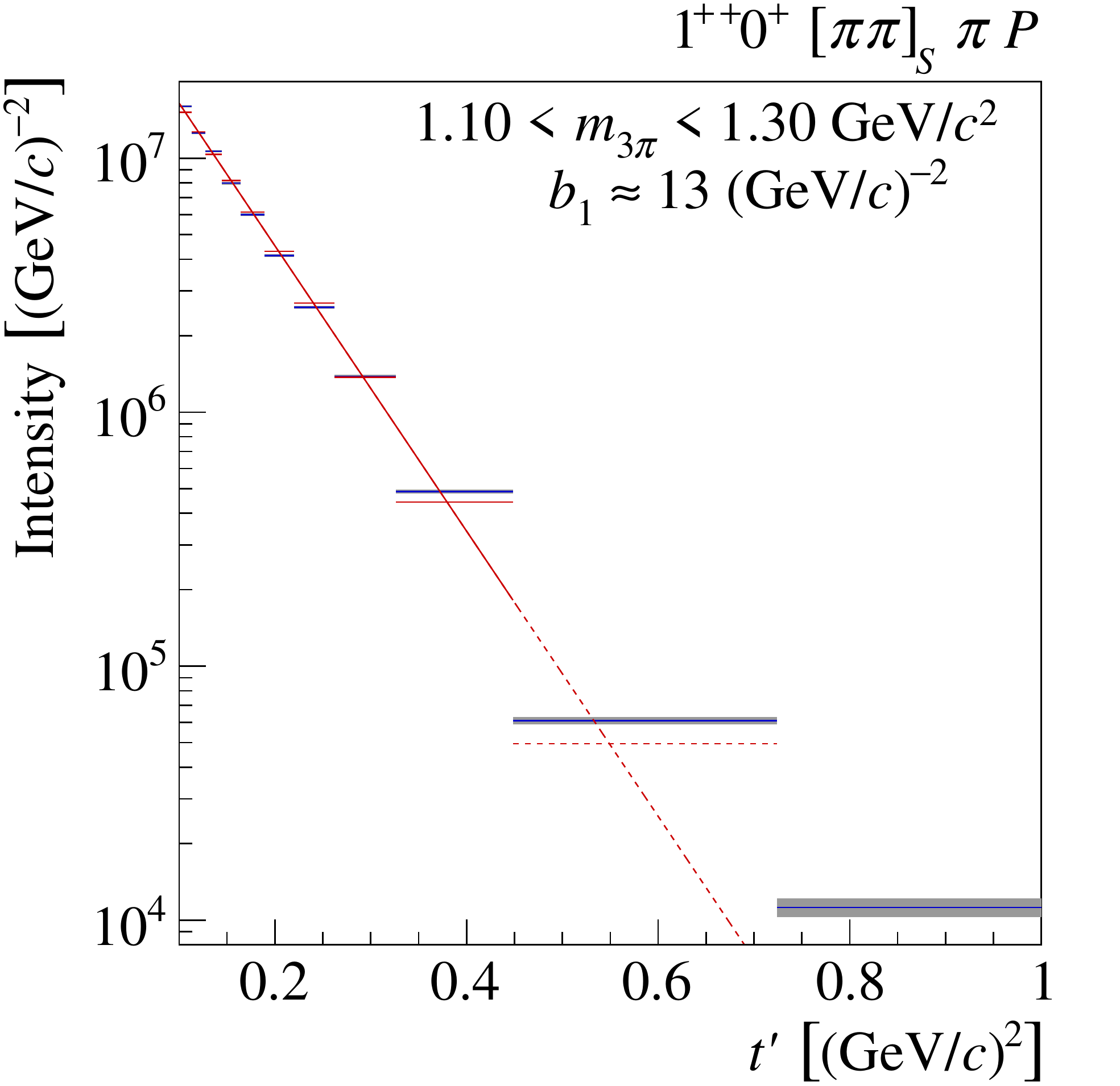}%
  }%
  \newLineOrHspace{\twoPlotSpacing}%
  \subfloat[][]{%
    \label{fig:t_pi_f0980_a1}%
    \includegraphics[width=\twoPlotWidth]{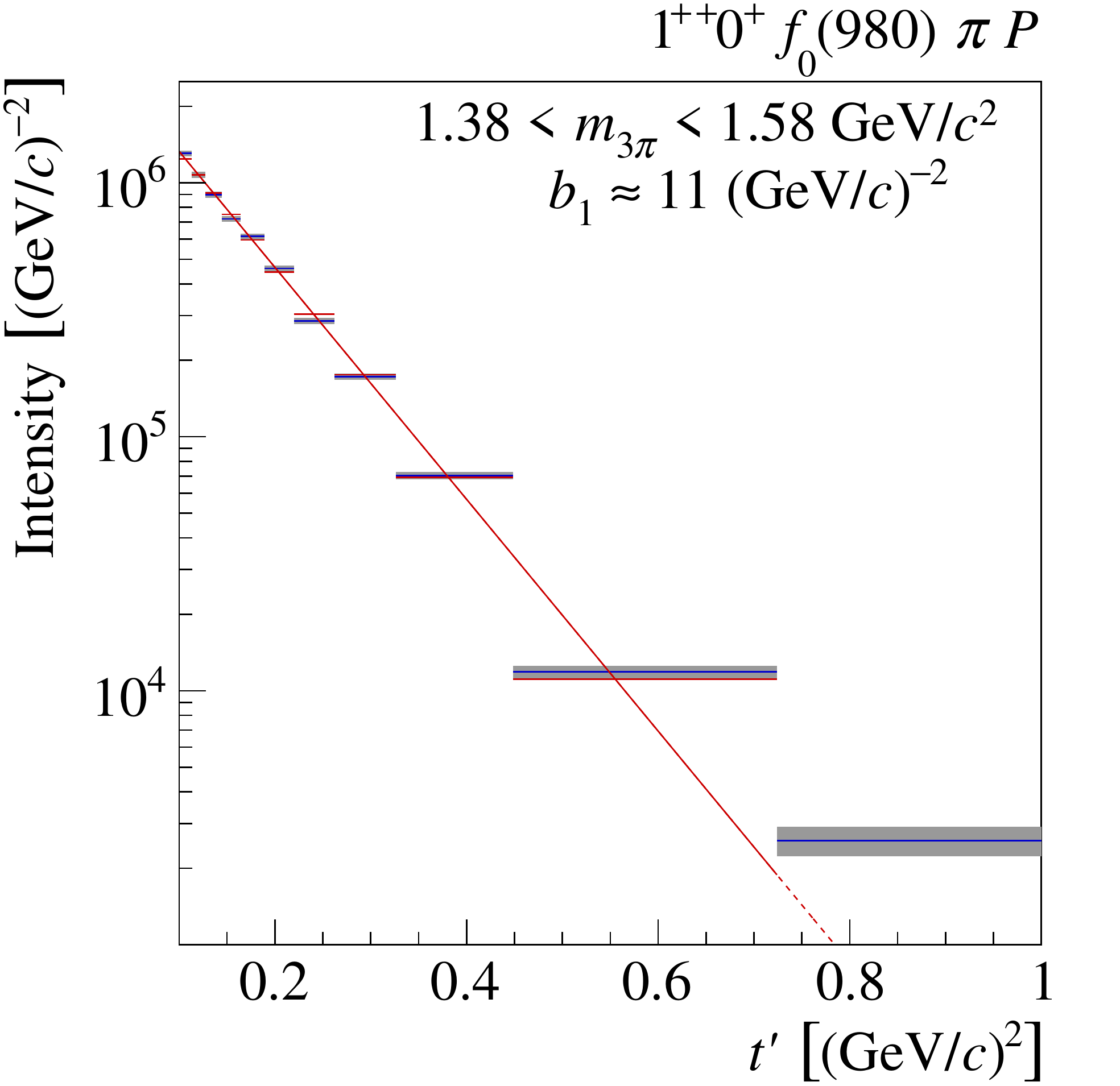}%
  }%
  \caption{\colorPlot The \tpr dependences of the intensities of the
    $1^{++}\,0^+$ waves with \pipiS~(a) and \PfZero[980]~(b) isobars
    taken around the peak region in the respective wave as indicated
    by the shaded bands in \cref{fig:1pp_pipiS,fig:1pp_f0980}.  The
    red curves represent single-exponential fits using
    \cref{eq:tprim-dependence} with $A_2 = 0$.  The fitted \tpr ranges
    are indicated by the solid curves; the extrapolations are shown as
    dashed curves.  See text for details.}
  \label{fig:t_dependence_PIPIS_pi_1pp}
\end{figure}

\paragraph{$\JPC = 1^{++}$:}
The mass region around the \PaOne peak contains both resonant and
nonresonant contributions, the latter ones dominated by the Deck
process.  Using a single slope, we obtain
$b \approx \SI{12}{\perGeVcsq}$ for the \PaOne region in the
\wave{1}{++}{0}{+}{\Prho}{S} wave, a similar value of
$b \approx \SI{13}{\perGeVcsq}$ in the \wave{1}{++}{0}{+}{\pipiS}{P}
wave (see \cref{fig:t_pi_S_m1_1pp}), and a steeper slope of
$b \approx \SI{16}{\perGeVcsq}$ for the \wave{1}{++}{1}{+}{\Prho}{S}
wave (\cref{fig:t_a1_m1}).  The \tpr distribution of the $M = 0$ wave
is much better described by two slopes with nearly equal intensity
(\cref{fig:a1_total_m0_t}).  The one for the $M = 1$ wave exhibits a
dip at approximately \SI{0.5}{\GeVcsq}.  However, the slope $b_1$ of
the steep component in the double-exponential fit of the $M = 0$ wave
is similar to that of the $M = 1$ wave extracted using the
single-exponential model in the region of lower \tpr.  If we interpret
the components with the steep slopes to be of nonresonant origin, we
would conclude that it contributes about \SI{50}{\percent} to the
$M = 0$ intensity and that it dominates the $M = 1$ wave.  The
intensity around the peak in the \wave{1}{++}{0}{+}{\PfTwo}{P} wave at
\SI{1.8}{\GeVcc} exhibits a nearly single-exponential \tpr spectrum
with a slope similar to that of the \PaTwo and \PaFour regions in the
respective $2^{++}$ and $4^{++}$ waves.  The \PaOne[1420] peak in the
\wave{1}{++}{0}{+}{\PfZero[980]}{P} wave is well described by a single
exponential (see \cref{fig:t_pi_f0980_a1}) and has a slope parameter
of $b \approx \SI{11}{\perGeVcsq}$ similar to that of the \Ppi[1800]
in the $0^{-+}$ waves with \PfZero[980] and \pipiS isobar.  This
finding is consistent with a slope parameter of
$b \approx \SI{10}{\perGeVcsq}$ that was extracted for the
\PaOne[1420] in a mass-dependent fit~\cite{Adolph:2015pws}.

\paragraph{$\JPC = 2^{++}$:}
The waves containing the \PaTwo are best described using two
exponentials and show similar behavior regardless of the type of the
isobar and the orbital angular momentum~$L$ in the decay.  A shallow
component with $b_2 \approx \SI{6}{\perGeVcsq}$ is accompanied by a
steeper component of comparable magnitude with
$b_1 \approx \SI{12}{\perGeVcsq}$.  Different spin projections~$M$ are
equally well described (see \cref{fig:t_dependence_2pp}).  In the
$2^{++}$ waves, the steep components cannot be directly identified
with nonresonant contributions, because they are small.  It cannot be
excluded that the two components are caused by the interference of the
low-mass tails of excited \PaTwo* states with the ground state, which
may contribute to the \tpr spectra with different slopes.

\paragraph{$\JPC = 2^{-+}$:}
This \JPC is studied in four partial waves containing \Prho and \PfTwo
isobars and two waves containing \pipiS and \PfZero[980] isobars.  The
latter two show striking interference effects and are discussed
further in \cref{sec:results_free_pipi_s_wave}.  The \tpr spectra are
studied in two different mass intervals: one containing the \PpiTwo,
the other the \PpiTwo[1880].  The observed pattern is rather
irregular.  Single-exponential fits yield slope values from about
\SIrange{6}{11}{\perGeVcsq}.  For the mass region around \PpiTwo, the
\tpr spectra fall into two classes: \one distributions that are
single-exponential or have only small contributions from a second
slope [\wave{2}{-+}{0}{+}{\Prho}{F}, \wave{2}{-+}{1}{+}{\PfTwo}{S},
\wave{2}{-+}{0}{+}{\PfTwo}{D}, and \wave{2}{-+}{0}{+}{\PfZero}{D}
waves] and \two distributions that need two exponentials
[\wave{2}{-+}{0}{+}{\PfTwo}{S} and \wave{2}{-+}{0}{+}{\pipiS}{D}
waves].  The latter waves have a shallower component with a slope of
around \SI{7}{\perGeVcsq} accompanied by a steep component of similar
magnitude.  The slopes of the single-exponential spectra vary
considerably.

The pattern is different for the \PpiTwo[1880] mass region.  Here, the
\wave{2}{-+}{0}{+}{\Prho}{F}, \wave{2}{-+}{0}{+}{\PfZero}{D}, and
\wave{2}{-+}{0}{+}{\pipiS}{D} waves are nearly single-exponential.
However, the latter has a steeper slope of \SI{11}{\perGeVcsq}
compared to $b_1 \approx \SI{7}{\perGeVcsq}$ for the former two.  The
\wave{2}{-+}{0}{+}{\PfTwo}{D} wave requires two slopes, where the
steep slope is about \SI{11}{\perGeVcsq} and the shallow one
approximately \SI{5.9}{\perGeVcsq}.

\paragraph{$\JPC = 4^{++}$:}
The waves containing the \PaFour are studied in decays into two
different isobars.  The \tpr spectra follow the pattern observed for
$\JPC = 2^{++}$, with one slope of $b_2 \approx \SI{6}{\perGeVcsq}$
and a steeper component described by
$b_1 \approx \SIrange{13}{14}{\perGeVcsq}$ of about equal strength.

In summary, for single-exponential fits of the \tpr spectra, we find a
general trend of shallower slopes with increasing mass.  Waves with
dominant resonant contributions, like \eg the $2^{++}$ and $4^{++}$
waves, have slopes in the range from \SIrange{7}{11}{\perGeVcsq},
which are equal for different decay modes.  In contrast, waves with
large nonresonant contributions, like \eg the $1^{++}\,\Prho\,\pi$
waves, show typically steeper slopes in the range of
\SIrange{12}{16}{\perGeVcsq}.  Many waves are better described by a
two-exponential model.  However, in general the two components do not
seem to separate nonresonant from resonant contributions.  This may be
due to possible large interference effects or contributions from
excited states.  Signs of such interferences are observed in the \tpr
spectra of some waves, which exhibit a dip around
$\tpr \approx \SIrange{0.3}{0.6}{\GeVcsq}$ and thus can be described
by the single-exponential function only in a limited \tpr interval.

Our results of the fits using single exponentials can to some extent
be compared to earlier analyses done on this topic.  The single-slope
parameters in the mass region around the \PaTwo of
\SIrange{7.5}{8.9}{\perGeVcsq} agree well with the results obtained
for the $\eta\,\pi^-$ and $\eta'\, \pi^-$ channels studied at the same
incident energy.  In the former channel, which is dominated by the
\PaTwo, the slope parameter is
\SI{8.45}{\perGeVcsq}~\cite{Adolph:2014rpp}.  In the $\eta'\, \pi^-$
channel, a slope parameter of \SI{8.2}{\perGeVcsq} is found in the
\PaTwo mass region~\cite{Schluter:2011dt}.  As in the present case,
natural-parity transfer ($M = 1$) is strongly dominant.  Hence, all
\PaTwo production characteristics are consistent with being
independent on the decay channels, as required for true resonances.
For \PaFour production, the $3\pi$ and $\eta'\, \pi^-$ results (see
\cref{tab:mass_independent:single-slope-parameters} and
\refCite{Schluter:2011dt}, respectively) are consistent with this
requirement as well.

The ACCMOR
collaboration~\cite{Daum:1979sx,Daum:1979iv,daum:1979ix,Daum:1980ay}
has pioneered such \tpr fits for selected waves in the $3\pi$ mass
region between \SIlist{0.8;1.9}{\GeVcc}, describing the \tpr spectra
in the range $\tpr < \SI{1.0}{\GeVcsq}$.  For the
\wave{1}{++}{0}{+}{\Prho}{S} wave, which contains the \PaOne[1260],
the authors quote an overall slope parameter of
$b = \SI{10.1 (9)}{\perGeVcsq}$, which is similar to our data.  For
the \wave{2}{++}{1}{+}{\Prho}{D} wave, they observe
$b = \SI{7.3 (1)}{\perGeVcsq}$ as compared to
$b \approx \SI{8.9}{\perGeVcsq}$ quoted in
\cref{tab:mass_independent:single-slope-parameters}.  Finally, for the
waves \wave{2}{-+}{0}{+}{\PfTwo}{S} and \wave{2}{-+}{0}{+}{\pipiS}{S},
they have extracted values of $b = \SI{8.5 (3)}{\perGeVcsq}$ and
$b = \SI{10.7 (11)}{\perGeVcsq}$, respectively, while selecting a mass
window from \SIrange{1.6}{1.7}{\GeVcc}.  Both values are in good
agreement with our findings (see
\cref{tab:mass_independent:single-slope-parameters}).  The authors
concluded that owing to strong nonresonant effects, the true values
for $b$ in direct resonance production might be around
$b = \SI{7.5}{\perGeVcsq}$.  This value agrees with our findings of
$b$ being in the range of \SIrange{7}{11}{\perGeVcsq} for the
$2^{++}$, $4^{++}$, as well as for the \wave{0}{-+}{0}{+}{\PfZero}{S}
and \wave{1}{++}{0}{+}{\PfZero}{P} waves, which we also ascribe to
resonant production.

Results from BNL E852 originate from a mass-independent fit in 12~bins
of \tpr and are shown in \refCite{Dzierba:2005jg} for the waves
\wave{2}{-+}{0}{+}{\PfTwo}{S}, \wave{2}{++}{1}{+}{\Prho}{D},
\wave{4}{++}{1}{+}{\Prho}{G}, and \wave{1}{-+}{1}{+}{\Prho}{P} for
\SIvalRange{0.1}{\tpr}{0.5}{\GeVcsq}, but without discussing a
functional description of the \tpr dependence in detail.
 %
%
%

\section{Determination of \pipi $S$-Wave Amplitudes}
\label{sec:results_free_pipi_s_wave}

As shown in \cref{sec:results_pwa_massindep_pipiS_waves}, the
$\JPC = 0^{++}$ isobars decaying into \twoPi in an $S$-wave are
important intermediate states in $3\pi$ meson decays.  In the
considered $3\pi$ mass range, they consist of \one a broad continuum,
which is usually described by a parametrization extracted from \pipiSW
elastic-scattering data, and \two at least two distinct resonances,
\PfZero[980] and \PfZero[1500].  The much debated \PfZero[1370] was
not included as a separate isobar in the analysis described in
\cref{sec:results_pwa_massindep}.  The key issue is: to what extent
the information extracted from \pipi elastic scattering can be used to
describe spectral shapes and phases of the two-pion $0^{++}$ isobars
in many-body decay amplitudes?

As discussed above, the \PaOne[1420] appears only in the
$\PfZero[980]\,\pi$ $P$-wave and its strength (but not its shape)
reveals some dependence on the detailed parametrization used for the
\PfZero[980], \ie a Breit-Wigner or a Flatt\'e amplitude (see
\cref{sec:pwa_massindep_systematic_studies,sec:appendix_syst_studies_massindep_isobar_param}).
This section addresses in particular the question whether the observed
\PaOne[1420] is truly related to the narrow \PfZero[980] or whether it
is an artifact of the isobar parametrizations employed in the fit.
This is relevant for the significance of the new observation as well
as for the interpretation of the \PaOne[1420].

\subsection{Method of extracting Isobar Amplitudes from Data}
\label{sec:results_free_pipi_s_wave_method}

The conventional isobar model uses fixed amplitudes for the
description of the \twoPi intermediate states $\xi$ (see
\cref{sec:pwa_method_isobar_parametrization}).  However, we cannot
exclude that the fit results are biased by the isobar parametrizations
used.  This is particularly true for the $0^{++}$ \twoPi isobars,
where we have separated a broad \pipiSW component from the
\PfZero[980] and \PfZero[1500] resonances.  In order to solve this
problem, a novel method inspired by \refCite{Aitala:2005yh} was
developed.  It allows us to determine the overall amplitude of the
\zeroPP \twoPi isobars directly from the data.

For selected isobars, the new method abandons the fixed description of
the mass-dependent amplitudes $\Delta_\xi(m_\xi)$, which appear in the
two-body isobar decay amplitude of
\cref{eq:decay_amplitude_isobar_dyn} and are part of the full decay
amplitude of the state $X^-$ defined in
\cref{eq:decay_amplitude_before_sym,eq:decay_amplitude_norm}.  The
latter amplitude factorizes into a part, $\mathcal{K}_a^\refl$, which
depends on the spherical angles
$\anglesGJ \equiv (\cosThetaGJ, \phiGJ)$ in the Gottfried-Jackson
frame as well as $\anglesHF \equiv (\cosThetaHF, \phiHF)$ in the
helicity frame, and a second part, $\Delta_a$, that is the
mass-dependent isobar amplitude.  Taking into account the Bose
symmetrization according to \cref{eq:decay_amplitude}, we write for a
particular \mThreePi bin
\begin{multlineOrEq}
  \label{eq:deisobarred_ansatz}
  \Psi_a^\refl(\tau_{13}, \tau_{23}) =
    \mathcal{K}_a^\refl(\Omega_{13}^\text{GJ}, \Omega_{13}^\text{HF})\, \Delta_a(m_{13}) \newLineOrNot
  + \mathcal{K}_a^\refl(\Omega_{23}^\text{GJ}, \Omega_{23}^\text{HF})\, \Delta_a(m_{23}).
\end{multlineOrEq}
The two terms represent the two possible \twoPi combinations of the
$\pi_1^-\pi_2^-\pi_3^+$ system.  The index~$a$ defined in
\cref{eq:wave_notation} represents the quantum numbers of the $3\pi$
partial wave.  This includes the quantum numbers of the \twoPi
subsystem $\xi$.

In our new \emph{freed-isobar} method, we replace the fixed
parametrizations for $\Delta_a(m_\xi)$ by a set of piecewise constant
functions that fully cover the allowed mass range for $m_\xi$.  The
isobar line shape is rewritten as:
\begin{equation}
  \label{eq:decay_amp_step_func}
  \Delta_a(m_\xi) = \sum_k \mathscr{T}_{a, k}\, \Pi_{k, \xi}(m_\xi),
\end{equation}
where the index $k$ runs over \twoPi mass bins.  These bins are
defined by sets of window functions $\cbrk{\Pi_{k, \xi}(m_\xi)}$ that
are non-zero only in a narrow $m_\xi$ interval in the isobar spectrum
given by the bin borders $\cbrk{m_{k, \xi}}$:
\begin{equation}
  \label{eq:dynamic_step_func}
  \Pi_{k, \xi}(m_\xi) = \begin{cases}
    1 & \text{if}~m_{k, \xi} \leq m_\xi < m_{{k + 1}, \xi}, \\
    0 & \text{otherwise}.
  \end{cases}
\end{equation}
Here, the bin width $\delta m_\xi = m_{{k + 1}, \xi} - m_{k, \xi}$ may
depend on the mass region of the \twoPi system considered.  The
$\mathscr{T}_{a, k}$ are unknown complex numbers that determine the
\emph{binned} amplitude $\Delta_a(m_\xi)$.

The intensity distribution in a given $3\pi$ mass bin, as defined in
\cref{eq:intensity_bin}, contains terms of the form
\begin{multline}
  \label{eq:deisobarred_intensity_term}
  \mathcal{T}_a^\refl\, \Psi_a^\refl(\tau_{13}, \tau_{23}) \\
  \begin{aligned}
    = \mathcal{T}_a^\refl\,
         \bigg[ \alignOrNot\mathcal{K}_a^\refl(\Omega_{13}^\text{GJ}, \Omega_{13}^\text{HF})\, \sum_k \mathscr{T}_{a, k}\, \Pi_{k, \xi}(m_{13}) \newLineOrNot
    +\> \alignOrNot\mathcal{K}_a^\refl(\Omega_{23}^\text{GJ}, \Omega_{23}^\text{HF})\, \sum_k \mathscr{T}_{a, k}\, \Pi_{k, \xi}(m_{23}) \bigg].
  \end{aligned}
\end{multline}
Absorbing the unknown isobar amplitudes $\mathscr{T}_{a, k}$ into the
transition amplitude $\mathcal{T}_a^\refl$ via
\begin{equation}
  \label{eq:deisobarred_prod_amp}
  \mathcal{T}_{a, k}^\refl \equiv \mathcal{T}_a^\refl\, \mathscr{T}_{a, k},
\end{equation}
the $m_\xi$ bins appear in the intensity via the index~$k$ that is
summed over coherently, in the same way as the partial-wave index~$a$
is:
\begin{equation}
  \label{eq:deisobarred_intensity_bin}
  \mathcal{I}(\tau)
  = \sum_{\refl = \pm 1}
  \bigg| \sum_a \sum_k \mathcal{T}_{a, k}^\refl\, \Psi_{a, k}^\refl(\tau_{13}, \tau_{23}) \bigg|^2 + \,\mathcal{I}_\text{flat},
\end{equation}
where
\begin{multlineOrEq}
  \Psi_{a, k}^\refl(\tau_{13}, \tau_{23}) =
    \mathcal{K}_a^\refl(\Omega_{13}^\text{GJ}, \Omega_{13}^\text{HF})\, \Pi_{k, \xi}(m_{13}) \newLineOrNot
  + \mathcal{K}_a^\refl(\Omega_{23}^\text{GJ}, \Omega_{23}^\text{HF})\, \Pi_{k, \xi}(m_{23}).
\end{multlineOrEq}
This means that each $2\pi$ mass bin can be treated like an
independent partial wave.  In this way, the same procedure as for the
standard mass-independent fit can be used.  The fits in $3\pi$ mass
bins yield transition amplitudes $\mathcal{T}_a^\refl$ that now depend
on \mThreePi and \mTwoPi.  According to
\cref{eq:deisobarred_prod_amp}, these amplitudes contain information
on the $3\pi$ system as well as on the \twoPi subsystem.  It should be
noted that the method is restricted to rank~1.  Therefore, the rank
index was omitted in the above formulas.  It was discussed in
\cref{sec:pwa_massindep_model} that rank~1 is sufficient for the
positive-reflectivity waves.

In the ansatz for the decay amplitude in \cref{eq:deisobarred_ansatz},
the isobar mass-dependent amplitude $\Delta_a(m_\xi)$ depends on the
$3\pi$ partial-wave index~$a$, \ie the model permits different isobar
amplitudes for different intermediate states $X^-$.  This is in
contrast to the conventional approach, which uses the \emph{same}
isobar parametrization in different partial waves.

The reduced model dependence of the new method and the additional
information about the \twoPi subsystem lead to a considerable increase
in the number of fit parameters in the mass-independent fit.  Thus,
even for large data sets, the freed-isobar approach can only be
applied to a subset of partial waves.  In the analysis presented here,
we replace the fixed parametrizations of the set of $\JPC = 0^{++}$
isobars, which consists of \pipiS, \PfZero[980], and \PfZero[1500], by
a set of single piecewise constant functions representing the overall
dynamical amplitude of all $0^{++}$ isobars as defined in
\cref{eq:decay_amp_step_func,eq:dynamic_step_func}.  In the following,
we shall denote the freed $0^{++}$ isobar amplitude by \pipiSF.

We determine the \pipiSF amplitudes simultaneously for the waves
\wave{0}{-+}{0}{+}{\pipiSF}{S}, \wave{1}{++}{0}{+}{\pipiSF}{P}, and
\wave{2}{-+}{0}{+}{\pipiSF}{D}, which are the dominant waves with
$0^{++}$ isobars.  These partial waves replace a set of seven waves
with conventional isobar parametrizations (see \cref{tab:waveset} in
\cref{sec:appendix_wave_set}).  For all other amplitudes, we keep the
isobar parametrizations as discussed in
\cref{sec:pwa_method_isobar_parametrization}.  The fits are performed
in \mThreePi bins with \SI{40}{\MeVcc} width, \ie twice as wide as
used in the conventional analysis.  Each fit results in an \Argand for
the \pipiSF amplitude ranging in the two-pion mass from $2 m_\pi$ to
$\mThreePi - m_\pi$.  The bin width in the \twoPi subsystem mass is
\SI{40}{\MeVcc}, except for the region
$\SIvalRange{920}{\mTwoPi}{1080}{\MeVcc}$ around the
\PfZero[980].\footnote{Also the first \mTwoPi bin, which covers the
  mass range from $2 m_\pi$ to \SI{320}{\MeVcc}, has a slightly
  different width.}  Here, finer bins of \SI{10}{\MeVcc} width are
chosen in order to better resolve the resonance structure.  In total,
62~two-pion mass bins are used.  In order to obtain reasonable
statistical accuracy, we perform this analysis in only four bins of
\tpr, which are listed in \cref{tab:t_binning_2pi_analysis}.

\begin{table}[tbp]
  \sisetup{%
    round-mode = places,
    round-precision = 3
  }
  \caption{Borders of the four non-equidistant \tpr bins, in which the
    partial-wave analysis with freed isobars is performed.  The intervals
    are chosen such that each bin contains approximately \num[round-mode =
    places, round-precision = 1]{11.5e6} events.}
  \label{tab:t_binning_2pi_analysis}
  \renewcommand{\arraystretch}{1.2}
  \newcolumntype{Z}{%
    >{\Makebox[0pt][c]\bgroup}%
    c%
    <{\egroup}%
  }
  \setlength{\tabcolsep}{0pt}  %
  \centering
  \begin{tabular}{l@{\extracolsep{12pt}}c@{\extracolsep{6pt}}Z*{4}{cZ}cc}
   \toprule
   \textbf{Bin} && 1 && 2 && 3 && 4 \\
   \midrule
   \textbf{\tpr [\si{\GeVcsq}]} &
   \num{0.10000} &&
   \num{0.14077} &&
   \num{0.19435} &&
   \num{0.32617} &&
   \num{1.00000} \\
   \bottomrule
  \end{tabular}
\end{table}

As in the conventional analysis, multiple fit attempts are performed
with randomly chosen starting values for the decay amplitudes
$\mathcal{T}_{a, k}^\refl$ in \cref{eq:deisobarred_intensity_bin}.
Here, the fit with the highest likelihood is selected from a set of
50~attempts.  For $3\pi$ masses below about \SI{1}{\GeVcc}, the fits
tend to be trapped in local maxima that deviate from each other only
by a few units of log-likelihood.  Such a behavior is also observed in
the fixed-isobar fits (see
\cref{sec:pwa_massindep_systematic_studies}).

\subsection{Comparison with the Fixed-Isobar Method}
\label{sec:results_free_pipi_s_wave_comparison}

In order to compare the new freed-isobar method with the conventional
analysis scheme, the fixed-isobar fit was repeated with the coarse
binning in $3\pi$ mass and \tpr.  Based on this fit, the amplitudes of
partial waves with the same $X^-$ quantum numbers but different
$0^{++}$ isobars, \ie\pipiS, \PfZero[980], and \PfZero[1500], are
added coherently.  For $\JPC = \zeroMP$, \onePP, and \twoMP of the
$3\pi$ system, the resulting intensities are shown in
\cref{fig:waves_with_s*} as blue data points in two \tpr regions
chosen as examples.  These spectra are related to those in
\cref{fig:pipi_s-wave}, which show the intensity distributions
separately for the \pipiS and \PfZero[980] isobars, integrated over
the full range of \tpr.  The striking interference effects observed in
the $2^{-+}\,0^+$ $\PfZero[980]\, \pi$ and $\pipiS\, \pi$ isobaric
waves are washed out in the coherent sum of the two.

\begin{figure*}[htbp]
  \centering
  \subfloat[][]{%
    \label{fig:0mp_mass_pi_s_lowt}%
    \includegraphics[width=\twoPlotWidth]{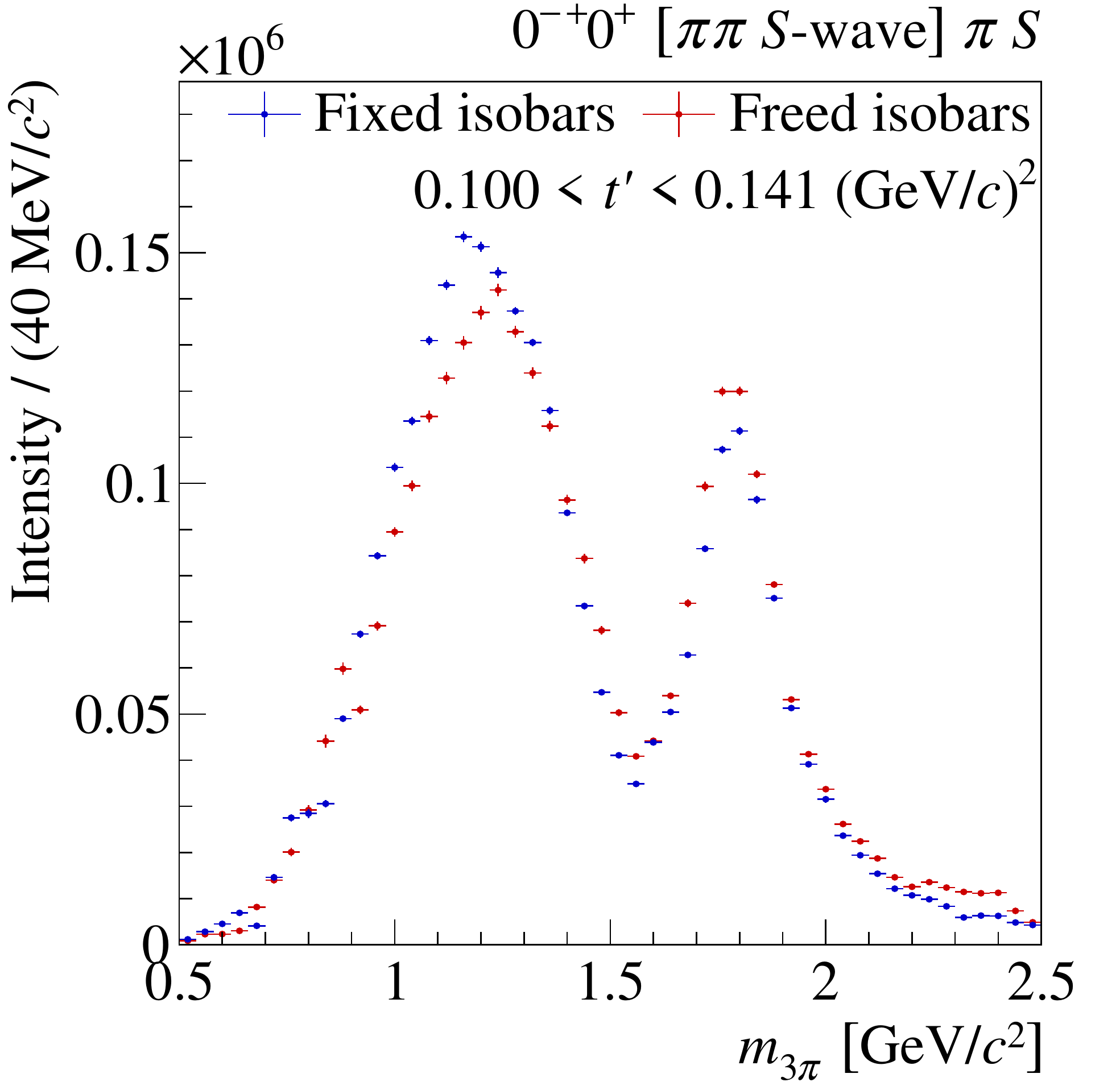}%
  }%
  \hspace*{\twoPlotSpacing}
  \subfloat[][]{%
    \label{fig:0mp_mass_pi_s_hight}%
    \includegraphics[width=\twoPlotWidth]{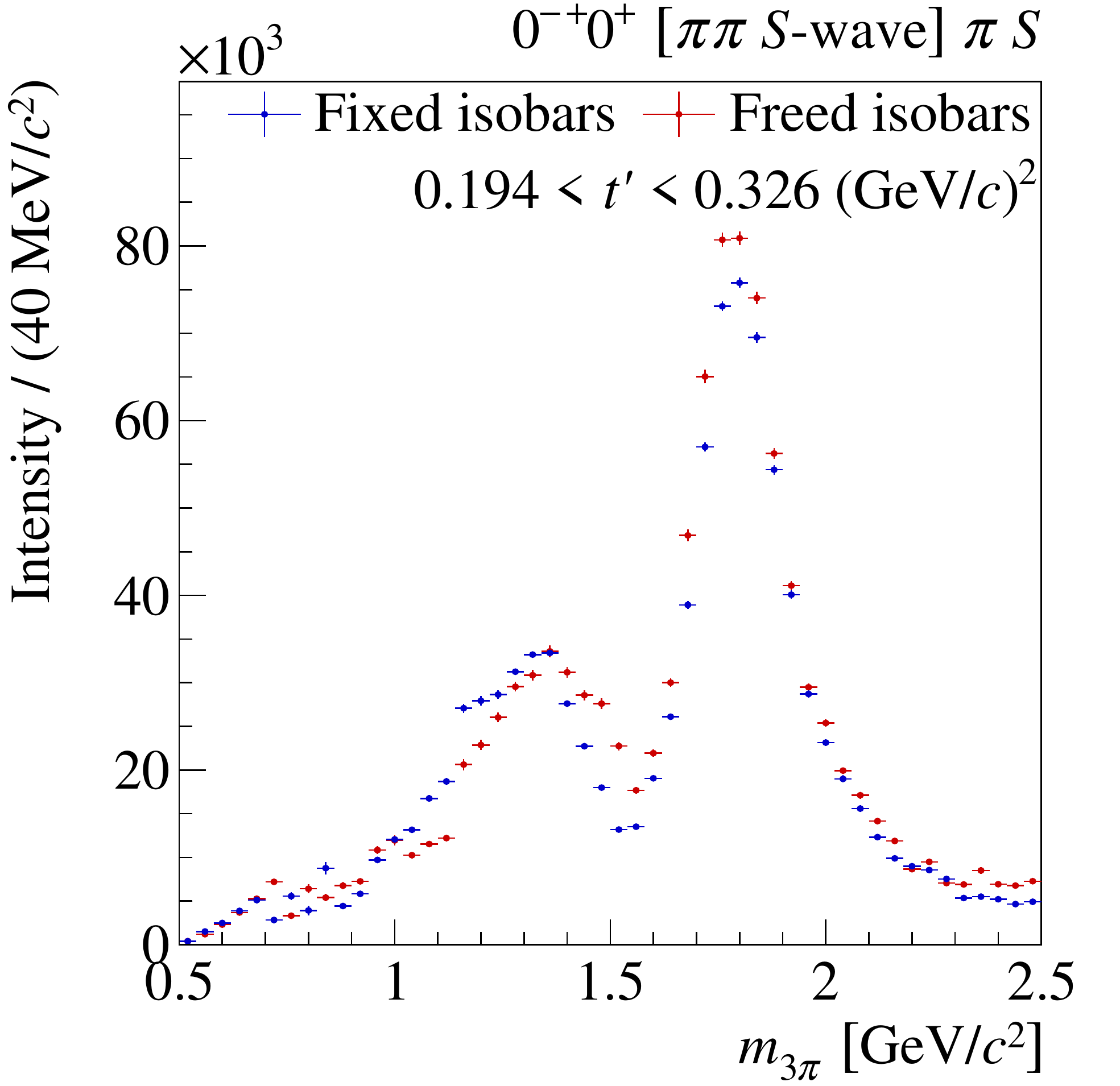}%
  }%
  \\
  \subfloat[][]{%
    \label{fig:1pp_mass_pi_s_lowt}%
    \includegraphics[width=\twoPlotWidth]{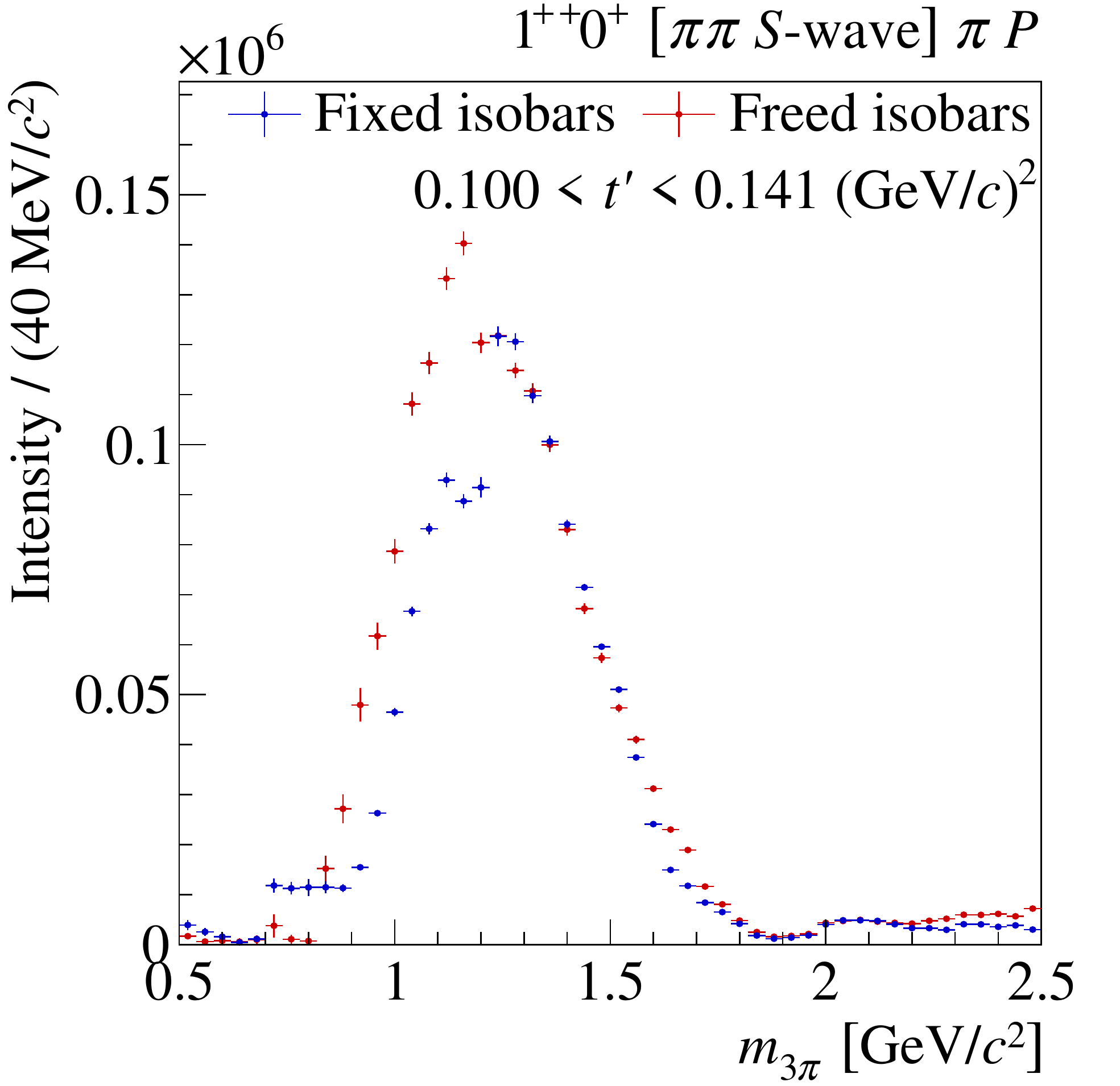}%
  }%
  \hspace*{\twoPlotSpacing}
  \subfloat[][]{%
    \label{fig:1pp_mass_pi_s_hight}%
    \includegraphics[width=\twoPlotWidth]{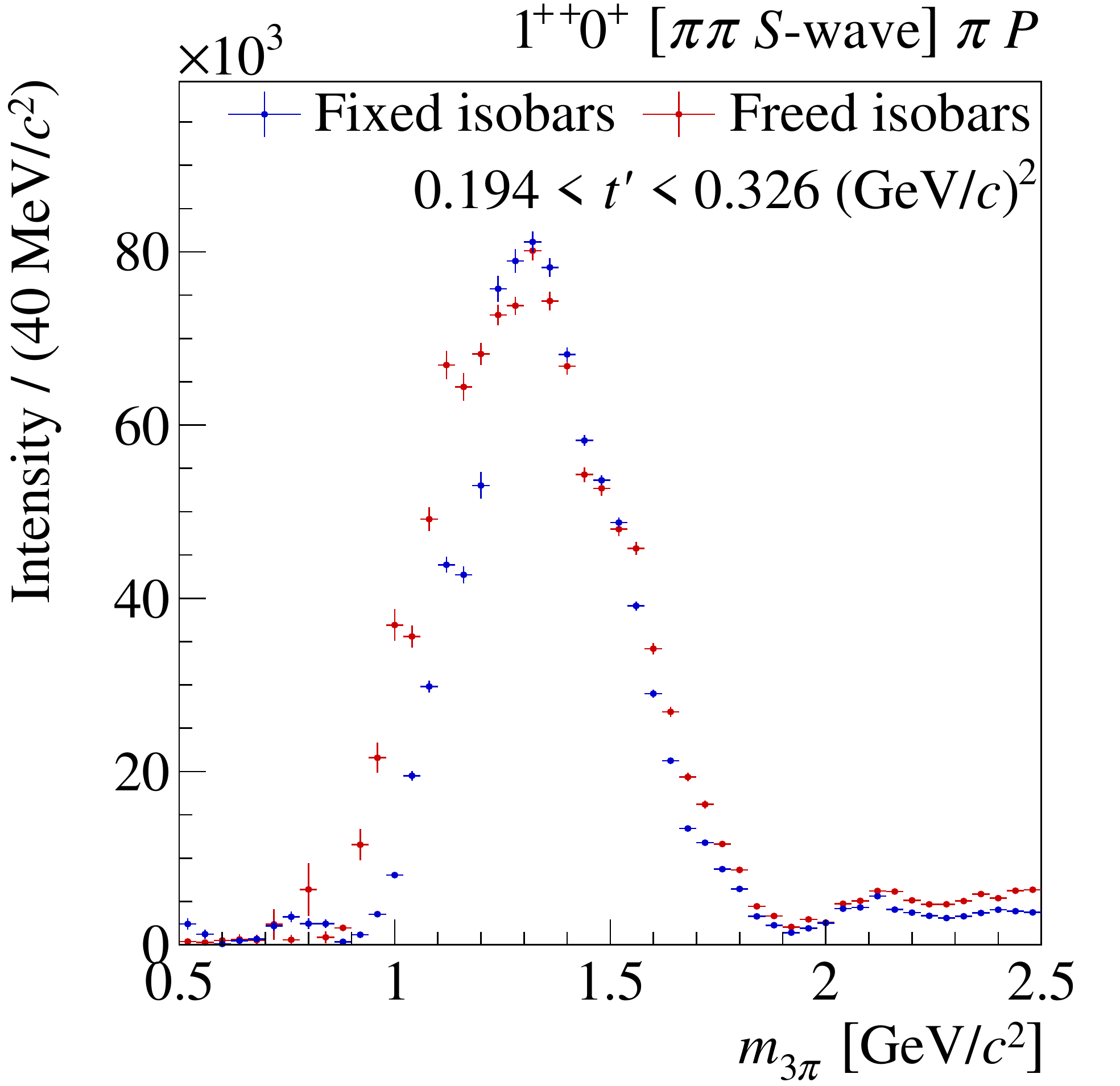}%
  }%
  \\
  \subfloat[][]{%
    \label{fig:2mp_mass_pi_s_lowt}%
    \includegraphics[width=\twoPlotWidth]{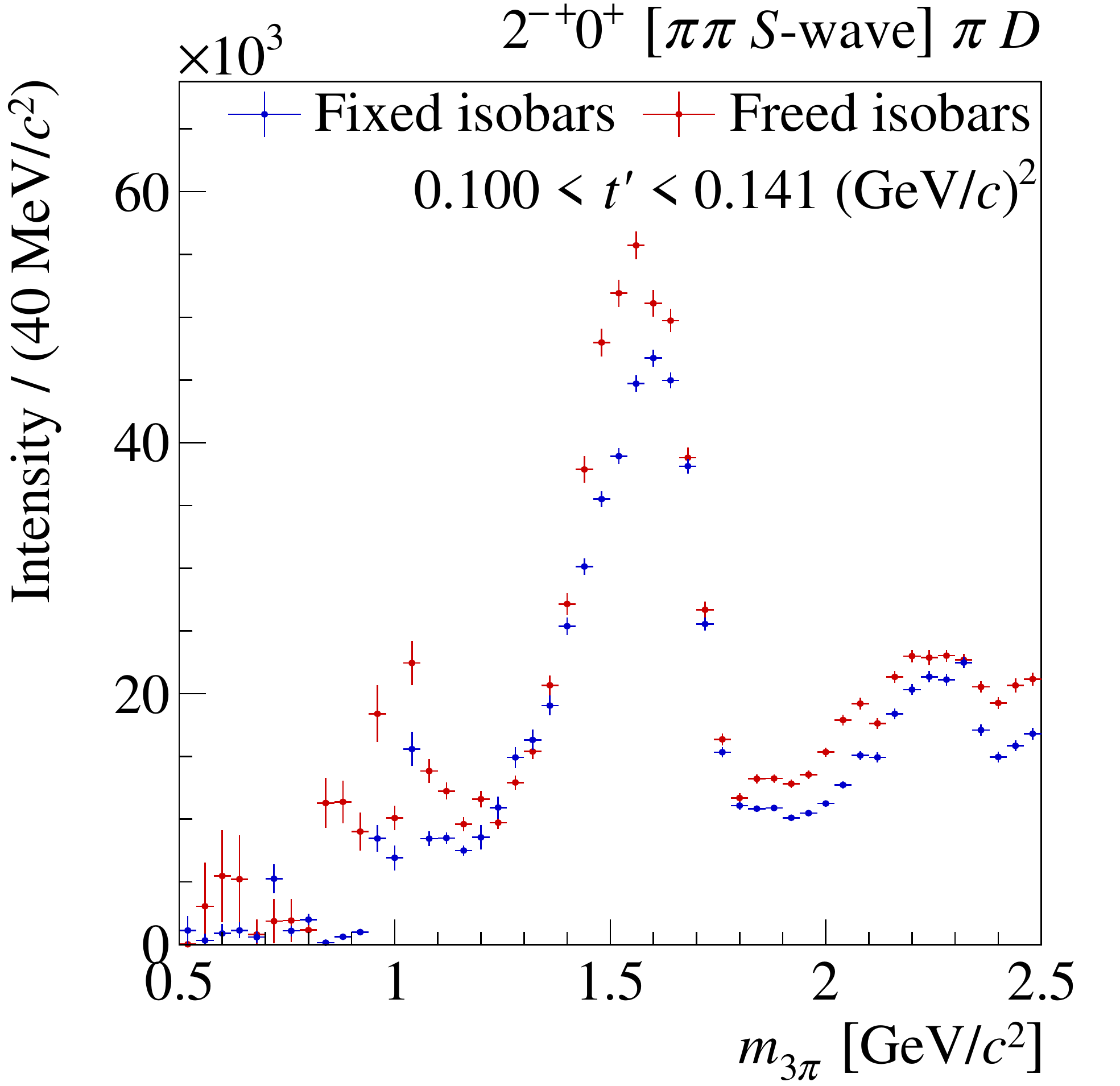}%
  }%
  \hspace*{\twoPlotSpacing}
  \subfloat[][]{%
    \label{fig:2mp_mass_pi_s_hight}%
    \includegraphics[width=\twoPlotWidth]{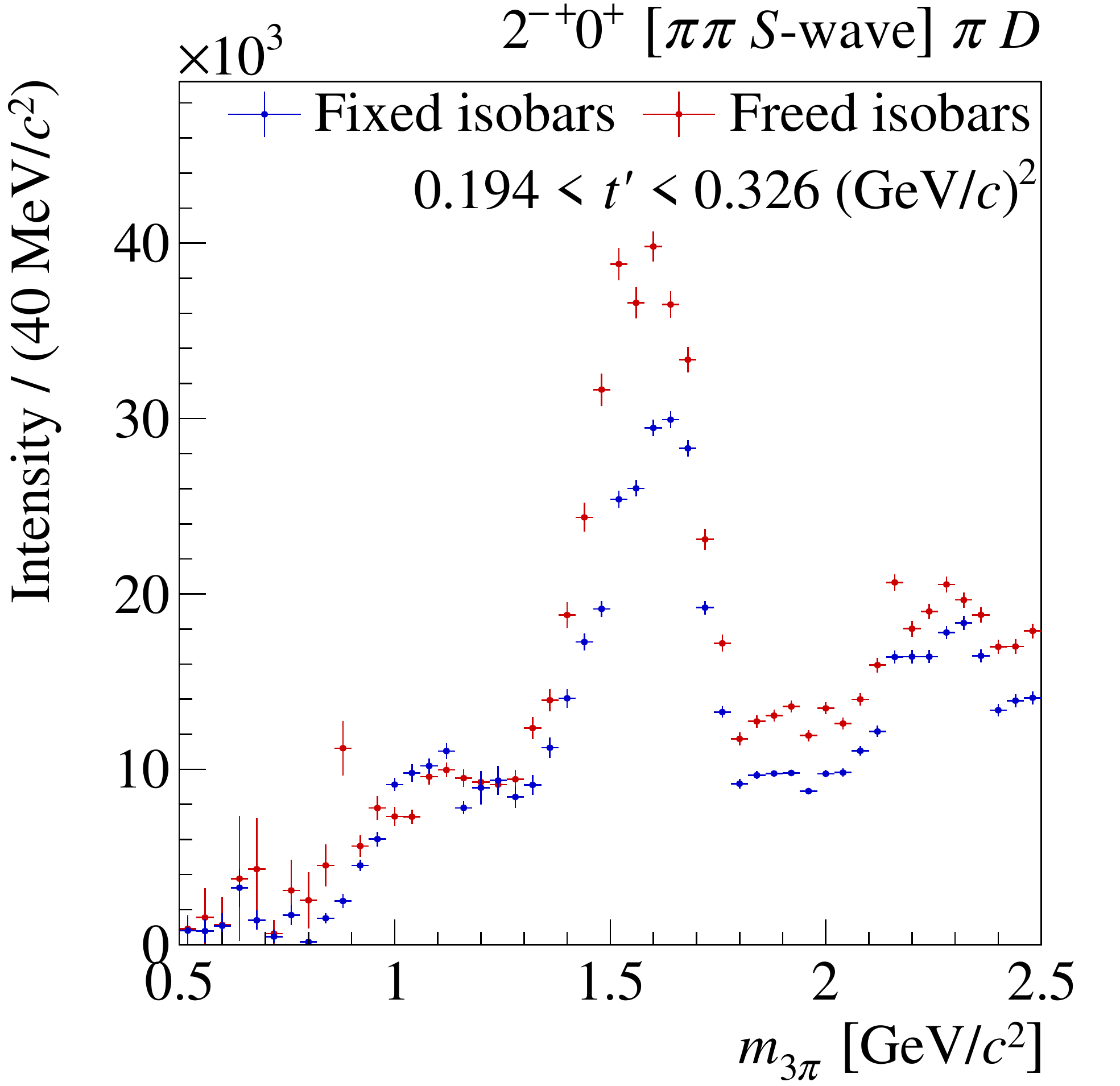}%
  }%
  \caption{\colorPlot Intensities of the coherent sum of partial waves
    with the same quantum numbers but different $0^{++}$ isobars, as
    obtained in the conventional analysis scheme with fixed isobar
    amplitudes (blue), and the corresponding intensities obtained from
    the freed-isobar fit (red).  Left column: low \tpr; Right column:
    high \tpr.}
  \label{fig:waves_with_s*}
\end{figure*}

In our novel approach, we do not separate the different $0^{++}$
isobar components but obtain the overall $0^{++}$ amplitude in bins of
\mTwoPi and \mThreePi.  This also implies that the correlation of the
relative phases between the components across the \mTwoPi spectrum is
not predetermined.  The red data points in \cref{fig:waves_with_s*}
show the $3\pi$ mass spectra for $\JPC = \zeroMP$, \onePP, and \twoMP,
obtained by coherently summing over all two-pion mass slices
[represented by index $k$ in \cref{eq:decay_amp_step_func}] in the two
chosen \tpr bins.  These intensity distributions can be compared
directly to those obtained by coherently summing over the $0^{++}$
isobars using the conventional analysis method shown as blue data
points in \cref{fig:waves_with_s*}.  The agreement is good in general,
in particular the \Ppi[1800] region in the $0^{-+}$ wave matches well.
In the $1^{++}$ wave, the region $\mThreePi < \SI{1.2}{\GeVcc}$ is
enhanced in the fit result for the freed isobars except for the
highest \tpr bin.  This is partly due to the fact that waves with
freed isobars have no $3\pi$ mass thresholds in the new fit, whereas
in the conventional fit, the \wave{1}{++}{0}{+}{\PfZero[980]}{P} wave
was limited to the region of $\mThreePi > \SI{1180}{\MeVcc}$ (see
\cref{tab:waveset} in \cref{sec:appendix_wave_set}).  The largest
differences appear in the $2^{-+}$ wave, where we observe in the
freed-isobar fit an increased intensity in the region around the
\PpiTwo[1670] across all \tpr bins.  Systematic studies indicate that
imperfections in the description of the other isobars used in the PWA
fit have an influence on the \zeroPP sector.  The agreement between
the results of the two methods validates the parametrizations of the
\pipiSW isobars that are employed in the simpler fixed-isobar fit.

\subsection{Correlation of $2\pi$ and $3\pi$ Mass Spectra for freed
  \pipi $S$-Wave Isobars}
\label{sec:results_free_pipi_s_wave_int_correlations}

It is instructive to look at the correlation of the \pipiSF mass
spectrum with the $3\pi$ mass spectrum in different partial waves.
The examples shown in
\cref{fig:pipis_2D_0mp,,fig:pipis_2D_1pp,,fig:pipis_2D_2mp} are
extracted from the \wave{0}{-+}{0}{+}{\pipiSF}{S},
\wave{1}{++}{0}{+}{\pipiSF}{P}, and \wave{2}{-+}{0}{+}{\pipiSF}{D}
waves, respectively.  The $z$ axis of the two-dimensional
representations (left columns) is given by
$\abs{\mathcal{T}_{a}^\refl(\mTwoPi, \mThreePi)}^2$, which is
normalized such that it can be interpreted as the number of events per
unit in \mTwoPi.  The apparent dependence of the shape of the $2\pi$
mass distribution on \mThreePi and on \JPC of the 3$\pi$ system
reveals the different coupling of $3\pi$ resonances to the various
$0^{++}$ components of the $2\pi$ subsystem.  In the following, we
will discuss the features for each three-pion \JPC in detail.

\begin{figure*}[htbp]
  \centering
  \subfloat[][]{%
    \includegraphics[width=\threePlotWidthTwoD]{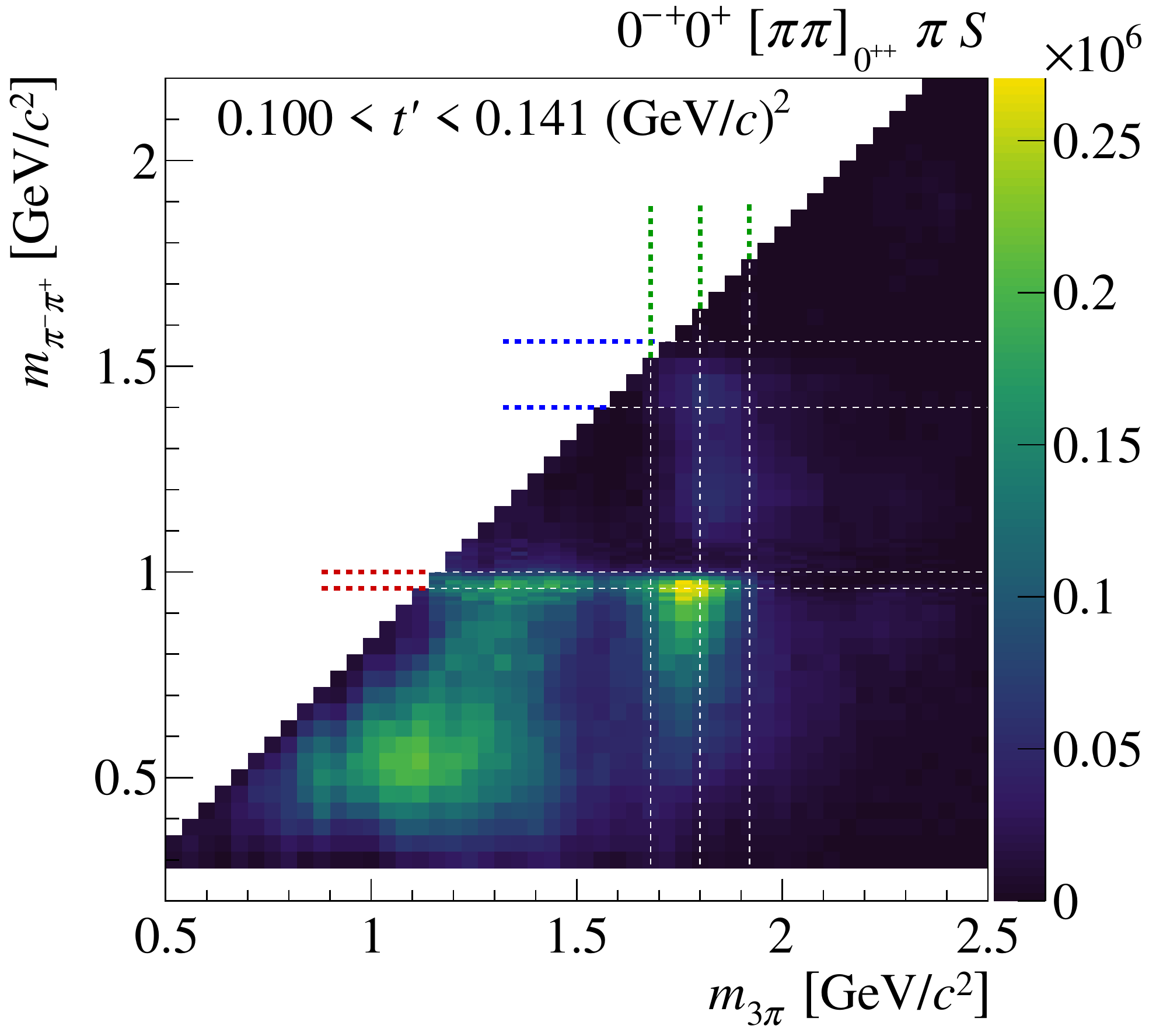}%
  }%
  \subfloat[][]{%
    \label{fig:PIPIS_0mp_ll_980}%
    \includegraphics[width=\threePlotWidth]{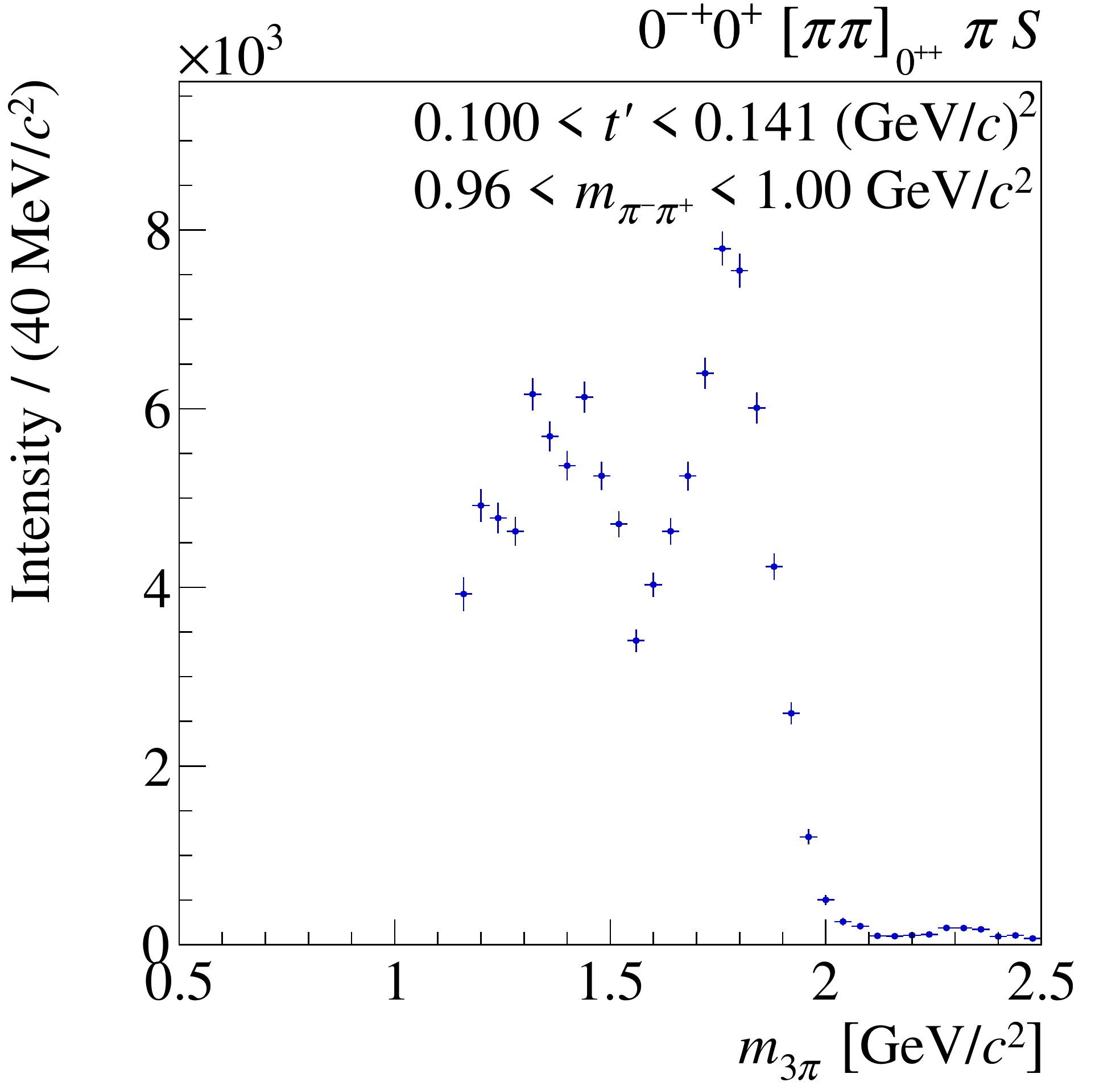}%
  }%
  \subfloat[][]{%
    \includegraphics[width=\threePlotWidth]{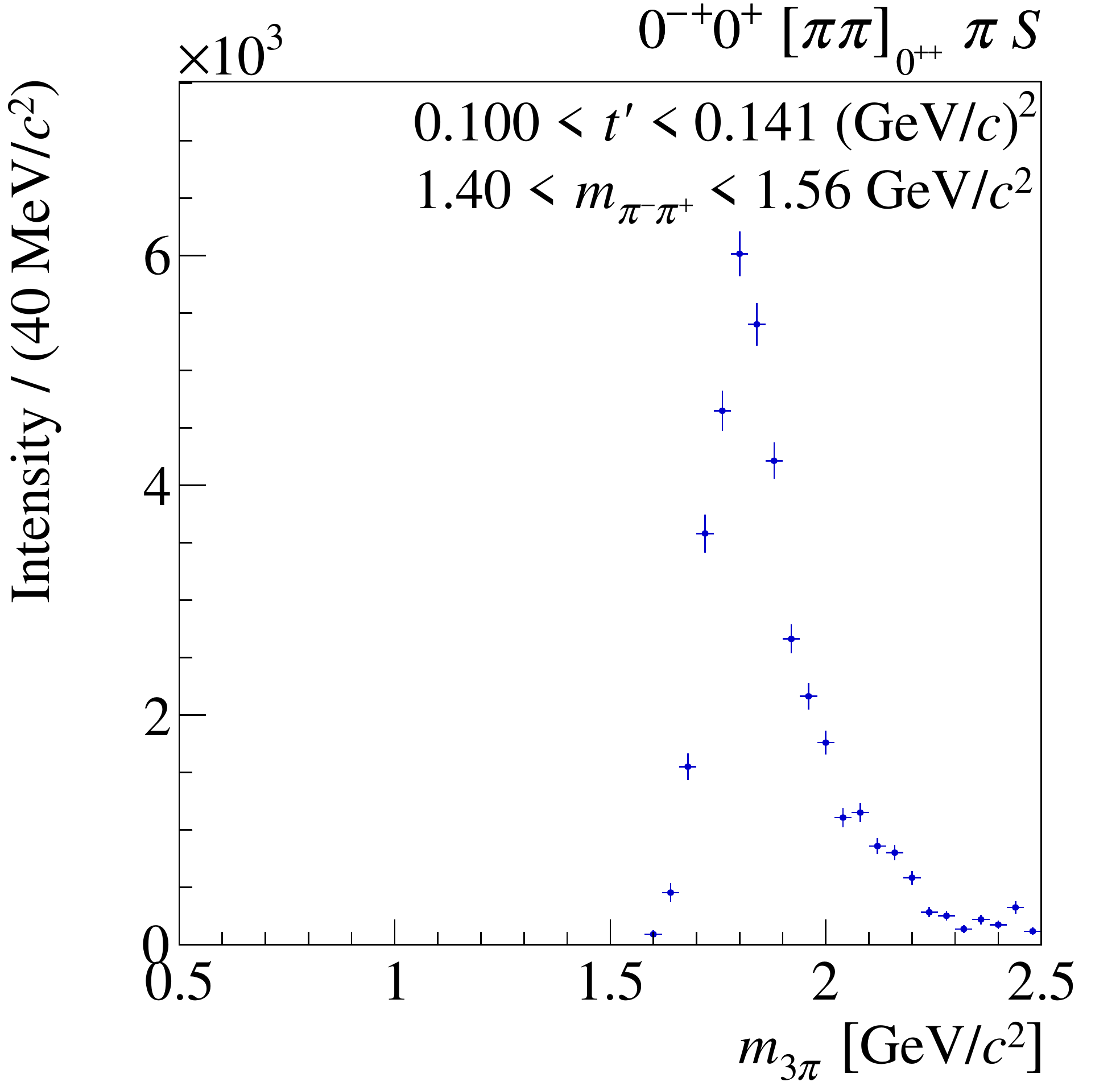}%
  }%
  \\
  \subfloat[][]{%
    \includegraphics[width=\threePlotWidthTwoD]{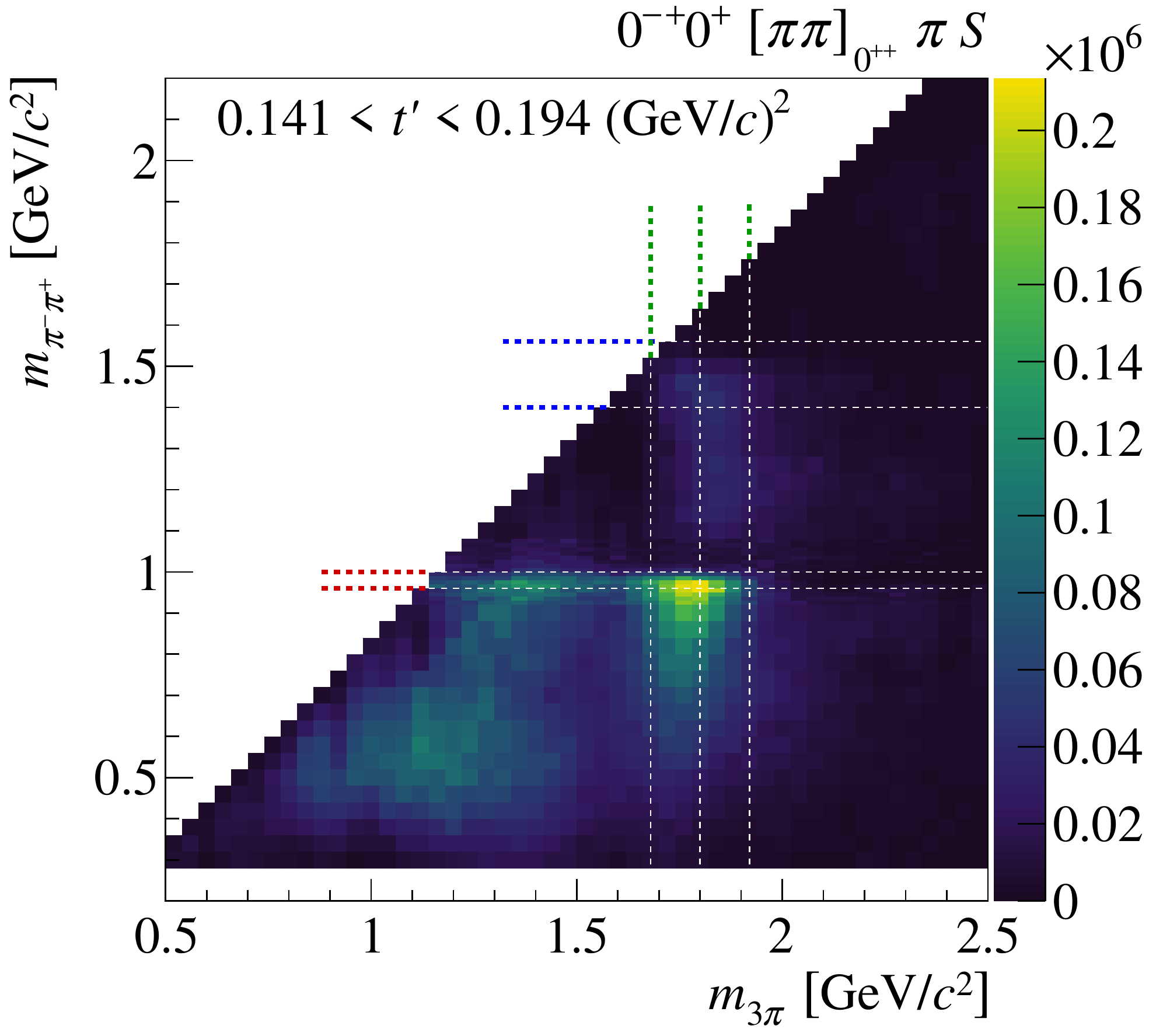}%
  }%
  \subfloat[][]{%
    \label{fig:PIPIS_0mp_lh_980}%
    \includegraphics[width=\threePlotWidth]{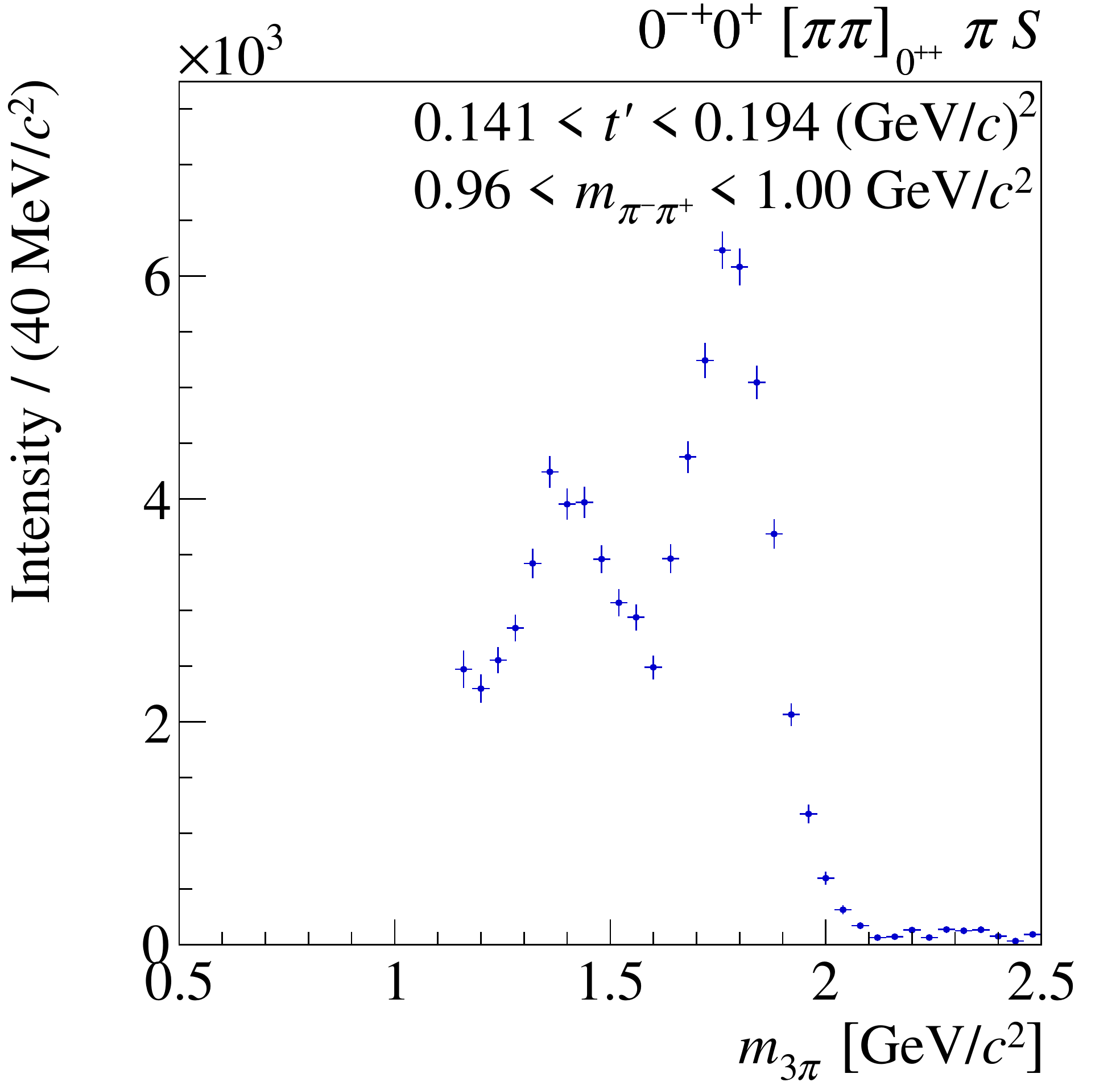}%
  }%
  \subfloat[][]{%
    \includegraphics[width=\threePlotWidth]{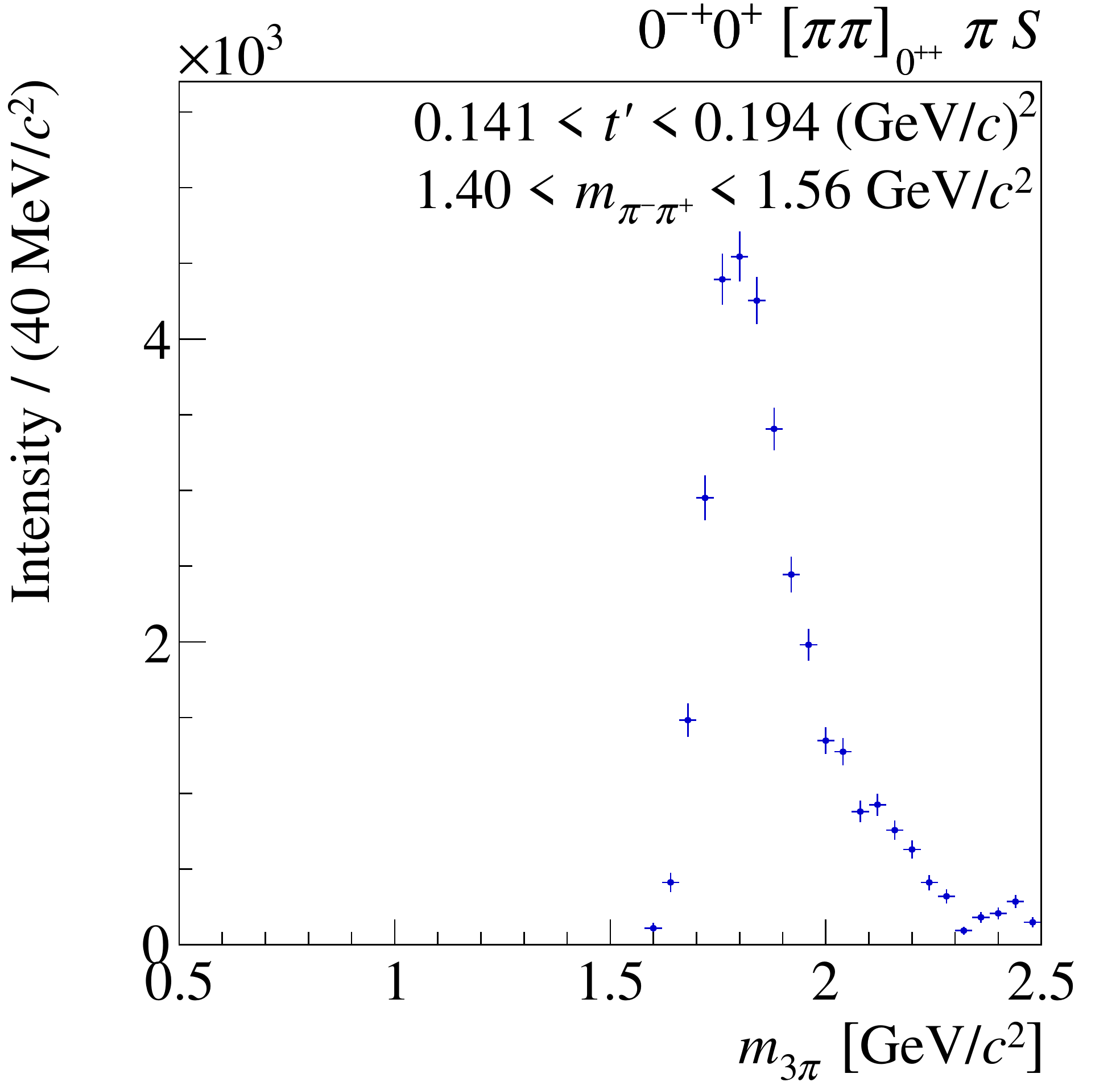}%
  }%
  \\
  \subfloat[][]{%
    \includegraphics[width=\threePlotWidthTwoD]{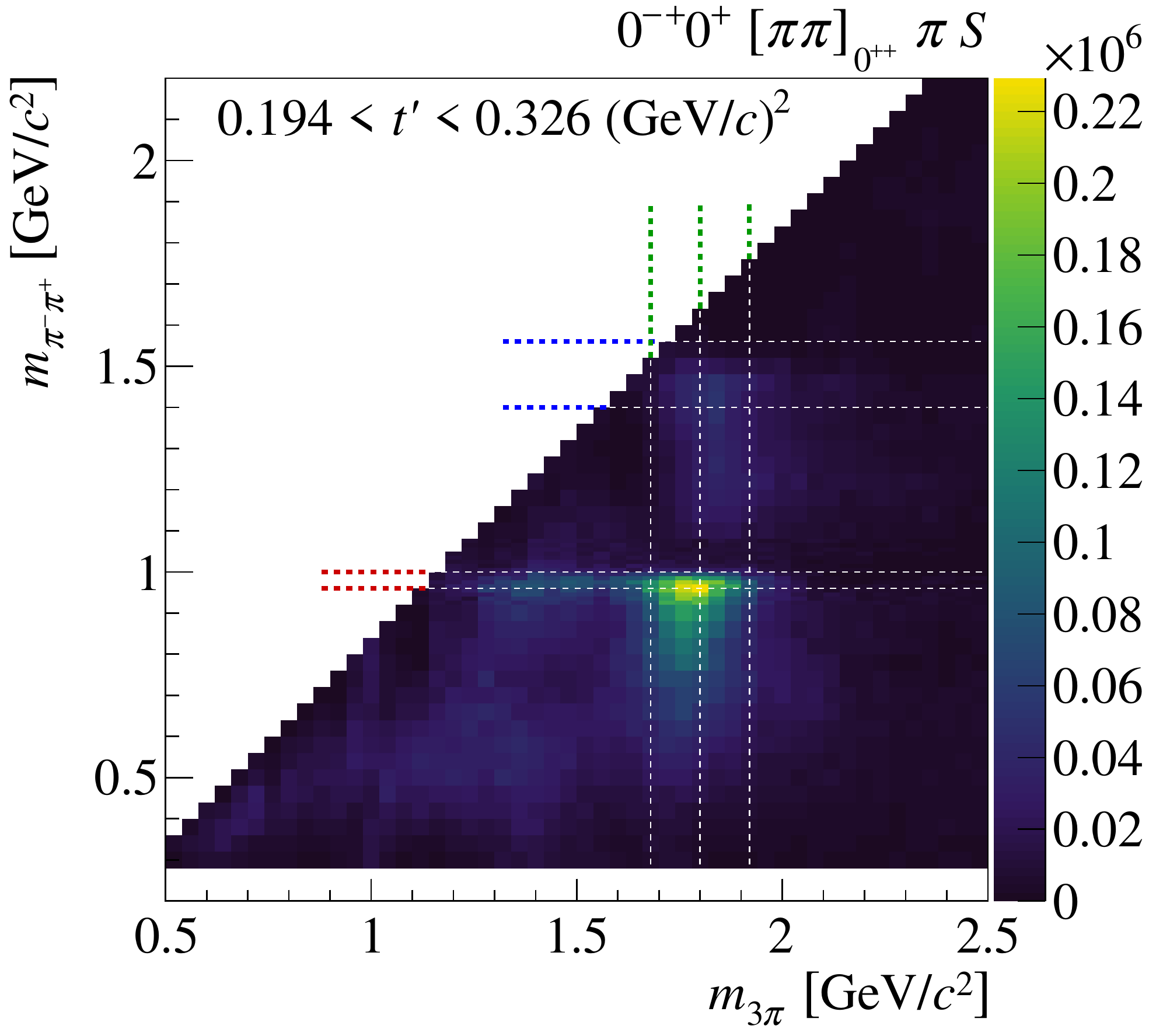}%
  }%
  \subfloat[][]{%
    \label{fig:PIPIS_0mp_hl_980}%
    \includegraphics[width=\threePlotWidth]{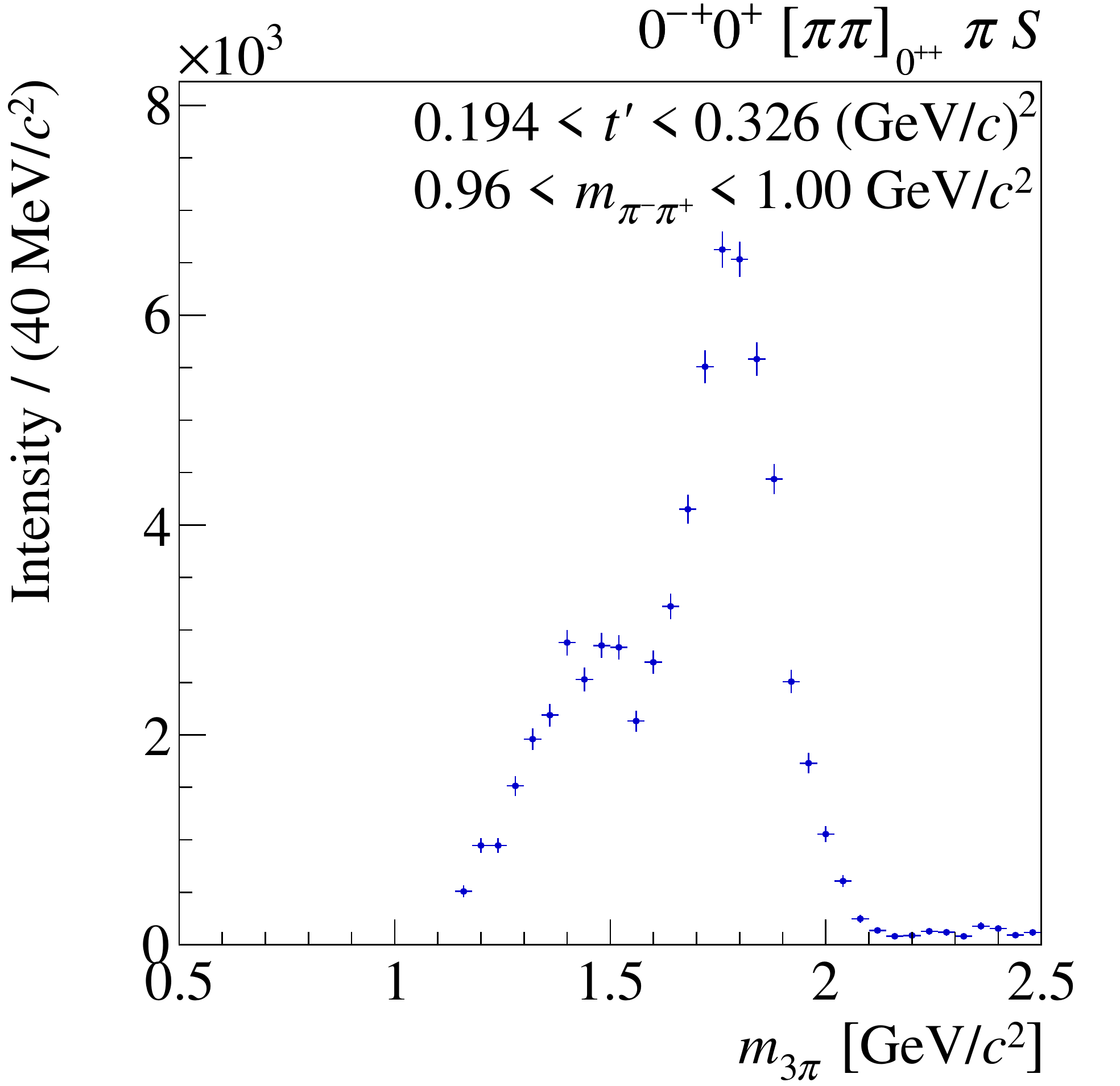}%
  }%
  \subfloat[][]{%
    \includegraphics[width=\threePlotWidth]{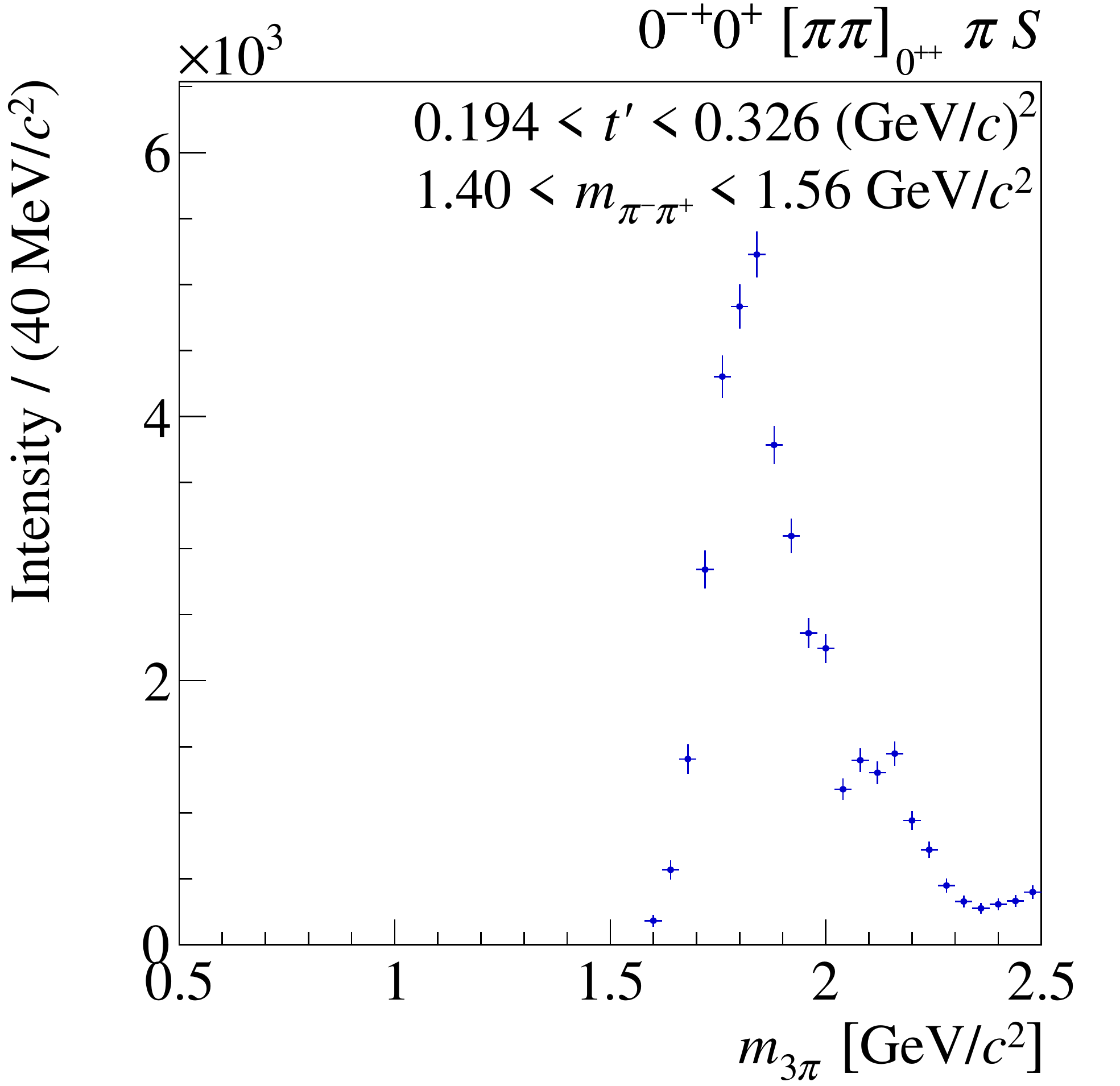}%
  }%
  \\
  \subfloat[][]{%
    \includegraphics[width=\threePlotWidthTwoD]{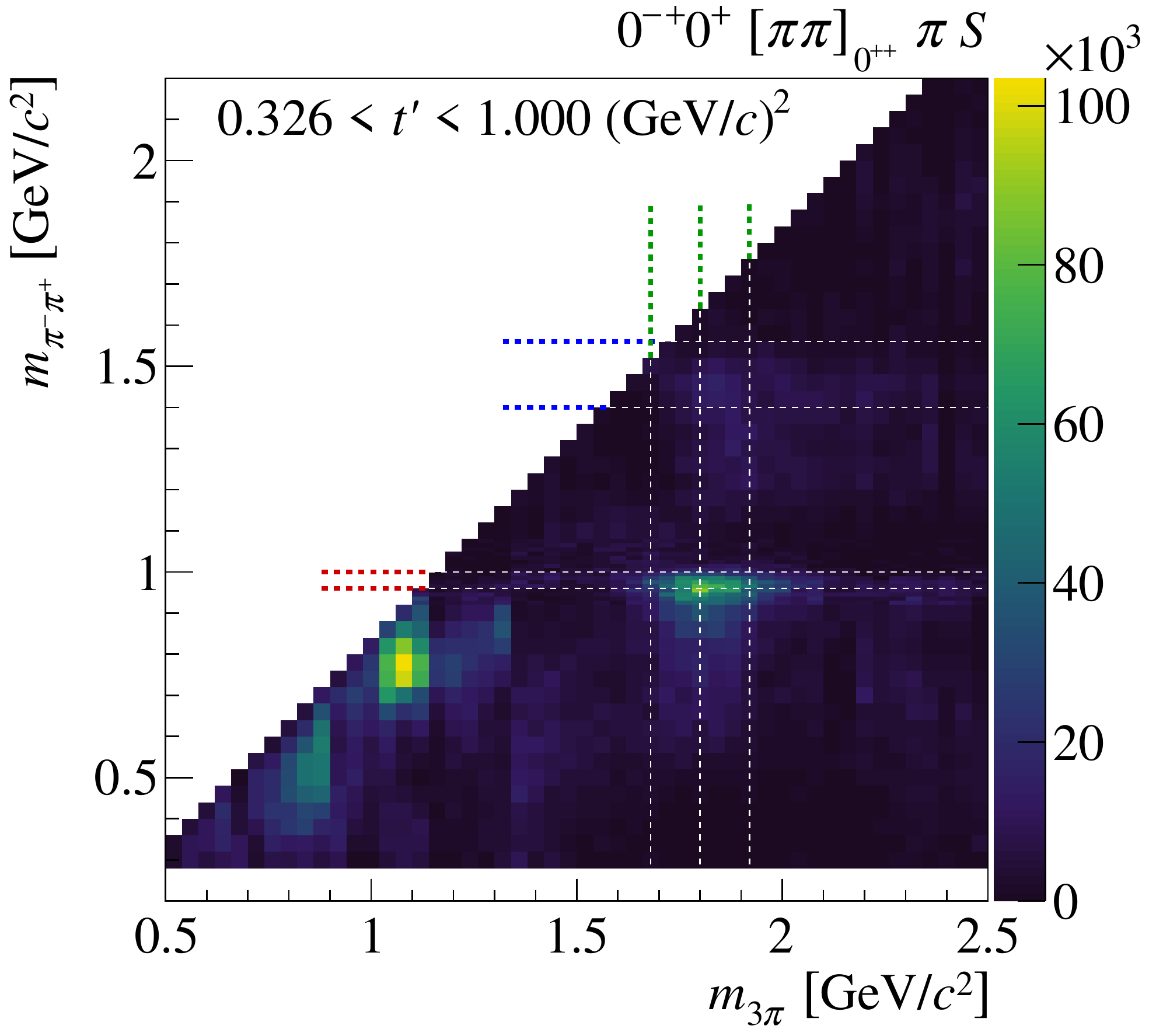}%
  }%
  \subfloat[][]{%
    \label{fig:PIPIS_0mp_hh_980}%
    \includegraphics[width=\threePlotWidth]{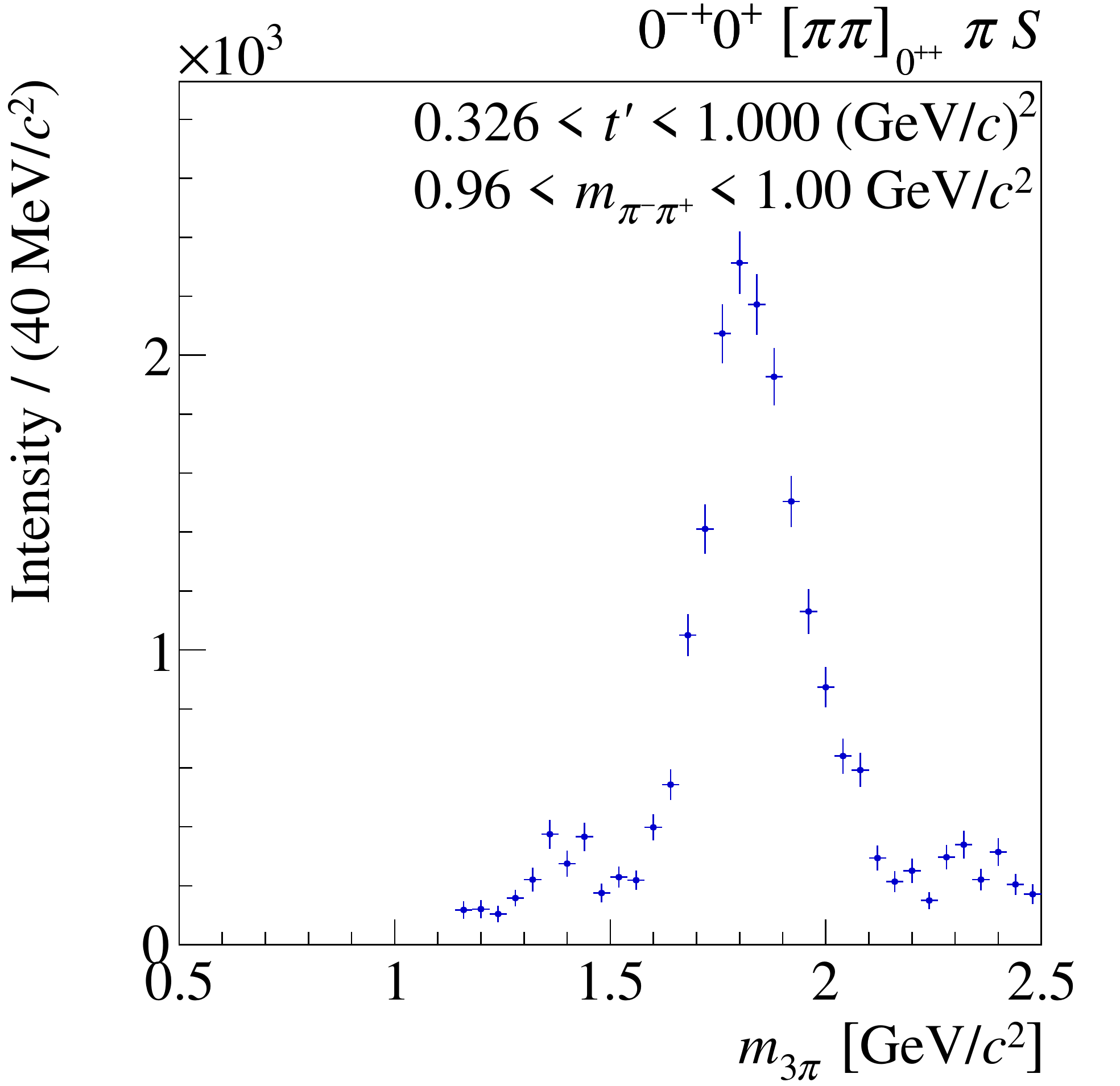}%
  }%
  \subfloat[][]{%
    \includegraphics[width=\threePlotWidth]{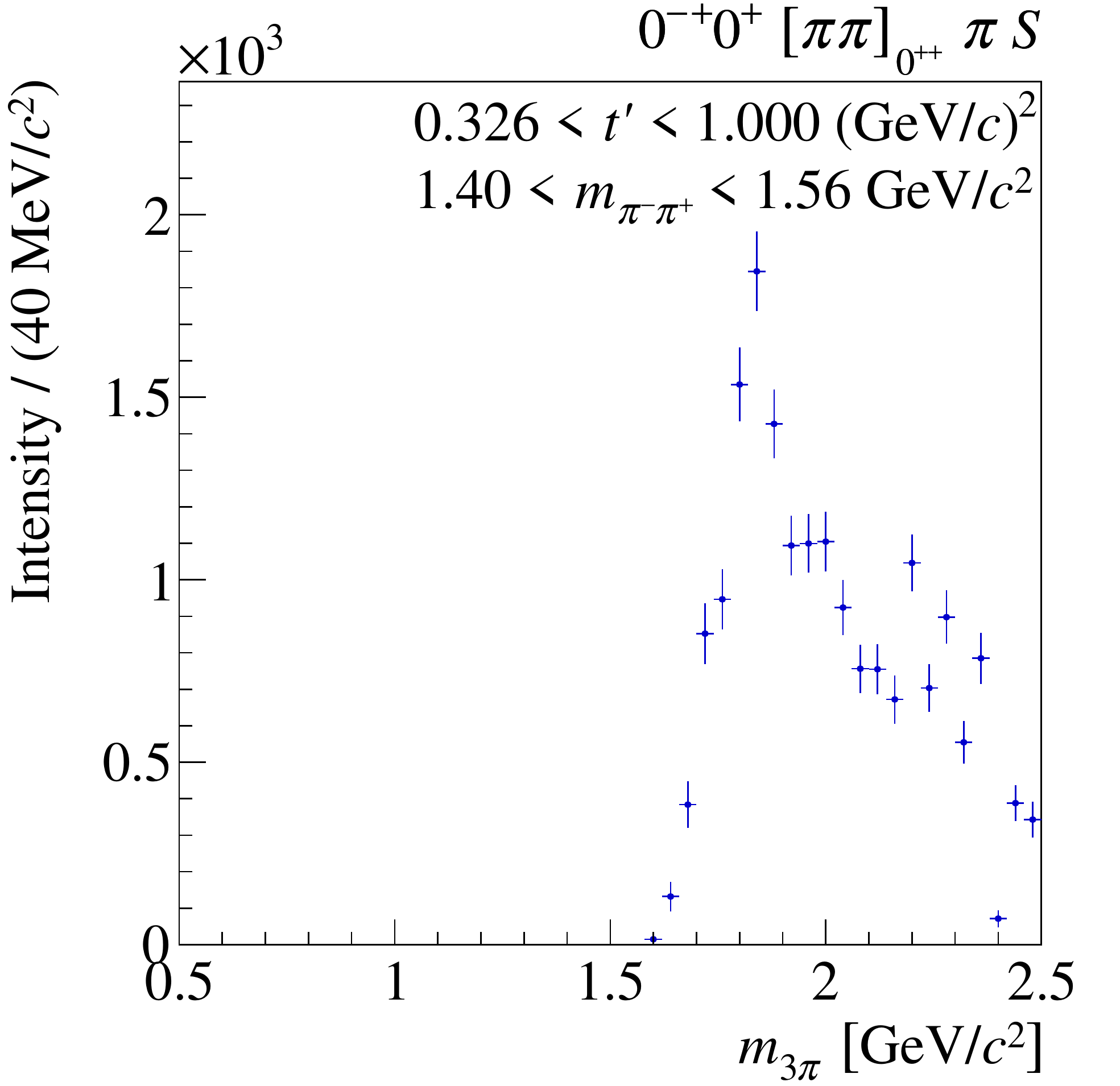}%
  }%
  \caption{\colorPlot Intensity of the $\pipiSF\, \pi$ $S$-wave
    component of the $\JPCMrefl = 0^{-+}\, 0^+$ amplitude resulting
    from the freed-isobar fits in four \tpr bins.  Left column:
    two-dimensional representation of the intensity of the
    \wave{0}{-+}{0}{+}{\pipiSF}{S} wave as a function of \mTwoPi and
    \mThreePi.  Central and right columns: intensity as a function of
    \mThreePi summed over selected \mTwoPi intervals around the
    \PfZero[980] (center) and around the \PfZero[1500] (right) as
    indicated by pairs of horizontal dashed lines in the left column.
    The vertical dashed lines indicate the centers of the \mThreePi
    bins discussed in \cref{sec:results_free_pipi_s_wave_argands}.}
  \label{fig:pipis_2D_0mp}
\end{figure*}

\begin{figure*}[htbp]
  \centering
  \subfloat[][]{%
    \includegraphics[width=\threePlotWidthTwoD]{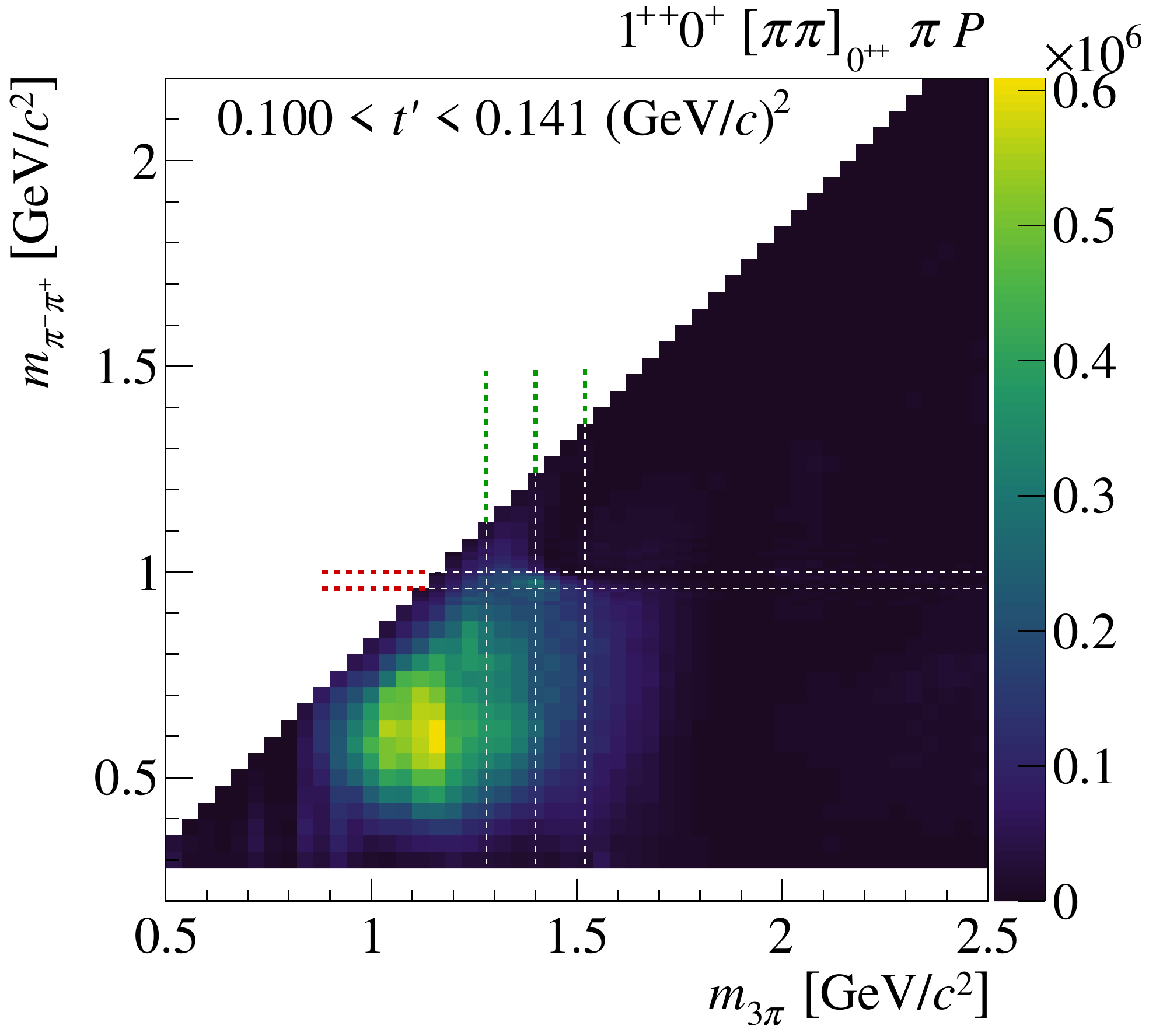}%
  }%
  \subfloat[][]{%
    \label{fig:PIPIS_1pp_ll_980}%
    \includegraphics[width=\threePlotWidth]{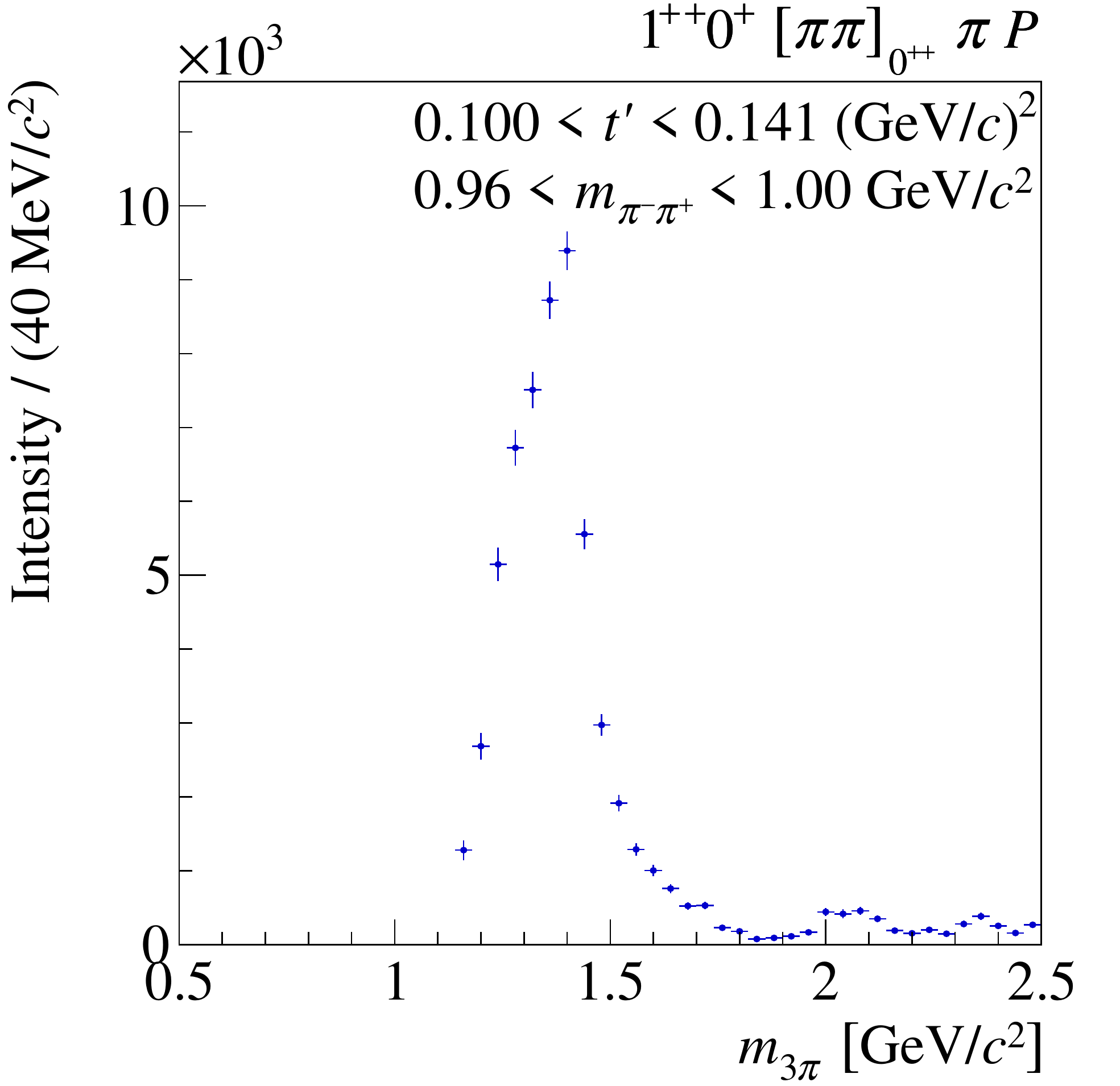}%
  }%
  \\
  \subfloat[][]{%
    \includegraphics[width=\threePlotWidthTwoD]{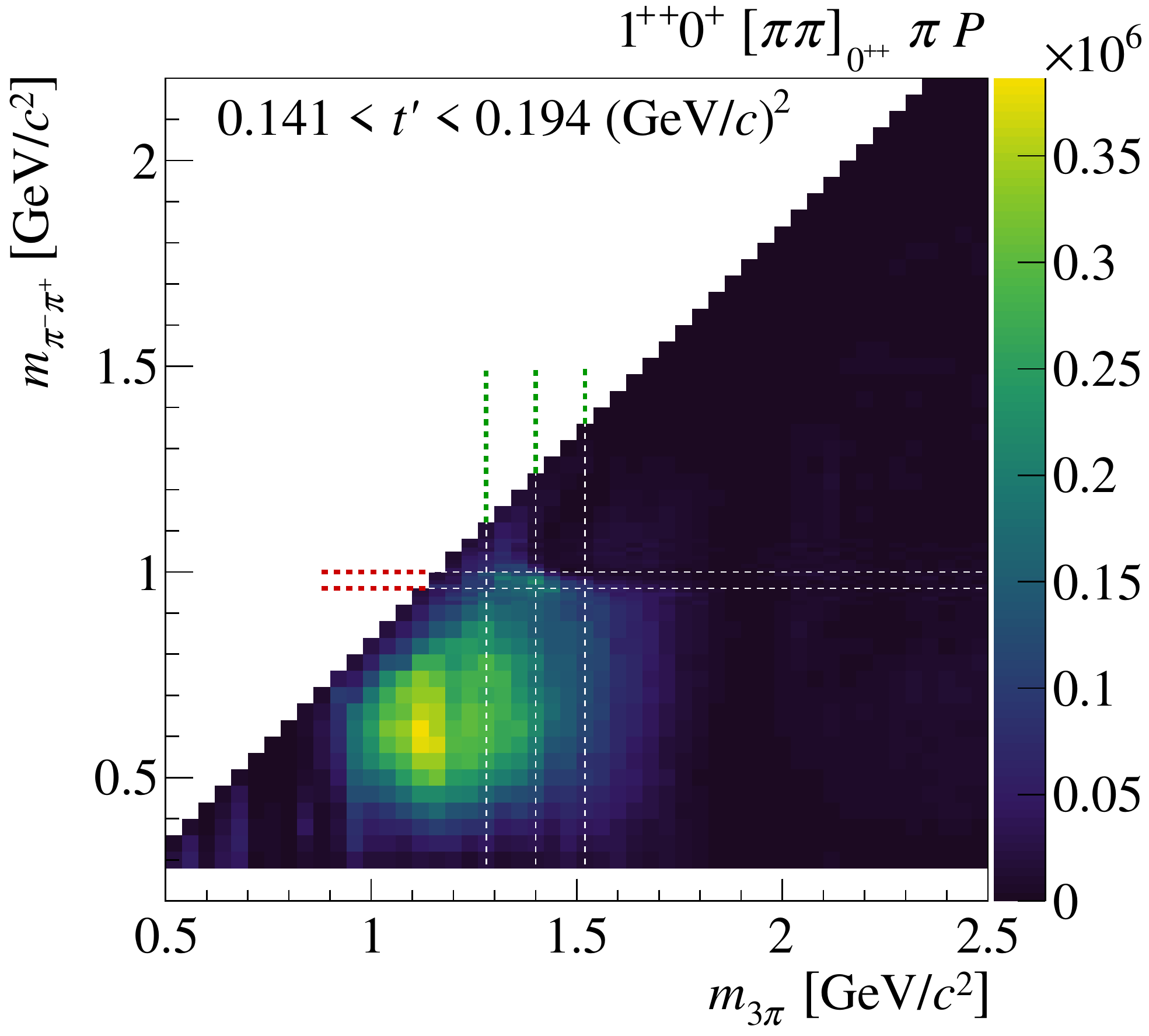}%
  }%
  \subfloat[][]{%
    \label{fig:PIPIS_1pp_lh_980}%
    \includegraphics[width=\threePlotWidth]{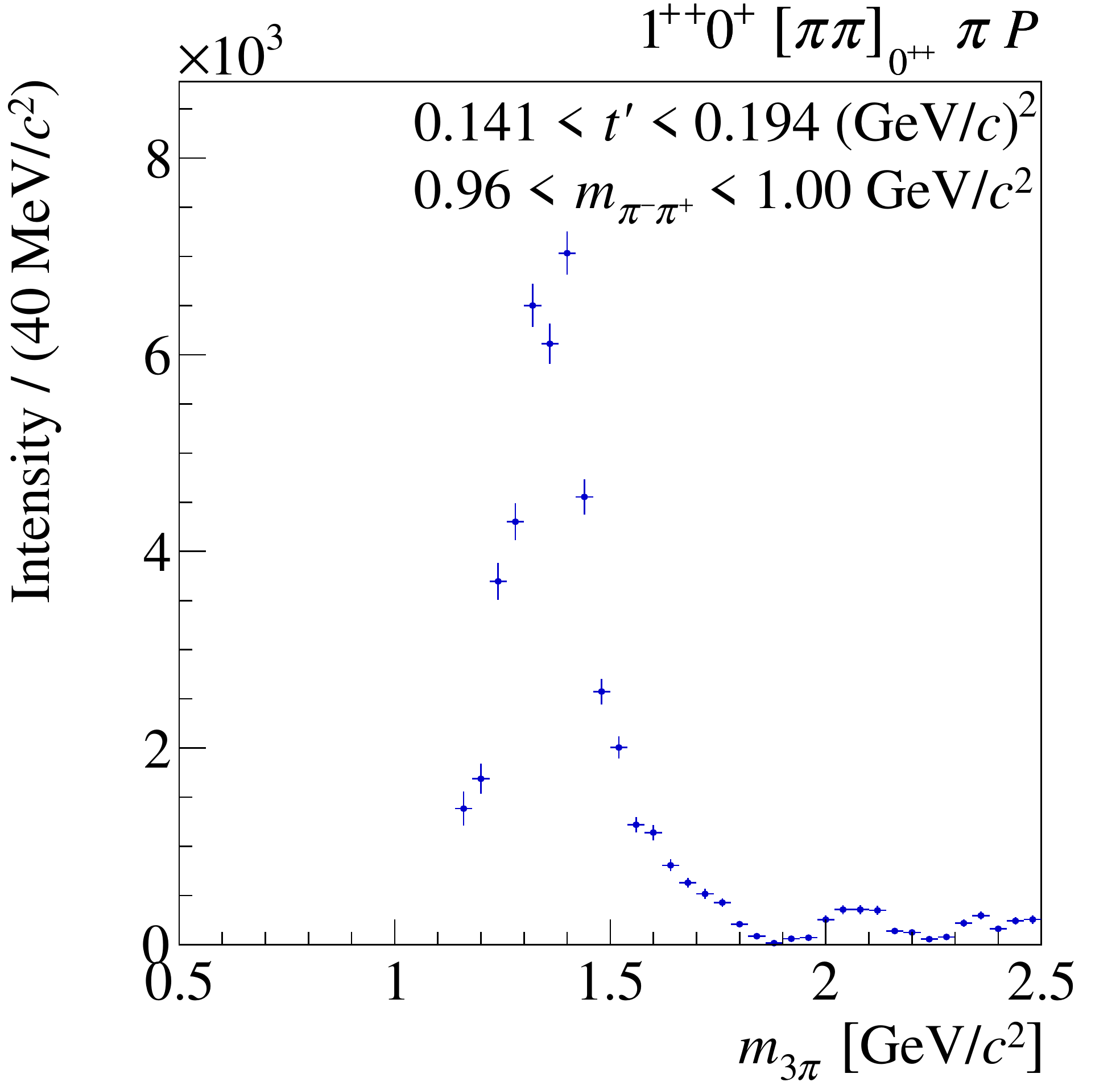}%
  }%
  \\
  \subfloat[][]{%
    \includegraphics[width=\threePlotWidthTwoD]{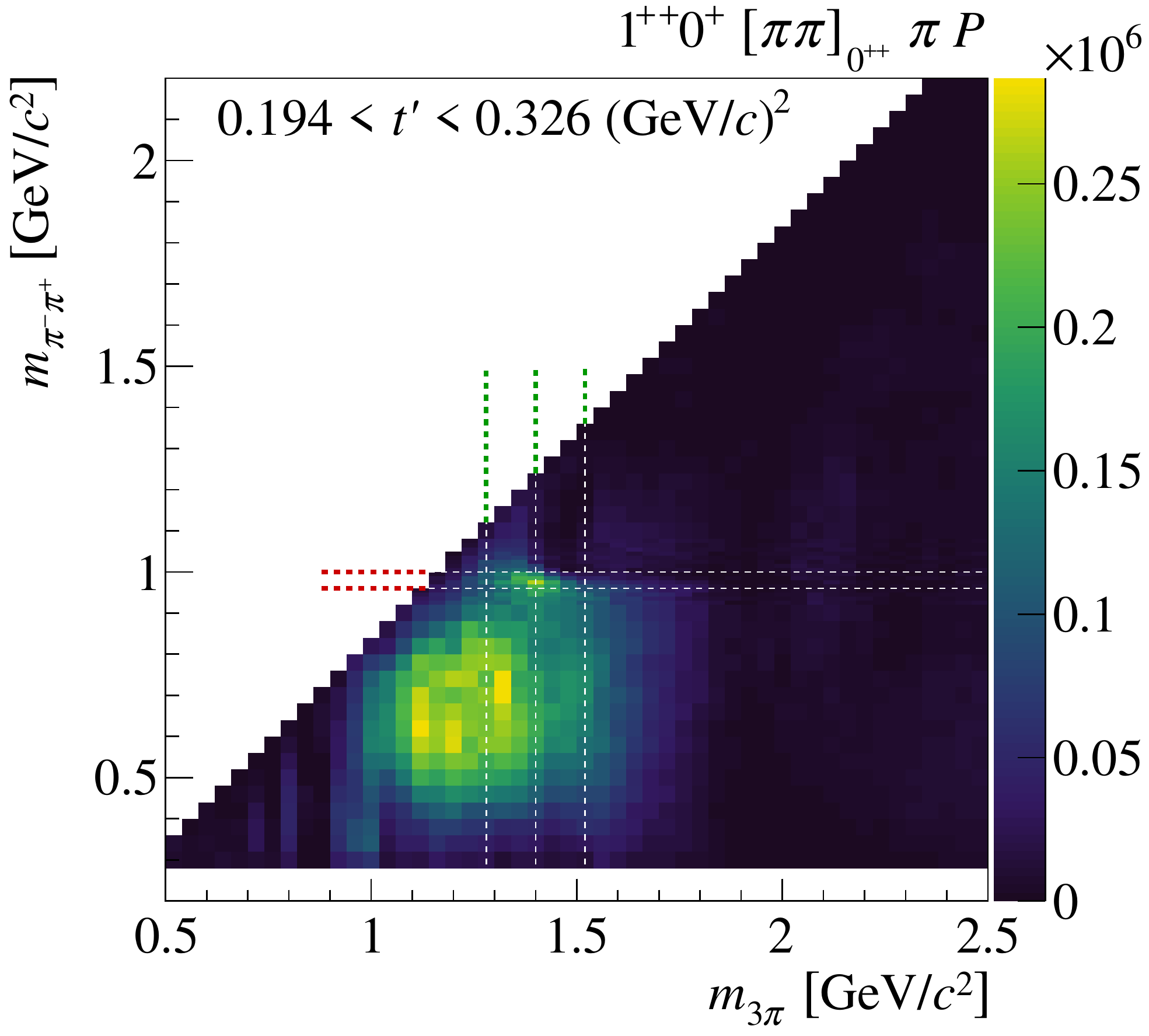}%
  }%
  \subfloat[][]{%
    \label{fig:PIPIS_1pp_hl_980}%
    \includegraphics[width=\threePlotWidth]{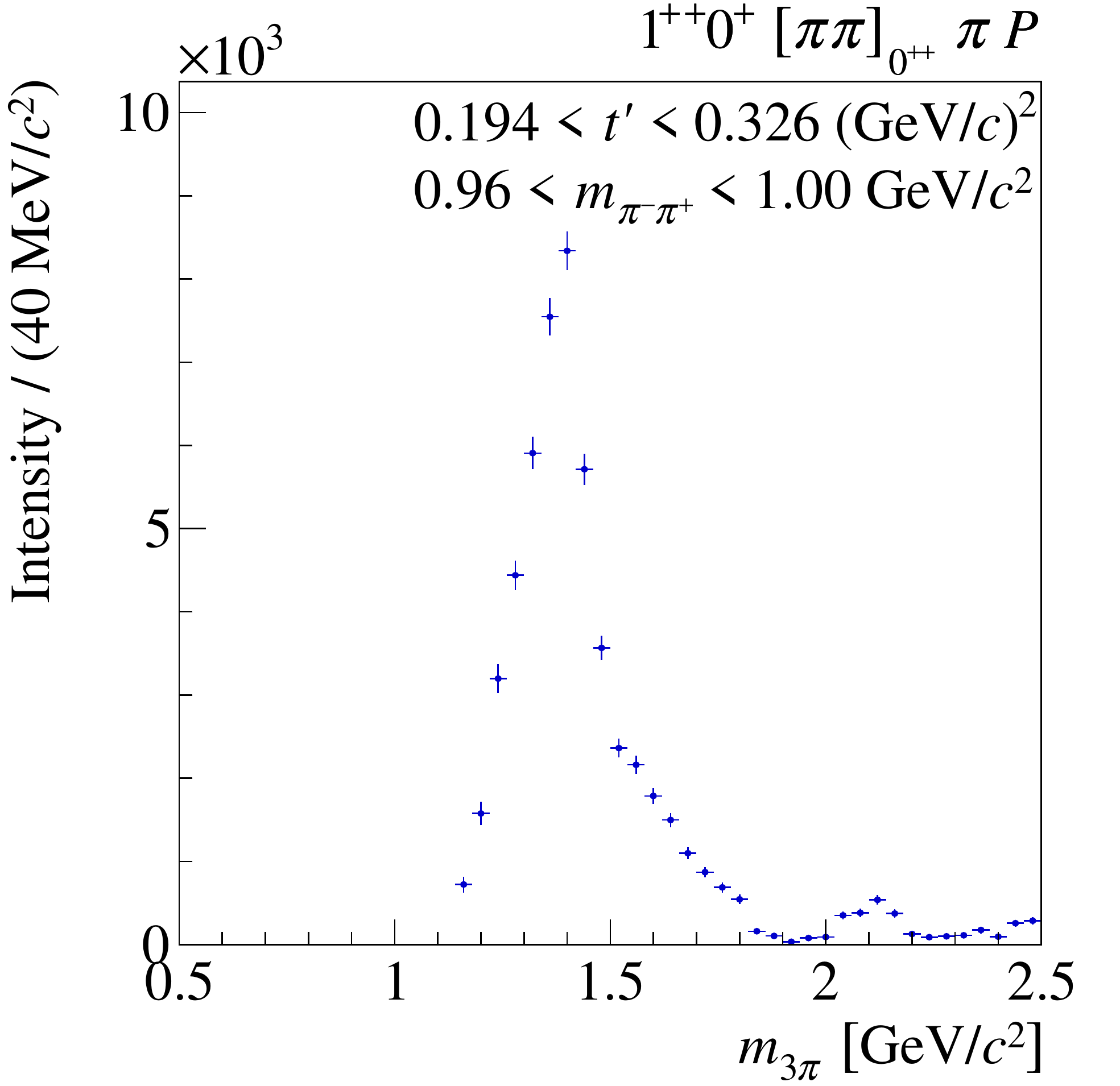}%
  }%
  \\
  \subfloat[][]{%
    \includegraphics[width=\threePlotWidthTwoD]{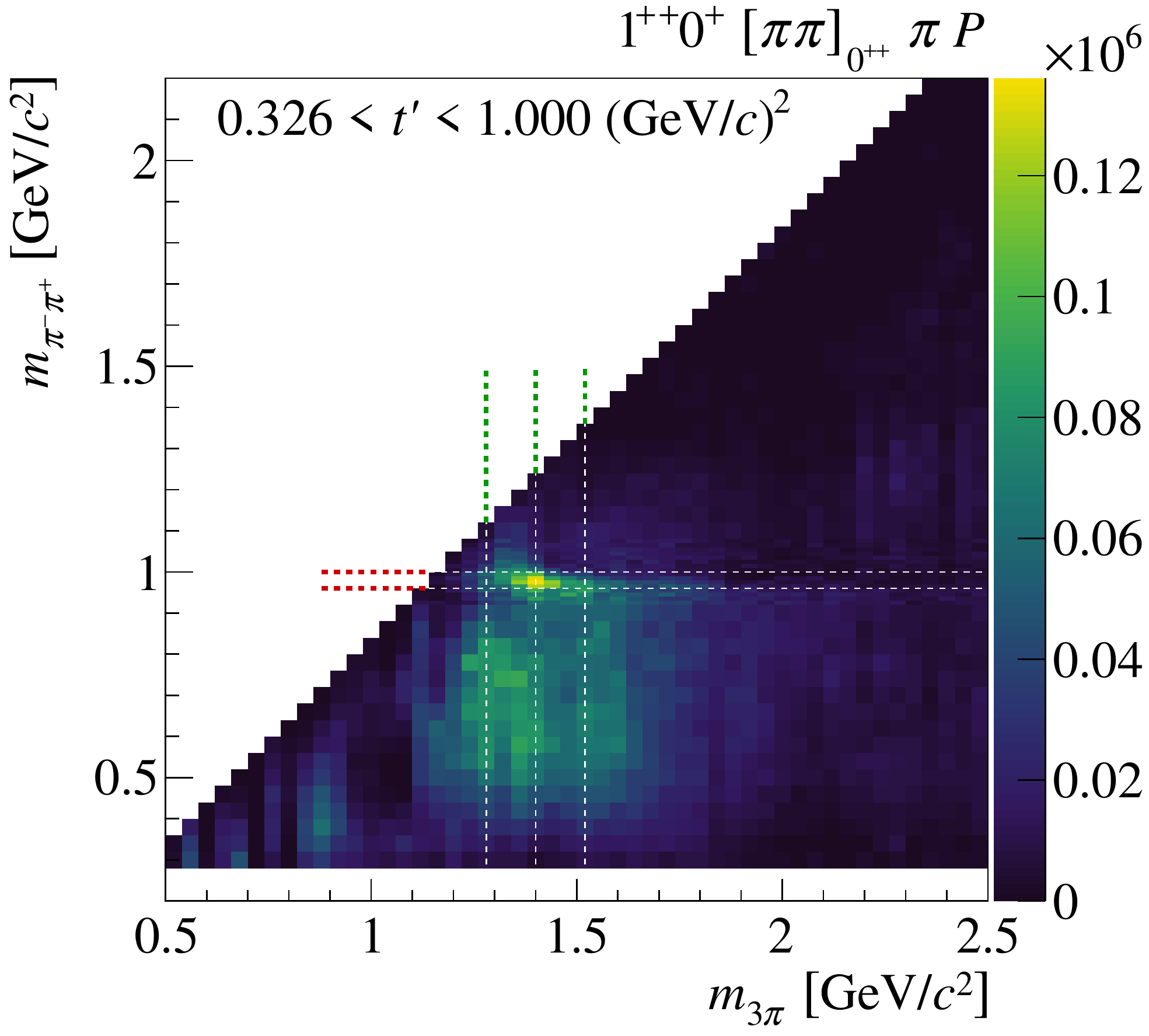}%
  }%
  \subfloat[][]{%
    \label{fig:PIPIS_1pp_hh_980}%
    \includegraphics[width=\threePlotWidth]{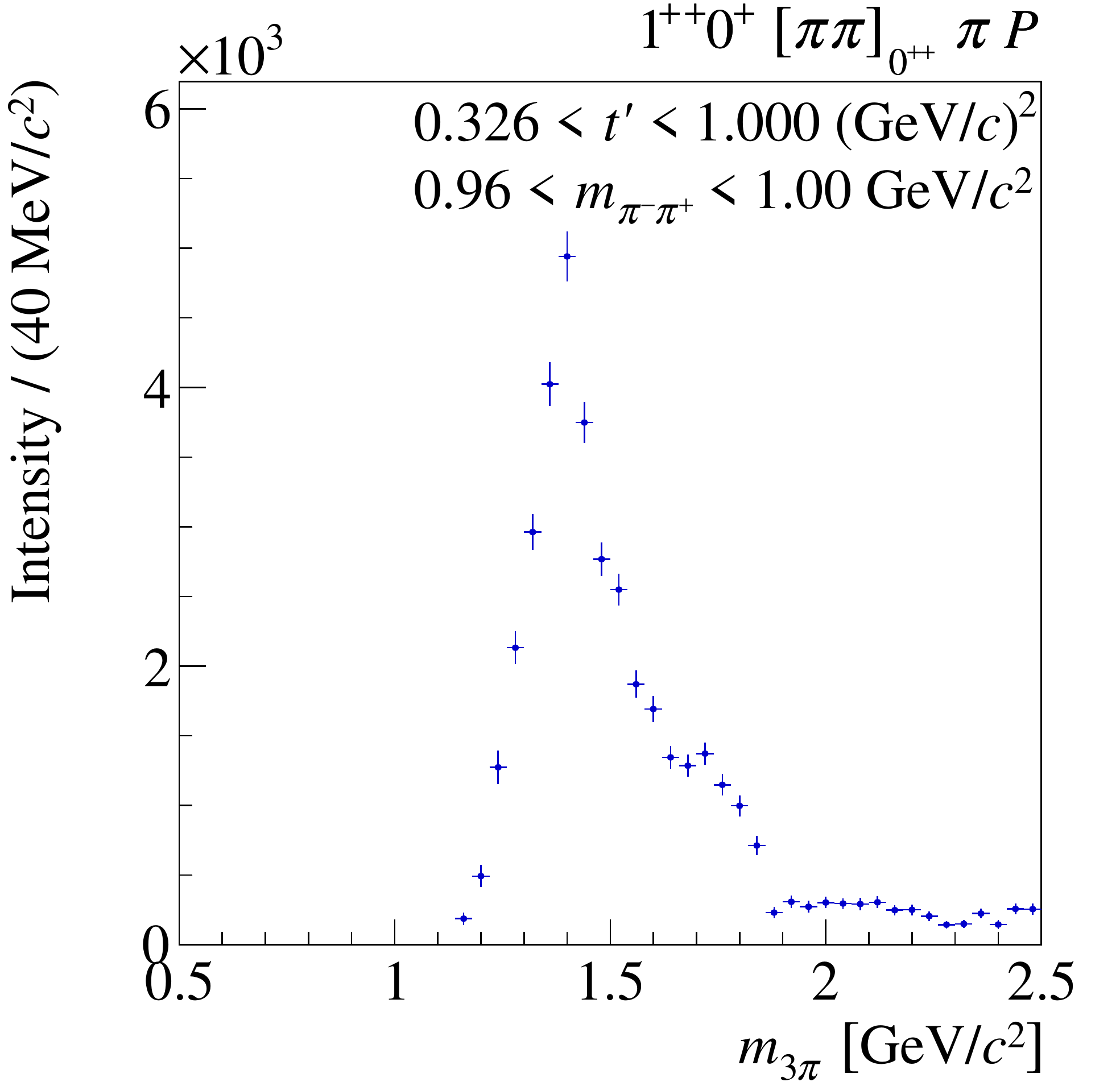}%
  }%
  \caption{\colorPlot Same as \cref{fig:pipis_2D_0mp}, but for the
    \wave{1}{++}{0}{+}{\pipiSF}{P} wave, except that the right column
    shows the \mTwoPi interval around the \PfZero[980].}
  \label{fig:pipis_2D_1pp}
\end{figure*}

\begin{figure*}[htbp]
  \centering
  \subfloat[][]{%
    \includegraphics[width=\threePlotWidthTwoD]{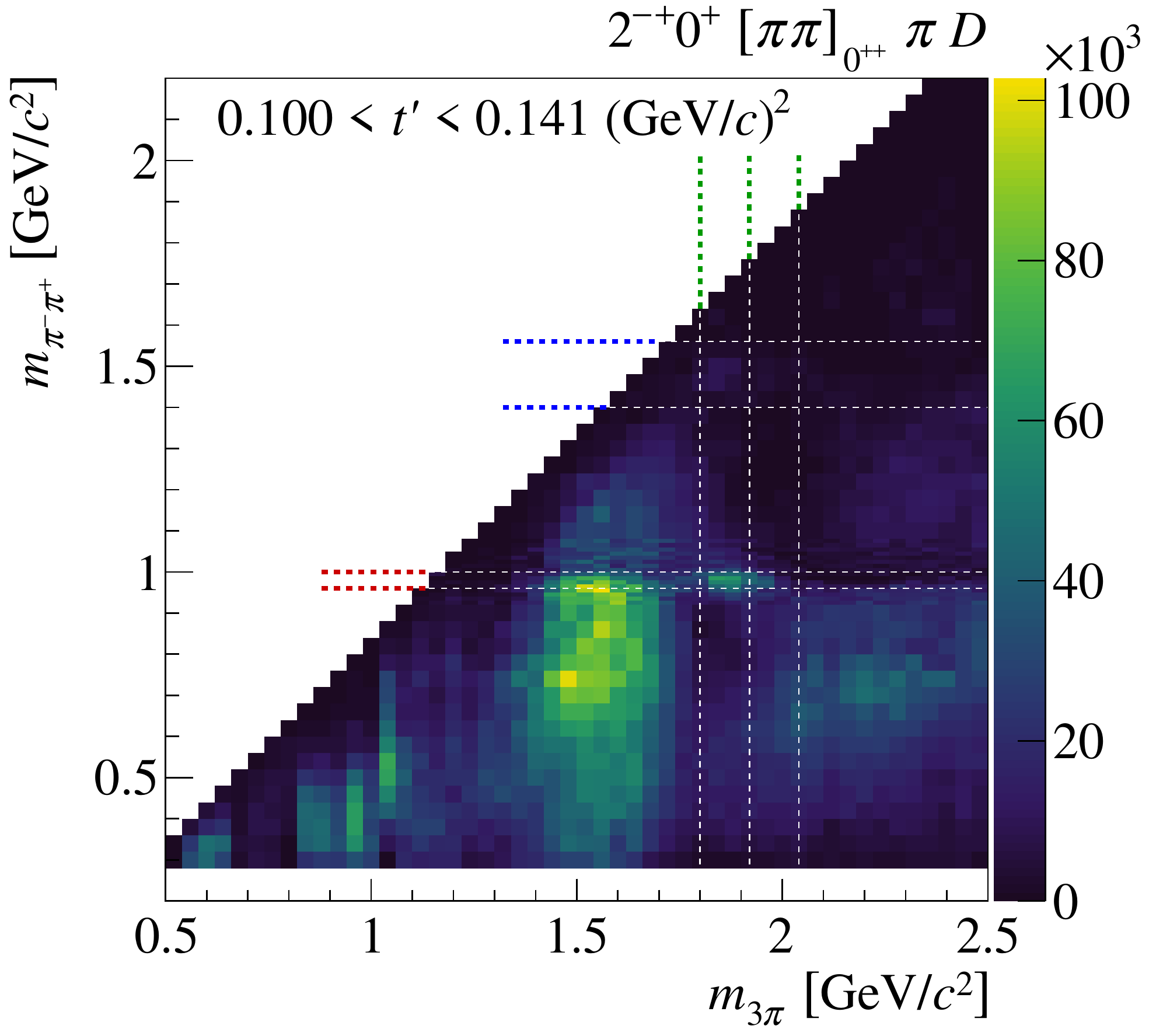}%
  }%
  \subfloat[][]{%
    \label{fig:PIPIS_2mp_ll_980}%
    \includegraphics[width=\threePlotWidth]{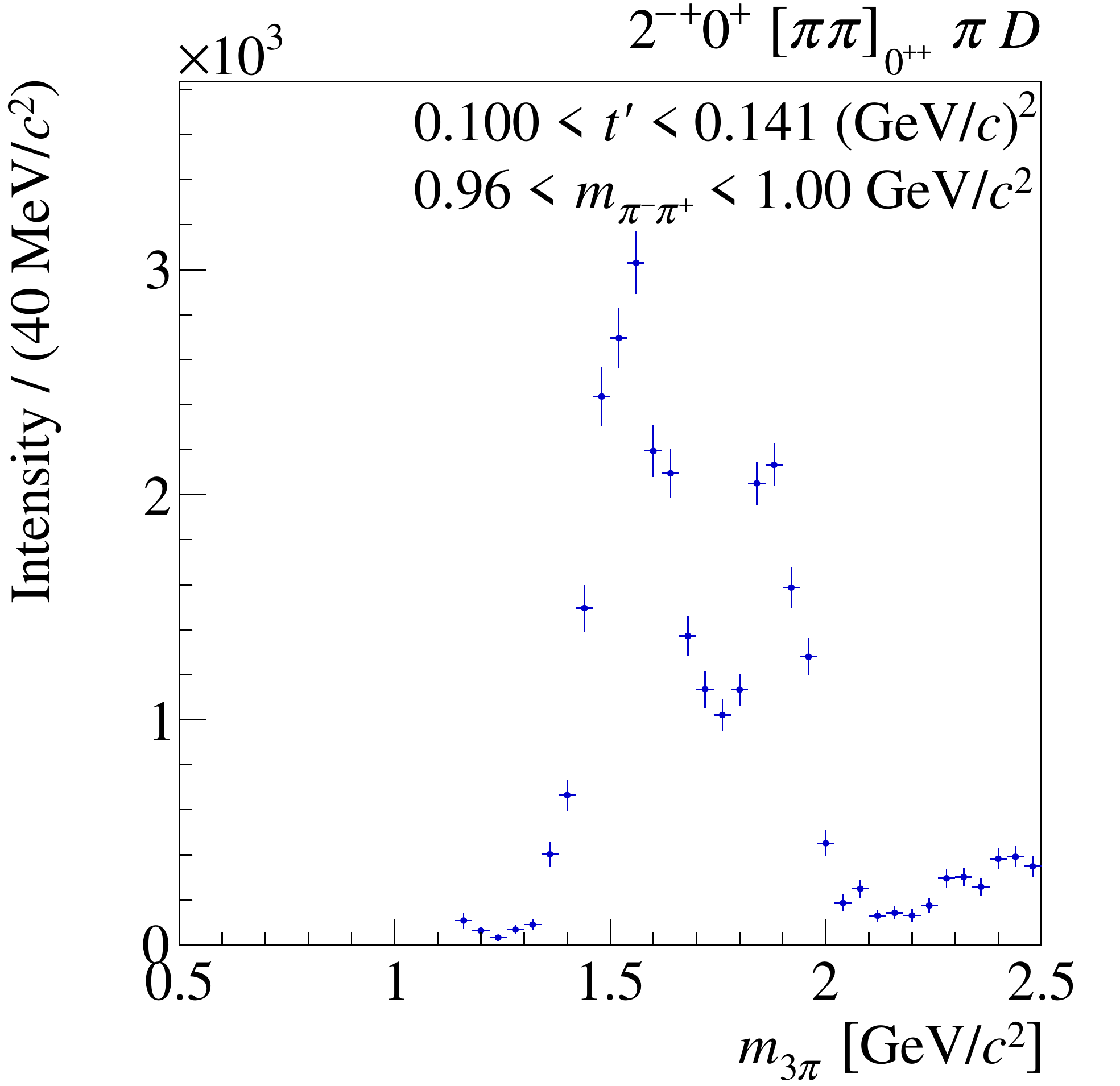}%
  }%
  \subfloat[][]{%
    \includegraphics[width=\threePlotWidth]{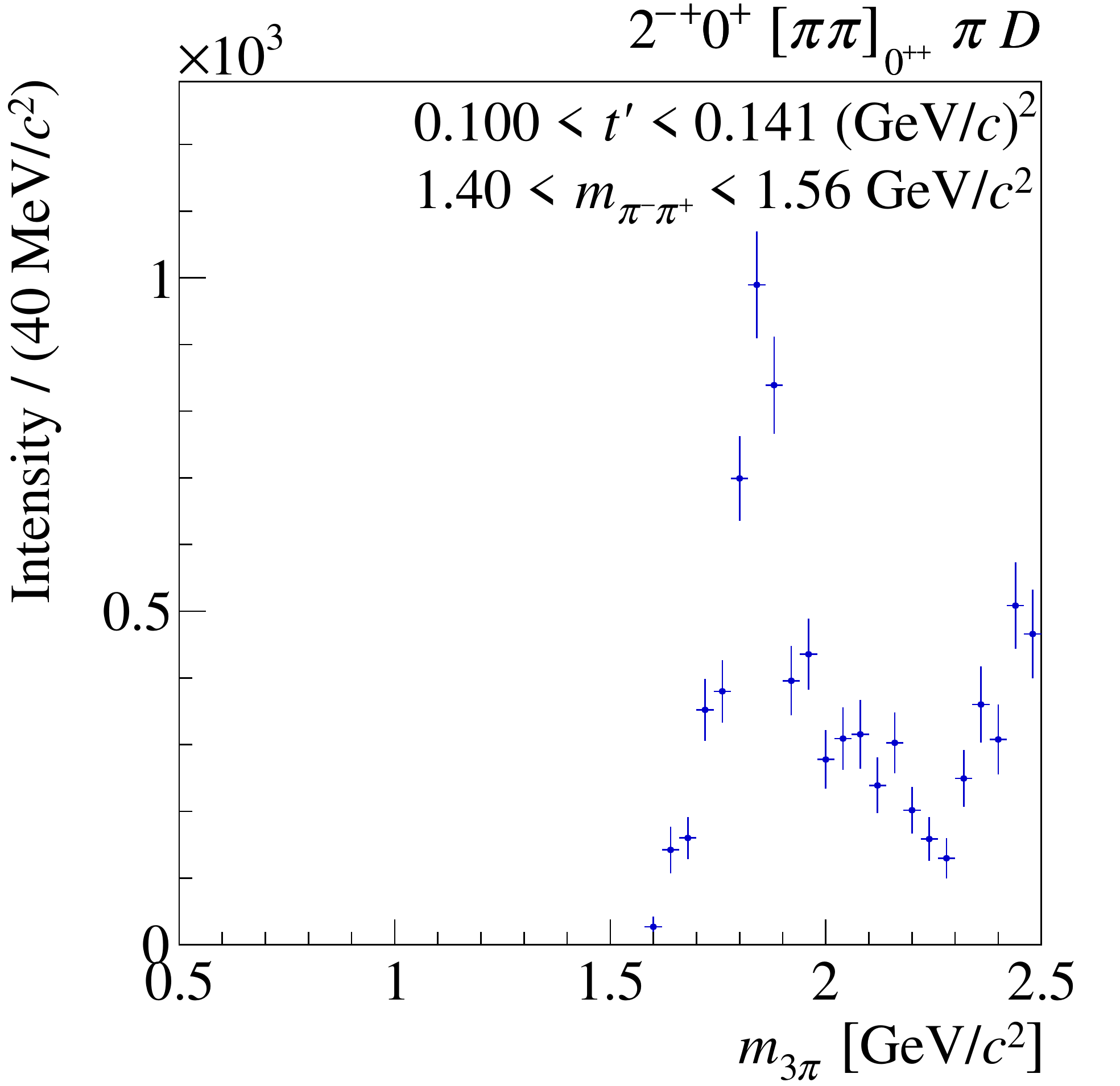}%
  }%
  \\
  \subfloat[][]{%
    \includegraphics[width=\threePlotWidthTwoD]{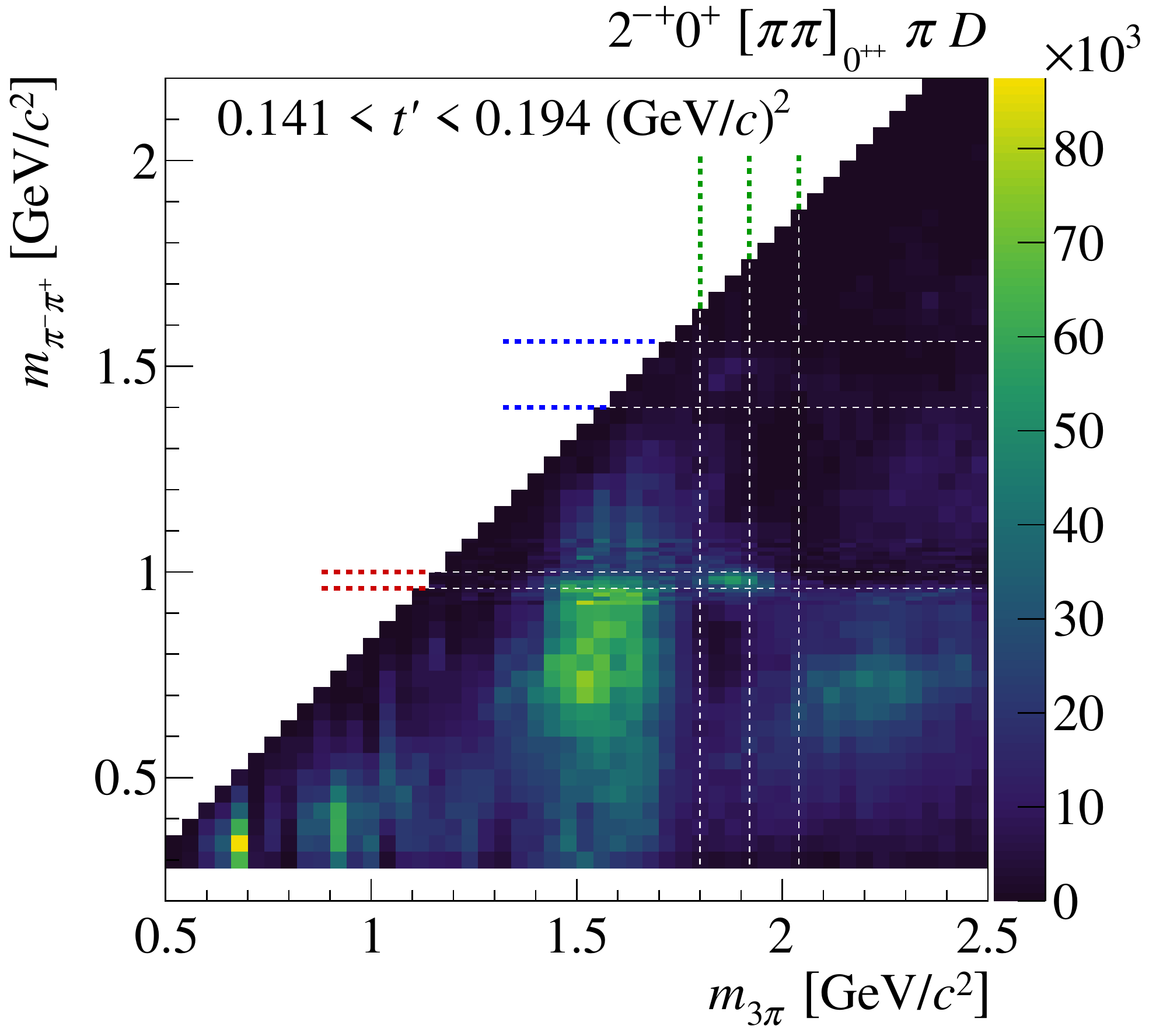}%
  }%
  \subfloat[][]{%
    \label{fig:PIPIS_2mp_lh_980}%
    \includegraphics[width=\threePlotWidth]{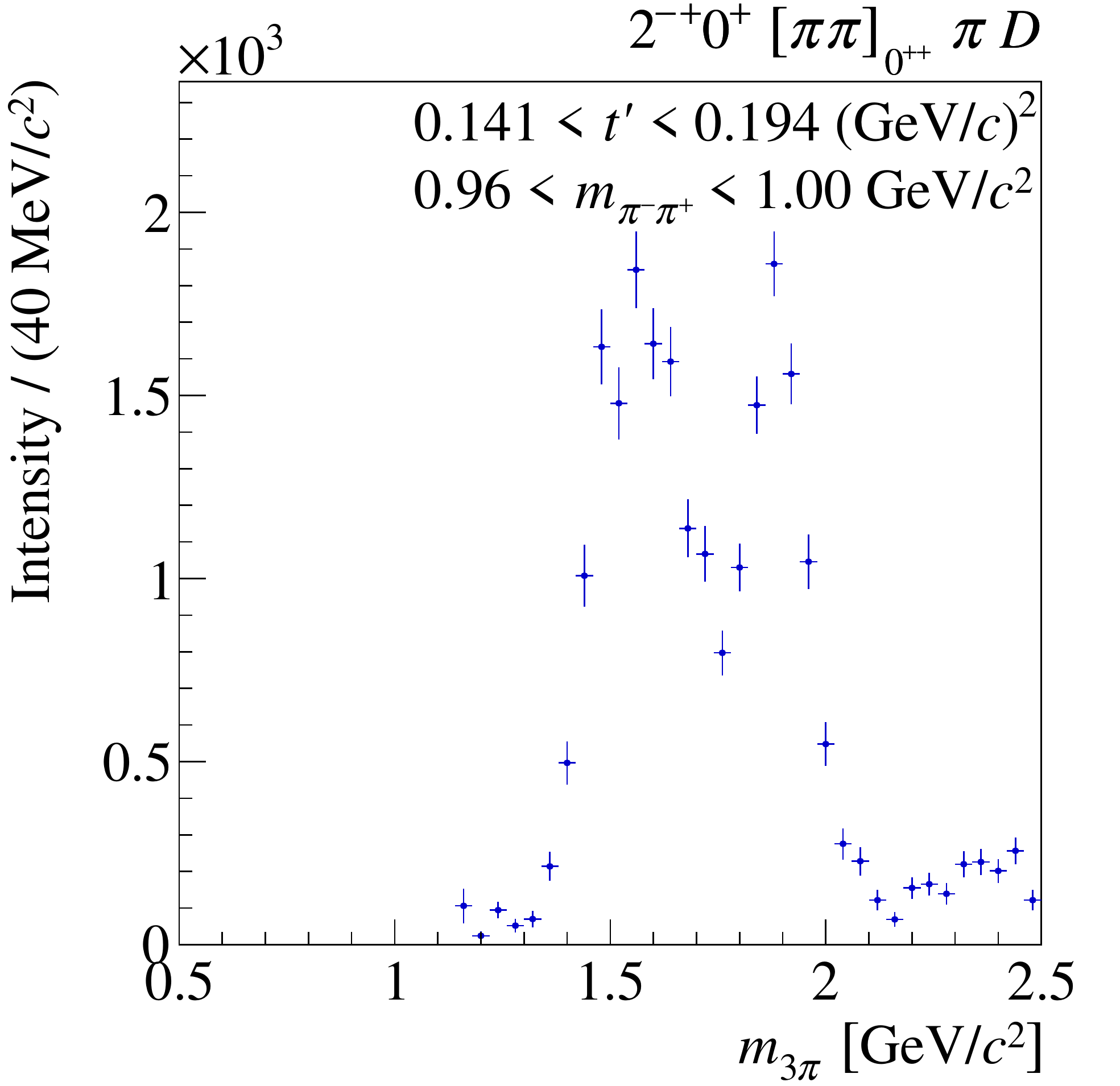}%
  }%
  \subfloat[][]{%
    \includegraphics[width=\threePlotWidth]{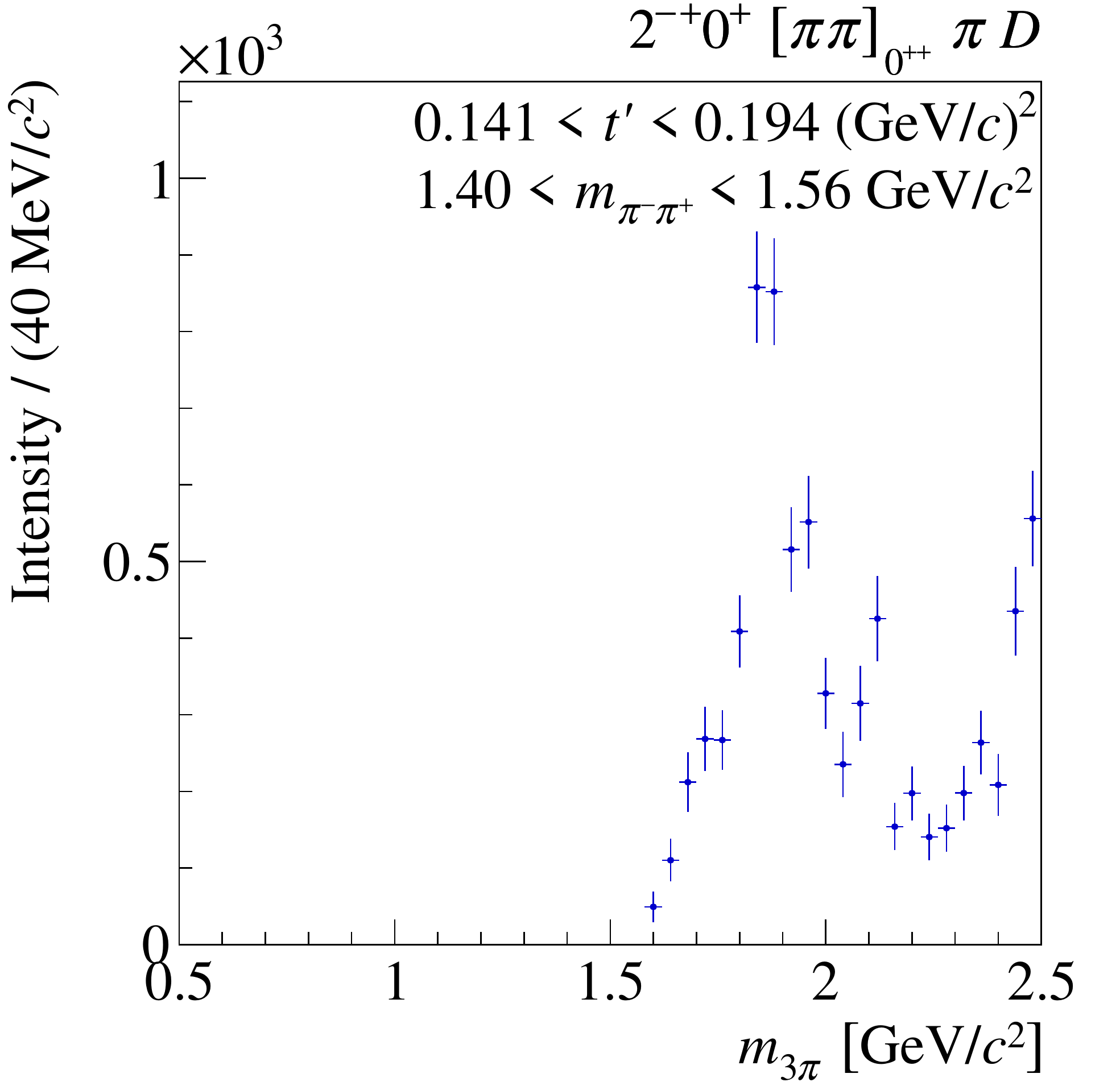}%
  }%
  \\
  \subfloat[][]{%
    \includegraphics[width=\threePlotWidthTwoD]{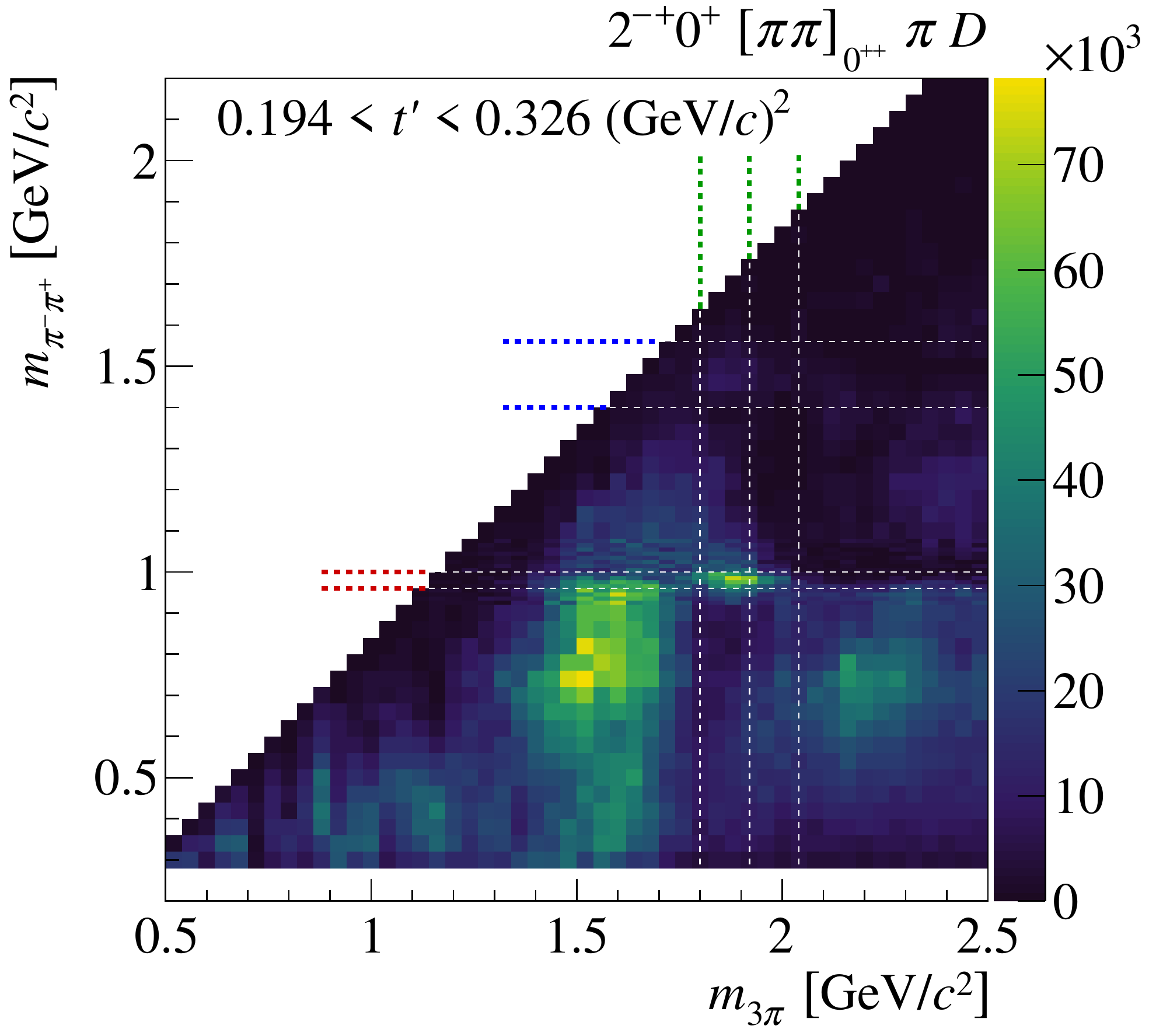}%
  }%
  \subfloat[][]{%
    \label{fig:PIPIS_2mp_hl_980}%
    \includegraphics[width=\threePlotWidth]{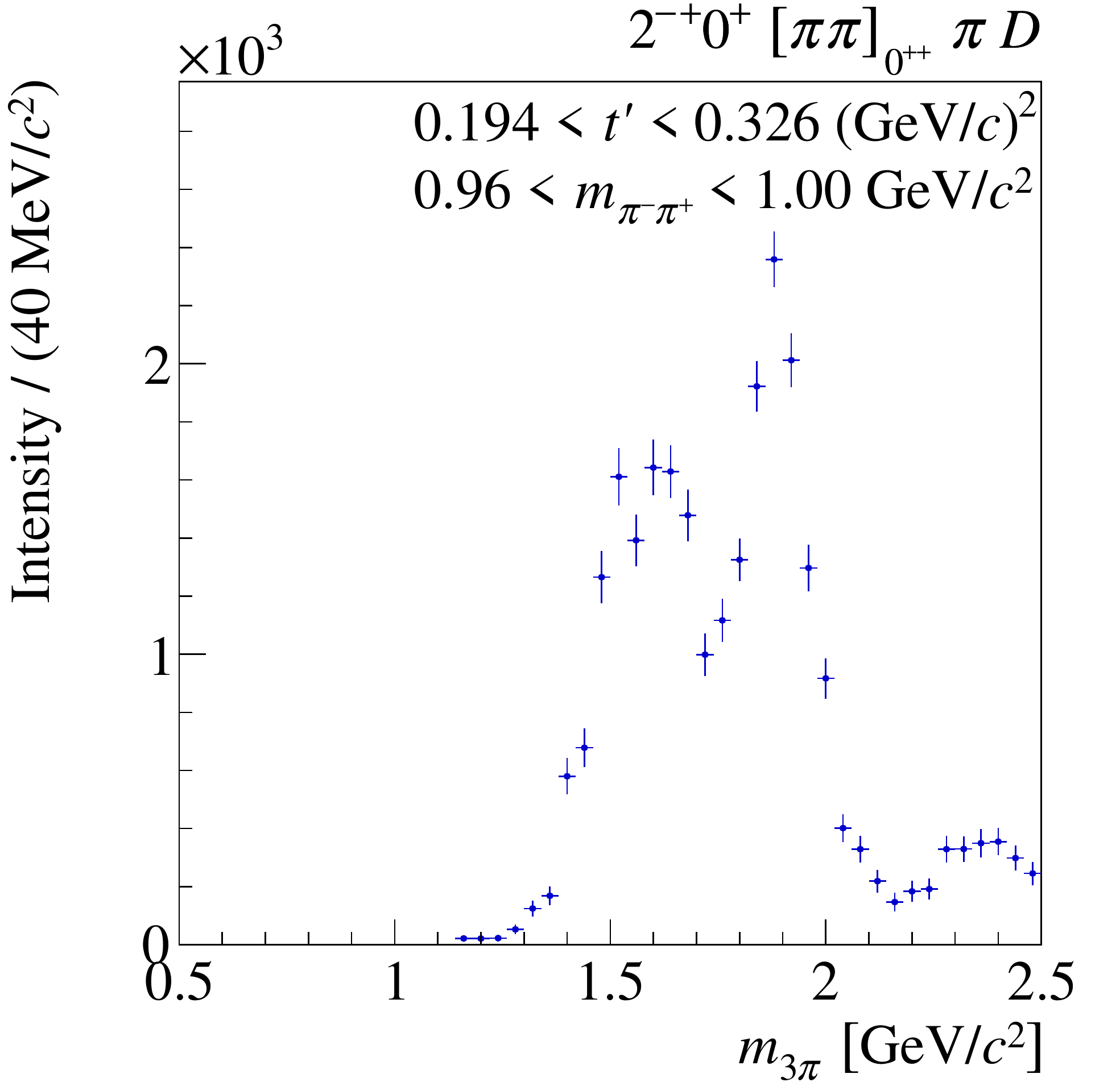}%
  }%
  \subfloat[][]{%
    \includegraphics[width=\threePlotWidth]{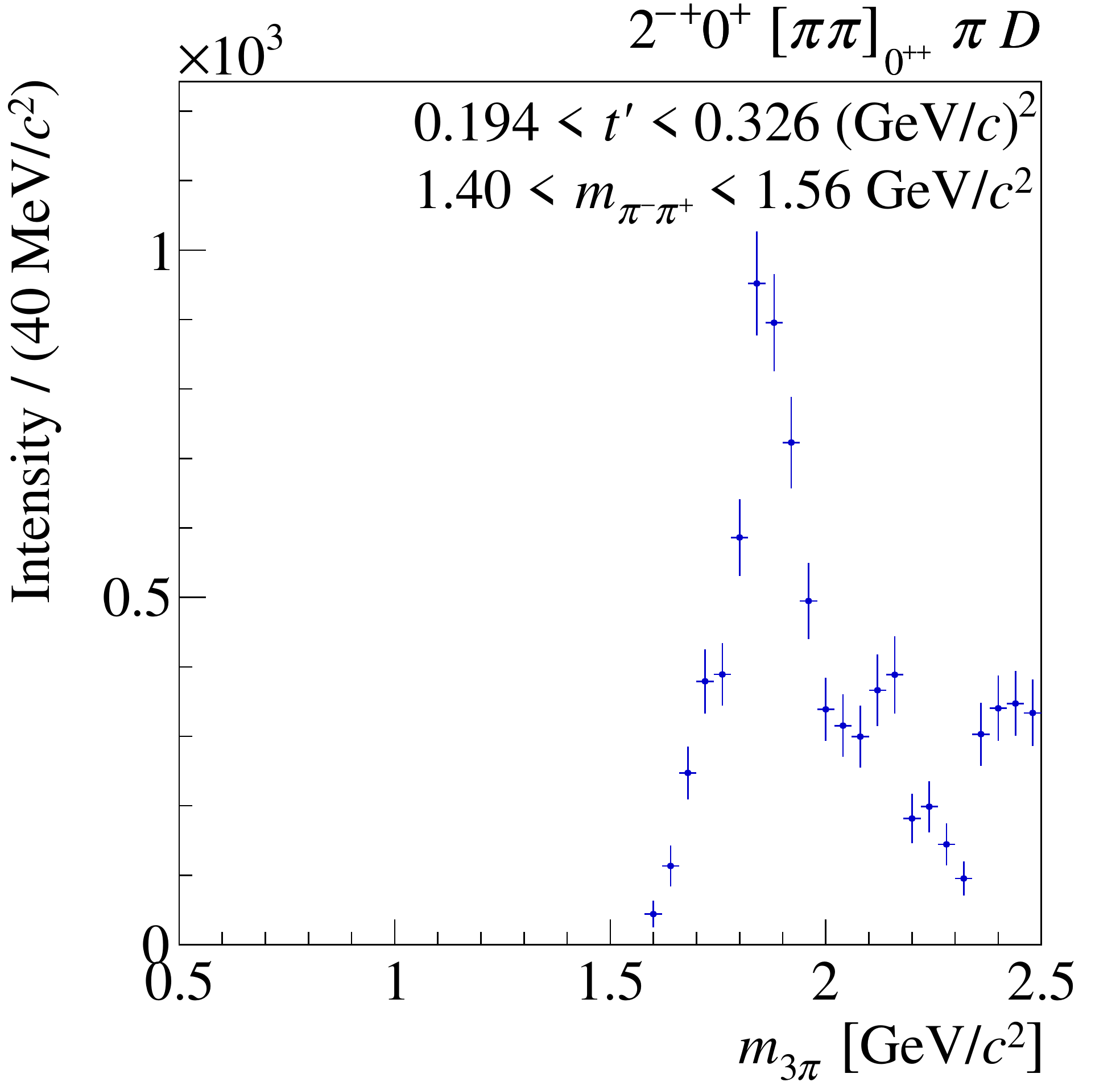}%
  }%
  \\
  \subfloat[][]{%
    \includegraphics[width=\threePlotWidthTwoD]{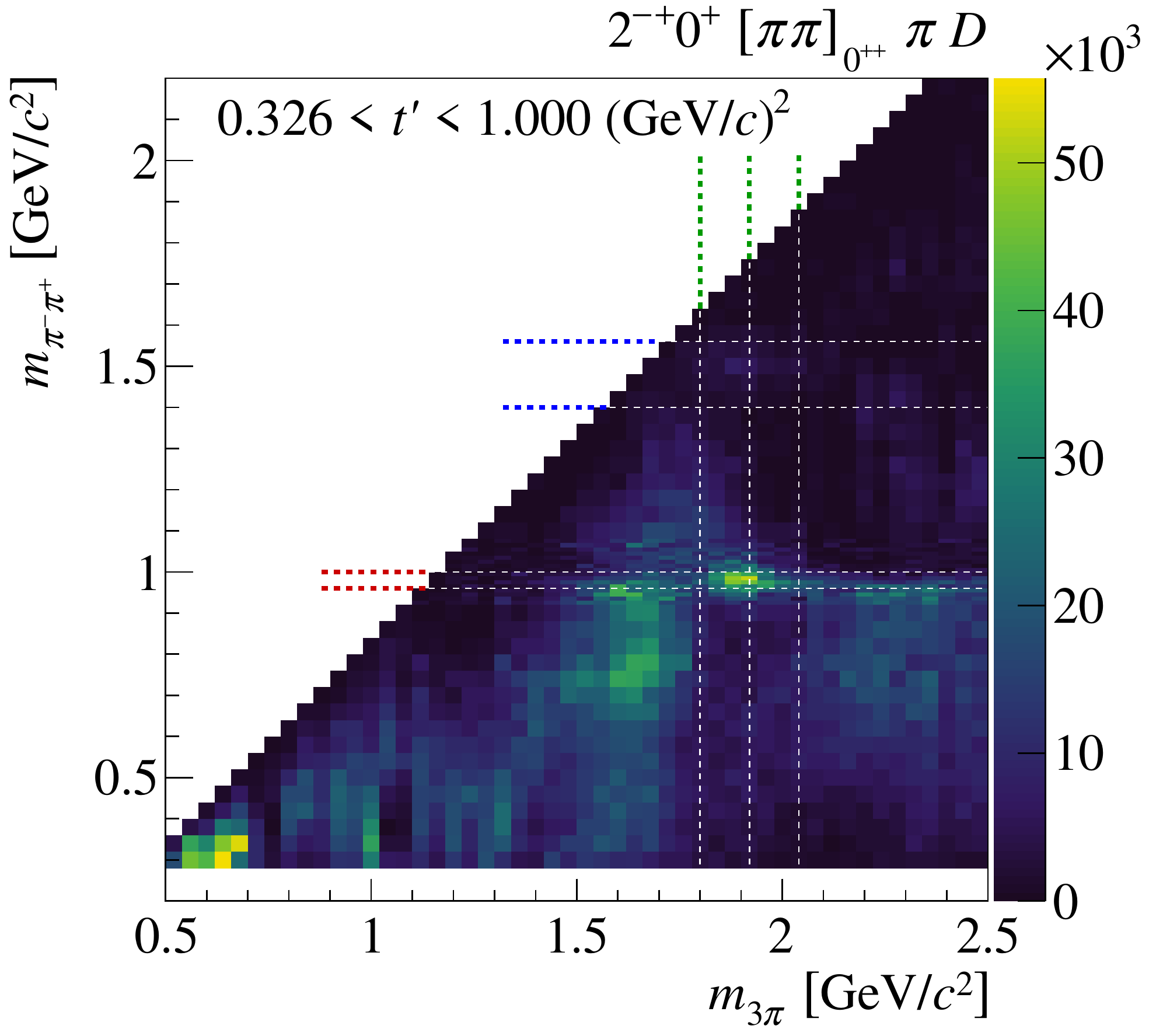}%
  }%
  \subfloat[][]{%
    \label{fig:PIPIS_2mp_hh_980}%
    \includegraphics[width=\threePlotWidth]{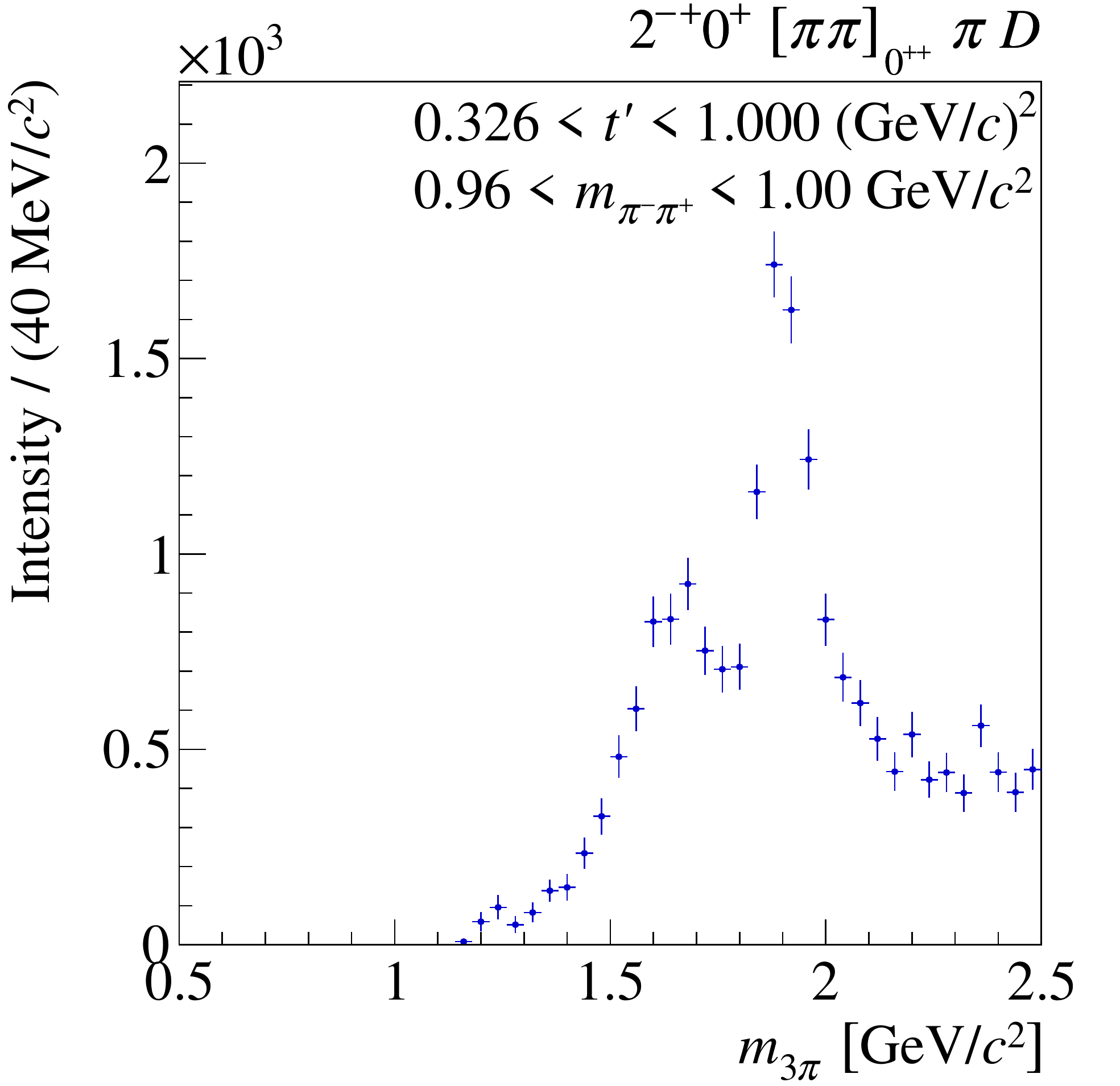}%
  }%
  \subfloat[][]{%
    \includegraphics[width=\threePlotWidth]{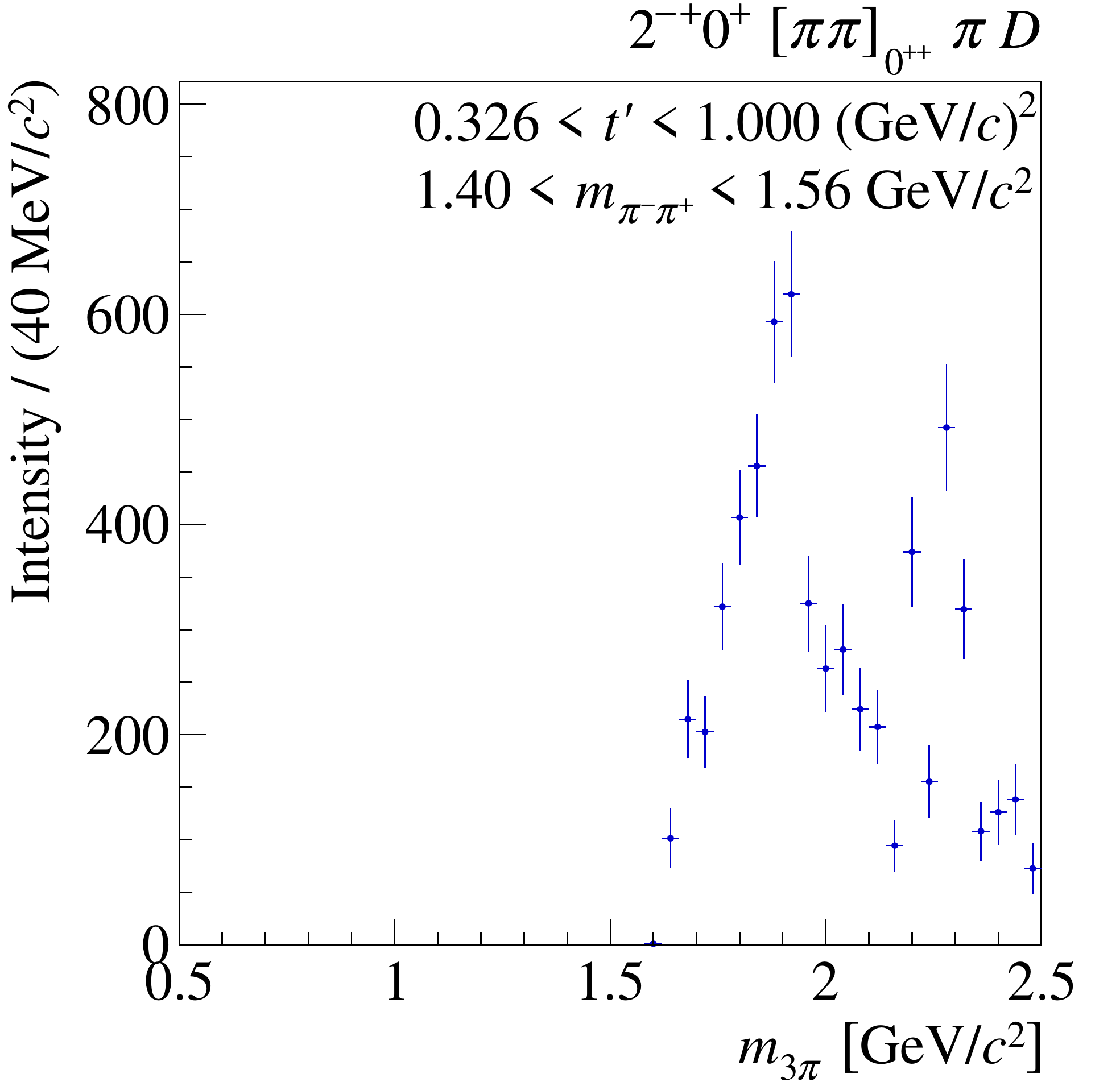}%
  }%
  \caption{\colorPlot Same as \cref{fig:pipis_2D_0mp}, but for the
    \wave{2}{-+}{0}{+}{\pipiSF}{D} wave.}
  \label{fig:pipis_2D_2mp}
\end{figure*}

\paragraph{$\JPC = 0^{-+}$:}
In the conventional analysis, the coherent sum of the waves with fixed
$0^{++}$ isobars exhibits two peaks in the intensity that may be
identified with the \Ppi[1300] and the \Ppi[1800].  These two peaks
appear very similar in the fit with freed $0^{++}$ isobars (see
\cref{fig:0mp_mass_pi_s_lowt,fig:0mp_mass_pi_s_hight}).  Since all
three pions are in a relative $S$-wave in the
\wave{0}{-+}{0}{+}{\pipiSF}{S} wave, it is very sensitive to
nonresonant $3\pi$ contributions.  In particular, in the wave with the
fixed \pipiS isobar, the region around $\mThreePi = \SI{1.3}{\GeVcc}$
seems to have nonresonant components (see
\cref{fig:0mp_pipiS,fig:0mp_pipiS_t_bins,fig:t_pi_S_m1}).  Also in the
freed-isobar fit, the shape of the \pipiSF intensity and its
considerable \tpr dependence in the \Ppi[1300] region suggest this to
be mostly nonresonant (see left column of \cref{fig:pipis_2D_0mp}).
These observations are in accordance with quark-model calculations for
the first radial excitation of the pion~\cite{barnes:1996ff} which
predict a strong suppression of the $\pipiS\,\pi$ decay mode as
compared to $\Prho\,\pi$.  Apparent enhancements of this wave in the
\Ppi[1300] region, which were observed in diffractive pion scattering
by the VES and BNL~E852 experiments~\cite{amelin:1995gu,Chung:2002pu},
are consistent with our observations and were attributed to the Deck
mechanism~\cite{klempt:2007cp,barnes:1996ff}.

In order to study the role of \PfZero[980] for the \zeroMP wave, we
sum up the intensity in the mass region
\SIvalRange{0.96}{\mTwoPi}{1.00}{\GeVcc}, which contains almost the
full \PfZero[980] and which is indicated by a pair of red horizontal
lines in the left column of \cref{fig:pipis_2D_0mp}.  The resulting
\mThreePi intensity spectra are shown in the central column of
\cref{fig:pipis_2D_0mp} for the four \tpr bins.  This simple method
does not take into account the interference of the \PfZero[980] with
the broad \pipiSW component.  The separation of amplitudes would only
be possible by fitting the \mTwoPi and \mThreePi dependences of the
amplitudes.  The \mThreePi intensity distributions exhibit a clear
signal for the \Ppi[1800].  In contrast, no clear correlation with a
possible \Ppi[1300] can be identified.  We observe the low-mass
structures in \mThreePi to vanish with growing \tpr.  This indicates
the existence of considerable nonresonant contributions in this $3\pi$
mass region.  In a similar way, the role of the \PfZero[1500] is
investigated by summing the intensity over the range
\SIvalRange{1.44}{\mTwoPi}{1.56}{\GeVcc} as indicated by a pair of
blue horizontal lines in the left column and shown in the right column
of \cref{fig:pipis_2D_0mp}.  Again, a clear correlation with the
\Ppi[1800] is observed.

\paragraph{$\JPC = 1^{++}$:}
The intensity correlations shown in the left column of
\cref{fig:pipis_2D_1pp} are dominated by a broad maximum between
\SIlist{0.6;0.8}{\GeVcc} in \mTwoPi.  For increasing \tpr, this
structure shifts from $\mThreePi = \SIrange{1.2}{1.4}{\GeVcc}$, almost
reaching the \PaOne[1420] region for the highest \tpr bin.  This
behavior suggests the existence of large nonresonant contributions,
which obstruct the observation of a possible coupling of the \PaOne to
the broad component of the \pipiSW.  The right column of
\cref{fig:pipis_2D_1pp} shows the intensity summed over the
\PfZero[980] mass region of \SIvalRange{0.96}{\mTwoPi}{1.00}{\GeVcc}.
For all \tpr bins, the mass spectra show a clear \PaOne[1420] peak and
no contribution of the \PaOne.  This demonstrates that the observed
\PaOne[1420] signal in the $\PfZero[980]\,\pi$ channel is not an
artifact of the $0^{++}$ isobar parametrizations used in the
conventional analysis method.

\paragraph{$\JPC = 2^{-+}$:}
The $2^{-+}$ intensity correlations shown in the left column of
\cref{fig:pipis_2D_2mp} exhibit a vertical band around
$\mThreePi = \SI{1.5}{\GeVcc}$ [below the \PpiTwo].  The \mTwoPi
distribution peaks below the \PfZero[980].  This structure changes its
shape and relative strength with \tpr.  The role of the \PfZero[980]
and \PfZero[1500] isobars is again illustrated by summing the
intensities over the respective $2\pi$ mass ranges, which are shown in
the central and right columns of \cref{fig:pipis_2D_2mp}.  For both
\twoPi mass regions, we observe a clear signal for the \PpiTwo[1880].
The intensity maximum around $\mThreePi = \SI{1.6}{\GeVcc}$ in the
\PfZero[980] slice changes its shape and position with \tpr, and hence
looks different from the \PpiTwo peak, as it is, for example, observed
in the $\PfTwo\,\pi$ decay mode and shown in
\cref{fig:pi2_t_bin_low,fig:pi2_t_bin_high,fig:pi2_total_m0_2,fig:pi2_total_m1_2}.
In addition, the position of the peak is \tpr dependent, which
indicates a nonresonant contribution.  The VES
experiment~\cite{amelin:1995gu} has reported on an excited \PpiTwo*
resonances at \SI{2.09}{\GeVcc}.  It was observed as a
\SI{520}{\MeVcc} broad enhancement in the $\pipiS\, \pi$ and
$\PfZero[980]\, \pi$ waves.  We also observe a similarly broad
structure in the $3\pi$ system at $\mThreePi \approx \SI{2.2}{\GeVcc}$
correlated with a broad bump at $2\pi$ masses of approximately
\SI{750}{\MeVcc}.  The shape in \mTwoPi seems to change as a function
of \tpr.  With the present analysis, we cannot confirm the resonance
interpretation of this structure.  While a corresponding peak is
observed around $\mThreePi = \SI{2.2}{\GeVcc}$ in the conventional
fixed-isobar fit in the \wave{2}{-+}{0}{+}{\PfZero[980]}{D} wave (see
\cref{fig:2mp_f0980}), no pronounced correlation with the $2\pi$
system in the \PfZero[980] mass region is seen in the freed-isobar
fit.

\subsection{Argand Diagrams and $2\pi$ Mass Spectra for freed \pipi
  $S$-Wave Isobars}
\label{sec:results_free_pipi_s_wave_argands}

The previous section discussed mainly the correlation of the \pipiSW
and the $3\pi$ partial-wave intensities.  The two-dimensional
transition amplitudes extracted from the data furthermore contain
information about the \mThreePi and the \mTwoPi dependences of the
relative phases.  These phases are measured \wrt the
\wave{1}{++}{0}{+}{\Prho}{S} anchor wave as a function of $2\pi$ mass.
They give insight into the composition of the \pipiSW amplitude.  In
order to study the influence of the $3\pi$ system, we look at the
$2\pi$ invariant mass spectra for three $3\pi$ mass bins, chosen
below, at, and above clear $3\pi$ resonance signals.  The centers of
the \mThreePi bins are indicated by green vertical lines in the left
columns of \cref{fig:pipis_2D_0mp,fig:pipis_2D_1pp,fig:pipis_2D_2mp}.

\paragraph{$\JPC = 0^{-+}$:}
The wave with the freed \pipiSF isobar shows a clear signal for the
\Ppi[1800] coupling to $\PfZero[980]\,\pi$ and $\PfZero[1500]\,\pi$.
The left column of \cref{fig:pipi_s_wave_0mp_highT} shows the \pipiSF
intensity as a function of \mTwoPi for three different values of
\mThreePi, \ie below, at, and above the \Ppi[1800] resonance for the
region of larger \tpr, where the resonance signal is clearer.  The
three \mTwoPi spectra are similar.  We observe prominent signals for
\PfZero[980].  Because of phase space, the \PfZero[1500] peak appears
only in the two higher $3\pi$ mass intervals.  The enhancement of both
states relative to the broad \pipiSW component is strongest at the
\Ppi[1800] mass.  The right column of \cref{fig:pipi_s_wave_0mp_highT}
shows the corresponding \Argands measured \wrt the
\wave{1}{++}{0}{+}{\Prho}{S} wave.  For a fixed mass of the $3\pi$
system, the \Argand describes magnitude and phase of the $2\pi$
amplitude.  The phase of the \pipiSF amplitude spans nearly two full
circles about the origin in the $2\pi$ mass range from threshold to
about \SI{1.6}{\GeVcc}.  This reflects the resonance character of the
\PfZero[980] and \PfZero[1500].  The positions of \PfZero[980] and
\PfZero[1500] (marked by the blue line segments in the \Argands)
rotate counterclockwise with increasing \mThreePi, reflecting the
growing phase of the \Ppi[1800] \wrt the anchor wave.  We conclude a
clear coupling of \Ppi[1800] to both \PfZero* states, which is more
pronounced than that to the broad component of the \pipiSW.  There is
no evidence for the \PfZero[1370] in this wave.  The observed behavior
of the \pipiSF phases corroborates the conclusions drawn from the
intensity correlations in \cref{fig:pipis_2D_0mp}.  In the \Ppi[1800]
region, the gross features of the \pipiSF phase motion are similar to
those observed by the \babar experiment in the $2\pi$ subsystem of
$D_s^+ \to \pi^+\pi^-\pi^+$ decays~\cite{Aubert:2008ao}.  Differences
in details are probably due to different nonresonant contributions in
the two processes.

\begin{figure*}[htbp]
  \centering
  \subfloat[][]{%
    \includegraphics[width=\twoPlotWidth]{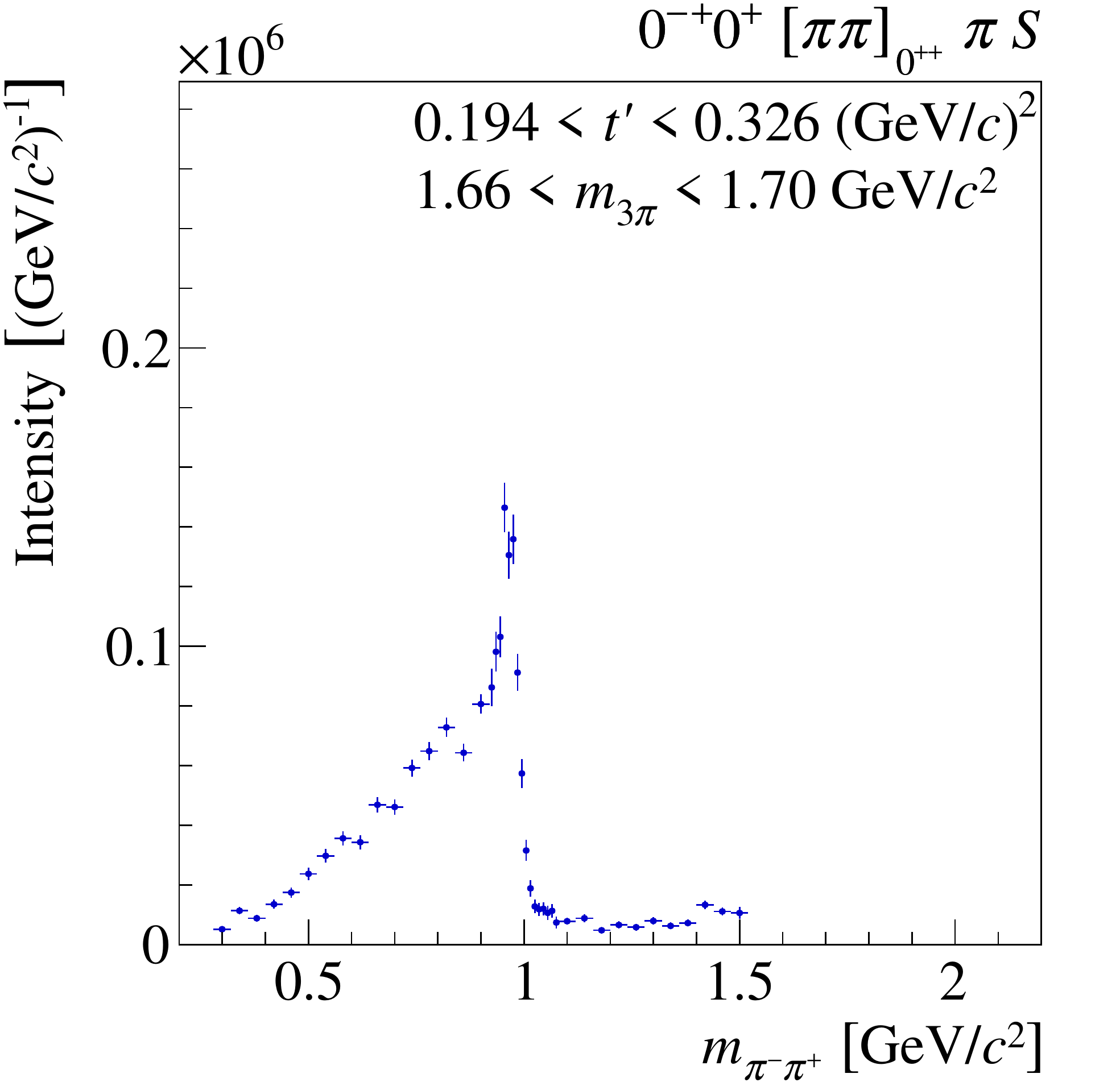}%
  }%
  \hspace*{\twoPlotSpacing}
  \subfloat[][]{%
    \includegraphics[width=\twoPlotWidth]{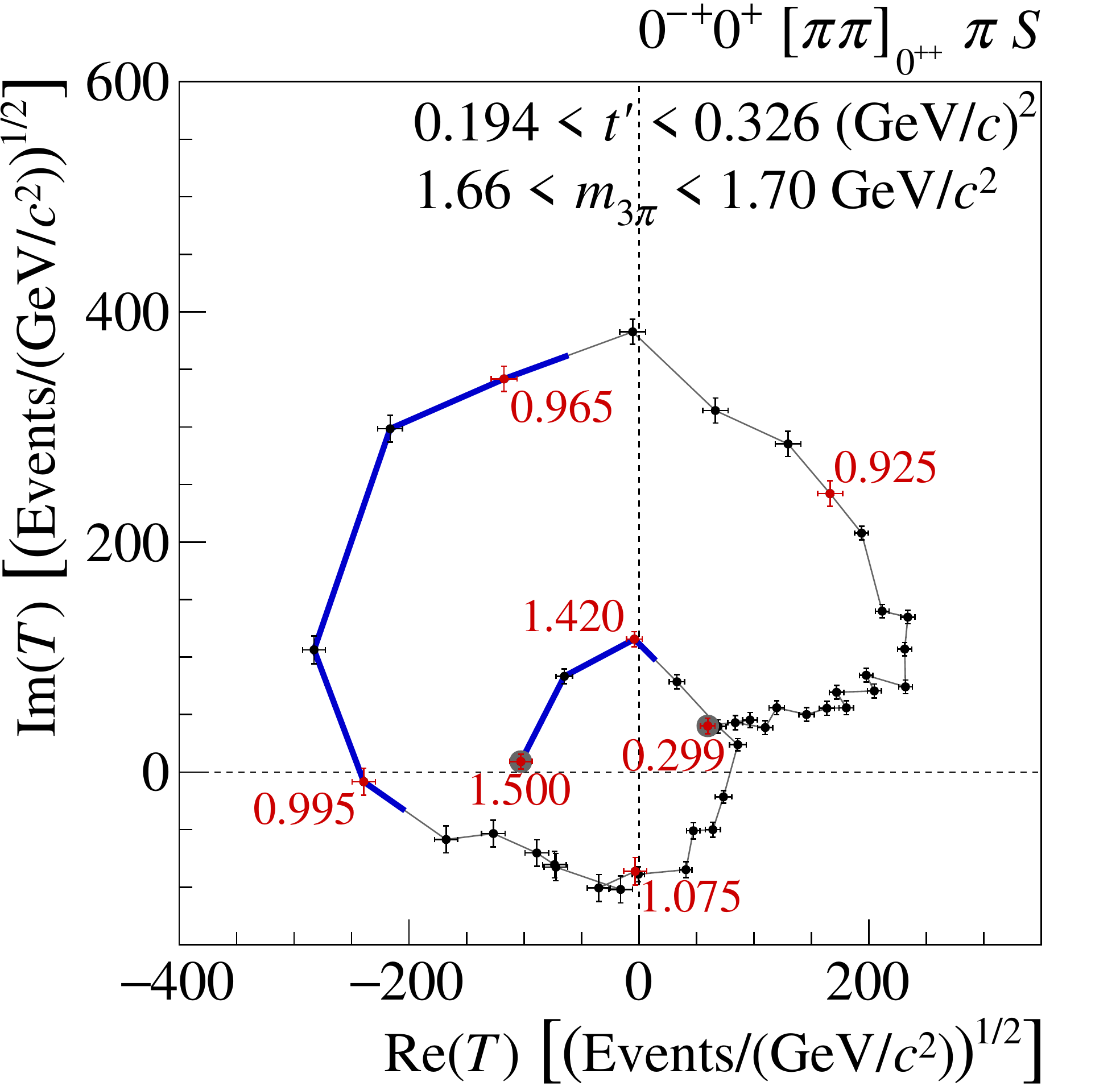}%
  }%
  \\
  \subfloat[][]{%
    \includegraphics[width=\twoPlotWidth]{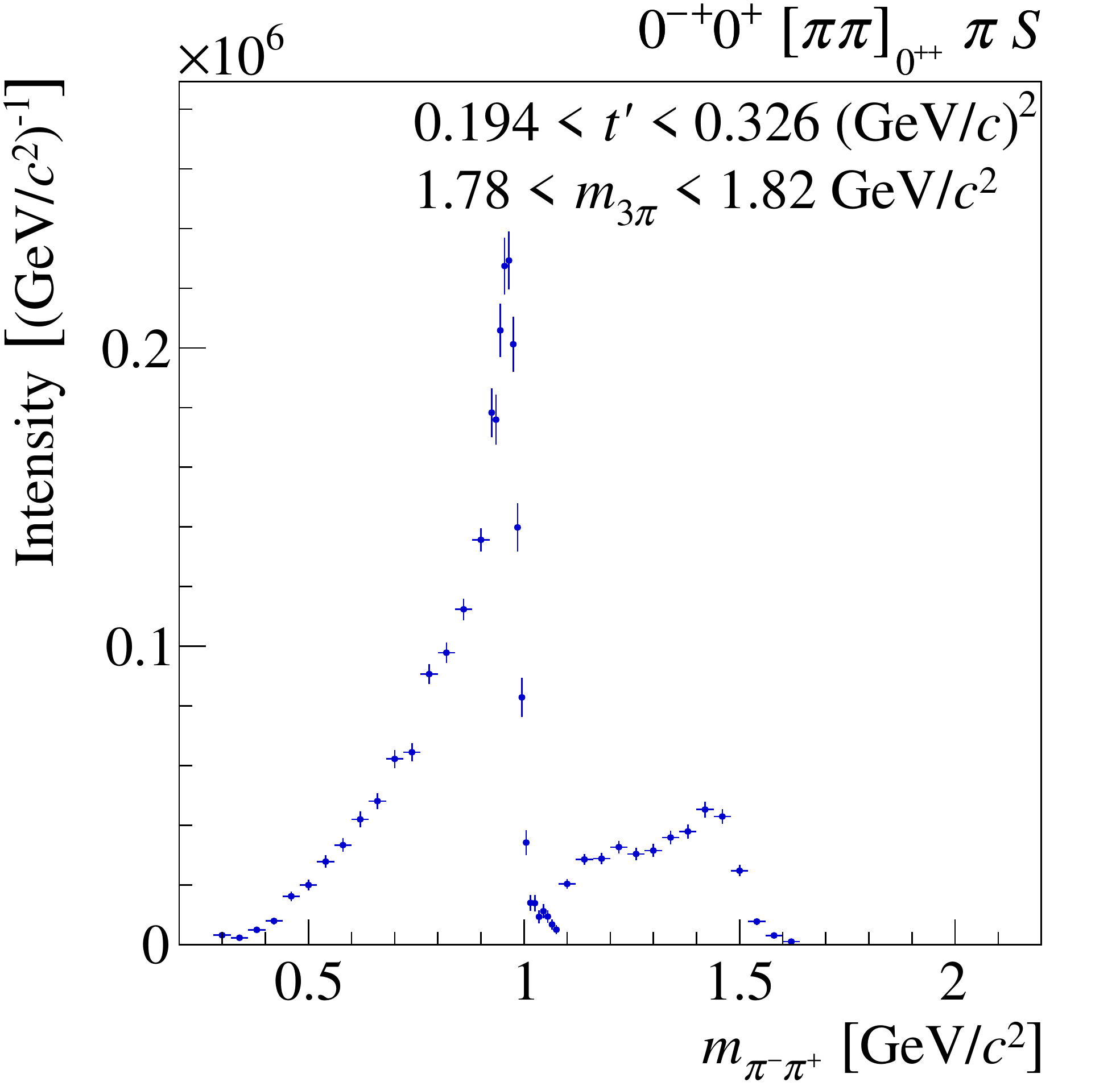}%
  }%
  \hspace*{\twoPlotSpacing}
  \subfloat[][]{%
    \includegraphics[width=\twoPlotWidth]{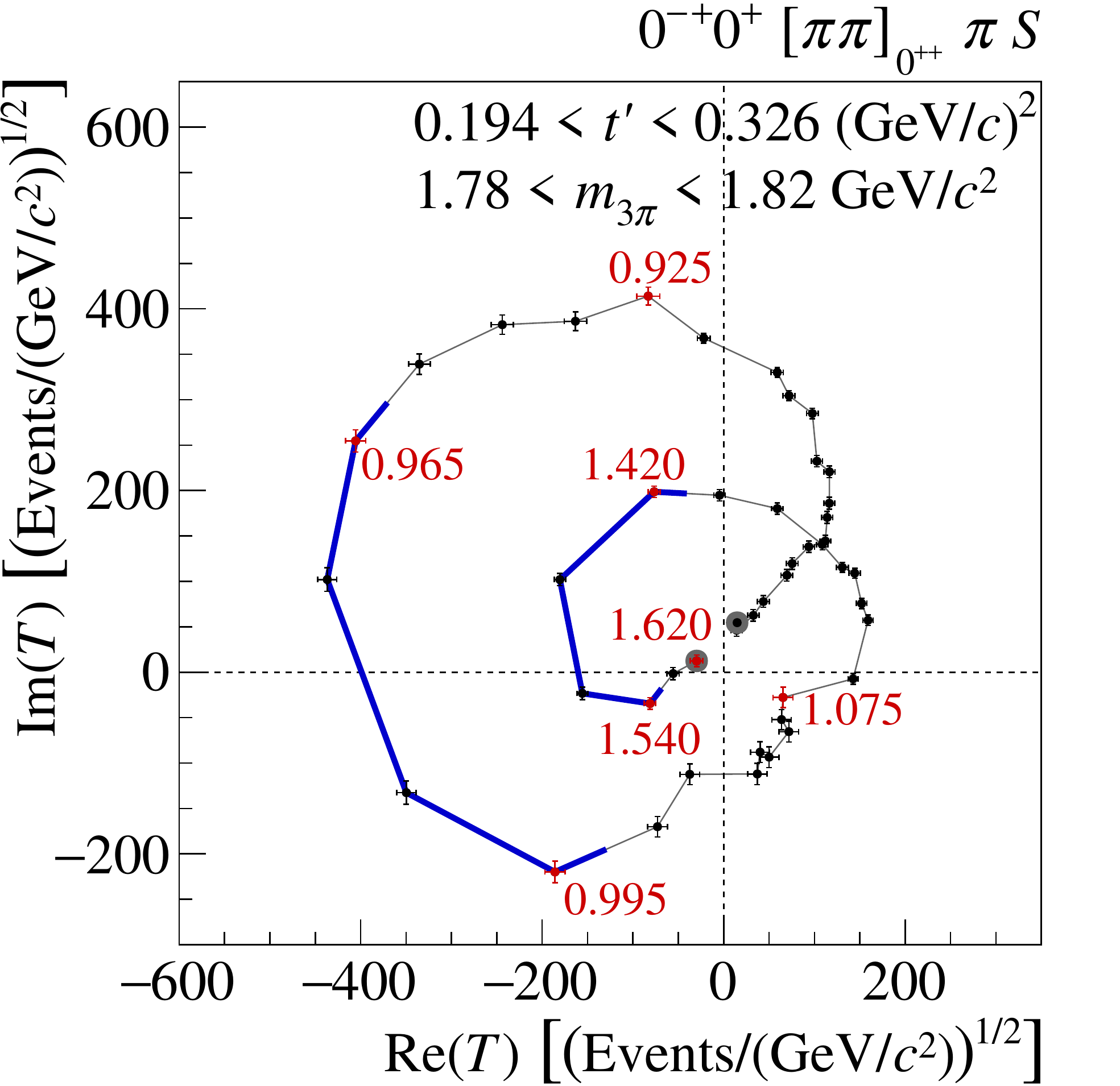}%
  }%
  \\
  \subfloat[][]{%
    \includegraphics[width=\twoPlotWidth]{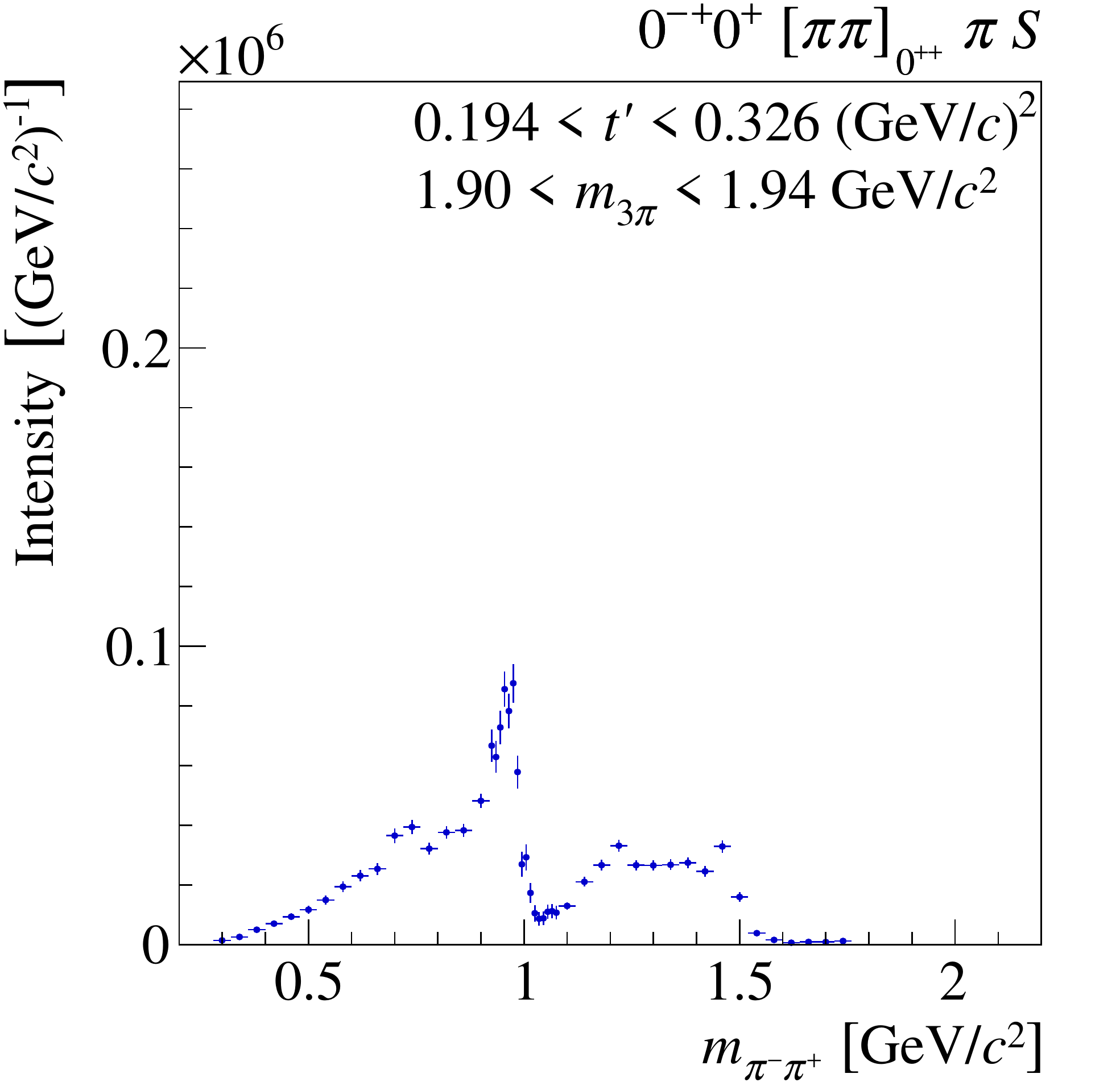}%
  }%
  \hspace*{\twoPlotSpacing}
  \subfloat[][]{%
    \includegraphics[width=\twoPlotWidth]{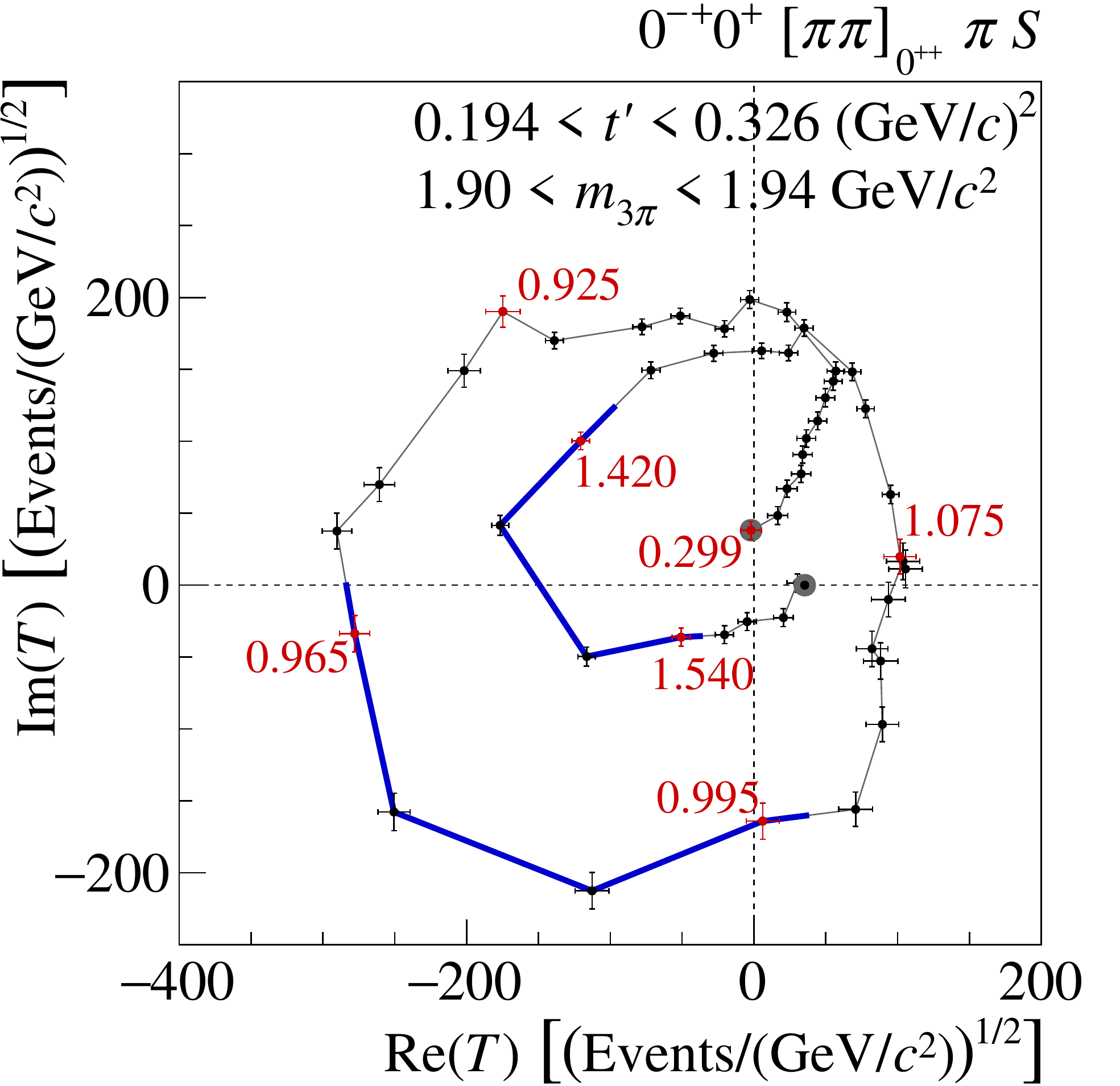}%
  }%
  \caption{\colorPlot The freed \pipiSF amplitude in the
    \wave{0}{-+}{0}{+}{\pipiSF}{S} wave in the range
    \SIvalRange{0.194}{\tpr}{0.326}{\GeVcsq} for three intervals in
    \mThreePi; below (top row), at (center row), and above the
    \Ppi[1800] (bottom row).  Left column: intensities as a function
    of \mTwoPi; Right column: \Argands.  The crosses with error bars
    are the result of the mass-independent fit.  The numbers in the
    \Argands show the corresponding mass value of the \twoPi system.
    The data points are connected by lines in order to indicate the
    order.  The line segments highlighted in blue correspond to the
    \mTwoPi ranges around the \PfZero[980] from
    \SIrange{960}{1000}{\MeVcc} and, if phase space permits, around
    the \PfZero[1500] from \SIrange{1400}{1560}{\MeVcc}.  The $2\pi$
    mass is binned in \SI{10}{\MeVcc} wide intervals around the
    \PfZero[980] and in \SI{40}{\MeVcc} wide slices elsewhere.  The
    phase of the \Argands is fixed by the \wave{1}{++}{0}{+}{\Prho}{S}
    wave.}
  \label{fig:pipi_s_wave_0mp_highT}
\end{figure*}

\begin{figure*}[htbp]
  \centering
  \subfloat[][]{%
    \includegraphics[width=\twoPlotWidth]{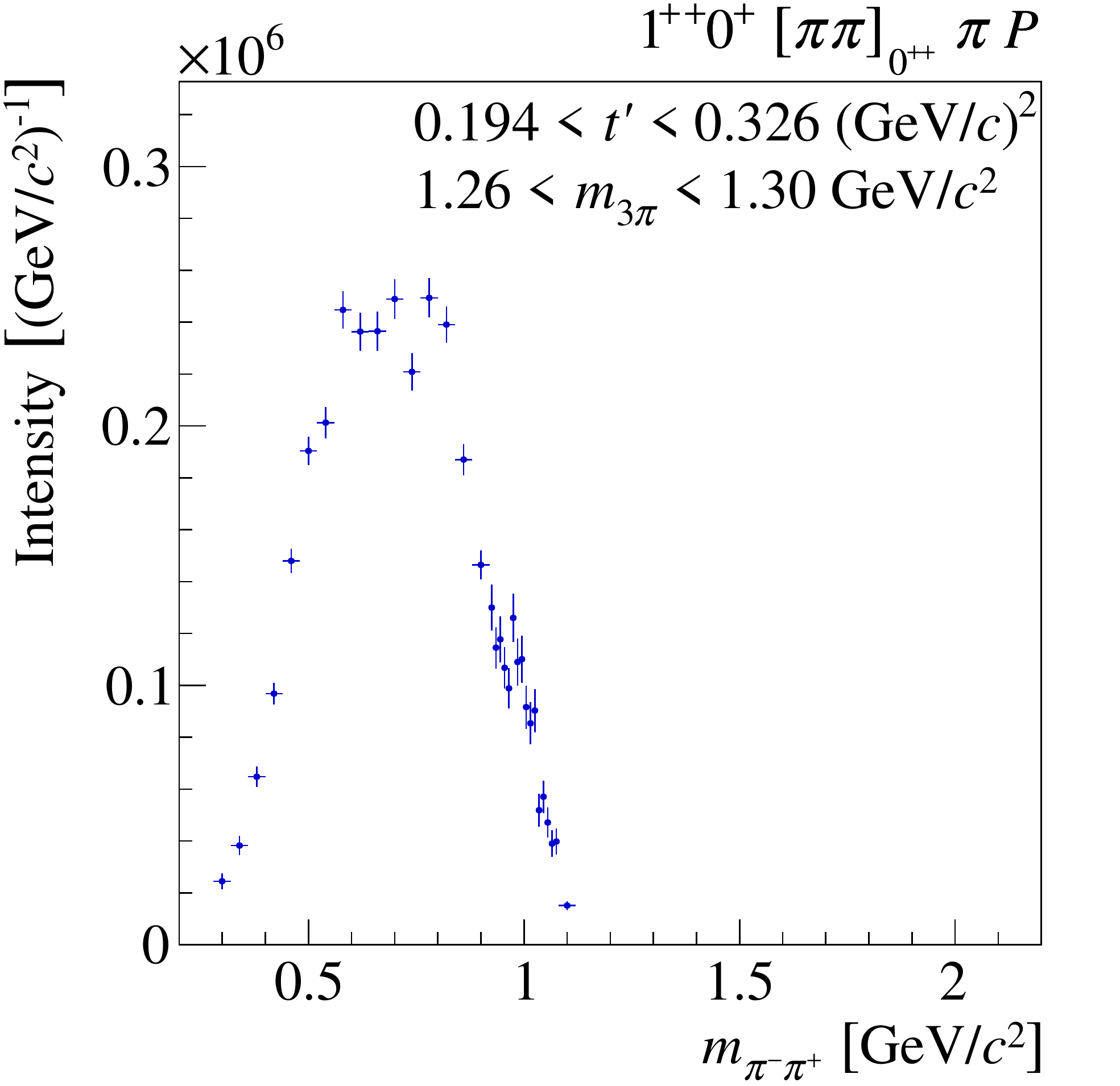}%
  }%
  \hspace*{\twoPlotSpacing}
  \subfloat[][]{%
    \includegraphics[width=\twoPlotWidth]{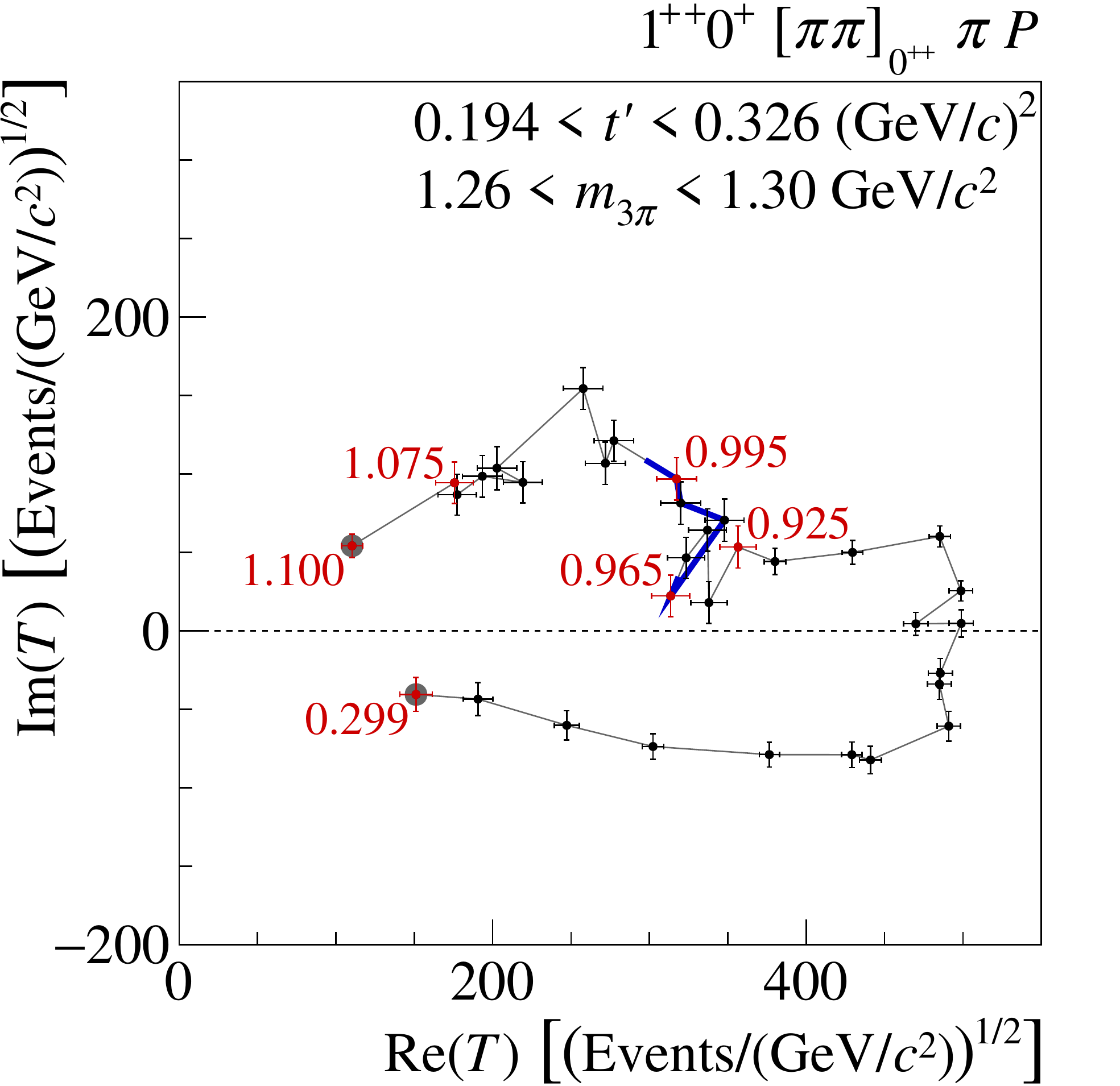}%
  }%
  \\
  \subfloat[][]{%
    \includegraphics[width=\twoPlotWidth]{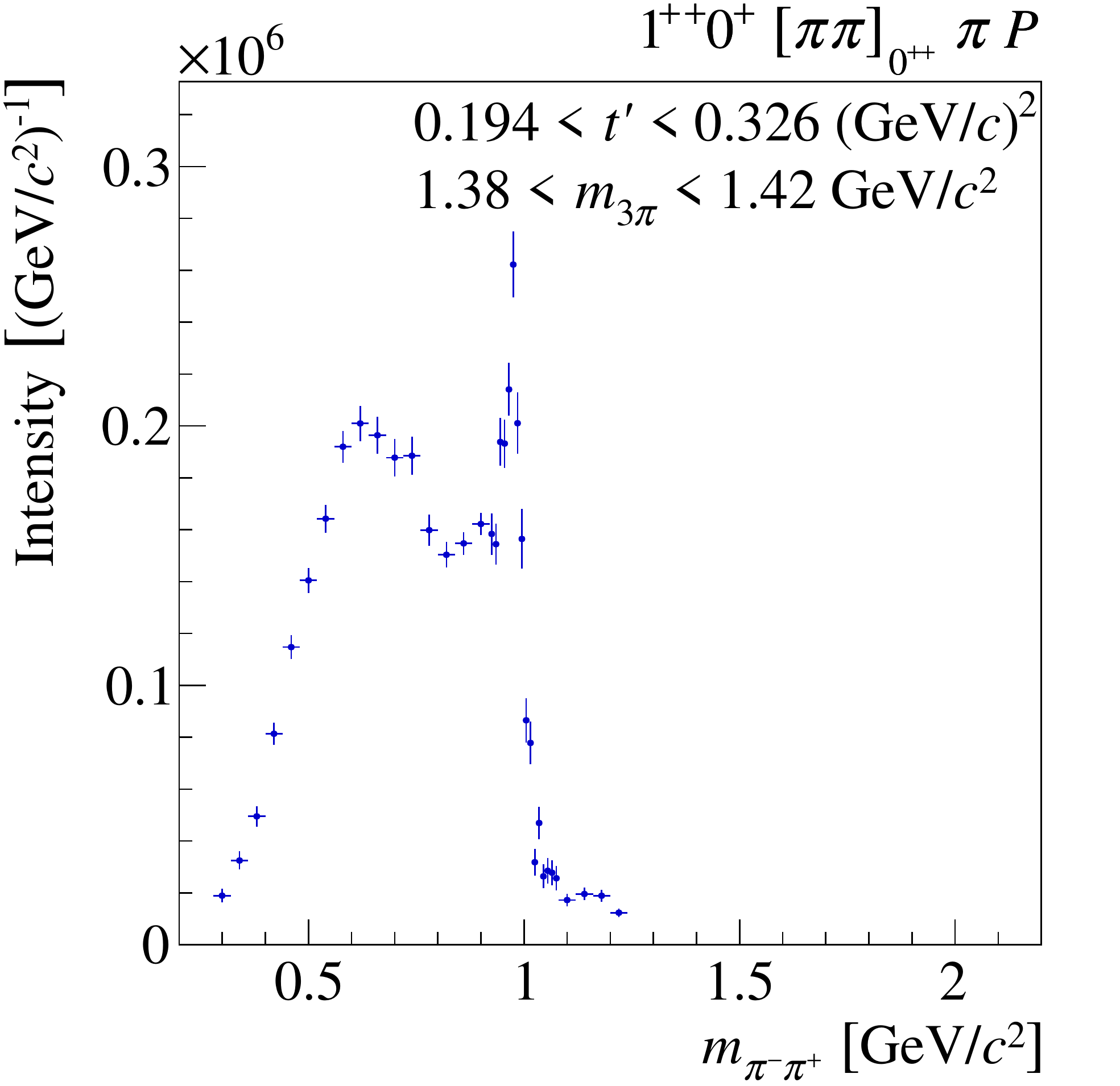}%
  }%
  \hspace*{\twoPlotSpacing}
  \subfloat[][]{%
    \label{fig:pipi_s_wave_1pp_highT_argand_at_res}%
    \includegraphics[width=\twoPlotWidth]{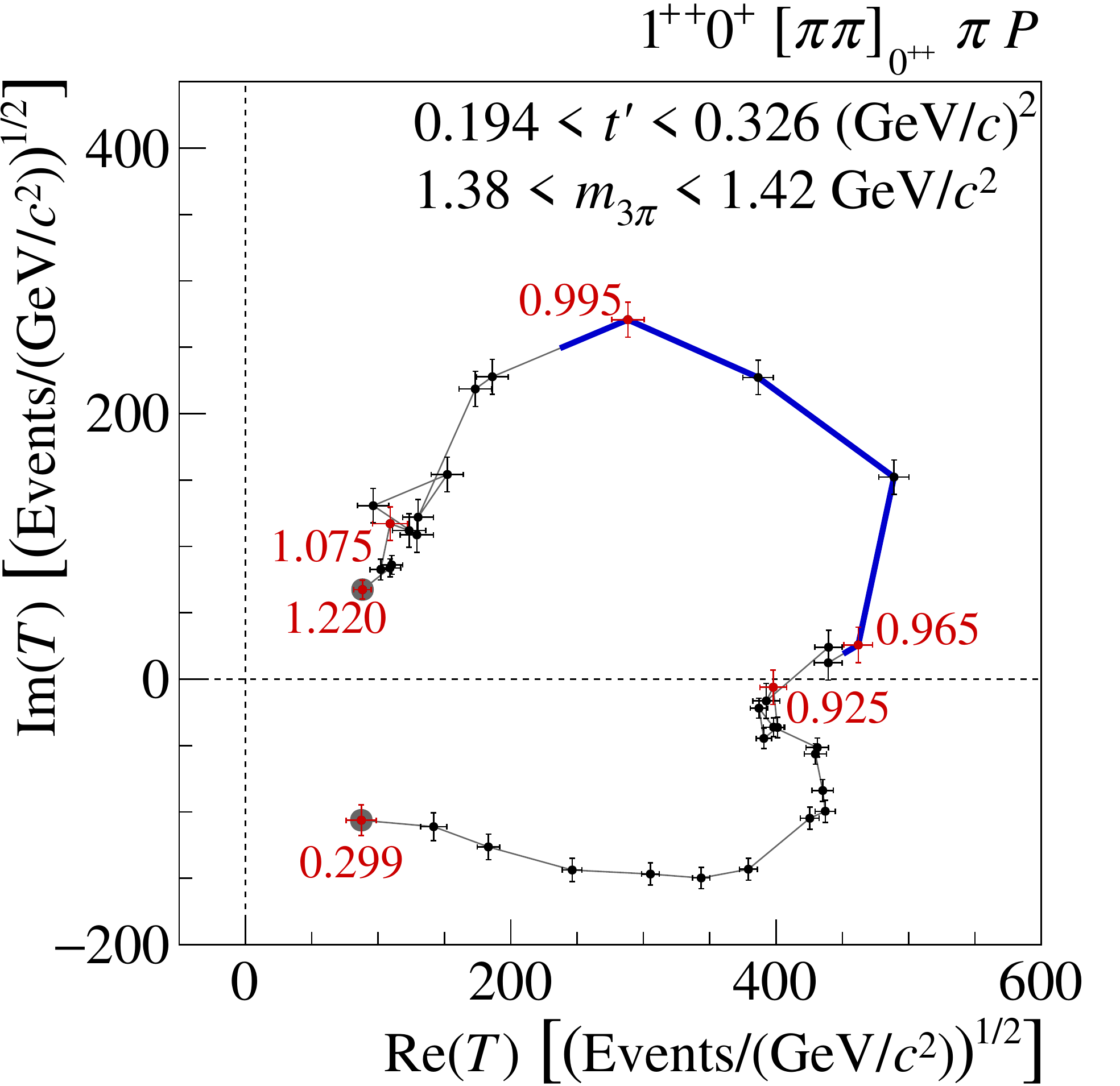}%
  }%
  \\
  \subfloat[][]{%
    \includegraphics[width=\twoPlotWidth]{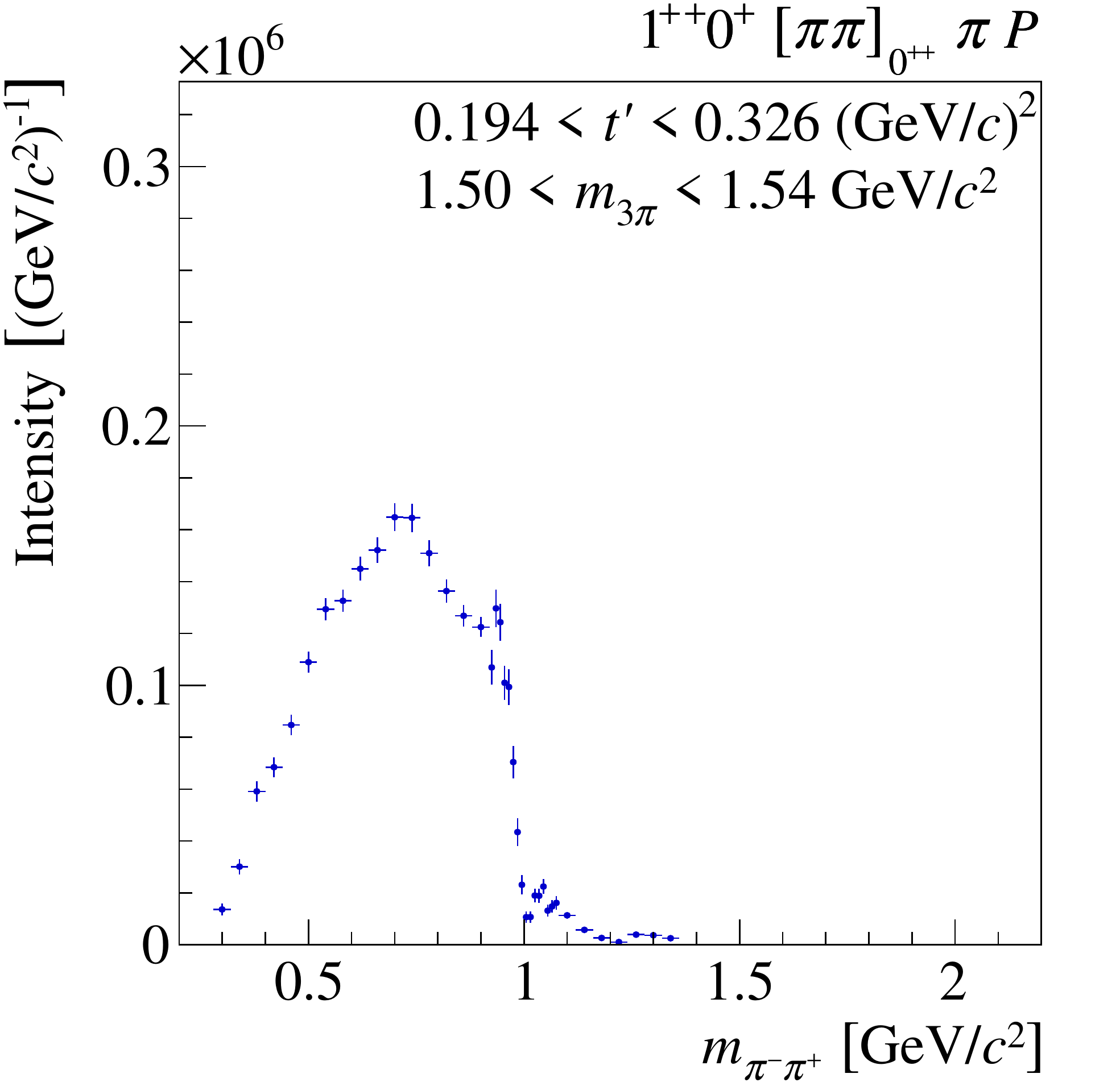}%
  }%
  \hspace*{\twoPlotSpacing}
  \subfloat[][]{%
    \label{fig:pipi_s_wave_1pp_highT_argand_above_res}%
    \includegraphics[width=\twoPlotWidth]{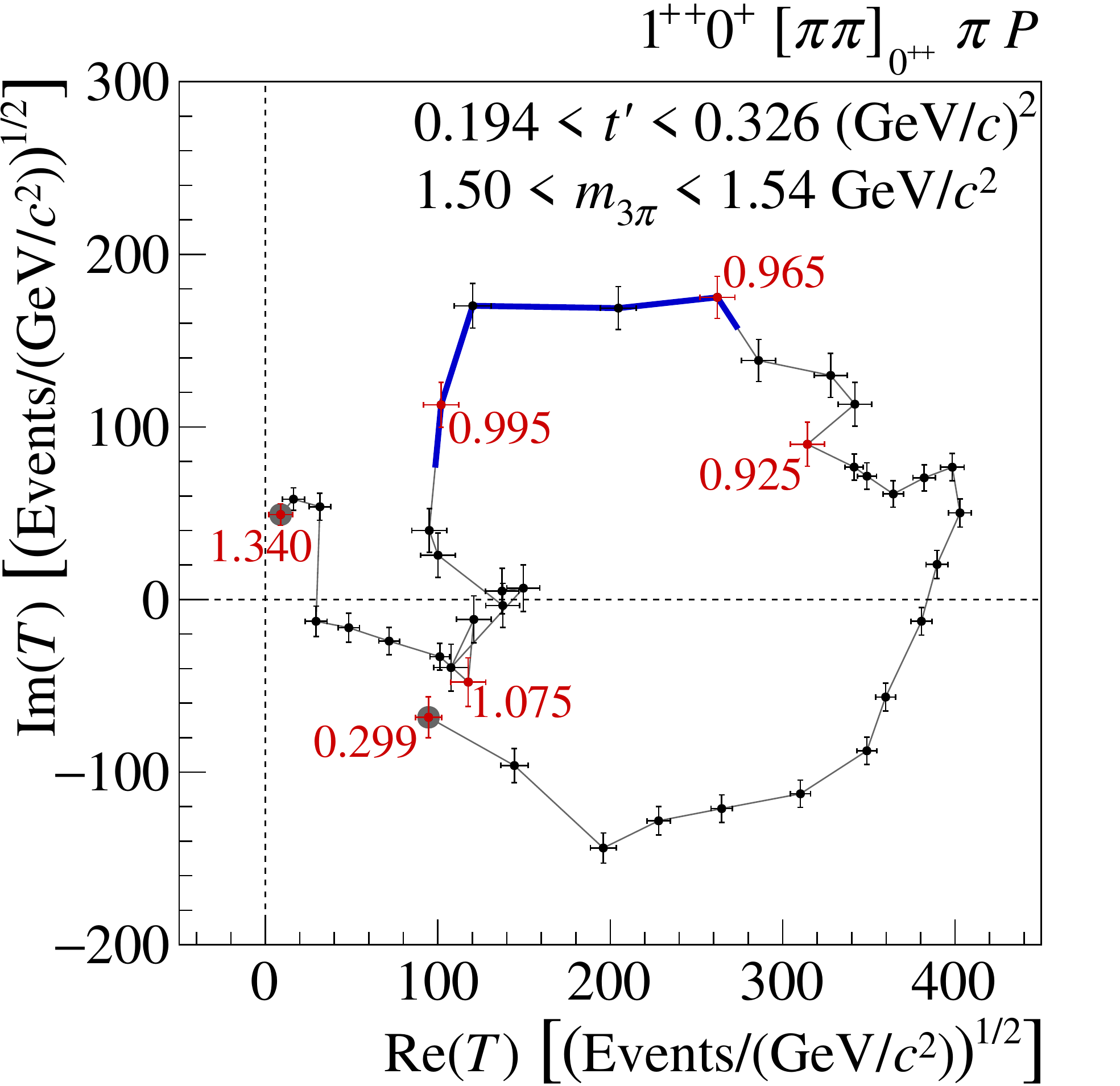}%
  }%
  \caption{\colorPlot Same as \cref{fig:pipi_s_wave_0mp_highT}, but
    for the \wave{1}{++}{0}{+}{\pipiSF}{P} wave in three \mThreePi
    bins around the \PaOne[1420].}
  \label{fig:pipi_s_wave_1pp_highT}
\end{figure*}

\begin{figure*}[htbp]
  \centering
  \subfloat[][]{%
    \includegraphics[width=\twoPlotWidth]{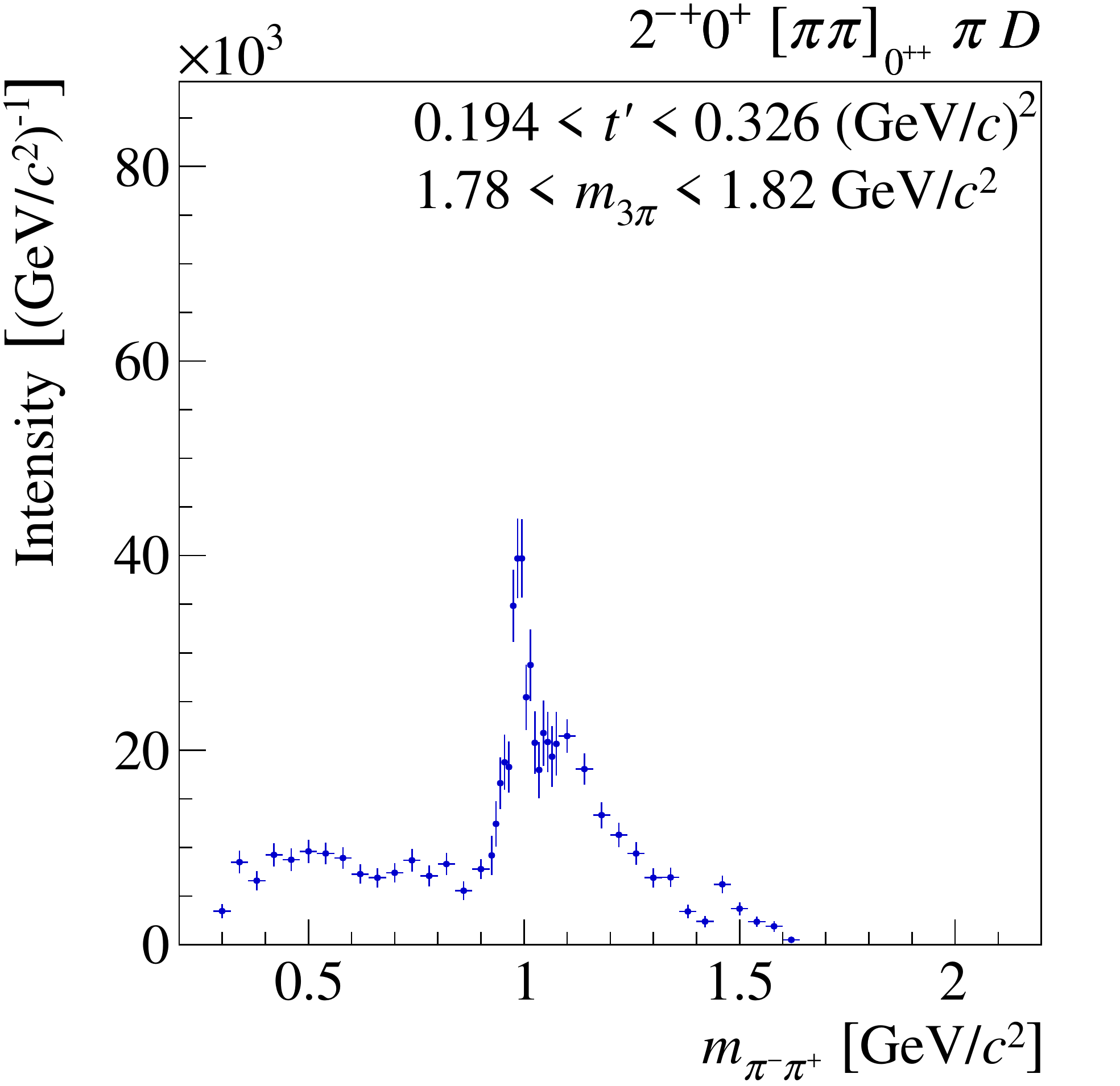}%
  }%
  \hspace*{\twoPlotSpacing}
  \subfloat[][]{%
    \includegraphics[width=\twoPlotWidth]{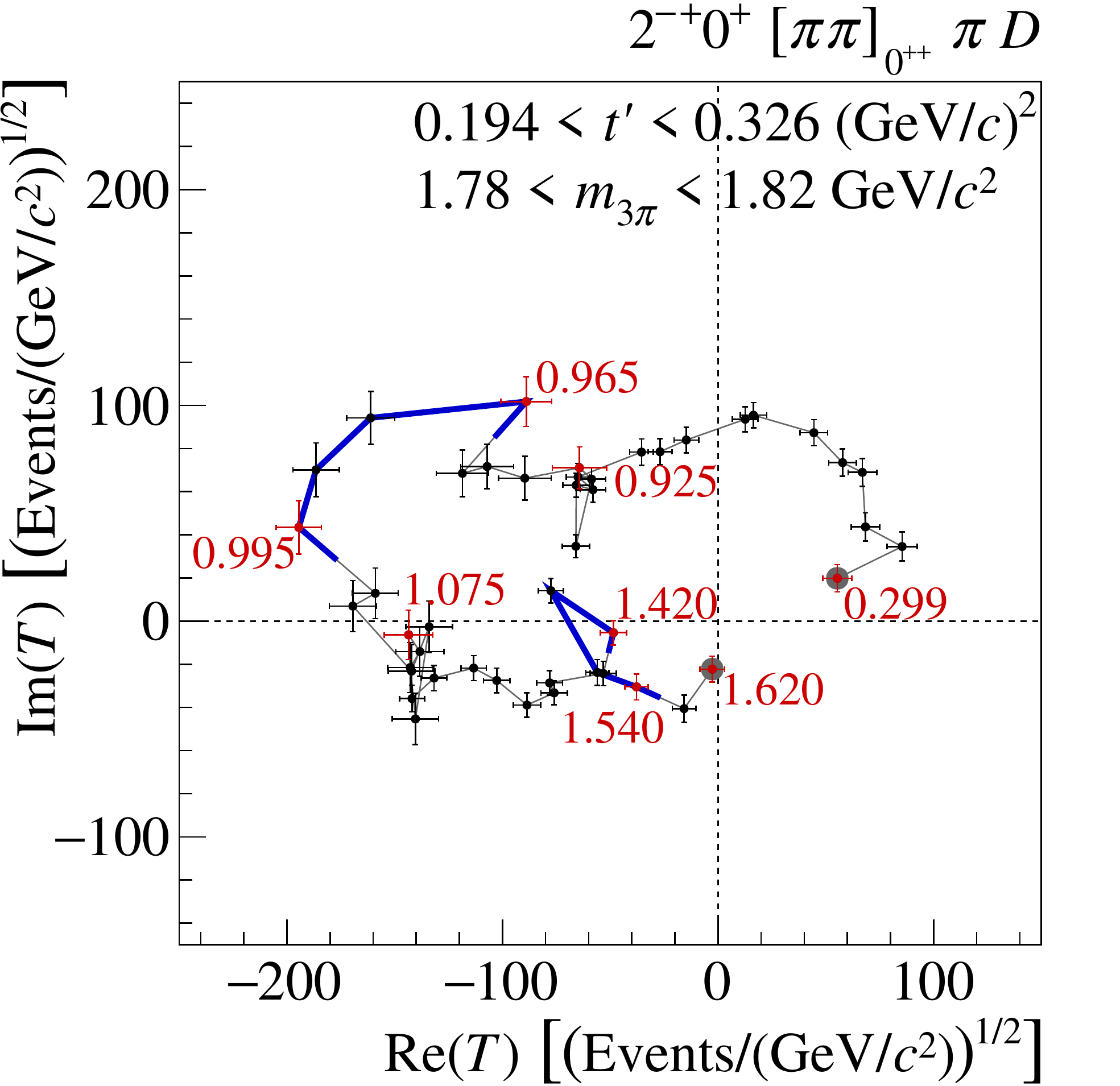}%
  }%
  \\
  \subfloat[][]{%
    \includegraphics[width=\twoPlotWidth]{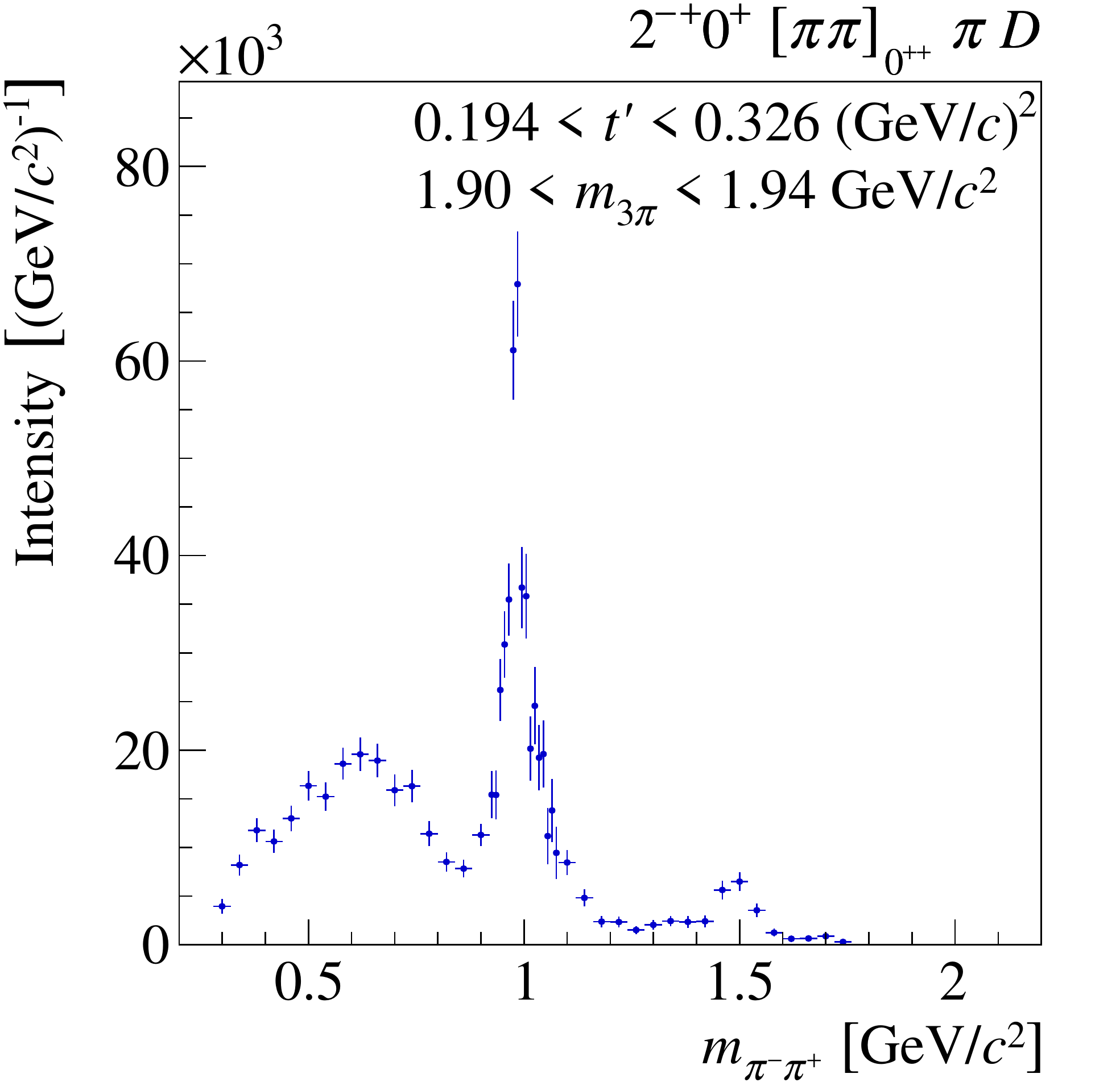}%
  }%
  \hspace*{\twoPlotSpacing}
  \subfloat[][]{%
    \includegraphics[width=\twoPlotWidth]{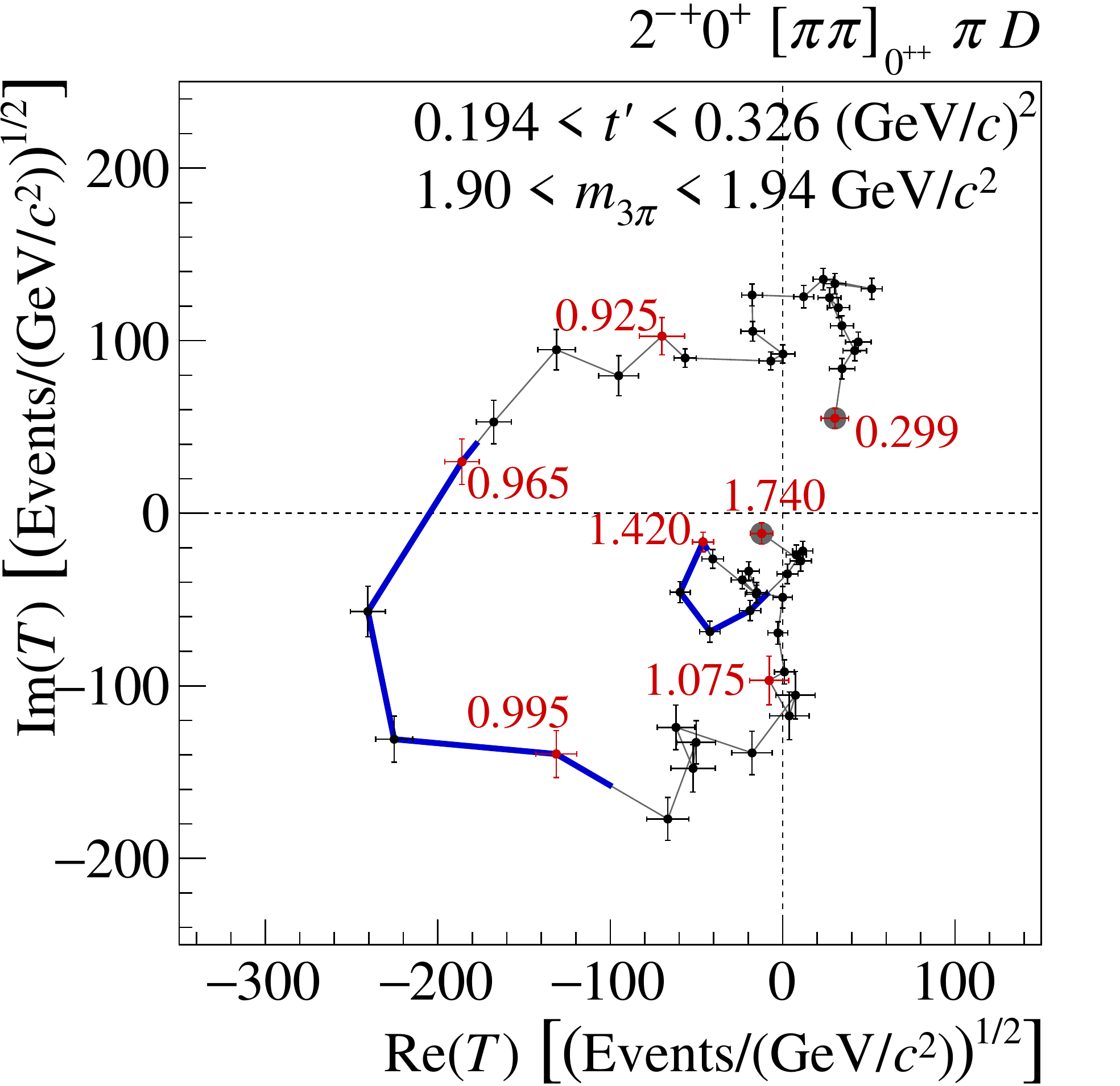}%
  }%
  \\
  \subfloat[][]{%
    \includegraphics[width=\twoPlotWidth]{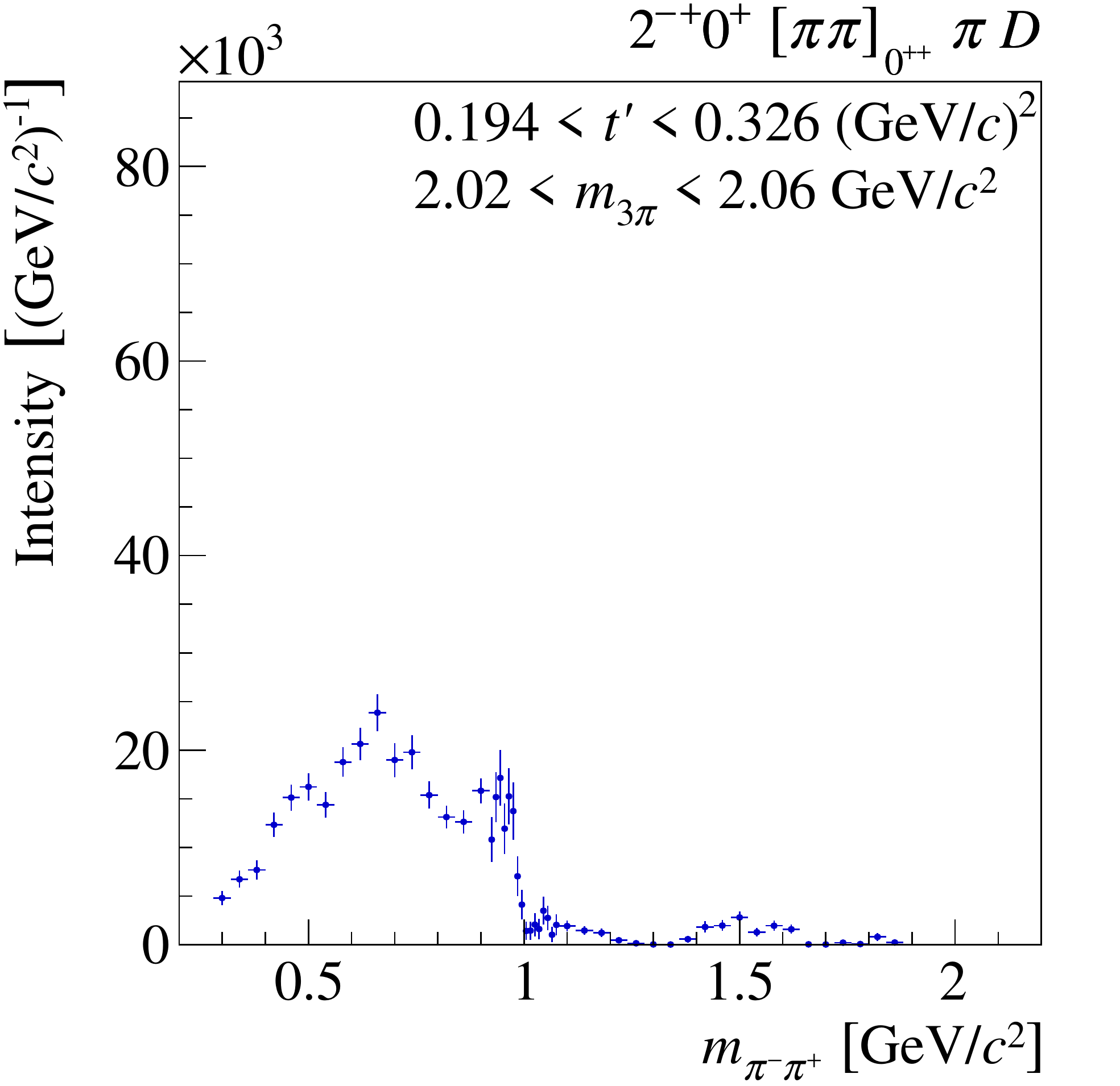}%
  }%
  \hspace*{\twoPlotSpacing}
  \subfloat[][]{%
    \includegraphics[width=\twoPlotWidth]{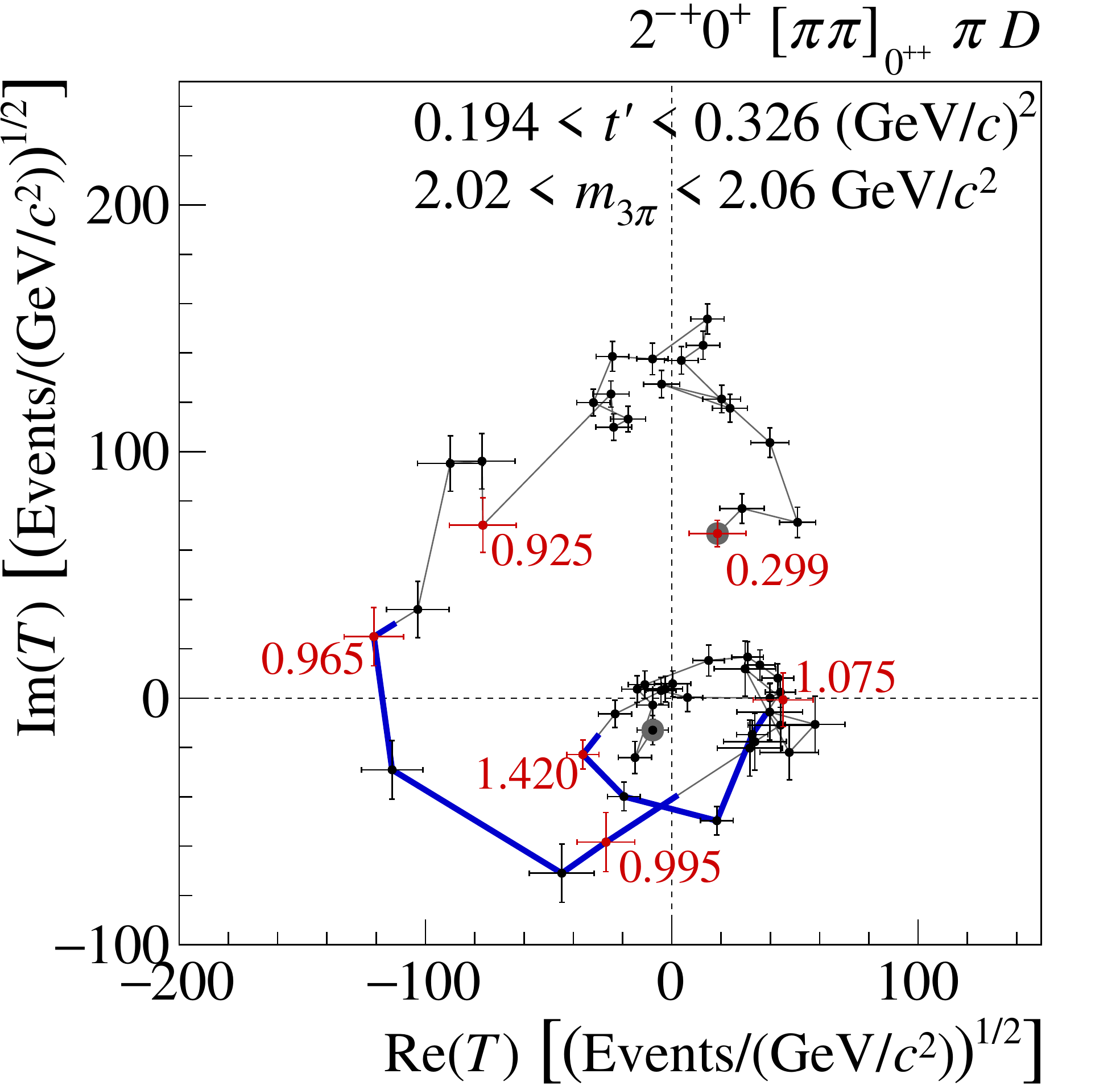}%
  }%
  \caption{\colorPlot Same as \cref{fig:pipi_s_wave_0mp_highT}, but
    for the \wave{2}{-+}{0}{+}{\pipiSF}{D} wave in three \mThreePi
    bins around the \PpiTwo[1880].}
  \label{fig:pipi_s_wave_2mp_highT}
\end{figure*}

\paragraph{$\JPC = 1^{++}$:}
In this wave, the \PaOne[1420] region is most interesting.  The
extracted \pipiSF intensities are shown in the left column of
\cref{fig:pipi_s_wave_1pp_highT} for three different values of
\mThreePi in a similar way as explained above.  For \mThreePi values
around \SI{1.4}{\GeVcc}, a signal for the \PfZero[980] appears sitting
above a broad \pipiSW structure.  The complicated shape of the \twoPi
amplitude is again illustrated by the \Argands shown in the right
column of \cref{fig:pipi_s_wave_1pp_highT}.  Here, the \PfZero[980]
contribution can be identified by the semicircle-like structure with a
shifted origin, which appears at the \PaOne[1420] resonance.
Comparing
\cref{fig:pipi_s_wave_1pp_highT_argand_at_res,fig:pipi_s_wave_1pp_highT_argand_above_res},
a significant counterclockwise rotation of this structure by about
\SI{90}{\degree} \wrt its center is observed above the \PaOne[1420].
This confirms the resonance interpretation of the \PaOne[1420] in the
$\PfZero[980]\,\pi$ decay.  For $\mTwoPi < \SI{0.8}{\GeVcc}$, the
amplitude does not change much \wrt \mThreePi.  As a consequence, the
relative phase of the \PfZero[980] component \wrt the broad \pipiSW
structure also changes by about \SI{90}{\degree}.  Hence the
interference pattern changes from partly constructive at the
\PaOne[1420] to partly destructive at the higher \mThreePi bin.  This
results in a sharp drop of the intensity in the $2\pi$ mass spectrum
above the \PfZero[980], which is followed by low intensity at higher
\mTwoPi.

\paragraph{$\JPC = 2^{-+}$:}
The clearest signal in this wave is the \PpiTwo[1880] that couples to
$\PfZero[980]\,\pi$ and $\PfZero[1500]\,\pi$.  We again study the
$2\pi$ subsystem in three $3\pi$ mass regions.  The intensity
distributions and \Argands are shown in
\cref{fig:pipi_s_wave_2mp_highT}.  Also here we find semicircle-like
structures with shifted origin that correspond to the \PfZero[980],
which is observed as a clear peak in the corresponding \mTwoPi
distributions.  At higher values of \mTwoPi, we observe an indication
of another small circular structure in the \Argand, which is
correlated with a rise of the intensity in the $2\pi$ mass
distribution attributable to \PfZero[1500].  The shape of the broad
\pipiSW component at low $2\pi$ masses is somewhat different from that
of the other waves, exhibiting more intensity close to the threshold.
 %
%
%

\section{Summary and Conclusions}
\label{sec:conclusions}

We have presented a detailed partial-wave analysis based on the
world's currently largest data set of the exclusive \threePi final
state from diffractive pion scattering off a proton target.  The PWA
was performed independently in 100~bins of the $3\pi$ mass \mThreePi,
each of which was subdivided into eleven slices of the reduced
four-momentum transfer squared \tpr.  We refer to this as
\emph{mass-independent fit}.  It is based on the largest wave set used
so far in a PWA of this final state, which contains in total 80~waves
with positive reflectivity, 7~with negative reflectivity, and one
incoherent isotropic wave representing three uncorrelated pions (see
\cref{sec:pwa_massindep_model,tab:waveset} in
\cref{sec:appendix_wave_set}).  In this paper, a subset of 18~partial
waves with positive reflectivity accounting for \SI{75.8}{\percent} of
the total intensity was studied in detail.

From the study of the general characteristics of partial-wave
intensities, two classes of waves can be identified: for some waves,
the shape of the mass spectrum shows little dependence on \tpr (see
\eg \cref{fig:a2_t_bin_low,fig:a2_t_bin_high}), while for others
moderate (see \eg \cref{fig:pi2_t_bin_low,fig:pi2_t_bin_high}) or even
large variations (see \eg \cref{fig:a1_t_bin_low,fig:a1_t_bin_high})
are seen.  These variations indicate the presence of weak or strong
nonresonant contributions that do interfere with the resonant
components and may have a characteristically different dependence on
\tpr.  Among the stable waves, where the peak positions do not
significantly depend on \tpr, we find:
\ifMultiColumnLayout{}{\begin{multicols}{2}}
  \begin{itemize}
    \renewcommand{\labelitemi}{}
  \item \wave{0}{-+}{0}{+}{\PfZero[980]}{S}
  \item \wave{1}{++}{0}{+}{\PfZero[980]}{P}
  \item \wave{2}{++}{1}{+}{\Prho}{D}
  \item \wave{2}{++}{2}{+}{\Prho}{D}
  \item \wave{2}{-+}{1}{+}{\PfTwo}{S}
  \item \wave{2}{-+}{0}{+}{\PfTwo}{D}
  \item \wave{4}{++}{1}{+}{\Prho}{G}
  \item \wave{4}{++}{1}{+}{\PfTwo}{F}
  \end{itemize}
\ifMultiColumnLayout{}{\end{multicols}}
\noindent
The following waves show significant peak shifts or large variations
of the shapes of their mass distribution as a function of \tpr:
\ifMultiColumnLayout{}{\begin{multicols}{2}}
  \begin{itemize}
    \renewcommand{\labelitemi}{}
  \item \wave{0}{-+}{0}{+}{\pipiS}{S}
  \item \wave{1}{++}{0}{+}{\pipiS}{P}
  \item \wave{1}{++}{0}{+}{\Prho}{S}
  \item \wave{1}{++}{1}{+}{\Prho}{S}
  \item \wave{1}{++}{0}{+}{\PfTwo}{P}
  \item \wave{2}{++}{1}{+}{\PfTwo}{P}
  \item \wave{2}{-+}{0}{+}{\pipiS}{D}
  \item \wave{2}{-+}{0}{+}{\Prho}{F}
  \item \wave{2}{-+}{0}{+}{\PfZero[980]}{D}
  \item \wave{2}{-+}{0}{+}{\PfTwo}{S}
  \end{itemize}
\ifMultiColumnLayout{}{\end{multicols}}

In the subset of 18~waves, clear resonance peaks are found in the
partial-wave intensities of the following decay modes:
\ifMultiColumnLayout{}{\begin{multicols}{2}}
  \begin{itemize}
    \renewcommand{\labelitemi}{}
  \item $\Ppi[1800] \to \PfZero\,\pi$ $S$-wave
  \item $\Ppi[1800] \to \pipiS\,\pi$ $S$-wave
  \item $\PaOne \to \Prho\,\pi$ $S$-wave
  \item $\PaOne[1420] \to \PfZero[980]\,\pi$ $P$-wave
  \item $\PaTwo \to \Prho\,\pi$ $D$-wave
  \item $\PpiTwo \to \PfTwo\,\pi$ $S$-wave
  \item $\PpiTwo[1880] \to \PfTwo\,\pi$ $D$-wave
  \item $\PaFour \to \Prho\,\pi$ $G$-wave
  \item $\PaFour \to \PfTwo\,\pi$ $F$-wave
  \end{itemize}
\ifMultiColumnLayout{}{\end{multicols}}
\noindent
The new \PaOne[1420], which was presented in \refCite{Adolph:2015pws},
is only seen in the \wave{1}{++}{0}{+}{\PfZero[980]}{P} wave.  No
evidence for a corresponding resonance structure is observed in
\wave{1}{++}{0}{+}{\pipiS}{P}, nor in waves containing other isobars.
The \PpiTwo and \PpiTwo[1880] seem to have different couplings to
various decay modes.  A peak attributable to the \PpiTwo[1670]
appears, for example, dominantly in \wave{2}{-+}{0}{+}{\PfTwo}{S},
while a \PpiTwo[1880] peak is dominant in
\wave{2}{-+}{0}{+}{\PfTwo}{D}.  Both states seem to couple to the
\wave{2}{-+}{0}{+}{\Prho}{F} wave, however, with different apparent
strength as a function of \tpr.  In turn, only the \PpiTwo[1880] shows
a clear coupling to the $\PfZero[980]\, \pi$ decay mode.  The shape of
the structure observed in this decay mode around
$\mThreePi = \SI{1.6}{\GeVcc}$ changes as a function of \tpr (see
\cref{sec:results_free_pipi_s_wave_int_correlations}).

\paragraph{\tpr Dependences}
We have investigated the production characteristics of the \threePi
final state by studying the \tpr dependence for the overall data
sample as a function of \mThreePi as well as for individual
partial-wave intensities in $3\pi$ mass regions around known
resonances.  The fits to the overall \tpr spectra require two
exponential functions in order to describe the fall-off with \tpr (see
\cref{fig:tprim_spectrum_lowM,fig:tprim_spectrum_highM}).  The slopes
of the two exponentials and their relative contributions change with
increasing $3\pi$ mass, leveling off for
$\mThreePi \gtrsim \SI{1.3}{\GeVcc}$ (see
\cref{fig:t-slopes_overall}).

The slope parameters for individual waves in $3\pi$ mass regions
around resonances exhibit a complex pattern.  Qualitatively, mass
regions with strong nonresonant contributions are characterized by a
steep drop-off with \tpr and thus larger values for the slope
parameter up to \SI{22}{\perGeVcsq}.  Considerable deviations from the
single-exponential behavior are observed for mass regions around the
\Ppi[1300] in the \wave{0}{-+}{0}{+}{\pipiS}{S} wave and around the
\PaOne in \wave{1}{++}{1}{+}{\Prho}{S}.  In these two waves, we find a
minimum of the intensity at values of \tpr of about \SI{0.4}{\GeVcsq}
(\cref{fig:t_pi_S_m1}) and \SI{0.6}{\GeVcsq} (\cref{fig:t_a1_m1}),
respectively, which may be attributed to interference effects of
different production processes.  Other distributions can be described
well by only a single exponential, a parametrization also employed for
the fit of the low-\tpr region in the case of dip structures.  Mass
regions dominated by resonances show typically a shallower drop-off
with slope parameters between \SIlist{7;11}{\perGeVcsq}.  However,
these regions are often better described by a double-exponential
model.  Hence the observation of a steep component does not exclude a
dominant resonant contribution.

For mass regions with clear resonance signals, \eg \PaTwo and \PaFour,
slope parameters are found to be similar for different waves belonging
to the same \JPC, even with different spin projections $M$.  We have
studied the production of waves with $\JPC = \onePP$, \twoMP, and
\twoPP with different $M$.  We observe a reduction in their production
rate by about an order of magnitude with every unit of $M$.  For the
\PaTwo, the intensity ratio for the two spin projections is in good
agreement with the one observed in the $\eta\,\pi^-$ decay
channel~\cite{Adolph:2014rpp}.  At the same time, we confirm that the
\tpr dependences follow the theoretically expected suppression factor
$(\tpr)^M$ at small values of \tpr for $M = \numlist{0;1;2}$ (see
\cref{fig:t_dependence_m0,fig:t_dependence_2pp}).  This observation
points to the spin characteristics of the Pomeron exchange, which is
dominant here.

The \wave{1}{++}{0}{+}{\PfZero[980]}{P} wave is of particular
interest.  In the mass region of the new \PaOne[1420], this wave
exhibits a nearly exponential \tpr spectrum with a slope parameter of
about \SI{11}{\perGeVcsq}, which is similar to that of the \Ppi[1800]
in the same decay mode.  This supports the resonance interpretation of
the \PaOne[1420] signal.  The slope is in agreement with the slope
parameter of about \SI{10}{\perGeVcsq} that was extracted for the
\PaOne[1420] in a mass-dependent fit~\cite{Adolph:2015pws}.

The \twoMP waves show no clear pattern.  Single-exponential fits in
different decay modes around the \PpiTwo and \PpiTwo[1880] give slope
values between \SIlist{6;11}{\perGeVcsq}.  For those $2^{-+}$ waves
that are better described by two exponentials, the dominant slope has
a similar range.

\paragraph{\pipi $S$-Wave Amplitudes}
For the first time, a detailed study of the amplitude of the \twoPi
$S$-wave isobar with $\IGJPC = 0^+\,0^{++}$ in the decay of the
\threePi system was performed.  This was achieved by using the
\emph{freed-isobar} technique (see
\cref{sec:results_free_pipi_s_wave}).  The $2\pi$ amplitudes are
extracted independently for different $3\pi$ partial waves in each bin
of \mThreePi and \tpr.  We have presented the correlations of the
intensities of the independent freed \twoPi isobar amplitudes with
those of the $3\pi$ system for $0^{-+}$, $1^{++}$, and $2^{-+}$
three-pion \JPC quantum numbers.  These correlations reveal a
selective coupling of $3\pi$ resonances to the scalar isobars
\PfZero[980] and \PfZero[1500] and less clear correlations with a
broad \pipiSW component.  The new method does not only yield the
two-dimensional intensity distribution, but also provides information
about the full $2\pi$ amplitude for each \mThreePi bin.  In the
corresponding \Argands, signals for \PfZero[980] and \PfZero[1500]
show up as semicircular structures with rapid counter-clockwise motion
with increasing \mTwoPi.  In the three waves studied, there is no
evidence for a distinct \PfZero[1370] resonance in the \twoPi
subsystem.

For $\JPC = \zeroMP$ and \onePP of the $3\pi$ system, the \mThreePi
spectra connected to the broad component of the \pipiSW show
enhancements around $\mThreePi = \SI{1.2}{\GeVcc}$, which might
na\"ively be interpreted as \PaOne and \Ppi[1300], respectively.  These
structures significantly change their shape as a function of \tpr,
thereby suggesting that they are influenced by nonresonant processes.
For the \Ppi[1800], we observe a coupling to $\PfZero[980]\, \pi$ and
a somewhat weaker one to $\PfZero[1500]\, \pi$.  The \Argand shows
clear semicircular structures corresponding to the \PfZero[980] and
\PfZero[1500].  Similarly, in the $2^{-+}$ $3\pi$ wave, the coupling
of the \PpiTwo[1880] to $\PfZero[980]\,\pi$ and $\PfZero[1500]\,\pi$
is seen.  For the $3\pi$ wave with $\JPC = \onePP$, we observe a clear
correlation of the \PfZero[980] isobar with the new \PaOne[1420]
resonance~\cite{Adolph:2015pws} in all bins of \tpr.  This is in
contrast to the broad component of the \pipiSW, which shows a strongly
\tpr-dependent correlation with \mThreePi and a shift of the intensity
maximum towards higher values of \mThreePi with increasing \tpr.  A
possible explanation of this shift is the Deck process.  At large
values of \tpr, the rapidity gap between the \pipiS system and the
bachelor pion is increased (see \cref{fig:deck_process}), which leads
to higher $3\pi$ masses.  The shift of intensity across the $3\pi$
mass spectrum with \tpr could explain the complicated behavior of some
\tpr spectra (see \cref{sec:tprim_dependence}).

Based on the analysis described in this paper, we extracted the
properties of resonances and of nonresonant contributions as well as
their production characteristics, which will be described in detail in
a forthcoming paper~\cite{COMPASS_3pi_mass_dep_fit}.

%
%
\clearpage
\appendix
%
%
%

\section{Wave Set}
\label{sec:appendix_wave_set}

\Cref{tab:waveset} lists the wave set used for the mass-independent
fit.  Note that in the reflectivity basis $\JPC = (\text{even})^{++}$
waves with $\Mrefl = 0^+$ are mathematically forbidden [see
\cref{eq:d_func_refl_basis} on page~\pageref{eq:d_func_refl_basis}].

\begin{table*}[p]
  \centering
  \caption{Wave set used for mass-independent fit: 80~waves with
    positive reflectivity, 7~with negative, plus an incoherent
    isotropic wave.}
  \label{tab:waveset}
  \ifMultiColumnLayout{}{\begin{scriptsize}}
    \subfloat{%
      \renewcommand{\arraystretch}{1.2}
      \begin{tabular}[t]{cccc}
        \toprule
        \textbf{\JPCMrefl} & \textbf{Isobar} & \textbf{$L$} & \textbf{Threshold [\si{\MeVcc}]} \\
        \midrule

        $0^{-+}\,0^+$ & \pipiS        & $S$ & --- \\
        $0^{-+}\,0^+$ & \Prho         & $P$ & --- \\
        $0^{-+}\,0^+$ & \PfZero       & $S$ & 1200 \\
        $0^{-+}\,0^+$ & \PfTwo        & $D$ & --- \\
        $0^{-+}\,0^+$ & \PfZero[1500] & $S$ & 1700 \\

        \midrule

        $1^{++}\,0^+$ & \pipiS     & $P$ & --- \\
        $1^{++}\,1^+$ & \pipiS     & $P$ & 1100 \\
        $1^{++}\,0^+$ & \Prho      & $S$ & --- \\
        $1^{++}\,1^+$ & \Prho      & $S$ & --- \\
        $1^{++}\,0^+$ & \Prho      & $D$ & --- \\
        $1^{++}\,1^+$ & \Prho      & $D$ & --- \\
        $1^{++}\,0^+$ & \PfZero    & $P$ & 1180 \\
        $1^{++}\,1^+$ & \PfZero    & $P$ & 1140 \\
        $1^{++}\,0^+$ & \PfTwo     & $P$ & 1220 \\
        $1^{++}\,1^+$ & \PfTwo     & $P$ & --- \\
        $1^{++}\,0^+$ & \PfTwo     & $F$ & --- \\
        $1^{++}\,0^+$ & \PrhoThree & $D$ & --- \\
        $1^{++}\,0^+$ & \PrhoThree & $G$ & --- \\

        \midrule

        $1^{-+}\,1^+$ & \Prho & $P$ & --- \\

        \midrule

        $2^{++}\,1^+$ & \Prho      & $D$ & --- \\
        $2^{++}\,2^+$ & \Prho      & $D$ & --- \\
        $2^{++}\,1^+$ & \PfTwo     & $P$ & 1000 \\
        $2^{++}\,2^+$ & \PfTwo     & $P$ & 1400 \\
        $2^{++}\,1^+$ & \PrhoThree & $D$ & 800 \\

        \midrule

        $2^{-+}\,0^+$ & \pipiS     & $D$ & --- \\
        $2^{-+}\,1^+$ & \pipiS     & $D$ & --- \\
        $2^{-+}\,0^+$ & \Prho      & $P$ & --- \\
        $2^{-+}\,1^+$ & \Prho      & $P$ & --- \\
        $2^{-+}\,2^+$ & \Prho      & $P$ & --- \\
        $2^{-+}\,0^+$ & \Prho      & $F$ & --- \\
        $2^{-+}\,1^+$ & \Prho      & $F$ & --- \\
        $2^{-+}\,0^+$ & \PfZero    & $D$ & 1160 \\
        $2^{-+}\,0^+$ & \PfTwo     & $S$ & --- \\
        $2^{-+}\,1^+$ & \PfTwo     & $S$ & 1100 \\
        $2^{-+}\,2^+$ & \PfTwo     & $S$ & --- \\
        $2^{-+}\,0^+$ & \PfTwo     & $D$ & --- \\
        $2^{-+}\,1^+$ & \PfTwo     & $D$ & --- \\
        $2^{-+}\,2^+$ & \PfTwo     & $D$ & --- \\
        $2^{-+}\,0^+$ & \PfTwo     & $G$ & --- \\
        $2^{-+}\,0^+$ & \PrhoThree & $P$ & 1000 \\
        $2^{-+}\,1^+$ & \PrhoThree & $P$ & 1300 \\

        \midrule

        $3^{++}\,0^+$ & \pipiS     & $F$ & 1380 \\
        $3^{++}\,1^+$ & \pipiS     & $F$ & 1380 \\
        $3^{++}\,0^+$ & \Prho      & $D$ & --- \\
        $3^{++}\,1^+$ & \Prho      & $D$ & --- \\

        \bottomrule
      \end{tabular}
    }%
    \ifMultiColumnLayout{\hspace{0.15\textwidth}}{\hspace{0.02\textwidth}}
    \subfloat{%
      \centering
      \renewcommand{\arraystretch}{1.2}
      \begin{tabular}[t]{cccc}
        \toprule
        \textbf{\JPCMrefl} & \textbf{Isobar} & \textbf{$L$} & \textbf{Threshold [\si{\MeVcc}]} \\
        \midrule

        $3^{++}\,0^+$ & \Prho      & $G$ & --- \\
        $3^{++}\,1^+$ & \Prho      & $G$ & --- \\
        $3^{++}\,0^+$ & \PfTwo     & $P$ & 960 \\
        $3^{++}\,1^+$ & \PfTwo     & $P$ & 1140 \\
        $3^{++}\,0^+$ & \PrhoThree & $S$ & 1380 \\
        $3^{++}\,1^+$ & \PrhoThree & $S$ & 1380 \\
        $3^{++}\,0^+$ & \PrhoThree & $I$ & --- \\

        \midrule

        $3^{-+}\,1^+$ & \Prho  & $F$ & --- \\
        $3^{-+}\,1^+$ & \PfTwo & $D$ & 1340 \\

        \midrule

        $4^{++}\,1^+$ & \Prho      & $G$ & --- \\
        $4^{++}\,2^+$ & \Prho      & $G$ & --- \\
        $4^{++}\,1^+$ & \PfTwo     & $F$ & --- \\
        $4^{++}\,2^+$ & \PfTwo     & $F$ & --- \\
        $4^{++}\,1^+$ & \PrhoThree & $D$ & 1700 \\

        \midrule

        $4^{-+}\,0^+$ & \pipiS & $G$ & 1400 \\
        $4^{-+}\,0^+$ & \Prho  & $F$ & --- \\
        $4^{-+}\,1^+$ & \Prho  & $F$ & --- \\
        $4^{-+}\,0^+$ & \PfTwo & $D$ & --- \\
        $4^{-+}\,1^+$ & \PfTwo & $D$ & --- \\
        $4^{-+}\,0^+$ & \PfTwo & $G$ & 1600 \\

        \midrule

        $5^{++}\,0^+$ & \pipiS     & $H$ & --- \\
        $5^{++}\,1^+$ & \pipiS     & $H$ & --- \\
        $5^{++}\,0^+$ & \Prho      & $G$ & --- \\
        $5^{++}\,0^+$ & \PfTwo     & $F$ & 980 \\
        $5^{++}\,1^+$ & \PfTwo     & $F$ & --- \\
        $5^{++}\,0^+$ & \PfTwo     & $H$ & --- \\
        $5^{++}\,0^+$ & \PrhoThree & $D$ & 1360 \\

        \midrule

        $6^{++}\,1^+$ & \Prho  & $I$ & --- \\
        $6^{++}\,1^+$ & \PfTwo & $H$ & --- \\

        \midrule

        $6^{-+}\,0^+$ & \pipiS     & $I$ & --- \\
        $6^{-+}\,1^+$ & \pipiS     & $I$ & --- \\
        $6^{-+}\,0^+$ & \Prho      & $H$ & --- \\
        $6^{-+}\,1^+$ & \Prho      & $H$ & --- \\
        $6^{-+}\,0^+$ & \PfTwo     & $G$ & --- \\
        $6^{-+}\,0^+$ & \PrhoThree & $F$ & --- \\

        \midrule

        $1^{++}\,1^-$ & \Prho & $S$ & --- \\

        \midrule

        $1^{-+}\,0^-$ & \Prho & $P$ & --- \\
        $1^{-+}\,1^-$ & \Prho & $P$ & --- \\

        \midrule

        $2^{++}\,0^-$ & \Prho  & $D$ & --- \\
        $2^{++}\,0^-$ & \PfTwo & $P$ & 1180 \\
        $2^{++}\,1^-$ & \PfTwo & $P$ & 1300 \\

        \midrule

        $2^{-+}\,1^-$ & \PfTwo & $S$ & --- \\

        \midrule

        Flat & & & --- \\

        \bottomrule
      \end{tabular}
    }%
  \ifMultiColumnLayout{}{\end{scriptsize}}
\end{table*}
 %
%
%

\section{Systematic Studies of Partial-Wave Analysis Model}
\label{sec:appendix_syst_studies_massindep}

\subsection{Rank of Spin-Density Matrix}
\label{sec:appendix_syst_studies_massindep_rank}

As pointed out in \cref{sec:pwa_decomp}, ranks $N_r > 1$ of the
spin-density matrix provide a way of modeling incoherences between
partial waves.  This is done by introducing additional sets of
transition amplitudes.  These sets are assumed to correspond to
different noninterfering production processes, each with its own
production phase.  By performing the analysis in bins of \tpr, it was
found that $N_r = 1$ is sufficient for positive-reflectivity waves.
This also leads to higher stability of the mass-independent fits.

In \cref{fig:major_waves_syst_rank2}, we show in red the intensities of
selected partial waves obtained from fits with rank~2 for the positive
and negative-reflectivity waves.  This is compared to the standard fit
(blue data points), where rank~2 was used only for waves with
$\refl = -1$.  In the rank-2 fit, the flat wave disappears completely
and the intensity of the negative-reflectivity waves is approximately
halved.  Slight modifications of the shape of resonance structures are
observed in some partial waves.  Several partial waves exhibit
artificial peak structures in the \SIrange{1.0}{1.3}{\GeVcc} $3\pi$
mass region, like \eg shown in \cref{fig:pipi_waves_syst_rank2_4pp}.
Altogether, we prefer to use rank~1 for the positive reflectivity
waves.

\begin{figure*}[htbp]
  \centering
  \subfloat[][]{%
    \includegraphics[width=\twoPlotWidth]{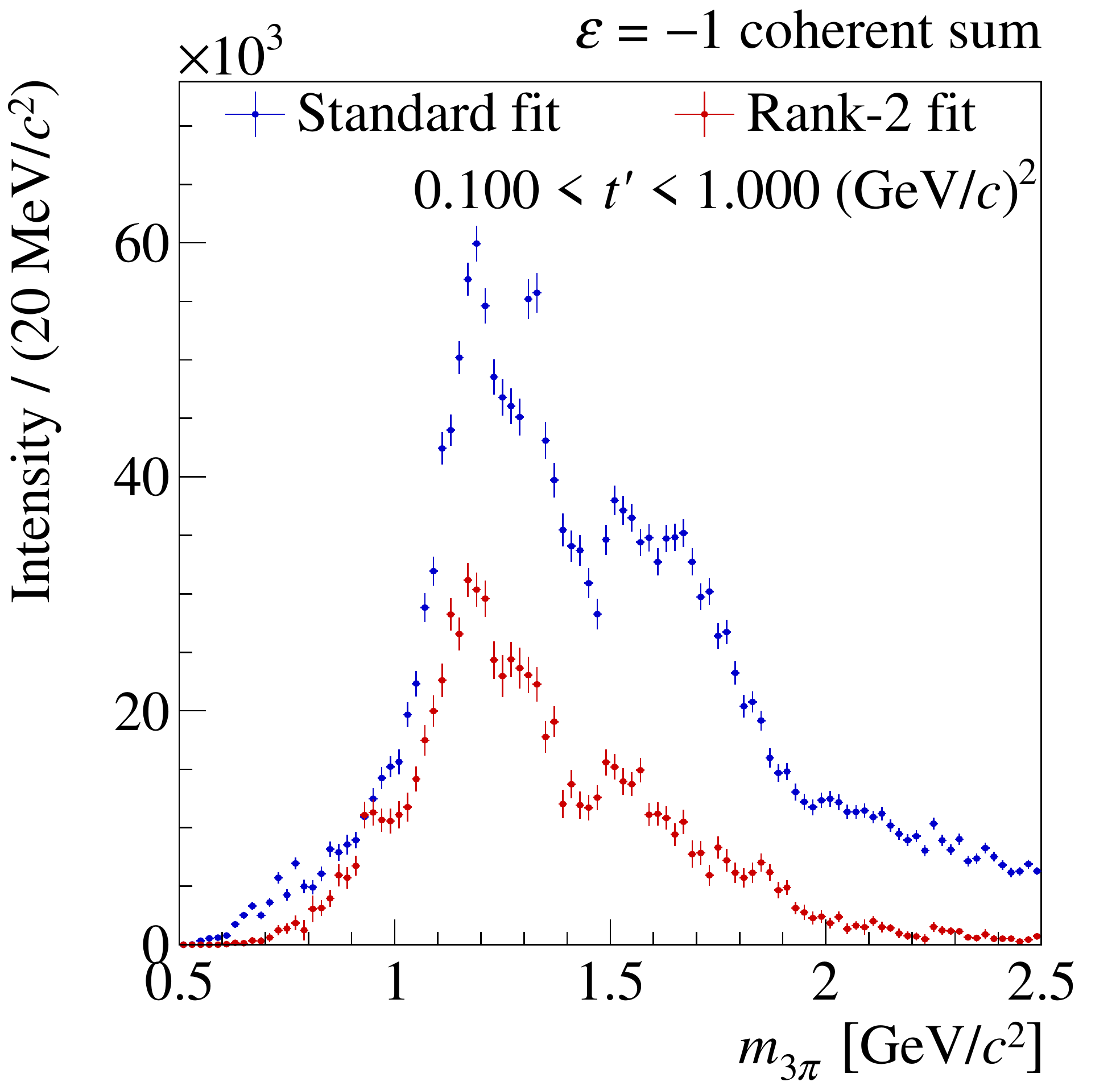}%
  }%
  \hspace*{\twoPlotSpacing}
  \subfloat[][]{%
    \includegraphics[width=\twoPlotWidth]{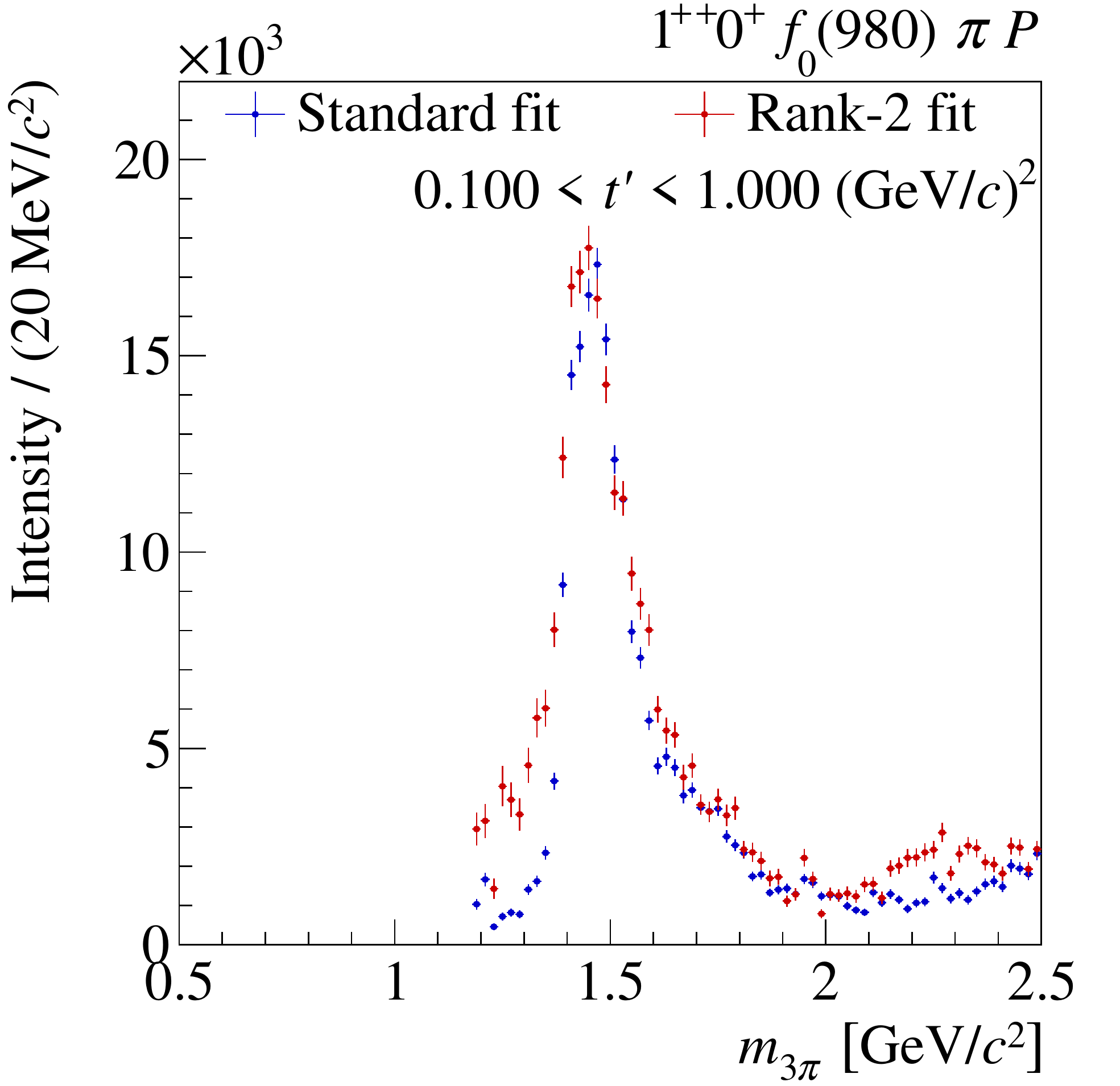}%
  }%
  \\
  \subfloat[][]{%
    \includegraphics[width=\twoPlotWidth]{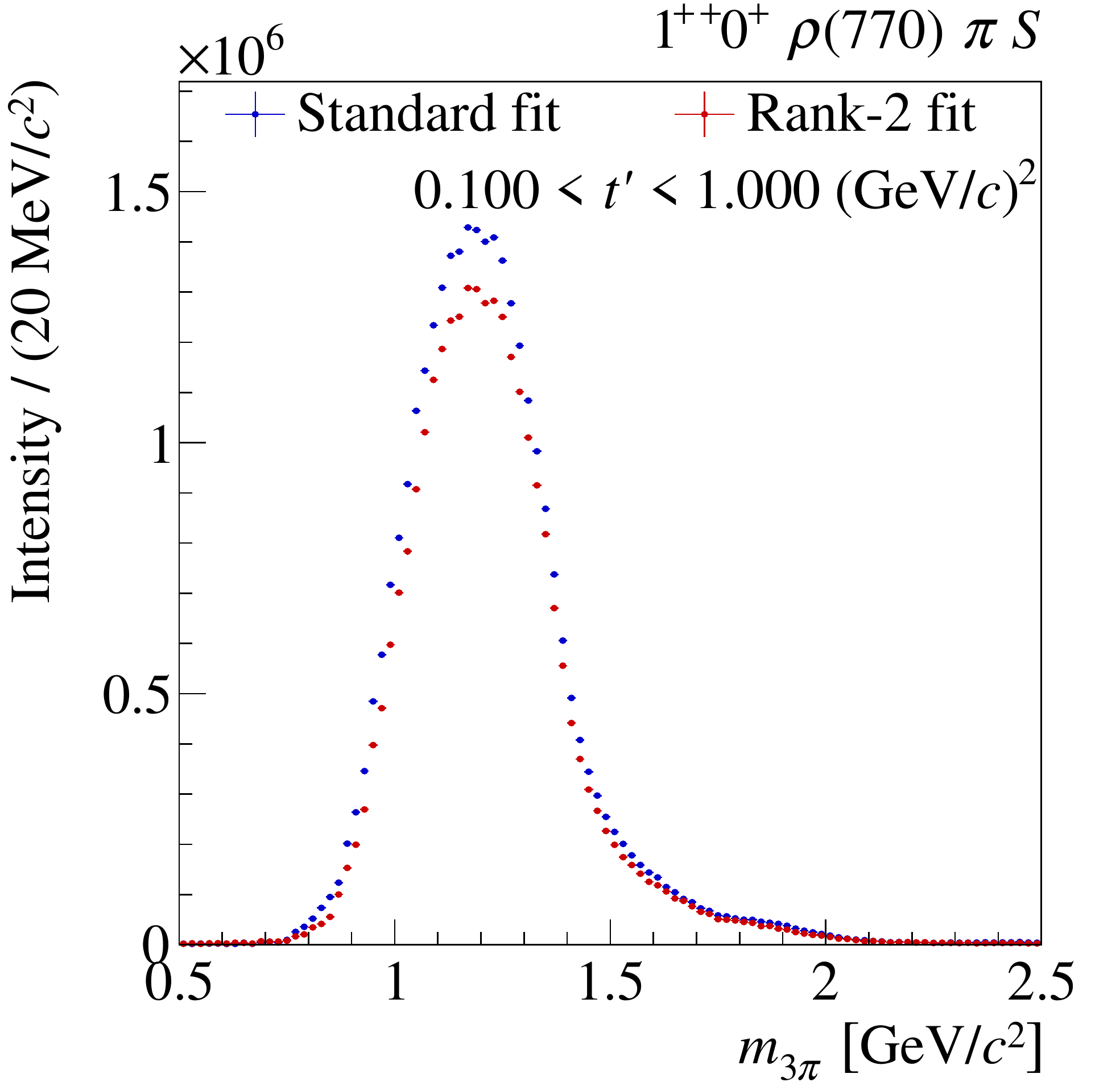}%
  }%
  \hspace*{\twoPlotSpacing}
  \subfloat[][]{%
    \label{fig:pipi_waves_syst_rank2_4pp}%
    \includegraphics[width=\twoPlotWidth]{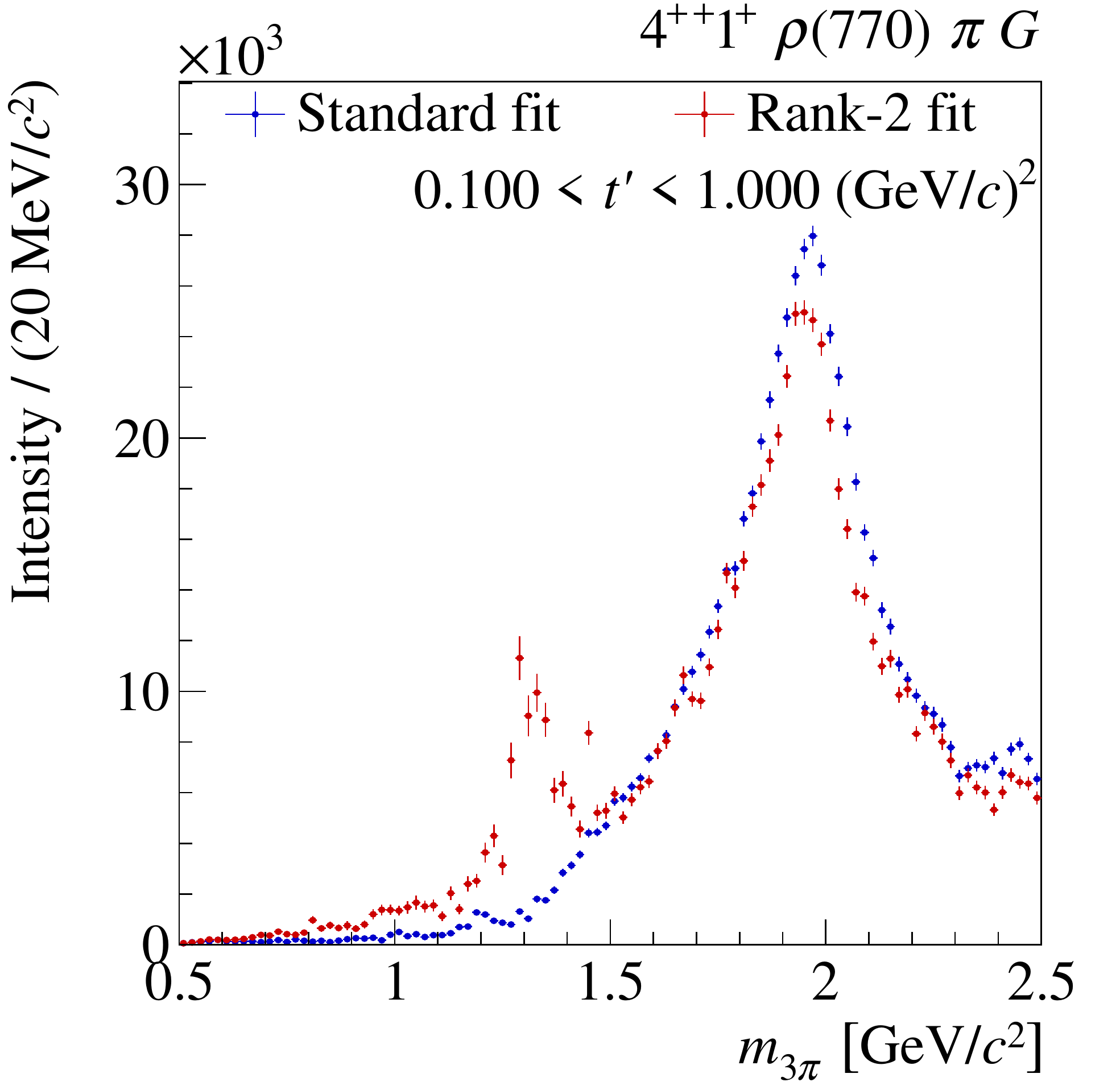}%
  }%
  \caption{\colorPlot Comparison of \tpr-summed partial-wave
    intensities obtained from the standard mass-independent fit with
    rank $N_r = 1$ of the spin-density matrix for waves with
    $\refl = +1$ and rank~2 for the ones with $\refl = -1$
    (blue/black) with the intensities from a fit with rank~2 for both
    sectors (red/gray).}
  \label{fig:major_waves_syst_rank2}
\end{figure*}

\subsection{Omission of Waves with Negative Reflectivity}
\label{sec:appendix_syst_studies_massindep_neg_refl}

The PWA model defined in \cref{sec:pwa_method} has two types of
incoherent contributions, rank and reflectivity [see
\cref{eq:intensity_bin}].  The latter one is determined by the
naturality of the exchange particle (Regge-trajectory) mediating the
scattering process.  Including $\refl = -1$ partial waves, we
effectively allow for the exchange of Reggeons other than the Pomeron,
which is expected to be suppressed at beam energies of \SI{190}{\GeV}.
In our PWA model, we have included seven waves with negative
reflectivity (see \cref{tab:waveset} in \cref{sec:appendix_wave_set}).
Negative and positive-reflectivity waves have different angular
distributions.  In order to study how well the fit is able to separate
the two sectors, we have performed fits without any $\refl = -1$
waves.  The result is shown in red in \cref{fig:waves_syst_noNegWaves}
for two selected waves.  With the exception of the flat wave, the
intensities of all waves stay practically unaltered.  This
demonstrates that the positive and negative-reflectivity sectors are
well separated by the analysis method.

\begin{figure*}[htbp]
  \centering
  \subfloat[][]{%
    \includegraphics[width=\twoPlotWidth]{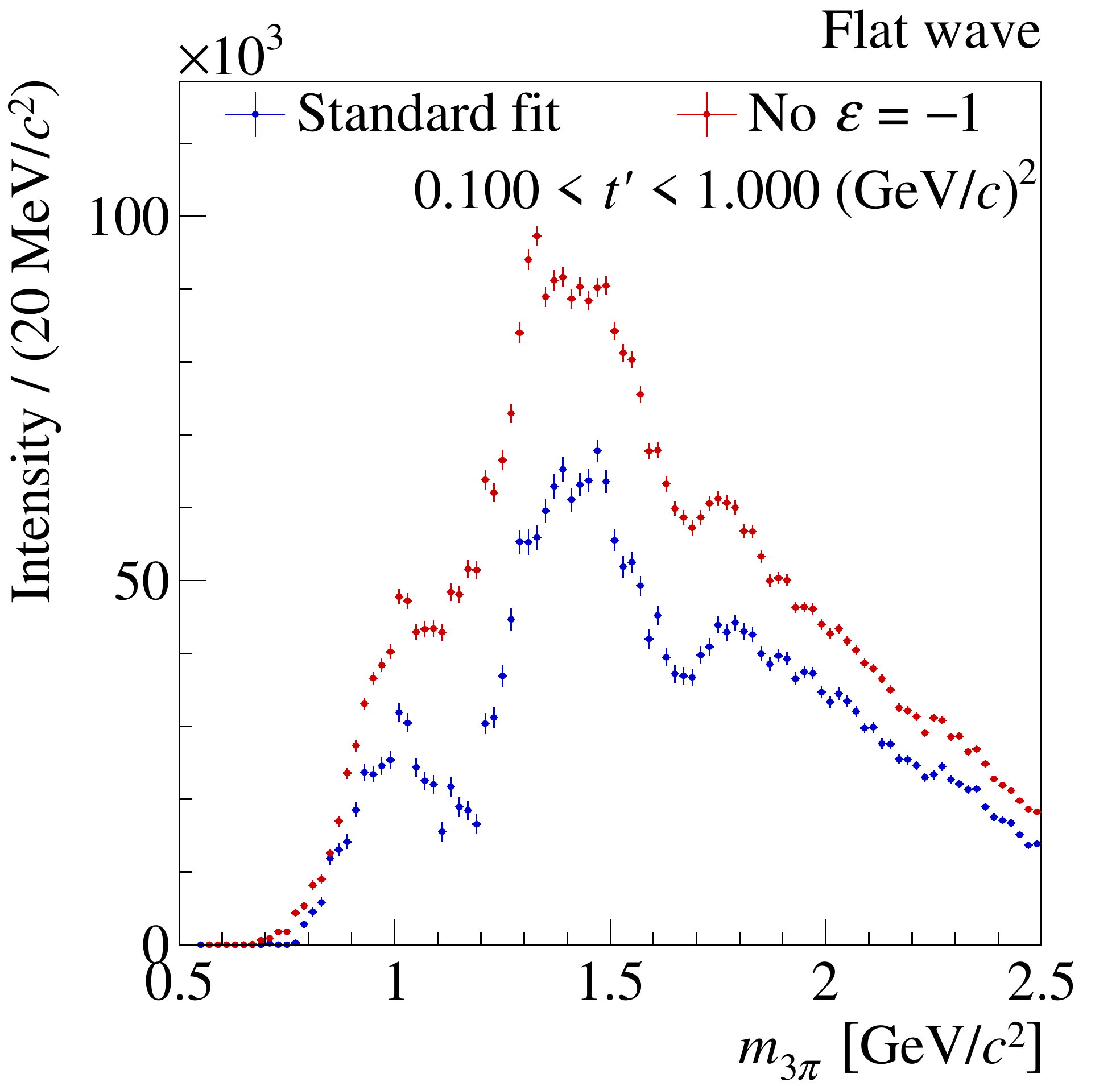}%
  }%
  \hspace*{\twoPlotSpacing}%
  \subfloat[][]{%
    \includegraphics[width=\twoPlotWidth]{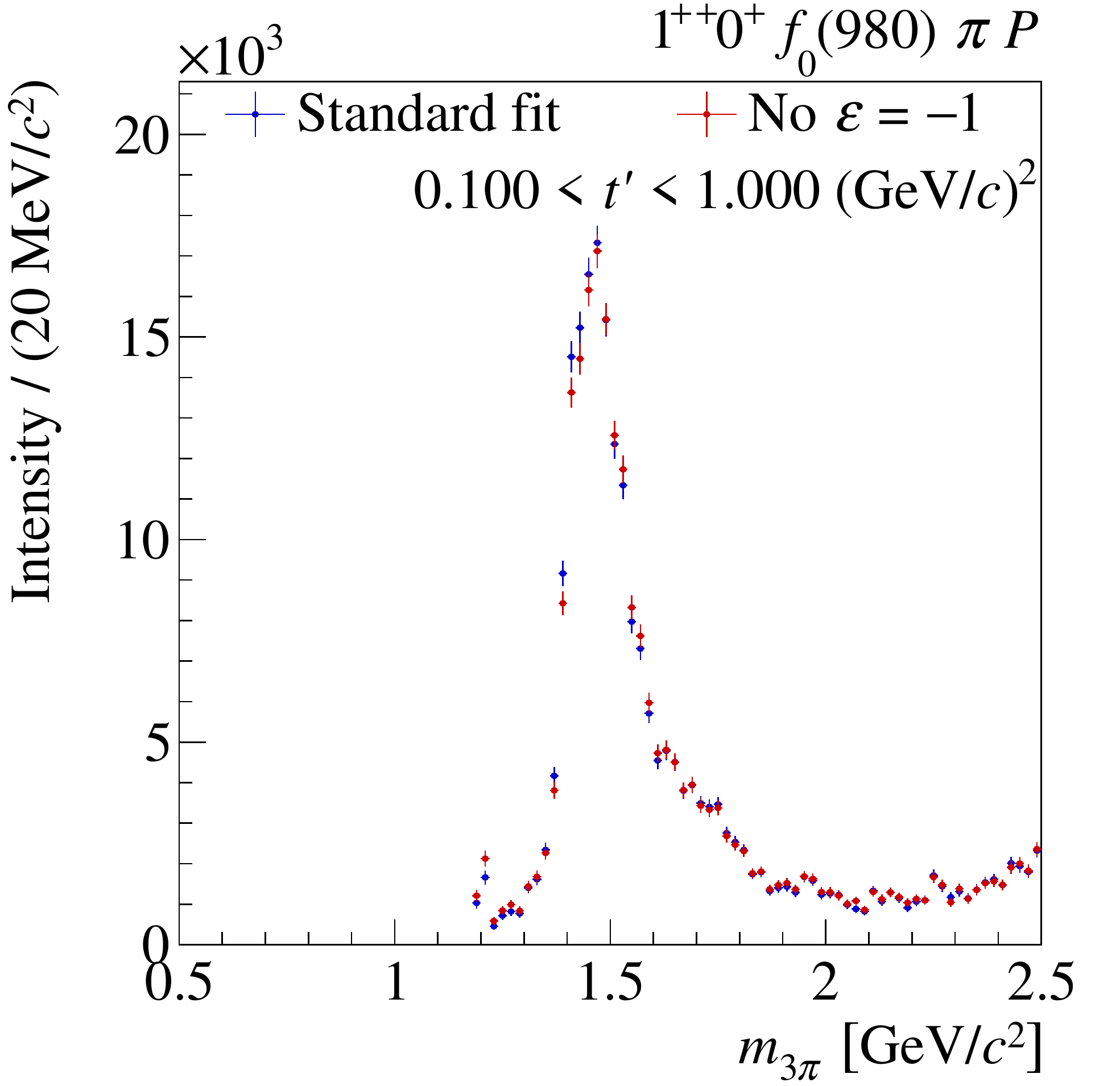}%
  }%
  \caption{\colorPlot Comparison of \tpr-summed partial-wave
    intensities obtained from the mass-independent fits with
    (blue/black) and without (red/gray) $\refl = -1$ waves.}
  \label{fig:waves_syst_noNegWaves}
\end{figure*}

\subsection{Variation of the Isobar Parametrization}
\label{sec:appendix_syst_studies_massindep_isobar_param}

In the employed PWA method, the $X^-$ decay amplitudes
$\Psi_a^\refl(\tau)$ [see \cref{eq:intensity_bin}] are not allowed to
have any free parameter.  Fixed parametrizations for the isobar
amplitudes $\Delta_\xi(m_\xi)$ [see
\cref{eq:decay_amplitude_isobar_dyn}] have to be used, which are taken
from literature (see \cref{tab:isobar_param}).  While eventually these
parametrizations could be extracted from our data following the
analysis scheme outlined in \cref{sec:results_free_pipi_s_wave}, for
this paper we still use the conventional approach.

The \Prho is the dominant isobar.  As discussed in
\cref{sec:pwa_method_isobar_parametrization}, different Breit-Wigner
parametrizations exist for the \Prho.  Using \cref{eq:massDepWidth}
instead of \cref{eq:massDepWidthRho} for the mass-dependent width
$\Gamma(m)$ of the \Prho changes the intensity of the structure in
the \PaOne mass region in the \wave{1}{++}{0}{+}{\pipiS}{P} wave and
that of the \PaTwo signal in the \wave{2}{++}{1}{+}{\PfTwo}{P} wave
(see \cref{fig:waves_syst_rho}).  Both structures seem to be
contaminated by model leakage from the respective dominant
$\Prho\,\pi$ decay modes.  The other 16~waves listed in
\cref{tab:selected-waves} remain practically unchanged.  When using
in addition the PDG averages for the \Prho parameters of
$m_0 = \SI{775.26}{\MeVcc}$ and
$\Gamma_0 = \SI{149.1}{\MeVcc}$~\cite{Agashe:2014kda}, the
log-likelihood values, summed over the 11~\tpr bins, decrease by more than
\SI{1000}{units} in the mass range between
\SIlist{0.95;1.35}{\GeVcc}.

\begin{figure*}[htbp]
  \centering
  \subfloat[][]{%
    \includegraphics[width=\twoPlotWidth]{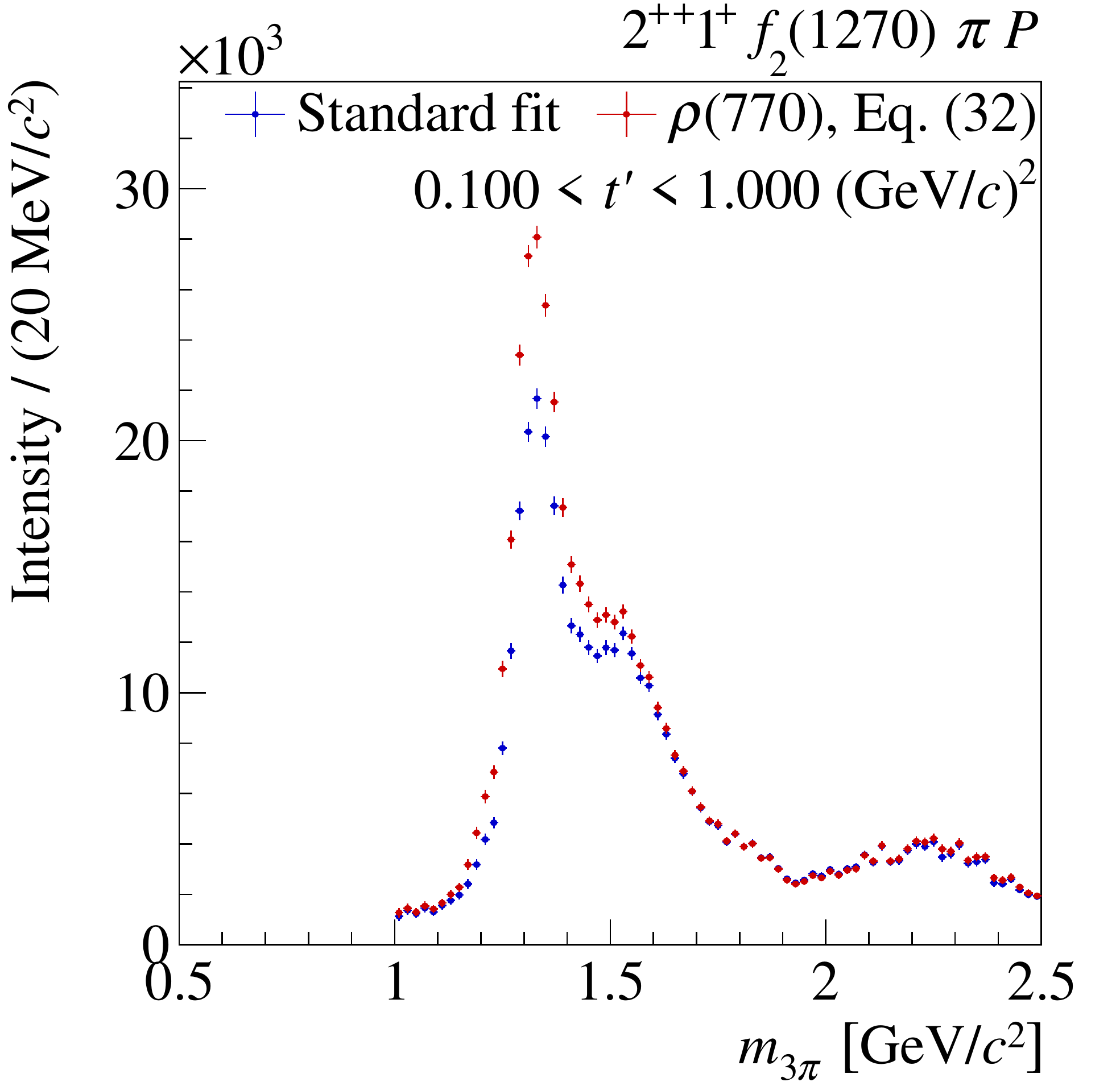}%
  }%
  \hspace*{\twoPlotSpacing}
  \subfloat[][]{%
    \includegraphics[width=\twoPlotWidth]{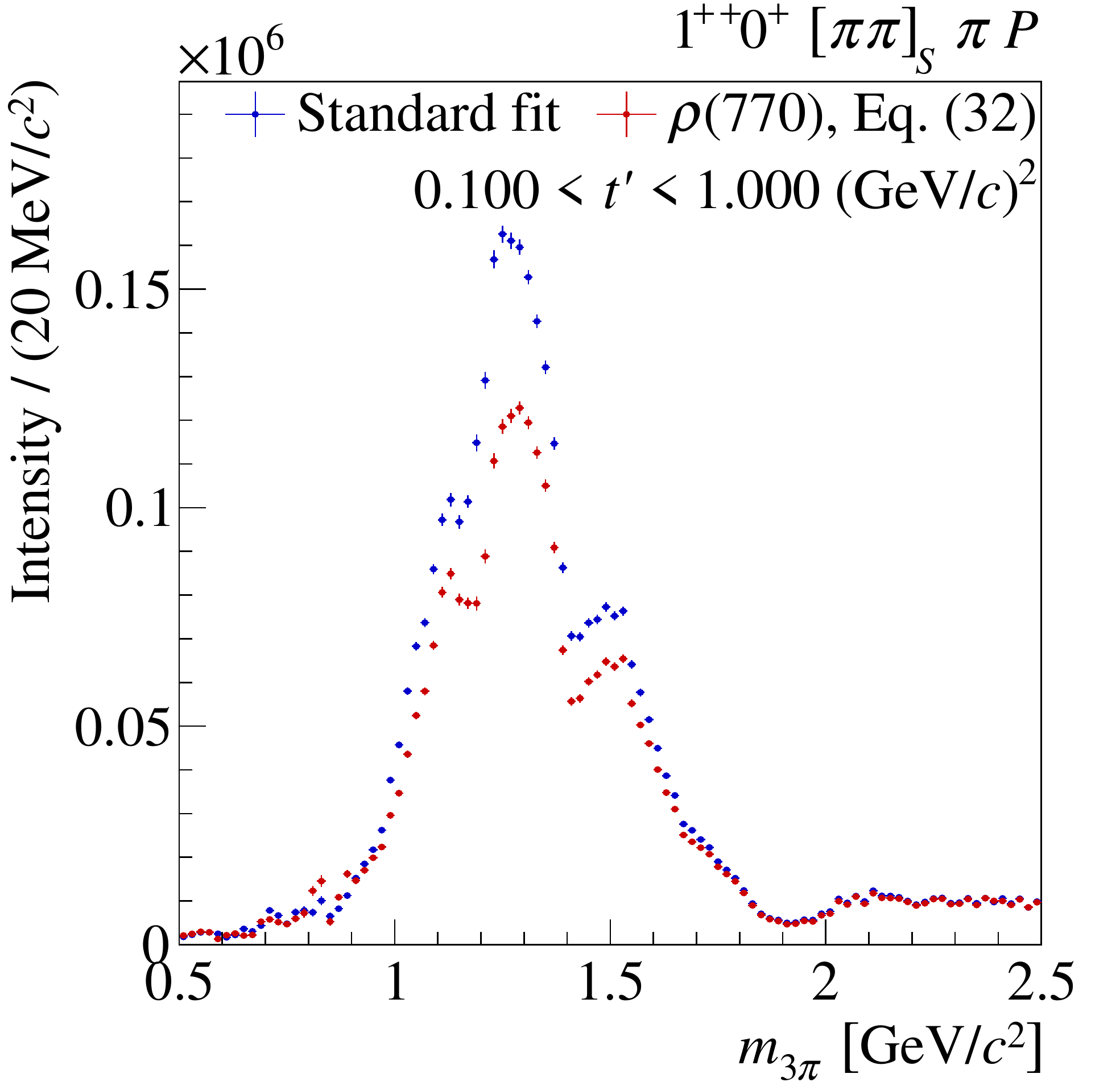}%
  }
  \caption{\colorPlot Comparison of \tpr-summed partial-wave
    intensities obtained from mass-independent fits using two
    different parametrizations for the \Prho isobar amplitude: a
    Breit-Wigner with \cref{eq:massDepWidthRho} (blue/black)
    and one with \cref{eq:massDepWidth} (red/gray).}
  \label{fig:waves_syst_rho}
\end{figure*}

We have also
investigated the sensitivity of the PWA result \wrt changes in the
parametrization of the \PfZero[980] and \pipiS isobars.
For the \PfZero[980], the Flatt\'e parametrization used in the standard
analysis is replaced by a modified $S$-wave
Breit-Wigner amplitude of the form:
\begin{equation}
  \label{eq:relBW_f0980_study}
  \Delta_{\PfZero[980]}(m; m_0, \Gamma_0) = \frac{m\, \Gamma_0}{m_0^2 - m^2 - i\, m_0\, \Gamma(m)},
\end{equation}
where
\begin{equation}
  \label{eq:massDepWidth_f0980_study}
  \Gamma(m) = \Gamma_0\, \frac{q}{q_0}.
\end{equation}
The \PfZero[980] parameters are  $m_0 = \SI{980}{\MeVcc}$
and $\Gamma_0 = \SI{40}{\MeVcc}$.  \Cref{fig:waves_syst_f0980BW} shows in red some selected
partial-wave intensities from this study.  The Breit-Wigner
parametrization has less pronounced tails and covers a narrower $2\pi$
mass range.  This leads to nearly a factor of two lower intensities in
the $\PfZero[980]\, \pi$ partial waves.  The shapes of the resonance
structures in these waves remain unaltered.  Interestingly, the
\Ppi[1800] peak in the \wave{0}{-+}{0}{+}{\pipiS}{S} wave also
decreases when the Breit-Wigner parametrization is used for the
\PfZero[980].  On the level of the mass-independent fit, this behavior
cannot be explained.  In contrast to the \Ppi[1800] peak, the
structure in the \Ppi[1300] region remains unaltered.  The fit with
the Flatt\'e parametrization has a higher likelihood than the fit with
the Breit-Wigner one.

\begin{figure*}[htbp]
  \centering
  \subfloat[][]{%
    \includegraphics[width=\twoPlotWidth]{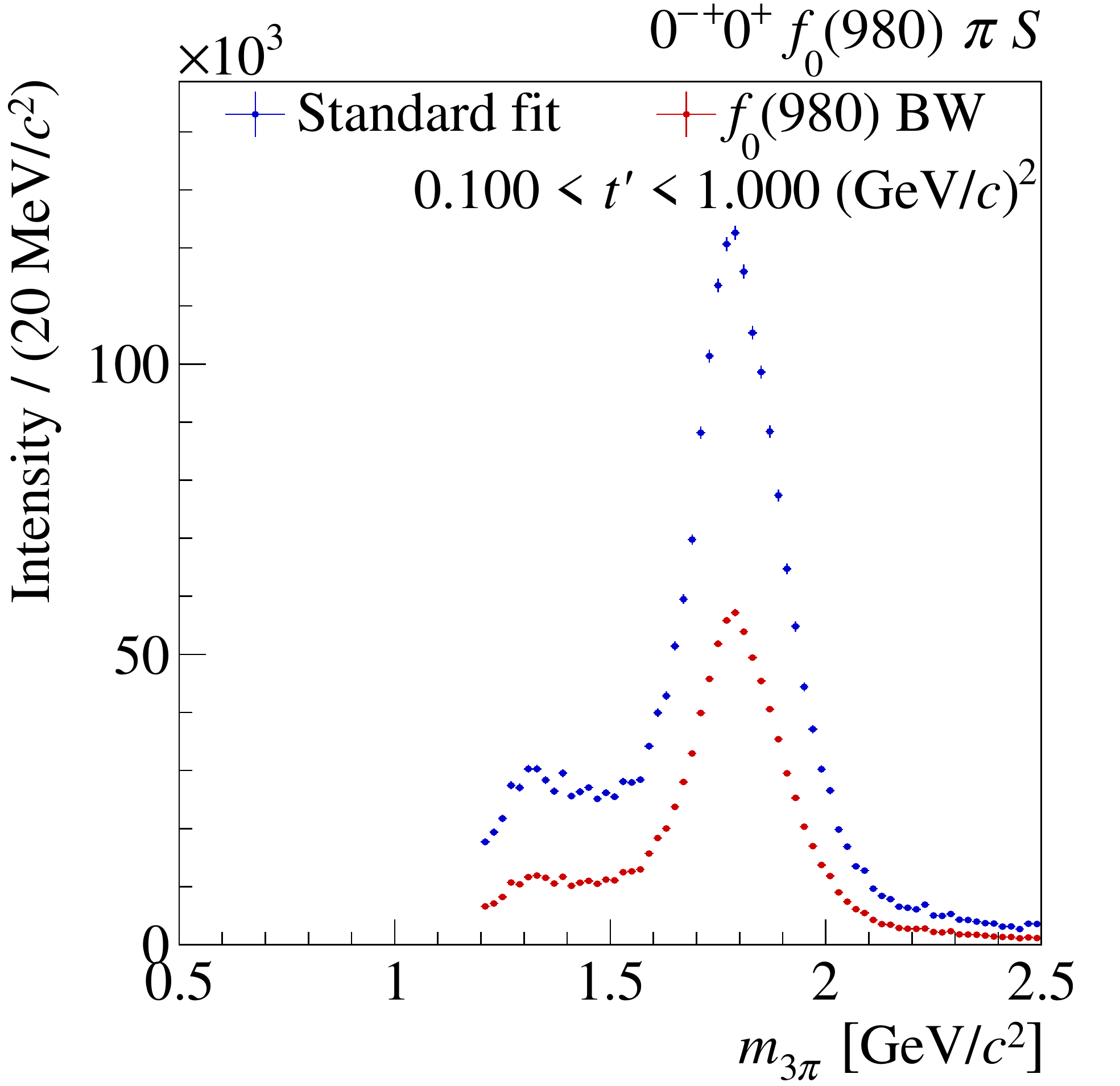}%
  }%
  \hspace*{\twoPlotSpacing}
  \subfloat[][]{%
    \includegraphics[width=\twoPlotWidth]{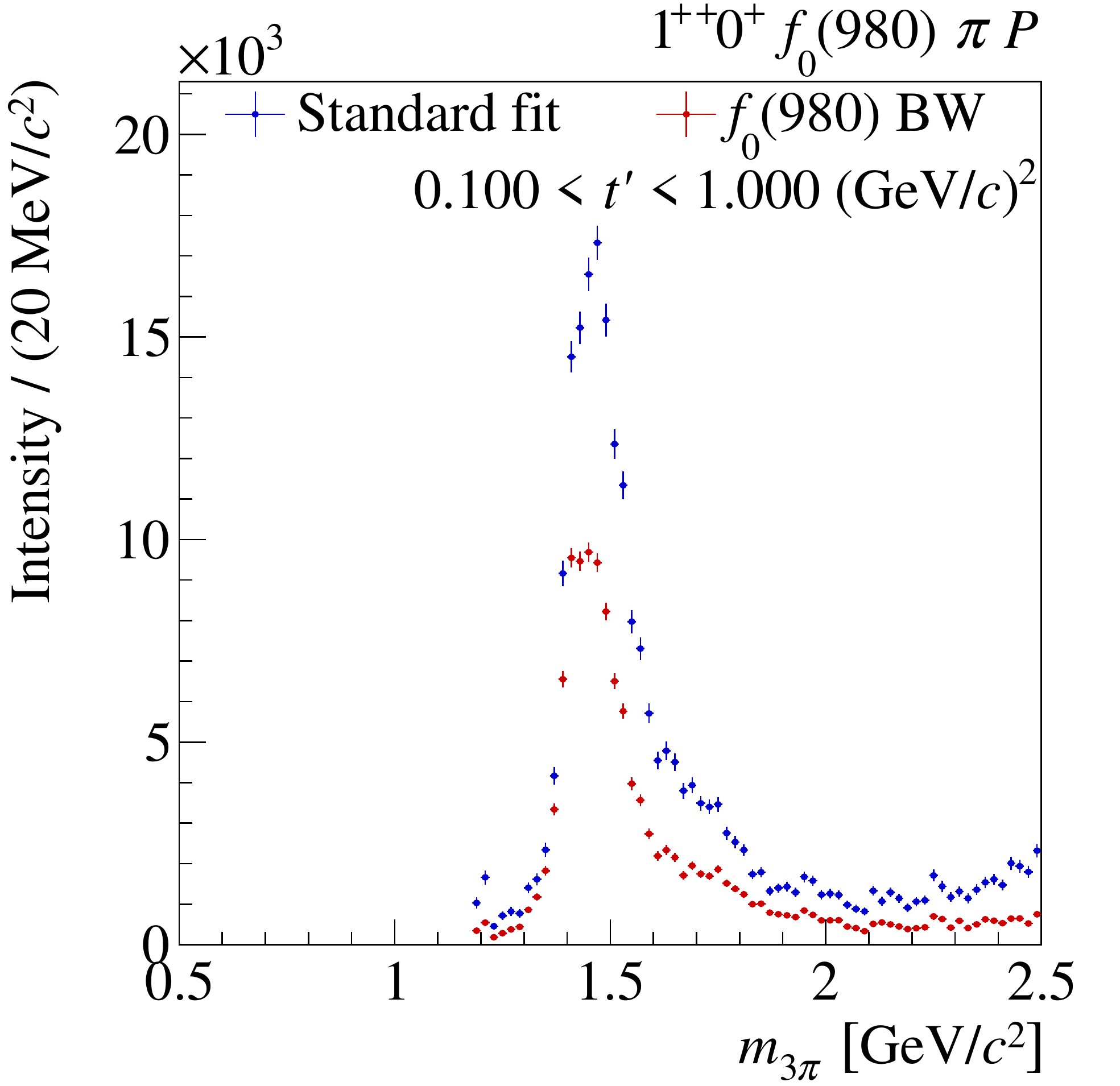}%
  }
  \\
  \subfloat[][]{%
    \includegraphics[width=\twoPlotWidth]{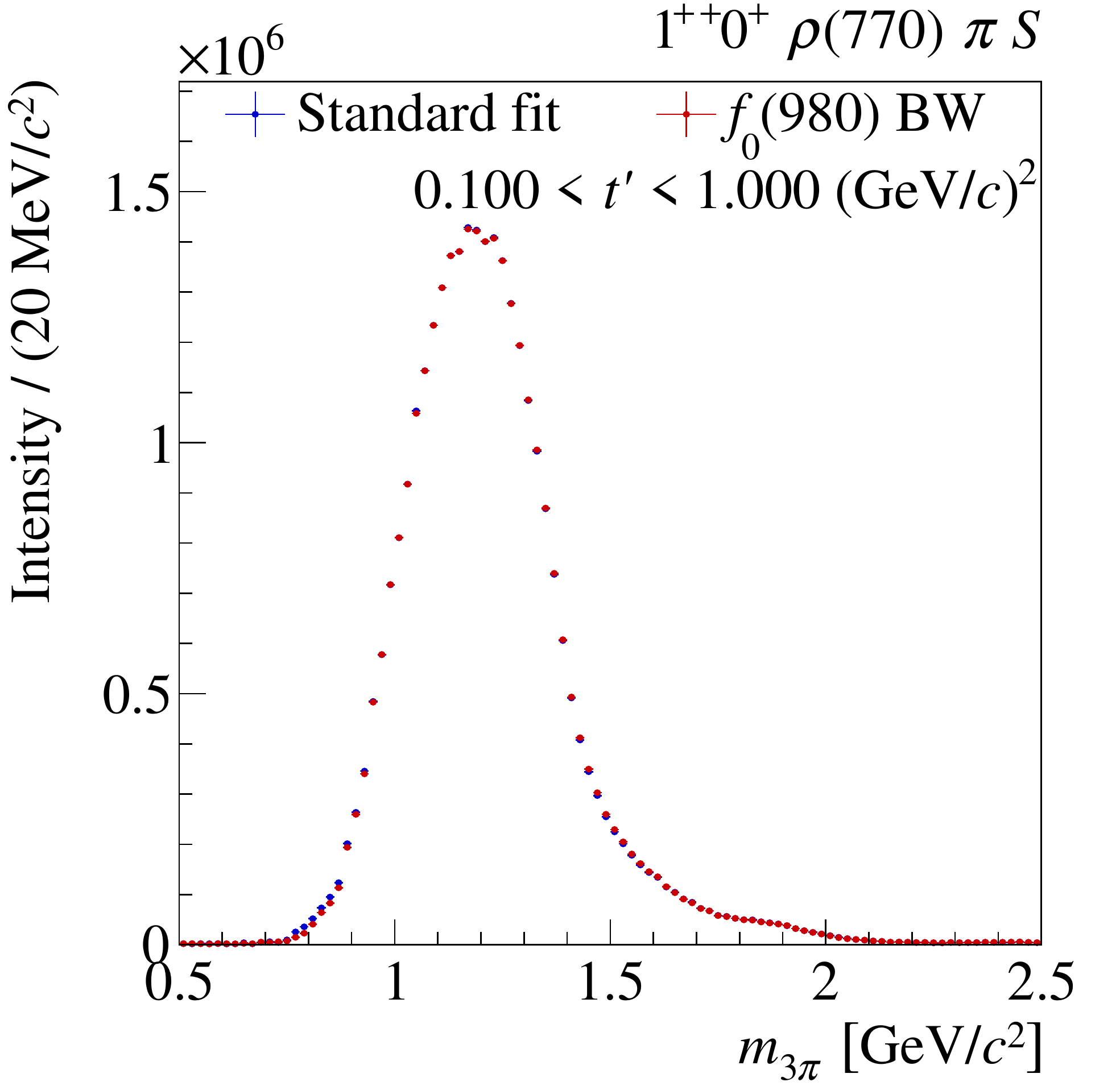}%
  }%
  \hspace*{\twoPlotSpacing}
  \subfloat[][]{%
    \includegraphics[width=\twoPlotWidth]{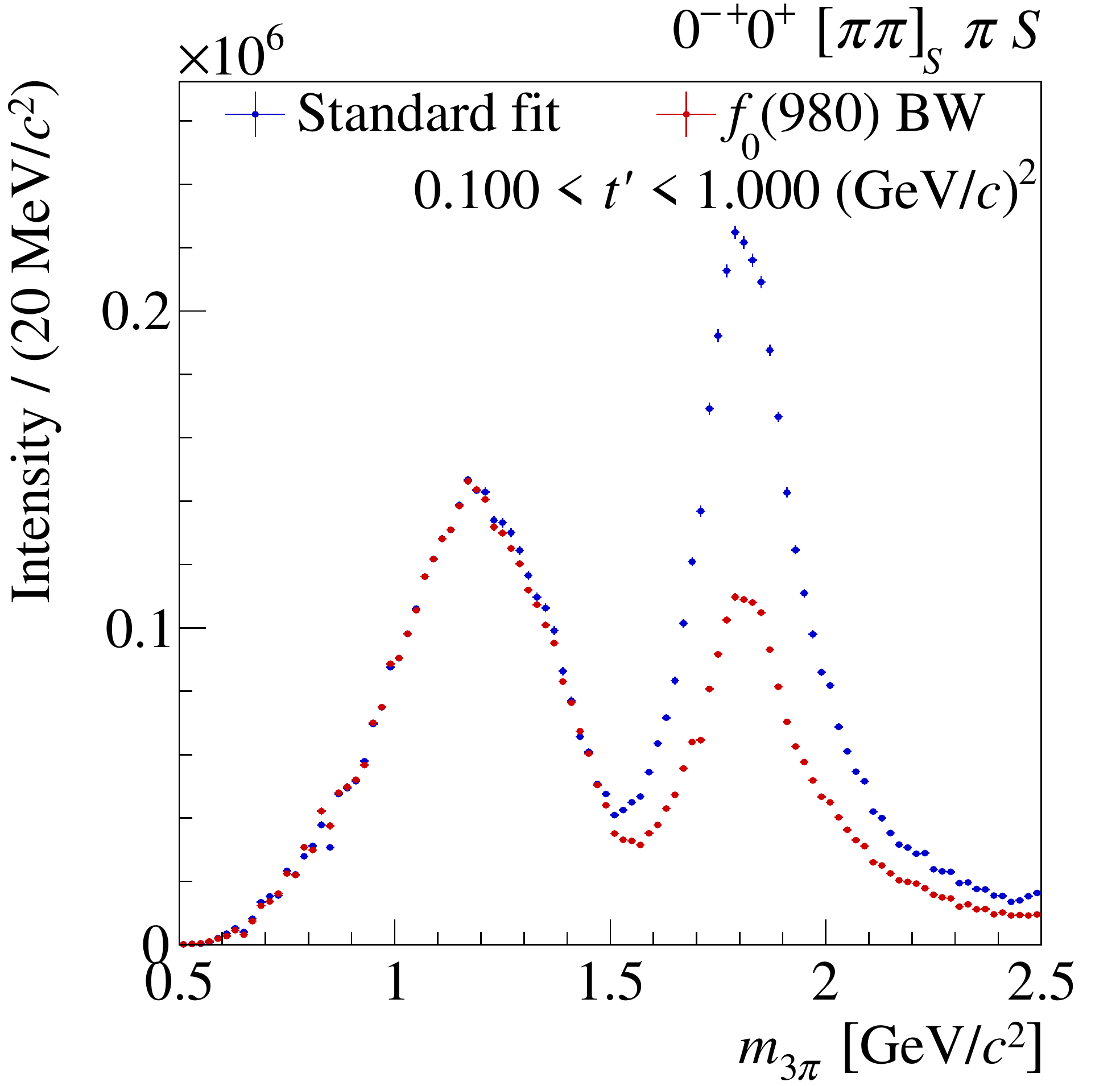}%
  }%
  \caption{\colorPlot Comparison of \tpr-summed partial-wave
    intensities obtained from mass-independent fits using two
    different parametrizations for the \PfZero[980] isobar amplitude:
    Flatt\'e parametrization [\cref{eq:f0980_flatte}, blue/black] and modified $S$-wave
    Breit-Wigner [\cref{eq:relBW_f0980_study} with \cref{eq:massDepWidth_f0980_study}, red/gray].}
  \label{fig:waves_syst_f0980BW}
\end{figure*}

Also for the broad component of the \pipiSW amplitude various
parametrizations exist.  In addition to the modified $M$~solution from
\refCite{au:1986vs}, we tried the $K_1$~solution from
\refCite{au:1986vs} with the \PfZero[980] pole subtracted, using the
modified $S$-wave Breit-Wigner amplitude of
  \cref{eq:relBW_f0980_study} with
  \cref{eq:massDepWidth_f0980_study}~\cite{amelin:1995gu}.
In order to be consistent, the same \PfZero[980] amplitude was also
used for the partial waves with the \PfZero[980] isobar.  The result is very similar to the
  one of the fit with the modified $M$~solution for
the \pipiS and the $S$-wave Breit-Wigner amplitude for the
\PfZero[980] isobar discussed above.

\subsection{Variation of Event Selection}
\label{sec:appendix_syst_studies_massindep_selection_cuts}

In order to study the potential influence of backgrounds from kaon
diffraction, kaon pairs in the final state, and central-production
reactions, the mass-independent fit was performed on a data sample, in
which the particle identification in the beam and spectrometer was not
used and the rejection of central-production events as described in
\cref{sec:event_selection_cuts} was not applied.  Therefore, possible
background contributions are expected to be enhanced in this data
sample.  The effect is shown in \cref{fig:waves_syst_event_selection}
for selected waves.  The data sample with looser cuts contains
approximately \SI{20}{\percent} more events in the analyzed range of
\mThreePi and \tpr.  Hence, the partial-wave intensities are larger
and typically scale proportionally to the sample size.  The peak
shapes of resonances are in general unaffected by the different event
selection.  The most noteworthy effects of the looser cuts are an
over-proportional increase of the flat-wave intensity by nearly a
factor of two and an enhancement of structures at large $3\pi$ masses
in some waves.

\begin{figure*}[htbp]
  \centering
  \subfloat[][]{%
    \includegraphics[width=\twoPlotWidth]{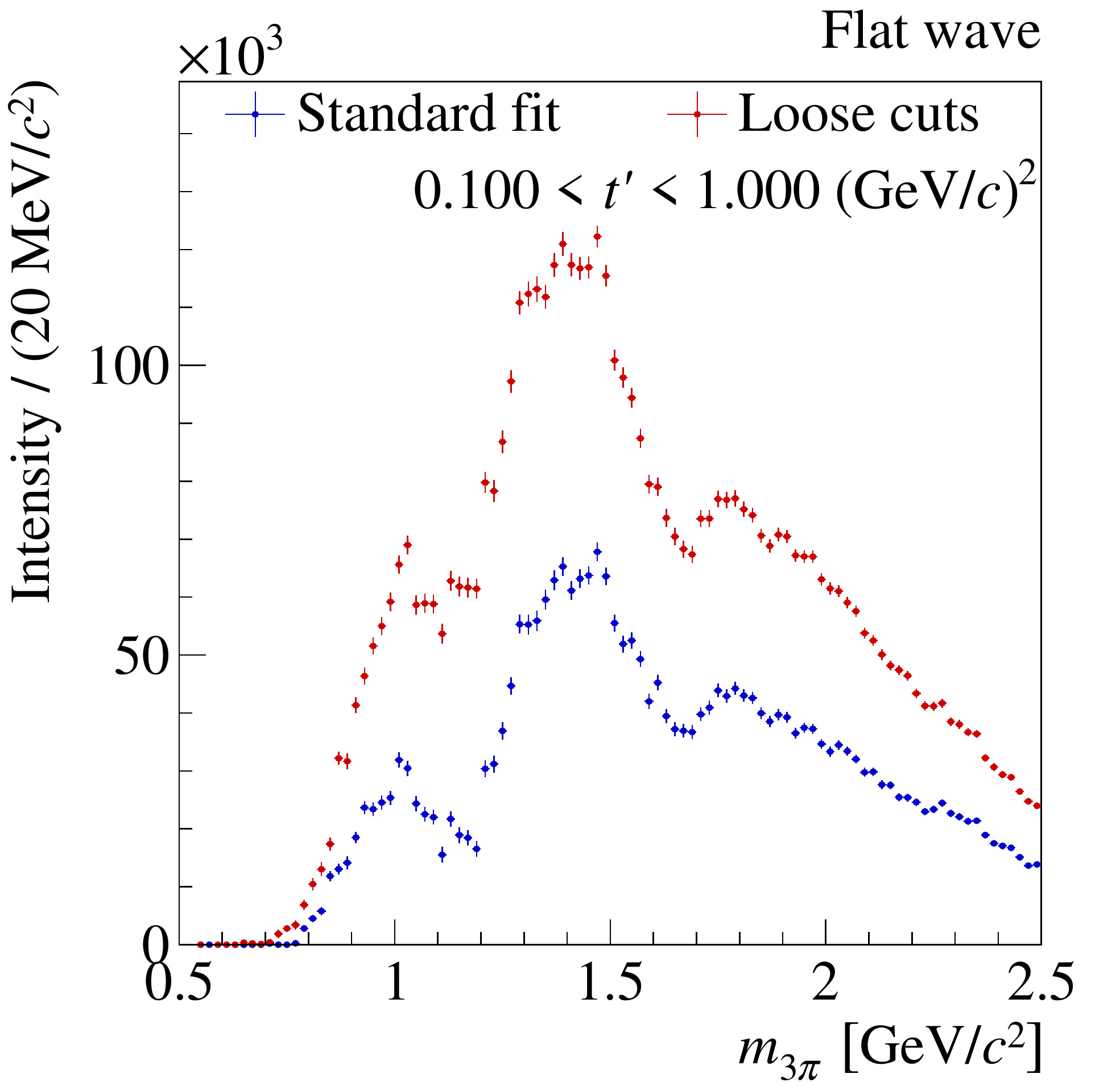}%
  }%
  \hspace*{\twoPlotSpacing}
  \subfloat[][]{%
    \includegraphics[width=\twoPlotWidth]{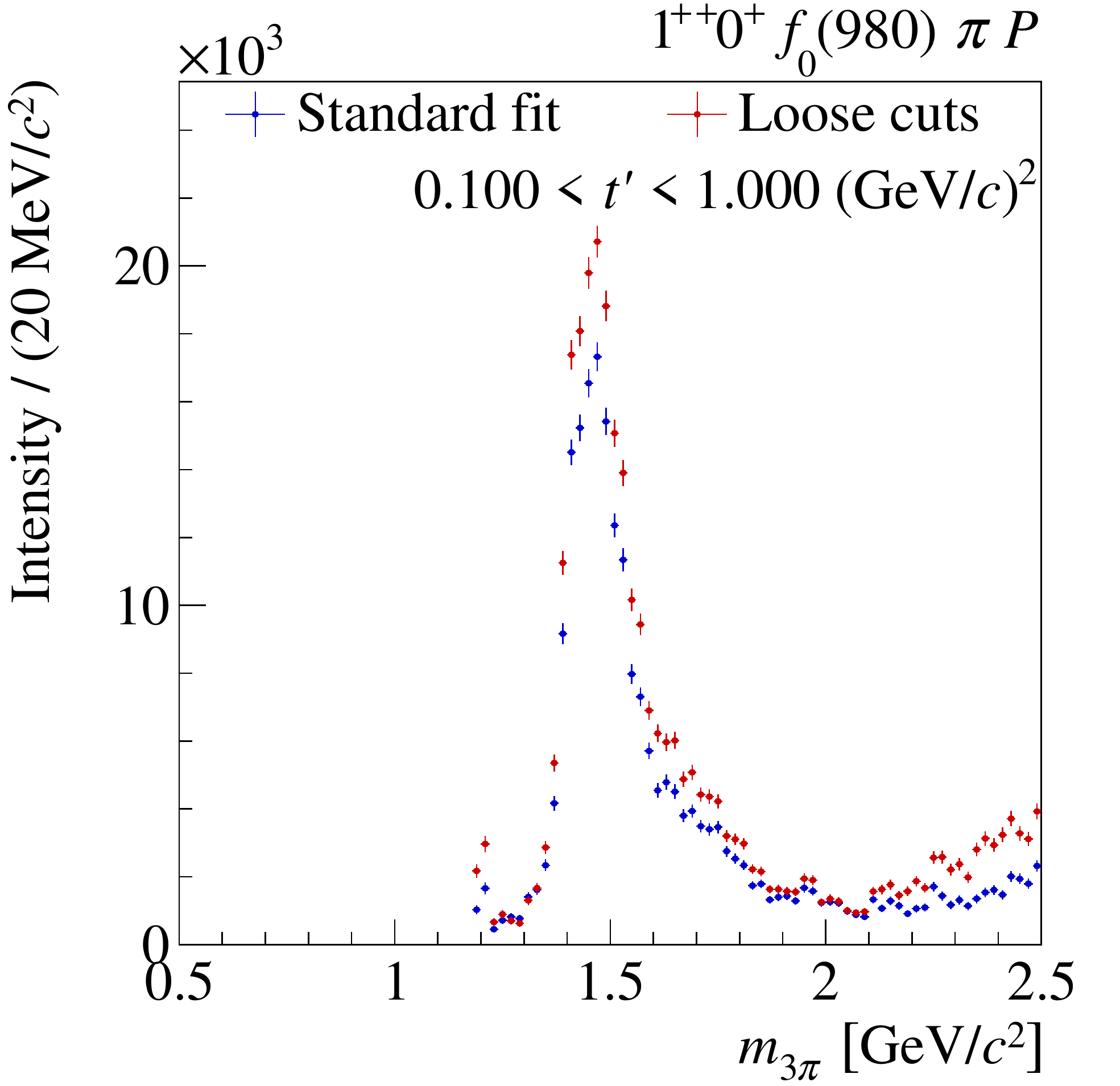}%
  }%
  \\
  \subfloat[][]{%
    \includegraphics[width=\twoPlotWidth]{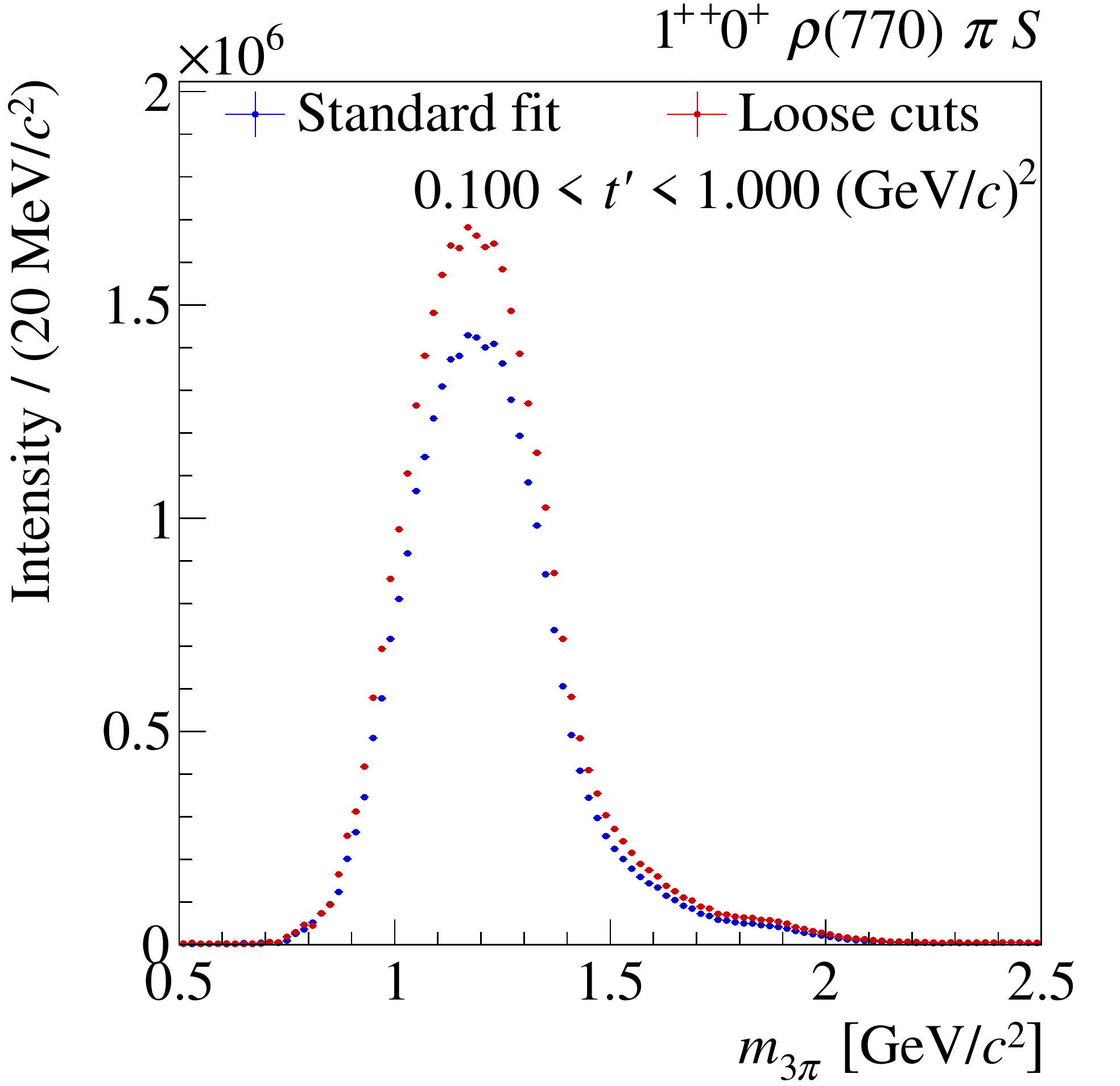}%
  }%
  \hspace*{\twoPlotSpacing}
  \subfloat[][]{%
    \includegraphics[width=\twoPlotWidth]{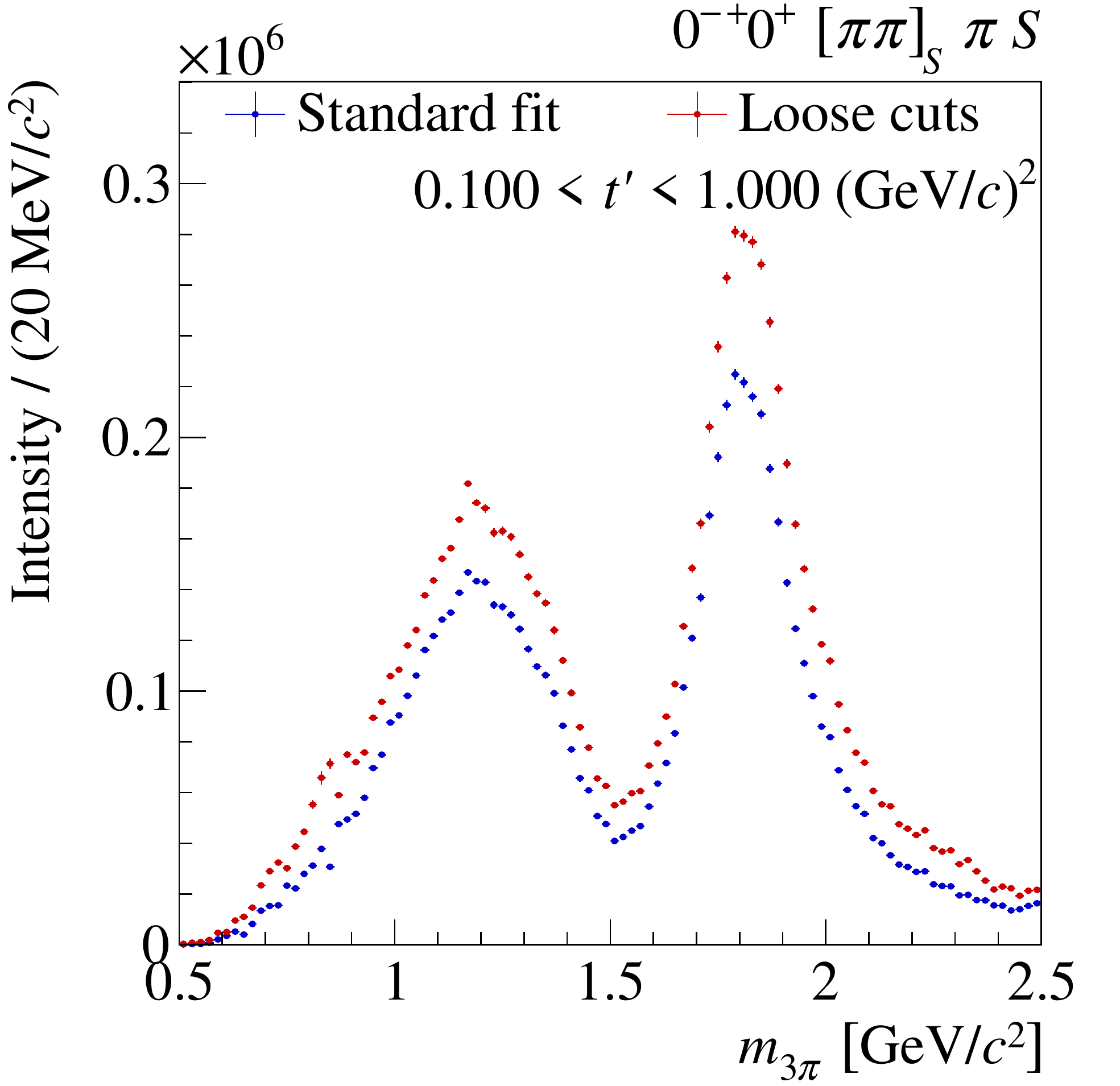}%
  }%
  \caption{\colorPlot Comparison of \tpr-summed partial-wave
    intensities obtained from mass-independent fits for standard event
    selection (blue/black) and a looser event selection (red/gray),
    where particle identification and central-production rejection
    have not been applied.}
  \label{fig:waves_syst_event_selection}
\end{figure*}

\subsection{Variation of \tpr Binning}
\label{sec:appendix_syst_studies_massindep_t_binning}

In order to study, whether the chosen binning in the four-momentum
transfer squared \tpr has any effect on the partial-wave analysis, the
\tpr bins defined in \cref{tab:t-bins} and shown in
\cref{fig:t_vs_m_binning} where halved, yielding in total 22~bins.
The finer \tpr binning has practically no effect on the partial-wave
intensities (see \cref{fig:waves_syst_t_binning}).  Only the flat wave
has lower intensity over the full \mThreePi range in the case of finer
\tpr bins.

\begin{figure*}[htbp]
  \centering
  \subfloat[][]{%
    \includegraphics[width=\twoPlotWidth]{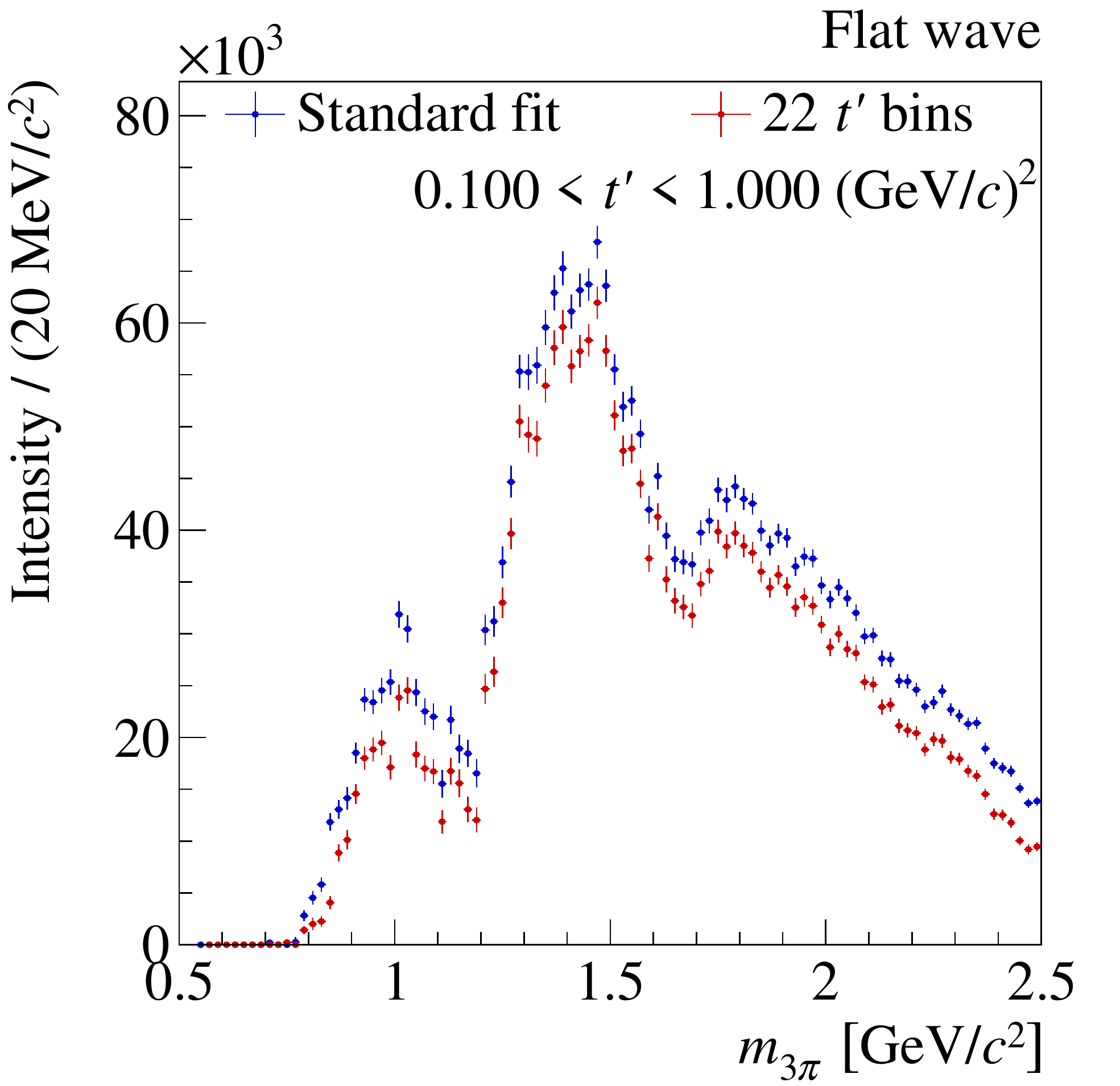}%
  }%
  \hspace*{\twoPlotSpacing}%
  \subfloat[][]{%
    \includegraphics[width=\twoPlotWidth]{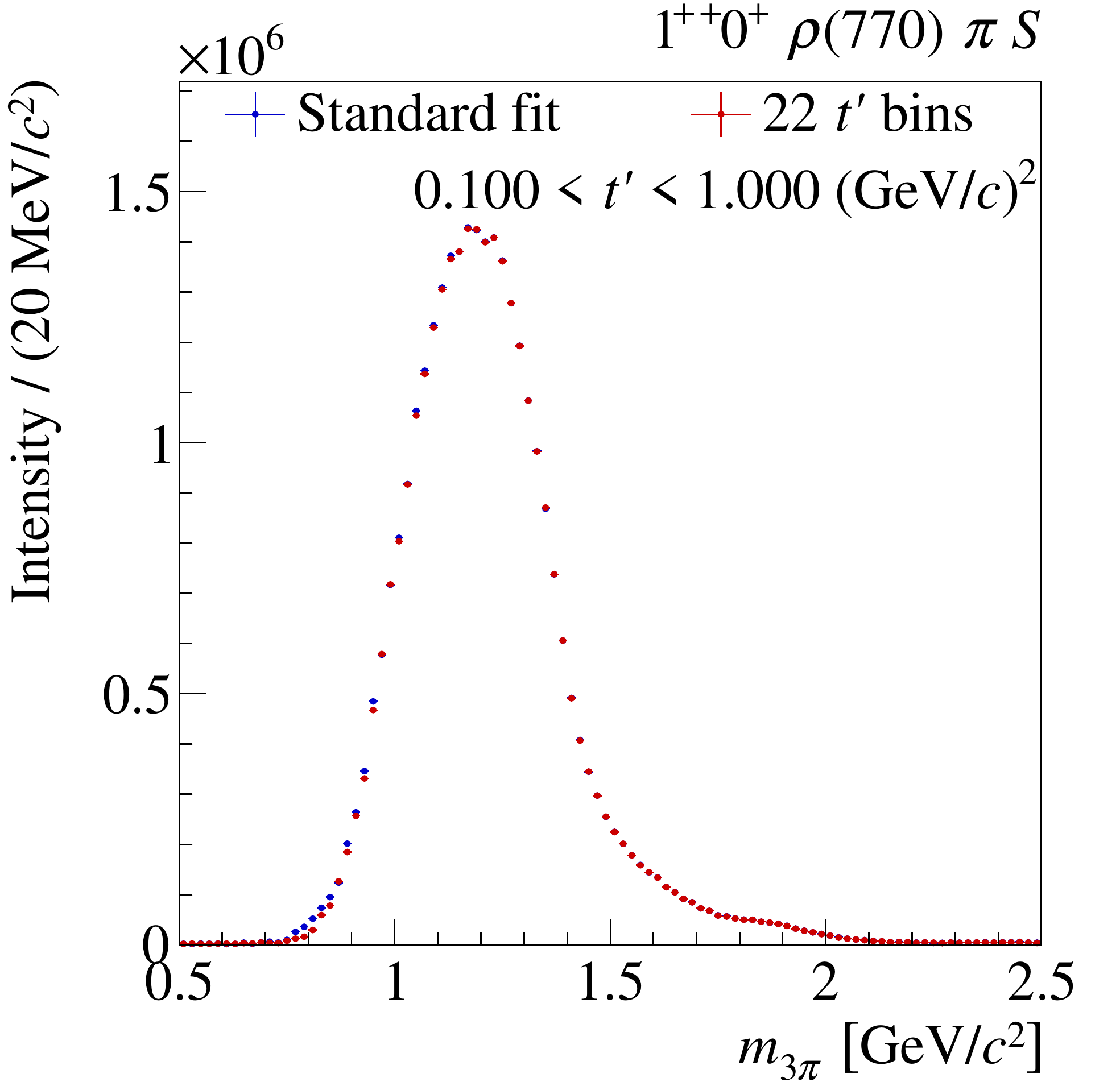}%
  }%
  \caption{\colorPlot Comparison of \tpr-summed partial-wave
    intensities obtained from mass-independent fits using
      two different \tpr binnings: 11~\tpr
    bins (blue/black) and  22~\tpr bins
    (red/gray).}
  \label{fig:waves_syst_t_binning}
\end{figure*}
 %
%
%

\section{Acceptance}
\label{sec:acc_res}

In the following, we describe the \threePi detection efficiency of the
COMPASS apparatus in absolute terms.  For this we have generated
$3\pi$ events distributed isotropically in phase space and passed them
through the COMPASS detector simulation and reconstruction chain.  The
same selection cuts were applied as used for the real data.  For fixed
values of \mThreePi and \tpr, the acceptance is a five-dimensional
function, of which we show only projections.

\Cref{fig:acceptance_phase_space_GJ} shows the \threePi detection
efficiency as a function of the two angles, \cosThetaGJ and \phiGJ, of
the isobar in the Gottfried-Jackson frame (see \cref{sec:isobar_model}
for the definition).  The acceptance is shown in four regions of
\mThreePi and \tpr.  The Monte Carlo data show a rather flat
acceptance with a small dip for in-plane events at forward angles.
This structure exhibits some dependence on \tpr and mass.  The
corresponding distributions in the helicity frame are shown in
\cref{fig:acceptance_phase_space_HF}.  Here, no significant structures
are visible.

\begin{figure*}
  \centering
  \includegraphics[width=\textwidth]{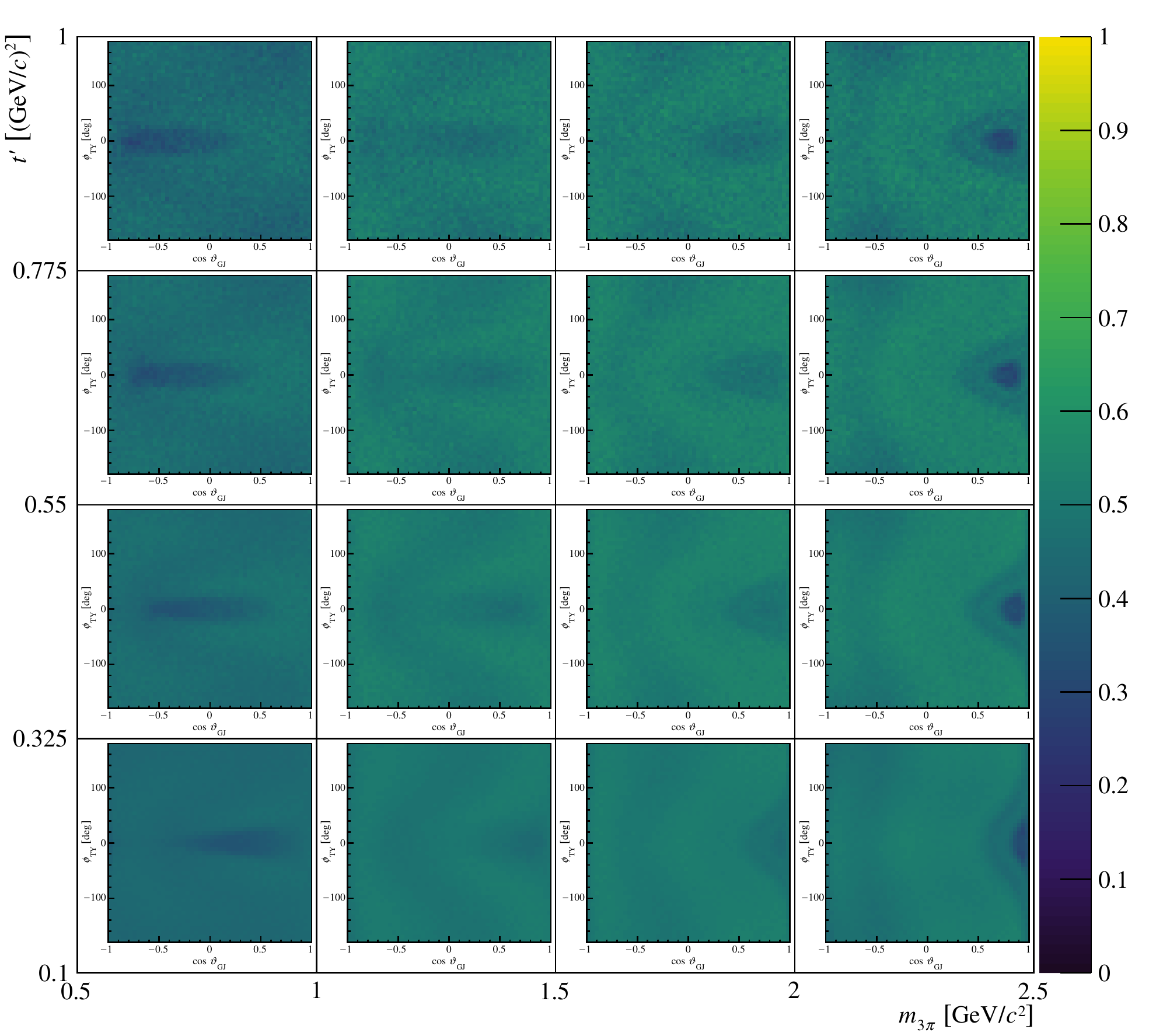}
  \caption{\colorPlot Detection efficiency for \threePi phase-space
    events in different regions of \mThreePi and \tpr.  Each graph
    shows the detection efficiency as a function of the angles,
    \cosThetaGJ (abscissa, from $-1$ to $+1$) and \phiGJ (ordinate,
    from \SI{-180}{\degree} to $+\SI{180}{\degree}$), of the isobar in
    the Gottfried-Jackson frame.}
  \label{fig:acceptance_phase_space_GJ}
\end{figure*}

\begin{figure*}
  \centering
  \includegraphics[width=\textwidth]{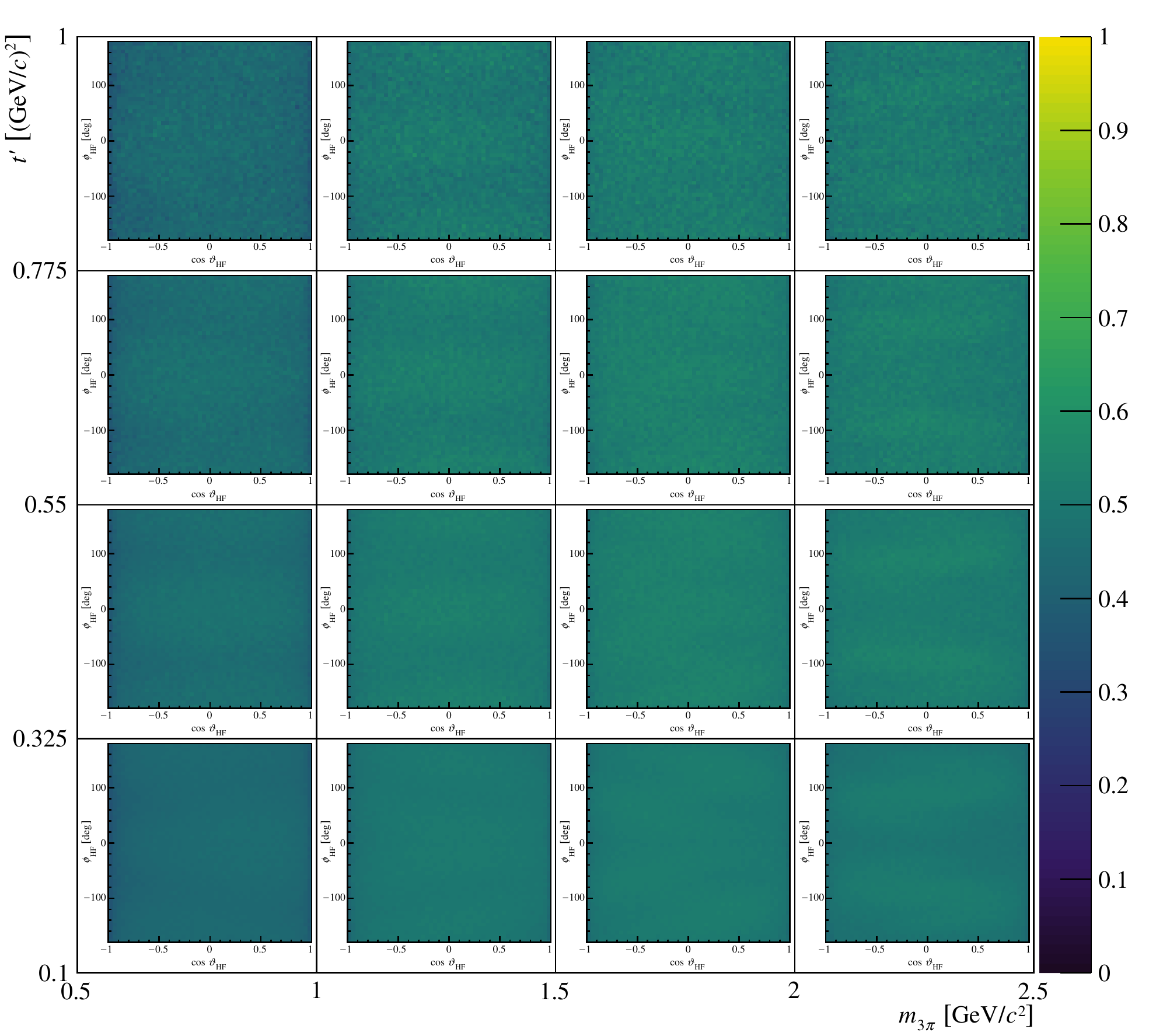}
  \caption{\colorPlot Same as \cref{fig:acceptance_phase_space_GJ},
    but for the detection efficiency as a function of the angles,
    \cosThetaHF and \phiHF, of the $\pi^-$ in the helicity frame.}
  \label{fig:acceptance_phase_space_HF}
\end{figure*}
 %
%
%

\section*{Acknowledgements}
\label{sec:acknowledgements}

We have received many suggestions and input during a series of PWA
workshops: a joint COMPASS-JLab-GSI Workshop on Physics and Methods in
Meson Spectroscopy (Garching/2008), Workshops on Spectroscopy at
COMPASS held 2009 and 2011 in Garching, and in the context of the
ATHOS workshop series (Camogli/2012, Kloster Seeon/2013, and
Ashburn/2015).  We are especially indebted to V.~Mathieu, W.~Ochs, J.~Pelaez,
M.~Pennington, and A.~Szczepaniak for their help and suggestions.
S.U.~Chung would like to thank the IAS at the TU M\"unchen and together
with D.~Ryabchikov the Excellence Cluster \enquote{Universe} for
supporting many visits to Munich during the last years.

We gratefully acknowledge the support of the CERN management and staff
as well as the skills and efforts of the technicians of the
collaborating institutions.  This work is supported by MEYS (Czech
Republic); \enquote{HadronPhysics2} Integrating Activity in FP7
(European Union); CEA, P2I and ANR (France); BMBF, DFG cluster of
excellence \enquote{Origin and Structure of the Universe}, the
computing facilities of the Computational Center for Particle and
Astrophysics (C2PAP), IAS-TUM, and Humboldt foundation (Germany); SAIL
(CSR) (India); ISF (Israel); INFN (Italy); MEXT, JSPS, Daiko, and
Yamada Foundations (Japan); NRF (Rep. of Korea); NCN (Poland); FCT
(Portugal); CERN-RFBR and Presidential Grant NSh-999.2014.2 (Russia).

%
%
\clearpage
\section*{\textsc{Supplemental Material}}
\addtocontents{toc}{\protect\contentsline{section}{\textsc{Supplemental Material}}{}{}}
\clearpage
%
%
%

\section{Additional Partial-Wave Intensities}
\label{sec:additional_waves}

In this section, we present the intensities of the remaining 69~waves
of the 88-wave PWA model that are not discussed in the
paper\ifMultiColumnLayout{~\cite{Adolph:2015tqa}}{}.  The waves are listed in
\cref{tab:remaining-waves} together with their relative intensities as
defined in Section~IV~C.  Out of the 69~waves, 64 have relative
intensities below \SI{1}{\percent}.  The relative intensities of the
69~waves add up to a total of \SI{26.4}{\percent}.

\begin{table*}[p]
  \sisetup{%
    round-mode = places,
    round-precision = 1,
    table-space-text-pre = ${<}$~
  }
  \caption{
    List of the 69~waves of the 88-wave PWA model that are not discussed in the
    paper\ifMultiColumnLayout{~\cite{Adolph:2015tqa}}{}.  The intensities are evaluated as a sum
    over the 11~\tpr bins and are normalized to the total number of
    acceptance-corrected events.  They do not include interference effects
    between the waves.
  }
  \centering
  \label{tab:remaining-waves}
  \ifMultiColumnLayout{}{\begin{scriptsize}\begin{adjustbox}{max width=\textwidth}}
    \subfloat{%
      \renewcommand{\arraystretch}{1.2}
      \begin{tabular}[t]{cccSl}
        \toprule
        \textbf{\JPCMrefl} &
        \textbf{Isobar} &
        \textbf{$L$} &
        \textbf{Relative} &
        \textbf{Shown in} \\
        & & & \textbf{intensity [\si{\percent}]} & \\
        \midrule

        $0^{-+}\,0^+$ & \Prho         & $P$ & 3.545      & \cref{fig:int_0mp0p_rho_P} \\
        $0^{-+}\,0^+$ & \PfTwo        & $D$ & 0.220      & \cref{fig:int_0mp0p_f2_D} \\
        $0^{-+}\,0^+$ & \PfZero[1500] & $S$ & 0.103      & \cref{fig:int_0mp0p_f01500_S} \\

        \midrule

        $1^{++}\,1^+$ & \pipiS        & $P$ & 0.185      & \cref{fig:int_1pp1p_pipiS_P} \\
        $1^{++}\,0^+$ & \Prho         & $D$ & 0.902      & \cref{fig:int_1pp0p_rho_D} \\
        $1^{++}\,1^+$ & \Prho         & $D$ & 0.557      & \cref{fig:int_1pp1p_rho_D} \\
        $1^{++}\,1^+$ & \PfZero[980]  & $P$ & 0.081      & \cref{fig:int_1pp1p_f0980_P} \\
        $1^{++}\,1^+$ & \PfTwo        & $P$ & 0.491      & \cref{fig:int_1pp1p_f2_P} \\
        $1^{++}\,0^+$ & \PfTwo        & $F$ & 0.145      & \cref{fig:int_1pp0p_f2_F} \\
        $1^{++}\,0^+$ & \PrhoThree    & $D$ & 0.116      & \cref{fig:int_1pp0p_rho3_D} \\
        $1^{++}\,0^+$ & \PrhoThree    & $G$ & ${<}$~ 0.1 & \cref{fig:int_1pp0p_rho3_G} \\

        \midrule

        $1^{-+}\,1^+$ & \Prho         & $P$ & 0.849      & \cref{fig:int_1mp1p_rho_P} \\

        \midrule

        $2^{++}\,2^+$ & \PfTwo        & $P$ & ${<}$~ 0.1 & \cref{fig:int_2pp2p_f2_P} \\
        $2^{++}\,1^+$ & \PrhoThree    & $D$ & ${<}$~ 0.1 & \cref{fig:int_2pp1p_rho3_D} \\

        \midrule

        $2^{-+}\,1^+$ & \pipiS        & $D$ & 0.382      & \cref{fig:int_2mp1p_pipiS_D} \\
        $2^{-+}\,0^+$ & \Prho         & $P$ & 3.825      & \cref{fig:int_2mp0p_rho_P} \\
        $2^{-+}\,1^+$ & \Prho         & $P$ & 3.327      & \cref{fig:int_2mp1p_rho_P} \\
        $2^{-+}\,2^+$ & \Prho         & $P$ & 0.163      & \cref{fig:int_2mp2p_rho_P} \\
        $2^{-+}\,1^+$ & \Prho         & $F$ & 0.303      & \cref{fig:int_2mp1p_rho_F} \\
        $2^{-+}\,2^+$ & \PfTwo        & $S$ & 0.112      & \cref{fig:int_2mp2p_f2_S} \\
        $2^{-+}\,1^+$ & \PfTwo        & $D$ & 0.196      & \cref{fig:int_2mp1p_f2_D} \\
        $2^{-+}\,2^+$ & \PfTwo        & $D$ & 0.080      & \cref{fig:int_2mp2p_f2_D} \\
        $2^{-+}\,0^+$ & \PfTwo        & $G$ & 0.076      & \cref{fig:int_2mp0p_f2_G} \\
        $2^{-+}\,0^+$ & \PrhoThree    & $P$ & 0.226      & \cref{fig:int_2mp0p_rho3_P} \\
        $2^{-+}\,1^+$ & \PrhoThree    & $P$ & 0.115      & \cref{fig:int_2mp1p_rho3_P} \\

        \midrule

        $3^{++}\,0^+$ & \pipiS        & $F$ & 0.228      & \cref{fig:int_3pp0p_pipiS_F} \\
        $3^{++}\,1^+$ & \pipiS        & $F$ & 0.319      & \cref{fig:int_3pp1p_pipiS_F} \\
        $3^{++}\,0^+$ & \Prho         & $D$ & 0.888      & \cref{fig:int_3pp0p_rho_D} \\
        $3^{++}\,1^+$ & \Prho         & $D$ & 0.989      & \cref{fig:int_3pp1p_rho_D} \\
        $3^{++}\,0^+$ & \Prho         & $G$ & 0.361      & \cref{fig:int_3pp0p_rho_G} \\
        $3^{++}\,1^+$ & \Prho         & $G$ & 0.118      & \cref{fig:int_3pp1p_rho_G} \\
        $3^{++}\,0^+$ & \PfTwo        & $P$ & 0.432      & \cref{fig:int_3pp0p_f2_P} \\
        $3^{++}\,1^+$ & \PfTwo        & $P$ & 0.449      & \cref{fig:int_3pp1p_f2_P} \\
        $3^{++}\,0^+$ & \PrhoThree    & $S$ & 0.434      & \cref{fig:int_3pp0p_rho3_S} \\
        $3^{++}\,1^+$ & \PrhoThree    & $S$ & 0.149      & \cref{fig:int_3pp1p_rho3_S} \\
        $3^{++}\,0^+$ & \PrhoThree    & $I$ & ${<}$~ 0.1 & \cref{fig:int_3pp0p_rho3_I} \\

        \bottomrule
      \end{tabular}
    }%
    \ifMultiColumnLayout{\hspace{0.15\textwidth}}{}
    \subfloat{%
      \renewcommand{\arraystretch}{1.2}
      \begin{tabular}[t]{cccSl}
        \toprule
        \textbf{\JPCMrefl} &
        \textbf{Isobar} &
        \textbf{$L$} &
        \textbf{Relative} &
        \textbf{Shown in} \\
        & & & \textbf{intensity [\si{\percent}]} & \\
        \midrule

        $3^{-+}\,1^+$ & \Prho         & $F$ & 0.052      & \cref{fig:int_3mp1p_rho_F} \\
        $3^{-+}\,1^+$ & \PfTwo        & $D$ & ${<}$~ 0.1 & \cref{fig:int_3mp1p_f2_D} \\

        \midrule

        $4^{++}\,2^+$ & \Prho         & $G$ & ${<}$~ 0.1 & \cref{fig:int_4pp2p_rho_G} \\
        $4^{++}\,2^+$ & \PfTwo        & $F$ & ${<}$~ 0.1 & \cref{fig:int_4pp2p_f2_F} \\
        $4^{++}\,1^+$ & \PrhoThree    & $D$ & ${<}$~ 0.1 & \cref{fig:int_4pp1p_rho3_D} \\

        \midrule

        $4^{-+}\,0^+$ & \pipiS        & $G$ & 0.283      & \cref{fig:int_4mp0p_pipiS_G} \\
        $4^{-+}\,0^+$ & \Prho         & $F$ & 0.981      & \cref{fig:int_4mp0p_rho_F} \\
        $4^{-+}\,1^+$ & \Prho         & $F$ & 0.379      & \cref{fig:int_4mp1p_rho_F} \\
        $4^{-+}\,0^+$ & \PfTwo        & $D$ & 0.299      & \cref{fig:int_4mp0p_f2_D} \\
        $4^{-+}\,1^+$ & \PfTwo        & $D$ & 0.138      & \cref{fig:int_4mp1p_f2_D} \\
        $4^{-+}\,0^+$ & \PfTwo        & $G$ & ${<}$~ 0.1 & \cref{fig:int_4mp0p_f2_G} \\

        \midrule

        $5^{++}\,0^+$ & \pipiS        & $H$ & 0.131      & \cref{fig:int_5pp0p_pipiS_H} \\
        $5^{++}\,1^+$ & \pipiS        & $H$ & 0.084      & \cref{fig:int_5pp1p_pipiS_H} \\
        $5^{++}\,0^+$ & \Prho         & $G$ & 0.330      & \cref{fig:int_5pp0p_rho_G} \\
        $5^{++}\,0^+$ & \PfTwo        & $F$ & 0.114      & \cref{fig:int_5pp0p_f2_F} \\
        $5^{++}\,1^+$ & \PfTwo        & $F$ & 0.087      & \cref{fig:int_5pp1p_f2_F} \\
        $5^{++}\,0^+$ & \PfTwo        & $H$ & ${<}$~ 0.1 & \cref{fig:int_5pp0p_f2_H} \\
        $5^{++}\,0^+$ & \PrhoThree    & $D$ & ${<}$~ 0.1 & \cref{fig:int_5pp0p_rho3_D} \\

        \midrule

        $6^{++}\,1^+$ & \Prho         & $I$ & ${<}$~ 0.1 & \cref{fig:int_6pp1p_rho_I} \\
        $6^{++}\,1^+$ & \PfTwo        & $H$ & ${<}$~ 0.1 & \cref{fig:int_6pp1p_f2_H} \\

        \midrule

        $6^{-+}\,0^+$ & \pipiS        & $I$ & 0.107      & \cref{fig:int_6mp0p_pipiS_I} \\
        $6^{-+}\,1^+$ & \pipiS        & $I$ & 0.055      & \cref{fig:int_6mp1p_pipiS_I} \\
        $6^{-+}\,0^+$ & \Prho         & $H$ & 0.699      & \cref{fig:int_6mp0p_rho_H} \\
        $6^{-+}\,1^+$ & \Prho         & $H$ & 0.139      & \cref{fig:int_6mp1p_rho_H} \\
        $6^{-+}\,0^+$ & \PfTwo        & $G$ & 0.055      & \cref{fig:int_6mp0p_f2_G} \\
        $6^{-+}\,0^+$ & \PrhoThree    & $F$ & ${<}$~ 0.1 & \cref{fig:int_6mp0p_rho3_F} \\

        \midrule

        $1^{++}\,1^-$ & \Prho         & $S$ & 0.296      & \cref{fig:int_1pp1m_rho_S} \\

        \midrule

        $1^{-+}\,0^-$ & \Prho         & $P$ & 0.256      & \cref{fig:int_0mp0m_rho_P} \\
        $1^{-+}\,1^-$ & \Prho         & $P$ & 0.670      & \cref{fig:int_0mp1m_rho_P} \\

        \midrule

        $2^{++}\,0^-$ & \Prho         & $D$ & 0.312      & \cref{fig:int_2pp0m_rho_D} \\
        $2^{++}\,0^-$ & \PfTwo        & $P$ & 0.184      & \cref{fig:int_2pp0m_f2_P} \\
        $2^{++}\,1^-$ & \PfTwo        & $P$ & 0.327      & \cref{fig:int_2pp1m_f2_P} \\

        \midrule

        $2^{-+}\,1^-$ & \PfTwo        & $S$ & 0.183      & \cref{fig:int_2mp1m_f2_S} \\

        \bottomrule
        \multicolumn{3}{r}{\textbf{Intensity Sum}} & 26.4 & \\  %
      \end{tabular}
    }%
  \ifMultiColumnLayout{}{\end{adjustbox}\end{scriptsize}}
\end{table*}

The employed analysis method has several known limitations.  Some of
them are related to presently open questions concerning the analysis
of diffractively produced multi-body final states.  Therefore, the
systematic uncertainties induced by the analysis method are in general
difficult to quantify.  In Section~IV~F and Appendix~B of the paper,
we discuss systematic effects that affect the 18~selected partial
waves.  In the following, we focus on the discussion of systematic
effects that are potentially affecting the remaining 69~waves.  Since
many of these waves have only small intensities, they are
more susceptible to systematic effects.

There are a number of systematic effects related to the truncation of
the partial-wave expansion [see Eq.~(24) in Section~III~C], \ie to the
choice of the wave set.  There is currently no generally accepted
objective method to determine from the data which partial waves
actually contribute.  As discussed in Section~IV~F, 17~of the
18~selected partial waves have been found to be fairly insensitive to
changes of the wave set.  This was assessed based on fits performed
with a significantly smaller set of 53~waves.  Mutual reshuffling of
intensity (leakage) was observed, which, however, was shown to have no
significant effect on 17 of the 18 selected waves.  Typical examples
of such leakage effects are unphysical enhancements below
\SI{1}{\GeVcc} as can be seen \eg in
\cref{fig:int_1pp1p_f2_P,fig:int_2mp2p_f2_S}.

Another class of artifacts that are related to the truncation of the
partial-wave expansion are discontinuities due to the thresholds
applied to 27~waves in the PWA model (see Table~IX in Appendix~A).
These thresholds gradually reduce the waves set towards lower
\mThreePi for $\mThreePi < \SI{1.7}{\GeVcc}$ and are required to
stabilize the PWA fit (see also discussion below).

Novel data-driven approaches to find an adequate wave set for a given
data sample, like the one proposed by the authors of
\refCite{Guegan:2015mea}, which is based on additional penalty terms
in the likelihood function, might help to solve the issues discussed above and
are currently under study.

Non-resonant contributions, like \eg the Deck effect, represent a
continuum of waves including waves with very high spin.  Thus, by
truncating the partial-wave expansion, the non-resonant contributions
are not fully accounted for.  This might induce additional leakage.
At the present state of the field, more theoretical work is required
in order to treat the non-resonant contributions.

The Bose symmetrization of the two final-state $\pi^-$ breaks the
orthogonality of the angular part of the decay amplitudes.  In
particular, amplitudes with the same \JPCMrefl quantum numbers but
different decay channels may develop large overlap integrals,
indicating that the corresponding phase-space distributions are
similar.  This typically leads to unstable maximum-likelihood
estimates for the intensities of these waves.  Such instabilities are
most pronounced at low \mThreePi, where only the low-mass tails of the
isobars contribute to the decay amplitudes.  Since the isobar
amplitudes exhibit similar behavior at low \mTwoPi, it becomes
difficult to distinguish them.  In the current approach, this is
counteracted by applying thresholds to some waves, thereby reducing
the wave set in the region of low \mThreePi (see Table~IX in
Appendix~A).

The employed isobar model uses predefined parametrizations for the
mass-dependent amplitudes of all isobars.  Systematic effects due to
possible imperfections in the isobar parametrizations are discussed in
Sections IV~F and VI and in Appendix~B~3.  Additional effects may be
caused by small contributions from heavier isobars, like \eg
\Prho[1450], \Prho[1700], or higher \PfZero* excitations, which are
currently neglected.  Including such excited states as isobars leads
to unstable fits because the waves with the excited isobars have large
overlap integrals with waves with the ground-state isobars, like \eg
\Prho or \pipiS (see also discussion of the \PfZero[1500] isobar in
Section~IV~B).

Another effect that is neglected by the isobar model
is a possible distortion of the mass-dependent isobar amplitudes due
to final-state interactions.  The above issues can be addressed in
future analyses by extending the freed-isobar method that currently
includes only $\JPC = 0^{++}$ isobars (see Section~VI) to also include
dominant partial waves with $\JPC = 1^{--}$ and $2^{++}$ isobars.

First studies of the relativistic corrections to the decay amplitudes
in the partial-wave analysis show that the 18~selected waves exhibit
only small changes (see Sections III~B and IV~F).  However, for some
of the remaining 69~waves these corrections are not small and will
have to be taken into account in future analyses.

For completeness and for future reference, we show in the following
for each of the 69~waves in \cref{tab:remaining-waves} the intensity
distribution summed over the 11~\tpr bins.  The percent numbers given
in the mass spectra are the relative intensities of the particular
partial wave shown.  In view of the systematic uncertainties possibly
induced by the effects discussed above, we refrain from interpreting
the intensity distributions here.

\ifMultiColumnLayout{\onecolumngrid}{}
\subsection{Waves with Positive Reflectivity}

\subsubsection{$\JPC = 0^{-+}$ Waves}

\begin{figure}[H]
  \centering
  \subfloat[][]{%
    \label{fig:int_0mp0p_rho_P}%
    \includegraphics[width=\threePlotWidth]{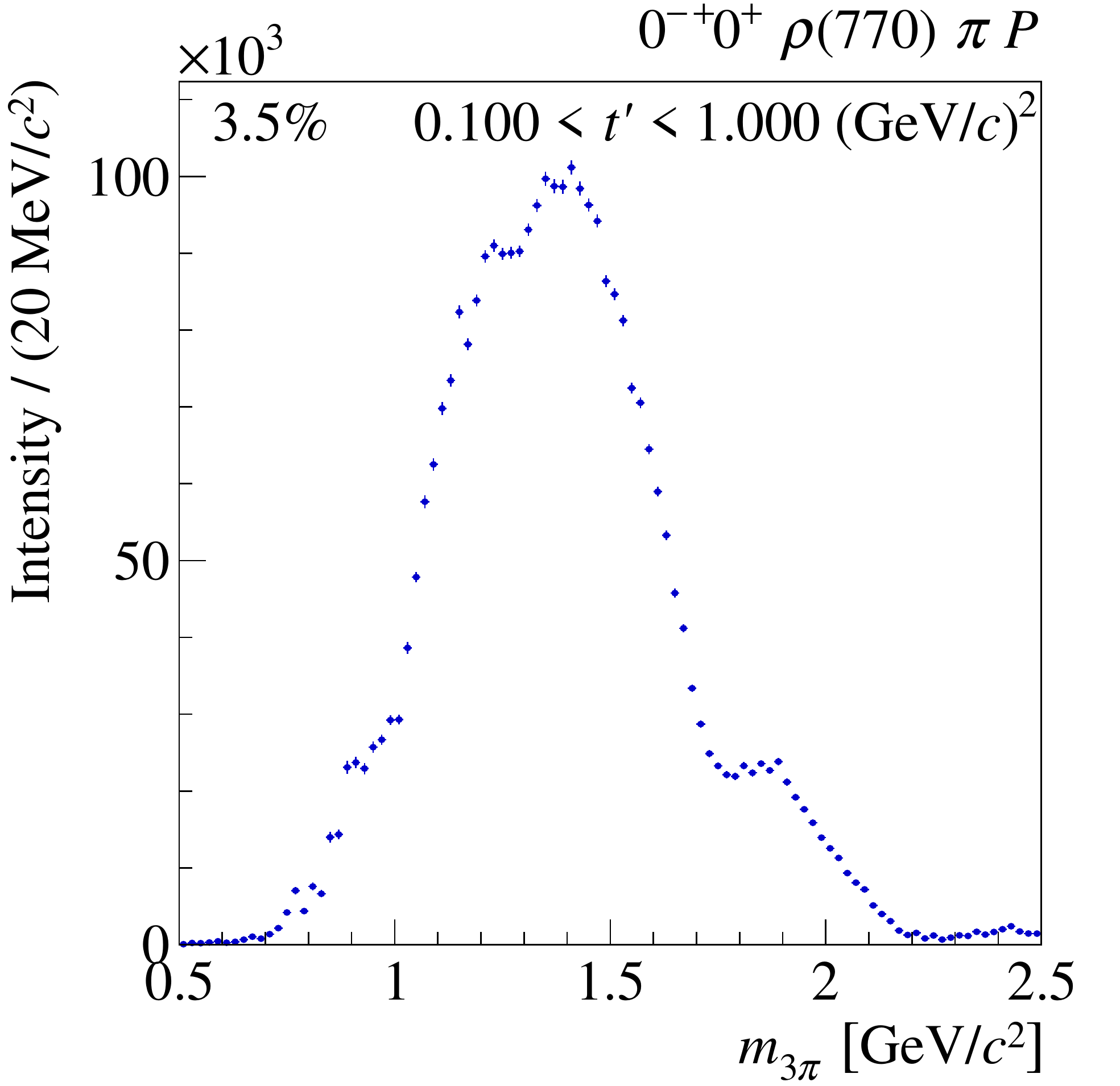}%
  }%
  \subfloat[][]{%
    \label{fig:int_0mp0p_f2_D}%
    \includegraphics[width=\threePlotWidth]{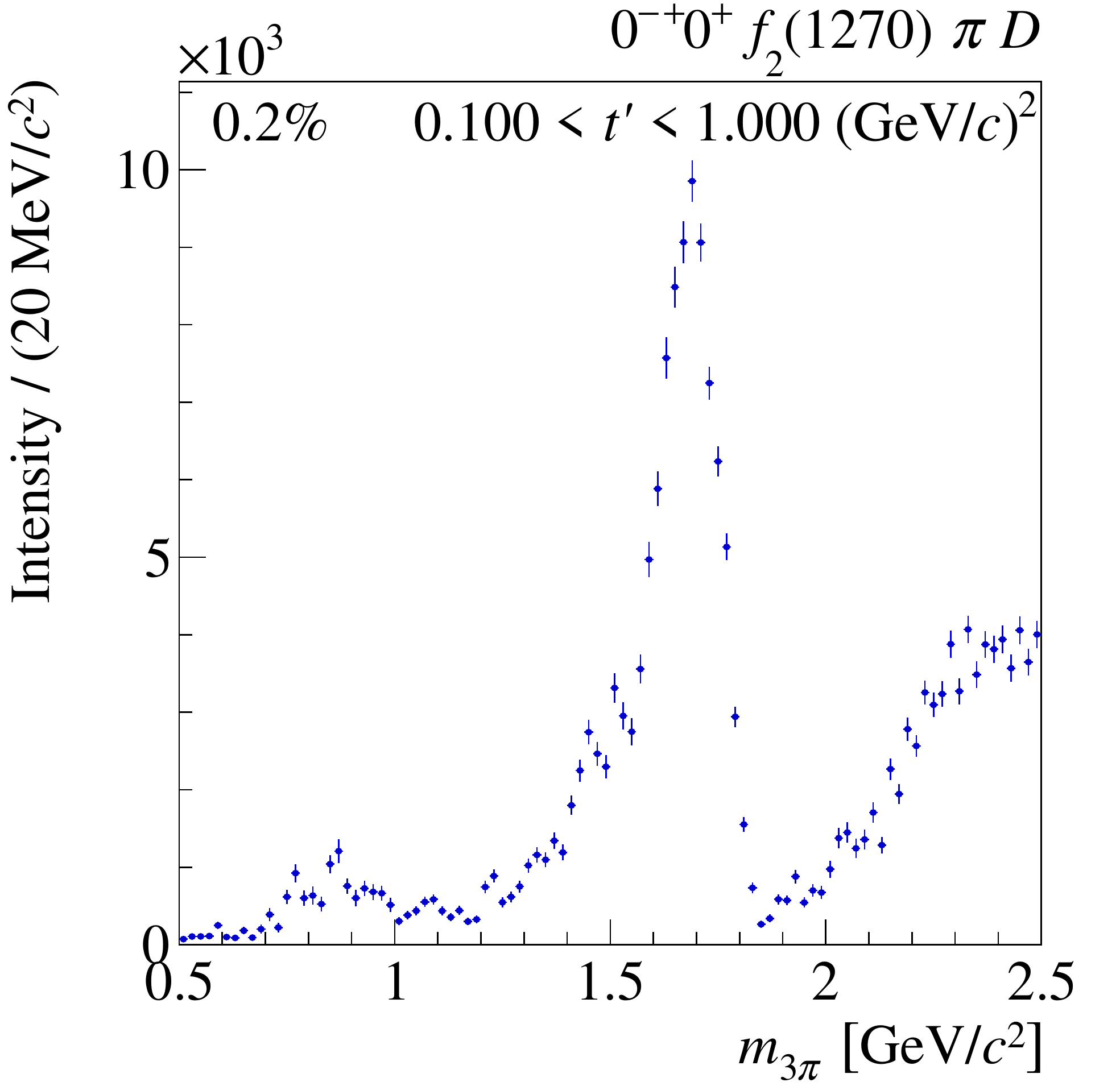}%
  }%
  \subfloat[][]{%
    \label{fig:int_0mp0p_f01500_S}%
    \includegraphics[width=\threePlotWidth]{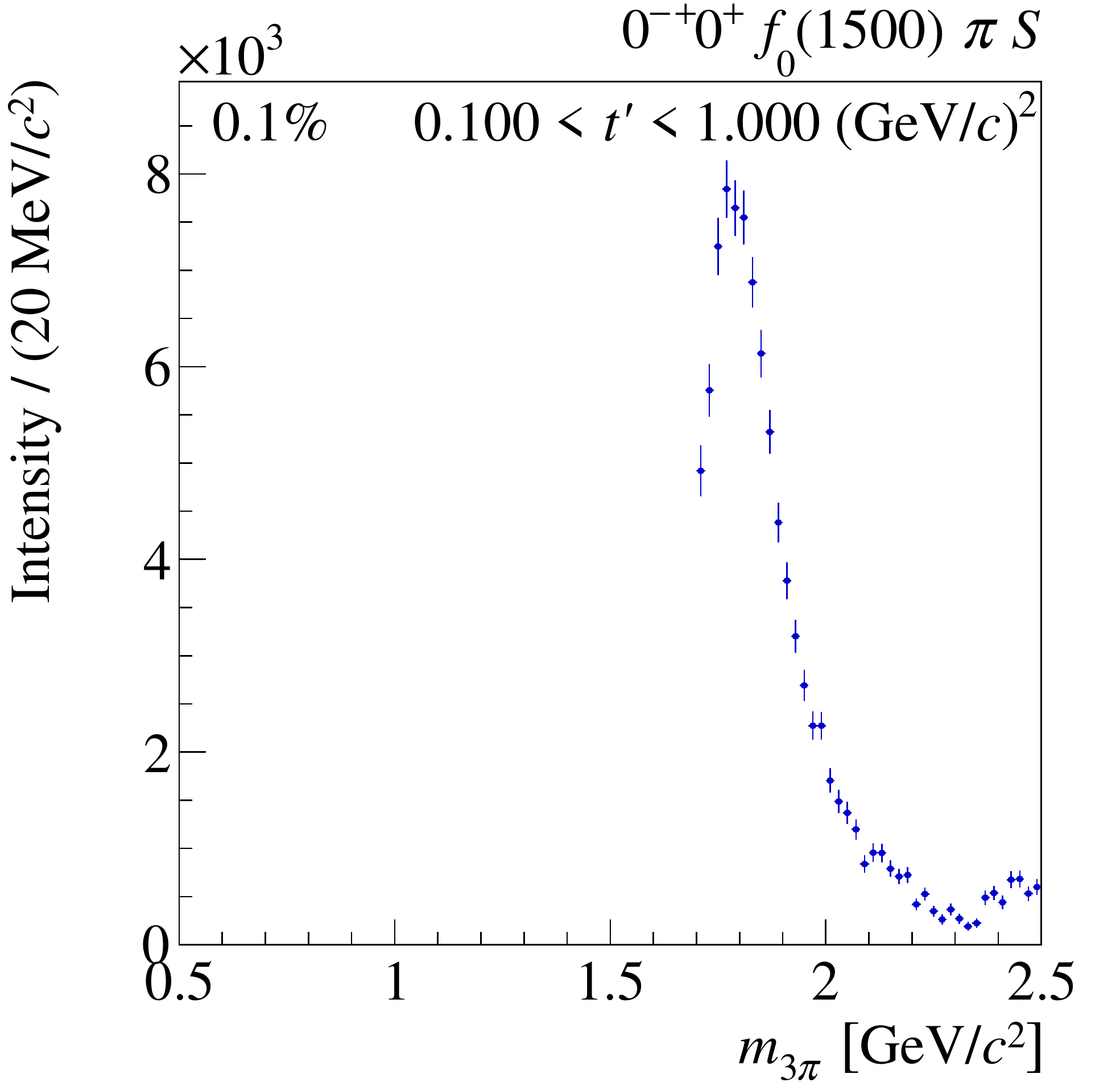}%
  }%
  \caption{The \tpr-summed intensities of partial waves with $\JPC =
    0^{-+}$ and positive reflectivity.}
  \label{fig:intensities_0mp}
\end{figure}

\subsubsection{$\JPC = 1^{++}$ Waves}

\begin{figure}[H]
  \centering
  \subfloat[][]{%
    \label{fig:int_1pp1p_pipiS_P}%
    \includegraphics[width=\threePlotWidth]{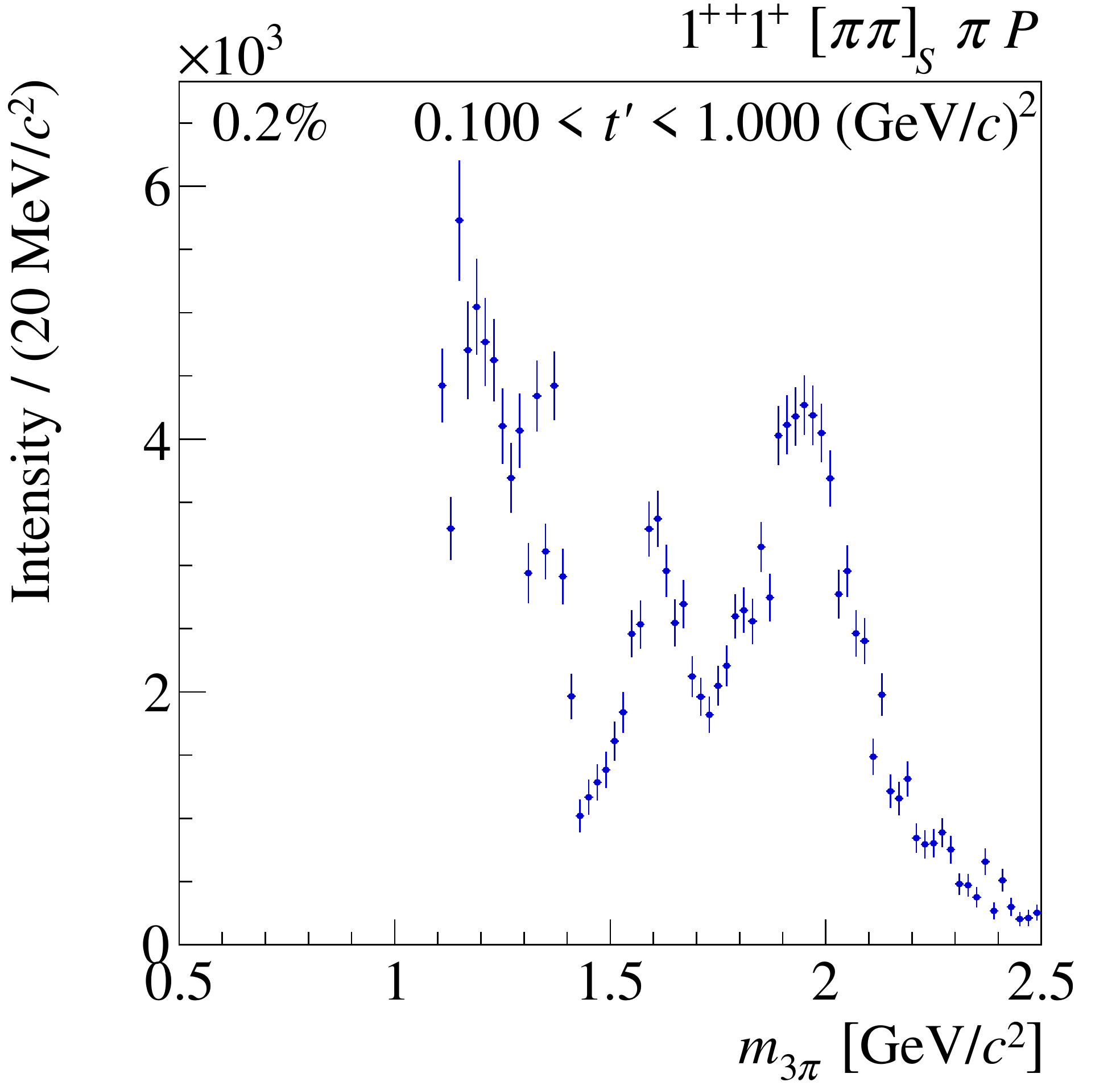}%
  }%
  \subfloat[][]{%
    \label{fig:int_1pp0p_rho_D}%
    \includegraphics[width=\threePlotWidth]{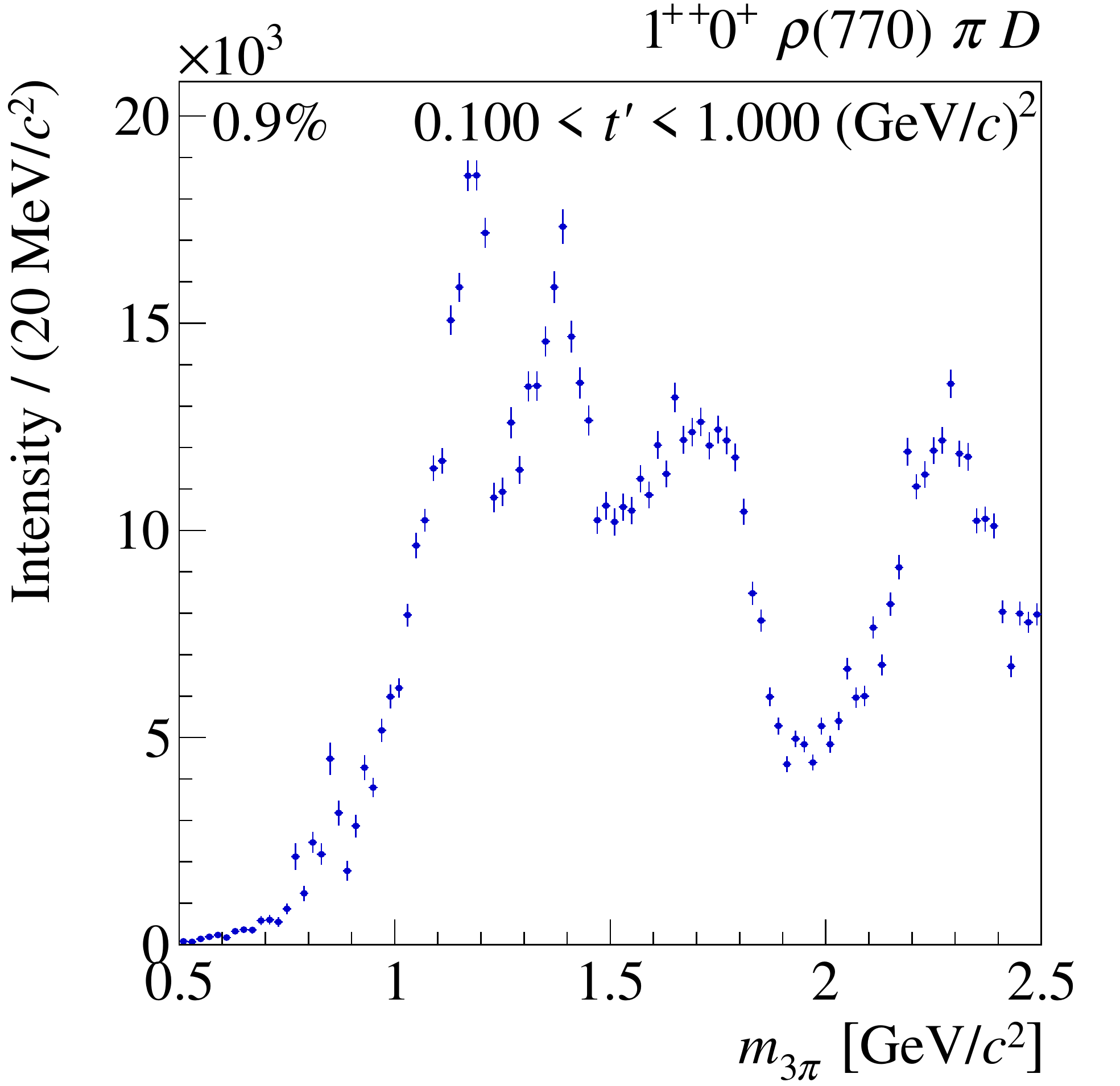}%
  }%
  \subfloat[][]{%
    \label{fig:int_1pp1p_rho_D}%
    \includegraphics[width=\threePlotWidth]{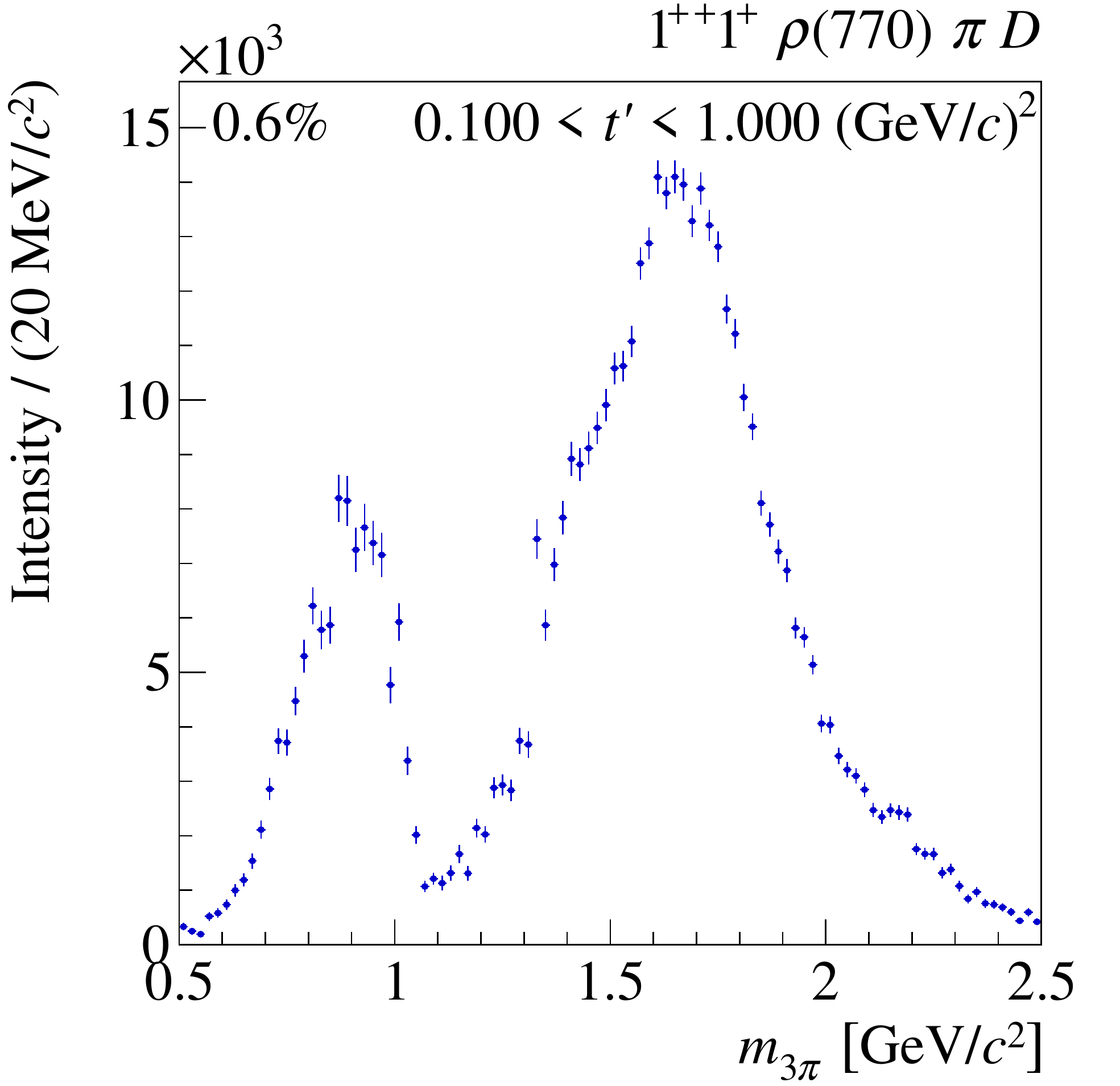}%
  }%
  \caption{The \tpr-summed intensities of partial waves with $\JPC =
    1^{++}$ and positive reflectivity.}
  \label{fig:intensities_1pp_1}
\end{figure}

\clearpage
\begin{figure}[H]
  \centering
  \subfloat[][]{%
    \label{fig:int_1pp1p_f0980_P}%
    \includegraphics[width=\threePlotWidth]{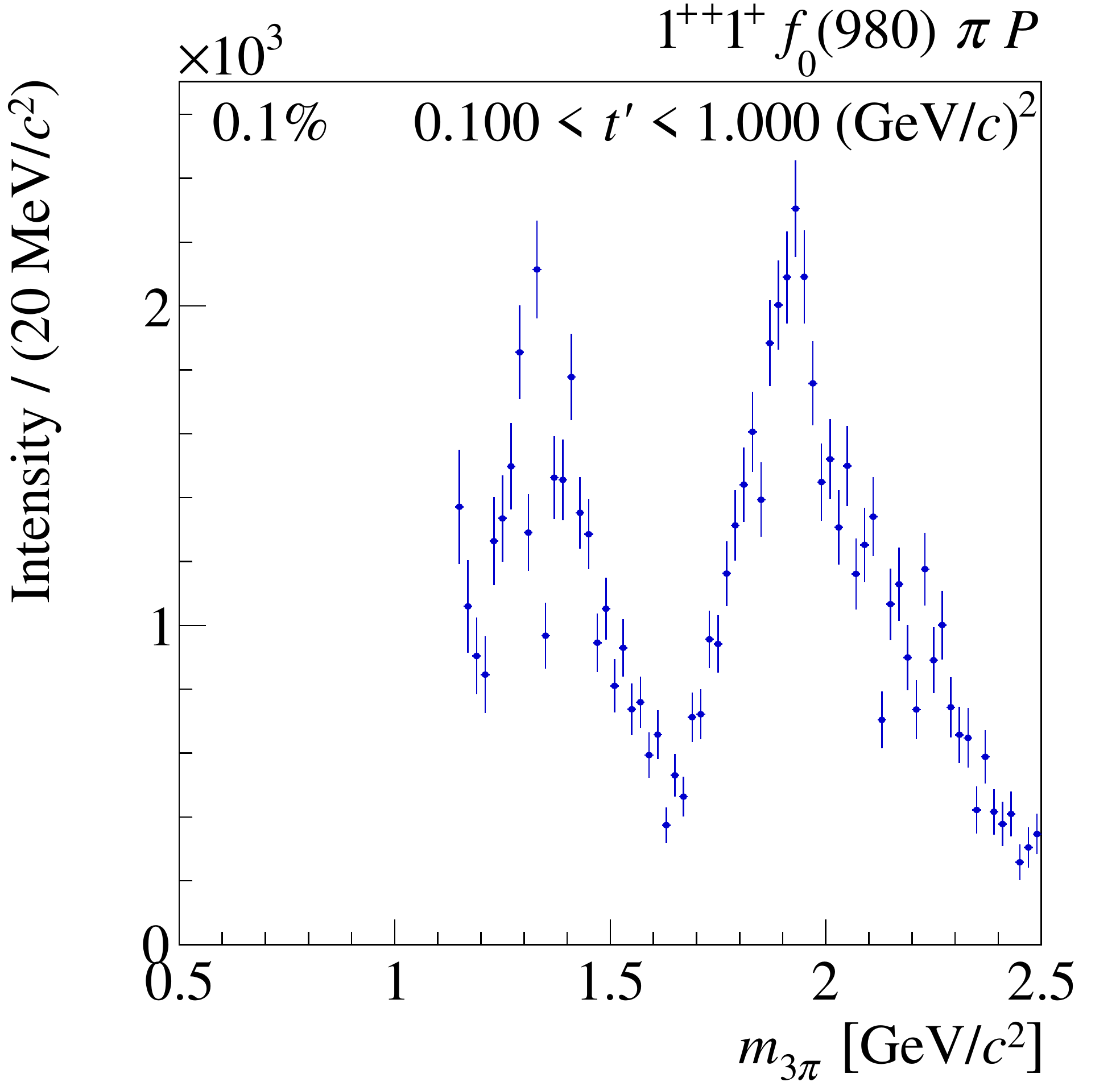}%
  }%
  \subfloat[][]{%
    \label{fig:int_1pp1p_f2_P}%
    \includegraphics[width=\threePlotWidth]{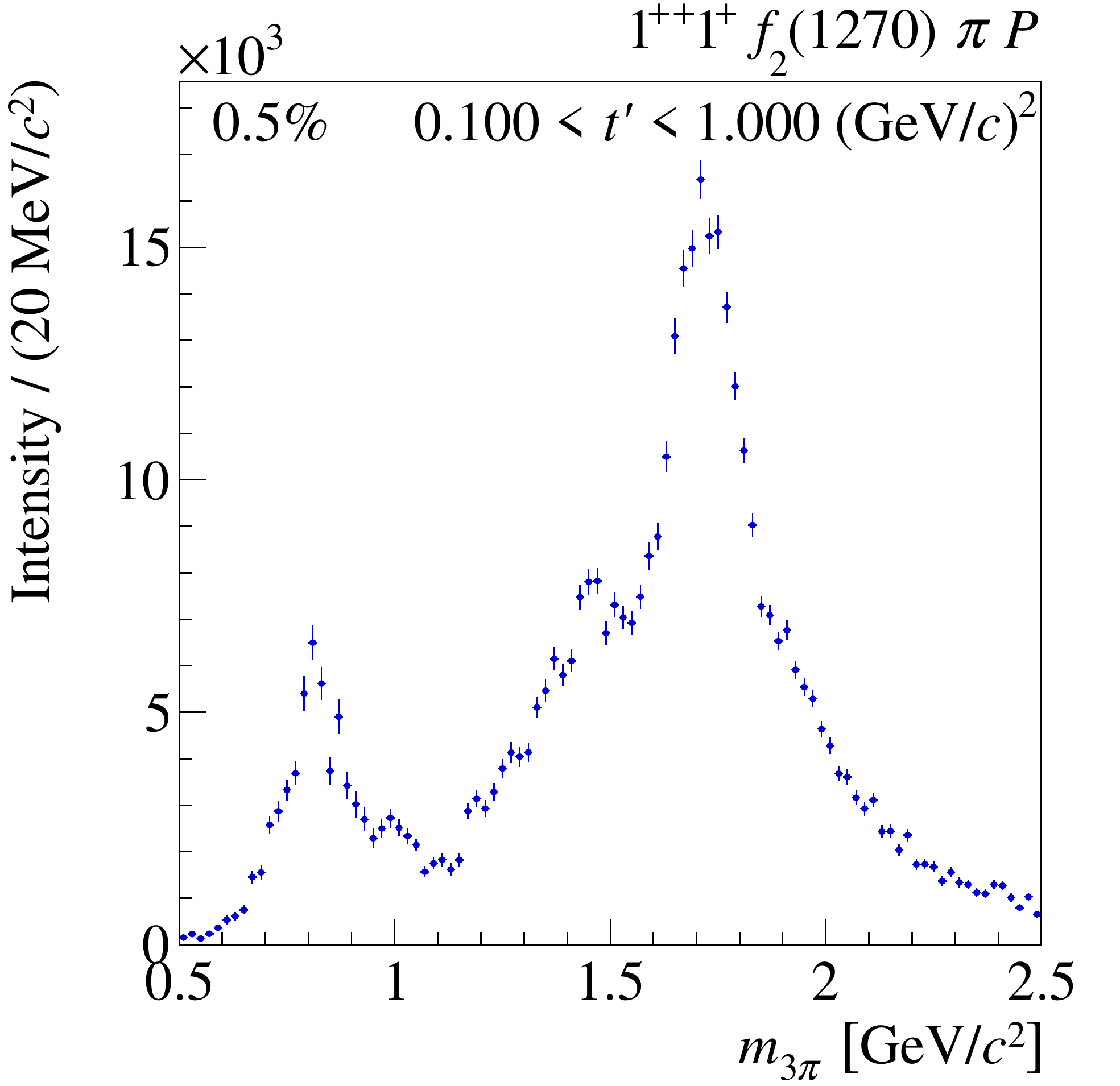}%
  }%
  \subfloat[][]{%
    \label{fig:int_1pp0p_f2_F}%
    \includegraphics[width=\threePlotWidth]{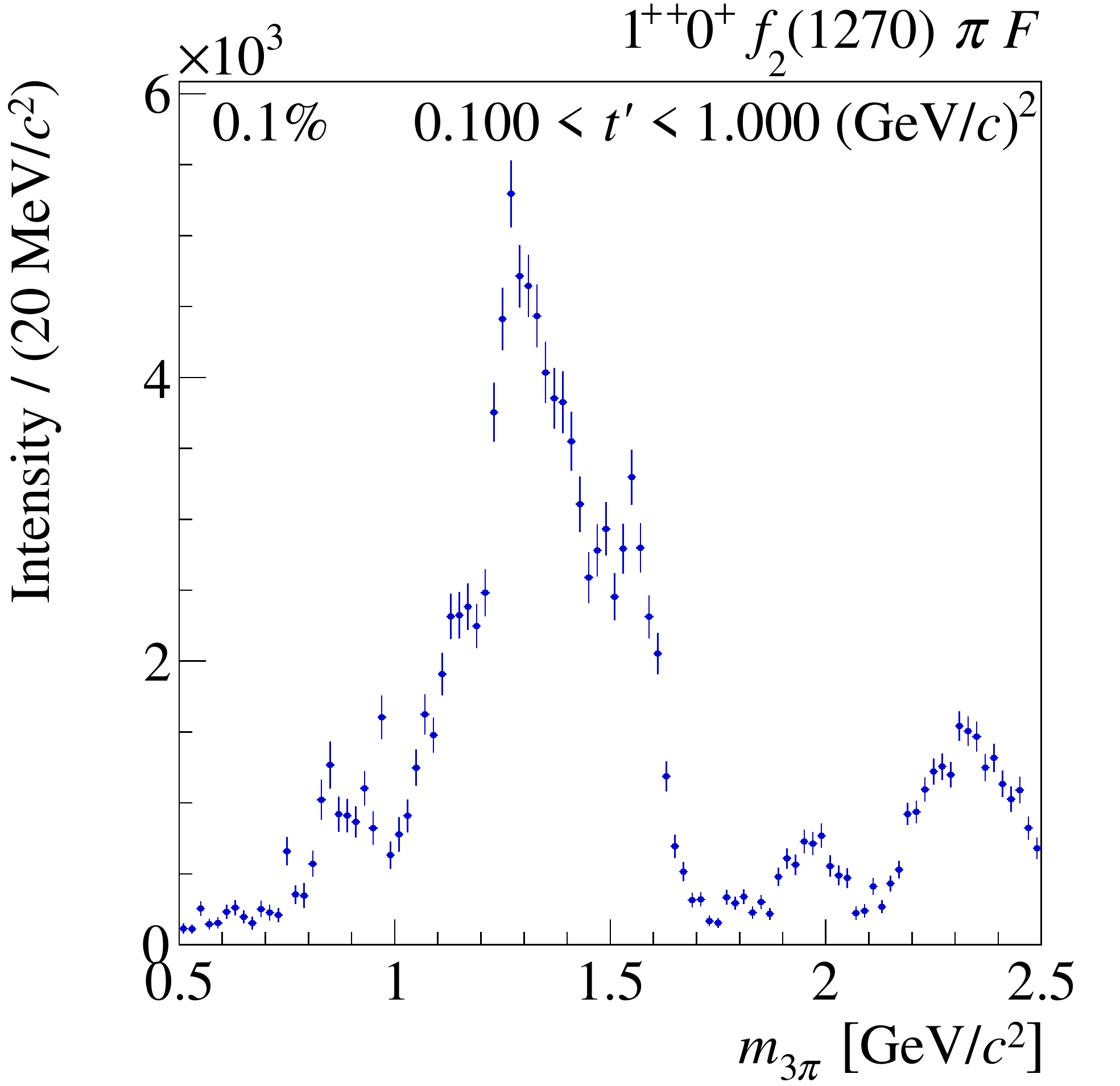}%
  }%
  \\
  \subfloat[][]{%
    \label{fig:int_1pp0p_rho3_D}%
    \includegraphics[width=\threePlotWidth]{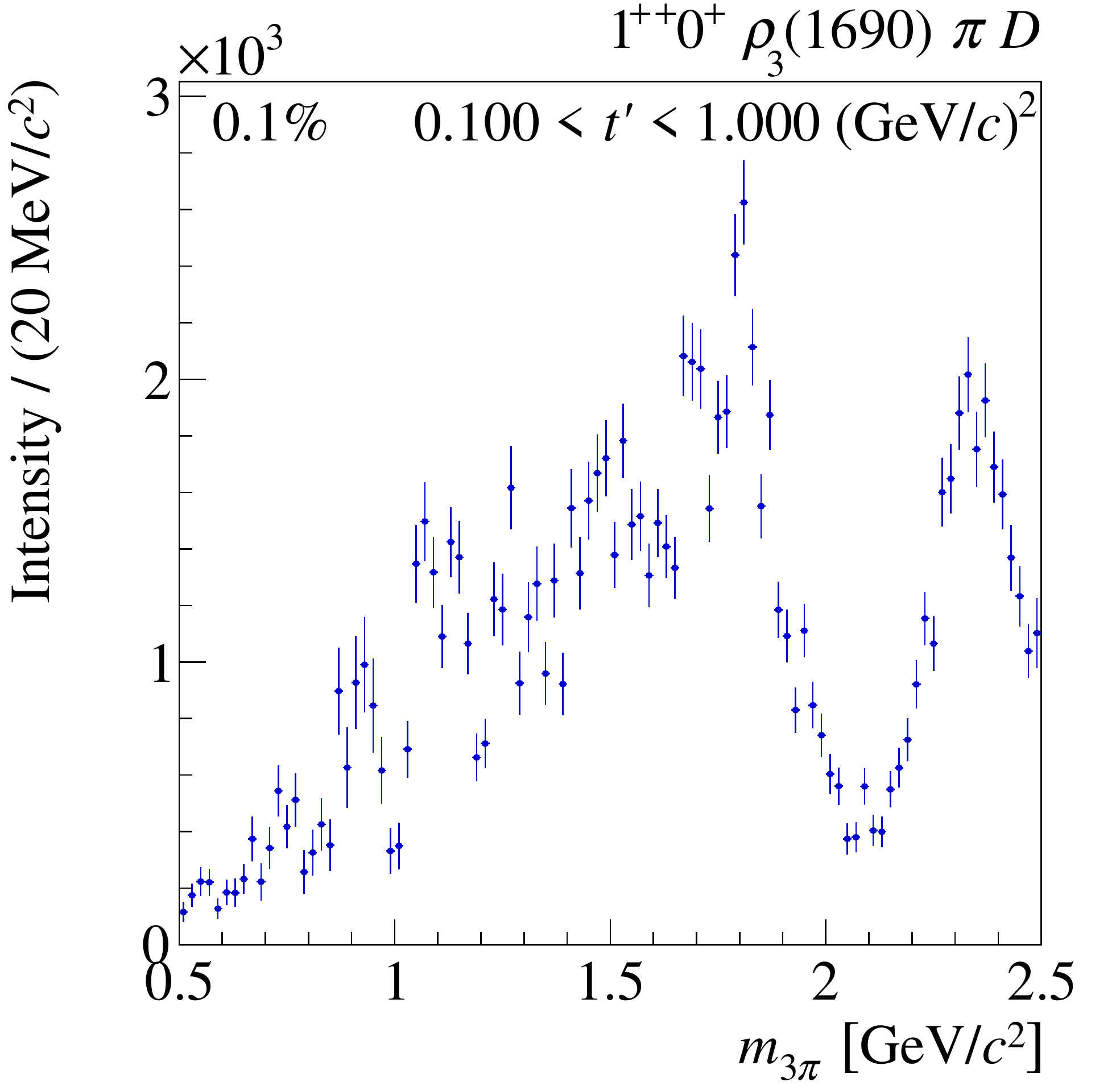}%
  }%
  \subfloat[][]{%
    \label{fig:int_1pp0p_rho3_G}%
    \includegraphics[width=\threePlotWidth]{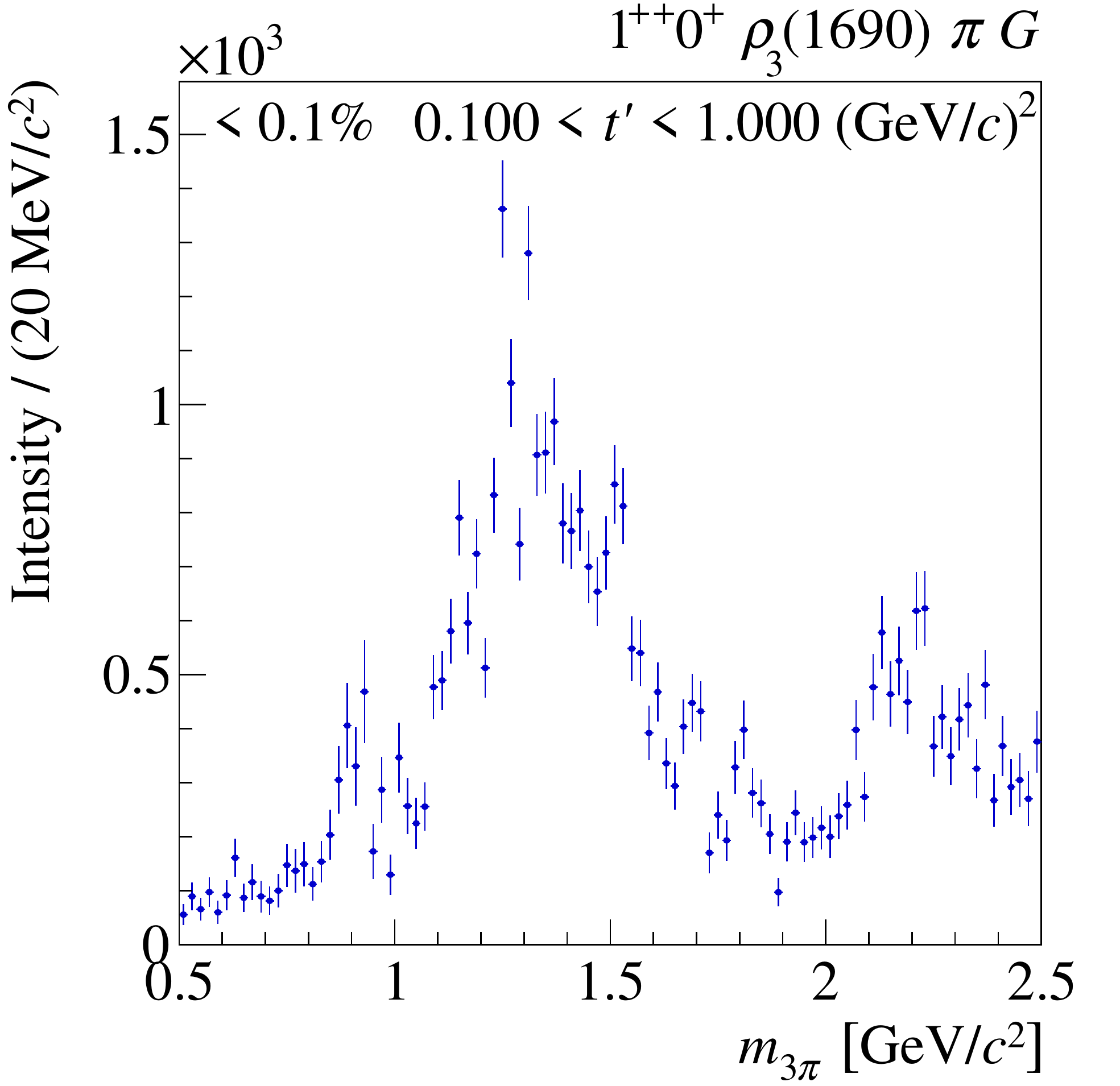}%
  }%
  \caption{The \tpr-summed intensities of partial waves with $\JPC =
    1^{++}$ and positive reflectivity.}
  \label{fig:intensities_1pp_2}
\end{figure}

\subsubsection{$\JPC = 1^{-+}$ Wave}

\begin{figure}[H]
  \centering
  \subfloat[][]{%
    \label{fig:int_1mp1p_rho_P}%
    \includegraphics[width=\threePlotWidth]{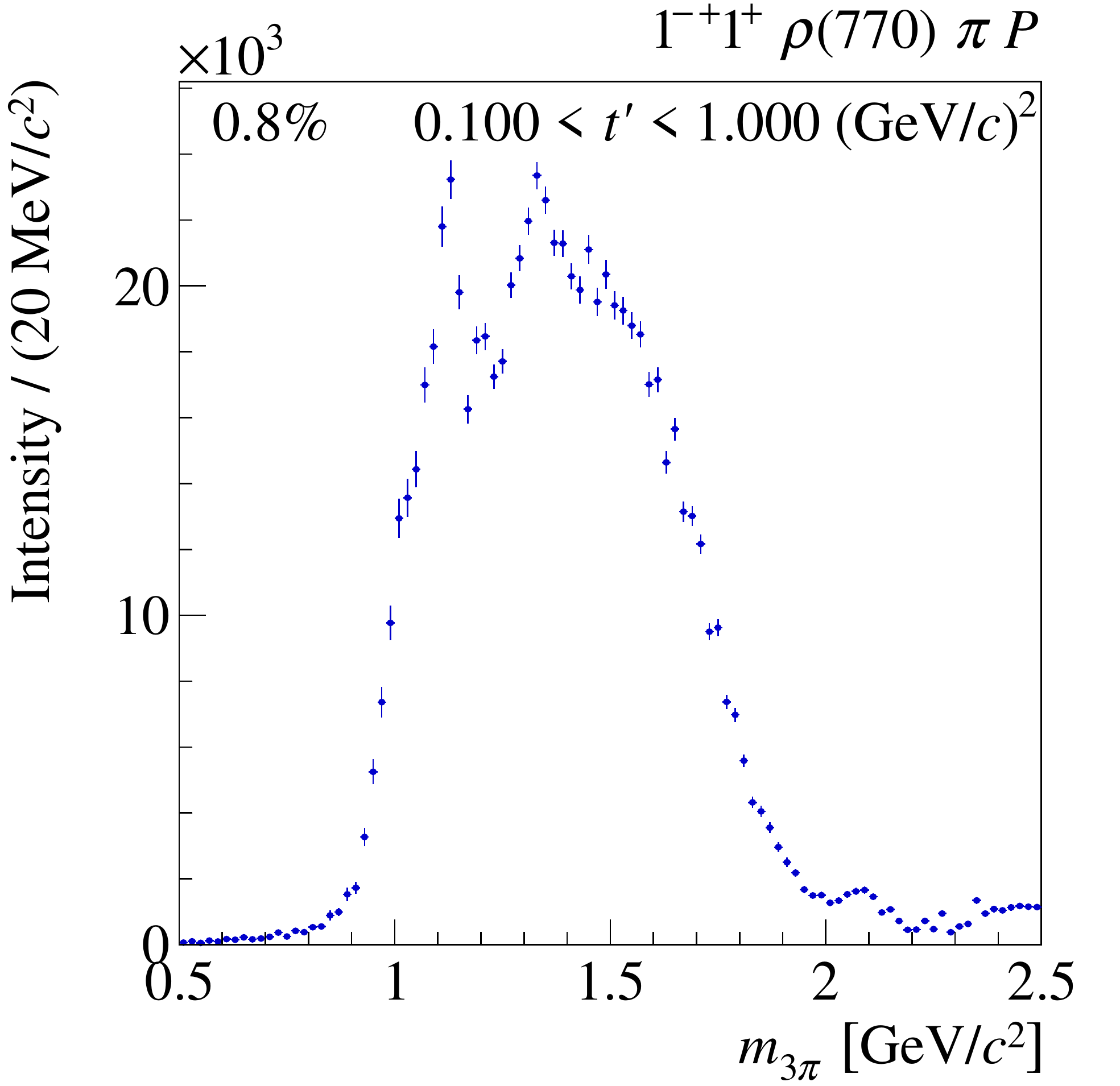}%
  }%
  \caption{The \tpr-summed intensity of the partial wave with
    spin-exotic $\JPC = 1^{-+}$ and positive reflectivity.}
  \label{fig:intensities_1mp}
\end{figure}

\clearpage
\subsubsection{$\JPC = 2^{++}$ Waves}

\begin{figure}[H]
  \centering
  \subfloat[][]{%
    \label{fig:int_2pp2p_f2_P}%
    \includegraphics[width=\threePlotWidth]{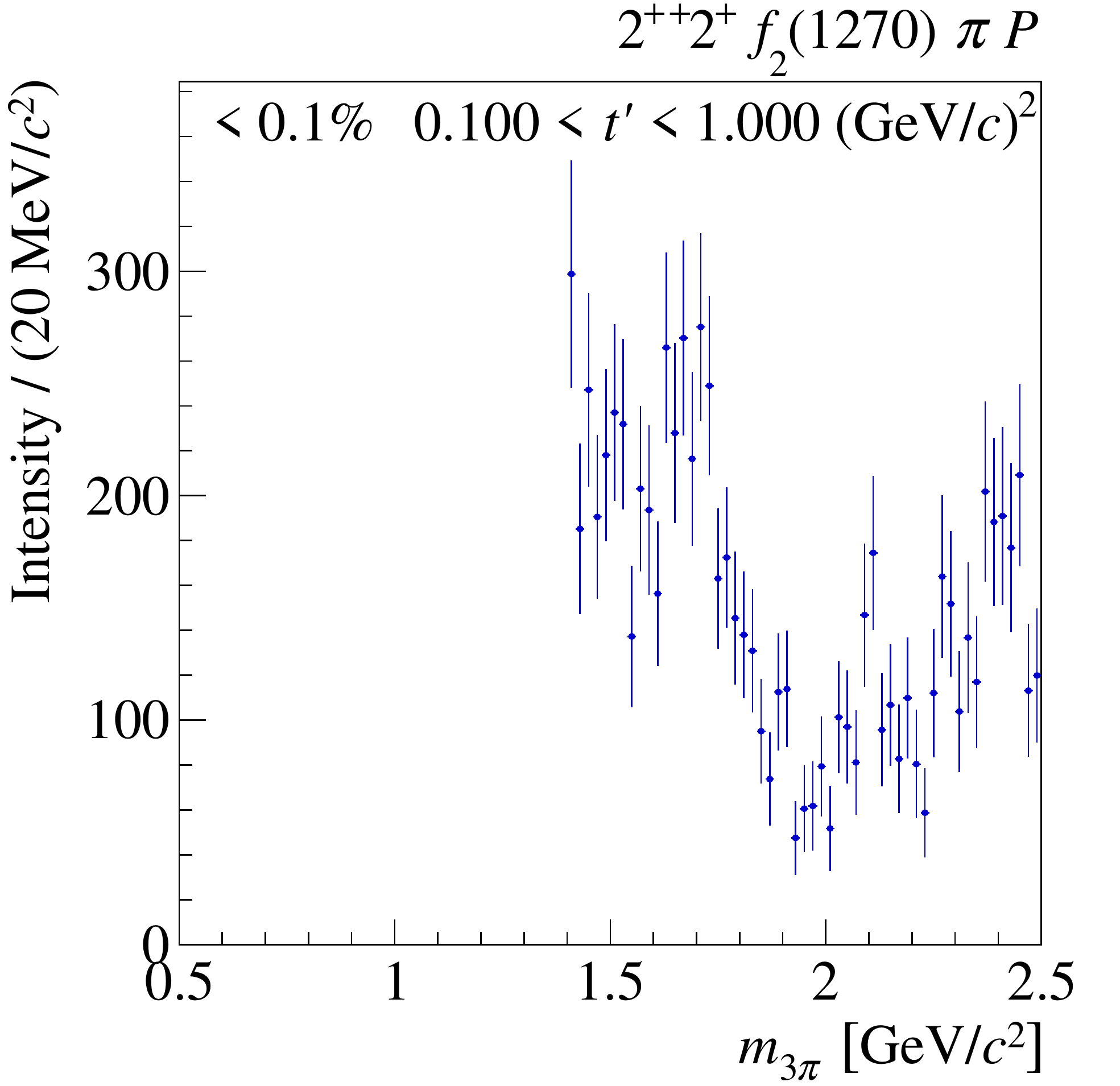}%
  }%
  \subfloat[][]{%
    \label{fig:int_2pp1p_rho3_D}%
    \includegraphics[width=\threePlotWidth]{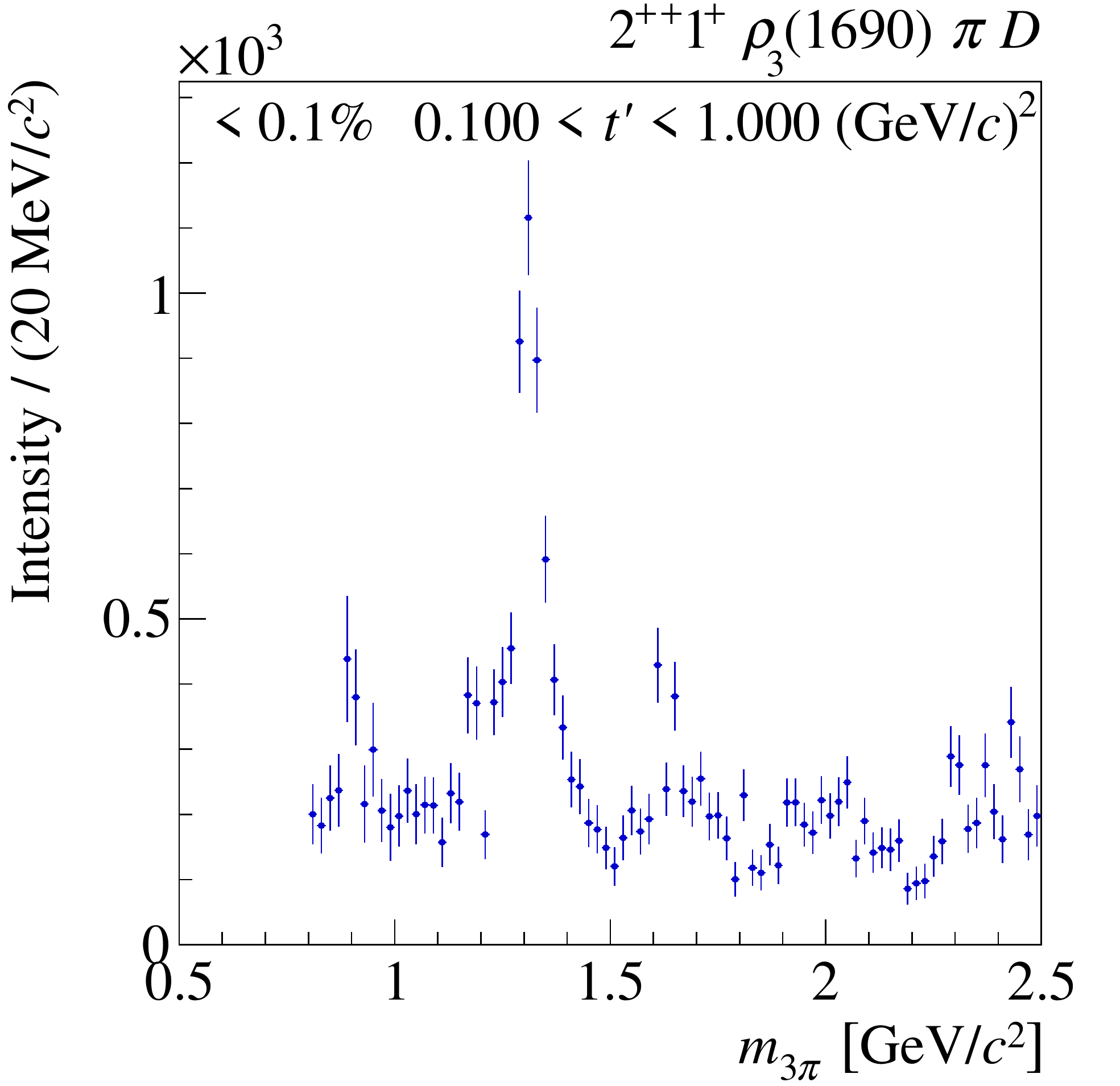}%
  }%
  \caption{The \tpr-summed intensities of partial waves with $\JPC =
    2^{++}$ and positive reflectivity.}
  \label{fig:intensities_2pp}
\end{figure}

\subsubsection{$\JPC = 2^{-+}$ Waves}

\begin{figure}[H]
  \centering
  \subfloat[][]{%
    \label{fig:int_2mp1p_pipiS_D}%
    \includegraphics[width=\threePlotWidth]{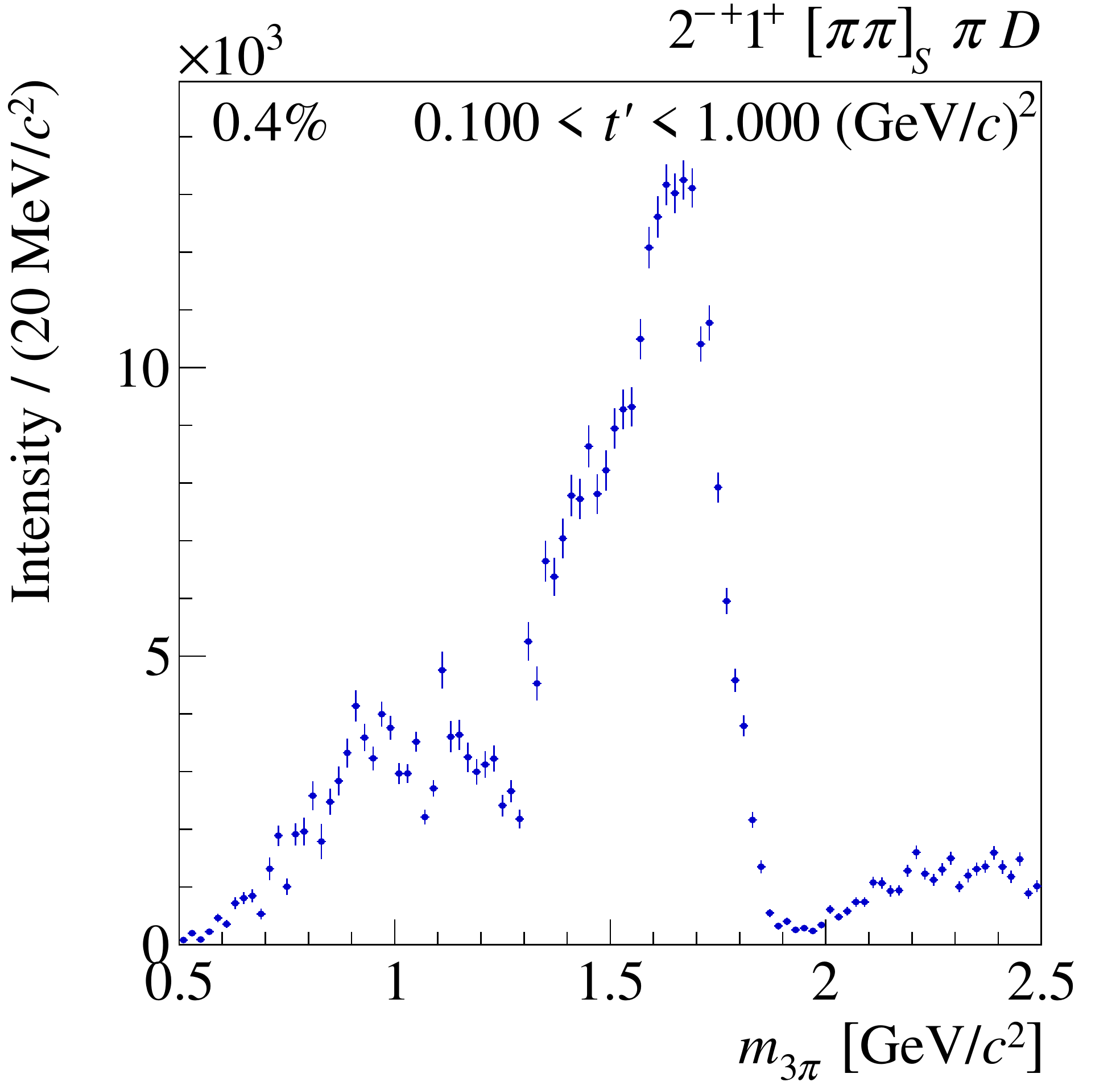}%
  }%
  \subfloat[][]{%
    \label{fig:int_2mp0p_rho_P}%
    \includegraphics[width=\threePlotWidth]{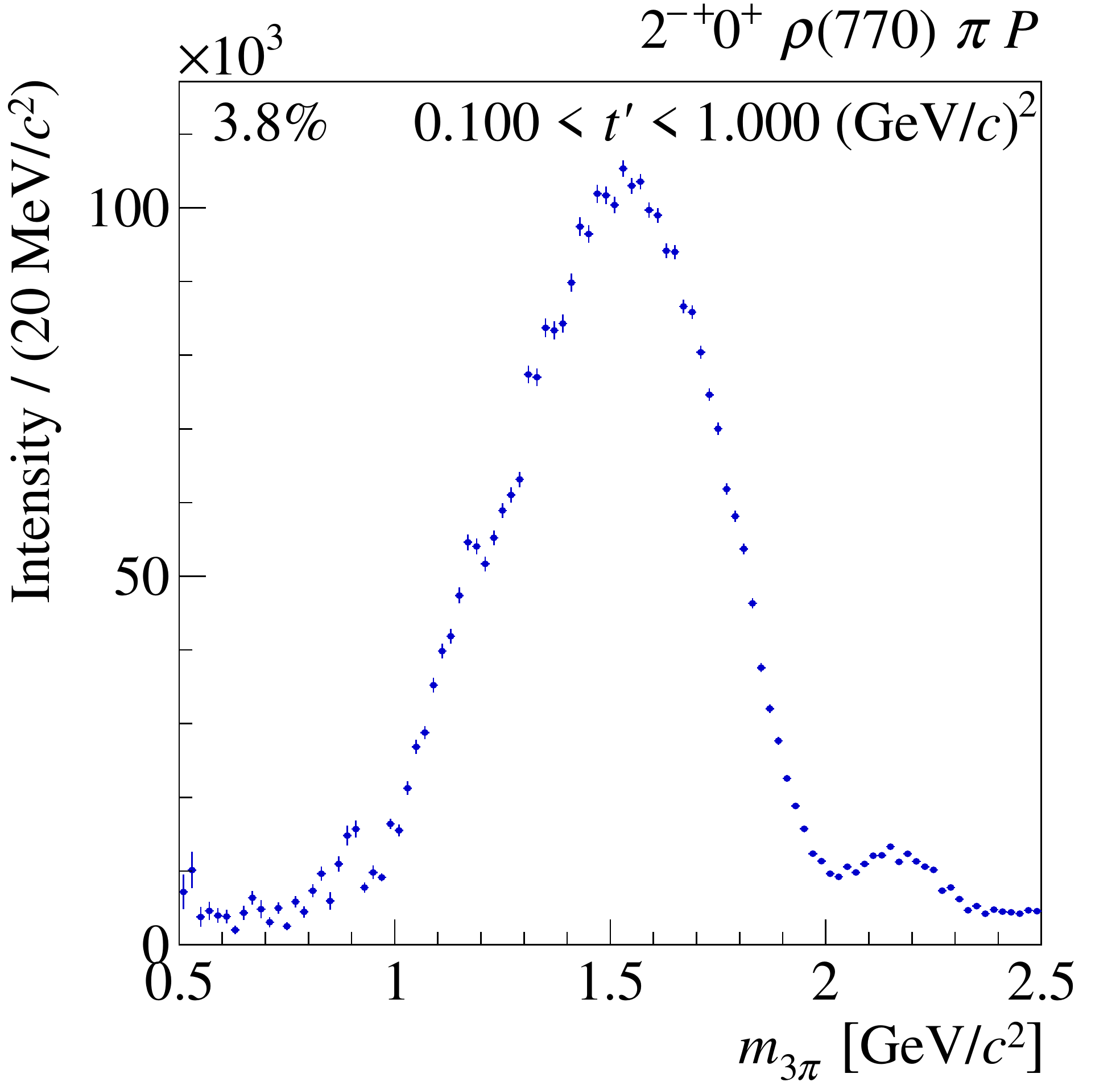}%
  }%
  \subfloat[][]{%
    \label{fig:int_2mp1p_rho_P}%
    \includegraphics[width=\threePlotWidth]{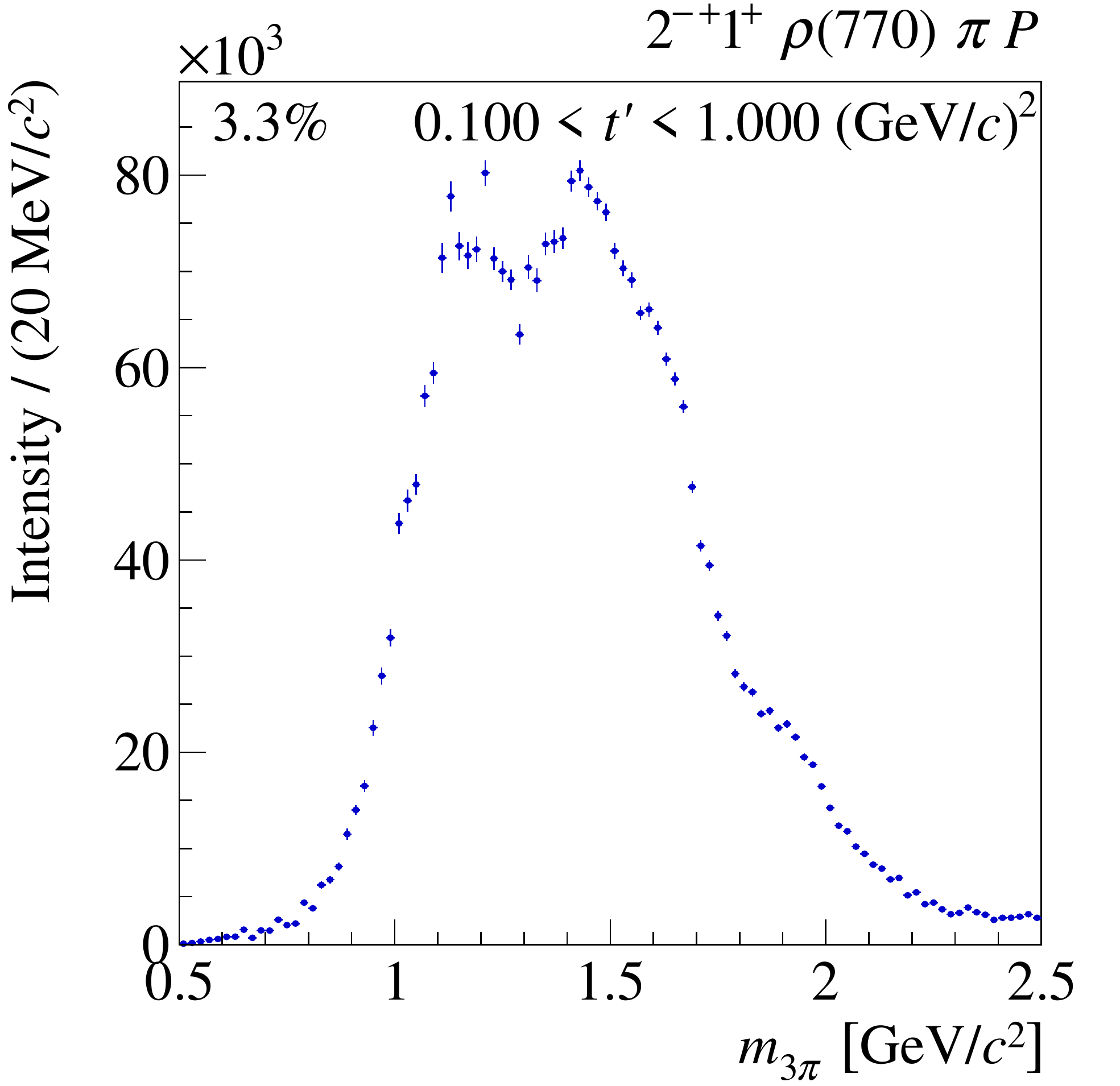}%
  }%
  \\
  \subfloat[][]{%
    \label{fig:int_2mp2p_rho_P}%
    \includegraphics[width=\threePlotWidth]{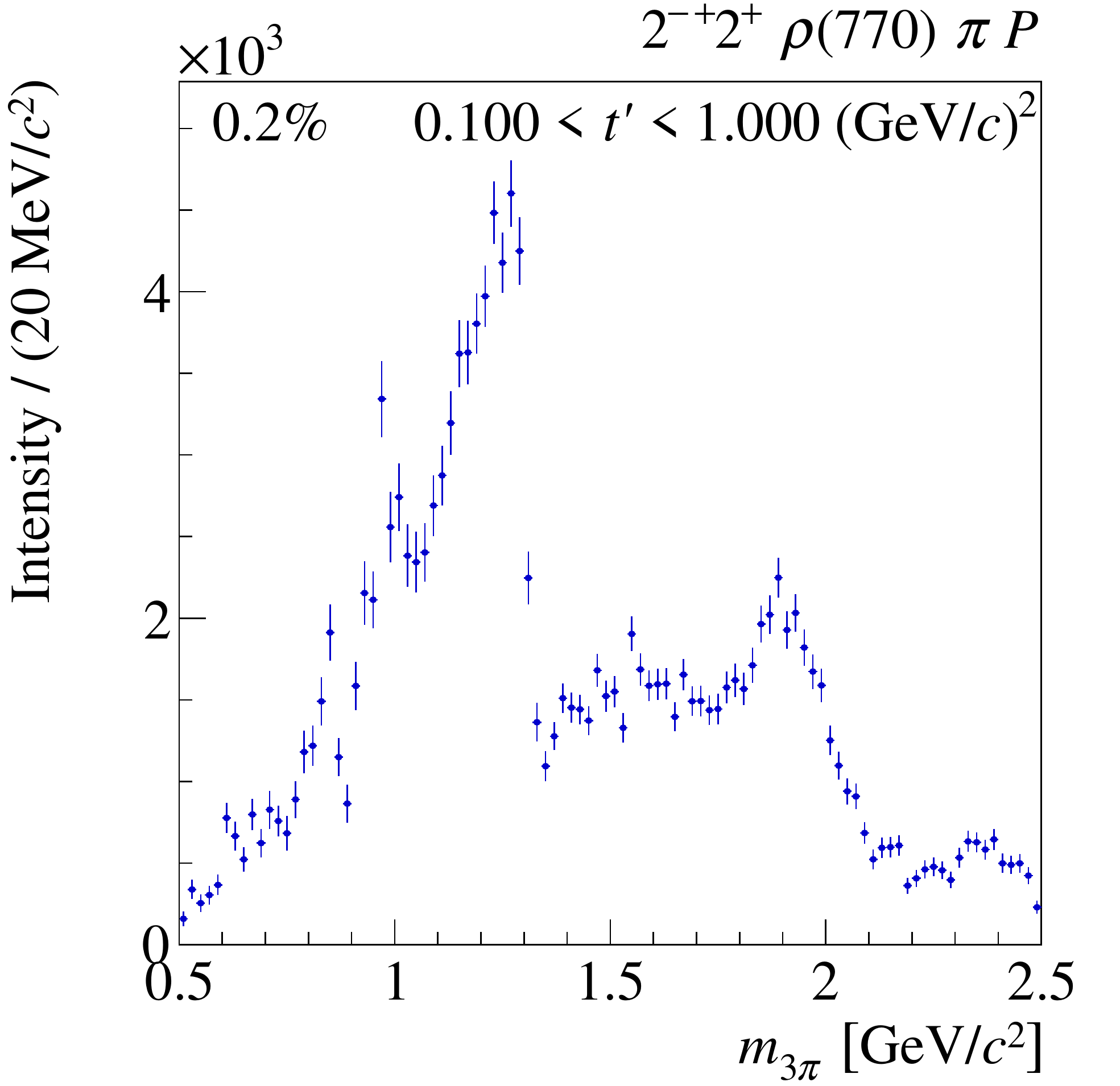}%
  }%
  \subfloat[][]{%
    \label{fig:int_2mp1p_rho_F}%
    \includegraphics[width=\threePlotWidth]{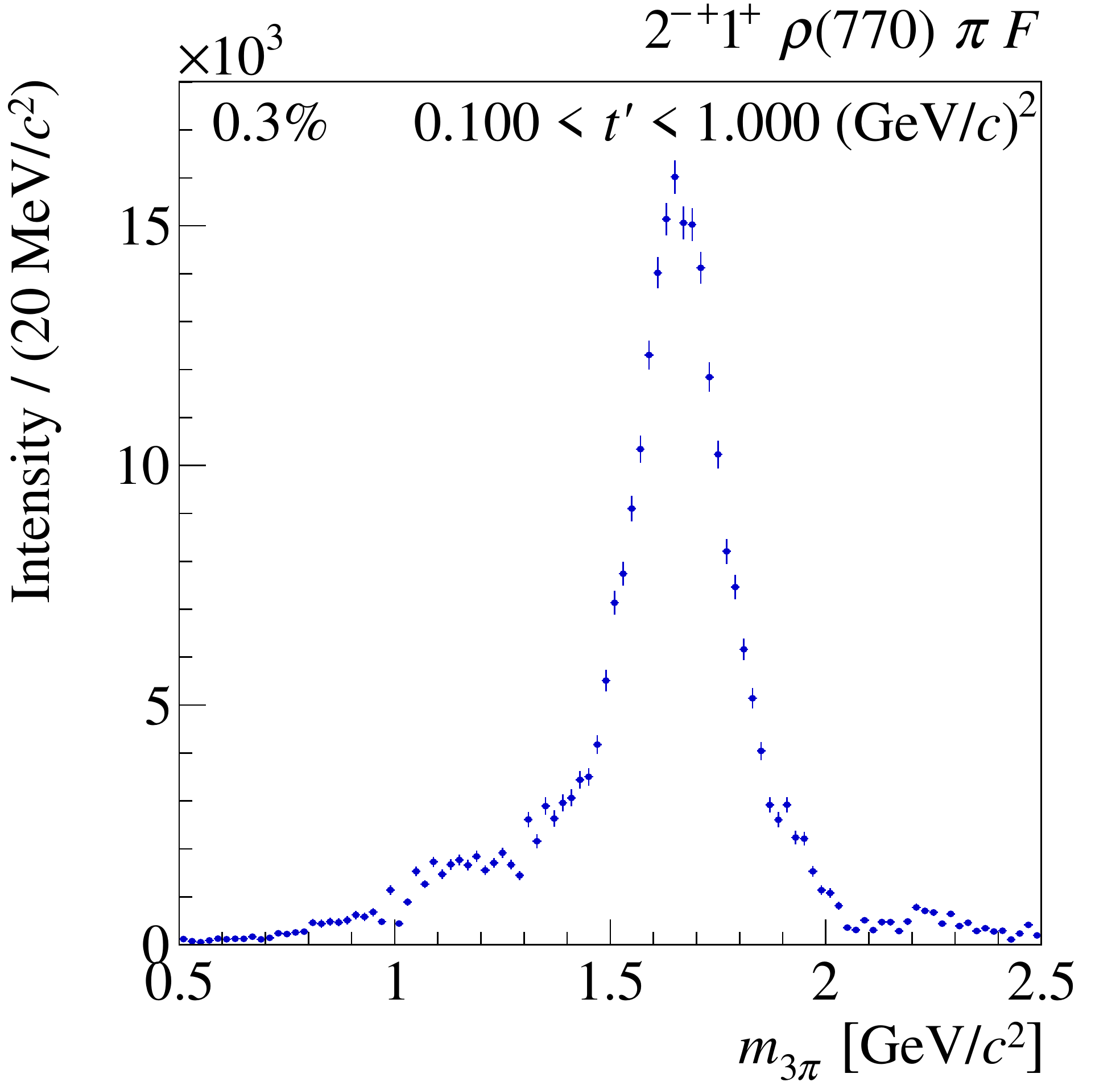}%
  }%
  \subfloat[][]{%
    \label{fig:int_2mp2p_f2_S}%
    \includegraphics[width=\threePlotWidth]{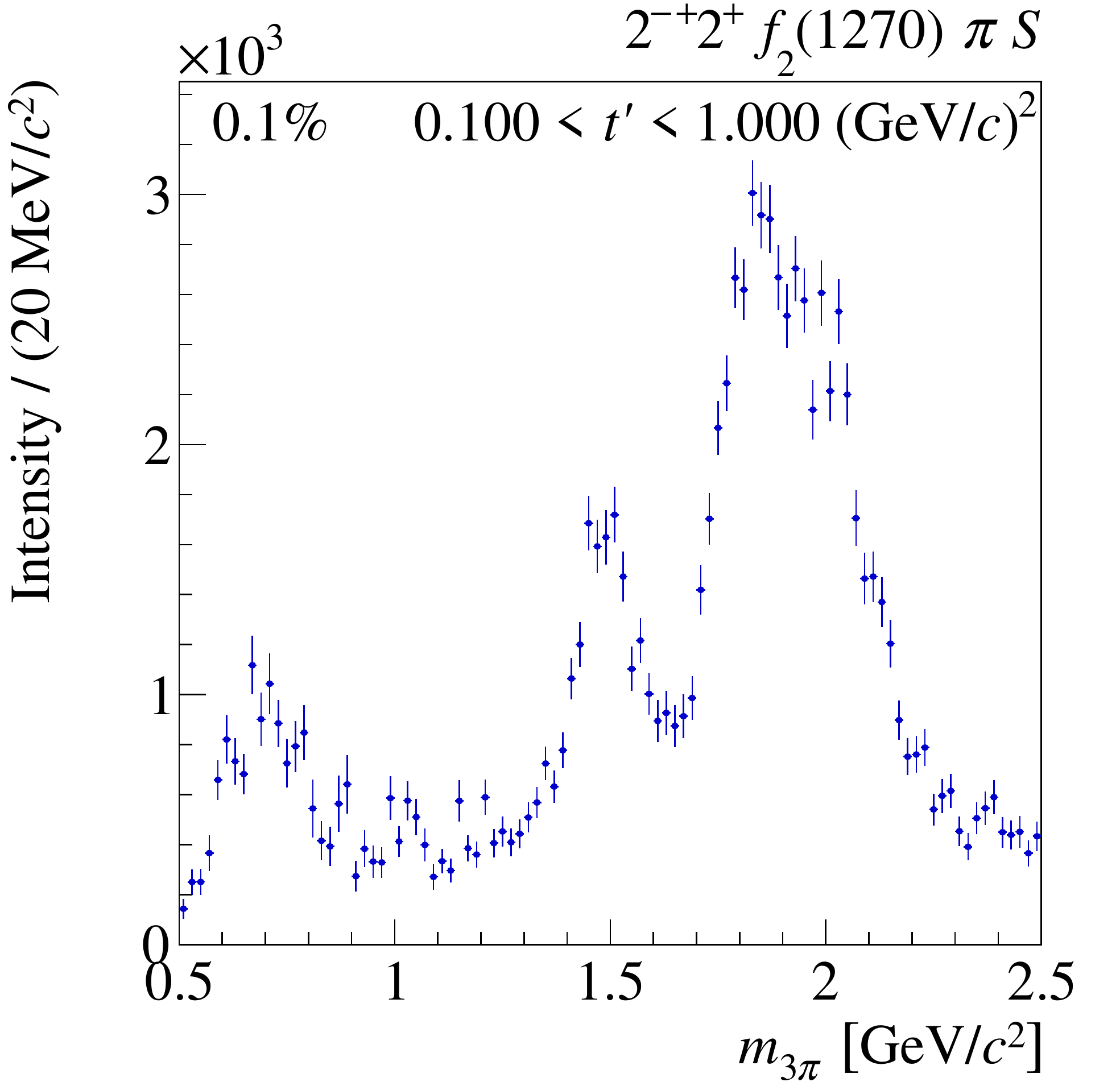}%
  }%
  \caption{The \tpr-summed intensities of partial waves with $\JPC =
    2^{-+}$ and positive reflectivity.}
  \label{fig:intensities_2mp_1}
\end{figure}

\clearpage
\begin{figure}[H]
  \centering
  \subfloat[][]{%
    \label{fig:int_2mp1p_f2_D}%
    \includegraphics[width=\threePlotWidth]{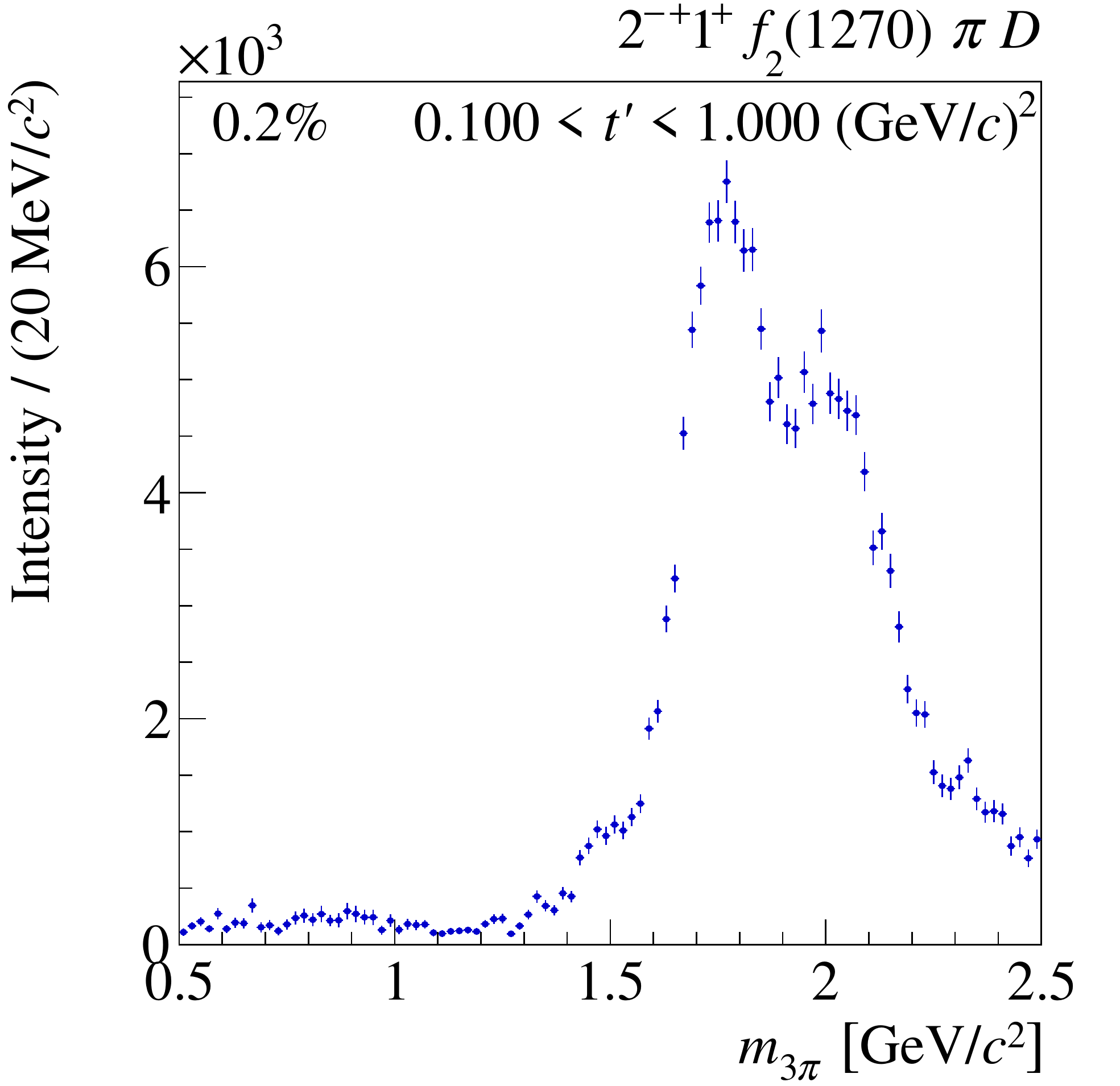}%
  }%
  \subfloat[][]{%
    \label{fig:int_2mp2p_f2_D}%
    \includegraphics[width=\threePlotWidth]{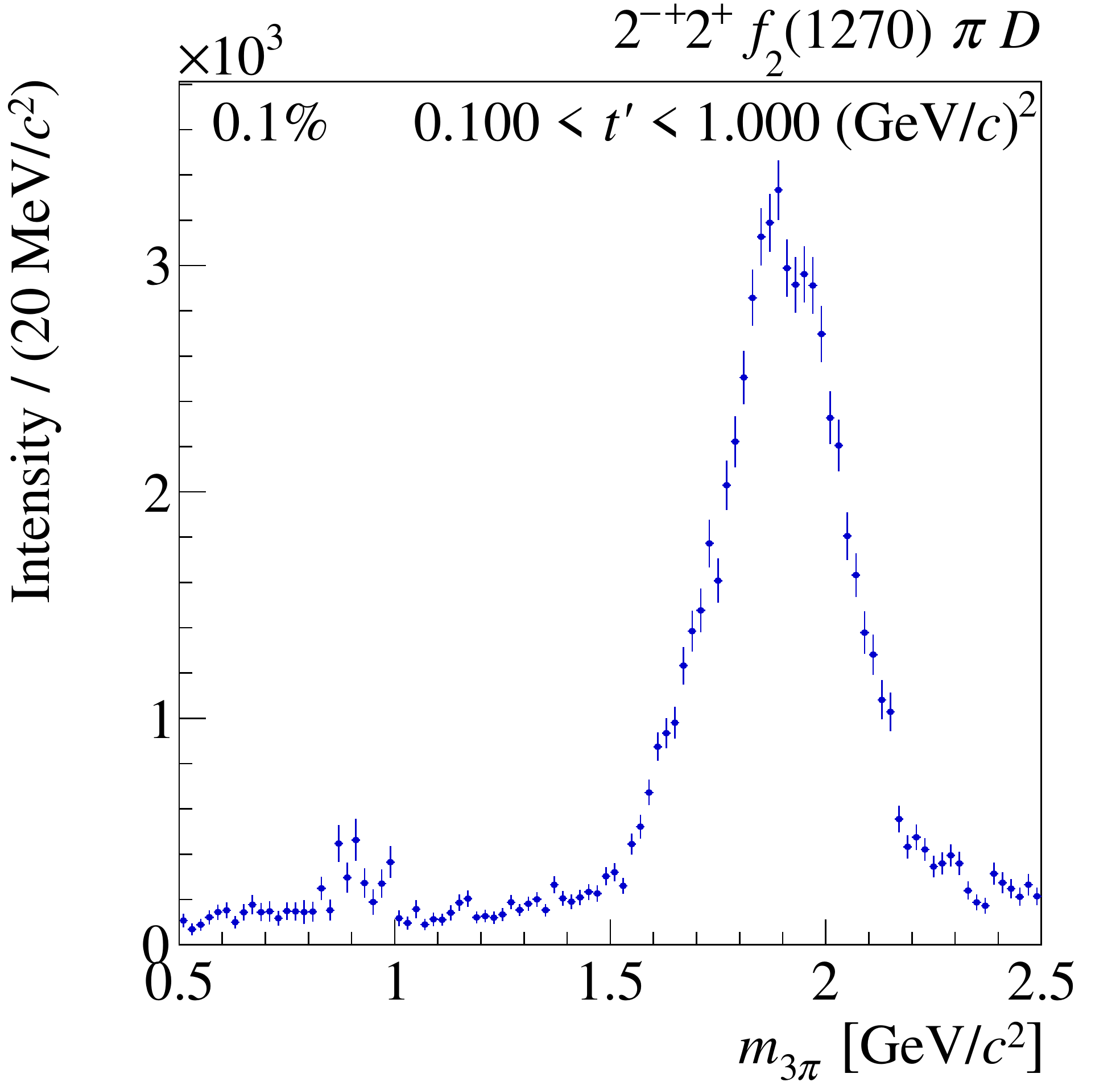}%
  }%
  \subfloat[][]{%
    \label{fig:int_2mp0p_f2_G}%
    \includegraphics[width=\threePlotWidth]{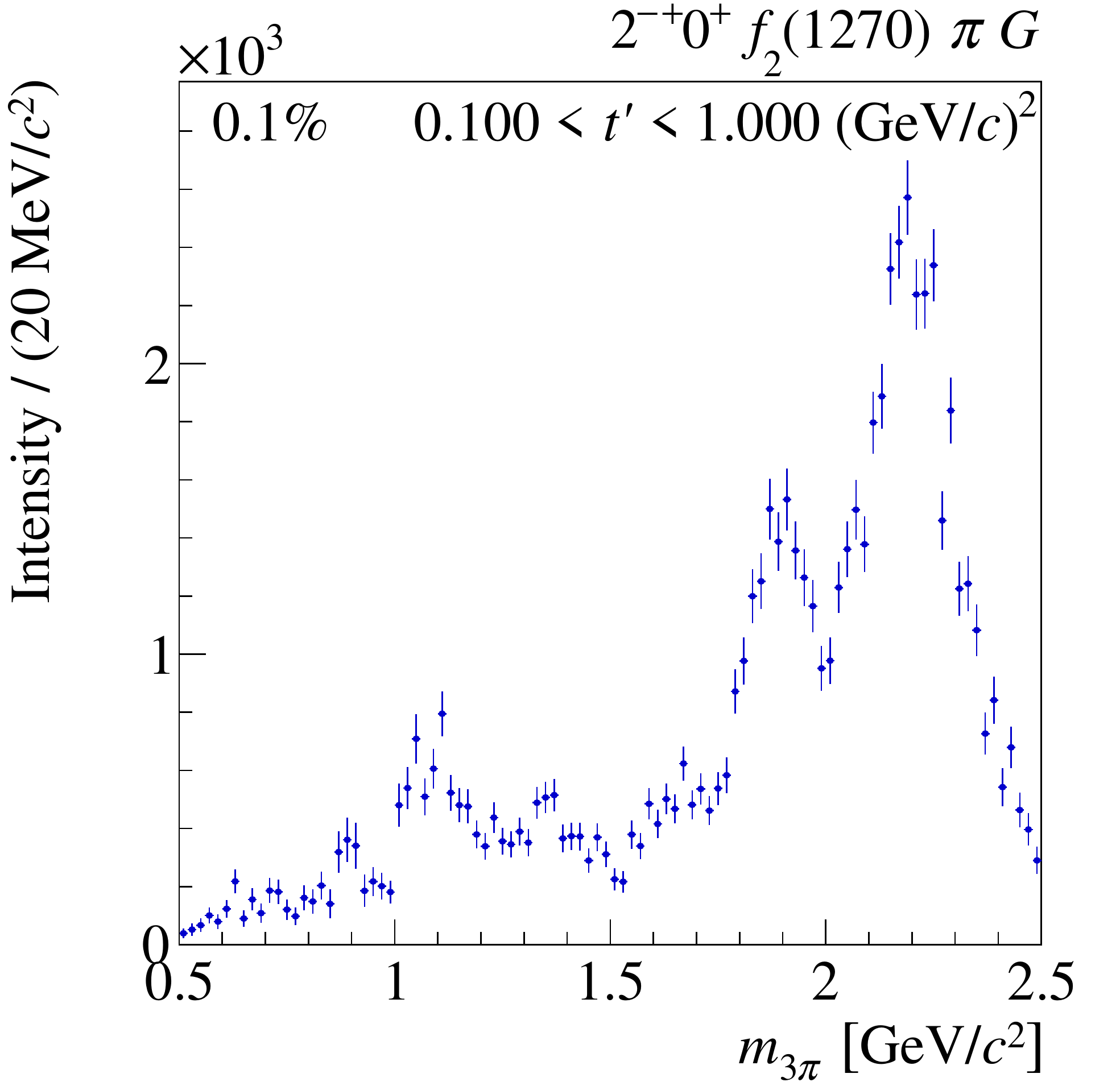}%
  }%
  \\
  \subfloat[][]{%
    \label{fig:int_2mp0p_rho3_P}%
    \includegraphics[width=\threePlotWidth]{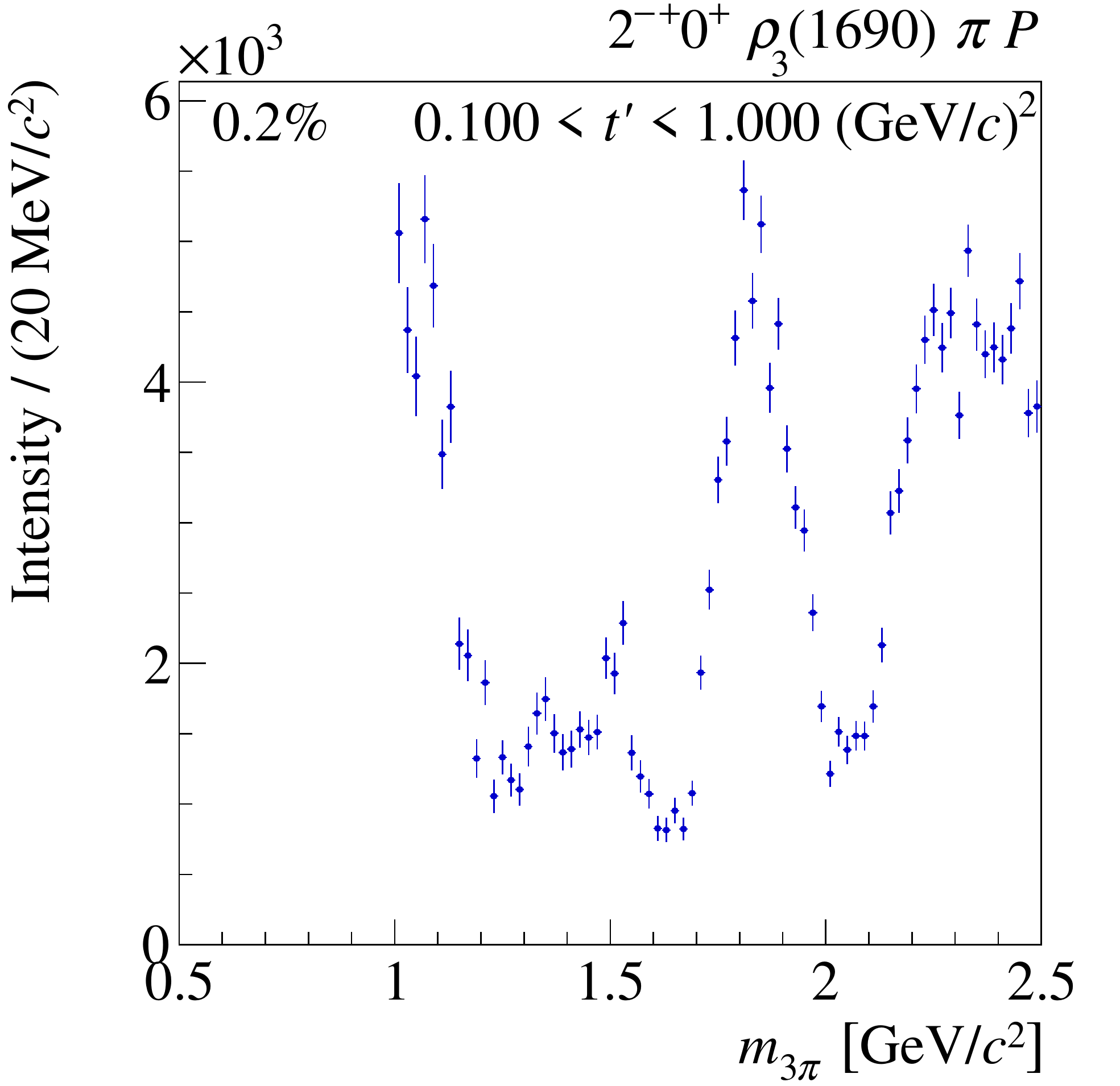}%
  }%
  \subfloat[][]{%
    \label{fig:int_2mp1p_rho3_P}%
    \includegraphics[width=\threePlotWidth]{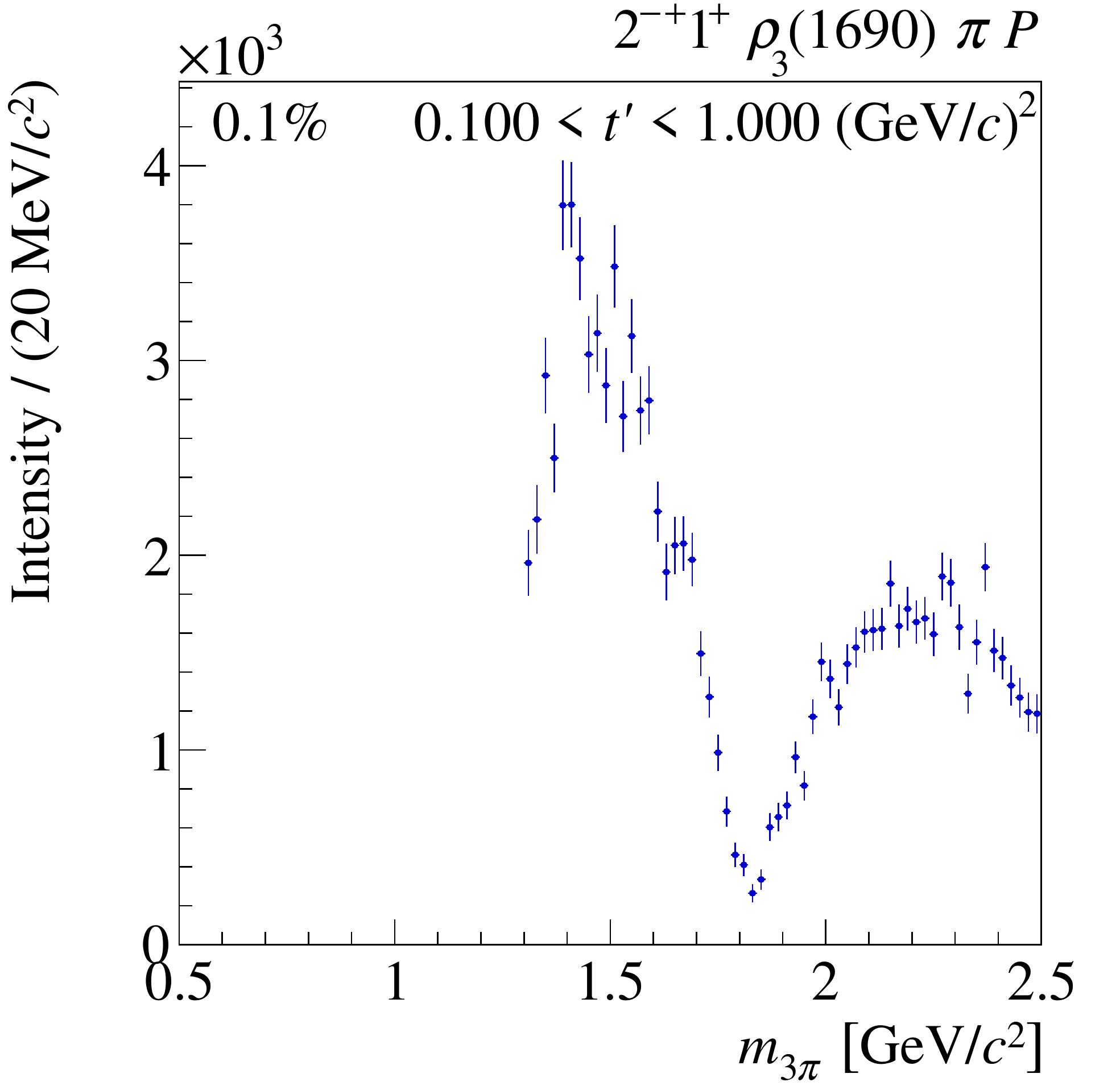}%
  }%
  \caption{The \tpr-summed intensities of partial waves with $\JPC =
    2^{-+}$ and positive reflectivity.}
  \label{fig:intensities_2mp_2}
\end{figure}

\subsubsection{$\JPC = 3^{++}$ Waves}

\begin{figure}[H]
  \centering
  \subfloat[][]{%
    \label{fig:int_3pp0p_pipiS_F}%
    \includegraphics[width=\threePlotWidth]{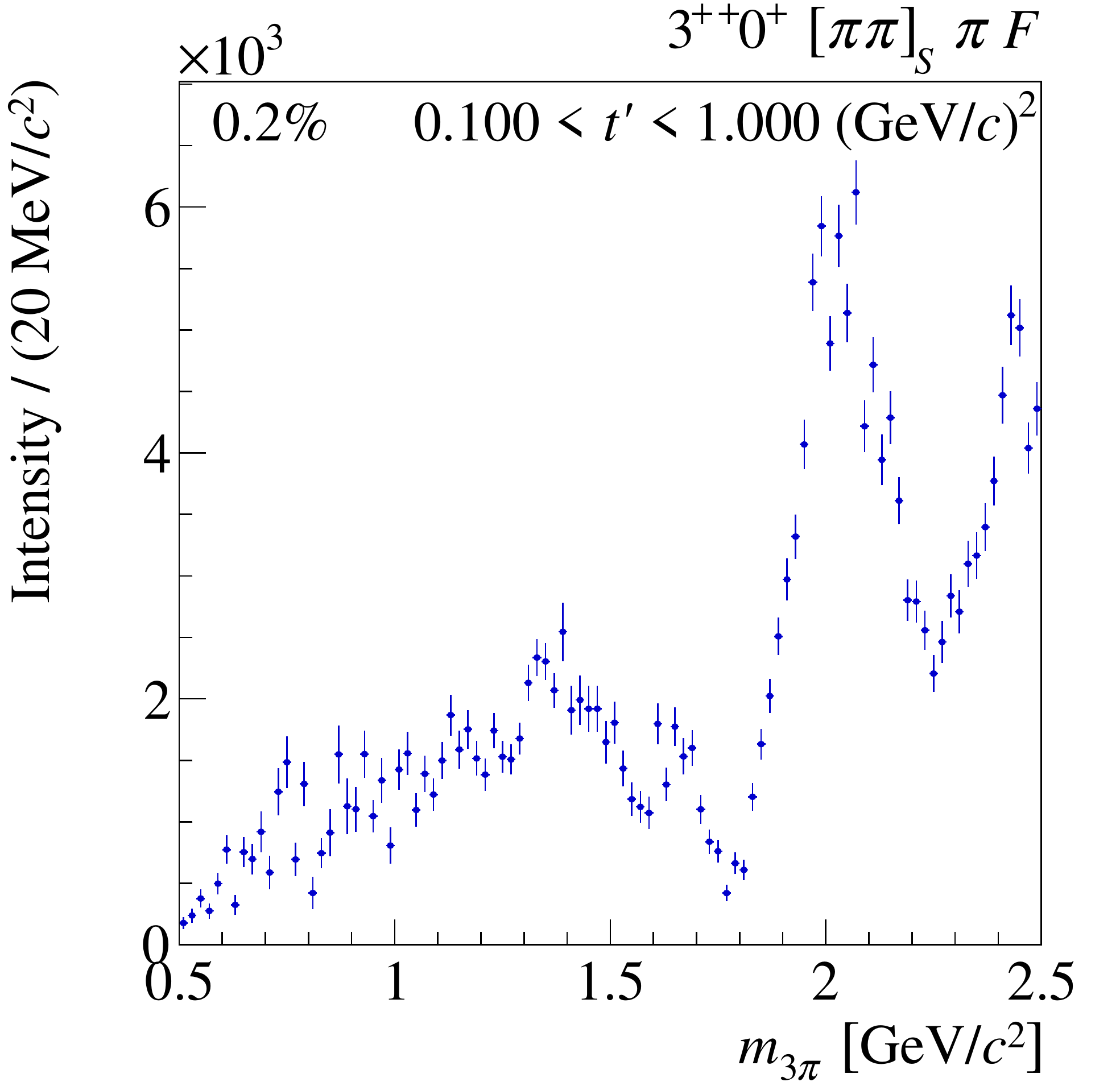}%
  }%
  \subfloat[][]{%
    \label{fig:int_3pp1p_pipiS_F}%
    \includegraphics[width=\threePlotWidth]{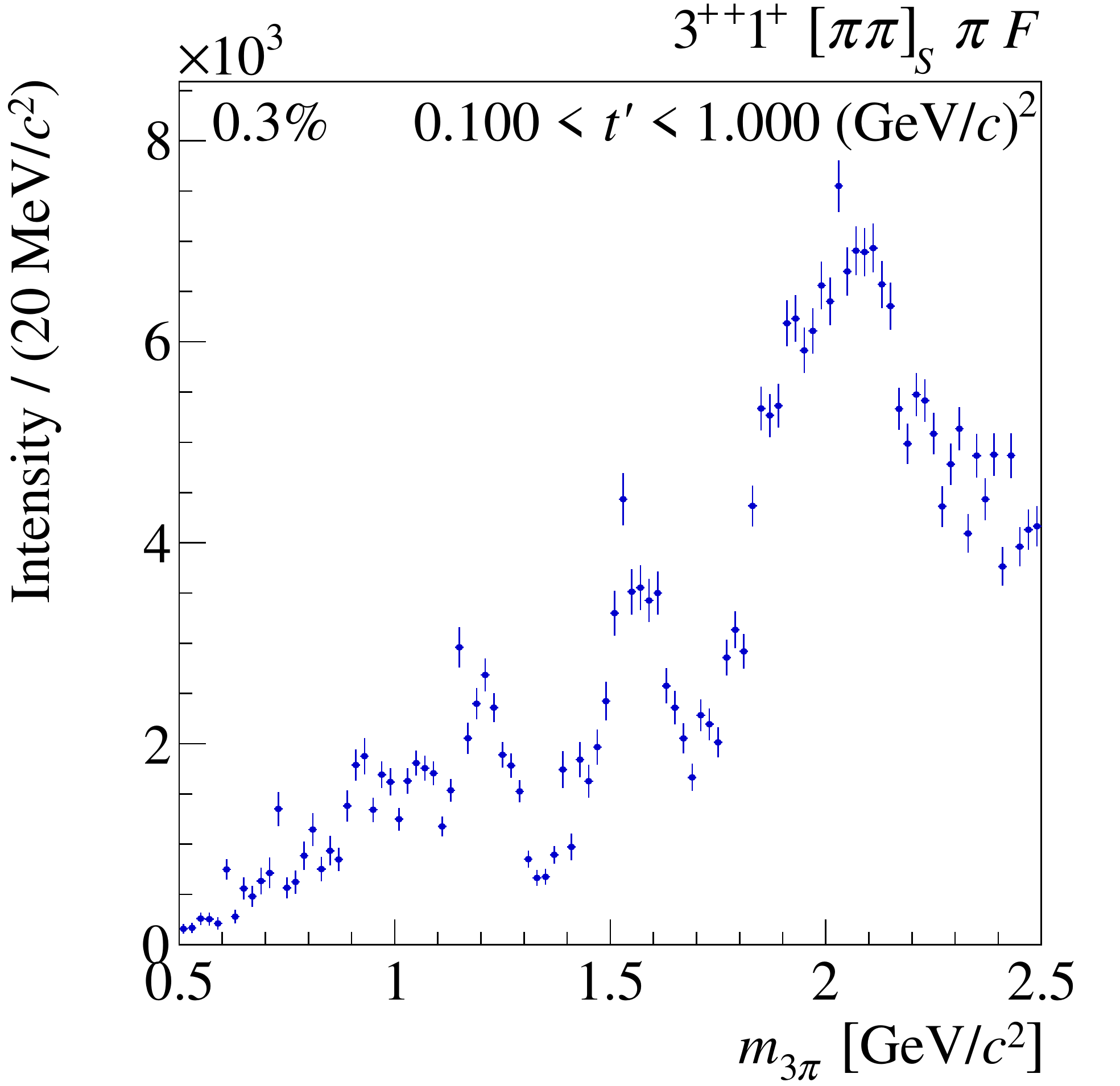}%
  }%
  \subfloat[][]{%
    \label{fig:int_3pp0p_rho_D}%
    \includegraphics[width=\threePlotWidth]{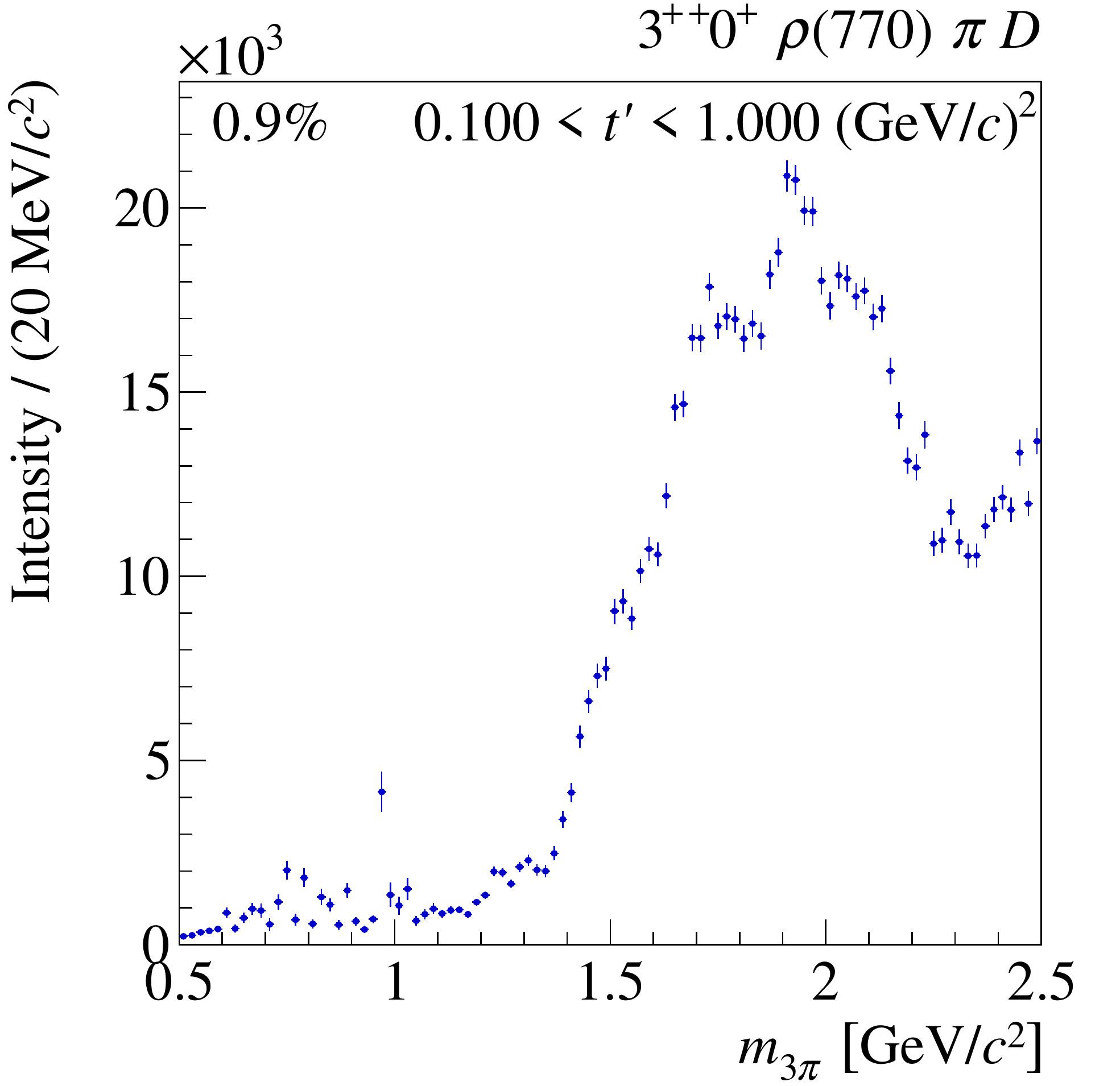}%
  }%
  \caption{The \tpr-summed intensities of partial waves with $\JPC =
    3^{++}$ and positive reflectivity.}
  \label{fig:intensities_3pp_1}
\end{figure}

\clearpage
\begin{figure}[H]
  \centering
  \subfloat[][]{%
    \label{fig:int_3pp1p_rho_D}%
    \includegraphics[width=\threePlotWidth]{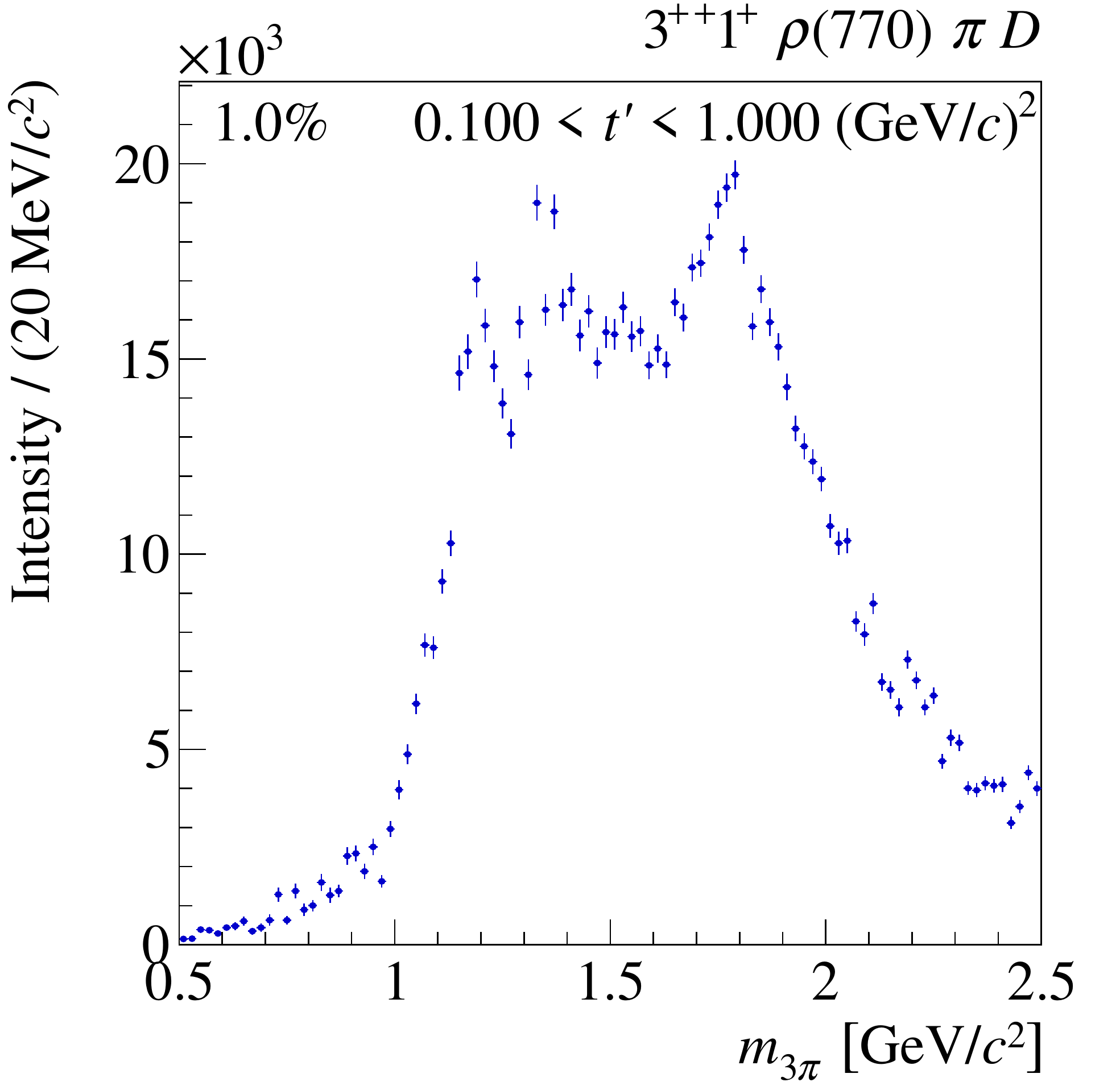}%
  }%
  \subfloat[][]{%
    \label{fig:int_3pp0p_rho_G}%
    \includegraphics[width=\threePlotWidth]{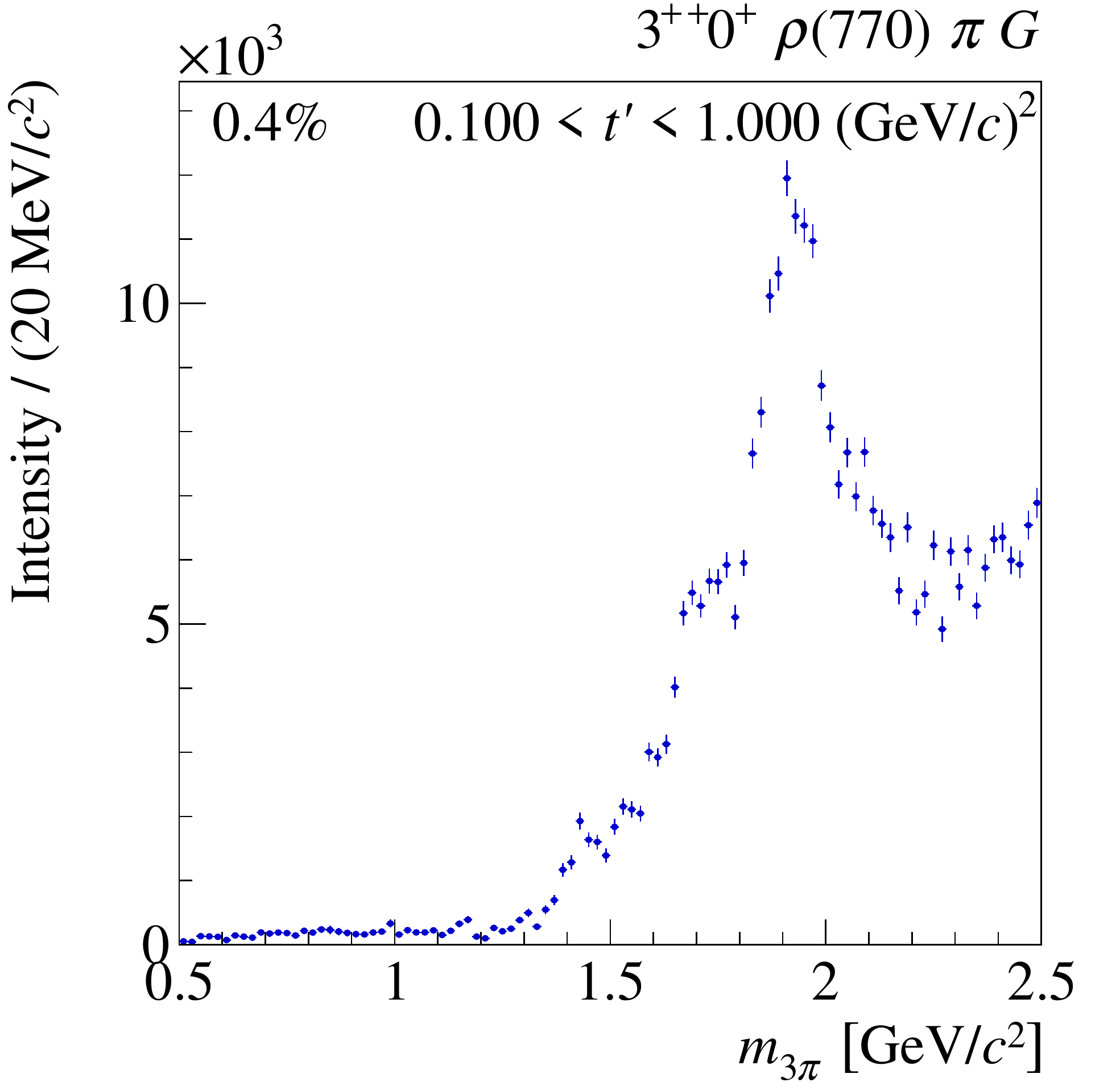}%
  }%
  \subfloat[][]{%
    \label{fig:int_3pp1p_rho_G}%
    \includegraphics[width=\threePlotWidth]{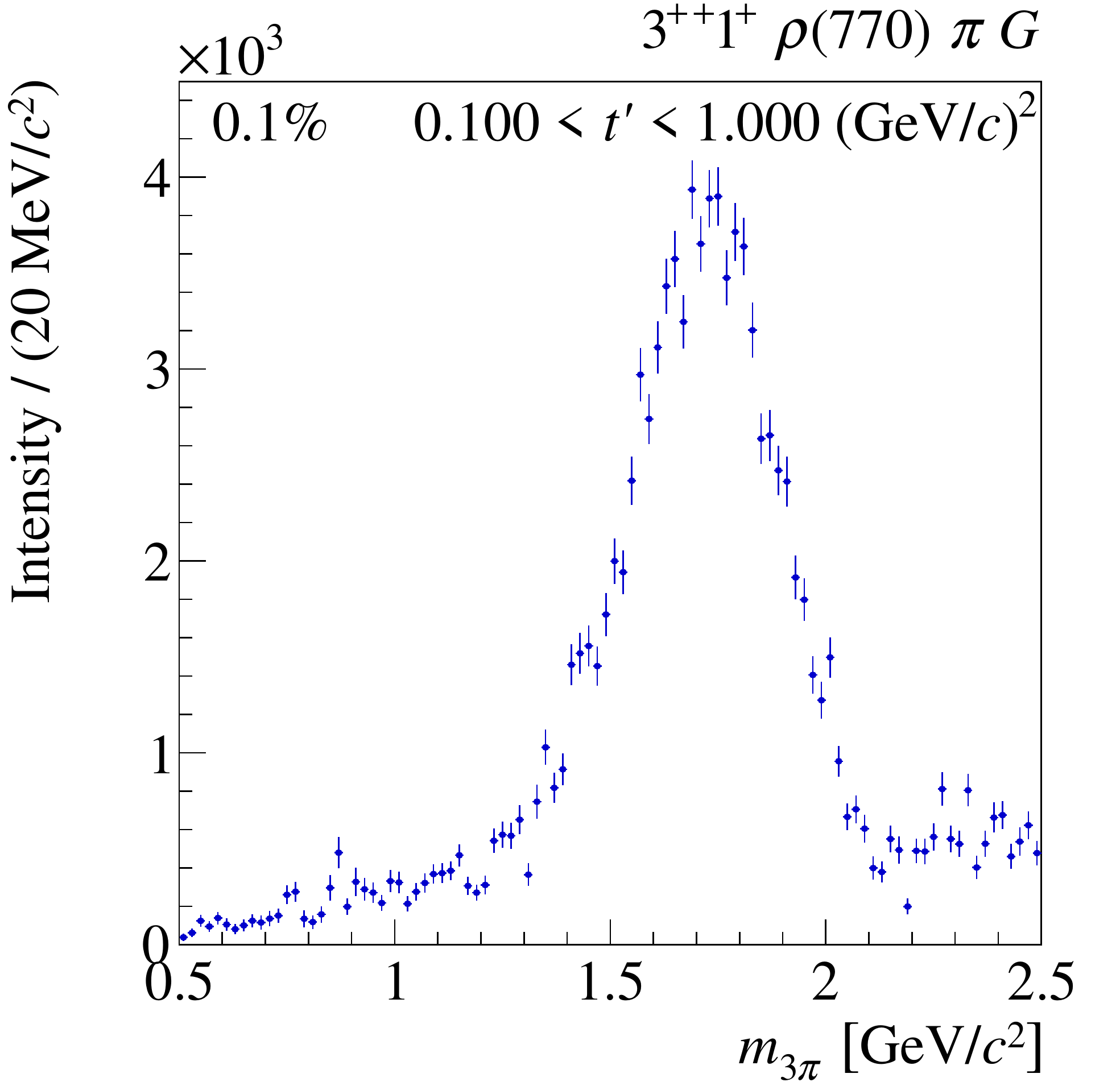}%
  }%
  \\
  \subfloat[][]{%
    \label{fig:int_3pp0p_f2_P}%
    \includegraphics[width=\threePlotWidth]{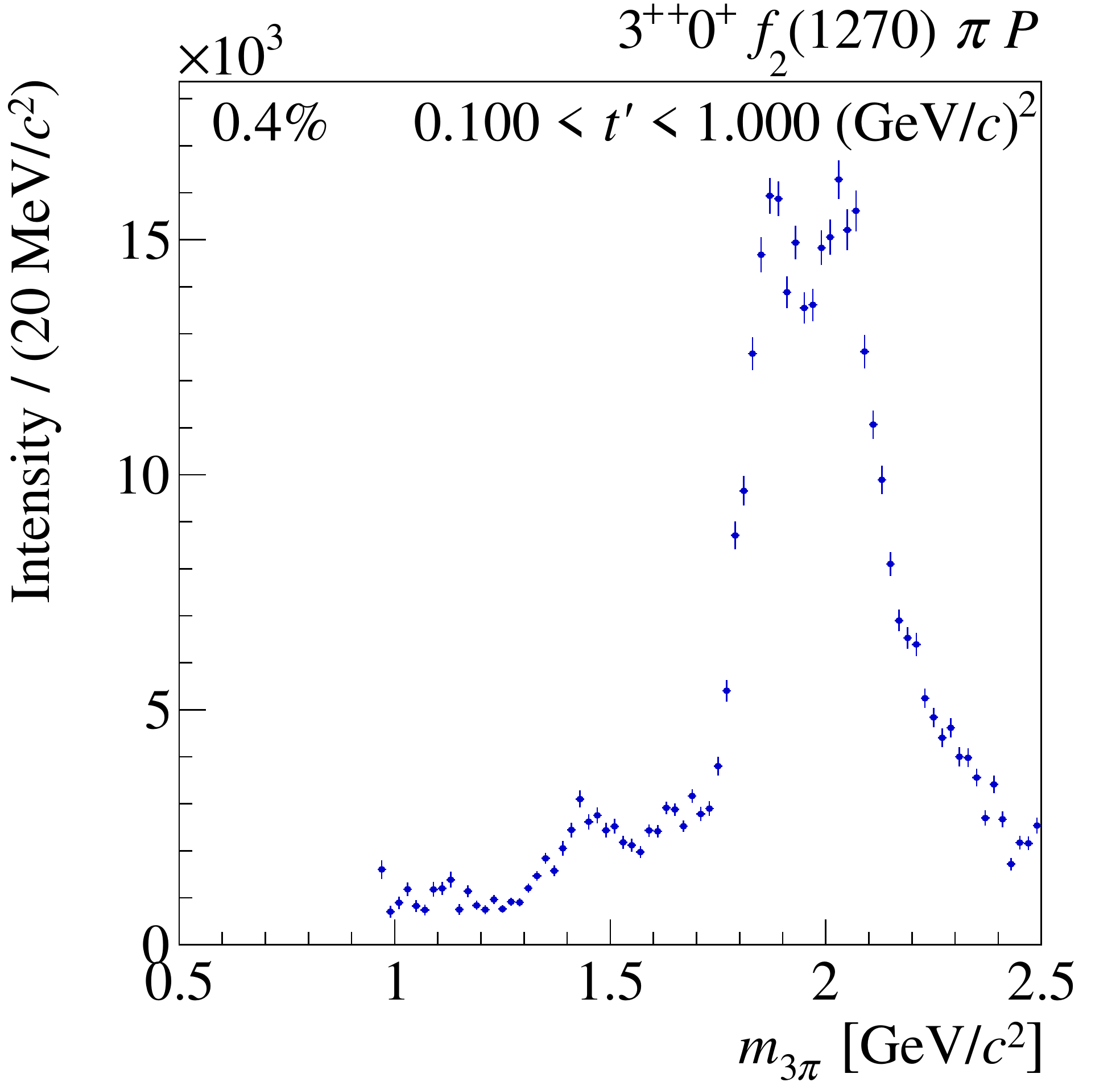}%
  }%
  \subfloat[][]{%
    \label{fig:int_3pp1p_f2_P}%
    \includegraphics[width=\threePlotWidth]{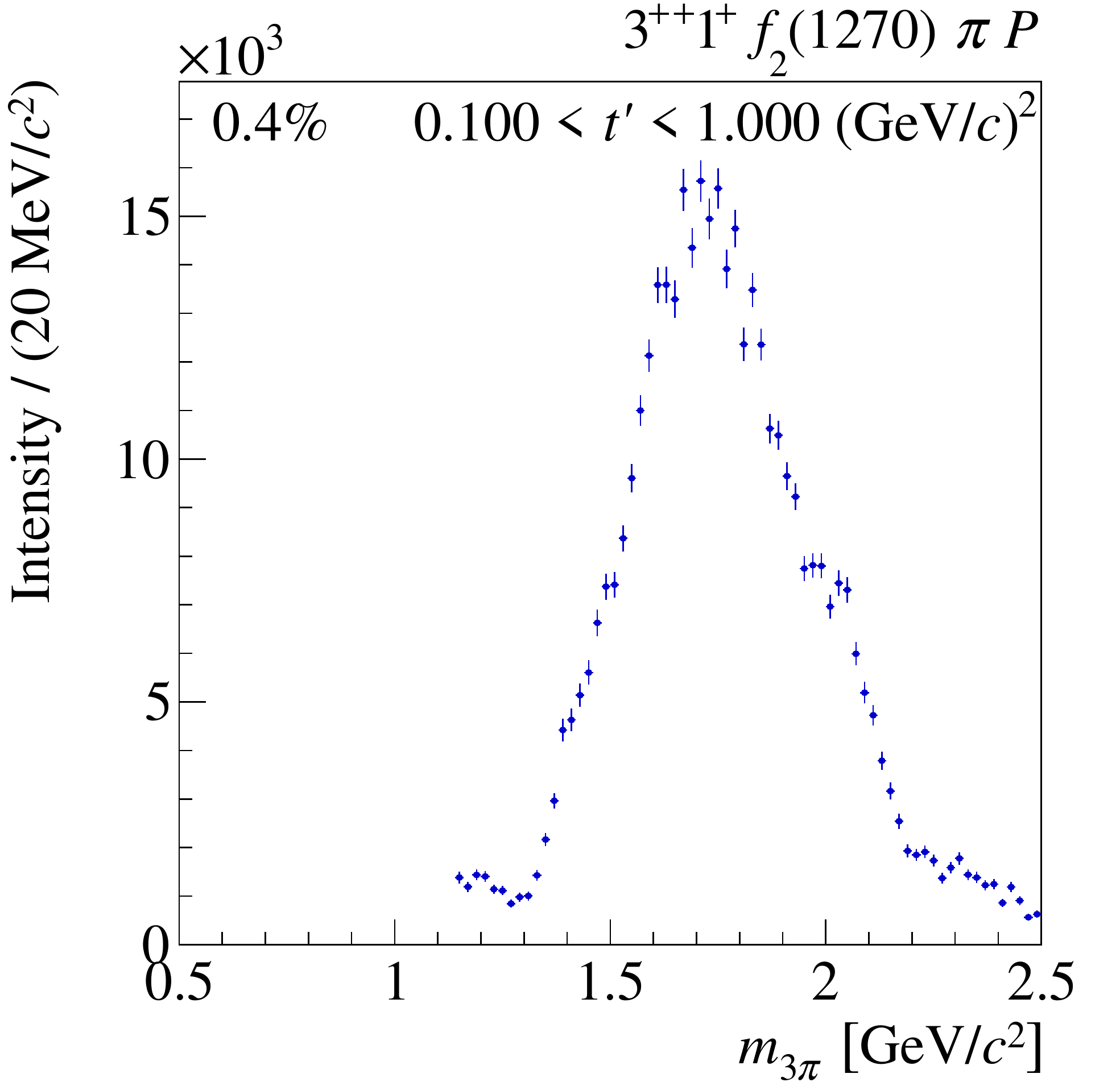}%
  }%
  \subfloat[][]{%
    \label{fig:int_3pp0p_rho3_S}%
    \includegraphics[width=\threePlotWidth]{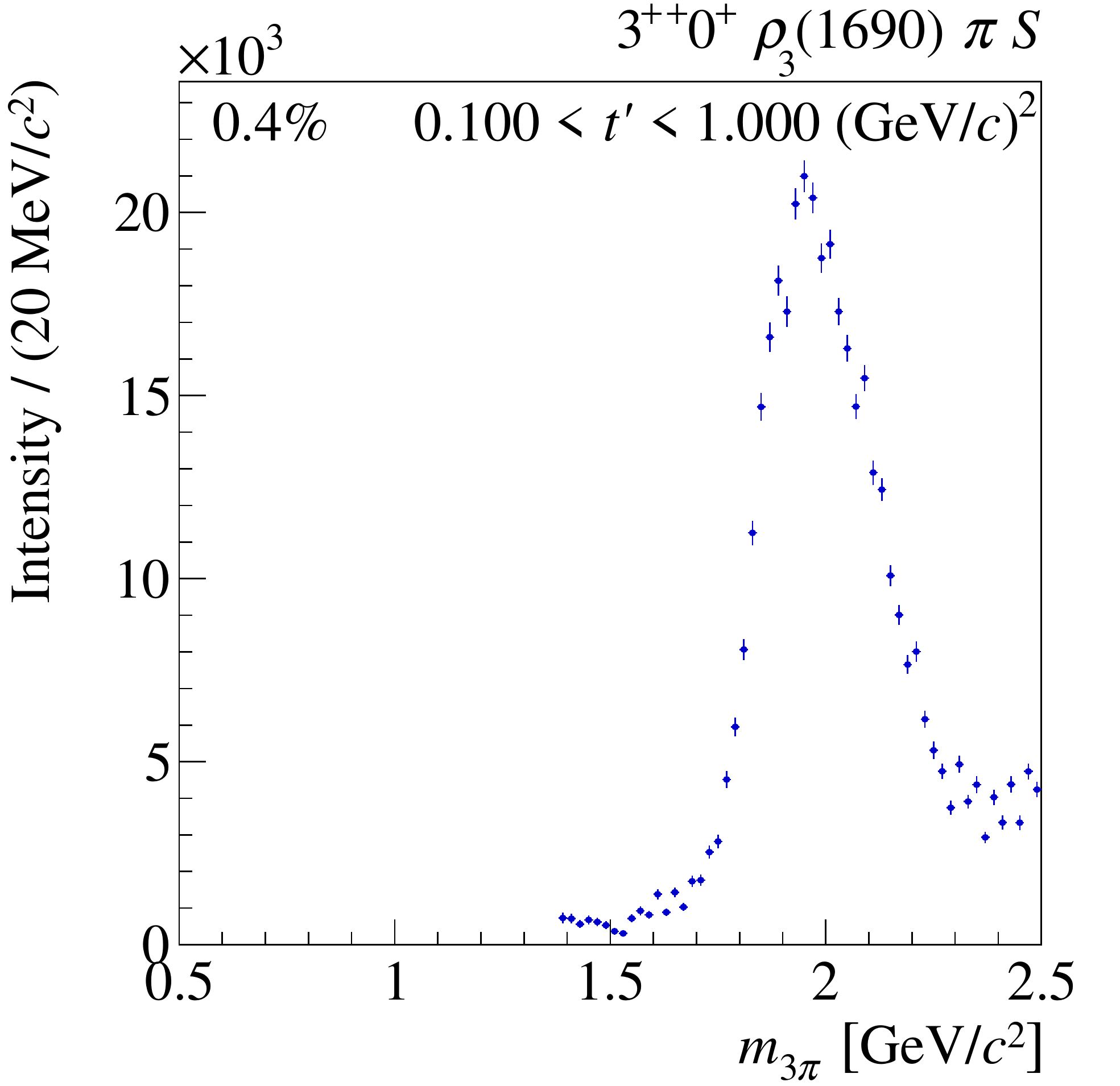}%
  }%
  \\
  \subfloat[][]{%
    \label{fig:int_3pp1p_rho3_S}%
    \includegraphics[width=\threePlotWidth]{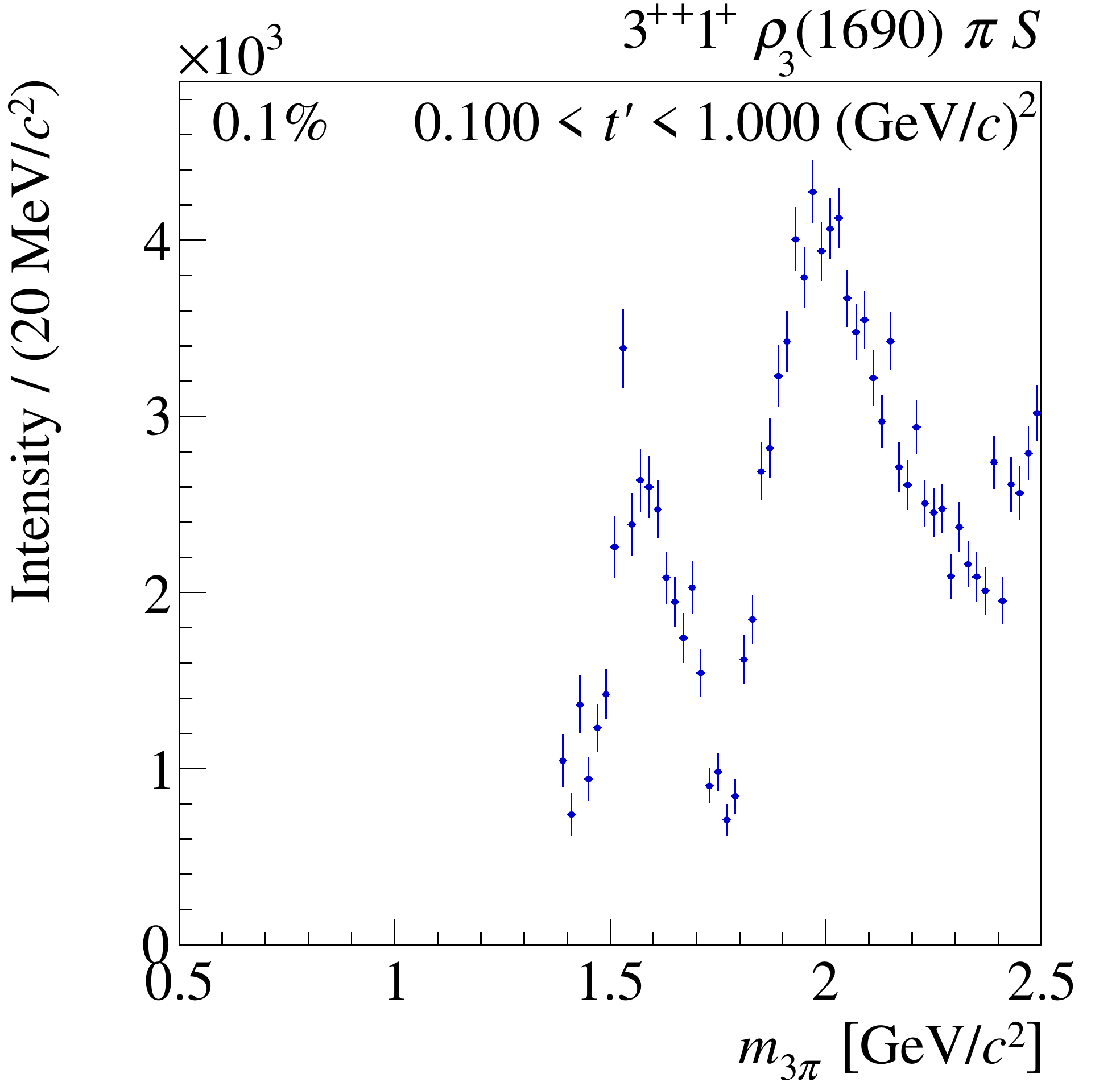}%
  }%
  \subfloat[][]{%
    \label{fig:int_3pp0p_rho3_I}%
    \includegraphics[width=\threePlotWidth]{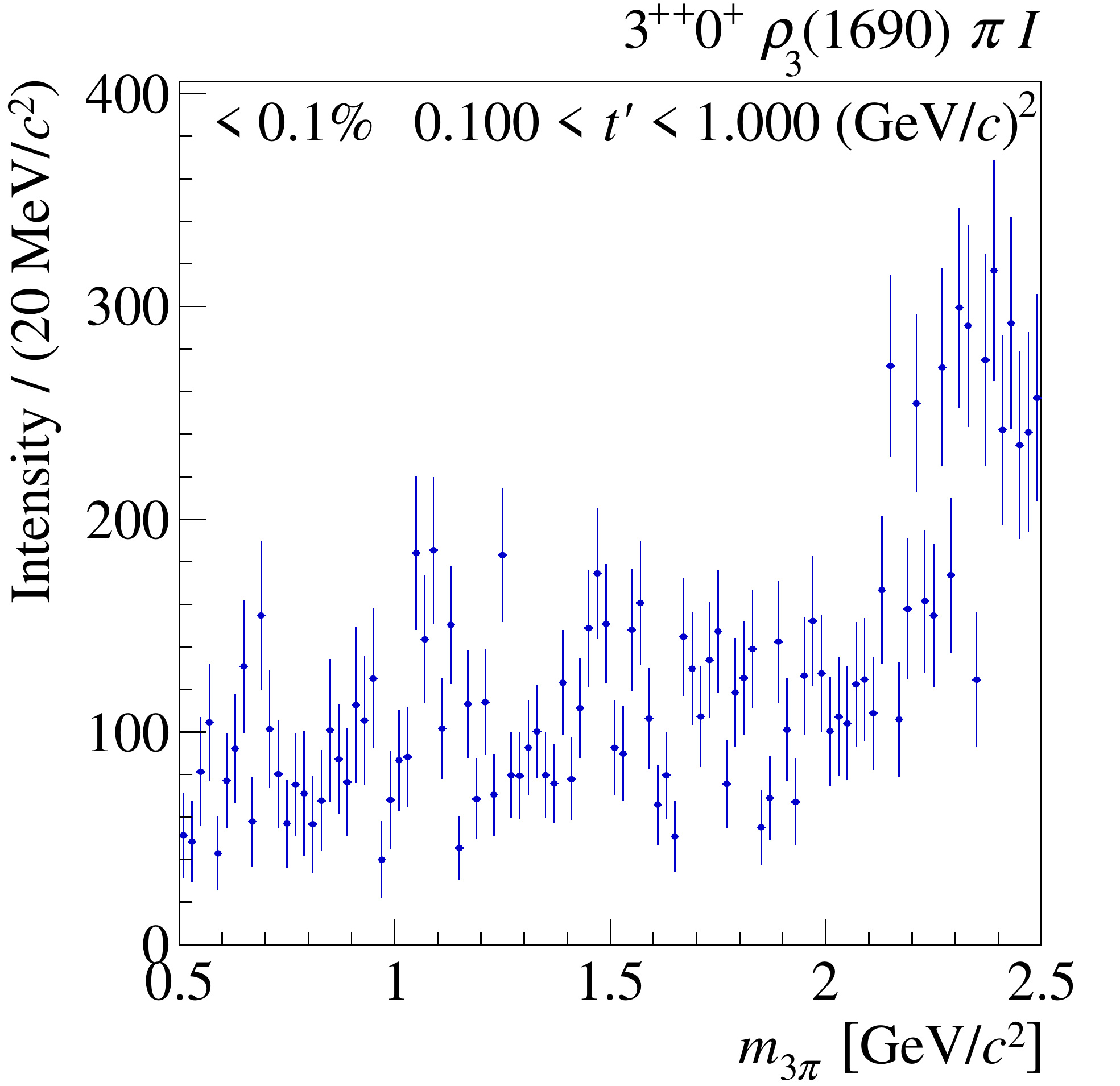}%
  }%
  \caption{The \tpr-summed intensities of partial waves with $\JPC =
    3^{++}$ and positive reflectivity.}
  \label{fig:intensities_3pp_2}
\end{figure}

\clearpage
\subsubsection{$\JPC = 3^{-+}$ Waves}

\begin{figure}[H]
  \centering
  \subfloat[][]{%
    \label{fig:int_3mp1p_rho_F}%
    \includegraphics[width=\threePlotWidth]{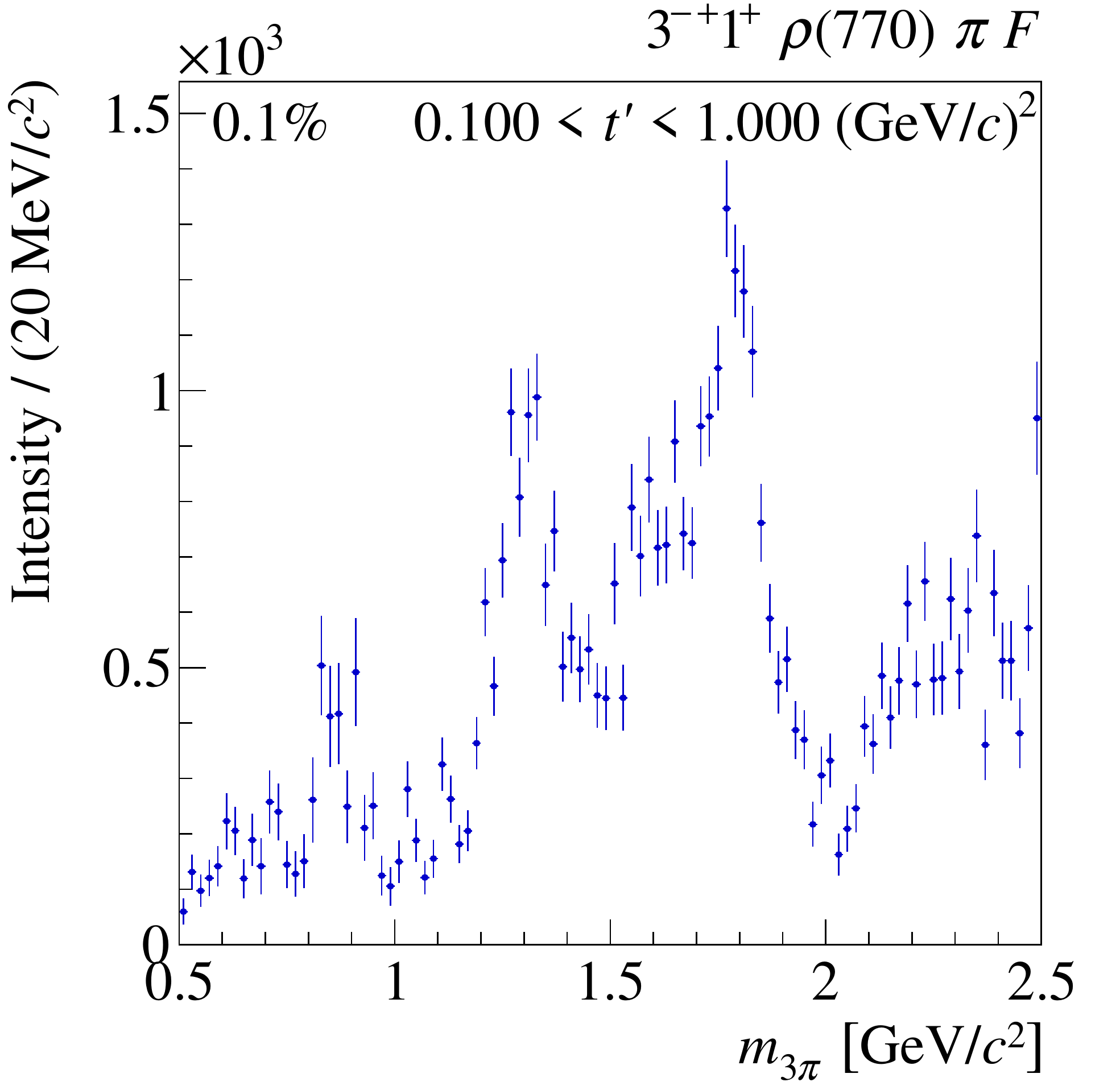}%
  }%
  \subfloat[][]{%
    \label{fig:int_3mp1p_f2_D}%
    \includegraphics[width=\threePlotWidth]{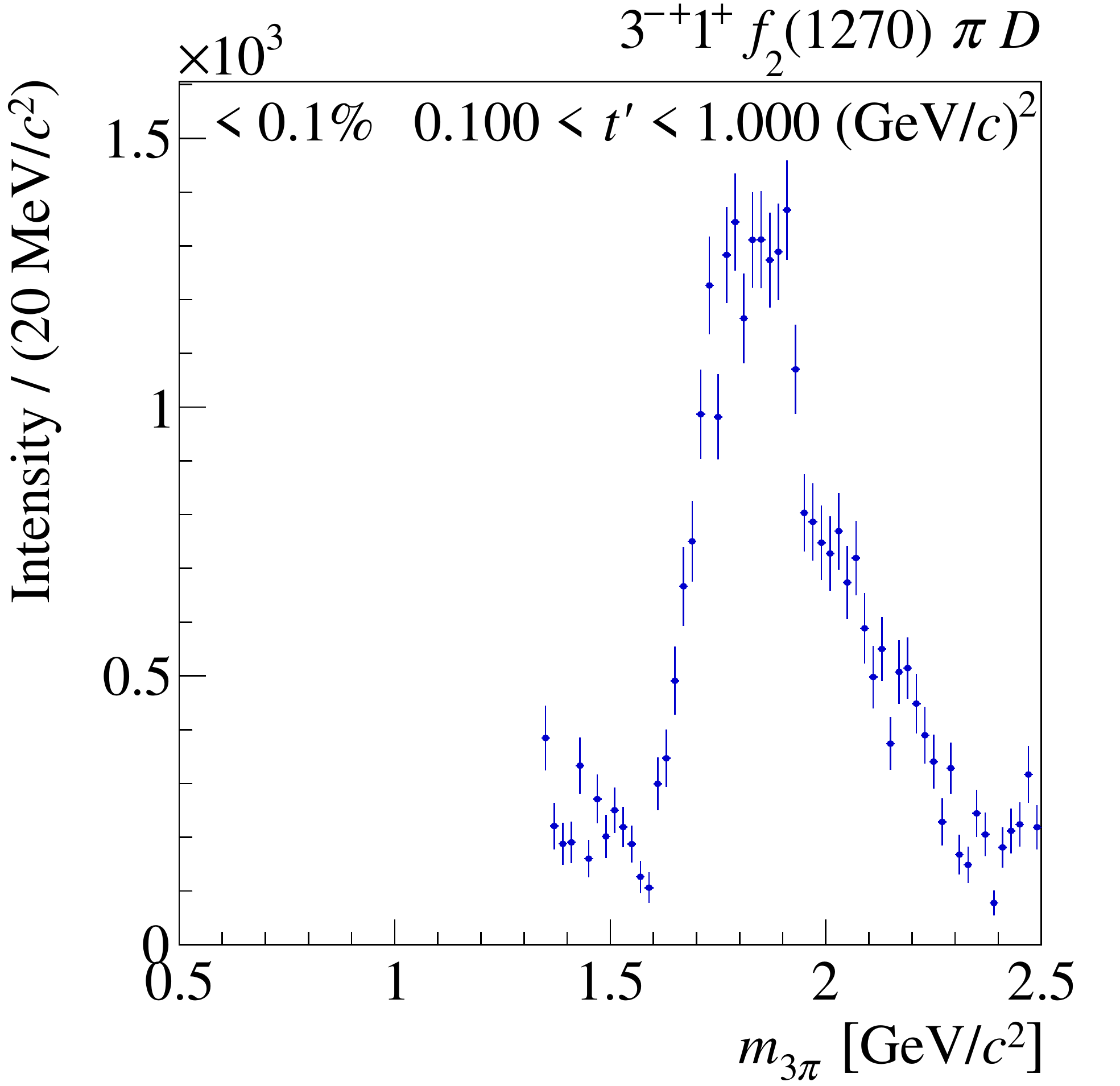}%
  }%
  \caption{The \tpr-summed intensities of partial waves with
    spin-exotic $\JPC = 3^{-+}$ and positive reflectivity.}
  \label{fig:intensities_3mp}
\end{figure}

\subsubsection{$\JPC = 4^{++}$ Waves}

\begin{figure}[H]
  \centering
  \subfloat[][]{%
    \label{fig:int_4pp2p_rho_G}%
    \includegraphics[width=\threePlotWidth]{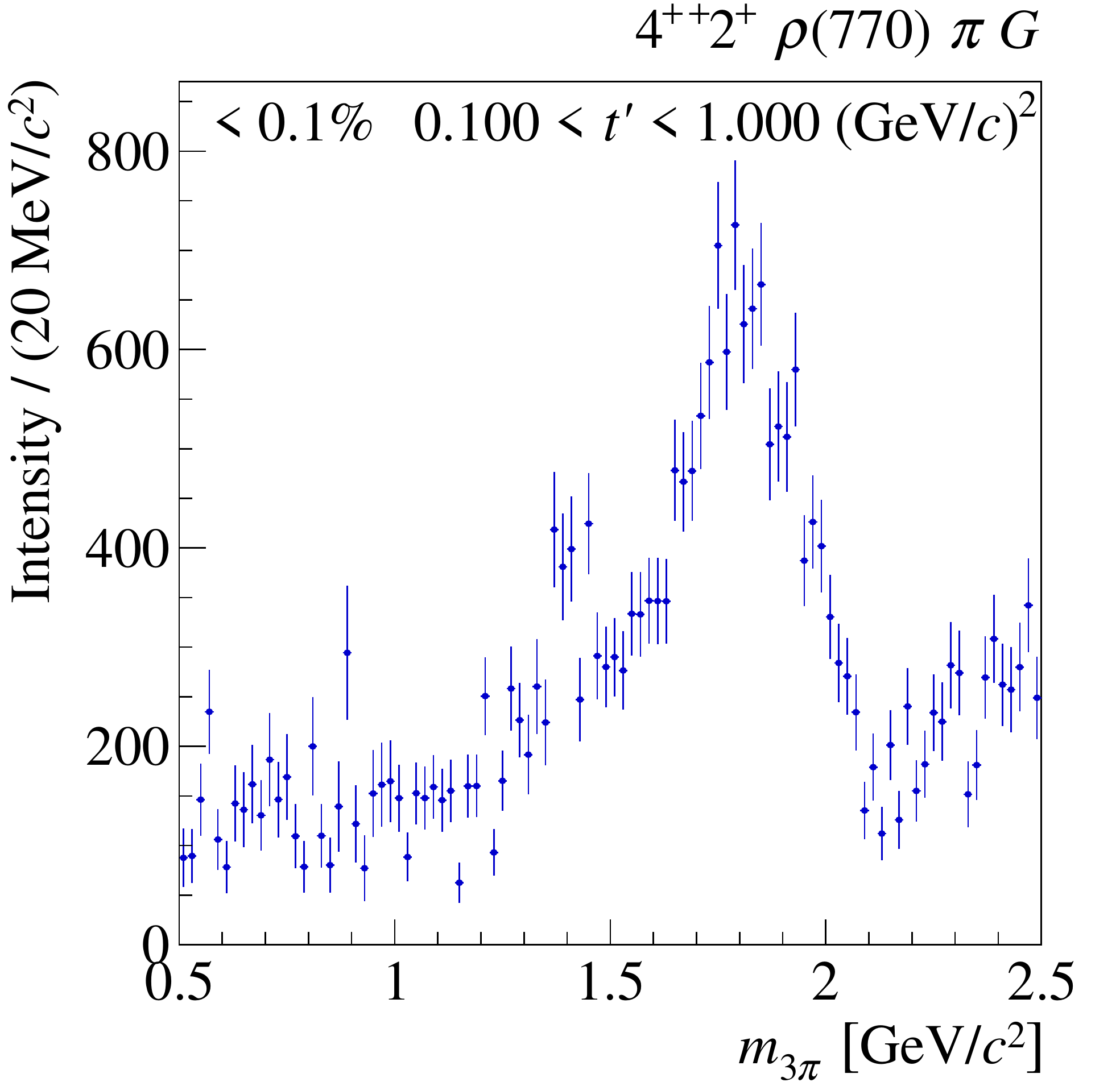}%
  }%
  \subfloat[][]{%
    \label{fig:int_4pp2p_f2_F}%
    \includegraphics[width=\threePlotWidth]{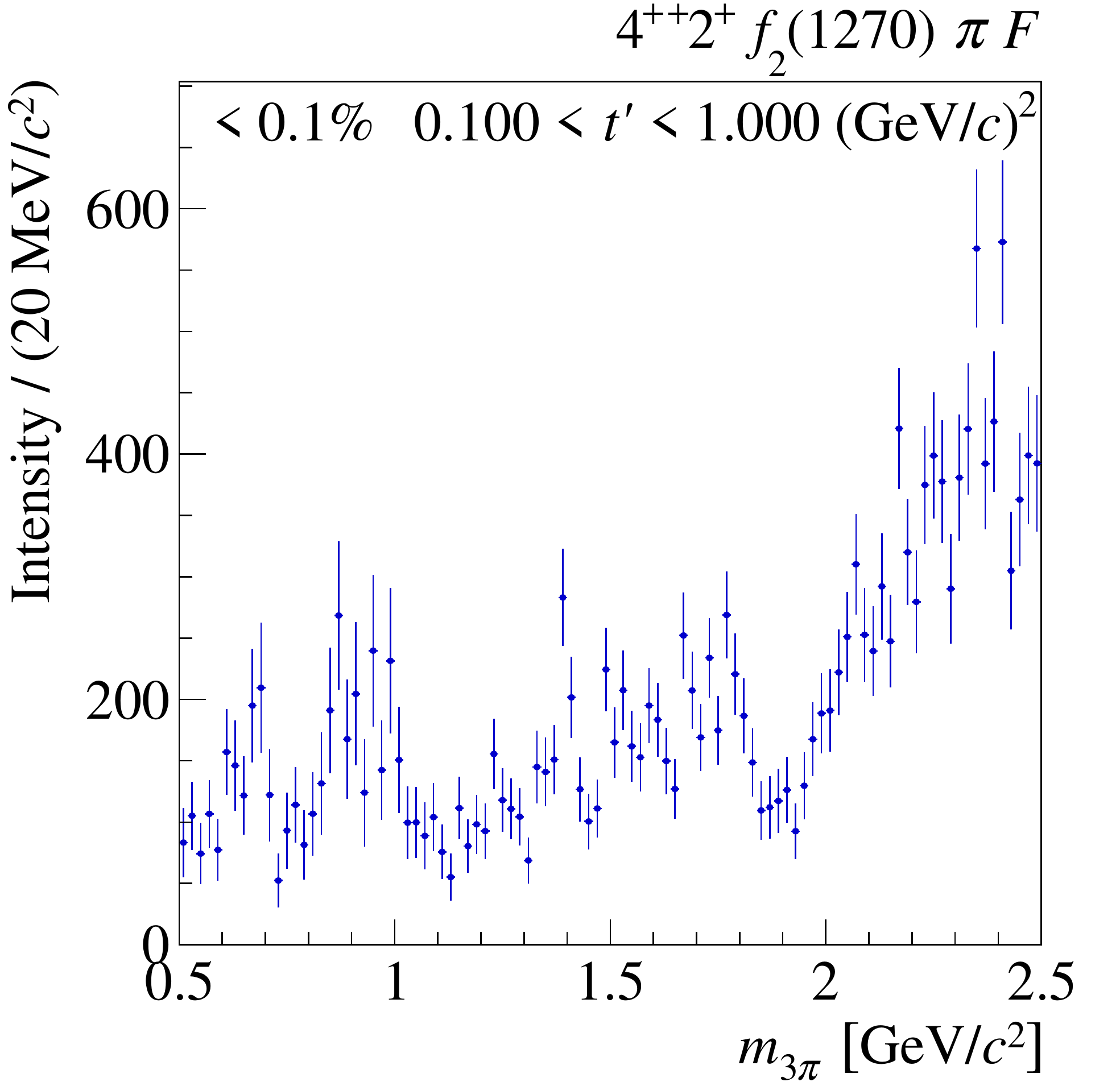}%
  }%
  \subfloat[][]{%
    \label{fig:int_4pp1p_rho3_D}%
    \includegraphics[width=\threePlotWidth]{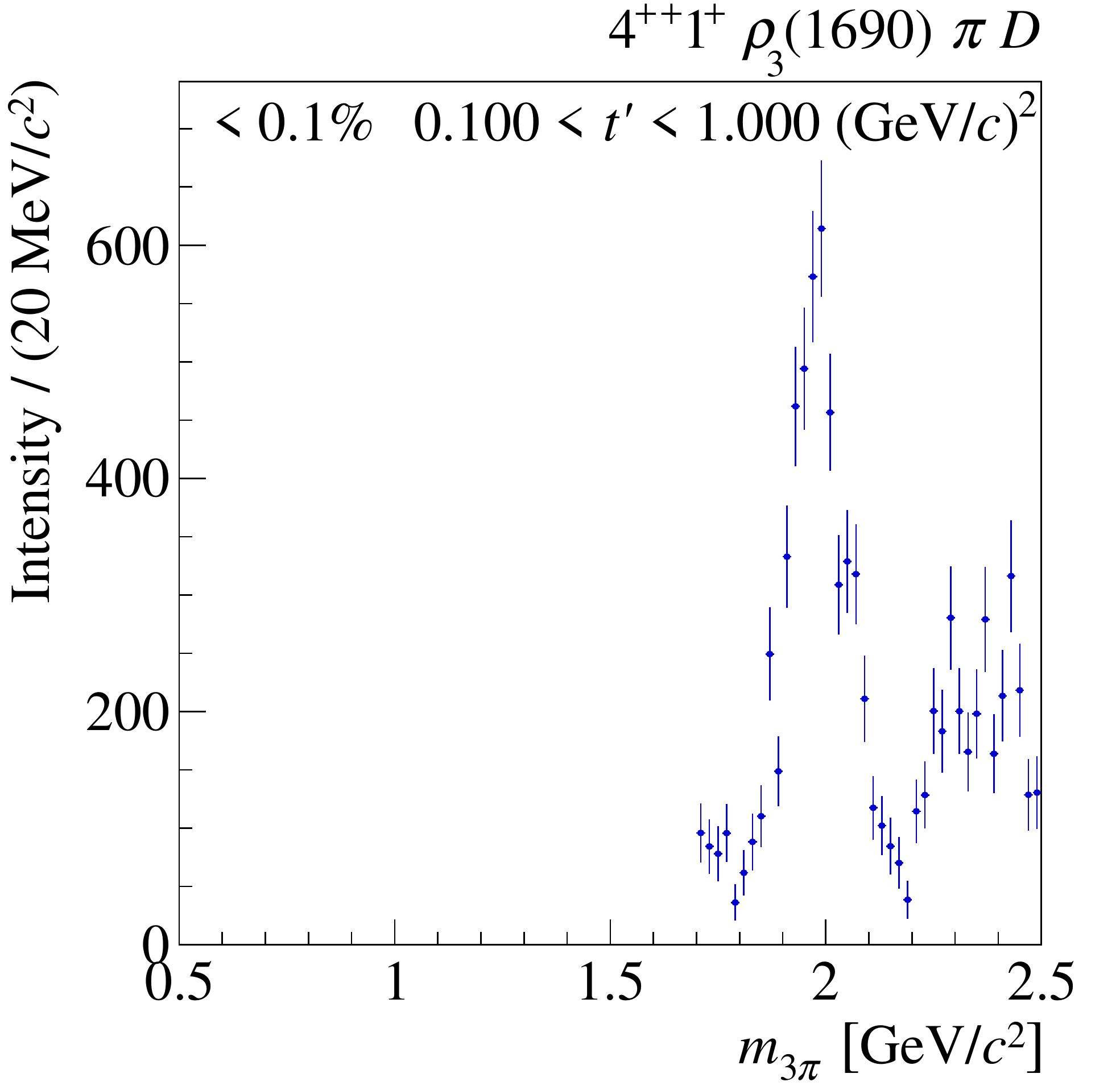}%
  }%
  \caption{The \tpr-summed intensities of partial waves with $\JPC =
    4^{++}$ and positive reflectivity.}
  \label{fig:intensities_4pp}
\end{figure}

\clearpage
\subsubsection{$\JPC = 4^{-+}$ Waves}

\begin{figure}[H]
  \centering
  \subfloat[][]{%
    \label{fig:int_4mp0p_pipiS_G}%
    \includegraphics[width=\threePlotWidth]{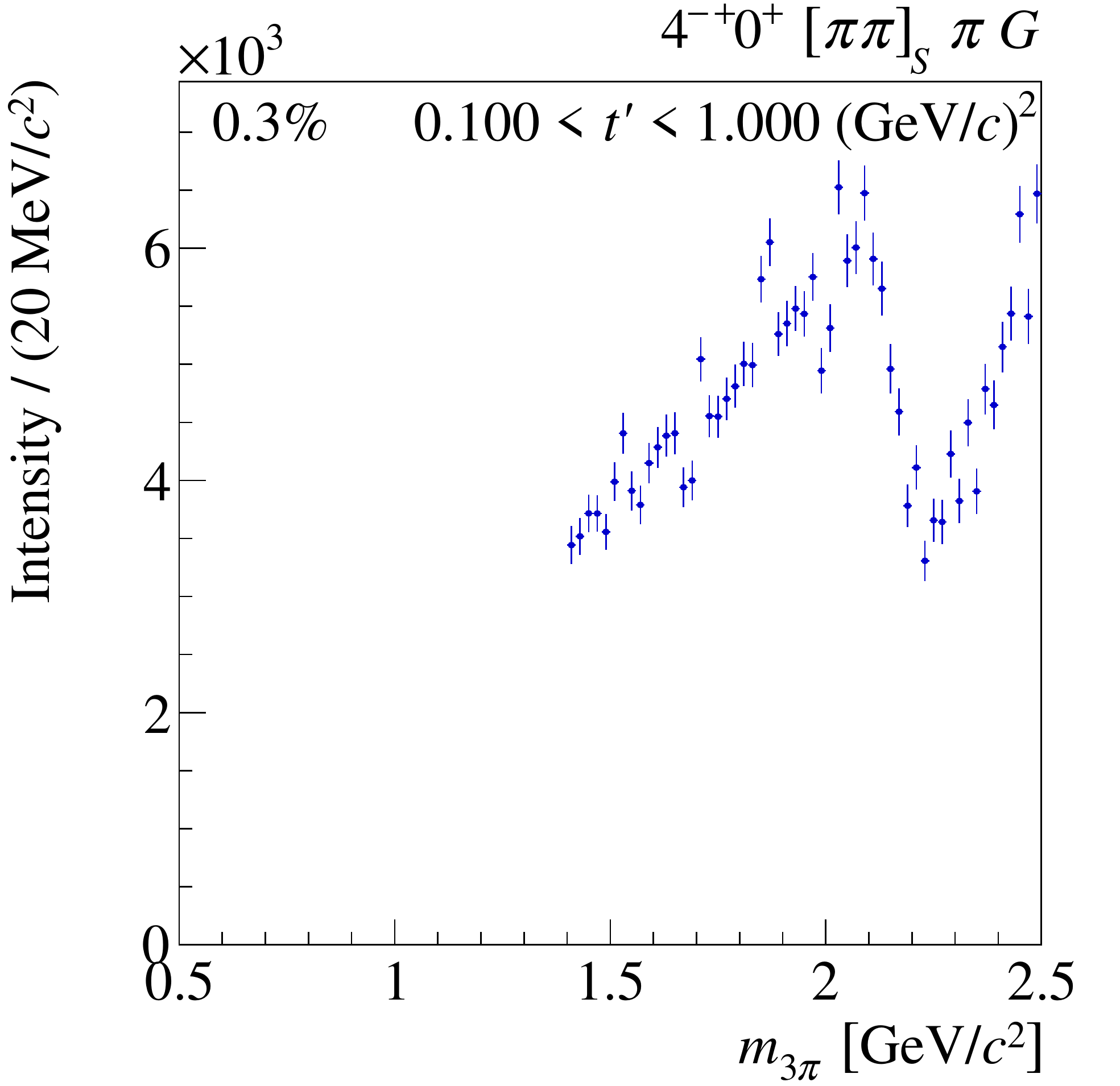}%
  }%
  \subfloat[][]{%
    \label{fig:int_4mp0p_rho_F}%
    \includegraphics[width=\threePlotWidth]{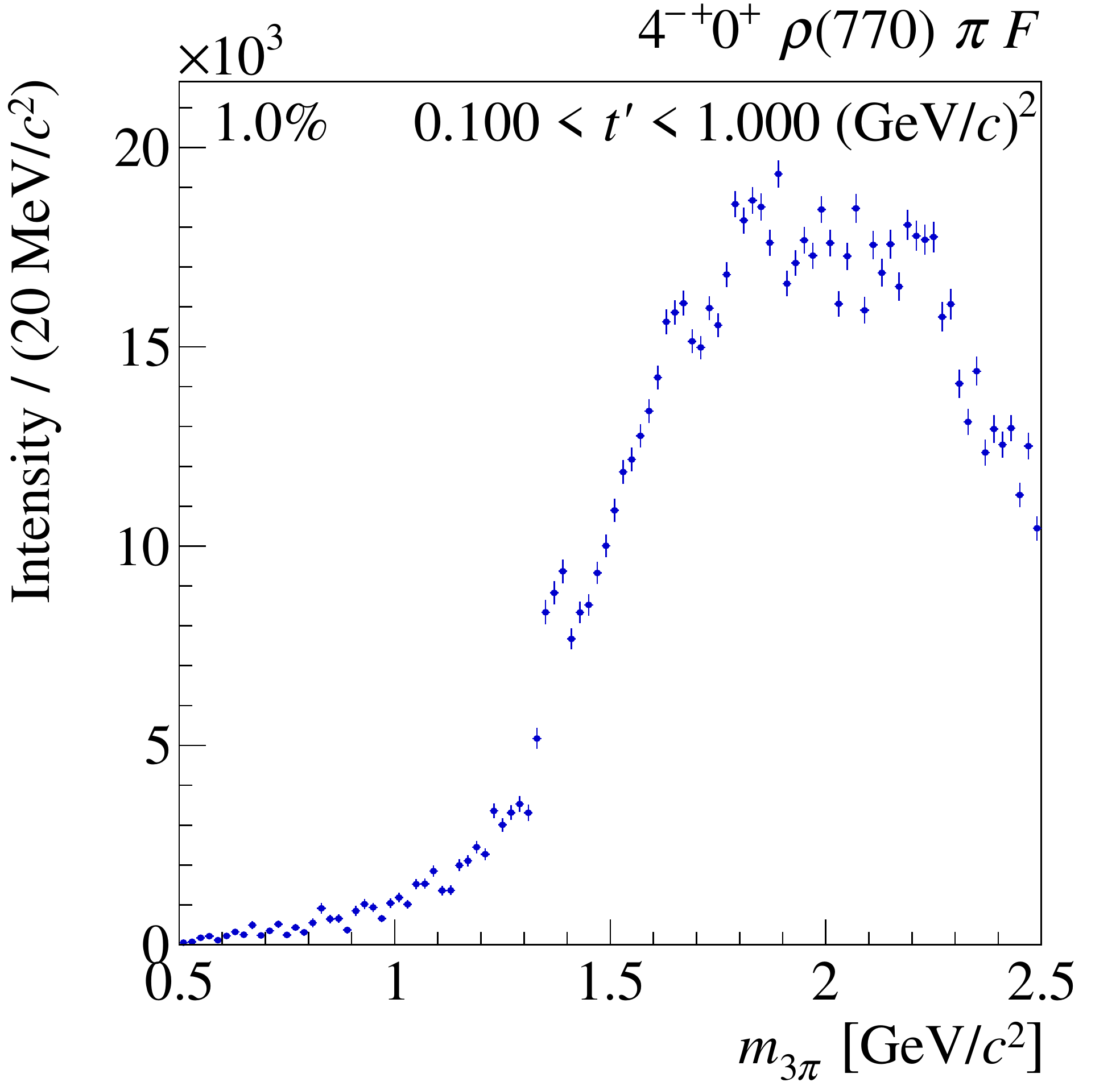}%
  }%
  \subfloat[][]{%
    \label{fig:int_4mp1p_rho_F}%
    \includegraphics[width=\threePlotWidth]{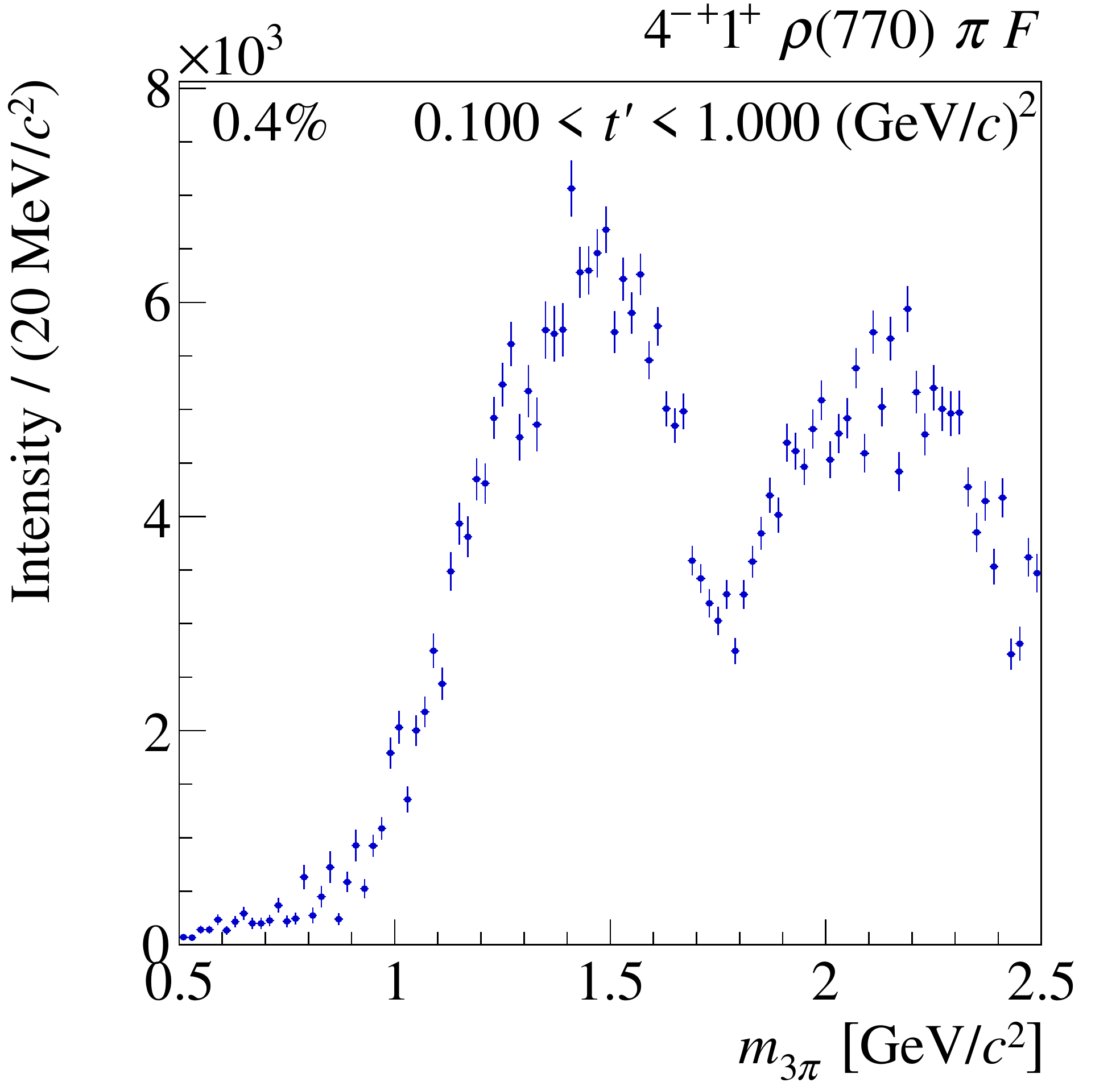}%
  }%
  \\
  \subfloat[][]{%
    \label{fig:int_4mp0p_f2_D}%
    \includegraphics[width=\threePlotWidth]{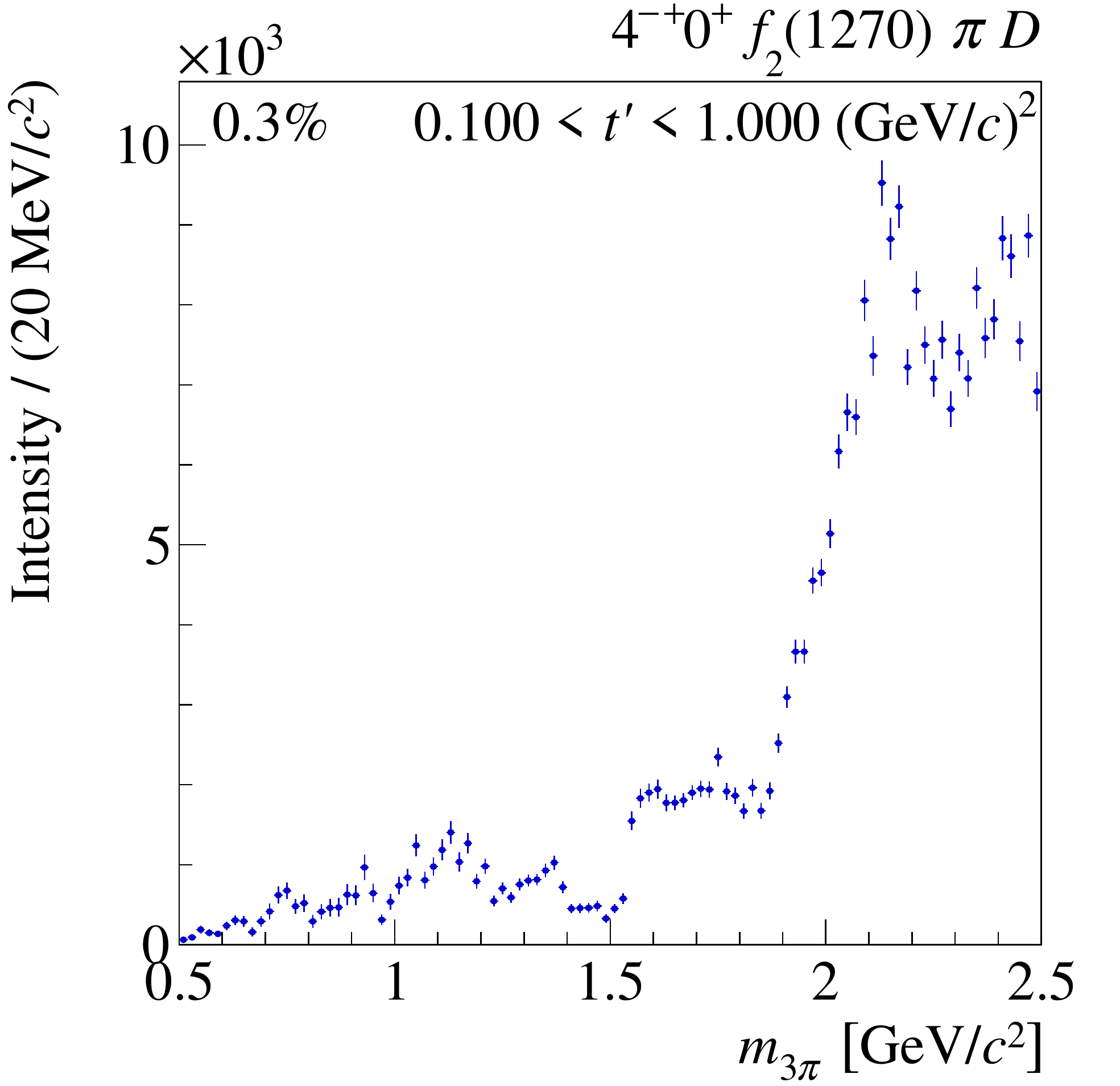}%
  }%
  \subfloat[][]{%
    \label{fig:int_4mp1p_f2_D}%
    \includegraphics[width=\threePlotWidth]{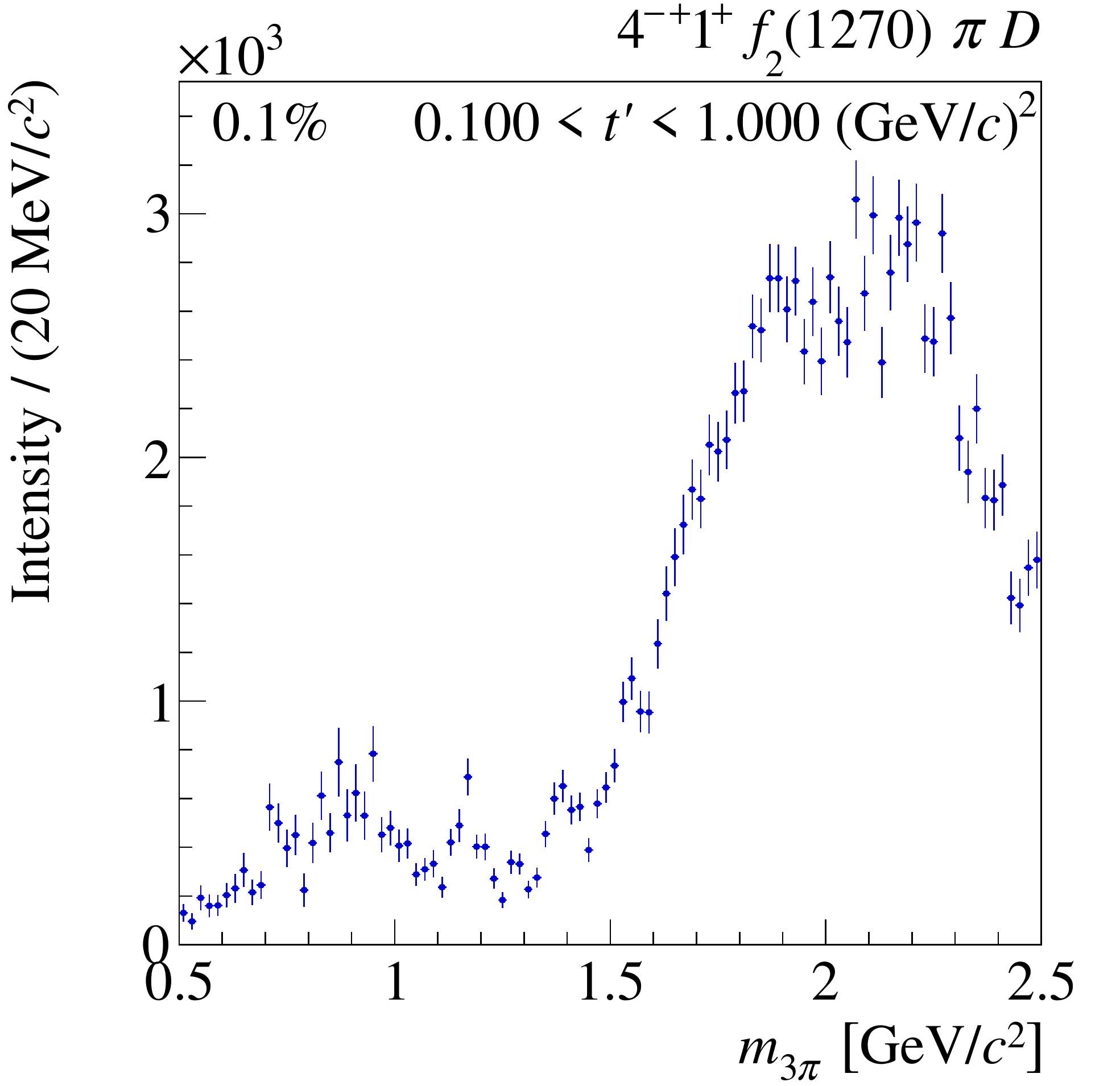}%
  }%
  \subfloat[][]{%
    \label{fig:int_4mp0p_f2_G}%
    \includegraphics[width=\threePlotWidth]{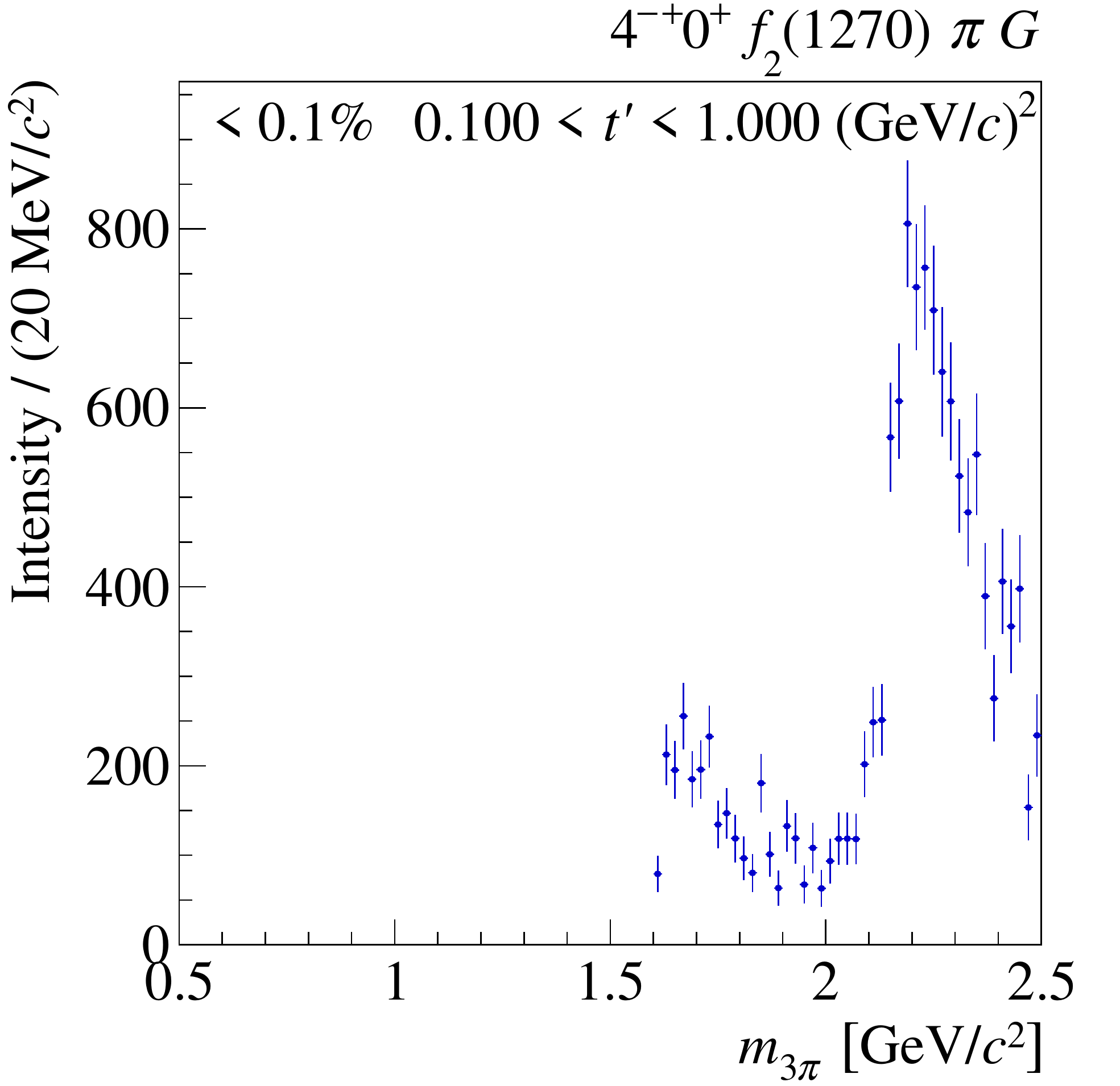}%
  }%
  \caption{The \tpr-summed intensities of partial waves with $\JPC =
    4^{-+}$ and positive reflectivity.}
  \label{fig:intensities_4mp}
\end{figure}

\subsubsection{$\JPC = 5^{++}$ Waves}

\begin{figure}[H]
  \centering
  \subfloat[][]{%
    \label{fig:int_5pp0p_pipiS_H}%
    \includegraphics[width=\threePlotWidth]{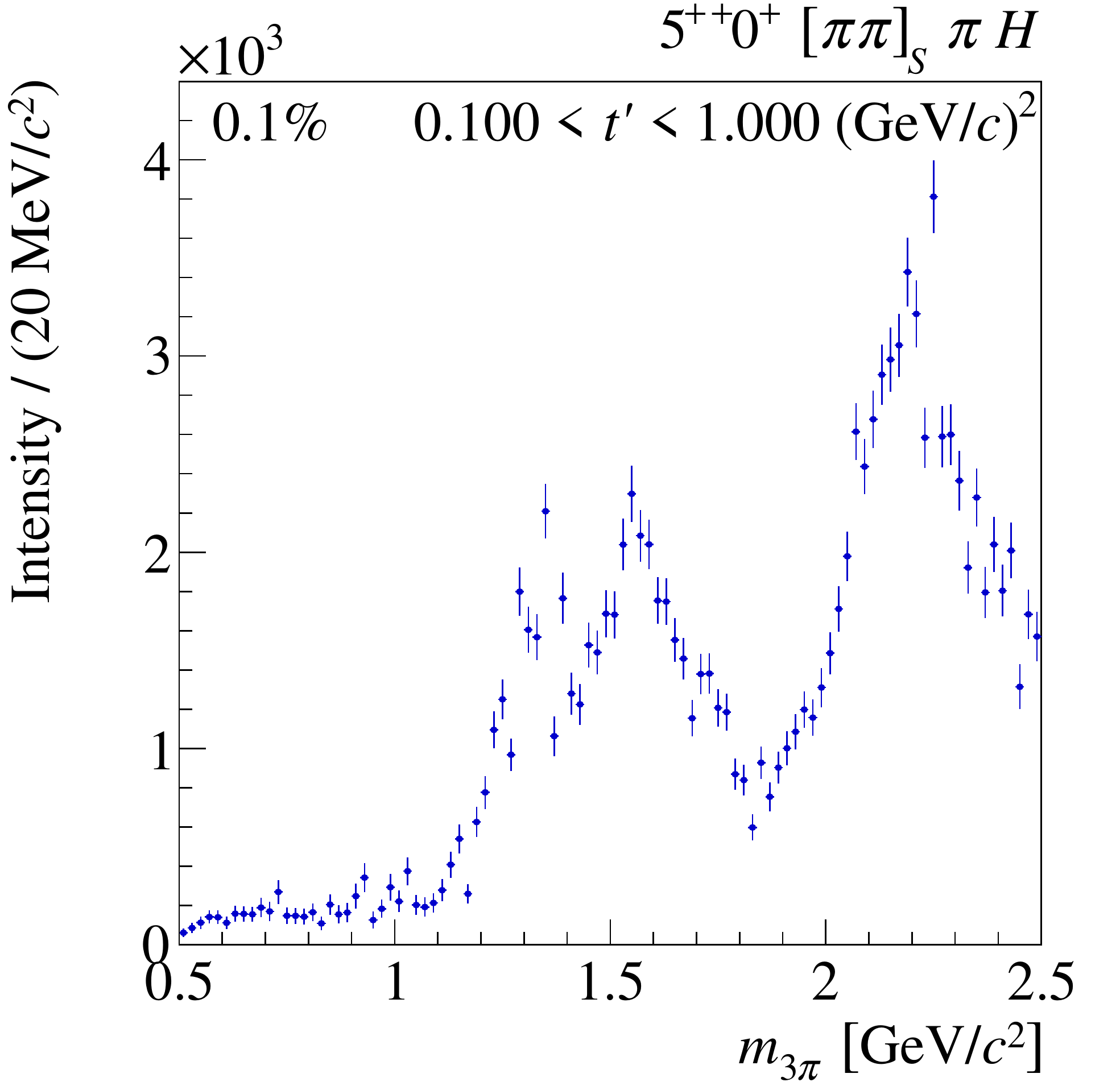}%
  }%
  \subfloat[][]{%
    \label{fig:int_5pp1p_pipiS_H}%
    \includegraphics[width=\threePlotWidth]{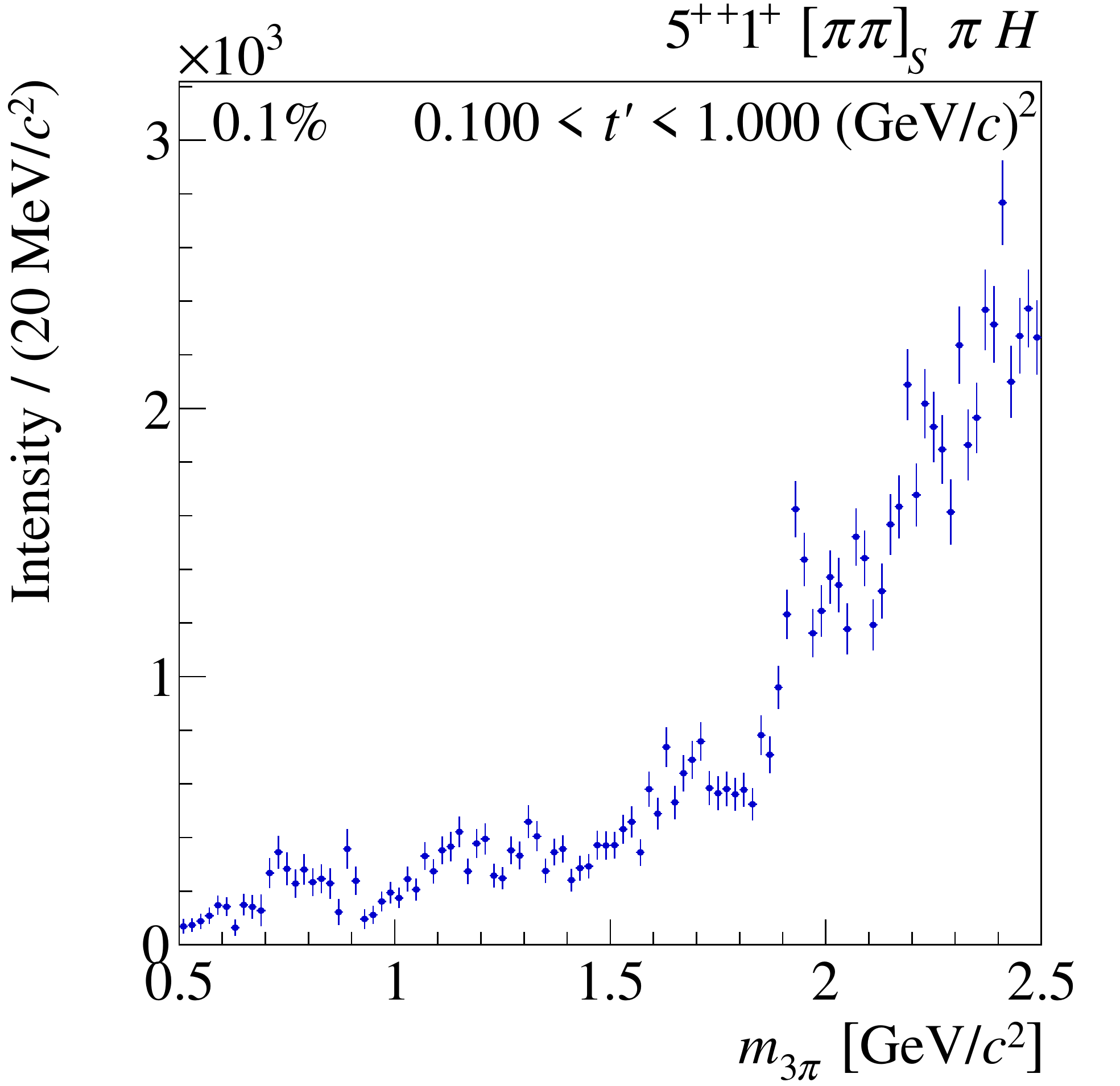}%
  }%
  \subfloat[][]{%
    \label{fig:int_5pp0p_rho_G}%
    \includegraphics[width=\threePlotWidth]{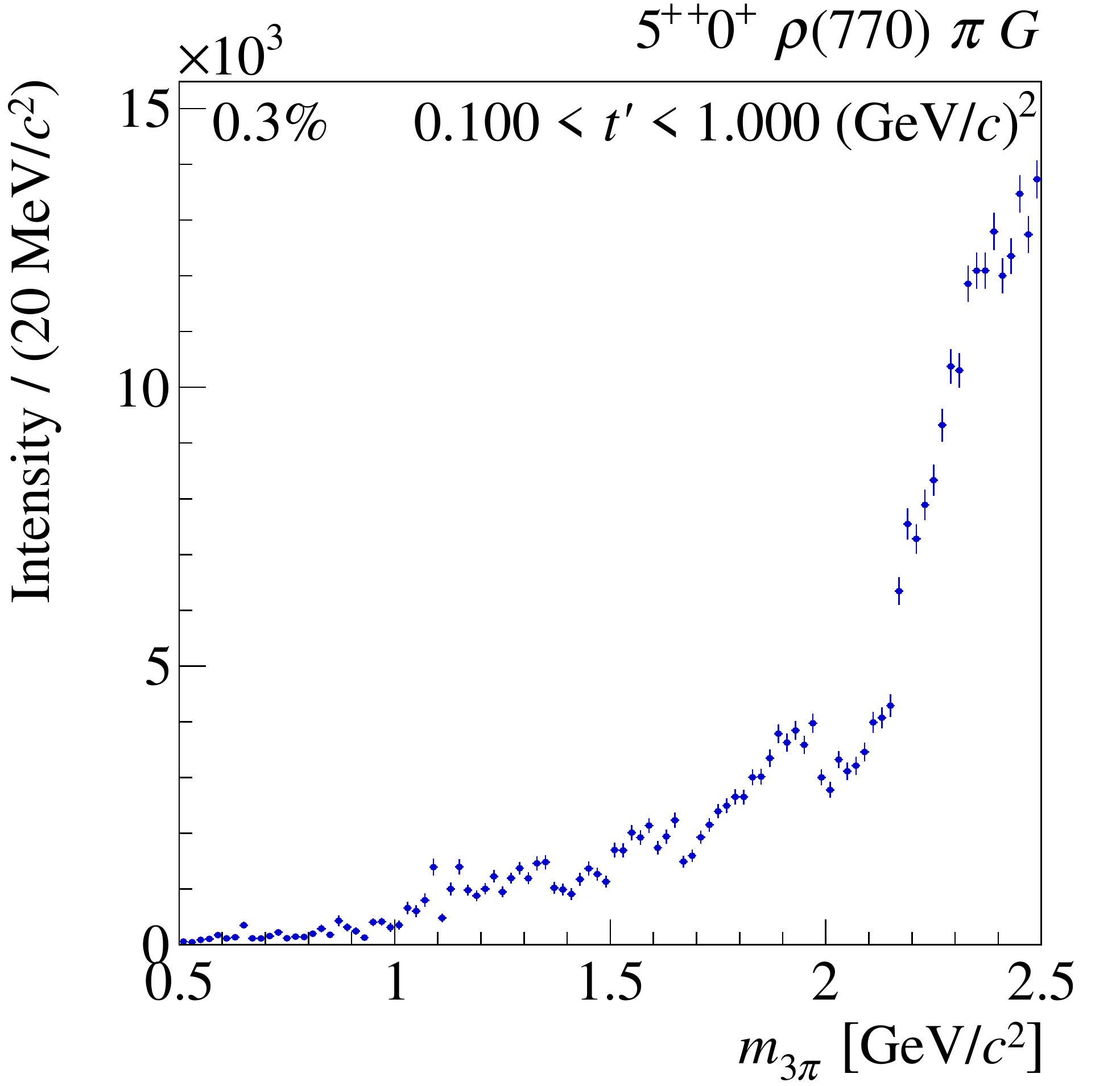}%
  }%
  \caption{The \tpr-summed intensities of partial waves with $\JPC =
    5^{++}$ and positive reflectivity.}
  \label{fig:intensities_5pp_1}
\end{figure}

\clearpage
\begin{figure}[H]
  \centering
  \subfloat[][]{%
    \label{fig:int_5pp0p_f2_F}%
    \includegraphics[width=\threePlotWidth]{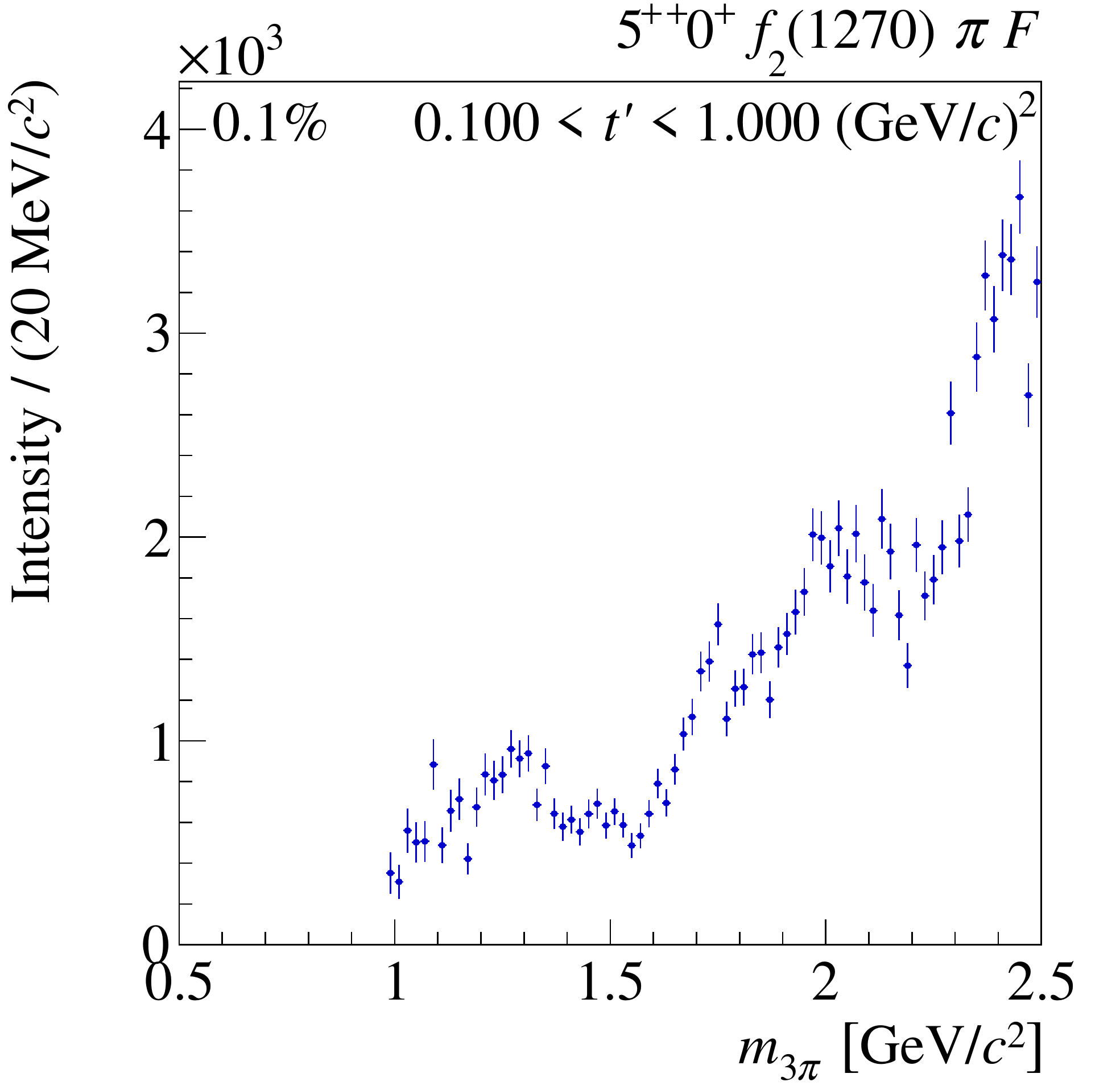}%
  }%
  \subfloat[][]{%
    \label{fig:int_5pp1p_f2_F}%
    \includegraphics[width=\threePlotWidth]{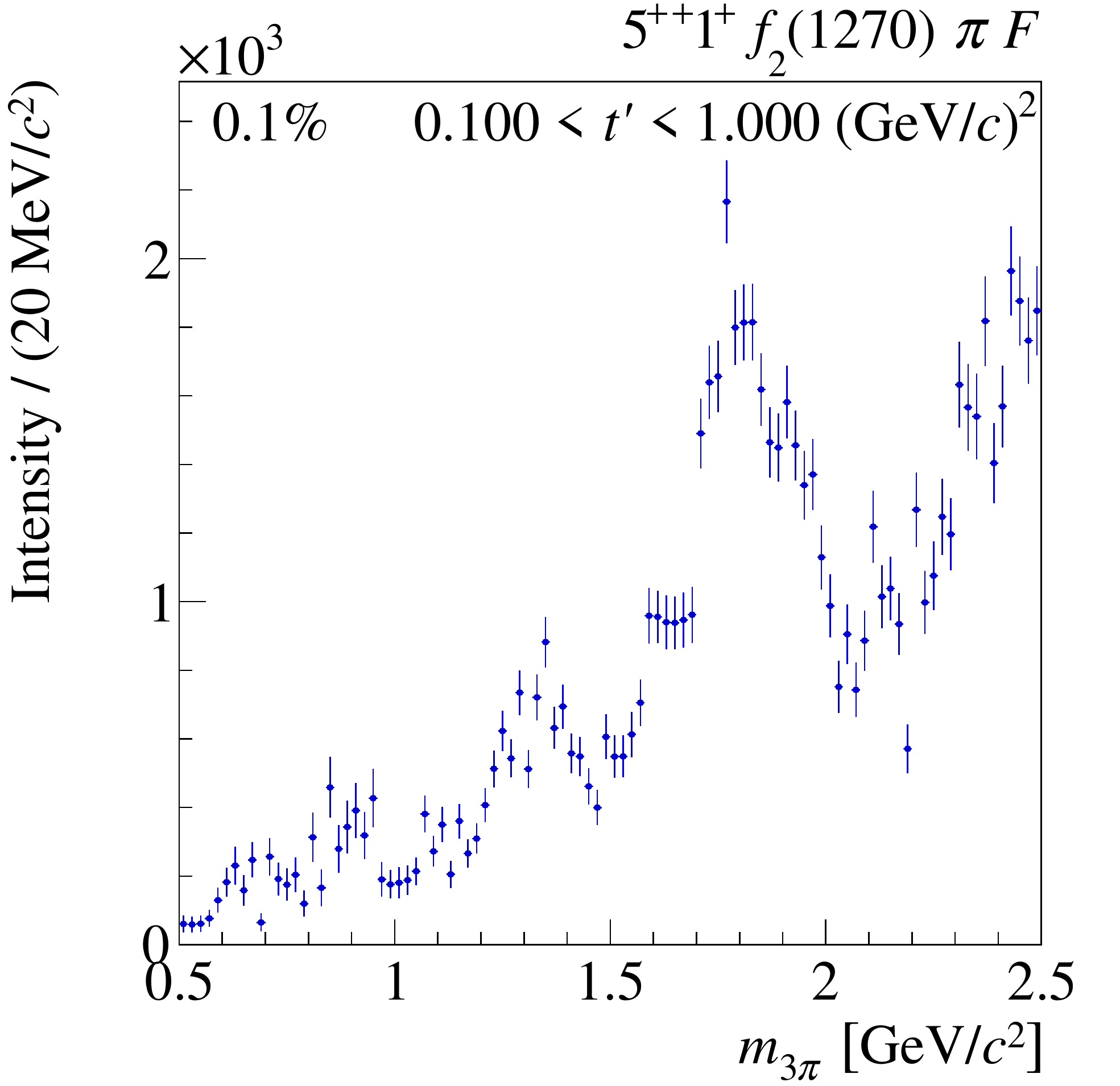}%
  }%
  \\
  \subfloat[][]{%
    \label{fig:int_5pp0p_f2_H}%
    \includegraphics[width=\threePlotWidth]{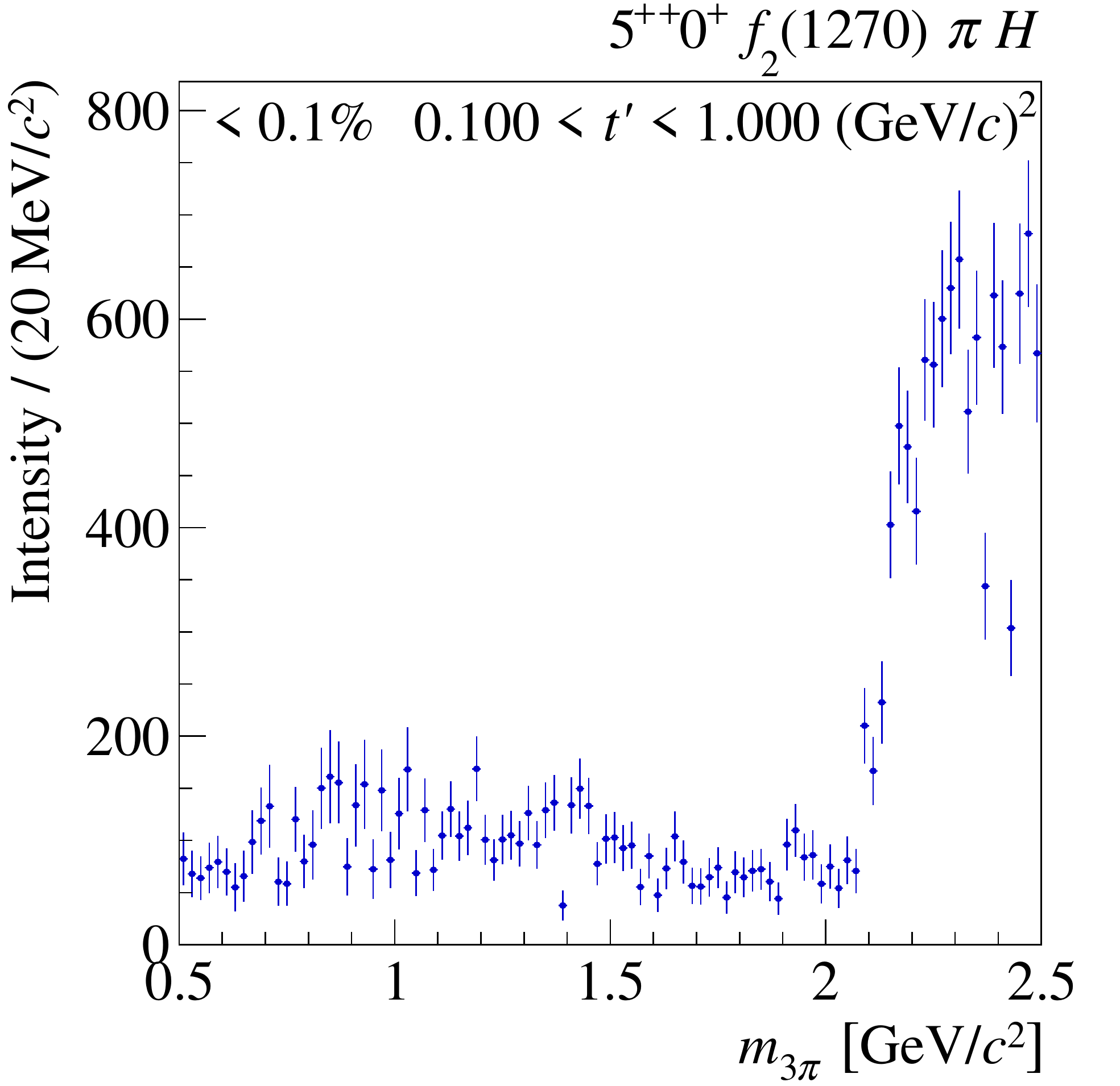}%
  }%
  \subfloat[][]{%
    \label{fig:int_5pp0p_rho3_D}%
    \includegraphics[width=\threePlotWidth]{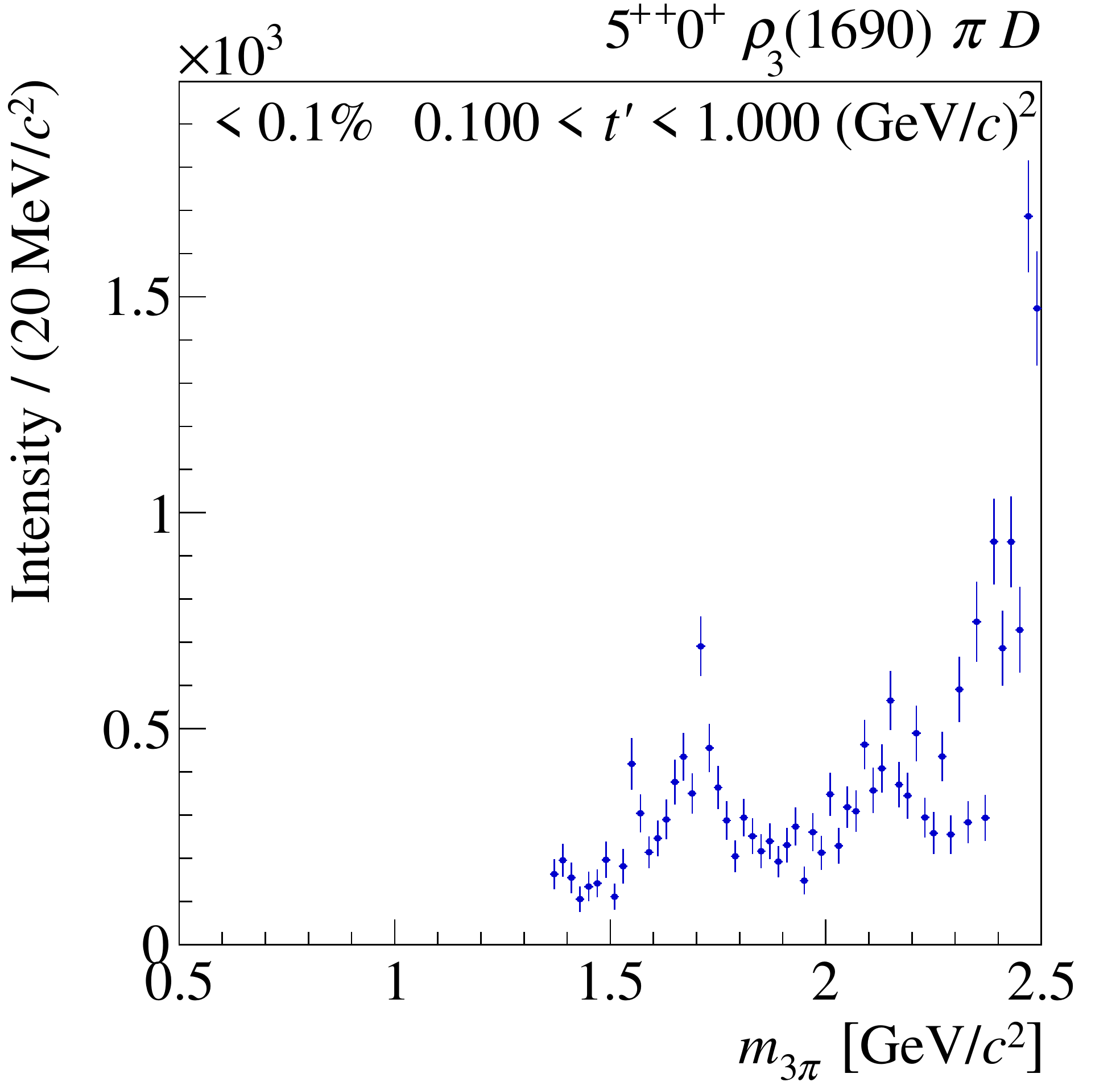}%
  }%
  \caption{The \tpr-summed intensities of partial waves with $\JPC =
    5^{++}$ and positive reflectivity.}
  \label{fig:intensities_5pp_2}
\end{figure}

\subsubsection{$\JPC = 6^{++}$ Waves}

\begin{figure}[H]
  \centering
  \subfloat[][]{%
    \label{fig:int_6pp1p_rho_I}%
    \includegraphics[width=\threePlotWidth]{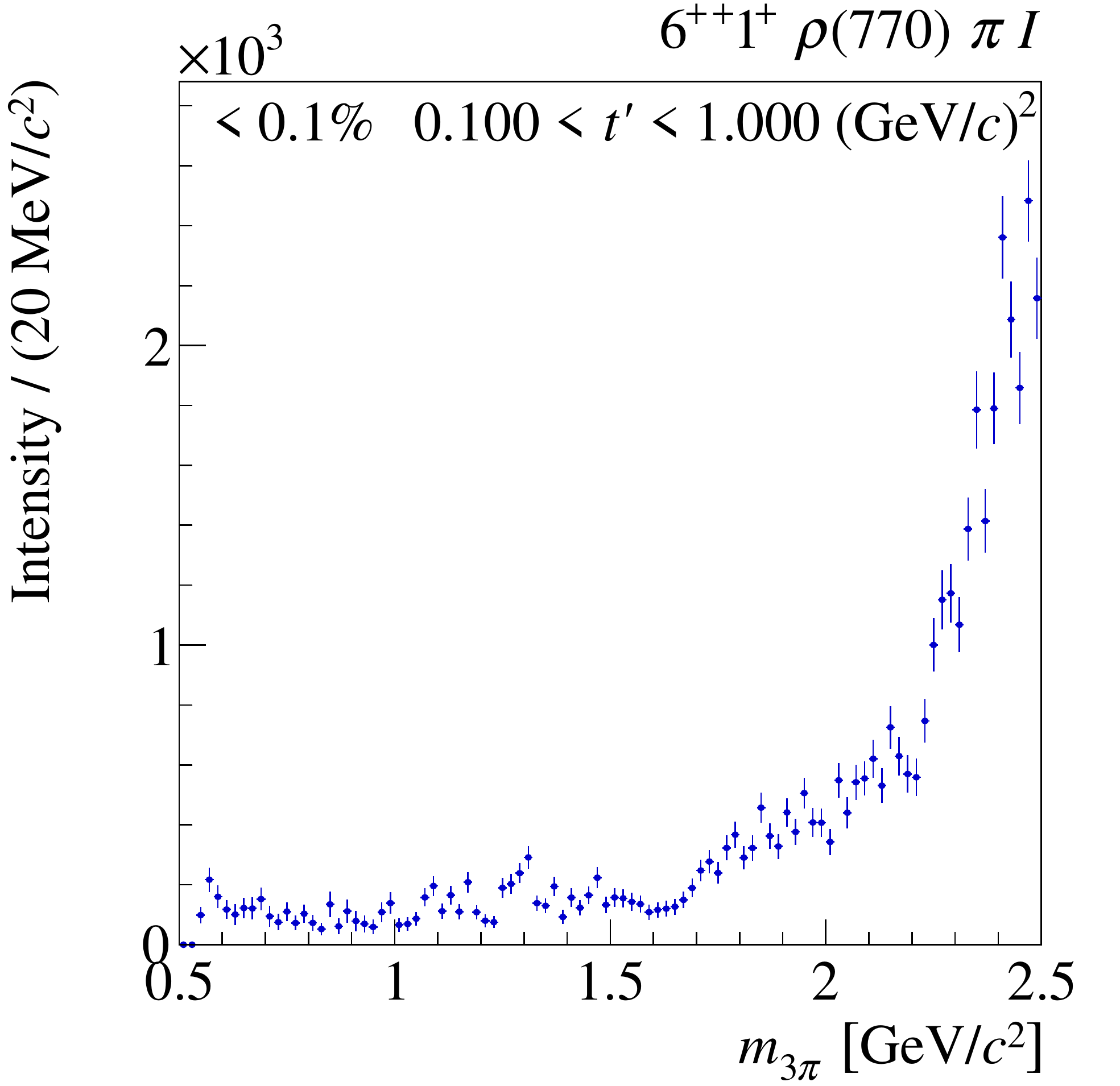}%
  }%
  \subfloat[][]{%
    \label{fig:int_6pp1p_f2_H}%
    \includegraphics[width=\threePlotWidth]{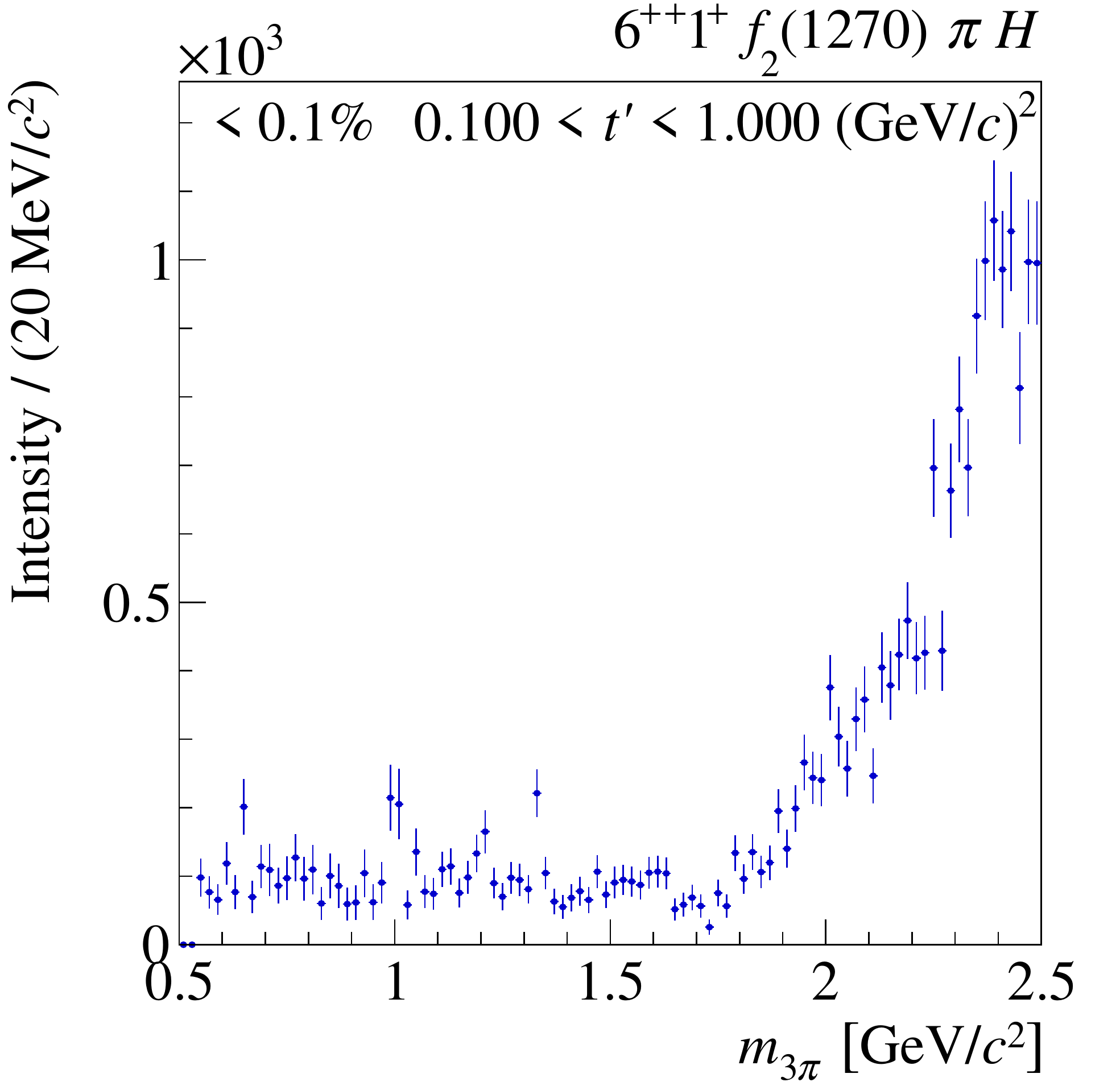}%
  }%
  \caption{The \tpr-summed intensities of partial waves with $\JPC =
    6^{++}$ and positive reflectivity.}
  \label{fig:intensities_6pp}
\end{figure}

\clearpage
\subsubsection{$\JPC = 6^{-+}$ Waves}

\begin{figure}[H]
  \centering
  \subfloat[][]{%
    \label{fig:int_6mp0p_pipiS_I}%
    \includegraphics[width=\threePlotWidth]{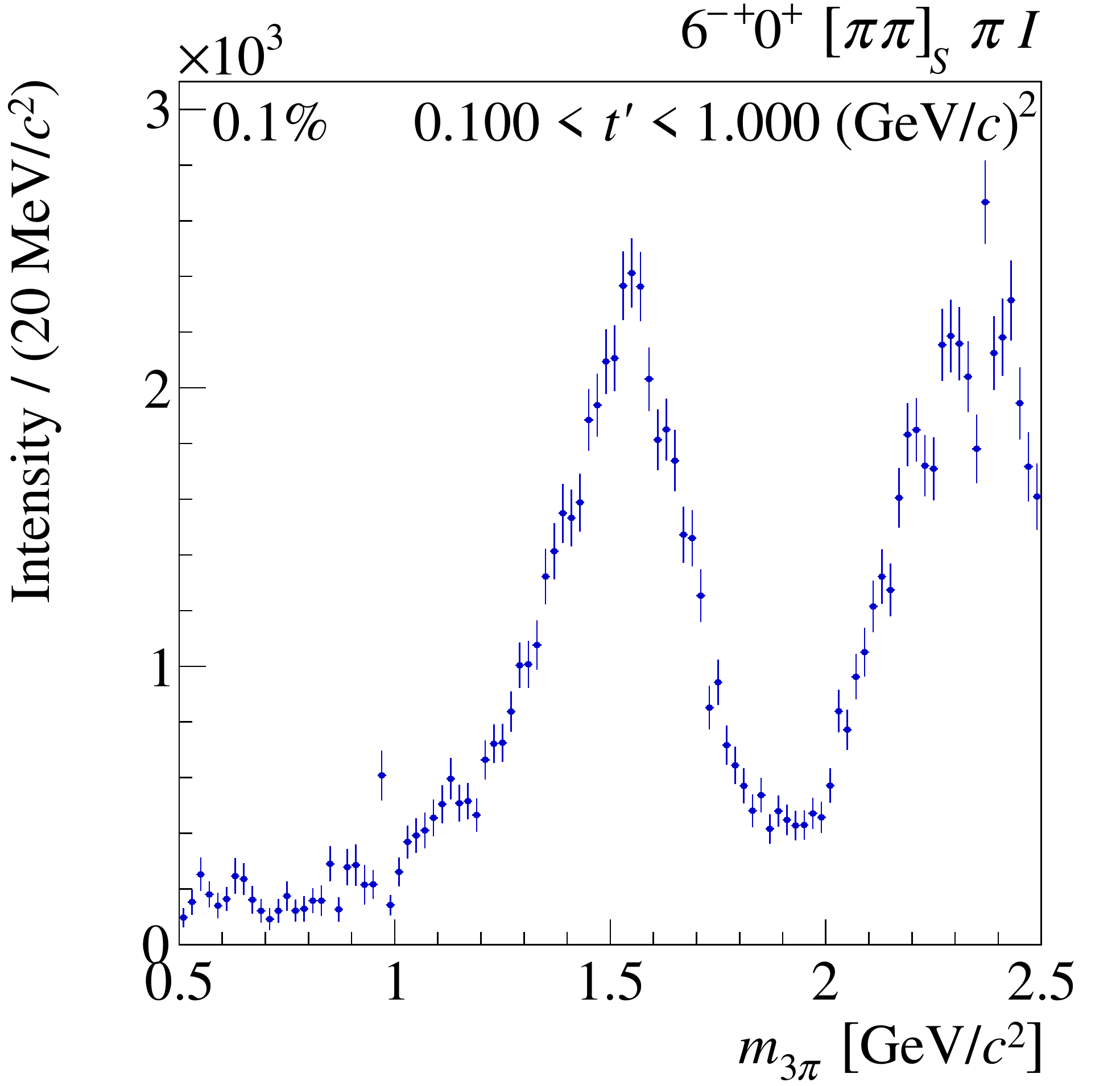}%
  }%
  \subfloat[][]{%
    \label{fig:int_6mp1p_pipiS_I}%
    \includegraphics[width=\threePlotWidth]{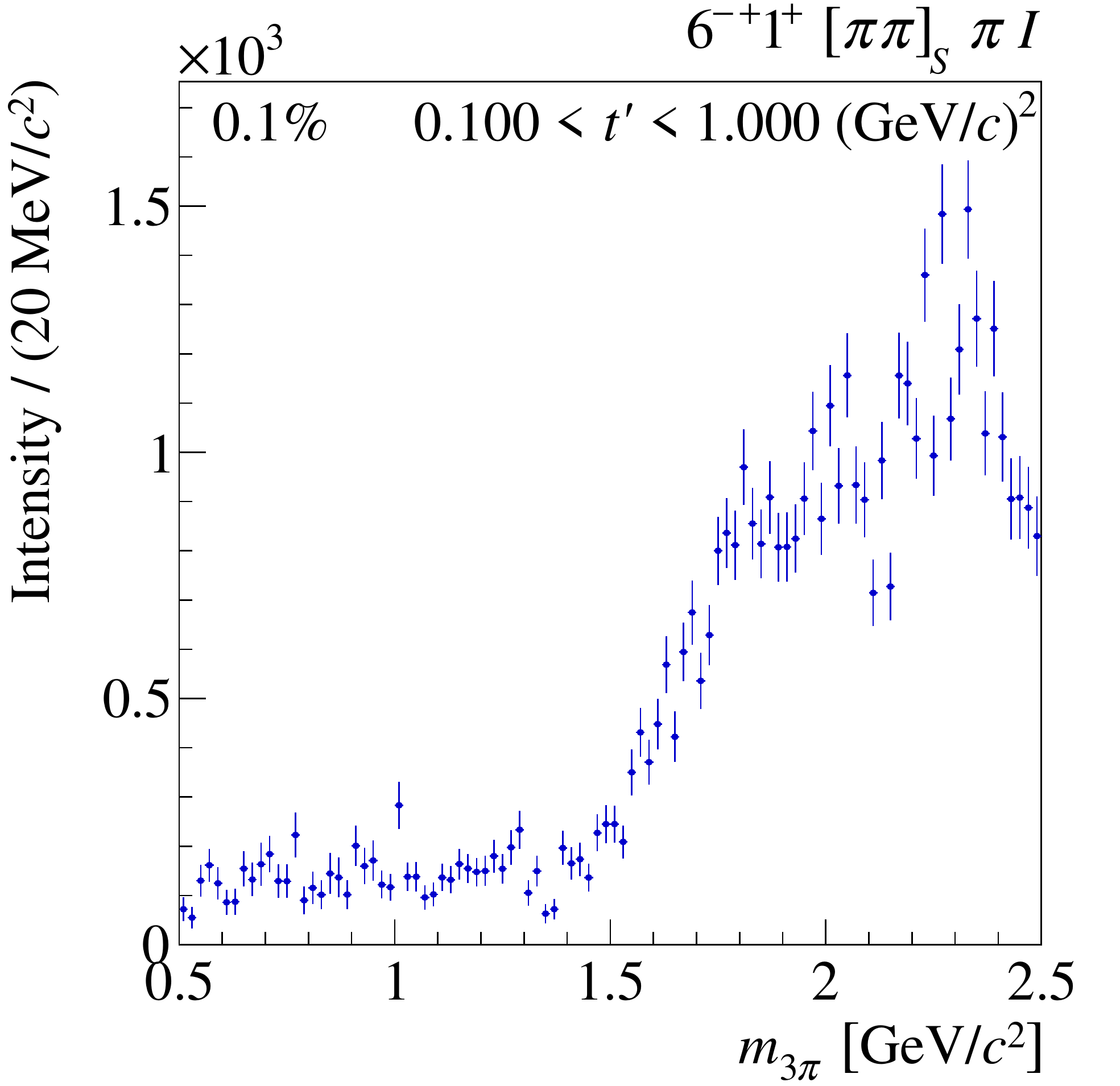}%
  }%
  \\
  \subfloat[][]{%
    \label{fig:int_6mp0p_rho_H}%
    \includegraphics[width=\threePlotWidth]{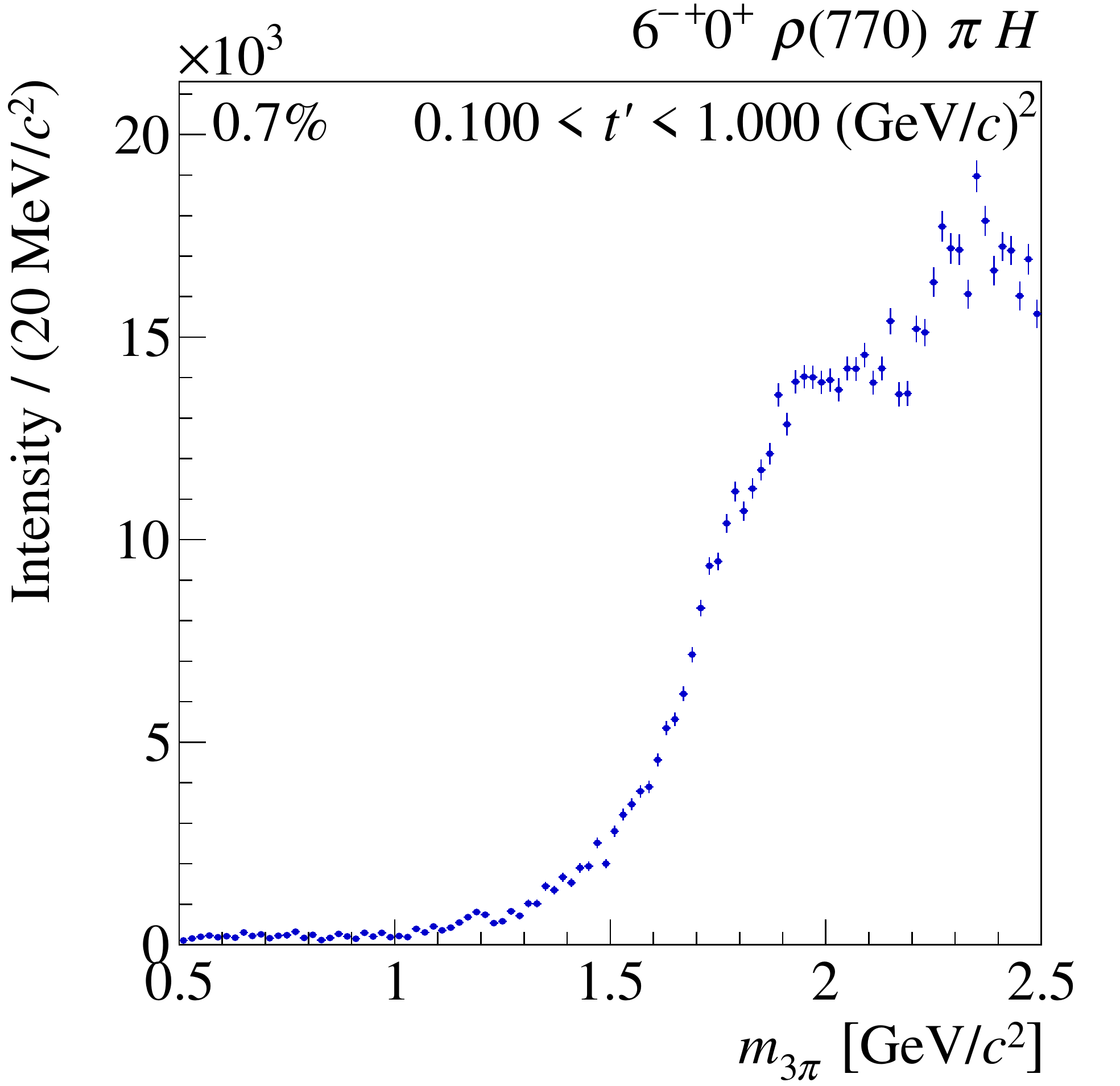}%
  }%
  \subfloat[][]{%
    \label{fig:int_6mp1p_rho_H}%
    \includegraphics[width=\threePlotWidth]{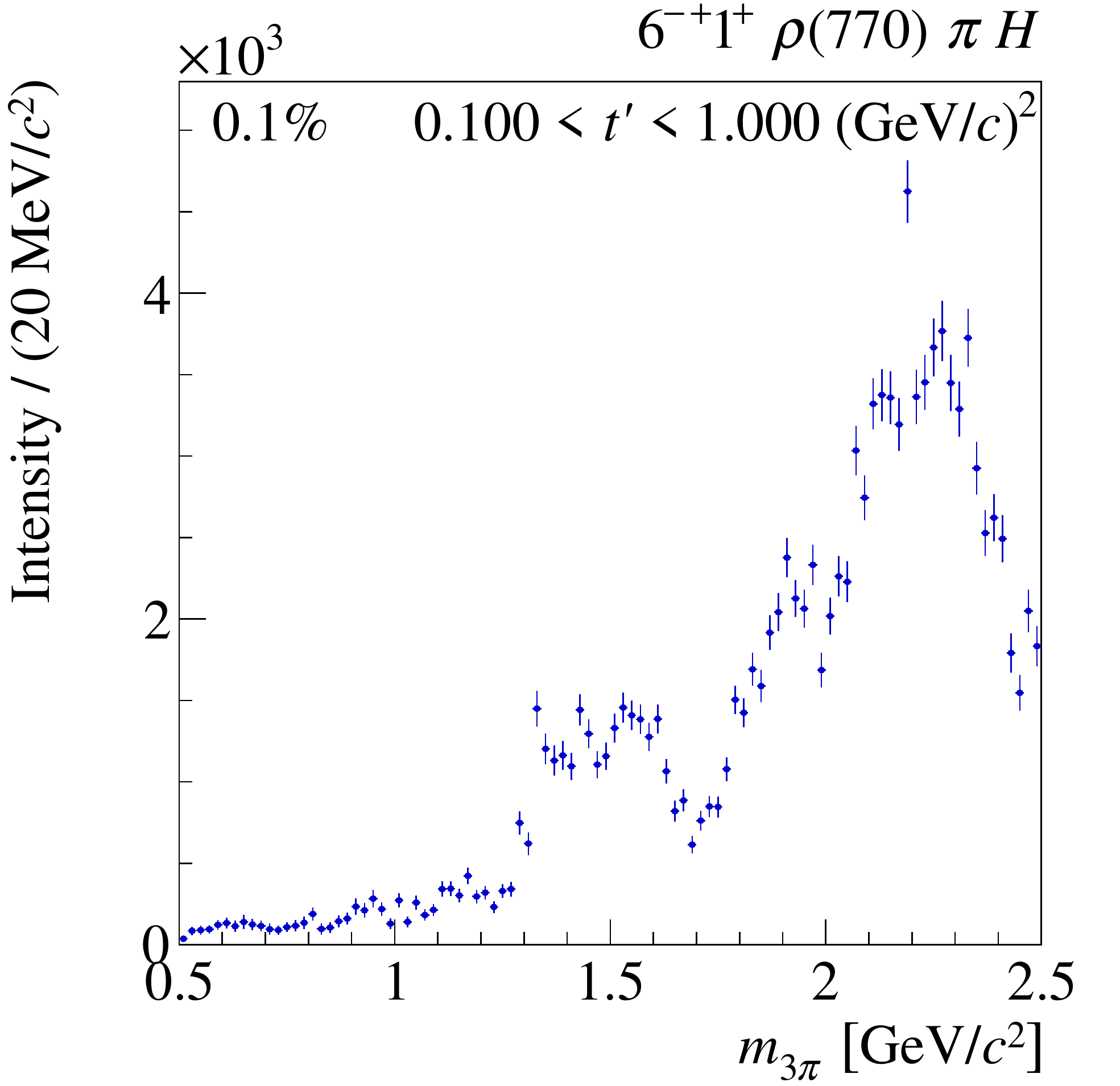}%
  }%
  \\
  \subfloat[][]{%
    \label{fig:int_6mp0p_f2_G}%
    \includegraphics[width=\threePlotWidth]{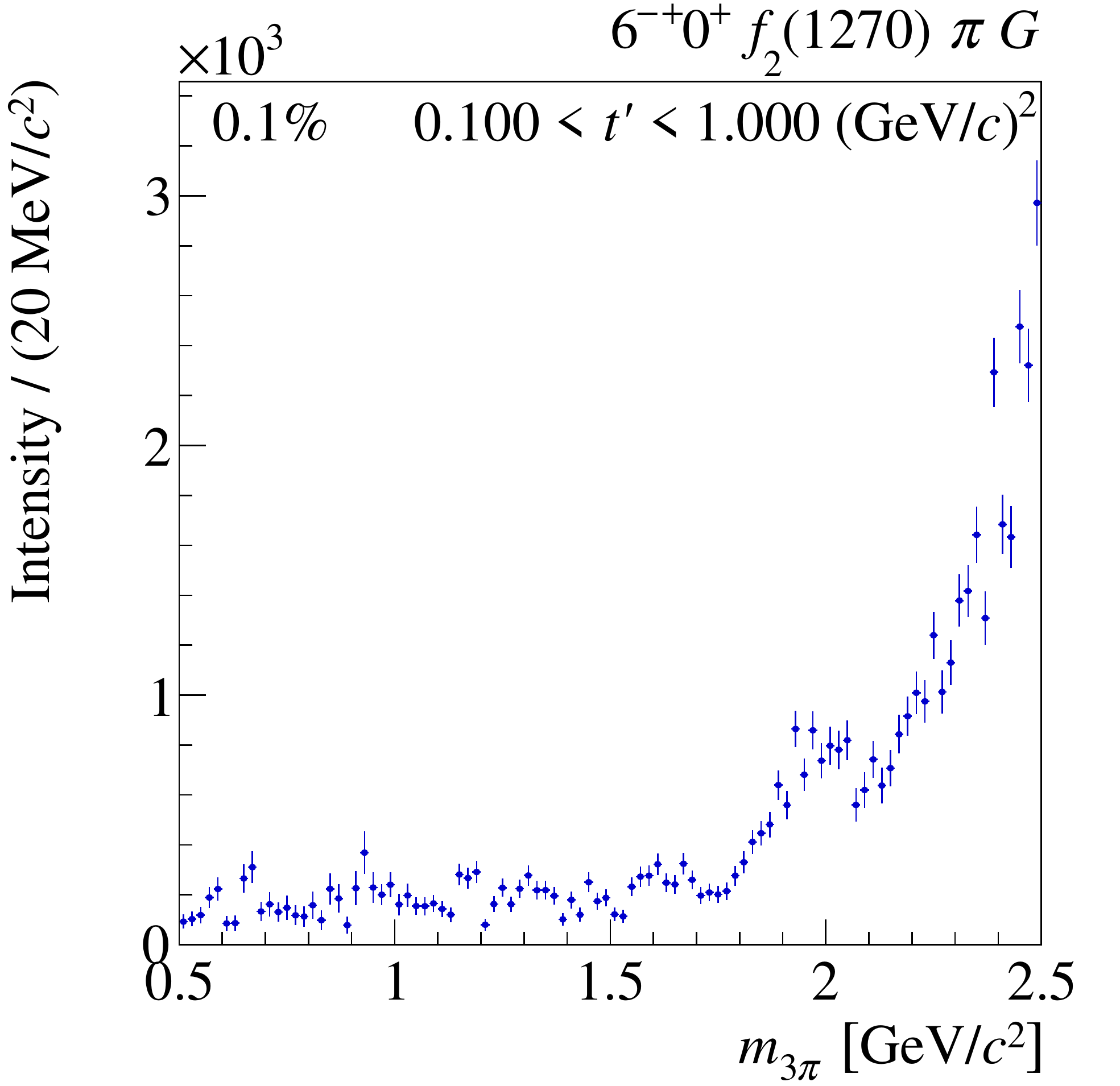}%
  }%
  \subfloat[][]{%
    \label{fig:int_6mp0p_rho3_F}%
    \includegraphics[width=\threePlotWidth]{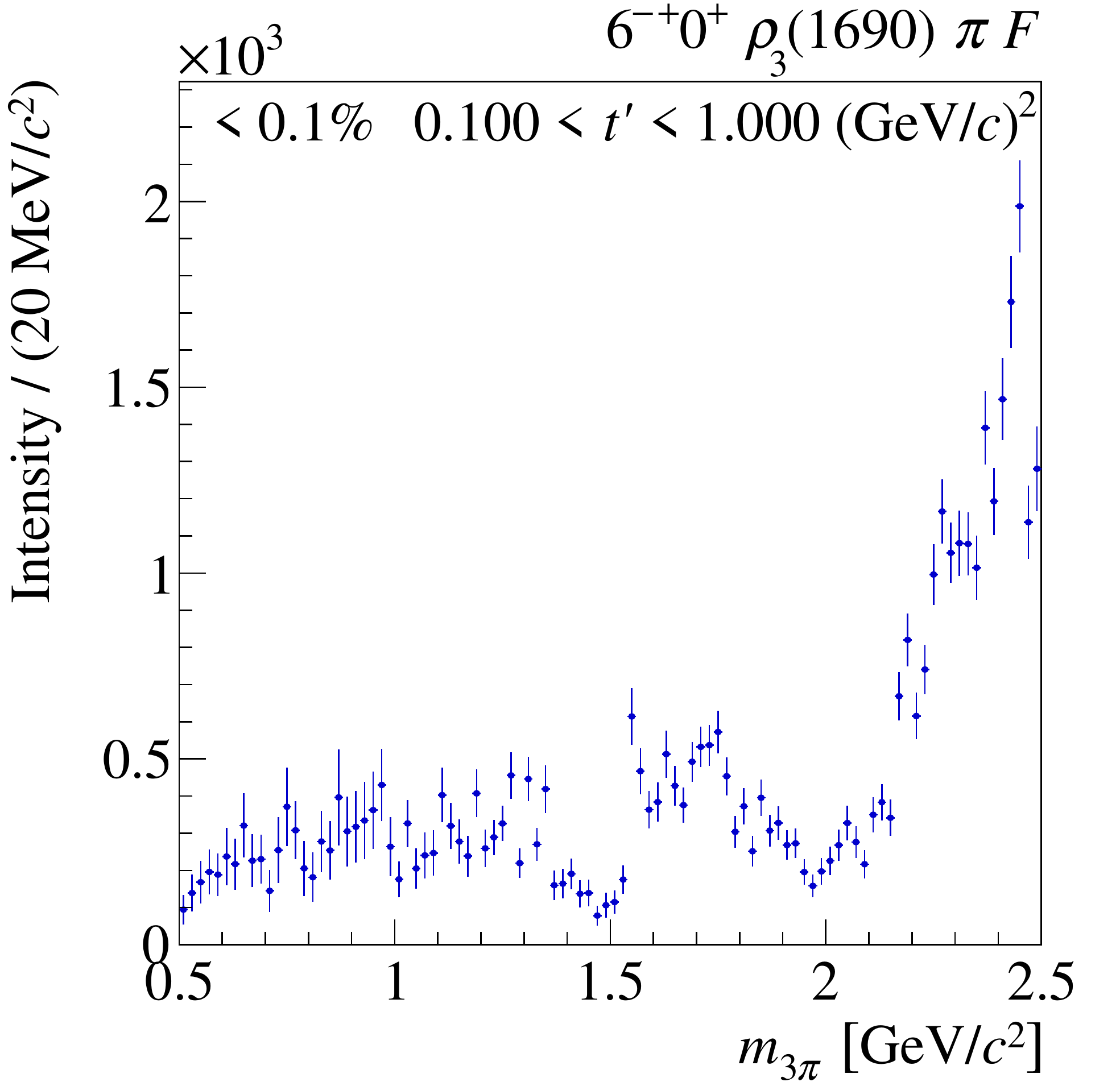}%
  }%
  \caption{The \tpr-summed intensities of partial waves with $\JPC =
    6^{-+}$ and positive reflectivity.}
  \label{fig:intensities_6mp}
\end{figure}

\clearpage
\subsection{Waves with Negative Reflectivity}

\begin{figure}[H]
  \centering
  \subfloat[][]{%
    \label{fig:int_1pp1m_rho_S}%
    \includegraphics[width=\threePlotWidth]{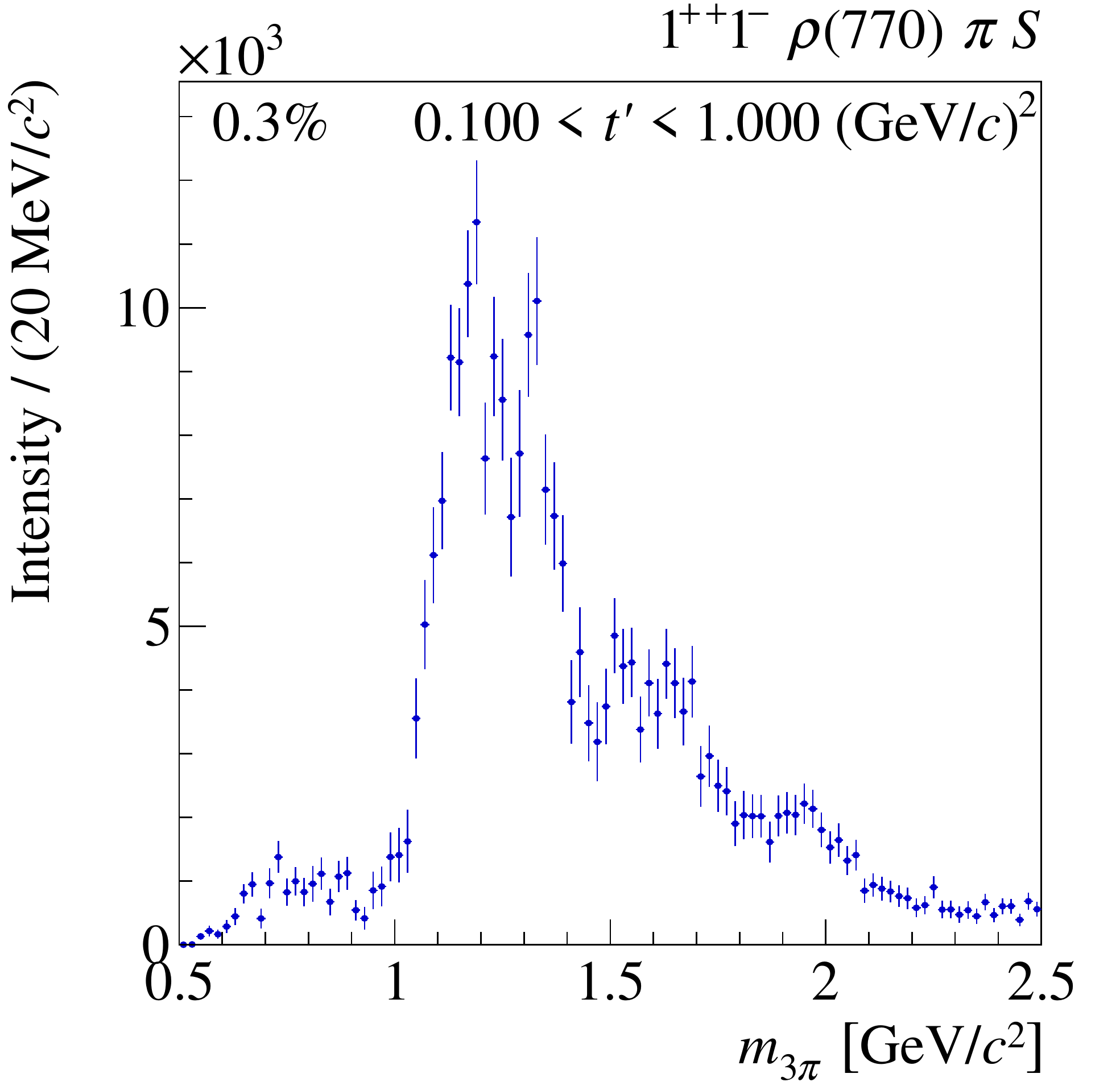}%
  }%
  \subfloat[][]{%
    \label{fig:int_0mp0m_rho_P}%
    \includegraphics[width=\threePlotWidth]{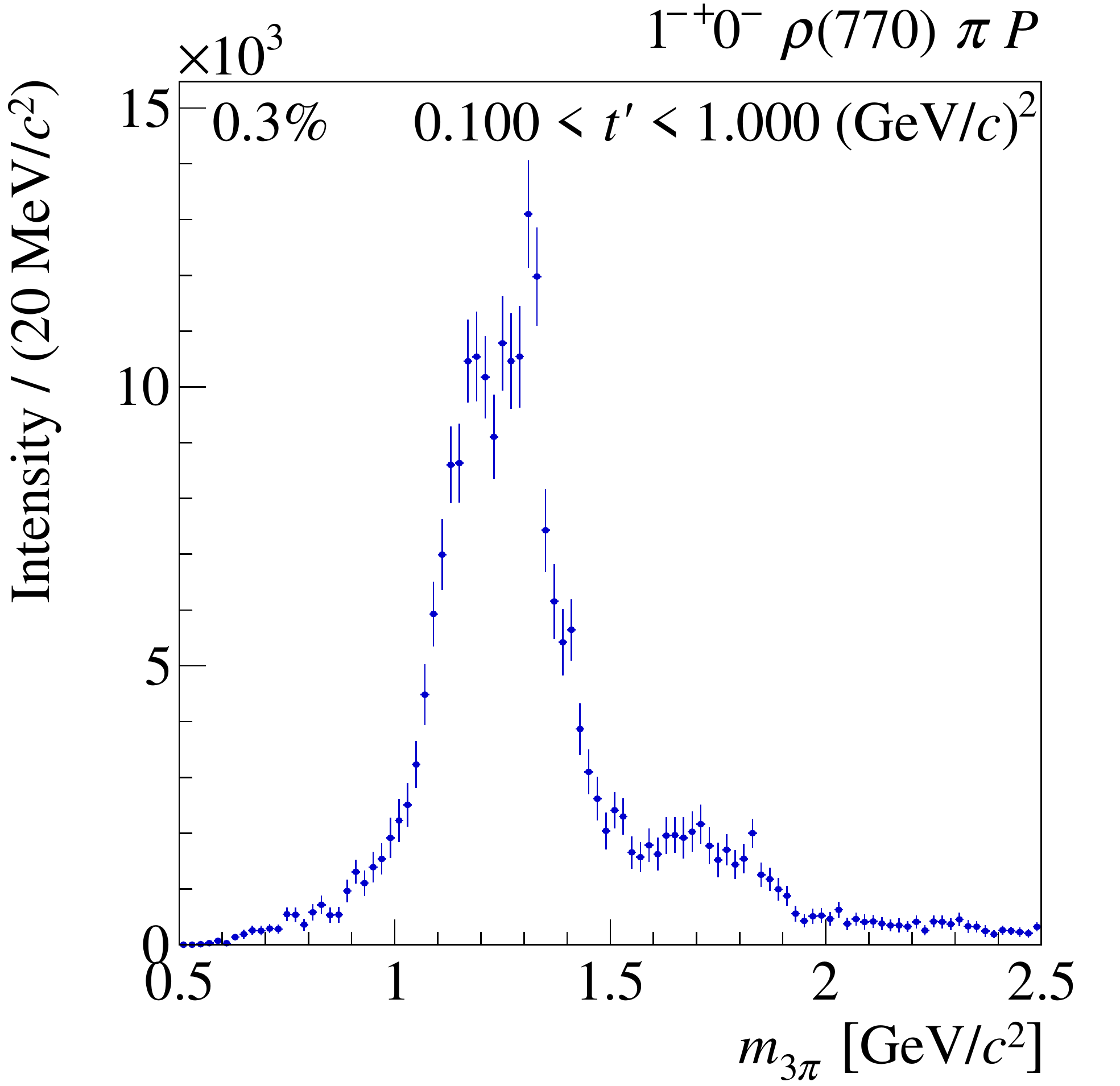}%
  }%
  \subfloat[][]{%
    \label{fig:int_0mp1m_rho_P}%
    \includegraphics[width=\threePlotWidth]{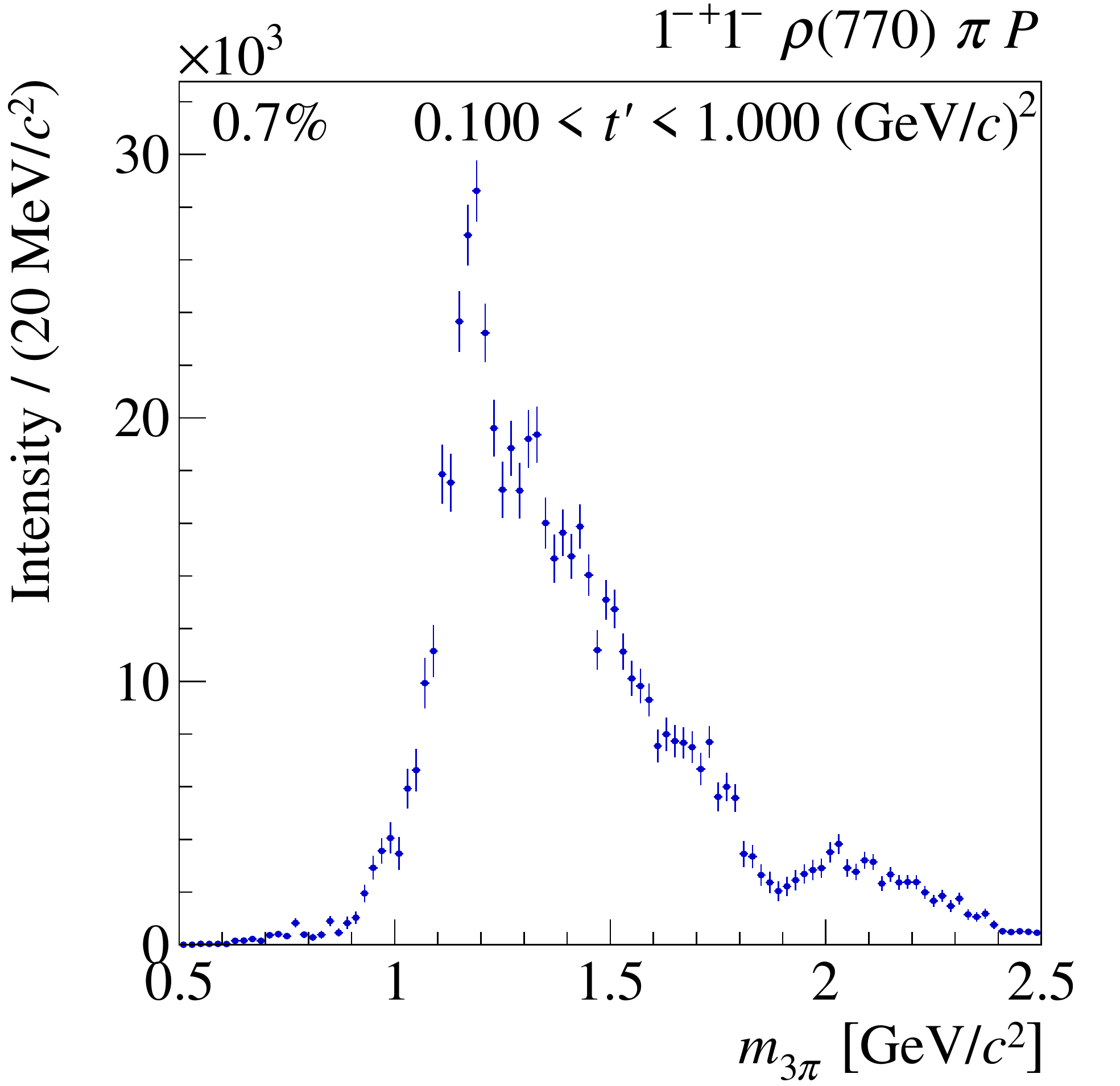}%
  }%
  \\
  \subfloat[][]{%
    \label{fig:int_2pp0m_rho_D}%
    \includegraphics[width=\threePlotWidth]{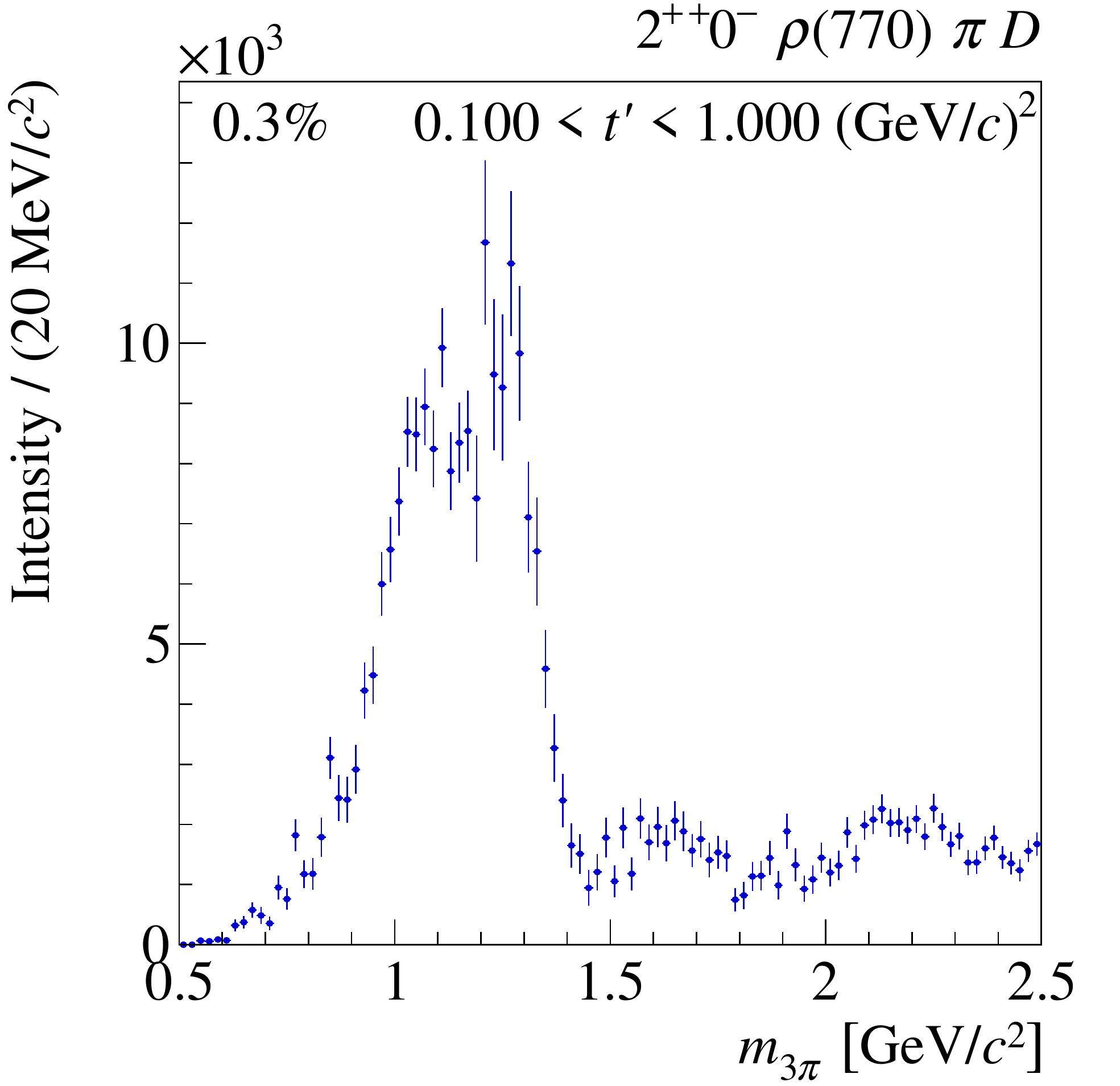}%
  }%
  \subfloat[][]{%
    \label{fig:int_2pp0m_f2_P}%
    \includegraphics[width=\threePlotWidth]{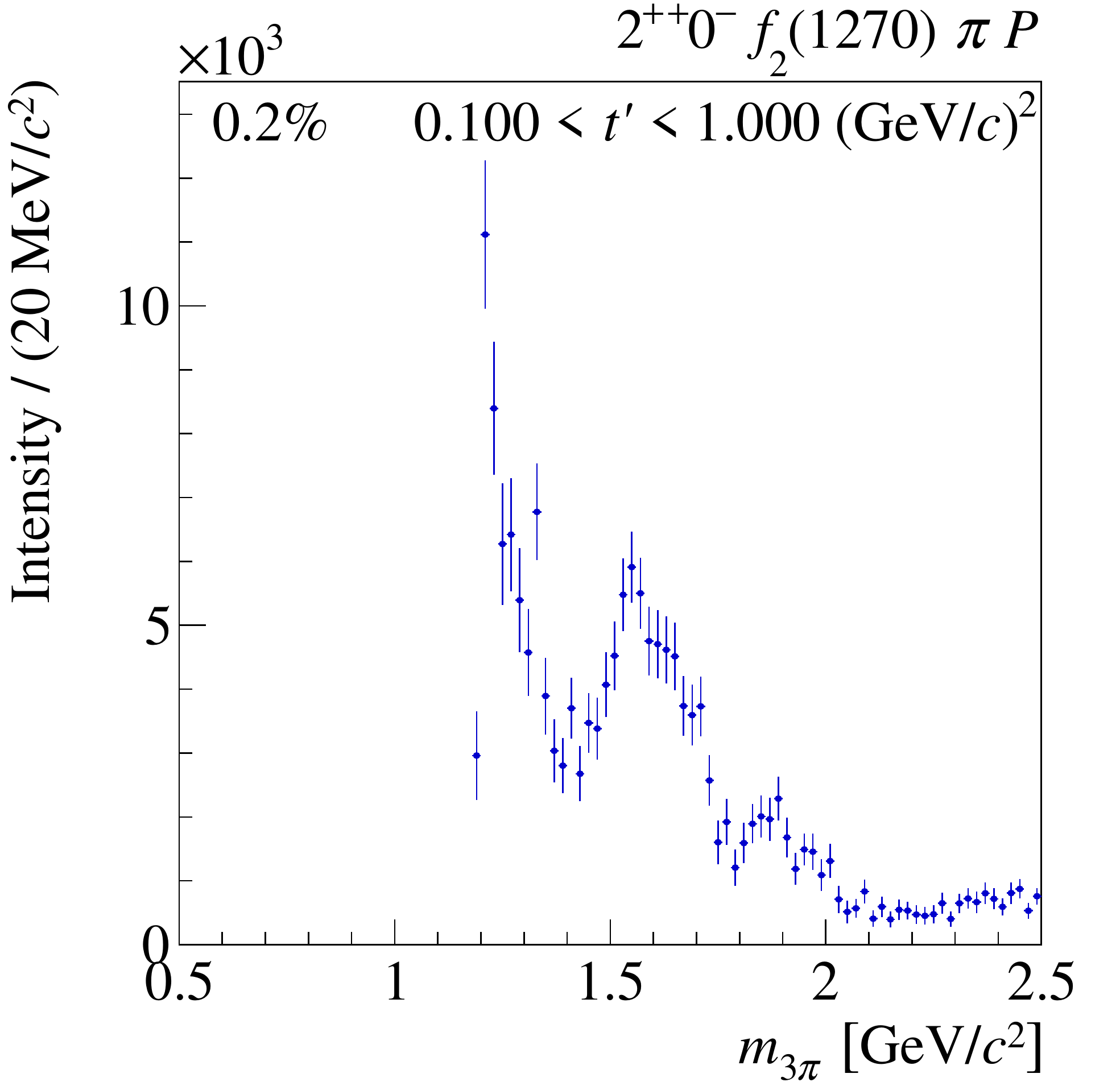}%
  }%
  \subfloat[][]{%
    \label{fig:int_2pp1m_f2_P}%
    \includegraphics[width=\threePlotWidth]{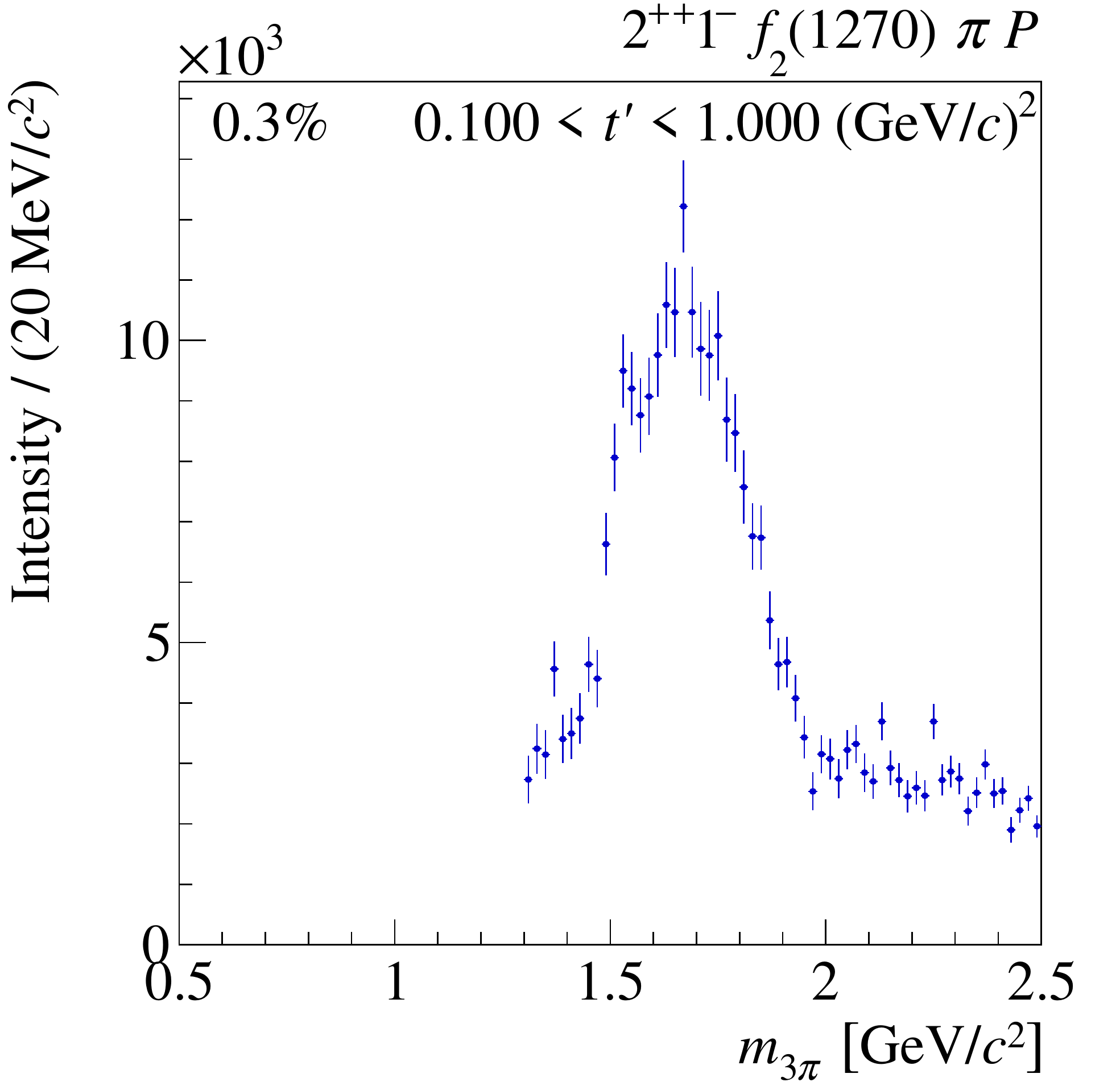}%
  }%
  \\
  \subfloat[][]{%
    \label{fig:int_2mp1m_f2_S}%
    \includegraphics[width=\threePlotWidth]{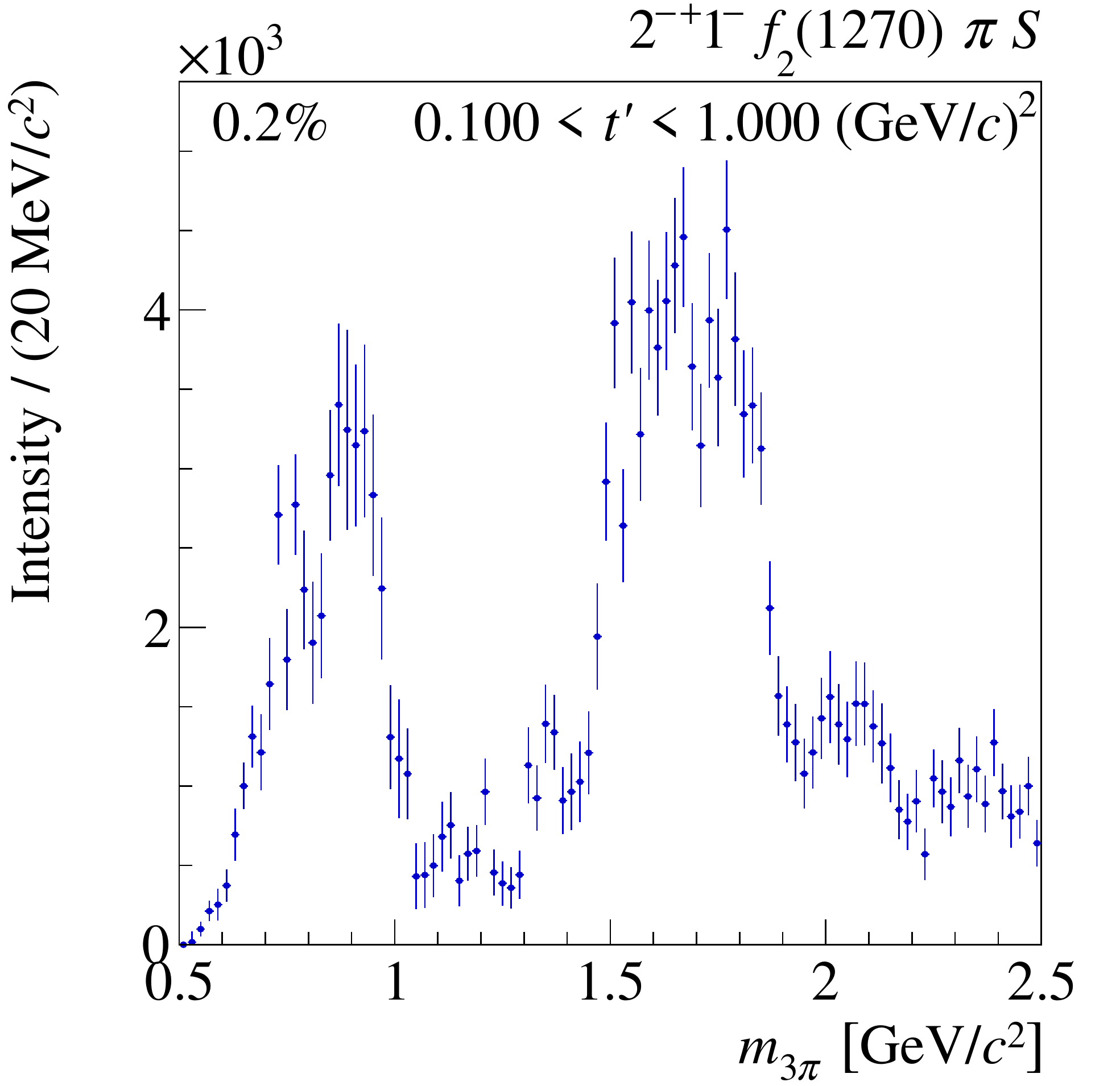}%
  }%
  \caption{The \tpr-summed intensities of partial waves with negative
    reflectivity.}
  \label{fig:intensities_neg_refl}
\end{figure}

%
%
\bibliographystyle{utphys_bgrube}
\providecommand{\href}[2]{#2}\begingroup\raggedright\endgroup

\end{document}